%% file: main.tex
\documentclass[phd]{luthesis}

\usepackage{hyperref}
\usepackage{booktabs}

\usepackage{natbib}
\usepackage{amssymb}
\usepackage{cite}
\usepackage{blkarray}
\usepackage{algorithmic}
\usepackage{graphicx}
\usepackage{textcomp}
\usepackage{booktabs}
\usepackage{tikz,pgfplots}
\usepackage[figuresright]{rotating}
\pgfplotsset{compat=1.18}
\usepackage{standalone}
\usepackage[listings,skins,breakable]{tcolorbox}
\usepackage{csquotes}
\usepackage{Figures/pgf-pie}
\usepackage{url}
\usepackage[tikz]{bclogo}
\usetikzlibrary{arrows.meta}
\usepackage{fancyhdr}
\usepackage{pifont}
\usepackage{makecell}
\usepackage{hyperref}
\usepackage{dingbat}
\usepackage[ruled, linesnumbered, vlined, commentsnumbered]{algorithm2e}
\usepackage{newtxmath}
\usepackage{multirow}
\usepackage{xcolor,colortbl}
\usepackage{graphicx}
\usepackage{longtable}
\usepackage{subcaption} 
\usepackage{tcolorbox}
\usepackage{booktabs}
\usepackage{adjustbox}
\usepackage{graphics}
\usepackage{amssymb,amsmath,mleftright,mathtools}

\definecolor{beaublue}{rgb}{0.74, 0.83, 0.9}

\definecolor{steel}{rgb}{0, 0.2, 0.9} 

\DeclareRobustCommand\dashed{\tikz[baseline=-0.6ex]\draw[thick,dashed] (0,0)--(0.54,0);}

\def\subfigsize{0.32}

\newtcolorbox{quotebox}{boxrule=0.4pt,colframe=black,fonttitle=\bfseries,top=2pt,bottom=2pt}

\newtcolorbox{answerbox}{colback=beaublue,boxrule=0.8pt,colframe=black,fonttitle=\bfseries,top=3pt,bottom=3pt}

\newcolumntype{P}[1]{>{\centering\arraybackslash}p{#1}}

\DeclareMathAlphabet\mathbfcal{OMS}{cmsy}{b}{n}

\SetKwInput{kwDeclare}{Declare}

\newcommand{\fixhd}[1]{%
  \smash[#1]{\vphantom{\Big|}}%
}

\newcommand{\Model}{\texttt{DaL}} 

\newcommand{\MetaModel}{\texttt{SeMPL}}

\definecolor{one}{HTML}{2b7bba}
\definecolor{two}{HTML}{d52221}

\newtheorem{property}{Property}

 \newcommand{\markone}[4]{\begin{adjustbox}{max width=.1\textwidth}\begin{picture}(20,5)
    {\linethickness{0.2mm}\color{one}\put(#1,3){\line(1,0){#2}}\color{one}\put(#3,3){\circle*{4}}}\end{picture}\end{adjustbox}}

    \newcommand{\marktwo}[4]{\begin{adjustbox}{max width=.1\textwidth}\begin{picture}(20,5)
    {\linethickness{0.2mm}\color{two}\put(#1,3){\line(1,0){#2}}\color{two}\put(#3,0.5){\large$\star$}}\end{picture}\end{adjustbox}}

\newcommand{\quart}[4]{\begin{adjustbox}{max width=.1\textwidth}\begin{picture}(100,5)
    {\color{black}\put(#1,5){\line(1,0){#2}}\color{blue!50}\put(#3,5){\circle*{7}}\color{black}\put(#3,5){\circle{7}}}\end{picture}\end{adjustbox}}

\newtcolorbox{insightbox}{
  sidebyside,sidebyside align=top,lower separated=true,lefthand width=0.5em,
  arc=0pt,
  boxsep=0pt,
  left=2pt,right=2pt,top=2pt,bottom=2pt,colframe=white,
  skin=bicolor,
  colback=black!50, 
  colupper=white,
  colbacklower=white,boxrule=0pt,colframe=white,
  sidebyside gap=5pt,
}

\title{Pushing the Boundary: Specialising Deep Configuration Performance Learning} 
\author{Jingzhi Gong}

\thesisDateDay{}
\thesisDateMonth{5}  
\thesisDateYear{2024}

\begin{document}

\include{ICPE_cover_letter}
\maketitle 

\frontmatter  
\include{abstract} 

\include{acknowledgements}

\tableofcontents
\listoffigures   
\listoftables    

\mainmatter   

\include{Chapter-introduction/introduction}

\include{Chapter-method/chapter-method}

\include{Chapter-survey/chapter-survey}

\include{Chapter-encoding/chapter-encoding.tex}

\include{Chapter-DAL/chapter-DAL.tex}

\include{Chapter-meta/chapter-meta.tex}
\include{Chapter-discussion/chapter-discussion}
\include{Chapter-conclusion/chapter-conclusion}


\addcontentsline{toc}{chapter}{References}
\bibliography{references}

\backmatter   
\appendix     

\include{Appendix/chap-meta-rq1-full}
\include{Appendix/chap-meta-rq2-full}
\include{Appendix/chap-meta-rq3-full}

\end{document}

%% file: abstract.tex
\chapter{Abstract}
Software systems often come with a multitude of configuration options that can be adjusted to adapt their performance (e.g., latency, execution time, and energy consumption) to various requirements. However, their combined influence on performance is often unknown, resulting in potential issues for software maintenance. Worse, the rapid growth in scale and complexity of modern software systems has made performance measurement increasingly resource-intensive, leaving limited datasets in most real-world scenarios. Consequently, it has become a major challenge to build accurate performance prediction models based on these limited measurements. To address this, deep learning approaches have gained popularity in recent years, due to their capabilities of capturing intricate representations and interactions even with only a few samples.

To facilitate this subject, this thesis starts by conducting a systematic literature review specialising the latest deep learning techniques employed for configuration performance modeling, covering 948 searched papers spanning six indexing services, based on which 85 primary papers were extracted and analyzed. The results disclose both positive and negative trends in the literature and reveal potential future directions to explore. Subsequently, three key knowledge gaps are incorporated and formalized as three objectives for this thesis to pursue.

Therein, the first knowledge gap observed is that, despite the presence of different encoding schemes, there is still little understanding of which is better and under what circumstances, which could be harmful to the community. To bridge this gap, this thesis performs an empirical study on three of the most popular encoding schemes for configuration performance learning, namely label, scaled label, and one-hot encoding. The results demonstrate that choosing the encoding scheme is non-trivial, and thereafter, a list of actionable suggestions is provided to enable more reliable decisions.

Meanwhile, the survey also reveals a crucial yet unaddressed knowledge gap, namely, the sparsity inherited from the configuration landscape. To handle this matter, this thesis presents a model-agnostic and sparsity-robust framework based on ``divide-and-learn'', dubbed \Model. To mitigate the sample sparsity, the samples from the configuration landscape are divided into distant divisions, for each of which a deep learning model, e.g., Hierarchical Interaction Neural Network, is built to deal with the feature sparsity. Experiment results from 12 real-world systems and five sets of training data reveal that \Model~performs better than the state-of-the-art approaches on 44 out of 60 cases with up to $1.61$ times improvement in accuracy.

Nonetheless, similar to the majority of the studies reviewed, \Model~is limited to predicting under static environments (e.g., hardware, version, and workload), which contradicts the dynamic nature of software. To address this concern, a sequential meta-learning framework is proposed in this thesis, named \MetaModel, which significantly enhances the prediction accuracy of state-of-the-art models in multi-environment scenarios. What makes it unique is that unlike common meta-learning frameworks (e.g., \texttt{MAML}) that train the meta environments in parallel, they are trained in a specialised order for deep neural networks. Through comparing with 15 state-of-the-art models under nine systems, it is demonstrated that \MetaModel~performs considerably better on $89\%$ of the systems with up to $99\%$ accuracy improvement.

Through the extensive studies conducted in this thesis, the critical knowledge gaps within the existing literature specialising deep performance learning have been identified and effectively addressed. As a result, the accuracy of performance learning has been significantly advanced to a new level of precision.
~\vspace{0.5cm}

\noindent \textbf{Keywords:} \emph{Highly Configurable Software, Configuration Performance Prediction, Configuration Performance Modeling, Configuration Performance Learning, Deep Learning, Software Engineering.}

%% file: acknowledgements.tex
\chapter{Acknowledgements}

First of all, I would like to express my deepest gratitude to my supervisor, Dr. Tao Chen, for your guidance and support throughout my doctoral study. Without your expertise and insightful feedback, the accomplishments I have achieved so far would not have been possible. 

I am also sincerely grateful to my parents, Peng Tang and Chuming Gong, who have given countless support to my doctoral study. Your belief in my abilities has given me the confidence to persevere and overcome obstacles along the way, and your love and encouragement have shaped me into the person I am today.

Furthermore, I would like to thank my life partner, Li Yan. Our paths crossed during my first year of PhD study in 2020, and in 2023, we embarked on the journey of marriage. I am grateful for your unwavering care during busy working periods and for your consistent support and encouragement during moments of tiredness and frustration.

Lastly, I extend my sincere appreciation to all my family and friends who have contributed to the completion of this thesis.






%% file: Chapter-introduction/introduction.tex
\chapter{Introduction}
\label{chap:introduction}


To meet the different performance requirements of users, modern configurable and adaptable software systems often permit possible configuration options to be adjusted at runtime. For example, around one-third of the configuration options of the database system \textsc{MySQL}\footnote{\url{https://dev.mysql.com/doc/refman/8.0/en/server-system-variable-reference.html}} can be changed at runtime, such as \texttt{max\_connections}; the remaining ones, e.g., \texttt{autocommit}, need to be fixed prior to the deployment.  

Given an environmental condition, these configuration options may have great impacts on the software performance~\citep{DBLP:journals/tosem/ChenLBY18,DBLP:conf/icse/Li023,DBLP:conf/esem/HanY16,DBLP:journals/corr/abs-2404-04744,DBLP:journals/tosem/ChenL23,DBLP:journals/tosem/ChenL23a}. Indeed, inappropriate configurations often cause serious performance bugs, which is the key reason why the users became frustrated enough to (threaten to) switch to another product~\citep{DBLP:conf/msr/ZamanAH12}. At the same time, simply using default configurations does little help. For instance,~\citet{DBLP:conf/cidr/HerodotouLLBDCB11} show that the default settings on \textsc{Hadoop} can actually result in the worst possible performance. In this regard, it is essential to understand the corresponding performance of a configuration, and a fundamental research problem is: 

\begin{displayquote}
\textit{Given a set of configurations, what is the performance of the software?}. 
\end{displayquote}

Indeed, one solution is to directly profile the software system for all possible configurations when needed. This, however, is impractical because (i) the number of possible configurations may be too high. For example, \textsc{MySQL} has more than millions of possible configurations. (ii) Even when such a number is small, the profiling of a single set of configurations can be too expensive.~\citet{DBLP:conf/sigsoft/WangHJK13} reports that it could take weeks of running time to benchmark and profile even a simple system. All these issues have called for a computational model, which is cheap to evaluate yet accurate, that captures the correlation between configuration options (and environmental factors, e.g., workload, hardware, \emph{etc}.) to a performance attribute.

Classic performance models have been relying on analytical methods~\citep{DBLP:conf/icse/Kumar0BB20,DBLP:journals/tse/GrassiDT92,DBLP:conf/wosp/DidonaQRT15,DBLP:conf/gecco/0001LY18,DBLP:journals/infsof/ChenLY19}, but soon they become ineffective due primarily to the soaring complexity of modern software systems. In particular, there are two key reasons which prevent the success of analytical methods: (i) analytical models often work on a limited set of configurations options~\citep{DBLP:conf/wosp/DidonaQRT15,DBLP:journals/tse/ChenB17}, but the number of configurations options and the complexity continues to increase. For example, \textsc{Hadoop} has only 17 configurations options in 2006, but it was increased by 9 times more to 173 at 2013~\citep{DBLP:conf/sigsoft/XuJFZPT15}; similarly, \textsc{MySQL} has 461 configuration options at 2014, in which around 50\% of them are of complex types. (ii) Their effectiveness is highly dependent on assumptions about the internal structure and the environment of the software being modeled. However, many modern scenarios, such as cloud-based systems, virtualized and multi-tenant software, intentionally hide such information to promote ease of use, which further reduces the reliability of the analytical methods~\citep{DBLP:conf/icse/Chen19b}.

As an alternative, machine learning-based configuration performance models have been explored over the past decade, e.g., liner regression~\citep{DBLP:conf/sigsoft/SiegmundGAK15}, decision tree~\citep{DBLP:journals/ese/GuoYSASVCWY18}, and random forest~\citep{DBLP:conf/msr/GongC22}, which work on arbitrary types of configuration options and do not rely on heavy human intervention~\citep{DBLP:journals/jss/PereiraAMJBV21, DBLP:journals/tse/WangHGGZFSLZN23}. Unlike analytical methods, machine learning is data-driven since it seeks to learn the patterns from the configuration data, hence generalizing the correlation between configuration and performance. This problem, namely configuration performance learning, has been gaining momentum in recent years~\citep{DBLP:conf/icse/HaZ19, DBLP:conf/sigsoft/Gong023, DBLP:conf/esem/ShuS0X20, DBLP:journals/tosem/ChengGZ23}. 

Among others, a particular type of data-driven configuration performance learning relies on deep learning, i.e., those that make use of deep neural networks~\citep{DBLP:journals/csur/YangXLG22, DBLP:journals/tosem/WatsonCNMP22}. Compared with statistical machine learning, deep learning often exhibits several advantages:

\begin{itemize}
    \item \textbf{Representation learning:} They are capable of extracting the underlying representations from the configuration options even without careful feature engineering, which allows them to learn directly from raw data, enabling end-to-end learning.
    \item \textbf{State-of-the-art accuracy:} They consist of multiple layers of interconnected neurons, enabling them to capture the nonlinear and complex relationships between the configuration options and performance.
    \item \textbf{Generalizability:} Deep learning models can leverage pre-trained knowledge from related tasks and quickly adapt to the target task through fine-tuning, hence they often have better generalizability.
\end{itemize}

Indeed, recent studies have demonstrated the benefits of deep learning for modeling configuration performance. For example, ~\citet{DBLP:conf/icse/HaZ19} propose \texttt{DeepPerf}, a deep neural network model combined with $L_1$ regularization to address the sparse performance functions, and~\citet{DBLP:journals/tosem/ChengGZ23} invent a hierarchical interaction neural network model called \texttt{HINNPerf} that achieves state-of-the-art accuracy.



\section{Configuration Performance Modeling with Deep Learning}
\label{chap-intro:deep_learning}

A typical example of datasets employed in deep learning studies for performance prediction can be viewed in Table~\ref{tb:data_example}, which exhibits a set of configurations and their corresponding performance of the video codec software \textsc{VP8}. Particularly, $x_1$ to $x_{n-1}$ are binary options (e.g., the option to enable or disable alternative reference frames), $x_n$ represents a numerical option (the number of parallel processing threads), and $\mathcal{P}$ stands for the runtime.

Without loss of generality, deep configuration performance learning seeks to build a function $f$ such that:
\begin{equation}
    y = f(\mathbf{c})
\end{equation}
whereby $y$ is the performance attribute that is of concern; $\mathbf{c}$ is a configuration consists of the values for $n$ configuration options, i.e., $\mathbf{c}=\{x_1,x_2,...,x_n\}$. Taking the simplest fully connected deep neural network as an example, $f$ is represented as multiple layers of interconnected neurons, where neurons are activated as:
\begin{equation}
    a_j^{l+1}=\sigma(\sum_i a_i^l w_{ij}^{l,l+1} + b_j^{l+1})
\end{equation}
where $\sum$ runs over all the lower-layer neurons that
are connected to neuron $j$. $i$ is the activation of a
neuron $i$ in the previous layer, and where $a_i^l w_{ij}^{l,l+1}$ is the contribution of neuron $i$ at layer $l$ to the activation of the neuron $j$ at layer $l + 1$. The function $\sigma$ is a nonlinear monotonously increasing activation function, e.g., a sigmoid function; $w_{ij}^{l,l+1}$ is the weight and $b_j^{l+1}$ is the bias term.

To build function $f$, the training in deep learning aims to find the set of weights for different neurons from all the layers such that a loss function can be minimized. For example, the mean squared error below is one possible loss function since configuration performance learning is essentially a regression:
\begin{equation}
    \mathbfcal{L}(\theta) =  {1 \over m} \sum^m_{i=1}{(f(\mathbf{c}_i) - y_i')^2}
\end{equation}
$f(\mathbf{c}_i)$ and $y_i'$ denote the predicted and actual performance values for the $i$th sample across $m$ data points in the training dataset.

\begin{table}[t!]
\centering
\footnotesize
\caption{An example of configurations and performance for \textbf{VP8}.}
\input{Tables/chap-encoding/data_example}
\label{tb:data_example}
\end{table}

\section{Aim, Objectives and Research Questions}
\label{chap-intro:aim}
Given the formal definition of the research problem, this section outlines the specific aim, objectives, and research questions that drive this thesis and guide the research conducted within it.

\subsection{Aim}

As mentioned above, the accuracy of performance models is of utmost importance as it directly impacts their reliability and effectiveness. Yet, since deep learning models heavily rely on training data for accurate predictions, and collecting performance data can often be costly, it becomes imperative to maximize the utilization of limited and complex performance data to achieve improved prediction outcomes. Recent studies like~\citep{DBLP:conf/icse/HaZ19, DBLP:conf/esem/ShuS0X20, DBLP:journals/tosem/ChengGZ23} have highlighted that the accuracy of deep performance models is limited by a range of unsolved challenges, such as capturing the intricate relationships between software configurations and performance characteristics, addressing the curse of dimensionality in high-dimensional configuration spaces, and dealing with the sparsity of performance functions. Therefore, the aim of this thesis is:

\begin{answerbox}
    \emph{\textbf{Aim:} To disclose the gaps and challenges in the literature concerning configuration performance modeling using deep learning techniques, contribute new knowledge and artifacts that can overcome the current limitations, and ultimately, push the boundaries of performance modeling using deep learning, enabling more accurate, robust, and reliable predictions of configurable software systems.}
\end{answerbox}



\subsection{Objevtives}
\label{chap-intro:objectives}
To accomplish the research aim, this thesis raises four key research objectives, each focusing on a specific aspect of the problem at hand, which are carefully examined and thoroughly discussed throughout the course of this study.

Firstly, although several reviews have been done in the field of performance modeling with machine learning~\citep{DBLP:journals/tse/HortKSH22} or deep learning for general software engineering~\citep{DBLP:journals/csur/YangXLG22, DBLP:journals/tosem/WatsonCNMP22, DBLP:journals/tse/WangHGGZFSLZN23}, to the best of my knowledge, none of them focuses specifically on deep learning techniques for performance modeling, which is one of the most popular and promising approaches in the recent years. Therefore, a systematic literature review (SLR) is necessary to summarize the application of techniques related to deep learning, identify any limitations and gaps of the current approaches, and analyze the reasons behind them. 

Secondly, given that the encoding scheme is a critical design decision for machine learning and deep learning models~\citep{DBLP:journals/corr/abs-1801-02175,DBLP:conf/sigsoft/SiegmundGAK15,DBLP:conf/sigsoft/NairMSA17}, it is surprised that the reviewed studies in the SLR (Chapter~\ref{chap:review}) hardly explain their motivation for choosing the encoding schemes. The lack of this knowledge will cause misunderstandings of model behaviors. Therefore, it is important to research the encoding schemes for performance learning.

Then, During the SLR conducted in this thesis, it was observed that a list of performance learning studies share the viewpoint that the performance models of numerous software systems exhibit sparsity~\citep{DBLP:conf/icse/HaZ19, DBLP:journals/jmlr/GlorotBB11,DBLP:journals/fgcs/LiLTWHQD19, DBLP:conf/mascots/GrohmannEEKKM19}, i.e., most of the configurable options have trivial influences on performance, and this sparsity could cause overfitting of the performance models and lead to poor accuracy of predictions. Yet, a major portion of studies specialising in deep learning fails to realize the sparsity problem. As such, developing DL models that can effectively overcome sparsity in performance learning holds promise for further improving prediction accuracy.

Furthermore, the systematic literature review (SLR) conducted in this study reveals that a majority of deep performance learning studies predominantly assume a static running environment. However, in real-world scenarios, software systems are often deployed in diverse environments, each exhibiting distinct performance behaviors. To mitigate this, the SLR highlights several studies on transfer learning~\citep{DBLP:conf/cloudcom/IorioHTA19, DBLP:journals/corr/abs-1803-03900, DBLP:conf/kbse/JamshidiSVKPA17}. While transfer learning models can leverage performance data from other environments, they often suffer from generalizability issues~\citep{torrey2010transfer}. Thugs, the last objective in this thesis is to utilize the performance data under different environments for better prediction results.

In summary, the primary objectives of this thesis are:

    \begin{itemize}
    \item[---] \textbf{Objective 1:} To systematically review and categorize the current literature in the discipline of configuration performance modeling using deep learning and identify possible knowledge gaps to fulfill.
    \item[---] \textbf{Objective 2:} To examine the impacts of encoding schemes for deep configuration performance learning and the behavior of the encoding schemes under different conditions and identify suggestions for future researchers.
    \item[---] \textbf{Objective 3:} To design and implement a deep learning model for addressing the sparsity problem in configuration performance prediction, such that the model is more accurate than the state-of-the-art models.
    \item[---] \textbf{Objective 4:} To design and implement a deep learning performance model that can leverage data from various environments, such that its accuracy is improved to be better than the state-of-the-art models in the target environment.
    \end{itemize}

\subsection{Research Questions}
\label{chap-intro:rq}
In order to reach the research objectives, several research questions (RQs) must be answered. Noteworthy, the RQs are not all proposed from the very beginning of this thesis, but through a sequential process model as specified in Chapter~\ref{chap:methodology}. For example, the first RQ is raised initially, the second and third are developed by conducting a systematic literature review (details in Chapter~\ref{chap:review}), and the last one is identified during the iterated research procedures.

Specifically, to address \textbf{objective 1}, the first RQ is formulated as:
\begin{answerbox}
\emph{\textbf{RQ1:} What techniques have been used in the discipline of deep learning for performance modeling, and what are the potential knowledge gaps to bridge for more accurate performance modeling?}
\end{answerbox}

To fully understand this RQ and cover techniques used throughout the four stages in the deep learning pipeline, the following sub-RQs are posted:

\begin{itemize}
\item \textbf{RQ1.1:} How to prepare performance data?
\item \textbf{RQ1.2:} How to train the performance model?
\item \textbf{RQ1.3:} How to evaluate the trained model?
\item \textbf{RQ1.4:} How to apply the performance model?
\end{itemize}

Following \textbf{objective 2}, the second RQ is set to be:

\begin{answerbox}
\emph{\textbf{RQ2:} Which encoding scheme is better for performance learning, and under what circumstances?}
\end{answerbox}

To research this question, an empirical study is conducted, and the below sub-RQs are asked in order to better understand the influence of encoding schemes:

\begin{itemize}
    \item \textbf{RQ2.1:} Is it practical to examine all encoding methods to find the best one under every system?
    \item \textbf{RQ2.2:} Which encoding scheme (paired with the model) helps to build a more accurate performance model?
    \item \textbf{RQ2.3:} Which encoding scheme (paired with the model) helps train a performance model faster?
    \item \textbf{RQ2.4:} What are the trade-offs between accuracy and training time when choosing the encoding and models?
\end{itemize}

Then, to achieve \textbf{objective 3}, it is essential to ask:

\begin{answerbox}
\emph{\textbf{RQ3:} Is it possible to improve prediction accuracy by addressing the sparsity problem in performance modeling with deep learning models?}
\end{answerbox}

In order to answer this RQ, this thesis proposes a novel DNN-based framework called \Model, which is systematically introduced in Chapter~\ref{chap:dal}. To evaluate \Model, the following sub-RQs are investigated:

\begin{itemize}
    \item \textbf{RQ3.1:} How accurate is \Model~compared with the state-of-the-art approaches, i.e., \texttt{HINNPerf}, \texttt{DeepPerf}, \texttt{Perf-AL}, \texttt{DECART}, and \texttt{SPLConqueror}, for configuration performance learning?

    \item \textbf{RQ3.2:} To what extent \Model~can improve different generic models (i.e., \texttt{HINNPerf}, regularized Deep Neural Network, \texttt{CART}, Linear Regression, and Support Vector Regression) when they are used locally therein for predicting configuration performance?

    \item \textbf{RQ3.3:} How do \Model~perform compared with the existing ensemble approaches such as \texttt{Bagging} and \texttt{Boosting}?

    \item \textbf{RQ3.4:} What is the benefit of the components in \Model? 
  
    \item \textbf{RQ3.5:} What is the sensitivity of \Model~to a fixed $d$ and how well does the adaptive mechanism perform in finding the optimal $d$ for each individual run?

    \item \textbf{RQ3.6:} What is the model building time for \Model?
\end{itemize}

Subsequently, to realize \textbf{objective 4}, the last research question is:

\begin{answerbox}
\emph{\textbf{RQ4:} Is it possible to improve the prediction accuracy of deep learning models under the target environments by utilizing data from other environments?}
\end{answerbox}

To that end, a sequential meta-learning approach with DNN as the base learner, referred to as \MetaModel, is developed in Chapter~\ref{chap:meta}, and a series of sub-RQs are identified for evaluating \MetaModel:

\begin{itemize}
    \item \textbf{RQ4.1:} How does \MetaModel~perform compared with existing single environment models?
  
    \item \textbf{RQ4.2:} How does \MetaModel~perform compared with the state-of-the-art models that handle multiple environments?
    
    \item \textbf{RQ4.3:} How effective is the sequence selection in \MetaModel?

    \item \textbf{RQ4.4:} What is the sensitivity of \MetaModel's accuracy to the sample size used in pre-training?
\end{itemize}

\section{Contributions}
Through answering the research questions, this thesis has accomplished a collection of significant contributions. Particularly, Table~\ref{chap1:contribution} summarizes the key contributions and the addressed research objectives and questions.

\input{Tables/contribution}

\input{Tables/publications}

Additionally, this thesis has resulted in a series of publications, including two accepted papers at top-level conferences in the field of software engineering, and three additional papers that have been submitted to one prominent SE conference and two leading journals in SE and CS. Detailed information about these publications can be found in Table~\ref{chap1:publication}.

\begin{figure}[ht]
\centering
\begin{adjustbox}{width=1.05\textwidth,center}
\includegraphics[width=\columnwidth]{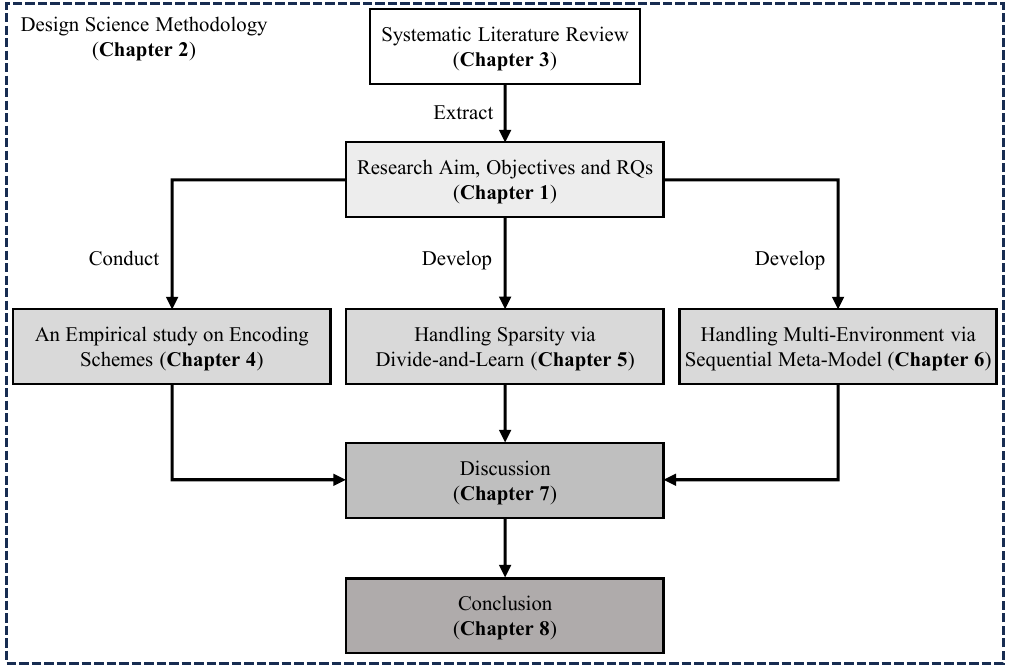}
\end{adjustbox}
\caption{Overview of the structure of this thesis.}
\label{fig:structure}
\end{figure}

\section{Thesis Outline}
Figure~\ref{fig:structure} captures the structure of this thesis, showcasing the logical flow and relations between the chapters. Specifically, all the studies conducted in this thesis follow a systematic and pragmatic research methodology as elaborated in Chapter~\ref{chap:methodology}. With the primary goal of establishing a concrete foundation and strong motivations, a systematic literature review is first performed in Chapter~\ref{chap:review}, from which the formal research aim, objectives, and research questions are extracted (Chapter~\ref{chap:introduction}). Following the objectives, dedicated artifacts are designed, implemented, and evaluated in Chapters~\ref{chap:encoding},~\ref{chap:dal} and~\ref{chap:meta}, respectively. Thereafter, the realization of the aim, the contributions, the recommendations, and the limitations of this thesis are discussed in Chapter~\ref{chap:discussion}. Finally, Chapter~\ref{chap:conclusion} revisits the key contents of the thesis. 

In a nutshell, the outline of the chapters is as follows:
    \begin{itemize}
    \item[---] \textbf{Chapter~\ref{chap:introduction}: Introduction} presents the backgrounds, aim, objectives, research questions, and contributions of this thesis.
    \item[---] \textbf{Chapter~\ref{chap:methodology}: Methodology} introduces the Design Science research methodology applied to conduct the studies in this thesis.
    \item[---] \textbf{Chapter~\ref{chap:review}: A Systematic Literature Review} surveys the current literature on deep learning for performance modeling, where statistics and taxonomy are presented. Justifications are made to reveal the positive and negative trends and, most importantly, the knowledge gaps to fill.
    \item[---] \textbf{Chapter~\ref{chap:encoding}: An Empirical Study on Encoding Schemes} includes the empirical study on the importance of choosing the encoding method and provides actionable suggestions for researchers to find the most appropriate encoding scheme.
    \item[---] \textbf{Chapter~\ref{chap:dal}: A `Divide-and-learn' Framework} covers the proposal of a model to solve the sparsity problem in performance models, along with the evaluations and justifications.
    \item[---] \textbf{Chapter~\ref{chap:meta}: A Sequential Meta Learning Framework.} In this chapter, a meta-learning framework is designed and implemented for the multi-environment problem, as well as the evaluations and discussions.
    \item[---] \textbf{Chapter~\ref{chap:discussion}: Discussion} analyzes the realization of the research aim and discusses the key contributions, recommendations, and limitations of this thesis.
    \item[---] \textbf{Chapter~\ref{chap:conclusion}: Conclusion.} The main contents of this report are summarised in this chapter.
    \end{itemize}

%% file: Tables/chap-encoding/data_example.tex
\begin{tabular}{lllllll||c}
\toprule
$x_{1}$ & $x_{2}$ & $x_{3}$ & $\cdots$ & $x_{n-2}$ & $x_{n-1}$ & $x_{n}$ & $y$ \\ \midrule
0 & 0 & 0 & $\cdots$ & 0 & 0 & 1 & 8190.6 seconds \\ 
0 & 1 & 0 & $\cdots$ & 0 & 0 & 2 & 6502.4 seconds \\ 
$\cdots$        & $\cdots$  & $\cdots$       & $\cdots$       & $\cdots$       & $\cdots$ & $\cdots$ & $\cdots$ \\ 
1 & 1 & 1 & $\cdots$ & 1 & 0 & 3 & 29102.2 seconds \\ 
1 & 1 & 1 & $\cdots$ & 1 & 1 & 4 & 25827.4 seconds \\ \bottomrule
\end{tabular}

%% file: Tables/contribution.tex
\begin{footnotesize} 
~\vspace{0.1cm}
\begin{longtable}{|p{0.65\columnwidth}|p{0.12\columnwidth}|p{0.12\columnwidth}|}
\caption{The contributions of this thesis, the addressed objectives, RQ, and the corresponding chapter.}
\label{chap1:contribution}
\\
\hline
\rowcolor[HTML]{C0C0C0} 
\textbf{Contribution}                        & \textbf{Objective \& RQ} & \textbf{Chapter}\\ \hline
A systematic literature review exclusively focused on deep learning for configuration performance is conducted, encompassing 948 papers from 2013 to 2023, sourced from six online repositories and 52 venues, where 85 prominent studies were extracted for data extraction and analysis. The examination covers the four stages within the deep configuration performance learning pipeline, including configuration data preparation, deep model training, accuracy evaluation, and model exploitation for configurable software systems.

&
  Objective 1 \& RQ1 & Chapter~\ref{chap:review} \\ \hline

Based on the review results, the following contributions are derived:

\begin{itemize}
    \item A taxonomy to categorize the techniques employed in each process of the deep configuration performance learning pipeline.
    
    \item A comprehensive summary of the positive and negative trends that are observed in the deep learning pipeline for modeling configuration performance.  
    
    \item A summary of the knowledge gaps in existing studies is presented, highlighting opportunities for further research in this specific area.
\end{itemize}

&
  Objective 1 \& RQ1 & Chapter~\ref{chap:review} \\ \hline
  
An empirical study that examines the performance of three widely used encoding schemes for software performance learning, namely label encoding, scaled label encoding, and one-hot encoding, covering five software systems and seven models.

&
  Objective 2 \& RQ2 & Chapter~\ref{chap:encoding} \\ \hline

The empirical study on encoding schemes yields significant findings:

\begin{itemize}
    \item The demonstration of the intricacy involved in selecting an encoding scheme for learning software performance and the subsequent costliness of conducting trial-and-error to identify optimal encoding schemes.
    \item A list of actionable suggestions for choosing the most suitable encoding scheme under various scenarios, enabling researchers and practitioners to make reliable and informative decisions.
\end{itemize}

&
  Objective 2 \& RQ2 & Chapterr~\ref{chap:encoding} \\ \hline

A systematic qualitative study is conducted to comprehend the SLR in Chapter~\ref{chap:review}, which analyzes the sparsity characteristics of configuration data, including both a literature review and an empirical study, hence better motivating the research on handling sparsity.

&
  Objective 3 \& RQ3 & Chapterr~\ref{chap:dal} \\ \hline
  
A `Divide-and-learn' framework to solve the sparsity problem, which divides the training dataset into distinct divisions where each is learned by a dedicated local deep learning model. 

Evaluations on 12 subject systems and 5 training sizes have demonstrated that it achieves no worse accuracy on 44 out of 60 cases, with 31 of them being significantly better. The improvements can be up to $1.61\times$ against the best counterpart.

&
  Objective 3 \& RQ3 & Chapter~\ref{chap:dal} \\ \hline

By formalizing the `divide-and-learn' framework, several notable contributions have been achieved: 

\begin{itemize}
    \item The formulation, on top of the regression problem of learning configuration performance, of a new classification problem without explicit labels.
    
    \item The extension of Classification and Regression Tree (\texttt{CART}) as a clustering algorithm to ``divide'' the samples into different divisions with similar characteristics, for each of which a local model is built.
    
    \item Newly given configurations would be assigned into a division inferred by a Random Forest classifier, which is trained using the pseudo-labeled data from the \texttt{CART}. The local model of the assigned division would be used for the final prediction thereafter.

    \item Experiments show that \Model~is sensitive to its only parameter $d$, which determines the number of divisions. This thesis proposes a novel adaptive mechanism that dynamically adapts the $d$ value to an appropriate level without additional training or profiling. 

    \item To determine the optimal $d$ value in the adaptation, a new indicator $\mu$HV is proposed, extending from the standard HV that is widely used for multi-objective evaluation, which can better reflect the goodness and balance between the ability to handle sample sparsity and the amount of data for learning in the divided configuration data.
\end{itemize}

&
  Objective 3 \& RQ3 & Chapter~\ref{chap:dal} \\ \hline
  
A meta-learning framework, dubbed \MetaModel, where data from multiple related software environments are learned in a specific order, such that the pre-trained model can be fine-tuned efficiently for the target environment. 

Based on nine systems with 3-10 meta environments and five training data sizes each, \MetaModel~(combined with DNN) is experimentally compared against 15 state-of-the-art models for single or multiple environments, where \MetaModel~significantly outperforms existing models in $89\%$ of the systems with up to $99\%$ accuracy improvement; it is also data-efficient with at most $3.86\times$ speedup.

&
  Objective 4 \& RQ4 & Chapter~\ref{chap:meta} \\ \hline

The justification for the suitability of the concept of sequential meta-learning in \MetaModel~for learning multi-environment configuration data, and the identification of three unique properties in \MetaModel:
    \begin{itemize}
        \item the sequence matters;
        \item train later contributes more;
        \item and using more meta environments are beneficial.
    \end{itemize}
&
  Objective 4 \& RQ4 & Chapter~\ref{chap:meta} \\ \hline

\end{longtable}
\end{footnotesize}

%% file: Tables/publications.tex
\begin{table*}[h]
\footnotesize
\vspace{0.5cm}
\caption{The publications resulted from this thesis, their status (until October 2024), and the addressed objective and RQ.}
\label{chap1:publication}
\begin{tabular}{|p{0.67\columnwidth}|p{0.10\columnwidth}|p{0.12\columnwidth}|}
\hline
\rowcolor[HTML]{C0C0C0} 
\textbf{Publication}                        & \textbf{Status} & \textbf{Objective \& RQ}\\ \hline
\underline{Jingzhi Gong}, Tao Chen and Rami Bahsoon, Dividable Configuration Performance Learning, \textit{IEEE Transactions on Software Engineering (TSE)} [An extended version of the publication in \textit{ESEC/FSE'23}].

&
  Accepted & Objective 3 \& RQ3\\ \hline
  
\underline{Jingzhi Gong}, Tao Chen, Deep Configuration Performance Learning: A Systematic Survey and Taxonomy, \textit{ACM Transactions on Software Engineering and Methodology (TOSEM)}.

&
  Accepted & Objective 1 \& RQ1\\ \hline
  
\underline{Jingzhi Gong} and Tao Chen, Predicting Configuration Performance in Multiple Environments with Sequential Meta-Learning, \textit{The ACM Joint European Software Engineering Conference and Symposium on the Foundations of Software Engineering (FSE'24)}, July 15-19, 2024, Porto de Galinhas, Brazil, 24 pages.

&
  Accepted & Objective 4 \& RQ4\\ \hline
  
\underline{Jingzhi Gong} and Tao Chen. 2023. Predicting Software Performance with Divide-and-Learn. \textit{In Proceedings of the 31st ACM Joint European Software Engineering Conference and Symposium on the Foundations of Software Engineering (ESEC/FSE’23)}, December 3–9, 2023, San Francisco, CA, USA., 13 pages. 

&
  Accepted & Objective 3 \& RQ3\\ \hline
\underline{Jingzhi Gong} and Tao Chen. 2022. Does Configuration Encoding Matter in Learning Software Performance? An Empirical Study on Encoding Schemes. \textit{In 19th IEEE/ACM International Conference on Mining Software Repositories (MSR'22)},
Pittsburgh, PA, USA, May 23-24, 2022. ACM, 482–494.

&
  Accepted & Objective 2 \& RQ2\\ \hline

\end{tabular}
\end{table*}

%% file: Chapter-method/chapter-method.tex
\chapter{Methodology}
\label{chap:methodology}
This chapter discusses the research methodology employed throughout this thesis, outlining its key components, objectives, and activities. It further explores how this methodology contributes to the overall success of the thesis and provides an overview of the research organization within this work.


\input{Chapter-method/introduction}

\input{Chapter-method/elements}

\input{Chapter-method/objectives}

\input{Chapter-method/activities}

\input{Chapter-method/summary}

%% file: Chapter-method/introduction.tex
\section{Introduction}
\label{sec:method-intro}

The methodology taken to carry out the studies in this thesis is a \textbf{\textit{Design Science Research Methodology (DSRM)}} in the field of information systems, as proposed by~\citet{DBLP:journals/jmis/PeffersTRC08}. To be specific, design science (DS) is a research paradigm that focuses on the design and development of artifacts with the explicit intention of solving a problem. In the meantime, information systems (IS) is an applied discipline where data is analyzed by computer science technologies and turned into pragmatic information for the decision-making of organizations. 

Therein, the DSRM is a widely used framework for DS research in IS. It covers the three fundamental components of a methodology~\citep{10.5555/2017212.2017217}, achieves three primary objectives of DS methodology, and contains six primitive activities.  

The utilization of this methodology is natural because this thesis aligns perfectly with the principles of DS and IS. Firstly, this research aims at building accurate predicting artifacts for solving the regression problem between the software configurations and the performance attribute, fulfilling the core objective of design science. Secondly, the performance prediction artifact can be utilized to help software engineers make developing decisions and help users to tune their software based on their specific performance requirements.

This chapter is organized as follows. Section~\ref{chap-method:elements} introduces the three key elements of the DSRM and how they are related to this study, Section~\ref{chap-method:objectives} explains the three objectives met by the methodology as well as their benefits to this study, and in Section~\ref{chap-method:activities}, the six main steps of the DSRM and the organizing details in this research are presented. At last, this chapter is summarised in Section~\ref{chap-method:summary}.



%% file: Chapter-method/elements.tex
\section{Key Elements}
\label{chap-method:elements}
According to~\citet{DBLP:journals/jmis/PeffersTRC08}, a DS research methodology is “principles, practices, and procedures applied to a specific branch of knowledge”. Therefore, any high-quality methodology for DS research should have these basic elements. The three elements provide significant guidance to this thesis. 

In this section, the three elements of the DSRM, as well as their benefits to this thesis, are introduced.

\subsubsection{Principles}
The principles of a DS research methodology define the conceptual meaning of the research. In the context of information systems, DSRM emphasizes the design, implementation, and improvement of artifacts that analyze and utilize data to support organizational decision-making~\citep{10.5555/2017212.2017217}. 

This principle aligns directly with the aim of this thesis, as outlined in Section~\ref{chap-intro:aim}, to develop advanced deep configuration performance models. By adopting this methodology, this thesis will empower software developers and users with their decision-making processes.


\subsubsection{Practice Rules}
According to~\citet{10.5555/2017212.2017217}, DSRM adheres to seven key practice rules to guide the design science research process: (1) the output of the research is an artifact, (2) the artifact is a solution to the research problem, (3) the effectiveness of the solution is evaluated, (4) the research output contributes to the current literature, (5) the development and evaluation methods of the artifact are rigorous, (6) search different techniques to meet the objective, and (7) the results are clearly presented to both the technical agents and business organizations.

This thesis has strictly adhered to all seven practice rules, ensuring the legitimacy and rigor of the research process. As a result, the thesis produces state-of-the-art performance models that effectively address the performance prediction problem (rules 1-4). Moreover, the models have been rigorously evaluated through comprehensive experiments and statistical tests, ensuring the significance of the results (rule 5). Additionally, the thesis explores various techniques and presents its findings clearly to both technical and non-technical audiences (rules 6-7).


\subsubsection{Procedures}

The final element of DSRM is the process model, which offers both a structured guide for research execution and a mental framework for presenting research findings~\citet{DBLP:journals/jmis/PeffersTRC08}. The specific procedures and implementation decisions made in this research are detailed in Section~\ref{chap-method:activities}.

By following the process model, the thesis progresses in a formal order, facilitating a clear understanding and presentation of the research results. Specifically, the thesis adheres to the process model outlined in Figure~\ref{fig:dsrm} and utilizes it as a mental model for presentation.


%% file: Chapter-method/objectives.tex
\section{Accomplished Objectives}
\label{chap-method:objectives}
The DSRM adopted in this research successfully achieves three objectives critical for a successful design science research, as outlined by~\citet{DBLP:journals/jmis/PeffersTRC08}. The following sections will explain how each objective is met and its contribution to this study.

\subsubsection{Providing a Nominal Process}

The DSRM offers a structured process for conducting design science research by assigning designated names to each research activity. This naming convention, as illustrated in Figure~\ref{fig:dsrm}, enhances the legitimacy and clarity of the research process.


\subsubsection{Building upon Existing Research}
The methodology integrates processes from previous successful design science research studies across various disciplines, such as information systems and engineering. By building upon the foundation of previous successful studies, this thesis increases its potential for impact and contribution to the field of software configuration performance prediction.


\subsubsection{Providing a Mental Model}
The DSRM procedures serve as a mental model, guiding the organization and presentation of this research. This model facilitates a clear understanding of the research processes and results, ensuring a well-structured and coherent thesis.

%% file: Chapter-method/activities.tex
\section{Sequential Activities}
\label{chap-method:activities}
\begin{figure}[t!]
  \centering
  \includegraphics[width=0.8\columnwidth]{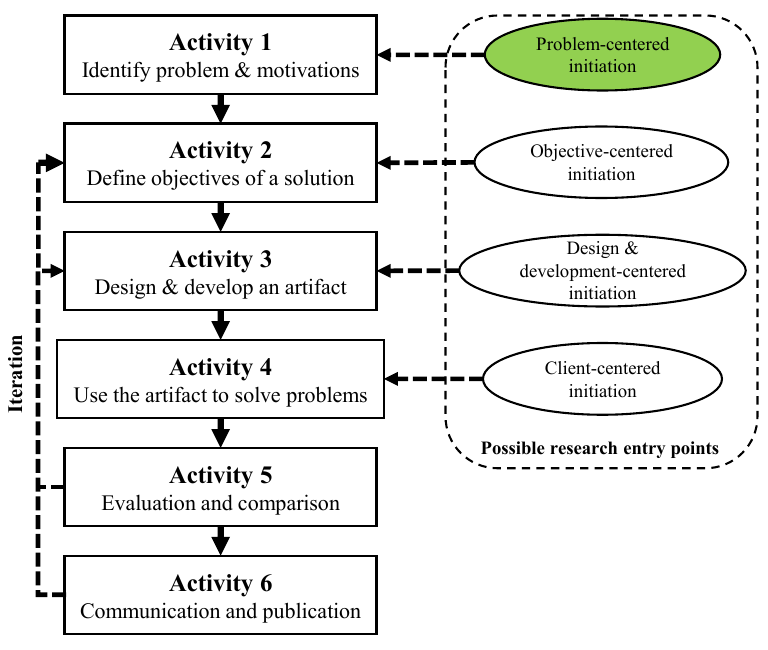}
  \caption{The process model of the design science research methodology used in this study.}
  \label{fig:dsrm}
\end{figure}
The procedures of this DSRM can be represented by the process model in Figure~\ref{fig:dsrm}. There are six key activities that should be conducted consecutively. Among the six activities, four entry points could be chosen according to the nature of the research, ranging from the first activity to the fourth. At the beginning of my doctoral study, the research problems are not identified. Therefore, the \textbf{problem-centered initiation} is chosen as the starting point. 

This section provides detailed explanations of each activity and outlines the organization of this thesis.

\subsubsection{Activity 1: Problem and Importance Identification}
This initial stage involves identifying the research problem and its importance. The research problem should be based on the current literature in the field and capture the knowledge gaps. Meanwhile, the importance of the research problem could motivate the researcher to conduct the studies and help the readers understand the problem better. 

In this thesis, the research problem and importance are formed by a preliminary study of the existing studies in the domain of software engineering and performance engineering. Particularly, with the growing complexity of modern software, it is often impractical to examine the performance of all the software configurations. Therefore, it is important to build a deep performance model that learns only a small subset of configurations and predicts the rest of them accurately, which can save a large amount of time and resources. 

Subsequently, a systematic literature review of the latest deep configuration performance learning approaches is conducted, as presented in Chapter~\ref{chap:review}, which facilitates the formalization of the research aim, which is to design and develop performance models that address the identified knowledge gaps in the community and enable better predictions, ultimately reducing measurement costs for software engineers.

    

\subsubsection{Activity 2: Define the Objectives for a Solution}

This phase aims at translating the identified research problem and aim into specific research objectives. These objectives serve as targets and milestones that must be achieved to solve the research problem. Objectives can be formulated either quantitatively, using specific metrics to measure the effectiveness of the artifact, or qualitatively, describing the expected research outcomes.

In this thesis, research objectives are derived from two sources. Firstly, the research aim specified in Activity 1 is analyzed, leading to the extraction of the first research question and its corresponding objective, which focuses on investigating the current literature and identifying knowledge gaps. Secondly, during the literature review, additional research questions and objectives are identified and added to the list.

Notably, some of these objectives require sequential completion, where they must be met in a specific order. Others are parallel, meaning they can be addressed by different models in any order. For instance, objectives 1 and 3 in Section~\ref{chap-intro:objectives}, which are to review the latest performance models and solve the sparsity problem, respectively, are sequential because objective 1 provides the foundation for the other. On the other hand, objectives 2, 3, and 4 address different gaps in configuration performance modeling and are met by different artifacts. Therefore, they can be addressed in parallel.

Furthermore, the research objectives include both quantitative and qualitative elements. For example, the first research objective is qualitative, as there is no specific metric to measure its completion. In contrast, objective 3, to make the accuracy of the deep performance model better than the state-of-the-art models by mitigating the sparsity issue, is quantitative, as the prediction accuracy can be measured.

\subsubsection{Activity 3: Design and Development}
This step focuses on designing and implementing artifacts that fulfill the research objectives. The design phase involves drafting the artifact's functionality and dataflow to ensure alignment with the desired objectives. A prototype is then implemented, trained with a set of training samples, and passed to the next stage. If the prototype fails to meet the objectives, it is iterated back to this activity for redesign and modification.

In this thesis, the design of the deep learning model for performance prediction involves reviewing the latest deep learning approaches from various domains and selecting the most promising one for the problem at hand. If a suitable model exists, its basic idea is used to draft the model structure, with adaptations made to fit the specific problem. Otherwise, a new model is designed based on the authors' domain knowledge.

The development phase involves creating a performance model that realizes the designed framework and training it with the provided configurations and their corresponding performance values. The model is then passed to the next activity for evaluation. If it fails to meet the designated objectives, the research returns to this activity via the iteration path, and the model is improved to meet the unfulfilled goals.

\subsubsection{Activity 4: Testing}
This activity seeks to demonstrate the artifact's ability to solve instances of the research problem. The artifact's outputs are then passed to the next activity for evaluation.

In this thesis, the research aim is to accurately predict software performance based on a given configuration. Therefore, the testing phase involves using the trained model to predict the performance of a set of testing configurations. The predicted performance values are then passed to the evaluation stage.

    

\subsubsection{Activity 5: Evaluation and Comparison}
This stage evaluates the accuracy of the model by comparing predicted results to actual values. The effectiveness of the model is measured using appropriate performance metrics, and the measured accuracy is then compared to an expected standard to determine whether the model meets the desired objectives. If not, the model is returned to the previous activity for further improvement. Otherwise, the evaluation results are passed to the next stage.

In this thesis, the predicted performance values are compared to the actual values, and the accuracy is measured using MAPE (or MRE). The computed MAPE is then compared to the accuracy of state-of-the-art performance modeling approaches such as \texttt{HINNPerf}\citep{DBLP:journals/tosem/ChengGZ23} and \texttt{DeepPerf}\citep{DBLP:conf/icse/HaZ19}. Statistical tests like the Scott-Knott test are used to ensure the statistical significance and effect size of the accuracy difference. If the proposed model does not meet the dedicated objectives, it is returned to Activity 3 for revision or reported to the supervisor in the next phase.


\subsubsection{Activity 6: Communication and Publication.}
This stage involves presenting the evaluation results and artifact to relevant professionals for discussion and feedback. If agreement on the artifact's accomplishment is reached, the research is concluded and prepared for publication in academic conferences or journals. Otherwise, iterations are carried out, and the research continues.

In this thesis, the accuracy of the performance model and comparison results are presented to the supervisor for discussion. If the model's accuracy is not statistically significantly better than existing approaches, it is returned to Activity 3 for further improvements. If new problems are identified during the research, the research returns to Activity 2 to update the objectives and either modify existing models or develop new solutions to meet these objectives. Finally, if the model achieves all specified objectives, a paper is prepared for publication.

This research has resulted in two models: the "Divide-and-Learn" (\Model) framework, which addresses the sparsity issue in performance data (details in Chapter~\ref{chap:dal}), and the meta-learning (\MetaModel) framework, which sequentially trains a meta-model with multiple meta-environment data, leading to more robust and accurate predictions (details in Chapter~\ref{chap:meta}). These models fulfill objectives 3 and 4 of this thesis, as outlined in Section~\ref{chap-intro:objectives}. Note that objective 2 is achieved through an empirical study on encoding schemes; however, as no artifact is designed in this research, the process does not fully follow the DSRM.

%% file: Chapter-method/summary.tex
\section{Chapter Summary}
\label{chap-method:summary}


This chapter introduces the design science research methodology employed in this thesis, which is a well-established approach in information systems research that provides a structured framework for conducting and presenting research.

The DSRM consists of three components: the principles that define the meaning of the research, the practice rules that provide conducting principles for this research, and the procedures which are step-by-step guidelines to conduct and present this study. 

Moreover, this methodology has three accomplishments: it provides a nominal process to legitimize this research, it is built on top of previous studies, which makes the DSRM a practical framework, and it presents a mental model for researchers to better organize the study and exhibit the results. 

Finally, the DSRM procedures follow an iterative process model with six sequential activities, as illustrated in Figure~\ref{fig:dsrm}. The model offers four entry points to begin the research, with an iteration loop between activities 5 and 6, allowing for adjustments based on evaluation results.

This thesis starts with the first activity, where a systematic literature review (details in Chapter~\ref{chap:review}) is conducted to identify the research problem and aim. Further research objectives are extracted in activity 2. Performance models are then designed and implemented in activity 3 to fulfill these objectives. In activities 4 and 5, the model is tested and evaluated using the MAPE accuracy metric. If the accuracy falls short of the requirements, the process returns to activity 3 for iteration. Finally, in activity 6, the evaluation results are presented to the supervisor and other relevant audiences.


Two models have been designed and produced following the methodology, which will be presented in the following Chapter~\ref{chap:dal} and~\ref{chap:meta}.

%% file: Chapter-survey/chapter-survey.tex
\chapter{Deep Configuration Performance Learning: A Systematic Survey and Taxonomy}
\label{chap:review}
The design science research methodology employed in this thesis involves two primary activities: identifying the research aim and defining the objectives (as discussed in Chapter~\ref{chap:methodology}). Through the primary studies, the research aim of this thesis is identified, which focuses on efficiently addressing the limitations in modeling the relationship between the software configuration and performance and enabling next-level prediction accuracy. Consequently, to realize this aim, it is crucial to investigate the behaviors documented in the literature and identify the current limitations that need to be overcome. Thereby, the first objective of the thesis, which coincides with the objective of the SLR, is defined as follows:

\begin{answerbox}
\emph{\textbf{Objective 1:} to systematically review the current literature in the discipline of performance modeling using deep learning, disclose both the positive and negative behaviors and identify the knowledge gaps that need to be addressed.}
\end{answerbox}

To achieve these, a systematic literature review (SLR) is conducted between 2013 and 2023\footnote{The original SLR was conducted from 2015 to 2020, coinciding with the start of my PhD study. In 2023, the review scope is expanded, and the review findings are updated to include a broader time range.}. In the end, the SLR classifies the techniques applied and concludes the positive and problematic practices in the different processes of deep learning for performance modeling, such as data preparation, model training, evaluation, and application, by answering a set of sub-research questions. Subsequently, the results are analyzed and discussed, leading to the disclosure of potential knowledge gaps that need to be further bridged by this thesis. Thereby, additional objectives and research questions were established, which will be introduced in detail in the forthcoming chapters of this thesis.

In this chapter, the methodology of the SLR, the statistics and analysis of the review, the findings of positive and negative behaviors, and the knowledge gaps identified in the SLR are explained.

\input{Chapter-survey/introduction}

\input{Chapter-survey/methodology}

\input{Chapter-survey/RQ1}
\input{Chapter-survey/RQ2}
\input{Chapter-survey/RQ3}
\input{Chapter-survey/RQ4}

\input{Chapter-survey/discussions}

\input{Chapter-survey/threats_to_validity}

\input{Chapter-survey/conclusion}

%% file: Chapter-survey/introduction.tex
\section{Introduction}
\label{chap-review:introduction}

\begin{figure}[!t]
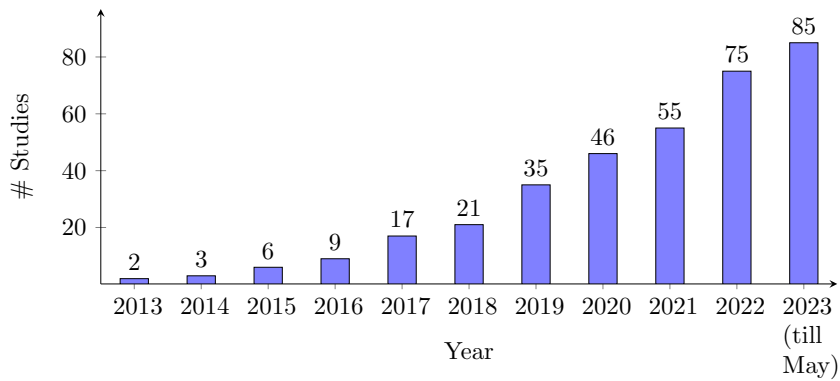

\centering
\includestandalone[width=0.75\columnwidth]{Figures/chap-review/year_publications_cumulative}
    \caption{Cumulative number of primary studies on deep configuration performance learning models.}
 \label{chap-review-fig:year_publications}
 \end{figure}

Recent studies have demonstrated the benefits of deep learning for modeling configuration performance. For example,~\citet{DBLP:conf/icse/HaZ19} propose \texttt{DeepPerf}, a DNN-based model combined with L1 regularization to address the sparse performance functions, and~\citet{DBLP:journals/tosem/ChengGZ23} invent a hierarchical interaction neural network model called \texttt{HINNPerf} that achieves state-of-the-art MRE. Yet, despite the importance of such research direction, to the best of our knowledge, there has been little work on a systematic survey that covers the full spectrum of deep configuration performance learning. The current reviews related to this topic mainly focus on either general machine learning models~\citep{DBLP:journals/jss/PereiraAMJBV21} or deep learning in the general context of software engineering~\citep{DBLP:journals/csur/YangXLG22, DBLP:journals/tosem/WatsonCNMP22, DBLP:journals/tse/WangHGGZFSLZN23}. Undoubtedly, systematically reviewing state-of-the-art studies on this particular research field can provide vast benefits, including summarizing the common categories, revisiting the important concepts, and more importantly, discussing novel perspectives on the positive and negative practices of the field, and providing insights for future opportunities.

To bridge such a gap, in this paper, a systematic literature review that covers 948 papers from six online repositories and 52 venues, published between 2013 and 2023 is conducted, based on which 85 prominent studies were extracted for data extraction and analysis. The results confirm that the significance of deep learning for configuration performance and its associated challenges has led to a notable increase in research efforts in this field. As shown in Figure~\ref{chap-review-fig:year_publications}, clearly, there is a growing trend of papers in this field. In particular, 75\% of the work has been published since 2019, indicating the emergence of this new paradigm for configuration performance learning.



\begin{figure}[t!]
  \centering
  \includegraphics[width=\columnwidth]{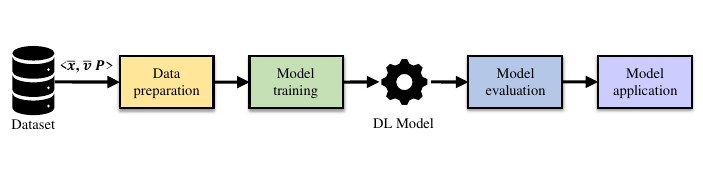}
  ~\vspace{-1cm}
  \caption{Deep learning pipeline for performance modeling.}
  \label{fig:pipeline}
\end{figure}

\subsection{Research Questions}
\label{chap-review-subsec:rq}
Following the DL pipeline in Figure~\ref{fig:pipeline}, the sub-research questions of this survey are derived. Specifically, the workflow of deep learning for learning software configurations and performance is formalized into 4 main stages. 

Particularly, the first step is to process the raw performance data collected from the full configuration space and make preparations for inputs to the deep learning models, including procedures like preprocessing, encoding, and sampling. Therefore, the first sub-questions is:

\begin{quotebox}
   \noindent
   \textit{\textbf{RQ1.1:} How to prepare performance data?}
\end{quotebox}

For RQ1.1, three key processes are examined:
        \begin{itemize}
        \item  What data preprocessing methods have been applied?
        \item  What encoding methods have been applied?
        \item  What data sampling methods have been applied?
        \end{itemize} 

Secondly, for learning the preprocessed data, various procedures are involved, including the handling of the specific challenges in performance prediction, the design, tuning, and optimization of the deep learning model. To survey this, the second sub-question is: 

\begin{quotebox}
   \noindent
   \textit{\textbf{RQ1.2:} How to train the performance model?}
\end{quotebox}

To answer this, four key processes need to be investigated:
        \begin{itemize}
        \item How to handle overfitting and sparsity problems?
        \item What Deep learning models have been utilized?
        \item How to tune the hyper-parameters?
        \item What optimization algorithms and activation functions have been used?
    \end{itemize}

Thirdly, the aim is to evaluate the trained performance model with a set of subject systems, a specific evaluation method, and a particular accuracy metric, and to examine the statistical significance of the results. Therefore, the question to ask is:

\begin{quotebox}
   \noindent
   \textit{\textbf{RQ1.3:} How to evaluate the trained model?}
\end{quotebox}

This incorporates Three key processes to review:
        \begin{itemize}
        \item How are the models evaluated and what are the metrics?
        \item How to ensure the statistical significance of the conclusions?
        \item How many subject software systems are used, and what are their domains?
        \end{itemize}

Finally, it is crucial for the performance model to be applied in a practical domain, adapt to the dynamic running environment, and enable readers to replicate and reproduce the findings. To that end, the last sub-RQ is:

\begin{quotebox}
   \noindent
   \textit{\textbf{RQ1.4:} How to apply the performance model?}
\end{quotebox}

Three key processes are related to this question:
        \begin{itemize}
        \item What are the application domains?
        \item Are dynamic environments considered?
        \item Are datasets and source codes available for replication?
        \end{itemize}

For each sub-RQ, the related techniques are investigated, and a taxonomy is made. More importantly, the positive practices that should be followed and the negative practices that would require immediate attention are summarized, and the future directions that can be addressed in this particular thread of research are justified. 

\subsection{Contributions}
\label{chap-survey:contributions}

To bridge the lack of understanding regarding deep learning models for software performance prediction, this chapter presents a systematic literature review that encompasses 948 papers sourced from six online repositories and 52 venues. The selected studies were published between 2013 and 2023, and 85 prominent studies were extracted for data extraction and analysis. The examination covers various stages within the deep configuration performance learning pipeline, including configuration data preparation, deep model training, accuracy evaluation, and model exploitation for configurable software systems.

In summary, the contributions of this chapter include:
\begin{itemize}
    \item A taxonomy that categorizes the techniques used and key concerns in deep configuration performance learning with up to 13 findings of the trends. 
    
    \item Comprehensive summaries of the key approaches used in the deep learning pipeline for configuration performance, including preparation, modeling, evaluation, and application, together with discussions on their benefits and shortcomings.

    \item Articulation on the good practices and bad smells observed from the findings.
    
    \item Gaps identified from existing studies, offering insights into the future opportunities for this particular thread of research.

    
    
\end{itemize}

In particular, the review results highlight several key observations that warrant attention in future studies:

\begin{itemize}
    \item More than half (45 out of 85) of studies have not applied any data pre-processing methods, which may limit the quality of the configuration data.
    \item The justification for the choice of encoding scheme is ignored by 65 (74\%) of the primary studies, which makes the influence and behavior of the encoding schemes unclear. Hence, it is crucial to conduct investigations specifically aimed at examining the influence of encoding schemes.
    \item Random sampling, which is inefficient in finding the most informative configurations, is the most commonly used method, as used in 66 studies. This indicates the need for exploring more informative sampling methods to enhance the quality of the collected data.
    \item A significant number of studies (33 out of 85) fail to address the issues of sparsity and overfitting, while 24 of the rest rely on inefficient manual feature selection methods. This emphasizes the importance of addressing sparsity issues and encourages researchers to explore alternative approaches.
    \item The majority of studies (52 out of 85) apply manual hyperparameter tuning, which relies heavily on human experts. Future research should investigate automatic and heuristic hyperparameter tuning techniques to reduce tuning costs and enhance efficiency.
    \item 46 Primary studies omit to introduce the optimization method and activation function for their deep learning model, which could be harmful to open science.
    \item A notable portion of studies neglect the usage of statistical (79 out of 85) and effect size tests (81 out of 85). To improve the reliability of evaluations and findings, future studies should incorporate statistical analysis.
    \item 56 of the primary studies (66\%) do not consider the challenge of dynamic environments. It is crucial for future research to address this aspect to strengthen the generalizability of performance models.
    \item Only a small percentage (14 out of 85) of the primary papers provide both source codes and datasets, which hinders open science. The sharing of source code and datasets in future works can facilitate replicability and reproducibility.
\end{itemize}

\subsection{Chapter Outline}
The chapter is organized as follows. Section~\ref{chap-review:methodology} presents the systematic review methodology. Section~\ref{chap-reivew:rq1},~\ref{chap-reivew:rq2},~\ref{chap-reivew:rq3}, and~\ref{chap-reivew:rq4} summarize the results of the review. Then, Section~\ref{chap-review:discussion} discusses the trends identified from the SLR, possible knowledge gaps in the current literature, and the research objectives extracted from the SLR for further studies of this thesis. Finally, Section~\ref{chap-reivew:threats} justifies the threats to validity, and Section~\ref{chap-review:conclusion} concludes this chapter.

%% file: Figures/chap-review/year_publications_cumulative.tex
\begin{tikzpicture}
\footnotesize
\begin{axis}[
    width=12cm,
    height=5.5cm,
    ybar,
    ymax = 95,
         axis x line  = bottom,
    axis y line  = left  ,
    bar width=0.4cm, 
   ylabel=\# Studies,
    enlarge y limits=0.02,
    enlarge x limits=0.05,
    y label style={at={(-0.08,0.5)}},
    x label style={at={(0.5,-0.18)}},
    xlabel=Year,
    symbolic x coords={A, B, C, D, E, F, G, H, I, J, K}, 
    xtick=data, 
     xticklabels={2013,2014,2015,2016,2017,2018,2019,2020,2021,2022,{2023\\(till May)}},
     xticklabel style={text width=0.6cm},
    nodes near coords, 
]
\addplot[fill=blue!50, draw=black] coordinates {(A, 2) (B, 3) (C, 6) (D, 9) (E, 17) (F, 21) (G, 35) (H, 46) (I, 55) (J, 75) (K, 85)};
\end{axis}
\end{tikzpicture}




%% file: Chapter-survey/methodology.tex
\section{Research Methodology}
\label{chap-review:methodology}

\begin{figure}[t!]
  \centering
  \begin{adjustbox}{width=1.3\textwidth,center}
  \includegraphics[width=\columnwidth]{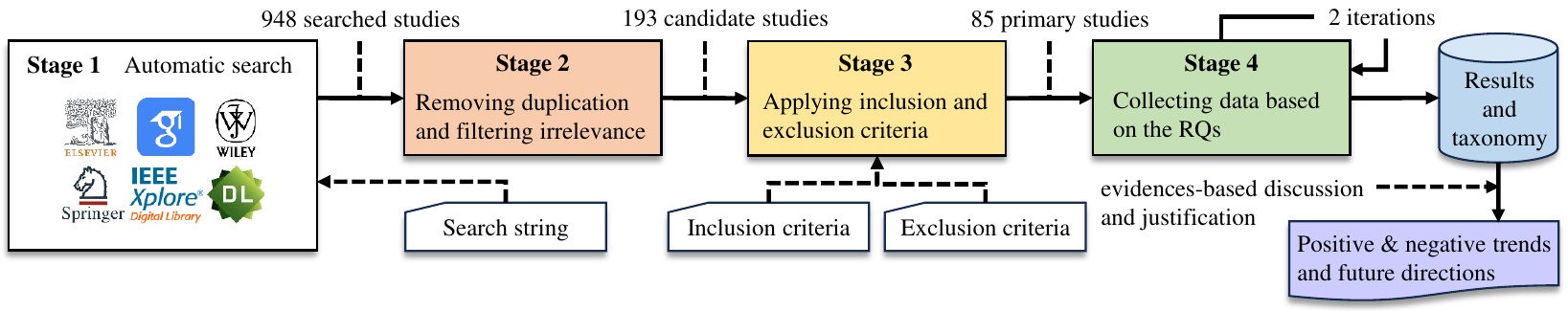}
  \end{adjustbox}
  \caption{Overview of the systematic literature review protocol.}
  \label{fig:protocol}
\end{figure}

This SLR covers the papers published between 2013 and 2023. This period was chosen because this thesis seeks to concentrate on the latest trends, avoiding noises from the old and disappeared practices in the field. In particular, the review methodology of this SLR follows the best practice of systematic literature review for software engineering~\citep{DBLP:journals/infsof/KitchenhamBBTBL09}, as shown in Figure~\ref{fig:protocol}.

\subsection{Stage 1: Automatic Search}
As can be seen in Figure~\ref{fig:protocol}, an automatic search over six highly influential indexing services was conducted, i.e., ACM Library, IEEE Xplore, Google Scholar, ScienceDirect, SpringerLink, and Wiley Online, as they have been used in a number of SLR in the field of software engineering and performance modeling~\citep{DBLP:journals/sigmetrics/NambiarKBSD16, DBLP:conf/wosp/PereiraA0J20, DBLP:journals/spe/HanYY23, DBLP:journals/tse/HortKSH22, DBLP:journals/tjs/Flores-Contreras21}.

The consideration behind formulating the search string was multi-faceted, as the aim was to identify primary studies that satisfy:
1) the inclusion of software systems; 2) the utilization of configurations as the input for performance models; 3) the application of performance models to predict the performance of software or systems; and 4) the incorporation of deep learning or neural network-based techniques in the learning models. 

To achieve this objective, different combinations of keywords in these four areas were explored to determine the optimal search string. This ensured that the search results excluded redundant and irrelevant works as much as possible, specifically retrieving papers that aligned with the research focus on the intersection of software performance prediction and deep learning approaches.

In a nutshell, the search was guided by the search string below:

\begin{tcolorbox}[breakable,left=2pt,right=2pt,top=2pt,bottom=2pt] 
\emph{(``software'' OR ``system'') AND ``configuration'' AND (``performance prediction'' OR ``performance modeling'' OR ``performance learning'' OR ``configuration tuning'') AND (``deep learning'' OR ``neural network'')}
\end{tcolorbox}

During the automatic search process, the search engines were configured with a time range filter of 2013-2023 to ensure that the retrieved results were limited to the specified period. 

The search resulted in 948 studies, accounting for duplicates and excluding non-English documents. Specifically, a total of 414 primary studies were obtained from Google Scholar, 172 studies from SpringerLink, 114 studies from Wiley Online, 98 studies from ACM Library, 94 studies from ScienceDirect, and 56 studies from IEEE Xplore.

\subsection{Stage 2: Removing Duplication and Filtering Irrelevance}
Next, in stage 2, the objective was to ensure that only unique and highly relevant studies were considered for further analysis in the review.

To achieve this, a careful examination of the titles of the identified papers was conducted to filter out any duplicates. This step ensured that each study included in the review was distinct and provided unique insights into the body of knowledge.

Subsequently, a brief evaluation was carried out to eliminate any documents that were clearly irrelevant to the specific research topic. For example, studies focused on human and student performance were excluded as the research did not directly address educational outcomes. Similarly, papers centered on the performance of physical systems, fuel systems, and football systems were disregarded as they were not directly aligned with the scope of the investigation.

As a result of these rigorous filtering procedures, 193 highly relevant candidate studies were identified, which have the potential to make significant contributions to the review.

\subsection{Stage 3: Applying Inclusion and Exclusion Criteria}
During stage 3 of the review process, various criteria were applied to extract a set of representative works.

To begin with, inclusion criteria were formulated to ensure that the selected studies closely aligned with the core themes of the research. Specifically, these criteria aimed to include studies that employed deep learning models for software performance learning, which constituted the central focus of the study. Additionally, the criteria emphasized the need for explicit details about the employed learning algorithms, as this was crucial for understanding, analyzing, and applying the models. Moreover, the inclusion of quantitative experiments was considered important to facilitate a robust evaluation process and enhance the depth and reliability of the selected studies.

Conversely, the exclusion criteria were designed to maintain the integrity and relevance of the study. By excluding papers unrelated to software or systems, the selected studies were ensured to relate to the research focus of the survey directly. Furthermore, the requirement for peer-reviewed publication enhanced the credibility of the included studies, while excluding certain types of works such as surveys, reviews, tutorials, case studies, or empirical studies helped filter out literature that may not align with the objective of investigating performance prediction using deep learning. Lastly, limitations on the length of papers were imposed to ensure that the selected studies possessed sufficient depth and detail for comprehensive analysis.

In a nutshell, a study is temporarily selected as a primary study if it meets all of the following inclusion criteria:

\begin{itemize}
\item The paper presents a performance modeling approach using deep learning algorithm(s).
\item The goal of the paper is to predict or analyze the performance of a software system.
\item The paper has at least one section that explicitly specifies the learning algorithm(s) used.
\item The paper contains quantitative experiments in the evaluation with details about how the results were obtained.
\end{itemize}

Subsequently, the following exclusion criteria are applied to the previously included study, which would be removed if it meets any: 
\begin{itemize}
\item The paper is not software or system engineering-related.
\item The paper is not published in a peer-reviewed public venue.
\item The paper is a survey, review, tutorial, case study, or an empirical type of work.
\item The paper is a short and work-in-progress work, i.e., shorter than 8 double-column or 15 single-column pages.
\end{itemize}

The above process was repeated for all candidate studies. Finally, 85 primary studies were identified for detailed data collection.

\subsection{Stage 4: Collecting and Extracting Data}
\subsubsection{Review elements}
After formulating the research questions, a comprehensive list of review elements has been developed for collection during the review process. The summary of these review elements is presented in Table~\ref{tb:data_items}, which includes a total of 22 review elements. Each element serves a specific purpose related to the corresponding research question. This section explains the design rationales behind these elements and clarifies the procedure for extracting and classifying the data from each element.

In the initial stage, review elements $E_1$ to $E_4$ are employed to gather meta-information about the reviewed studies, such as the title, author, and publication information. This meta-information enables accurate referencing and organization of the reviewed studies, establishing a robust foundation for further analysis and interpretation.

To address RQ1.1, review elements $E_5$ to $E_7$ are designed to extract the corresponding data preparation methods. These elements aim to examine the key processes in preparing configuration data, where it is common to pre-process the raw configuration data, encode the data into the most appropriate form, and select a subset of informative samples for model training. For instance, by using $E_5$ the pre-processing methods used in each primary study are extracted according to the fundamental types of methods. Methods such as min-max scaling and z-score, which are all used to normalize the scale of the configuration options, can be classified into the same category.

To answer RQ1.2, review element $E_8$ is used to review the methods used to deal with data sparsity in configuration data. Review elements $E_9$ to $E_{13}$ are then used to understand how the deep learning models are chosen and trained according to their most general type. For example, on $E_{11}$, hyperparameter tuning methods can be set into three categories:

\begin{itemize}
    \item using the default hyperparameter settings without any tuning;
    \item tuning the hyperparameters via manual effort and domain knowledge;
    \item or relying on automated heuristic methods in the tuning.
\end{itemize}

To examine the evaluation-related techniques in RQ1.3, review elements $E_{14}$ to $E_{19}$ are employed. These elements examine the evaluation procedure ($E_{14}$), metrics ($E_{15}$), methods of statistical validation ($E_{16}$ and $E_{17}$), the number of subject systems used ($E_{18}$), and the domain of subject systems used ($E_{19}$). These review elements ensure the examination of the statistical difference between the comparisons and the generalizability of the conclusions.

To address RQ1.4, review elements $E_{20}$ to $E_{22}$ are identified to examine the exploitation information regarding the deep configuration performance models. These elements examine the application domains of the performance models ($E_{20}$), the environmental conditions considered ($E_{21}$), and the data for replication and reproduction of the proposed models ($E_{22}$). These review elements ensure the examination of the generalizability, robustness, and reproducibility of the deep configuration performance models.

\input{Tables/chap-review/data_items}

\subsubsection{Data Collection}

Stage 4 of the review methodology aims to collect all review elements related to each sub-RQ presented in Section~\ref{chap-review-subsec:rq}. 

The data collection procedure involves a systematic examination of the primary studies to extract relevant information for addressing the research questions. Specifically, for RQ1.1 related to data preprocessing, the initial focus is on identifying keywords such as preprocessing, encoding, and sampling. Then, sections like ``Experiment setup'', ``Data processing'', or ``Implementation'' are carefully read to identify any potentially related techniques.

Subsequently, for the process of handling overfitting in RQ1.2, the examination begins with the introduction and justifications provided in the papers to determine if they mention sparsity and overfitting problems. The analysis then extends to sections that illustrate the learning model to identify techniques used to address these challenges. Similarly, strategies adopted to address processes like model selection, hyper-parameter tuning, and optimization algorithms are extracted from sections that introduce the deep learning model, such as ``Model structure'', ``Model implementation'', or ``Learning model''.

Regarding RQ1.3, which involves model evaluation and statistical significance, the primary focus is to understand how researchers assess model performance. Consequently, relevant data is primarily sought in sections that evaluate model performance, such as ``Evaluations'', ``Experiments'', and ``Validations''.

Lastly, for RQ1.4, which considers application domains, dynamic environments, and the availability of datasets and source codes, the data collection process consists of a comprehensive reading of the papers, particularly the backgrounds, the challenges addressed, and the replicability and reproducibility of the primary studies.



%% file: Tables/chap-review/data_items.tex
 \begin{table}[t!]
  \centering
\footnotesize
\caption{The data items considered in this survey.}
\begin{adjustbox}{width=\textwidth,center}
  \begin{subtable}[t]{0.45\textwidth}
    \centering
\footnotesize
\begin{tabular}{lll}
\toprule
\multicolumn{1}{c}{\textbf{ID}} & \multicolumn{1}{c}{\textbf{Review Element}} & \multicolumn{1}{c}{\textbf{RQ}} \\ \hline
$E_{1}$ & Authors & N/A \\ \hline
$E_{2}$ & Year & N/A \\ \hline
$E_{3}$ & Title & N/A \\ \hline
$E_{4}$ & Venue (conference or journal) & N/A \\ \hline
$E_{5}$ & Data preprocessing methods & RQ1.1 \\ \hline
$E_{6}$ & Data encoding sachems & RQ1.1 \\ \hline
$E_{7}$ & Data sampling strategies & RQ1.1 \\ \hline
$E_{8}$ & Sparsity handling mechanisms & RQ1.2 \\ \hline
$E_{9}$ & Deep learning models & RQ1.2 \\ \hline
$E_{10}$ & Reasons for model selection & RQ1.2 \\ \hline
$E_{11}$ & Hyperparameter tuning optimizers & RQ1.2 \\ 
\bottomrule
\end{tabular}
  \end{subtable}
  \hspace{1.6cm}
  \begin{subtable}[t]{0.45\textwidth}
    \centering
\footnotesize
\begin{tabular}{lll}
\toprule
\multicolumn{1}{c}{\textbf{ID}} & \multicolumn{1}{c}{\textbf{Review Element}} & \multicolumn{1}{c}{\textbf{RQ}} \\ \hline
$E_{12}$ & Training algorithms & RQ1.2 \\ \hline
$E_{13}$ & Activation functions & RQ1.2 \\ \hline
$E_{14}$ & Evaluation procedures & RQ1.3 \\ \hline
$E_{15}$ & Accuracy metrics & RQ1.3 \\ \hline
$E_{16}$ & Statistical test & RQ1.3 \\ \hline
$E_{17}$ & Effect size measurement & RQ1.3 \\ \hline
$E_{18}$ & Number of subject systems & RQ1.3 \\ \hline
$E_{19}$ & Domain of subject systems & RQ1.3 \\ \hline
$E_{20}$ & Application categories & RQ1.4 \\ \hline
$E_{21}$ & Handling of dynamic environments & RQ1.4 \\ \hline
$E_{22}$ & Availability of code and dataset & RQ1.4
\\ \bottomrule
\end{tabular}
  \end{subtable}

\end{adjustbox}
  \label{tb:data_items}
\end{table}


%% file: Chapter-survey/RQ1.tex
\section{RQ1.1: How to Prepare Performance Data? }
\label{chap-reivew:rq1}
As shown in Figure~\ref{fig:pipeline}, data preparation stands at the forefront of deep learning problems, serving as a critical step that is aimed at sampling, pre-processing, and encoding raw data to ensure its suitability for deep learning models. In order to improve the accuracy and reliability of deep learning models to the best, data preparation is necessary~\citep{DBLP:journals/infsof/HuangL015}. To comprehensively explore the current literature on data preparation methods for software performance prediction using deep learning, this section addresses three key processes in this stage.

\subsection{How to Pre-process raw Performance Data?}
\label{subsec: preprocess}

\input{Tables/chap-review/RQ-preprocessing_methods}

Data preprocessing methods play a fundamental role in preparing data for deep learning problems~\citep{DBLP:journals/infsof/HuangL015}, and understanding the preprocessing techniques employed in the context of performance learning is crucial for researchers to gain insights and enhance the quality and interpretability of the data. Hence, this section aims to investigate the diverse range of techniques utilized for processing raw performance data.

In Table~\ref{tb:preprocessing}, the taxonomy of the preprocessing methods and their corresponding studies are summarised. It is evident that while various methods have been employed, a significant number of authors (45 out of 85) opt to use the raw dataset without any preprocessing. On the other hand, normalization techniques such as min-max scaling have been applied in 32 studies to standardize numerical features. 15 studies have employed regularization techniques to reduce the impact of noisy or irrelevant features. Methods like Principal Component Analysis for dimensionality reduction have been used in 10 studies. 5 studies have explored embedding techniques for feature construction, while another 5 studies have focused on outlier detection using methods like isolation forest algorithm. Besides, imbalance processing approaches have been applied in 3 studies, and 1 study removes multicollinearity by addressing highly correlated columns. In the following, detailed explanations of these categories will be provided.

\subsubsection{Normalization.} 
Normalization techniques are fundamental in data preprocessing to ensure that different features are brought to a standardized range, thereby preventing any particular feature from dominating the analysis or modeling process due to its larger values~\citep{DBLP:journals/asc/SinghS20}. Several widely used normalization techniques include: \textbf{min-max scaling}~\citep{DBLP:journals/taco/MoolchandaniKS22, DBLP:conf/sigsoft/Gong023, DBLP:conf/icse/HaZ19, DBLP:conf/msr/GongC22}, used in 25 out of the 32 studies, is the most widely used technique that transforms numerical features to a specific range, typically between 0 and 1, by subtracting the minimum value and dividing by the range of the feature. \textbf{Z-score}~\citep{DBLP:journals/fgcs/LiLTWHQD19, DBLP:journals/mam/JooyaDB19} transforms numerical features by subtracting the mean and dividing by the standard deviation of the feature. This method results in a distribution with a mean of 0 and a standard deviation of 1, allowing for easier comparison and interpretation across different features. \textbf{Box-Cox transformation}~\citep{DBLP:conf/cluster/IsailaBWKLRH15} is a power transformation technique used to normalize skewed data by applying a suitable power transformation. It helps in achieving a more symmetric distribution and reducing the impact of outliers.

\subsubsection{Regularization.} 
Regularization techniques introduce additional constraints or penalties to the model's objective function, aiming to control complexity, reduce noise, and enhance generalization to new data. Several commonly used examples: \textbf{L1 regularization}~\citep{DBLP:journals/taco/WangLWB19, DBLP:conf/iccad/KimMMSR17, DBLP:journals/tosem/ChengGZ23, myung2021machine, DBLP:conf/esem/ShuS0X20, DBLP:conf/sigsoft/Gong023, DBLP:journals/access/ThraneZC20, DBLP:conf/icse/HaZ19, DBLP:journals/ese/VituiC21a}, also known as Lasso regularization, adds a penalty term proportional to the absolute value of the model's coefficients. It encourages sparsity in the model by driving some coefficients to zero, effectively performing feature selection and reducing the impact of irrelevant features. \textbf{L2 regularization}~\citep{DBLP:journals/access/ThraneZC20, DBLP:conf/im/JohnssonMS19, DBLP:journals/jsa/TangLLLZ22, DBLP:journals/ese/VituiC21a, myung2021machine}, also called Ridge regularization, adds a penalty term proportional to the squared magnitude of the model's coefficients. It encourages smaller coefficients and can help mitigate multicollinearity issues by reducing the impact of highly correlated features. \textbf{Laplacian regularization}~\citep{DBLP:journals/pvldb/ZhouSLF20} is a technique that incorporates a smoothness constraint on the model's parameters, which encourages the model to have similar parameter values for similar input samples, promoting smoother predictions and reducing the impact of noisy data.

\subsubsection{Dimension reduction.} Dimension reduction techniques help to simplify the data representation, remove redundant or irrelevant features, and overcome the curse of dimensionality, thereby improving computational efficiency, visualization, and interpretability of the data while retaining as much relevant information as possible. For example, \textbf{Principal Component Analysis (PCA)}~\citep{DBLP:conf/middleware/GrohmannNIKL19, DBLP:journals/mam/JooyaDB19, DBLP:journals/jss/ZhuYZZ21, DBLP:journals/concurrency/OuaredCD22, DBLP:conf/icccnt/KumarMBCA20, DBLP:conf/cluster/IsailaBWKLRH15} transforms the original features into a new set of uncorrelated variables called principal components, which are linear combinations of the original features and are ordered by the amount of variance they explain. \textbf{Graph embedding}~\citep{DBLP:journals/corr/abs-2304-13032, DBLP:journals/pvldb/ZhouSLF20} involves mapping the original complex graph like Abstract
Syntax Tree to a lower-dimensional space, where nodes or edges with similar properties are mapped close to each other.

\subsubsection{Anomaly detection.} Anomaly detection techniques aim at identifying data absences, errors, and outliers within a dataset, ensuring the integrity and reliability of deep learning tasks. Examples include: \textbf{Smoothing}~\citep{DBLP:journals/fgcs/LiLTWHQD19} aims at capturing the underlying trends, patterns, or regularities in the data while suppressing or filtering out the noise. \textbf{Isolation Forest}~\citep{DBLP:journals/tecs/TrajkovicKHZ22} algorithm constructs random decision trees to measure the average number of steps required to isolate an instance, allowing it to identify anomalies as instances that require fewer steps for isolation.

\subsubsection{Feature construction.} Feature construction, or featurization, is the process where raw data are transformed into more meaningful and practical features that capture important patterns or relationships in the data, which are crucial for more accurate modeling. For example,~\citet{DBLP:conf/splc/Acher0LBJKBP22} actively incorporate the number of
``y'' options values as an extra feature for the Linux kernel, because it explicitly captures the domain knowledge between the performance and the number of options. Besides, \textbf{alphanumeric cleaning}~\citep{DBLP:conf/wosp/CengizFAM23} removes non-alphanumeric characters from a text or string, retaining only letters and numbers, which are important to construct a meaningful feature. Moreover, \textbf{discretization}~\citep{DBLP:journals/fgcs/LiLTWHQD19} involves converting continuous options into discrete and representative categories, to save efforts on exploring the huge configuration space. Particularly, the authors discretize the reach rate of a Content Delivery Network by calculating its fifth power, such that the reach rate is projected into several discrete intervals, and each interval is assigned a unique value.

\subsubsection{Others.} Imbalance processing techniques are crucial in dealing with imbalanced datasets, where the number of instances in different classes is significantly uneven, such as \textbf{SMOTE (Synthetic Minority Over-sampling Technique)}~\citep{DBLP:conf/sigsoft/Gong023}, which addresses the class imbalance by generating synthetic samples of the minority class, and \textbf{SMOTUNED (SMOTE TUNED)}~\citep{DBLP:journals/jss/ZhuYZZ21}, an extension of SMOTE that incorporates an additional step of hyperparameter tuning to improve its performance and adaptability to different datasets. Besides, \textbf{Kendall’s rank correlation}~\citep{DBLP:conf/wosp/CengizFAM23}, which quantifies the strength and direction of the association between two ranked variables, is used to remove highly correlated columns and ensure that the predictors are independent or have minimal correlation, allowing for more reliable and meaningful estimation of their individual effects on the response variable.

\begin{figure}[!t]
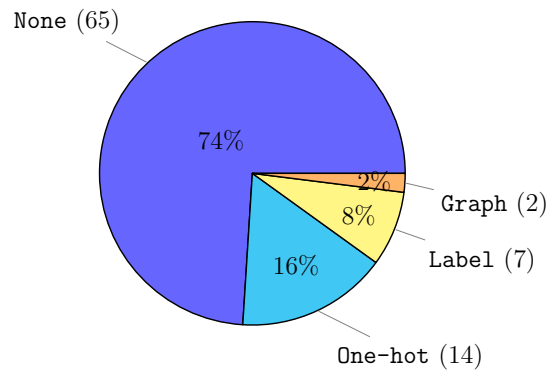

\centering
\includestandalone[width=0.5\columnwidth]{Figures/chap-review/RQ-encoding}
    \caption{Data on encoding schemes from primary studies.}
 \label{fig:encoding}
 \end{figure}

\subsection{How to Encode Performance Data?}
\label{chap-survey:how-to-encode}
After data preprocessing, the configuration data may still remains in the format of natural language, including binary options like ``yes/no'' or ``on/off'', as well as categorical options such as ``low/medium/high quality''. In such cases, it is often essential to apply suitable encoding schemes to further convert the configuration data into a more appropriate format for deep learning. Thereby, this process stands out as a fundamental step alongside other preprocessing methods. 

However, to the best of my knowledge, the justifications for choosing specific encoding schemes are often overlooked, leading to uncertainty regarding which technique is superior and under what conditions it performs best. Therefore, this sub-section seeks to make a comprehensive investigation of techniques employed for data encoding for performance learning, hoping to provide a clear overview of the current literature and enable researchers to make informed decisions.

Upon the summarized results in Figure~\ref{fig:encoding}, it is observed that a substantial portion, specifically 65 studies, opted to utilize default inputs directly from the previous step without applying any explicit encoding schemes. Among the utilized encoding methods, \textbf{one-hot encoding} emerges as the most prevalent, employed by 14 primary studies~\citep{DBLP:journals/taco/WangLWB19, DBLP:conf/kbse/BaoLWF19, DBLP:journals/access/ThraneZC20, DBLP:journals/taco/LiL22, DBLP:conf/icccnt/KumarMBCA20, DBLP:conf/sc/MaratheAJBTKYRG17}. It transforms categorical variables into binary vectors, where each category is represented as a binary feature, enabling deep learning models to appropriately interpret and capture relationships between different categories. On the other hand, \textbf{label encoding}~\citep{chai2023perfsage, DBLP:conf/splc/Acher0LBJKBP22, DBLP:journals/tecs/TrajkovicKHZ22, DBLP:journals/jsa/ZhangLWWZH18, DBLP:journals/taco/WangLWB19, DBLP:conf/kbse/BaoLWF19} assigns a unique numerical label to each category, converting them into integer representations. Label encoding preserves the ordinal relationship between categories but does not introduce additional dimensions like one-hot encoding. Moreover, it is found that 2 studies employed \textbf{graph encoding} methods, leveraging the graph representations in certain datasets, such as the computational graphs of neural networks~\citep{DBLP:conf/pldi/SinghHLS22, DBLP:journals/corr/abs-2304-13032}.

 \begin{figure}[!t]
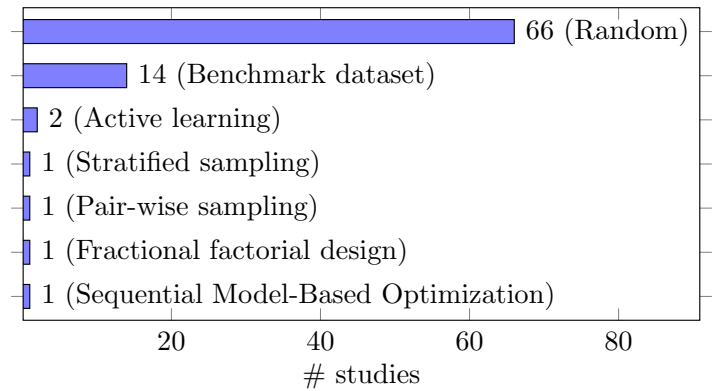

\centering
\includestandalone[width=0.65\columnwidth]{Figures/chap-review/RQ-sampling}
    \caption{Data on sampling methods from primary studies.}
 \label{fig:sampling}
 \end{figure}

\subsection{What Sampling Methods Have Been Used?}
Sampling, which is typically performed to guide the raw data collection, is also a critical step in data preparation, as the quality of the collected samples decides the prediction accuracy of the deep learning models~\citep{DBLP:conf/wosp/PereiraA0J20}. By investigating the various sampling approaches utilized in previous studies, valuable guidance is provided for selecting appropriate sampling strategies for future research and practical applications. 

Figure~\ref{fig:sampling} shows the categories of the sampling methods in the reviewed studies. First, an overwhelming majority of the studies surveyed, specifically 66 out of 85, employed \textbf{random sampling} as a prevalent data sampling technique in their research. Random sampling involves selecting data instances from the original dataset in an unbiased and random manner. The widespread adoption of random sampling highlights its efficiency and effectiveness in creating balanced and diverse datasets~\citep{DBLP:journals/jsa/ZhangLWWZH18, DBLP:conf/icse/GaoGZLY23, DBLP:journals/taco/WangLWB19, DBLP:conf/infocom/LiHZLT20, DBLP:journals/access/ThraneZC20, DBLP:journals/ese/VituiC21a}. 
Then, 14 of them followed a common practice of utilizing the default training and testing samples provided within the \textbf{benchmark} dataset~\citep{DBLP:journals/taco/MoolchandaniKS22, du2013performance, DBLP:conf/wosp/CengizFAM23, DBLP:journals/cee/KundanSA21, ding2021portable}, 
which allows for a direct comparison of their models' performance against the existing results, and 2 studies applied \textbf{active learning} to obtain the samples~\citep{DBLP:journals/corr/abs-2304-13032, DBLP:journals/taco/LiL22}, which seek to select the most informative samples to be measured for training, thereby accelerating the learning process and reducing the costs of data collection. Finally, 1 study adopted the following methods: 
\begin{itemize}
    \item \textbf{Stratified sampling}~\citep{DBLP:journals/pr/TranSWQ20} involves dividing the population into homogeneous subgroups based on specific characteristics and then sampling from each subgroup proportionally to its representation in the overall population.
    \item \textbf{Pair-wise sampling}~\citep{DBLP:journals/tsc/KumaraACMHT23} seek to sample data that achieve maximum coverage of interaction effects between pairs of features while minimizing the number of samples required to prevent biases.
    \item \textbf{Fractional factorial design}~\citep{DBLP:conf/cluster/IsailaBWKLRH15}, which provides a way to select a subset of factor combinations that represent the most important effects.
    \item \textbf{Sequential Model-Based Optimization (SMBO)}~\citep{wang2021morphling} builds a surrogate model capable of estimating uncertainty to effectively guide the search for optimal training configurations.
\end{itemize}

In summary, the answer to RQ1.1 is as follows:
\begin{quotebox}
   \noindent\textbf{RQ1.1:} In the preparation stage of performance learning, it is found that:
   \begin{enumerate}
       \item The default datasets, without any preprocessing, are used by most studies (45 out of 85). Among the rest of the preprocessing methods, normalization is the most popular, used in 32 of them.
       \item A majority of studies, 65 in particular, do not justify their encoding methods, while one-hot encoding is applied in 14 studies, being the most welcomed in the remaining ones.
       \item Random sampling is the most widely employed technique as observed in 66 studies, followed by the default sampling pipeline as in the benchmark in 14 works.
   \end{enumerate}

\end{quotebox}


%% file: Tables/chap-review/RQ-preprocessing_methods.tex
\begin{table}[t!]
\caption{Data on preprocessing methods from primary studies.}
\centering
\scriptsize
\begin{adjustbox}{width=1.1\linewidth,center}

\begin{tabular}{p{2cm}p{3cm}p{0.1cm}p{7cm}}
\toprule
\multicolumn{1}{c}\textbf{{Category}} &
  \multicolumn{1}{c}\textbf{{Examples}} &
  \multicolumn{1}{c}\textbf{{\# studies}} &
  \multicolumn{1}{c}\textbf{{References}} \\ \hline
None & N/A
 &
  45 &
  \citep{DBLP:journals/infsof/WartschinskiNVK22}, \citep{ding2021portable}, \citep{chokri2022performance}, \citep{DBLP:journals/cee/KundanSA21}, \citep{DBLP:journals/concurrency/CiciogluC22}, \citep{DBLP:journals/concurrency/RosarioSZNB23}, \citep{DBLP:journals/fgcs/KousiourisMKGV14}, \citep{DBLP:journals/comcom/AteeqAAK22}, \citep{DBLP:conf/infocom/LiHZLT20}, \citep{DBLP:journals/concurrency/YinH00C23}, \citep{DBLP:journals/access/NadeemAMFA19}, \citep{DBLP:conf/dsrt/DuongZCLZ16}, \citep{DBLP:conf/springsim/LuxWCBLXBBCH18}, \citep{DBLP:journals/tcc/PhamDF20}, \citep{DBLP:conf/ic2e/RahmanL19}, \citep{DBLP:journals/jsa/ChengCWX17}, \citep{DBLP:journals/pvldb/MarcusP19}, \citep{DBLP:journals/taco/LiL22}, \citep{chai2023perfsage}, \citep{du2013performance}, \citep{DBLP:journals/access/WangXTW20}, \citep{DBLP:conf/nsdi/LiangFXYLZYZ23}, \citep{wang2021morphling}, \citep{DBLP:journals/concurrency/FalchE17}, \citep{myung2021machine}, \citep{DBLP:conf/ipps/FalchE15}, \citep{DBLP:conf/IEEEcloud/MarosMSALGHA19}, \citep{DBLP:conf/mascots/KarniavouraM17}, \citep{DBLP:journals/tsc/KumaraACMHT23}, \citep{blott2018finn}, \citep{DBLP:conf/noms/SanzEJ22}, \citep{DBLP:conf/sc/MalikFP09}, \citep{DBLP:conf/wosp/DidonaQRT15}, \citep{DBLP:conf/msr/GongC22}, \citep{DBLP:conf/cf/LiuMCV20}, \citep{DBLP:conf/kbse/ChenHL022}, \citep{DBLP:conf/icse/HaZ19}, \citep{DBLP:conf/spects/KimK17}, \citep{DBLP:conf/sc/MalakarBVMK18}, \citep{DBLP:conf/splc/GhamiziCPT19}, \citep{DBLP:journals/access/LiLSJ20}, \citep{DBLP:journals/tompecs/MakraniSNDSMRH21}, \citep{DBLP:conf/kbse/BaoLWF19}, \citep{said2021accurate}, \citep{DBLP:conf/pldi/SinghHLS22} \\ \hline
Normalization &
  Min-max scaling, z-score, Box-Cox transformation, centering, making unit consistent, standardization &
  32 &
  \citep{DBLP:conf/sigsoft/Gong023}, \citep{DBLP:conf/nsdi/FuGMR21}, \citep{DBLP:journals/pvldb/ZhouSLF20}, \citep{DBLP:journals/taco/MoolchandaniKS22}, \citep{DBLP:journals/jiii/YuGLZIY22}, \citep{DBLP:journals/concurrency/JiZL22}, \citep{DBLP:conf/icse/GaoGZLY23}, \citep{DBLP:journals/jsa/ZhangLWWZH18}, \citep{zain2022software}, \citep{DBLP:journals/tse/ChenB17}, \citep{sabbeh2016performance}, \citep{DBLP:journals/ese/VituiC21a}, \citep{DBLP:journals/pr/TranSWQ20}, \citep{DBLP:journals/tosem/ChengGZ23}, \citep{DBLP:conf/icpp/DouWZC22}, \citep{DBLP:journals/tnse/CaoPC22}, \citep{DBLP:conf/icccnt/KumarMBCA20}, \citep{DBLP:conf/icst/PorresARLT20}, \citep{DBLP:conf/icpp/MadireddyBCLLRS19}, \citep{DBLP:conf/sbac-pad/NemirovskyAMNUC17}, \citep{DBLP:conf/middleware/GrohmannNIKL19}, \citep{DBLP:journals/mam/JooyaDB19}, \citep{DBLP:journals/access/TousiL22}, \citep{DBLP:conf/cluster/IsailaBWKLRH15}, \citep{DBLP:journals/mam/JooyaDB19}, \citep{DBLP:journals/fgcs/LiLTWHQD19}, \citep{DBLP:conf/cluster/IsailaBWKLRH15}, \citep{DBLP:conf/wosp/CengizFAM23}, \citep{DBLP:journals/jss/ZhuYZZ21}, \citep{DBLP:conf/sbac-pad/NemirovskyAMNUC17}, \citep{DBLP:conf/sc/MaratheAJBTKYRG17}, \citep{DBLP:conf/sigmod/ZhangLZLXCXWCLR19} \\ \hline
Regularization &
  Bayesian regularization, L1 regularization, L2 regularization, Laplacian regularization &
  15 &
  \citep{DBLP:conf/middleware/MahgoubWGMGHMGB17}, \citep{DBLP:conf/sigsoft/Gong023}, \citep{DBLP:journals/access/ThraneZC20}, \citep{DBLP:journals/taco/WangLWB19}, \citep{DBLP:conf/iccad/KimMMSR17}, \citep{DBLP:journals/access/ThraneZC20}, \citep{DBLP:journals/jsa/TangLLLZ22}, \citep{DBLP:conf/im/JohnssonMS19}, \citep{DBLP:journals/pvldb/ZhouSLF20}, \citep{DBLP:journals/concurrency/OuaredCD22}, \citep{DBLP:conf/esem/ShuS0X20}, \citep{DBLP:conf/icse/HaZ19}, \citep{DBLP:journals/tosem/ChengGZ23}, \citep{DBLP:journals/access/TousiL22}, \citep{DBLP:conf/ic2e/RahmanL19} \\ \hline
Dimension reduction &
  PCA, graph embedding &
  10 &
  \citep{DBLP:conf/cluster/IsailaBWKLRH15}, \citep{DBLP:conf/middleware/GrohmannNIKL19}, \citep{DBLP:journals/mam/JooyaDB19}, \citep{DBLP:journals/corr/abs-2304-13032}, \citep{DBLP:journals/concurrency/OuaredCD22}, \citep{DBLP:conf/icccnt/KumarMBCA20}, \citep{DBLP:journals/mam/JooyaDB19}, \citep{DBLP:journals/jss/ZhuYZZ21}, \citep{DBLP:conf/sc/MalikFP09}, \citep{DBLP:journals/pvldb/ZhouSLF20} \\ \hline
Anomaly detection &
  Outlier detection, smoothing, isolation forest algorithm &
  5 &
  \citep{DBLP:journals/mam/JooyaDB19}, \citep{DBLP:conf/nsdi/FuGMR21}, \citep{DBLP:conf/wosp/CengizFAM23}, \citep{DBLP:journals/tecs/TrajkovicKHZ22}, \citep{DBLP:journals/fgcs/LiLTWHQD19} \\ \hline
Feature construction &
  Discretization, making columns categorical, featurization, alphanumeric cleaning &
  5 &
  \citep{DBLP:conf/splc/Acher0LBJKBP22}, \citep{DBLP:conf/nsdi/FuGMR21}, \citep{DBLP:conf/wosp/CengizFAM23}, \citep{DBLP:journals/fgcs/LiLTWHQD19}, \citep{DBLP:conf/wosp/CengizFAM23} \\ \hline
Imbalance processing &
  SMOTUNED, SMOTE &
  3 &
  \citep{DBLP:conf/sigsoft/Gong023}, \citep{DBLP:journals/jss/ZhuYZZ21}, \citep{DBLP:journals/fgcs/LiLTWHQD19} \\ \hline
Multicollinearity processing&
  Kendall's rank correlation &
  1 &
  \citep{DBLP:conf/wosp/CengizFAM23, DBLP:conf/splc/Acher0LBJKBP22}

\\ \bottomrule
\end{tabular}

\end{adjustbox}
\label{tb:preprocessing}
\end{table}

%% file: Figures/chap-review/RQ-encoding.tex
\begin{tikzpicture}[scale=1]
\tikzstyle{every node}=[font=\large]
    \pie[
        /tikz/every pin/.style={align=center},
        text=pin,
        rotate=0,
        explode=0
        ]
        {
        74/\texttt{None} (65),
        16/\texttt{One-hot} (14),
        8/\texttt{Label} (7),
        2/\texttt{Graph} (2)
        }

    \end{tikzpicture}

%% file: Figures/chap-review/RQ-sampling.tex
\begin{tikzpicture}
\begin{axis}[
    xbar=0.05cm,
    width=10cm,
    height=5.5cm,
    xmax=90,
    bar width=0.3cm,
 enlarge y limits=0.09,
 enlarge x limits=0.01,
    legend style={at={(1.05,0.7)},anchor=north,legend columns=1,font=\tiny},
    legend cell align={left},
   xlabel={\# studies},
     x label style={at={(0.5,-0.1)},font=\footnotesize},
     nodes near coords,
          point meta=explicit symbolic,
          every node near coord/.append style={anchor=west,font=\footnotesize},
        symbolic y coords={
        SMBO,
        FFD,
        Coverage-based,
        Stratified,
        Active learning,
        Benchmark,
        Random
        },
         y tick label style={font=\footnotesize},
          x tick label style={font=\footnotesize},
          yticklabels={,,},
         ytick=data
    ]

\addplot[fill=blue!50, draw=black] coordinates {(66,Random)[66 (Random)] (14,Benchmark)[14 (Benchmark dataset)] (2,Active learning)[2 (Active learning)]  (1,Stratified)[1 (Stratified sampling)] (1,Coverage-based)[1 (Pair-wise sampling)] (1,FFD)[1 (Fractional factorial design)] (1,SMBO)[1 (Sequential Model-Based Optimization)]};




\end{axis}
\end{tikzpicture}

%% file: Chapter-survey/RQ2.tex
\section{RQ1.2: How to Model Software Performance? }
\label{chap-reivew:rq2}
In the second phase of the workflow in Figure~\ref{fig:pipeline}, the aim is to model the prepared data from the previous phase accurately. This section addresses four key processes in this phase to provide a comprehensive understanding of how deep learning can be effectively built to learn software performance, offering insights into best practices and potential problems in this field.

\subsection{How to Deal with Sparsity and Prevent Overfitting?}
\label{chap-review-subsec:sparsity}

\input{Tables/chap-review/RQ-overfitting}

As large-scale software systems generate vast amounts of data, sparsity arises when valuable information is scarce in features or unevenly distributed across the configuration space. Conversely, overfitting occurs when models excessively learn the noise or specific features of the training data, leading to poor generalization on unseen samples~\citep{DBLP:conf/icse/HaZ19, DBLP:conf/sc/MalikFP09, DBLP:conf/middleware/GrohmannNIKL19, DBLP:journals/corr/abs-2304-13032, DBLP:journals/pvldb/MarcusP19, DBLP:journals/tecs/TrajkovicKHZ22, DBLP:conf/splc/Acher0LBJKBP22, DBLP:conf/esem/ShuS0X20}. Understanding these methods is crucial for developing robust and reliable deep models. Yet, despite the significance of the sparsity problem, there is currently no comprehensive survey available that thoroughly examines this issue. In this sub-section, the investigation focuses on the number of studies that have addressed the sparsity problem and the methods and techniques employed by researchers to tackle these challenges.

Table~\ref{tb:overfitting} presents the taxonomy of methods used to address sparsity and overfitting. Notably, 33 out of 85 studies did not adopt any specific technique to handle sparsity, which is the largest category in the table. Among the others, 24 papers rely on human effort to select the most important features, while automatic methods like regularization (15 studies), dimension reduction (10 studies), and dropout (8 studies) are widely used to eliminate insignificant dimensions and parameters. Then, a range of other methods are utilized, including correlation-based selection (4 studies), tree-based feature selection (4 studies), filter-based feature selection (2 studies), NN-based feature selection (2 studies), wrapper-based feature selection (2 studies), and metaheuristic feature selection (1 study). Furthermore, a study introduces a framework known as `divide-and-learn' specifically designed to address sparsity issues in performance learning. Next, the categories will be introduced in detail. Note that regularization and dimension reduction have been previously discussed in Section~\ref{subsec: preprocess} and will be omitted in this section.

\subsubsection{Dropout.} \textbf{Dropout}~\citep{DBLP:journals/access/ThraneZC20, DBLP:conf/cf/LiuMCV20, DBLP:journals/infsof/WartschinskiNVK22, said2021accurate, zain2022software, DBLP:conf/esem/ShuS0X20, DBLP:journals/fgcs/LiLTWHQD19} approaches work by randomly deactivating a fraction of neurons during the training phase of a neural network. By doing so, dropout prevents the network from relying too heavily on specific neurons, forcing it to learn more robust and generalized features. In addition,~\citep{DBLP:conf/middleware/MahgoubWGMGHMGB17} applied \textbf{ensemble pruning} of networks to prune the top 30\% of the ensemble networks to prevent overfitting.

\subsubsection{Correlation-based feature selection.}
Correlation-based feature selection techniques such as \textbf{correlation analysis}~\citep{DBLP:journals/ese/VituiC21a, DBLP:conf/sc/MalikFP09, DBLP:journals/tecs/TrajkovicKHZ22, DBLP:journals/jss/ZhuYZZ21} leverage statistical measures to determine the strength of relationships between variables, which enable the selection of features with strong correlations, reducing sparsity and enhancing the model's ability to generalize and mitigate overfitting.

\subsubsection{Tree-based feature selection.}
Tree-based feature selection techniques inherently enable analyzing the structure and splits of the trees, which can determine the most informative features for prediction. \textbf{Extra trees regression}~\citep{DBLP:conf/icpp/MadireddyBCLLRS19} involves constructing multiple decision trees with randomized splits to estimate feature importance. Moreover,~\citep{DBLP:conf/middleware/GrohmannNIKL19} utilize the ensemble nature of \textbf{random forest} to rank features based on their contribution to the model's predictive power.

\subsubsection{Filter-based feature selection.}
Filter-based feature selection techniques evaluate each feature independently of the target variable and rank them based on predefined criteria. An example is the \textbf{Minimum Redundancy Maximum Relevance Feature Selection (mRMR)}~\citep{DBLP:journals/jsa/ZhangLWWZH18}, which seeks a balance between selecting features that have high relevance to the target variable while minimizing redundancy among selected features.

\subsubsection{NN-based feature selection.}
NN-based feature selection applies neural network layers to identify relevant features, which have the inherent ability to learn and extract informative features from high-level representations. For example, \textbf{CNN layers}~\citep{DBLP:journals/access/ThraneZC20} (Convolutional Neural Network) are commonly used in image-related tasks to extract local patterns and spatial hierarchies, and \textbf{GNN layers}~\citep{chai2023perfsage} (Graph Neural Networks) are particularly effective for structured data represented as graphs, which analyze relationships and dependencies among nodes in a graph to capture important patterns and interactions.

\subsubsection{Wrapper-based feature selection.}~\citep{DBLP:journals/access/TousiL22} uses \textbf{Recursive Feature Elimination} (RFE) to iteratively eliminate features with the least importance, while evaluating the model's performance using cross-validation. \textbf{Hybrid dual-learner} is proposed by~\citep{DBLP:journals/tse/ChenB17}, which combines the outputs of multiple learning algorithms to guide the feature selection process.


\subsubsection{Metaheuristic feature selection.} 
\citep{DBLP:journals/jss/ZhuYZZ21} proposed to combine Whale optimization algorithm (WOA) and Simulated annealing (SA) to construct an Enhanced Metaheuristic search-based feature selection algorithm named \textbf{EMWS}.

\begin{figure}[!t]
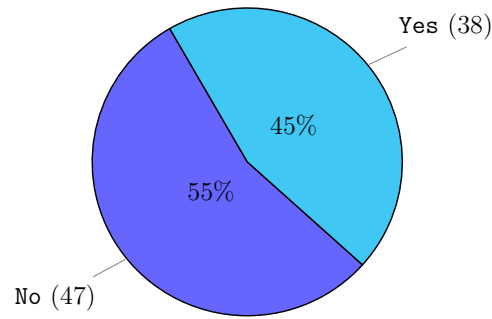

  \centering
    \centering
\includestandalone[width=0.45\columnwidth]{Figures/chap-review/reasons}
\caption{Data on DL model justification from primary studies.}
 \label{fig:reasons}
\end{figure}

\subsection{What Deep Learning Models have been Utilized?}
\label{chap-survey:what-deep-learning}

\input{Tables/chap-review/RQ-learning_model}

Reviewing the deep learning methods employed to model software performance is critical. Firstly, it allows researchers to understand the state-of-the-art techniques and advancements in the field, enabling them to build upon existing knowledge. Secondly, it facilitates the identification of the most effective models for specific tasks, saving time and resources by avoiding redundant experimentation. Lastly, reviewing the models helps researchers identify potential limitations or areas for improvement, driving further innovation in deep learning methodologies~\citep{DBLP:journals/tosem/LiuGXLGY22, DBLP:journals/tosem/WatsonCNMP22}.

The deep learning models applied in the reviewed studies are shown in Table~\ref{tb:learning}. In particular, Feedforward neural network is supported by most studies (61 out of 85). Additionally, recurrent neural networks and convolutional neural networks have each been studied in 9 research papers, followed by graph neural networks in 8 other papers. Graph Neural Networks have 5 supporting studies. Finally, adversarial Learning approaches are explored in 3 studies, and meta-learning and reinforcement learning are each used in only 1 study. In the following, the deep learning models will be introduced in detail.

\subsubsection{Feedforward neural network (FNN)} FNN refers to a general type of neural network where information flows through the network in a single direction, from the input layer to the output layer, without any loops or feedback connections. Among these, the most prevalent approach is Multilayer Perceptron (MLP) in 54 studies. MLP is a fundamental neural network consisting of interconnected neurons organized in multiple layers. These neurons apply activation functions to the weighted sum of their inputs, introducing non-linear transformations to the data.~\citet{DBLP:conf/icse/HaZ19} propose the utilization of regularization techniques in regularized Deep Neural Networks (rDNN), such that additional penalties are introduced during training to remove insignificant features, control feature sparsity of the model and thereby prevent overfitting.~\citet{DBLP:conf/sigsoft/Gong023} combines rDNN with the divide-and-learn framework to further address the sample sparsity issues. In addition, Kernel Extreme Learning Machines (KELM) are also used in two studies, aiming at simplifying the training process of neural networks, which randomly initializes the weights between the input and hidden layers, then applies a kernel function to obtain the transformed feature representation, and analytically determines the weights between the hidden and output layers based on the transformed features. A most recent work~\citep{DBLP:journals/tosem/ChengGZ23} employs a Hierarchical Interaction Neural Network (HINN) that leverages a hierarchical structure for performance learning, where lower-level layers typically capture low-level options, and higher-level layers learn to combine these low-level options into more complex representations. Further, dynamic neural networks and radial basis function neural networks are each utilized in one study.

\subsubsection{Recurrent Neural Networks (RNN)}
However, FNNs cannot handle sequential or temporal data effectively in many real-world tasks, such as natural language processing and time series prediction. To mitigate this, RNN are specifically designed to maintain an internal memory or hidden state that persists across time steps, allowing them to capture information from past inputs. Among our review scope, RNNs have been employed in nine works. For example, Long Short-Term Memory (LSTM) is a dedicated version of the recurrent neural network with internal memory cells, which allows them to selectively retain or forget information based on the input and previous context by using specialized units called gates, and therefore overcome the limitations of traditional RNNs in capturing and remembering long-term dependencies in sequential data. \textcolor{black}{For example,~\citet{DBLP:journals/jsa/TangLLLZ22} employ LSTM to tune the performance of file systems like \textsc{Ext4}, \textsc{F2FS}, and \textsc{PMFS}, where the operations and workloads are dynamically changing over time and it is crucial to apply time-related deep learning models to model the temporal data.}

\subsubsection{Convolutional Neural Networks (CNN)} CNNs leverage the hierarchical feature learning capability of the convolutional layers, allowing them to automatically learn and extract meaningful representations from the input data, which are especially effective in processing grid-like structured data, such as images and videos. Notably, a particular variant of CNNs, known as Residual Networks (ResNets), has been exploited by~\citet{DBLP:conf/wosp/CengizFAM23}, which incorporate novel architectural designs with residual connections to address the issue of vanishing gradients during training. \textcolor{black}{As an example,~\citet{DBLP:conf/cf/LiuMCV20} leverage the ability of CNN that can disclose the hidden and complex correlations among different options to predict the performance of high-performance computing systems like RISC-V.}

\subsubsection{Graph Neural Networks (GNN)} GNNs and their variant DAG-transformer have been used in six studies, which utilize graph convolutions and message-passing techniques to capture complex relationships and dependencies within interconnected entities, e.g., graphs, which are mathematical structures composed of nodes connected by edges. \textcolor{black}{For example,~\citet{chai2023perfsage} utilize GNN to predict the inference latency, energy, and memory footprint of DNN inference, because it can directly process arbitrary DNN TFlite graphs and has good generalizability.}

\subsubsection{Generative Adversarial Networks (GAN)} GAN is an \textbf{adversarial learning} approach that is primarily designed for generative modeling tasks, aiming at learning the underlying distribution of the training data and generating new samples that resemble the training data. It consists of two neural networks: a generator network and a discriminator network, which are trained in an adversarial manner, with the generator trying to produce more realistic samples to deceive the discriminator, and the discriminator trying to become better at distinguishing real from generated samples. \textcolor{black}{For instance, GAN is used by~\citet{DBLP:conf/kbse/BaoLWF19} to automatically generate configuration samples for different software systems like \textsc{Kafka}, \textsc{Spark}, and \textsc{MySQL} to save performance measurement costs.}

\subsubsection{Others.} 
\textbf{Meta-learning} and \textbf{reinforcement learning} methods are explored in one study, respectively. Specifically, Model-Agnostic Meta-Learning (MAML) pre-trains meta-model's parameters using the known software environments to facilitate fast adaptation to new environments, allowing for efficient generalization and rapid adaptation. On the other hand, reinforcement learning approaches like Q-learning networks are used to approximate the action-value function (Q-function), which represents the expected cumulative reward for taking a specific action from a given state and following a particular policy. \textcolor{black}{As an example, \citet{DBLP:journals/concurrency/YinH00C23} utilizes a Q-network to deal with the dynamic changes of workloads while predicting the configuration performance for multitier web systems like \textsc{RUBiS}.}

In addition to the statistics of learning methods, the survey also investigates whether the studies have provided an explanation for their choice of a specific learning method, which is captured in Figure~\ref{fig:reasons}. 

Among the 85 studies analyzed, it was found that only 37 of them provided explicit reasons for the utilization of a specific learning model, accounting for approximately 44\% of the total studies examined. On the other hand, a significant portion of the studies, 47 in total, did not offer any justification or rationale behind their choice of learning model.

\begin{figure}[!t]
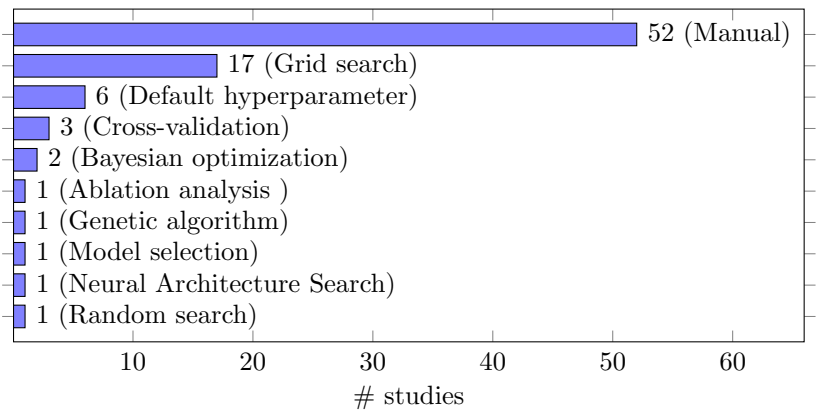

  \centering
  \begin{minipage}{0.75\textwidth}
    \centering
\includestandalone[width=1\columnwidth]{Figures/chap-review/RQ-hyperparameter}
  \caption{Data on hyperparameter tuning methods from primary studies.}
 \label{fig:hyperparameter}
  \end{minipage}
\end{figure}

\subsection{How to Tune Hyperparameters?}
\label{chap-review-subsec:hyperparameter}

Hyperparameter tuning plays a critical role in the performance and generalization ability of deep learning models. However, determining the optimal values for hyperparameters is a challenging task as the hyperparameter space is often huge, and it will be time-consuming. In this sub-section, the exploration focuses on the hyperparameter tuning methods that have been employed for deep learning models.

Figure~\ref{fig:hyperparameter} presents an analysis of the hyperparameter tuning methods utilized in the examined studies. It is most worth noting that a great portion of 51 studies rely on human experts and domain knowledge to tune the hyperparameters. \textbf{Grid search}, which exhaustively tests over a predefined grid of values to find the best configuration, is the second widely used technique and is applied in 17 studies~\citep{DBLP:conf/icse/HaZ19, DBLP:journals/corr/abs-2304-13032, DBLP:journals/tosem/ChengGZ23, DBLP:conf/pldi/SinghHLS22, DBLP:conf/mascots/KarniavouraM17, sabbeh2016performance}. 
In addition, 6 studies rely on \textbf{default} hyperparameter settings~\citep{DBLP:conf/icpp/MadireddyBCLLRS19, DBLP:journals/access/LiLSJ20, DBLP:journals/tecs/TrajkovicKHZ22, DBLP:conf/cluster/IsailaBWKLRH15, DBLP:conf/wosp/DidonaQRT15, said2021accurate}, where models are trained using pre-defined configurations without any explicit tuning. \textbf{Cross-validation}, which iteratively divides the dataset into multiple subsets for training, validation, and testing, is utilized in 3 studies~\citep{DBLP:conf/nsdi/FuGMR21, DBLP:conf/IEEEcloud/MarosMSALGHA19, DBLP:journals/access/TousiL22}. Then, \textbf{Bayesian optimization}, a method that uses probability models to efficiently search for optimal hyperparameters by considering the performance of previously evaluated configurations, is employed in 2 studies~\citep{DBLP:journals/pr/TranSWQ20, zain2022software}. Other less frequently used methods include ablation analysis~\citep{DBLP:journals/jss/ZhuYZZ21}, genetic algorithm~\citep{DBLP:conf/splc/GhamiziCPT19}, model selection~\citep{DBLP:journals/taco/WangLWB19}, neural architecture search~\citep{DBLP:journals/taco/LiL22}, and random search~\citep{DBLP:journals/access/ThraneZC20}, where each method is employed in 1 study.

\subsection{Optimization Methods and Activation Functions}
\label{chap-survey:optimization-and-activation}

\begin{figure}[!t]
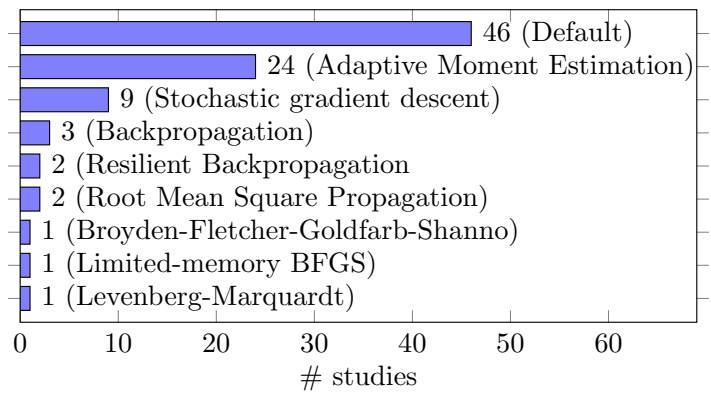

\centering
\includestandalone[width=0.65\columnwidth]{Figures/chap-review/RQ-optimization}
  \caption{Data on optimization methods from primary studies.}
 \label{subfig:optimization}
 \end{figure}

The effective training of deep learning models heavily relies on the choice of optimization methods and activation functions, as the former determines how the model's parameters are updated during training, while activation functions introduce non-linearity and enable complex representations within the neural network. In this sub-section, the question of what optimization methods and activation functions have been employed in the context of deep learning-based software performance prediction is examined in detail.

Figure~\ref{subfig:optimization} provides data on the optimization methods employed in the examined studies. A notable observation is that: a significant number of studies, 46 in total, {omitted} mentioning the specific optimization method employed both in their manuscript and the public repository (if available)~\citep{DBLP:conf/ic2e/RahmanL19, DBLP:journals/tsc/KumaraACMHT23, DBLP:conf/nsdi/LiangFXYLZYZ23, DBLP:conf/middleware/GrohmannNIKL19, DBLP:conf/splc/Acher0LBJKBP22, DBLP:journals/mam/JooyaDB19, DBLP:journals/concurrency/RosarioSZNB23}. Clearly, it is not possible to conjecture the actual optimizer simply according to their deep learning model, therefore, it is assumed these studies employ the \textbf{default} optimizer provided by their implementation libraries. 
Despite this omission, \textbf{Adaptive Moment Estimation (Adam)} optimizer emerges as the most prevalent optimization method, being utilized in 24 studies~\citep{zain2022software, DBLP:journals/access/TousiL22, DBLP:conf/icse/HaZ19, DBLP:journals/tosem/ChengGZ23, DBLP:conf/noms/SanzEJ22, DBLP:conf/icpp/MadireddyBCLLRS19}, 
which is an adaptive optimization algorithm that maintains adaptive learning rates based on the first and second moments of the gradients. \textbf{Stochastic gradient descent (SGD)}, which is an iterative optimization algorithm that updates the model parameters by computing gradients on small randomly sampled subsets of the training data, stands as the second most frequently used method, appearing in 9 studies~\citep{DBLP:conf/sbac-pad/NemirovskyAMNUC17, DBLP:journals/comcom/AteeqAAK22, DBLP:conf/IEEEcloud/MarosMSALGHA19, myung2021machine}.

 \begin{figure}[!t]
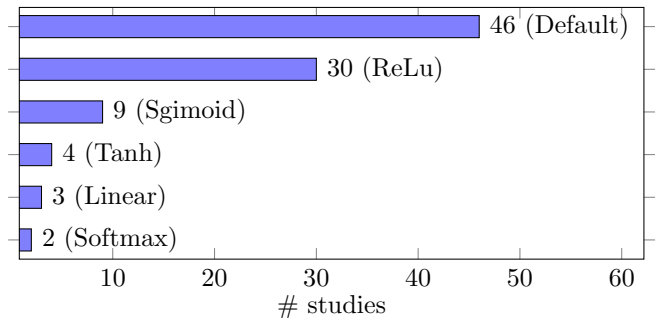

\centering
\includestandalone[width=0.6\columnwidth]{Figures/chap-review/RQ-activation}
  \caption{Data on activation functions from primary studies.}
 \label{subfig:activation}
 \end{figure}
 
In Figure~\ref{subfig:activation}, an overview of the activation functions employed in the surveyed studies is provided. Similar to the data of optimization methods, a significant number of studies, 46 out of 85, have neither explicitly reported the activation function employed in their paper nor provided the source codes, therefore, it is presumed that they utilize the \textbf{default} activation function in the implementation library. Despite this omission, \textbf{Rectified Linear Unit (ReLU)} emerges as the most prevalent, utilized in 30 studies~\citep{DBLP:conf/springsim/LuxWCBLXBBCH18, DBLP:journals/comcom/AteeqAAK22, DBLP:conf/sbac-pad/NemirovskyAMNUC17, DBLP:journals/jss/ZhuYZZ21, DBLP:journals/concurrency/RosarioSZNB23}, 
which is known for its simplicity and effectiveness in handling non-linearities. \textbf{Sigmoid} activation function follows as the second most commonly used, appearing in 9 studies~\citep{zain2022software, DBLP:journals/concurrency/FalchE17, DBLP:conf/dsrt/DuongZCLZ16, DBLP:journals/tse/ChenB17, du2013performance}, 
and it can map the output of a neural network to a probability-like value between 0 and 1. In addition, the \textbf{hyperbolic tangent (tanh)} activation function, which is a smooth, symmetric, and nonlinear function that maps the input to a continuous range between -1 and 1~\citep{DBLP:journals/access/TousiL22, DBLP:conf/nsdi/FuGMR21, DBLP:conf/sc/MalikFP09, DBLP:journals/taco/WangLWB19}, and \textbf{linear} activation functions are employed in 4 and 3 studies, respectively~\citep{DBLP:journals/concurrency/JiZL22, DBLP:journals/tsc/KumaraACMHT23, DBLP:conf/middleware/GrohmannNIKL19}, while \textbf{softmax} is used in 2 studies~\citep{DBLP:conf/splc/GhamiziCPT19, DBLP:conf/middleware/GrohmannNIKL19}, which ensures the output values lie in the range of 0 to 1 and sum up to 1, representing valid probabilities.

In conclusion, RQ1.2 can be answered by:
\begin{quotebox}
   \noindent\textbf{RQ1.2:} In the second step of performance learning, which involves model training, several important findings can be highlighted: 
   \begin{enumerate}
       \item The largest number of studies fail to handle the problem of sparsity and overfitting (33 out of 85). Among the studies that deal with this challenge, 24 of them rely on manual feature selection, which is the largest.
       \item Various deep learning models have been leveraged for performance modeling, while the most common one is the multilayer perception, chosen by 53 out of 85 works. However, it is worth noting that nearly 55\% of the primary studies do not provide explicit justifications for their choice of deep learning model.
       \item Manual hyperparameter tuning is preferred by the majority of studies (52 out of 85), followed by grid search (17 studies).
       \item Despite 46 of the primary studies omit the justifications of optimization and activation functions, it is observed that a number of different techniques, where the  Adam optimizer and ReLu activation function stand out as the most commonly applied methods, utilized in 24 and 30 papers, respectively.
   \end{enumerate}

\end{quotebox}

%% file: Tables/chap-review/RQ-overfitting.tex
\begin{table}[t!]
\caption{Data on sparsity handling methods from primary studies.}
\centering
\scriptsize
\begin{adjustbox}{width=1.1\linewidth,center}

\begin{tabular}{p{2cm}p{3cm}p{0.1cm}p{7cm}}
\toprule
\multicolumn{1}{l}{\textbf{Category}} &
  \multicolumn{1}{l}{\textbf{Examples}} &
  \multicolumn{1}{l}{\textbf{\# studies}} &
  \multicolumn{1}{l}{\textbf{References}} \\ \hline
None &
  N/A &
  33 &
  \citep{chokri2022performance}, \citep{DBLP:conf/dsrt/DuongZCLZ16}, \citep{DBLP:conf/msr/GongC22}, \citep{DBLP:conf/noms/SanzEJ22}, \citep{DBLP:conf/springsim/LuxWCBLXBBCH18}, \citep{DBLP:conf/wosp/CengizFAM23}, \citep{DBLP:journals/access/NadeemAMFA19}, \citep{DBLP:journals/cee/KundanSA21}, \citep{DBLP:journals/concurrency/CiciogluC22}, \citep{DBLP:journals/concurrency/JiZL22}, \citep{DBLP:journals/concurrency/RosarioSZNB23}, \citep{DBLP:journals/fgcs/KousiourisMKGV14}, \citep{DBLP:journals/jsa/ChengCWX17}, \citep{DBLP:journals/pvldb/MarcusP19}, \citep{DBLP:journals/taco/LiL22}, \citep{ding2021portable}, \citep{DBLP:conf/icst/PorresARLT20}, \citep{DBLP:conf/kbse/BaoLWF19}, \citep{DBLP:conf/kbse/ChenHL022}, \citep{DBLP:conf/mascots/KarniavouraM17}, \citep{DBLP:conf/nsdi/LiangFXYLZYZ23}, \citep{DBLP:conf/pldi/SinghHLS22}, \citep{DBLP:conf/sbac-pad/NemirovskyAMNUC17}, \citep{DBLP:conf/wosp/DidonaQRT15}, \citep{DBLP:journals/corr/abs-2304-13032}, \citep{wang2021morphling}, \citep{ DBLP:conf/icse/GaoGZLY23}, \citep{DBLP:conf/infocom/LiHZLT20}, \citep{DBLP:journals/taco/MoolchandaniKS22}, \citep{DBLP:journals/taco/WangLWB19}, \citep{du2013performance}, \citep{DBLP:journals/concurrency/YinH00C23}, \citep{DBLP:conf/sigmod/ZhangLZLXCXWCLR19} \\ \hline
Manual feature selection &
  N/A &
  24 &
  \citep{DBLP:conf/sc/MalakarBVMK18}, \citep{DBLP:journals/access/LiLSJ20}, \citep{DBLP:journals/tompecs/MakraniSNDSMRH21}, \citep{DBLP:journals/jiii/YuGLZIY22}, \citep{blott2018finn}, \citep{DBLP:conf/cf/LiuMCV20}, \citep{DBLP:conf/icpp/DouWZC22}, \citep{DBLP:conf/IEEEcloud/MarosMSALGHA19}, \citep{DBLP:conf/ipps/FalchE15}, \citep{DBLP:conf/nsdi/FuGMR21}, \citep{DBLP:conf/sc/MaratheAJBTKYRG17}, \citep{DBLP:conf/spects/KimK17}, \citep{DBLP:journals/access/WangXTW20}, \citep{DBLP:journals/comcom/AteeqAAK22}, \citep{DBLP:journals/concurrency/FalchE17}, \citep{DBLP:journals/jsa/TangLLLZ22}, \citep{DBLP:journals/pr/TranSWQ20}, \citep{DBLP:journals/tcc/PhamDF20}, \citep{DBLP:journals/tnse/CaoPC22}, \citep{myung2021machine}, \citep{sabbeh2016performance}, \citep{DBLP:journals/fgcs/LiLTWHQD19}, \citep{DBLP:conf/splc/GhamiziCPT19}, \citep{DBLP:journals/tsc/KumaraACMHT23} \\ \hline
Regularization &
  Bayesian regularization, L1 regularization, L2 regularization, Laplacian regularization &
  15 &
  \citep{DBLP:conf/middleware/MahgoubWGMGHMGB17}, \citep{DBLP:conf/sigsoft/Gong023}, \citep{DBLP:journals/access/ThraneZC20}, \citep{DBLP:journals/taco/WangLWB19}, \citep{DBLP:conf/iccad/KimMMSR17}, \citep{DBLP:journals/access/ThraneZC20}, \citep{DBLP:journals/jsa/TangLLLZ22}, \citep{DBLP:conf/im/JohnssonMS19}, \citep{DBLP:journals/pvldb/ZhouSLF20}, \citep{DBLP:journals/concurrency/OuaredCD22}, \citep{DBLP:conf/esem/ShuS0X20}, \citep{DBLP:conf/icse/HaZ19}, \citep{DBLP:journals/tosem/ChengGZ23}, \citep{DBLP:journals/access/TousiL22}, \citep{DBLP:conf/ic2e/RahmanL19} \\ \hline
Dimension reduction &
  PCA, CCA, Embedding &
  10 &
  \citep{DBLP:conf/cluster/IsailaBWKLRH15}, \citep{DBLP:conf/middleware/GrohmannNIKL19}, \citep{DBLP:journals/mam/JooyaDB19}, \citep{DBLP:journals/corr/abs-2304-13032}, \citep{DBLP:journals/concurrency/OuaredCD22}, \citep{DBLP:conf/icccnt/KumarMBCA20}, \citep{DBLP:journals/mam/JooyaDB19}, \citep{DBLP:journals/jss/ZhuYZZ21}, \citep{DBLP:conf/sc/MalikFP09}, \citep{DBLP:journals/pvldb/ZhouSLF20} \\ \hline
Dropout &
  Network pruning, Network dropout &
  8 &
  \citep{DBLP:journals/access/ThraneZC20}, \citep{DBLP:conf/cf/LiuMCV20}, \citep{DBLP:journals/infsof/WartschinskiNVK22}, \citep{said2021accurate}, \citep{zain2022software}, \citep{DBLP:conf/esem/ShuS0X20}, \citep{DBLP:journals/fgcs/LiLTWHQD19}, \citep{DBLP:conf/middleware/MahgoubWGMGHMGB17} \\ \hline
Correlation-based feature selection &
  Correlation analysis, Factor analysis &
  4 &
  \citep{DBLP:journals/ese/VituiC21a}, \citep{DBLP:conf/sc/MalikFP09}, \citep{DBLP:journals/tecs/TrajkovicKHZ22}, \citep{DBLP:journals/jss/ZhuYZZ21} \\ \hline
Tree-based feature selection &
  Extra trees, Random forest &
  4 &
  \citep{DBLP:conf/icpp/MadireddyBCLLRS19}, \citep{DBLP:conf/middleware/GrohmannNIKL19}, \citep{DBLP:conf/im/JohnssonMS19}, \citep{DBLP:conf/splc/Acher0LBJKBP22} \\ \hline
Filter-based feature selection &
  Minimum Redundancy Maximum Relevance Feature Selection (mRMR) &
  2 &
  \citep{DBLP:journals/jsa/ZhangLWWZH18}, \citep{DBLP:journals/concurrency/OuaredCD22} \\ \hline
NN-based feature selection &
  CNN layers, GNN layers &
  2 &
  \citep{DBLP:journals/access/ThraneZC20}, \citep{chai2023perfsage} \\ \hline
Wrapper-based feature selection &
  Recursive Feature Elimination (RFE), Hybrid dual-learner &
  2 &
  \citep{DBLP:journals/access/TousiL22}, \citep{DBLP:journals/tse/ChenB17} \\ \hline
Metaheuristic feature selection &
  EMWS &
  1 &
  \citep{DBLP:journals/jss/ZhuYZZ21}

\\ \bottomrule
\end{tabular}

\end{adjustbox}
\label{tb:overfitting}
\end{table}

%% file: Figures/chap-review/reasons.tex
\begin{tikzpicture}[scale=1]
\tikzstyle{every node}=[font=\large]
    \pie[
        /tikz/every pin/.style={align=center},
        text=pin,
        rotate=120,
        explode=0, 
        ]
        {
        55/\texttt{No} (47),
        45/\texttt{Yes} (38)
        }

    \end{tikzpicture}

%% file: Tables/chap-review/RQ-learning_model.tex
\begin{table}[t!]
  \caption{Data on deep learning models from primary studies.}
\centering
\footnotesize
\begin{adjustbox}{width=\linewidth,center}
\begin{tabular}{llp{4cm}lp{7cm}}
\toprule

\textbf{Category} & \textbf{Total \#} & \textbf{Example} & \textbf{\# Studies} & \textbf{References} \\ \hline
\multirow{6}{*}{Feedforward neural network} & \multirow{6}{*}{61} & Multilayer Perceptron & 54 & \citep{DBLP:conf/sbac-pad/NemirovskyAMNUC17}, \citep{DBLP:journals/tse/ChenB17}, \citep{du2013performance}, \citep{DBLP:conf/dsrt/DuongZCLZ16}, \citep{DBLP:journals/fgcs/KousiourisMKGV14}, \citep{DBLP:conf/ipps/FalchE15}, \citep{sabbeh2016performance}, \citep{DBLP:journals/concurrency/OuaredCD22}, \citep{DBLP:journals/concurrency/FalchE17}, \citep{DBLP:journals/concurrency/JiZL22}, \citep{DBLP:conf/middleware/MahgoubWGMGHMGB17}, \citep{DBLP:journals/access/ThraneZC20}, \citep{DBLP:conf/ic2e/RahmanL19}, \citep{DBLP:journals/taco/WangLWB19}, \citep{DBLP:journals/comcom/AteeqAAK22}, \citep{chokri2022performance}, \citep{DBLP:conf/wosp/DidonaQRT15}, \citep{DBLP:journals/tecs/TrajkovicKHZ22}, \citep{DBLP:journals/taco/MoolchandaniKS22}, \citep{DBLP:journals/cee/KundanSA21}, \citep{ding2021portable}, \citep{DBLP:journals/tsc/KumaraACMHT23}, \citep{DBLP:conf/springsim/LuxWCBLXBBCH18}, \citep{DBLP:journals/pvldb/ZhouSLF20}, \citep{DBLP:journals/tcc/PhamDF20}, \citep{DBLP:conf/wosp/CengizFAM23}, \citep{DBLP:journals/tnse/CaoPC22}, \citep{DBLP:journals/tompecs/MakraniSNDSMRH21}, \citep{DBLP:conf/sc/MalikFP09}, \citep{myung2021machine}, \citep{DBLP:conf/nsdi/LiangFXYLZYZ23}, \citep{DBLP:journals/ese/VituiC21a}, \citep{DBLP:conf/icst/PorresARLT20}, \citep{DBLP:journals/pr/TranSWQ20}, \citep{DBLP:journals/access/TousiL22}, \citep{DBLP:conf/im/JohnssonMS19}, \citep{DBLP:conf/noms/SanzEJ22}, \citep{DBLP:conf/icpp/DouWZC22}, \citep{DBLP:conf/nsdi/FuGMR21}, \citep{DBLP:conf/sc/MaratheAJBTKYRG17}, \citep{DBLP:conf/spects/KimK17}, \citep{DBLP:conf/cluster/IsailaBWKLRH15}, \citep{DBLP:conf/sc/MalakarBVMK18}, \citep{DBLP:conf/msr/GongC22}, \citep{DBLP:conf/icccnt/KumarMBCA20}, \citep{DBLP:conf/splc/Acher0LBJKBP22}, \citep{DBLP:conf/middleware/GrohmannNIKL19}, \citep{DBLP:conf/iccad/KimMMSR17}, \citep{DBLP:journals/taco/LiL22}, \citep{DBLP:journals/fgcs/LiLTWHQD19},  \citep{DBLP:journals/jsa/ChengCWX17}, \citep{DBLP:journals/pvldb/MarcusP19}, \citep{DBLP:journals/mam/JooyaDB19}, \citep{DBLP:conf/IEEEcloud/MarosMSALGHA19} \\ \cline{3-5} 
 &  & Regularized DNN & 2 & \citep{DBLP:conf/icse/HaZ19}, \citep{DBLP:conf/sigsoft/Gong023} \\ \cline{3-5} 
 &  & Kernel extreme learning machines & 2 & \citep{DBLP:journals/jss/ZhuYZZ21}, \citep{DBLP:journals/access/WangXTW20} \\ \cline{3-5} 
 &  & Hierarchical interaction neural network & 1 & \citep{DBLP:journals/tosem/ChengGZ23} \\ \cline{3-5} 
 &  & Dynamic neural networks & 1 & \citep{said2021accurate} \\ \cline{3-5} 
 &  & Radial basis function neural network & 1 & \citep{DBLP:journals/access/NadeemAMFA19} \\ \hline
Recurrent neural network & 9 & Long short-term memory & 9 & \citep{DBLP:journals/jsa/TangLLLZ22}, \citep{DBLP:journals/infsof/WartschinskiNVK22}, \citep{DBLP:conf/mascots/KarniavouraM17}, \citep{DBLP:journals/concurrency/CiciogluC22}, \citep{DBLP:conf/icpp/MadireddyBCLLRS19}, \citep{DBLP:conf/infocom/LiHZLT20}, \citep{DBLP:journals/fgcs/LiLTWHQD19}, \citep{DBLP:journals/ese/VituiC21a}, \citep{DBLP:journals/jsa/ZhangLWWZH18} \\ \hline
\multirow{2}{*}{Convolutional neural network} & \multirow{2}{*}{9} & CNN & 8 & \citep{zain2022software}, \citep{DBLP:conf/cf/LiuMCV20}, \citep{blott2018finn}, \citep{DBLP:conf/kbse/ChenHL022}, \citep{DBLP:conf/splc/GhamiziCPT19}, \citep{DBLP:journals/jss/ZhuYZZ21}, \citep{DBLP:conf/wosp/CengizFAM23}, \citep{DBLP:journals/ese/VituiC21a} \\ \cline{3-5} 
 &  & Residual neural network & 1 & \citep{DBLP:conf/wosp/CengizFAM23} \\ \hline
\multirow{2}{*}{Graph neural network} & \multirow{2}{*}{6} & GNN & 5 & \citep{DBLP:journals/concurrency/RosarioSZNB23}, \citep{DBLP:conf/icse/GaoGZLY23}, \citep{chai2023perfsage}, \citep{DBLP:conf/pldi/SinghHLS22}, \citep{DBLP:journals/corr/abs-2304-13032} \\ \cline{3-5} 
 &  & DAG-transformer & 1 & \citep{DBLP:journals/jiii/YuGLZIY22} \\ \hline
Adversarial learning & 3 & Generative adversarial network & 3 & \citep{DBLP:conf/esem/ShuS0X20}, \citep{DBLP:journals/access/LiLSJ20}, \citep{DBLP:conf/kbse/BaoLWF19} \\ \hline
Meta-learning & 1 & Model-Agnostic Meta-Learning & 1 & \citep{wang2021morphling} \\ \hline
Reinforcement learning & 1 & Q-learning network & 1 & \citep{DBLP:journals/concurrency/YinH00C23}
\\ \bottomrule
\end{tabular}

\end{adjustbox}
\label{tb:learning}
\end{table}

%% file: Figures/chap-review/RQ-hyperparameter.tex
\begin{tikzpicture}
\begin{axis}[
    xbar=0.05cm,
    width=12cm,
    height=6cm,
    xmax=65,
    bar width=0.3cm,
 enlarge y limits=0.09,
 enlarge x limits=0.015,
    legend style={at={(1.05,0.7)},anchor=north,legend columns=1,font=\tiny},
    legend cell align={left},
   xlabel={\# studies},
     x label style={at={(0.5,-0.1)},font=\footnotesize},
     nodes near coords,
          point meta=explicit symbolic,
          every node near coord/.append style={anchor=west,font=\footnotesize},
        symbolic y coords={random,NAS,MS,GA,ablation,BO,cross,default,grid, manual},
         y tick label style={font=\footnotesize},
          x tick label style={font=\footnotesize},
          yticklabels={,,},
         ytick=data
    ]

\addplot[fill=blue!50, draw=black] coordinates {(52,manual)[52 (Manual)] (17,grid)[17 (Grid search)] (6,default)[6 (Default hyperparameter)] (3,cross)[3 (Cross-validation)] (2,BO)[2 (Bayesian optimization)] (1,ablation)[1 (Ablation analysis
)] (1,GA)[1 (Genetic algorithm)] (1,MS)[1 (Model selection)] (1,NAS)[1 (Neural Architecture Search)] (1,random)[1 (Random search)]};




\end{axis}
\end{tikzpicture}

%% file: Figures/chap-review/RQ-optimization.tex
\begin{tikzpicture}
\begin{axis}[
    xbar=0.05cm,
    width=10cm,
    height=5.5cm,
    xmax=68,
    bar width=0.3cm,
 enlarge y limits=0.09,
 enlarge x limits=0.015,
    legend style={at={(1.05,0.7)},anchor=north,legend columns=1,font=\tiny},
    legend cell align={left},
   xlabel={\# studies},
     x label style={at={(0.5,-0.1)},font=\footnotesize},
     nodes near coords,
          point meta=explicit symbolic,
          every node near coord/.append style={anchor=west,font=\footnotesize},
     symbolic y coords={ LM, LBFGS, BFGS, RMSprop, RPROP, back, SGD, adam, not},
         y tick label style={font=\footnotesize},
          x tick label style={font=\footnotesize},
          yticklabels={,,},
         ytick=data
    ]
    

\addplot[fill=blue!50, draw=black] coordinates { (1,LM)[1 (Levenberg-Marquardt)] (1,LBFGS)[1 (Limited-memory BFGS)] (1,BFGS)[1 (Broyden-Fletcher-Goldfarb-Shanno)] (2,RMSprop)[2 (Root Mean Square Propagation)]  (2,RPROP)[2 (Resilient Backpropagation] (3,back)[3 (Backpropagation)] (9,SGD)[9 (Stochastic gradient descent)] (24,adam)[24 (Adaptive Moment Estimation)] (46,not)[46 (Default)]
};



\end{axis}
\end{tikzpicture}

%
%
%

%% file: Figures/chap-review/RQ-activation.tex
\begin{tikzpicture}
\begin{axis}[
    xbar=0.05cm,
    width=10cm,
    height=5cm,
    xmax=61,
    bar width=0.3cm,
 enlarge y limits=0.09,
 enlarge x limits=0.02,
    legend style={at={(1.05,0.7)},anchor=north,legend columns=1,font=\tiny},
    legend cell align={left},
   xlabel={\# studies},
     x label style={at={(0.5,-0.1)},font=\footnotesize},
     nodes near coords,
          point meta=explicit symbolic,
          every node near coord/.append style={anchor=west,font=\footnotesize},
     symbolic y coords={
     F,
     E,
     D,
     C,
     B,
     A},
         y tick label style={font=\footnotesize},
          x tick label style={font=\footnotesize},
          yticklabels={,,},
         ytick=data
    ]
    

\addplot[fill=blue!50, draw=black] coordinates {(2,F)[2 (Softmax)] (3,E)[3 (Linear)]
(4,D)[4 (Tanh)] (9,C)[9 (Sgimoid)] (30,B)[30 (ReLu)]  (46,A)[46 (Default)] 
};



\end{axis}
\end{tikzpicture}

%
%
%

%% file: Chapter-survey/RQ3.tex
\section{RQ1.3: How to Evaluate the Performance Model? }
\label{chap-reivew:rq3}
As shown in Figure~\ref{fig:pipeline}, the third step of the deep learning workflow is to accurately evaluate the performance of deep learning models. This step is crucial as it decides the effectiveness of the proposed model, and the evaluation results are important references for the decision-making of future researchers. In this section, three key processes for model evaluation are addressed. The first focuses on the evaluation methods and metrics used to measure the performance of software performance prediction models; the second investigates the statistical test methods employed to validate the significance and effect size of the obtained results; while the third is on the number and domain of subject systems used in the evaluations.

\subsection{What Evaluation Methods and Accuracy Metrics have been Applied?}
\label{chap-survey:evaluation-and-accuracy}

\begin{figure}[!t]
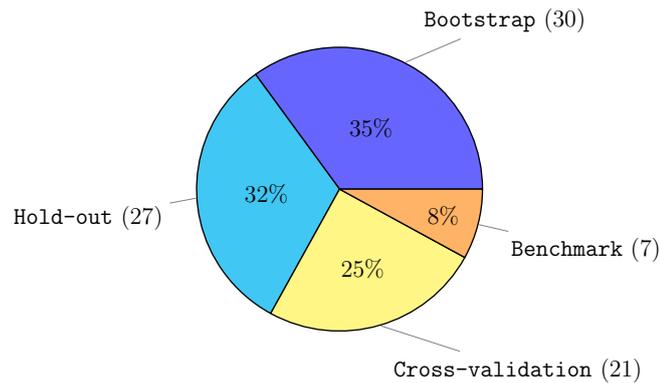

\centering
\includestandalone[width=0.6\columnwidth]{Figures/chap-review/RQ-evaluation}
  \caption{Data on evaluation methods from primary studies.}
 \label{subfig:evaluation}
 \end{figure}

By surveying 85 papers in the field, a diverse range of evaluation methods and metrics are identified, ensuring reliable and comprehensive performance model assessments and guiding future research efforts. 

In Figure~\ref{subfig:evaluation}, a summary of the analysis is provided. Among the identified evaluation methods, the most prevalent approach is \textbf{bootstrap}, utilized by 30 out of the 85 studies~\citep{DBLP:conf/nsdi/LiangFXYLZYZ23, DBLP:journals/jsa/ZhangLWWZH18, DBLP:conf/esem/ShuS0X20, DBLP:conf/spects/KimK17, said2021accurate, DBLP:conf/infocom/LiHZLT20}, 
which is a resampling technique that involves creating multiple datasets by randomly sampling from the original data, allowing for robust estimation of model performance. Yet, bootstrap samples may contain duplicate instances, which can result in overfitting. \textbf{Hold-out} evaluation, another widely used method, is employed in 26 papers~\citep{DBLP:journals/jiii/YuGLZIY22, DBLP:conf/wosp/DidonaQRT15, DBLP:journals/ese/VituiC21a, DBLP:journals/tsc/KumaraACMHT23, DBLP:journals/pvldb/MarcusP19, chai2023perfsage}. 
This approach involves splitting the dataset into training and testing sets based on a specific percentage. While hold-out is easy to implement and fast in evaluation overhead, it only evaluates the artifact once and is sensitive to the particular random split of the data, making the results bias and less reliable. On the other hand, \textbf{cross-validation}, which involves iteratively partitioning the dataset into multiple subsets for training and testing, is used by 21 studies~\citep{DBLP:journals/tompecs/MakraniSNDSMRH21, DBLP:conf/ic2e/RahmanL19, DBLP:journals/tecs/TrajkovicKHZ22, DBLP:journals/cee/KundanSA21, DBLP:journals/concurrency/OuaredCD22, DBLP:journals/concurrency/JiZL22}. To address the issues of hold-out, cross-validation makes multiple partitions like hold-out and computes the average performance, eventually utilizing the entire dataset for training and testing, which is efficient in the usage of data and offers better reliability.
Additionally, it is found that 7 studies opt to use the default evaluation pipeline provided by \textbf{benchmark} datasets, simplifying the evaluation process~\citep{DBLP:conf/middleware/MahgoubWGMGHMGB17, DBLP:journals/concurrency/YinH00C23, ding2021portable, DBLP:conf/im/JohnssonMS19}.

 \begin{figure}[!t]
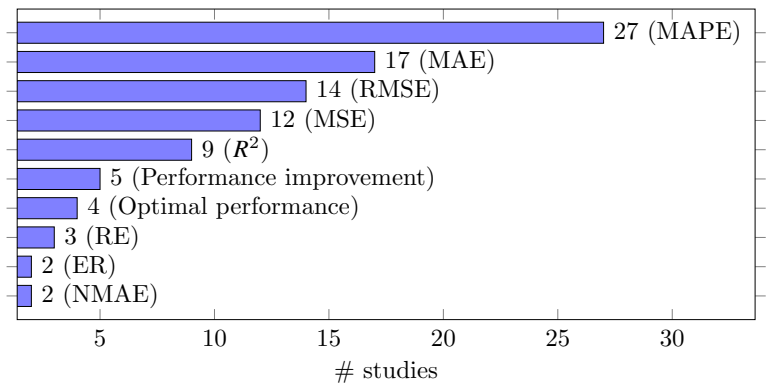

\centering
\includestandalone[width=0.7\columnwidth]{Figures/chap-review/metric}
  \caption{Data on top 10 accuracy metrics from primary studies.}
 \label{subfig:metric}
 \end{figure}
 
Figure~\ref{subfig:metric} presents the frequency of usage for the top 10 mostly used accuracy metrics. Among the identified metrics, \textbf{Mean Absolute Percentage Error (MAPE)}, also referred to as Mean Relative Error (MRE), emerges as the most commonly used in 27 studies~\citep{DBLP:journals/concurrency/FalchE17, DBLP:journals/concurrency/OuaredCD22, DBLP:journals/taco/LiL22, DBLP:journals/tosem/ChengGZ23, DBLP:conf/icse/HaZ19, DBLP:conf/icse/GaoGZLY23}. 
MAPE/MRE measures the percentage difference between the predicted and actual values. \textbf{Mean Absolute Error (MAE)} follows as the second most frequently employed metric, appearing in 17 studies~\citep{DBLP:journals/fgcs/LiLTWHQD19, said2021accurate, DBLP:journals/ese/VituiC21a, DBLP:conf/wosp/CengizFAM23, DBLP:journals/concurrency/CiciogluC22}, 
which calculates the average absolute difference between the predicted and actual values, offering a straightforward measure of model accuracy. \textbf{Root Mean Square Error (RMSE)}~\citep{ DBLP:conf/wosp/DidonaQRT15, DBLP:conf/cluster/IsailaBWKLRH15, DBLP:journals/concurrency/CiciogluC22} 
and \textbf{Mean Squared Error (MSE)}~\citep{DBLP:journals/jsa/TangLLLZ22, sabbeh2016performance, DBLP:journals/access/WangXTW20} 
are utilized in 14 and 12 studies, respectively, providing insights into the model's ability to capture both small and large errors. The \textbf{coefficient of determination (R$^{2}$)}, a metric indicating the proportion of variance in the dependent variable explained by the model, is used in 9 studies~\citep{DBLP:conf/wosp/CengizFAM23, DBLP:conf/icccnt/KumarMBCA20, DBLP:journals/tecs/TrajkovicKHZ22}. 
Additionally, performance improvement~\citep{DBLP:journals/concurrency/YinH00C23, DBLP:conf/kbse/ChenHL022, DBLP:conf/infocom/LiHZLT20, DBLP:conf/kbse/BaoLWF19, DBLP:conf/icpp/DouWZC22} and optimal performance~\citep{wang2021morphling, ding2021portable, DBLP:journals/jiii/YuGLZIY22, DBLP:conf/sigmod/ZhangLZLXCXWCLR19}, which are computed using the performance resulting from a selected configuration, are used by 5 and 4 works. Lastly, relative error (RE), error rate (ER)~\citep{du2013performance, DBLP:journals/concurrency/JiZL22, DBLP:conf/springsim/LuxWCBLXBBCH18}, and normalized mean absolute error (NMAE)~\citep{DBLP:journals/access/LiLSJ20, DBLP:journals/pvldb/ZhouSLF20} are employed in 3, 2 and 2 papers, respectively.

\begin{figure}[!t]
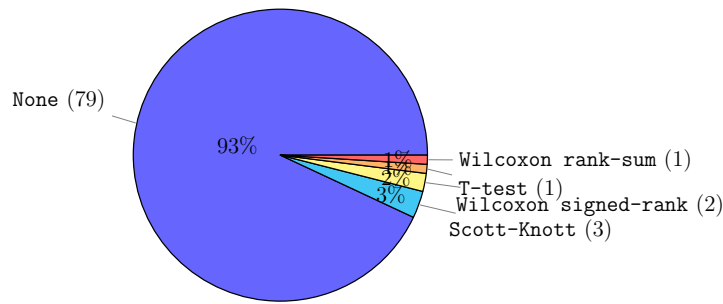

\centering
\includestandalone[width=0.65\columnwidth]{Figures/chap-review/RQ-statistical}
  \caption{Data on statistical test methods from primary studies.}
 \label{subfig:statistical}
 \end{figure}

\subsection{What Statistical and Effect Size Test Methods have been Applied?}
\label{chap-review-subsec:statistical}

As one of the inclusive criteria of this survey, nearly all of the primary studies performed comparisons between their proposed performance models and the state-of-the-art models, which puts a high requirement on accurate statistical analysis for ensuring the significance and effect size of the results, thereby drawing reliable conclusions and making valid inferences. In this sub-section, the aim is to answer the question of which statistical test and effect size test methods have been utilized in the surveyed studies.

In Figure~\ref{subfig:statistical}, the statistical test approaches are listed. Surprisingly, the majority of the studies, 79 out of 85, do not utilize any specific statistical test methods to measure the significance of their comparison results. Among the studies that do apply statistical tests, the \textbf{Scott-Knott Effect Size Difference (ESD)} test emerged as the most commonly used, appearing in 3 studies~\citep{DBLP:conf/sigsoft/Gong023, DBLP:conf/msr/GongC22, DBLP:journals/jss/ZhuYZZ21}, which is a non-parametric test that groups data into distinct subset with significant difference. Additionally, the \textbf{Wilcoxon signed-rank} test was employed in 2 studies~\citep{DBLP:journals/jss/ZhuYZZ21, DBLP:journals/tse/ChenB17}, providing a non-parametric test for comparing two paired sets of data. Furthermore, the \textbf{t-test}~\citep{DBLP:conf/icse/HaZ19} and the \textbf{Wilcoxon rank-sum}~\citep{DBLP:journals/tosem/ChengGZ23} test were each used in one study, offering comparisons for two independent data groups.

\begin{figure}[!t]
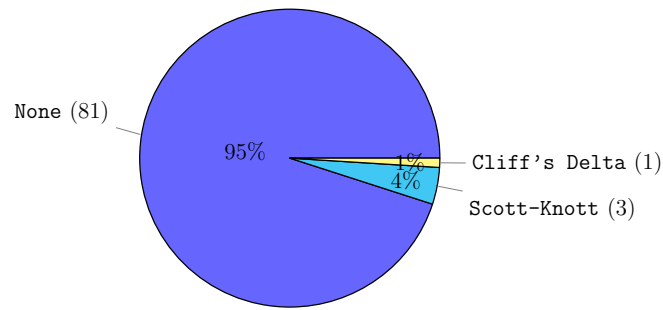

\centering
\includestandalone[width=0.6\columnwidth]{Figures/chap-review/effect_size}
  \caption{Data on effect size test methods from primary studies.}
 \label{subfig:effect_size}
 \end{figure}
 
Similarly, Figure~\ref{subfig:effect_size} reveals a predominant limitation in the utilization of effect size tests, where the majority of the studies, 80 out of 85, do not employ any specific effect size tests. Among the rest of the studies, the \textbf{Scott-Knott ESD} test is used in 3 studies~\citep{DBLP:conf/sigsoft/Gong023, DBLP:conf/msr/GongC22, DBLP:journals/jss/ZhuYZZ21}, which can identify significant differences on the effect size between groups. Additionally,~\citep{DBLP:journals/jss/ZhuYZZ21} utilized \textbf{Cliff's Delta}, a non-parametric effect size measure that quantifies the ordinal association between two variables.

\begin{figure}[!t]
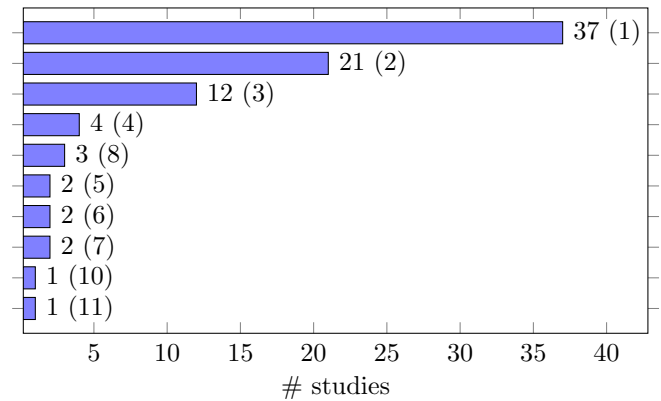

\centering
\includestandalone[width=0.6\columnwidth]{Figures/chap-review/RQ-subject_system}
  \caption{Data on \# subject systems from primary studies.}
 \label{subfig:num_subject}
 \end{figure}

\subsection{How Many Subject Systems have been Used, and What are Their Domains?}

Another key aspect that requires careful consideration in the evaluation phase is the number of subject systems. The choice of subject systems significantly impacts the generalizability and reliability of the models' performance predictions. Therefore, it is essential to study and understand the implications of the number of subject systems on the evaluation process. Additionally, reviewing the domains of the software systems allows for an understanding of the applicability and generalizability of the proposed deep learning methods across different software domains.

Figure~\ref{subfig:num_subject} provides the distribution of the number of studies used in the reviewer articles, among which 37 solely evaluate their models using a \textbf{single software}, and 21 studies evaluate \textbf{2 software systems}. Furthermore, 12 studies assess \textbf{3 systems}, 4 studies evaluate \textbf{4 subject systems}, and 3 studies evaluate \textbf{8 software}. Additionally, there were 2 studies each that evaluate \textbf{5, 6, and 7 software systems}, respectively, while 1 study each evaluates \textbf{10 and 11 subject systems}, respectively.

 \begin{figure}[!t]
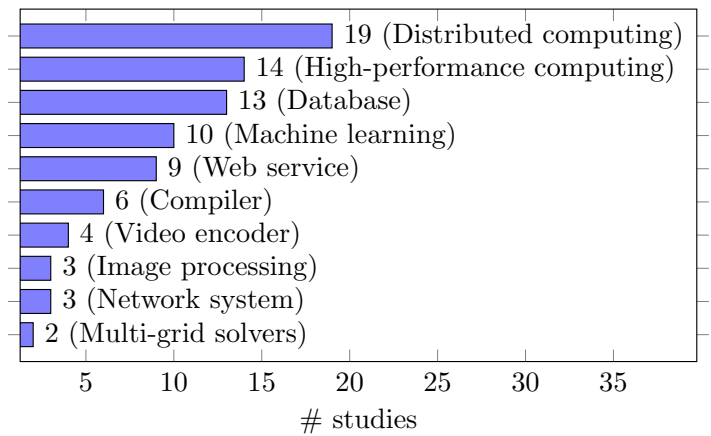

\centering
\includestandalone[width=0.65\columnwidth]{Figures/chap-review/domain_subject_system}
  \caption{Data on top 10 domains of subject systems from primary studies.}
 \label{subfig:domain_subject}
 \end{figure}
 
Moreover, the top 10 most popular domains of software systems are shown in Figure~\ref{subfig:domain_subject}. Among these domains, 19 primary studies apply \textbf{distributed computing} systems like Spark, which enables distributed processing of large datasets, particularly in cloud environments. \textbf{High-performance computing (HPC)} systems like Hadoop, which focus on utilizing powerful computing resources to solve big data processing problems efficiently, are investigated in 14 studies. In parallel, \textbf{Database} systems such as SQLite and MySQL, which pertain to the storage, retrieval, and management of structured information, are used in 13 studies. 10 studies focus on the modeling of \textbf{machine learning} systems, where predictive models like DNN are trained with specified datasets. Furthermore, \textbf{Web service} software like Apache HTTP server, which has been explored in 9 studies, involves the utilization of internet-based technologies to provide a platform for hosting websites. \textbf{Compilers}, such as SaC, examined in 6 studies, deal with translating high-level programming languages into machine code. Additionally, \textbf{Video encoder} like x264 and \textbf{image processing} software like HIPA$^{cc}$ are each used in 3 papers. \textbf{Network systems}, explored in 3 studies, enable the communication and exchange of data between devices within a network, e.g., Wireshark. Finally, \textbf{multi-grid solvers} like Dune, which are designed for solving partial differential equations (PDEs) using numerical methods, are used in 2 studies.

To sum up, it is disclosed in this section that:
\begin{quotebox}
   \noindent\textbf{RQ1.3:} The third phase of deep learning workflow involves evaluating the performance model, where it is observed that: 
   \begin{enumerate}
        \item Bootstrap is the most favorable evaluation method, which is utilized in 30 out of 85 studies, and the most popular accuracy metric is MAPE/MRE, with 27 studies used.
       \item A significant portion of 93\% and 95\% studies omit the usage of statistical and effect size tests, while the Scott-Knott ESD test is used in 3 studies, standing out as the most common method for both tests.
       \item The number of studies that only compare with one software system is the greatest (37 out of 85). Furthermore, among the various domains explored, distributed computing emerges as the most popular domain for subject systems, utilized in 19 papers. 
   \end{enumerate}

\end{quotebox}

%% file: Figures/chap-review/RQ-evaluation.tex
\begin{tikzpicture}[scale=1]
\tikzstyle{every node}=[font=\large]
    \pie[
        /tikz/every pin/.style={align=center},
        text=pin,
        rotate=0,
        explode=0, 
        ]
        {
        35/\texttt{Bootstrap} (30),
        32/\texttt{Hold-out} (27),
        25/\texttt{Cross-validation} (21),
        8/\texttt{Benchmark} (7)
        }

    \end{tikzpicture}

%% file: Figures/chap-review/metric.tex
\begin{tikzpicture}
\begin{axis}[
    xbar=0.05cm,
    width=12cm,
    height=6cm,
    xmax=33,
    bar width=0.3cm,
 enlarge y limits=0.09,
 enlarge x limits=0.02,
    legend style={at={(1.05,0.7)},anchor=north,legend columns=1,font=\tiny},
    legend cell align={left},
   xlabel={\# studies},
     x label style={at={(0.5,-0.1)},font=\footnotesize},
     nodes near coords,
          point meta=explicit symbolic,
          every node near coord/.append style={anchor=west,font=\footnotesize},
        symbolic y coords={A,B,C,D,E,F,G,H,I,J},
         y tick label style={font=\footnotesize},
          x tick label style={font=\footnotesize},
          yticklabels={,,},
         ytick=data
    ]

\addplot[fill=blue!50, draw=black] coordinates {(27,J)[27 (MAPE)] (17,I)[17 (MAE)] (14,H)[14 (RMSE)] (12,G)[12 (MSE)] (9,F)[9 ($R^2$)] (5,E)[5 (Performance improvement)] (4,D)[4 (Optimal performance)] (3,C)[3 (RE)] (2,B)[2 (ER)] (2,A)[2 (NMAE)]};




\end{axis}
\end{tikzpicture}

%% file: Figures/chap-review/RQ-statistical.tex
\begin{tikzpicture}[scale=1.1]
\tikzstyle{every node}=[font=\large]
    \pie[
        /tikz/every pin/.style={align=center},
        text=pin,
        rotate=0,
        explode=0, 
        ]
        {
        93/\texttt{None} (79),
        3/\texttt{Scott-Knott} (3),
        2/\texttt{Wilcoxon signed-rank} (2),
        1/\texttt{T-test} (1),
        1/\texttt{Wilcoxon rank-sum} (1)
        }

    \end{tikzpicture}

%% file: Figures/chap-review/effect_size.tex
\begin{tikzpicture}[scale=1.1]
\tikzstyle{every node}=[font=\large]
    \pie[
        /tikz/every pin/.style={align=center},
        text=pin,
        rotate=0,
        explode=0, 
        ]
        {
        95/\texttt{None} (81),
        4/\texttt{Scott-Knott} (3),
        1/\texttt{Cliff’s Delta} (1)
        }

    \end{tikzpicture}

%% file: Figures/chap-review/RQ-subject_system.tex
\begin{tikzpicture}
\begin{axis}[
    xbar=0.05cm,
    width=10cm,
    height=6cm,
    xmax=42,
    bar width=0.3cm,
 enlarge y limits=0.09,
 enlarge x limits=0.02,
    legend style={at={(1.05,0.7)},anchor=north,legend columns=1,font=\tiny},
    legend cell align={left},
   xlabel={\# studies},
     x label style={at={(0.5,-0.1)},font=\footnotesize},
     nodes near coords,
          point meta=explicit symbolic,
          every node near coord/.append style={anchor=west,font=\footnotesize},
     symbolic y coords={
     J,
     I,
     H,
     G,
     F,
     E,
     D,
     C,
     B,
     A},
         y tick label style={font=\footnotesize},
          x tick label style={font=\footnotesize},
          yticklabels={,,},
         ytick=data
    ]

\addplot[fill=blue!50, draw=black] coordinates {(1,J)[1 (11)] (1,I)[1 (10)] (2,H)[2 (7)] (2,G)[2 (6)] (2,F)[2 (5)] (3,E)[3 (8)]
(4,D)[4 (4)] (12,C)[12 (3)] (21,B)[21 (2)]  (37,A)[37 (1)] 
};

\end{axis}
\end{tikzpicture}

%% file: Figures/chap-review/domain_subject_system.tex
\begin{tikzpicture}
\begin{axis}[
    xbar=0.05cm,
    width=10cm,
    height=6cm,
    xmax=39,
    bar width=0.3cm,
 enlarge y limits=0.09,
 enlarge x limits=0.02,
    legend style={at={(1.05,0.7)},anchor=north,legend columns=1,font=\tiny},
    legend cell align={left},
   xlabel={\# studies},
     x label style={at={(0.5,-0.1)},font=\footnotesize},
     nodes near coords,
          point meta=explicit symbolic,
          every node near coord/.append style={anchor=west,font=\footnotesize},
     symbolic y coords={
     J,
     I,
     H,
     G,
     F,
     E,
     D,
     C,
     B,
     A},
         y tick label style={font=\footnotesize},
          x tick label style={font=\footnotesize},
          yticklabels={,,},
         ytick=data
    ]

\addplot[fill=blue!50, draw=black] coordinates {(2,J)[2 (Multi-grid solvers)] (3,I)[3 (Network system)] (3,H)[3 (Image processing)] (4,G)[4 (Video encoder)] (6,F)[6 (Compiler)] (9,E)[9 (Web service)] (10,D)[10 (Machine learning)] (13,C)[13 (Database)] (14,B)[14 (High-performance computing)]  (19,A)[19 (Distributed computing)] 
};

\end{axis}
\end{tikzpicture}

%% file: Chapter-survey/RQ4.tex
\section{RQ1.4: How to Apply the Performance Model? }
\label{chap-reivew:rq4}
The successful application of deep learning methods in software performance prediction relies on considering various factors, including application purposes, adaptability to dynamic software running environments, and the availability of resources for reproducibility. These analyses are crucial for researchers to effectively apply suitable performance models for specific tasks, adapt to dynamic scenarios, and reproduce the proposed models. In this section, these aspects are explored by investigating the techniques applied in these three processes. 

\input{Tables/chap-review/RQ-application}

\subsection{What are the Application Purposes of the Performance Model?}

Understanding the diverse application purposes of performance models is important for researchers to utilize the most suitable performance model for desired tasks. In this sub-section, the specific application tasks are investigated, aiming to contribute to a comprehensive overview of the application landscape for deep learning-based performance models in software performance learning.

Table~\ref{tb:application} shows that out of the 85 papers surveyed, 57 studies applied performance models specifically for prediction tasks. Performance tuning follows with 20 studies, and performance testing is represented by 6 studies. Additionally, scheduling and model design have fewer studies, with 2 and 1, respectively. Below, a brief overview of each application and its corresponding examples will be provided.

\subsubsection{Prediction.} The application of pure prediction involves learning and modeling the performance behaviors of systems or software without restrictions on the domain of tasks. Terms such as performance modeling, performance prediction, performance learning, and configuration learning are typically used to represent this practice.

\subsubsection{Tuning.} Tuning tasks aim at optimizing the performance by finding the optimal configuration. It involves tasks such as \textbf{configuration auto-tuning}, where software configurations are automatically adjusted to fit performance requirements, and \textbf{self-adaptive systems}, which dynamically adapt to varying conditions to achieve optimal performance.

\subsubsection{Testing.} Performance testing involves assessing the system's behavior for specific configurations or under different environments to identify potential bottlenecks or performance issues. For example, \textbf{defect prediction} seeks to identify potential defects or issues that are caused by incorrect performance. \textbf{Vulnerabilities detection} focuses on identifying security vulnerabilities that could affect system performance or compromise its functionality. 

\subsubsection{Others.} \textbf{Scheduling} seek to allocate the available resources, such as CPU, memory, or network bandwidth, in an optimal manner to maximize system performance. Meanwhile, applications on \textbf{development} involve identifying the best set of hyperparameters for developing models or software.

\begin{figure}[!t]
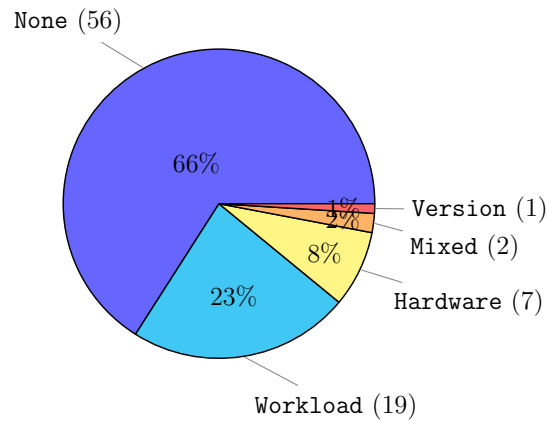

  \centering
\includestandalone[width=0.5\columnwidth]{Figures/chap-review/RQ-environments}
    \caption{Data on dynamic environment consideration from primary studies.}
 \label{fig:environments}
\end{figure}

\subsection{Does the Paper Consider Dynamic Environments?}
\label{chap-review-subsec:environment}


The running environments often influence the performance of software systems. However, it is natural that software systems are often run under dynamic conditions, such as different workloads, shifting system versions, and varying hardware availability, making it essential for performance models to adapt and provide accurate predictions in such scenarios. In this sub-section, the exploration focuses on whether the surveyed papers consider the challenges posed by dynamic software running environments when applying performance models.

The survey results in Figure~\ref{fig:environments} indicate that a significant number of studies do not explicitly consider the challenges posed by dynamic software running environments. Specifically, out of the 85 papers surveyed, 56 studies do not address the issue of dynamic environments in their application of performance models~\citep{DBLP:conf/icse/HaZ19, DBLP:conf/splc/GhamiziCPT19, DBLP:conf/sc/MaratheAJBTKYRG17, DBLP:conf/im/JohnssonMS19, DBLP:conf/sigsoft/Gong023, sabbeh2016performance}
. This suggests that a considerable portion of the research in deep learning-based software performance prediction has focused primarily on static or controlled settings without accounting for the real-world complexities of dynamic environments. Nevertheless, among the 35 studies that acknowledge and consider dynamic environments in their application of performance models, 19 studies exclusively focus on exploring the impact of different workloads on performance~\citep{DBLP:conf/mascots/KarniavouraM17, DBLP:journals/jsa/TangLLLZ22, DBLP:journals/concurrency/YinH00C23, chokri2022performance, DBLP:journals/pvldb/MarcusP19, DBLP:journals/jsa/ChengCWX17}. Additionally, 7 studies concentrate solely on investigating the influence of different hardware configurations~\citep{DBLP:conf/middleware/GrohmannNIKL19, DBLP:journals/concurrency/FalchE17, chai2023perfsage}. Furthermore, one study specifically examines the impact of different software versions~\citep{DBLP:conf/noms/SanzEJ22}. Lastly, 2 studies research the implications of mixed environment types, i.e., 1 considers 3 workloads combined with 2 hardware~\citep{DBLP:conf/IEEEcloud/MarosMSALGHA19} and 1 considers 12 workloads combined with 2 hardware~\citep{DBLP:conf/icpp/DouWZC22}.

\begin{figure}[!t]
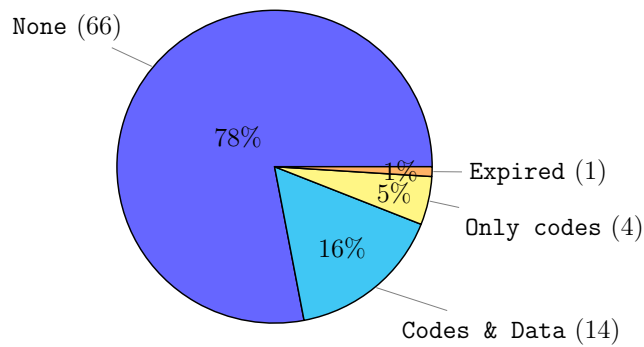

  \centering
\includestandalone[width=0.58\columnwidth]{Figures/chap-review/RQ-repo}
    \caption{Data on public repository availability from primary studies.}
 \label{fig:repo}
\end{figure}

\subsection{Does the Paper Provide a Public Repository?}
\label{chap-review-subsec:repo}


The availability of public repositories plays an essential role in promoting replicability, transparency, and development in research. In this sub-section, the investigation focuses on whether the surveyed papers provide a public repository containing the source codes and datasets for reproducing the experiments and results.

The results of the survey in Figure~\ref{fig:repo} reveal that a significant number of studies do not provide an open-access repository containing the necessary resources for replicability. In particular, out of the 85 papers surveyed, a total of 66 studies do not make their code, data, or models publicly available~\citep{DBLP:journals/ese/VituiC21a, DBLP:conf/icst/PorresARLT20, DBLP:journals/jss/ZhuYZZ21, zain2022software, DBLP:journals/infsof/WartschinskiNVK22}. This indicates a lack of emphasis on open science practices and hinders the ability of other researchers and practitioners to validate and build upon the findings. On the other hand, only 19 studies offer an open-access repository, of which 14 provide both source codes and datasets~\citep{DBLP:conf/splc/GhamiziCPT19, DBLP:conf/kbse/BaoLWF19, DBLP:conf/icst/PorresARLT20, DBLP:journals/concurrency/OuaredCD22, DBLP:conf/wosp/DidonaQRT15}, and 4 of them only provide source codes~\citep{DBLP:conf/sigmod/ZhangLZLXCXWCLR19, DBLP:conf/kbse/ChenHL022, wang2021morphling, DBLP:journals/infsof/WartschinskiNVK22}. Nevertheless, it is worth noting that in one study, the provided link has expired~\citep{DBLP:journals/tnse/CaoPC22}.

In short, RQ1.4 can be summarized with the following:
\begin{quotebox}
   \noindent\textbf{RQ1.4:} In the application step of the DL pipeline, it can be summarized that: 
   \begin{enumerate}
        \item The applications of performance models can be classified into 5 categories, where the greatest subset of 57 studies solely focus on the prediction of software performance.
        \item More than half of the 85 primary studies (66\%) do not handle the challenge of dynamic environments. Among the others, workload stands out as the most commonly considered factor(23\%). 
        \item Only 22\% of the examined papers offer public repositories, with an even lower percentage of 16\% providing both source codes and datasets.
   \end{enumerate}

\end{quotebox}

%% file: Tables/chap-review/RQ-application.tex
\begin{table}[t!]
  \caption{Data on model applications from primary studies.}
\centering
\scriptsize
\begin{adjustbox}{width=1.1\linewidth,center}

\begin{tabular}{p{2cm}p{3cm}p{0.1cm}p{7cm}}
\toprule
\multicolumn{1}{c}\textbf{{Category}} &
  \multicolumn{1}{c}\textbf{{Alternative terms}} &
  \multicolumn{1}{c}\textbf{{\# studies}} &
  \multicolumn{1}{c}\textbf{{References}} \\ \hline
Prediction &
  Performance prediction, performance modeling, performance learning, configuration learning &
  57 &
  \citep{DBLP:journals/cee/KundanSA21}, \citep{DBLP:journals/tsc/KumaraACMHT23}, \citep{DBLP:journals/jiii/YuGLZIY22}, \citep{DBLP:journals/taco/MoolchandaniKS22}, \citep{DBLP:journals/tcc/PhamDF20}, \citep{DBLP:conf/dsrt/DuongZCLZ16}, \citep{DBLP:conf/ic2e/RahmanL19}, \citep{DBLP:journals/fgcs/LiLTWHQD19}, \citep{DBLP:journals/pvldb/ZhouSLF20}, \citep{DBLP:journals/pvldb/MarcusP19}, \citep{DBLP:conf/pldi/SinghHLS22}, \citep{chai2023perfsage}, \citep{DBLP:journals/mam/JooyaDB19}, \citep{chokri2022performance}, \citep{DBLP:conf/icpp/MadireddyBCLLRS19}, \citep{DBLP:conf/spects/KimK17}, \citep{DBLP:conf/sc/MaratheAJBTKYRG17}, \citep{DBLP:conf/springsim/LuxWCBLXBBCH18}, \citep{said2021accurate}, \citep{DBLP:journals/access/ThraneZC20}, \citep{DBLP:journals/access/WangXTW20}, \citep{DBLP:journals/concurrency/CiciogluC22}, \citep{DBLP:conf/splc/Acher0LBJKBP22}, \citep{DBLP:conf/cluster/IsailaBWKLRH15}, \citep{DBLP:conf/sc/MalikFP09}, \citep{DBLP:conf/im/JohnssonMS19}, \citep{DBLP:conf/noms/SanzEJ22}, \citep{DBLP:journals/taco/WangLWB19}, \citep{DBLP:journals/concurrency/OuaredCD22}, \citep{DBLP:journals/jsa/ChengCWX17}, \citep{DBLP:journals/tecs/TrajkovicKHZ22}, \citep{DBLP:conf/wosp/DidonaQRT15}, \citep{DBLP:conf/middleware/GrohmannNIKL19}, \citep{DBLP:journals/corr/abs-2304-13032}, \citep{DBLP:conf/IEEEcloud/MarosMSALGHA19}, \citep{DBLP:journals/fgcs/KousiourisMKGV14}, \citep{DBLP:journals/concurrency/JiZL22}, \citep{DBLP:journals/access/NadeemAMFA19}, \citep{DBLP:conf/iccad/KimMMSR17}, \citep{DBLP:conf/nsdi/LiangFXYLZYZ23}, \citep{DBLP:conf/sc/MalakarBVMK18}, \citep{DBLP:conf/msr/GongC22}, \citep{DBLP:journals/jsa/ZhangLWWZH18}, \citep{DBLP:conf/nsdi/FuGMR21}, \citep{DBLP:conf/esem/ShuS0X20}, \citep{DBLP:conf/icccnt/KumarMBCA20}, \citep{DBLP:conf/sigsoft/Gong023}, \citep{du2013performance}, \citep{DBLP:conf/cf/LiuMCV20}, \citep{DBLP:journals/access/TousiL22}, \citep{DBLP:journals/tosem/ChengGZ23}, \citep{DBLP:conf/icse/HaZ19}, \citep{DBLP:conf/wosp/CengizFAM23}, \citep{DBLP:conf/icse/GaoGZLY23}, \citep{DBLP:conf/mascots/KarniavouraM17}, \citep{sabbeh2016performance}, \citep{myung2021machine} \\ \hline
Tuning &
  Configuration auto-tuning, self-adaptive system, performance tuning &
  20 &
  \citep{DBLP:journals/concurrency/FalchE17}, \citep{DBLP:journals/concurrency/FalchE17}, \citep{DBLP:journals/concurrency/RosarioSZNB23}, \citep{DBLP:conf/icpp/DouWZC22}, \citep{DBLP:conf/splc/GhamiziCPT19}, \citep{DBLP:journals/access/LiLSJ20}, \citep{DBLP:journals/comcom/AteeqAAK22}, \citep{DBLP:conf/infocom/LiHZLT20}, \citep{DBLP:conf/middleware/MahgoubWGMGHMGB17}, \citep{DBLP:conf/kbse/BaoLWF19}, \citep{DBLP:journals/tompecs/MakraniSNDSMRH21}, \citep{wang2021morphling}, \citep{ding2021portable}, \citep{DBLP:conf/ipps/FalchE15}, \citep{DBLP:journals/tse/ChenB17}, \citep{DBLP:conf/sigmod/ZhangLZLXCXWCLR19}, \citep{DBLP:journals/tnse/CaoPC22}, \citep{blott2018finn}, \citep{DBLP:journals/taco/LiL22}, \citep{DBLP:journals/jsa/TangLLLZ22} \\ \hline
Testing &
  Performance testing, defect prediction, vulnerabilities detection &
  6 &
  \citep{DBLP:journals/ese/VituiC21a}, \citep{DBLP:conf/icst/PorresARLT20}, \citep{DBLP:conf/kbse/ChenHL022}, \citep{DBLP:journals/jss/ZhuYZZ21}, \citep{zain2022software}, \citep{DBLP:journals/infsof/WartschinskiNVK22} \\ \hline
Scheduling &
  Resource allocation, CPU scheduling &
  2 &
  \citep{DBLP:journals/concurrency/YinH00C23}, \citep{DBLP:conf/sbac-pad/NemirovskyAMNUC17} \\ \hline
Development &
  Hyperparameter optimization &
  1 &
  \citep{DBLP:journals/pr/TranSWQ20}
\\ \bottomrule
\end{tabular}

\end{adjustbox}
\label{tb:application}
\end{table}

%% file: Figures/chap-review/RQ-environments.tex
\begin{tikzpicture}[scale=1]
\tikzstyle{every node}=[font=\large]
    \pie[
        /tikz/every pin/.style={align=center},
        text=pin,
        rotate=0,
        explode=0, 
        ]
        {
        66/\texttt{None} (56),
        23/\texttt{Workload} (19),
        8/\texttt{Hardware} (7),
        2/\texttt{Mixed} (2),
        1/\texttt{Version} (1)
        }

    \end{tikzpicture}

%% file: Figures/chap-review/RQ-repo.tex
\begin{tikzpicture}[scale=1]
\tikzstyle{every node}=[font=\large]
    \pie[
        /tikz/every pin/.style={align=center},
        text=pin,
        rotate=0,
        explode=0, 
        ]
        {
        78/\texttt{None} (66),
        16/\texttt{Codes \& Data} (14),
        5/\texttt{Only codes} (4),
        1/\texttt{Expired} (1)
        }

    \end{tikzpicture}

%% file: Chapter-survey/discussions.tex
\section{Answer to RQ1 and Discussion}
\label{chap-review:discussion}
The first objective of this thesis is to realize the knowledge gaps in the existing literature via a comprehensive survey. This chapter will provide an overview of the literature, give insights on the good trends that researchers can follow and bad trends that can be avoided, and discuss the potential areas for improvement in future studies, hence demonstrating the accomplishment of the research \textbf{objective 1}. Furthermore, the additional objectives will be derived from the identified knowledge gaps, followed by a discussion on the limitations, publication, and future work of this chapter.

\subsection{Positive and Negative Trends}
As depicted in Figure~\ref{fig:pipeline} and discussed in Section~\ref{chap-review-subsec:rq}, the workflow of deep learning for performance prediction of configurable software comprises four stages, each involving several key processes. In this sub-section, both positive trends and potential negative trends are revealed for each investigated process within the deep learning pipeline.

\subsubsection{Satge 1: Data preprocessing.} Table~\ref{tb:preprocessing} provides a comprehensive summary of the preprocessing techniques employed in the surveyed studies, revealing several noteworthy trends. 

\textbf{Positive trends:} A majority of studies have recognized the importance of data preprocessing in optimizing the performance of deep learning models. These preprocessing techniques play a non-trivial role in enhancing the effectiveness and efficiency of such models. For instance, it is commonly recognized that software features often exist in various scales and units, and unifying these attributes can lead to improved model performance. Additionally, correctly featurizing system attributes into appropriate types (binary, categorical, continuous) significantly influences prediction accuracy and time costs, as emphasized by~\citet{DBLP:conf/msr/GongC22}. Furthermore, software feature spaces are often high-dimensional and sparse, which puts high demands on dimension reduction and regularization techniques to eliminate irrelevant and redundant features. Therefore, utilizing different data preprocessing methods according to the characteristics of the dataset is beneficial and should be kept by researchers.

\textbf{Negative trends:} A significant portion of the surveyed studies (45 out of 85) have paid inadequate attention to the importance of preprocessing methods. Neglecting these crucial steps can severely limit the performance and effectiveness of deep learning models~\citep{DBLP:journals/jss/PereiraAMJBV21}. Therefore, it is imperative for future studies to recognize the significance of preprocessing techniques and incorporate them systematically to ensure optimal model performance.

\subsubsection{Stage 1: Encoding.} 
Figure~\ref{fig:encoding} presents a summary of the techniques used for encoding, revealing several noteworthy trends.

\textbf{Positive trends:} Encouragingly, different encoding schemes have been employed to encode the raw data. The choice of encoding scheme is significant since it profoundly impacts the prediction accuracy and time consumption of deep learning models~\citep{DBLP:conf/msr/GongC22}. For instance, one-hot encoding tends to yield the highest accuracy, albeit with longer training times compared to other schemes, while label encoding often leads to faster training times. Therefore, encoding the raw data using the most suitable encoding scheme is considered to be positive.

\textbf{Negative trends:} The influence of encoding schemes has been underestimated by a majority of the studies, which solely rely on the default dataset without providing justifications for the encoding schemes in their papers. This oversight can be detrimental to the literature, as it leaves new researchers unaware of which encoding method to use, potentially resulting in ambiguous data formats for deep learning tasks~\citep{DBLP:journals/jmlr/AlayaBGG19, DBLP:conf/kbse/BaoLWF19}.

\subsubsection{Stage 1: Sampling.} 
In Figure~\ref{fig:sampling}, a comprehensive summary of the sampling approaches used is presented, showcasing several notable trends.

\textbf{Positive trends:} In the current literature, a number of sampling methods that address different coverage criteria have been utilized. In software performance datasets, the configuration space is usually huge, while the budget for collecting training samples is often limited, so it is important to measure the most informative configurations to train performance models. As such, maintaining such a list of sampling methods to choose from according to specific requirements is a good practice.

\textbf{Negative trends:} Despite the availability of various sampling methods, a significant portion of studies tends to focus on the most straightforward approach, namely random sampling. While random sampling is effective in representing the distribution of the dataset, it can lead to overfitting in highly sparse systems, as many of the features may prove ineffective for software performance analysis~\citep{DBLP:conf/wosp/PereiraA0J20}. Therefore, it is important for more researchers to explore alternative sampling methods in such scenarios to enhance the accuracy of the performance models.

\subsubsection{Stage 2: Sparsity and overfitting handling.} 
Table~\ref{tb:overfitting} illustrates the techniques employed to address the issue of sparsity and mitigate the risk of overfitting, revealing certain patterns that have been observed.

\textbf{Positive trends:} Encouragingly, a majority of studies have recognized the significance of addressing sparsity and overfitting, which are critical for causing the degradation of predictive models, leading to the application of a diverse range of handling techniques~\citep{DBLP:conf/sigsoft/Gong023, DBLP:conf/icse/HaZ19, DBLP:conf/sc/MalikFP09, DBLP:conf/middleware/GrohmannNIKL19, DBLP:journals/corr/abs-2304-13032, DBLP:journals/pvldb/MarcusP19, DBLP:journals/tecs/TrajkovicKHZ22, DBLP:conf/splc/Acher0LBJKBP22, DBLP:conf/esem/ShuS0X20}. It is worth noting that most of these techniques are feature selection methods as the software features often have redundant information, and most of them have little influence on the software performance. Yet,~\citet{DBLP:conf/sigsoft/Gong023} discover that except for feature sparsity, which has been realized by most studies, the distribution of samples in terms of performance is also sparse, which is also a significant property in the performance learning problem, and the authors propose a `divide-and-learn' framework to learn handle the sample sparsity, which is worth further researching.

\textbf{Negative trends:} However, given that sparsity has been addressed by 52 out of 85 papers, 24 of them rely on human efforts to select the correct feature sets to reduce sparsity, which is of low efficiency and low generalizability. In addition, there are still 33 works that did nothing to prevent them. Moreover, there is only 1 study focused on solving the sparsity problem from the perspective of sample distributions, which could be a key limitation of the literature. 

\subsubsection{Stage 2: Deep learning models.}
Table~\ref{tb:learning} provides an overview of the learning models employed in the studies analyzed, and Figure~\ref{fig:reasons} surveys how many studies explain their decision to choose the specific learning model.

\textbf{Positive trends:} The table indicates that MLP is the most commonly used learning model, with a total of 53 studies employing it. This demonstrates the wide recognition and acceptance of MLP as an effective deep learning approach for learning software configurations and performance. In addition, the presence of multiple DL models suggests their suitability and effectiveness in modeling complex relationships. Notice that some models are improved versions of others, for example, rDNN~\citep{DBLP:conf/icse/HaZ19, DBLP:conf/sigsoft/Gong023} is an MLP armed with regularization techniques to address the problem of feature sparsity, and ResNet~\citep{DBLP:conf/wosp/CengizFAM23} is a type of neural network that incorporates residual connections to overcome the challenges of vanishing gradient. This is a positive practice as it demonstrates the potential of the deep learning models by addressing specific challenges and sheds light for future studies on improving the existing models.

\textbf{Negative trends:} On the other hand, in Figure~\ref{fig:reasons}, it can be seen that 55\% papers (47 out of 85) do not provide any justification for their choices of deep learning models, which is a concerning trend. It is essential for researchers to clarify their reasons for selecting specific models, as this enhances the understanding of the research motivations and contributions~\citep{DBLP:journals/tosem/LiuGXLGY22}. In the meantime, the lack of explicit explanations of model selection could be harmful to the community as researchers may miss out on important considerations or potential limitations.

\subsubsection{Stage 2: Hyperparameter tuning.}
Figure~\ref{fig:hyperparameter} presents an overview of the hyperparameter tuning methods employed in the analyzed studies, shedding light on both positive and negative trends that have been observed.

\textbf{Positive trends:} Encouragingly, the table reveals a variety of hyperparameter tuning methods for deep learning models. The diverse range of methods reflects the researchers' efforts to explore different approaches for optimizing hyperparameters and enhancing model performance, which should be kept by future researchers.

\textbf{Negative trends:} However, negative trends are also observed. Notably, a great portion (52 out of 85) of papers have chosen to tune the hyperparameters manually, which is time-consuming and not generalizable. Besides, a subset of 6 studies relies solely on default hyperparameters, which may not be optimal for all scenarios. This limitation may restrict the exploration of the optimal hyperparameter settings and hinder the potential for further optimization~\citep{DBLP:journals/csur/YangXLG22, DBLP:conf/sc2/FrankHLB17}. As such, it is important for researchers to consider automatic tuning methods in order to maximize the performance and robustness of deep learning models. 

\subsubsection{Stage 2: Model optimization.}
Figure~\ref{subfig:optimization} and Figure~\ref{subfig:activation} present comprehensive overviews of the optimization methods and activation functions employed in the studied deep learning models, respectively, providing both good and bad insights for researchers.

\textbf{Positive trends:}
The fact that different optimization methods and activation functions have been adopted is promising since the best optimization method under different conditions could be different. For example, problems with sparse gradients may benefit from adaptive optimization methods like Adam, while problems with non-convex loss surfaces may require optimization methods with strong global optimization properties. Similarly, choosing the appropriate activation function, such as ReLU or sigmoid, can impact the model's ability to effectively capture nonlinear relationships in the data. 

\textbf{Negative trends:} Surprisingly, more than half of the studies (46 out of 85) have omitted to provide information on their choices of optimization and activation approaches. This omission poses a significant challenge to understanding and reproducing the deep learning models' results and findings~\citep{DBLP:conf/msr/GongC22, DBLP:conf/icse/HaZ19, DBLP:conf/sigsoft/Gong023}. Therefore, justification for these decisions should be provided in future works, as it allows readers and researchers to comprehend the rationale behind the decision-making. 

\subsubsection{Stage 3: Evaluation methods and metrics.}
Figure~\ref{subfig:evaluation} and Figure~\ref{subfig:metric} present an analysis of the evaluation methods and metrics employed to assess the performance of deep learning models, which provide insights for researchers.

\textbf{Positive trends:} A variety of metrics for accuracy has been applied, which is a rather positive sign for the community, as~\citet{mathews1994towards} point out, no single measure gives an unambiguous indication of the modeling performance. More importantly, it has been discovered that all of the primary studies have utilized some form of repeated evaluation procedure. This is crucial to ensure that the results were not produced due to chance or biased samples.

\textbf{Negative trends:} It is found that MAPE is the most widely used metric due primarily to the fact that it is unit-free and easy to calculate. However, it is known that MAPE suffers the problem of being \emph{asymmetric} on the error above and below the actual value~\citep{goodwin1999asymmetry}, where the former receives a much greater penalty. As a result, such a strong bias towards MAPE tends to be problematic.

\subsubsection{Stage 3: Statistical and effect size tests.}
Figure~\ref{subfig:statistical} and Figure~\ref{subfig:effect_size} serve as summaries of the statistical test and effect size test methods employed to evaluate the results of deep learning models. These figures offer valuable trends identified during the survey.

\textbf{Positive trends:}
One positive trend observed in the literature is that almost all papers have conducted comparisons between their models and others, and a few of them have utilized statistical tests and effect size tests to ensure the statistical significance and effect size difference of the comparison. This practice is vital in deep learning research as it enhances the reliability and robustness of the findings.

\textbf{Negative trends:} 
However, a notable concern arises from the fact that a significant majority of studies, around 93\% (79 out of 85), did not conduct any statistical tests, and 95\% of studies did not measure the effect size. This represents a significant problem as it introduces a potential source of randomness and uncertainty in the comparison results. Such a bad practice can limit the ability to draw reliable conclusions and increase the risk of causing severe external threats to validity. This highlights the need for greater awareness and adoption of statistical analysis in deep learning research. 

\subsubsection{Stage 3: Subject systems.}
The number of subject systems and the most popular domains of software used in deep performance learning studies have been listed in Figure~\ref{subfig:num_subject} and~\ref{subfig:domain_subject}.

\textbf{Positive trends:}
A positive observation is that 48 studies have considered multiple subject systems in their experiments. This is significant because evaluating the proposed model on software from different domains, scales, and performance indicators demonstrates its generalizability and robustness~\citep{ DBLP:conf/icccnt/KumarMBCA20, DBLP:conf/noms/SanzEJ22, DBLP:conf/bigdataconf/GhoshalWPS19, icse2020-empirical, DBLP:conf/icse/HaZ19}.

Moreover, it is encouraging to see that the reviewed studies have examined a variety of domains. This diversity in domain coverage enhances the credibility and applicability of the findings across various contexts, providing broader insights for developers and practitioners.

\textbf{Negative trends:} 
Nonetheless, it is worth noting that a significant proportion of studies (44\% of 85) evaluate their models on only one software, which can be considered a potential threat to the validity of the results and findings of the study. 

Software systems of different characteristics can exhibit unique performance behavior due to factors such as architecture, complexity, and usage patterns. Therefore, it is important for future studies to evaluate their models on multiple software systems, spanning various domains, scales, and performance indicators, to ensure the robustness and learning ability of the proposed performance models.

\subsubsection{Stage 4: Applications.}
Table~\ref{tb:application} provides a comprehensive summary of the application purposes for deep learning models, highlighting both positive and negative behaviors identified in the literature.

\textbf{Positive trends:} The best practice in the model applications is their widespread adoption across various domains and for different tasks. This variety is critical as it showcases the potential of performance models in addressing diverse challenges and provides practical insights for researchers in different fields~\citep{DBLP:journals/jss/PereiraAMJBV21}.

\textbf{Negative trends:} Potential concern is that 57 studies have only considered pure prediction tasks. While these studies contribute valuable insights into performance prediction, they raise the question of whether these models can effectively be applied to other tasks beyond prediction. In other words, the applicability and generalizability of performance models across different domains and tasks remain unclear.

\subsubsection{Stage 4: Dynamic environments.}
Figure~\ref{fig:environments} provides a depiction of the consideration given to dynamic software environments in the literature. This sub-section encapsulates both positive aspects and patterns of concern.

\textbf{Positive trends:} The most encouraging trend is that 29 out of 85 studies have actively considered dynamic environments, demonstrating the robustness of models. This dynamism is a significant property as the behavior of software could be completely different across the environments~\citep{muhlbaueranalyzing,DBLP:conf/wcre/Chen22,DBLP:conf/wosp/PereiraA0J20}. By accounting for dynamic workloads, versions, and hardware in real-world scenarios, these studies contribute to a more comprehensive understanding of how performance models can be applied in practical settings.

\textbf{Negative trends:} It is a potential concern that 66\% of 85 primary studies have solely focused on datasets from a single environment without considering dynamic factors that can impact model performance, which means there is a risk of overestimating the model's capabilities or encountering unexpected issues when applied in real-world scenarios~\citep{muhlbaueranalyzing}. Therefore, it is essential for future research to address this limitation by actively incorporating dynamic environments into their evaluation processes.

\subsubsection{Stage 4: Open-access repository.}
Figure~\ref{fig:repo} depicts the availability of a public repository in the primary studies, highlighting the observed positive and negative patterns related to this question.

\textbf{Positive trends:} 19 out of the total studies have embraced the principles of open science by providing a public repository that includes source codes and datasets. This is a good practice as it allows researchers to replicate or reproduce the experiment results, build further research based on the existing knowledge, and collaborate with each other~\citep{DBLP:journals/tosem/LiuGXLGY22}.

\textbf{Negative trends:} It is important to note that a significant number of studies (78\%) still missed the opportunity to promote open science by not providing public repositories. This is a limitation as it restricts the replicability, reproducibility, and development of the community, which could be addressed by future studies. Further, among the 19 studies that offer an open-access repository, only 14 of them provide both source codes and datasets, which is important to fully replicate the results in the experiments. 

It is worth noting that although~\citet{DBLP:journals/tnse/CaoPC22} present a link in their paper, the link has already expired, and the repository is unavailable, which is rather a problematic practice. Therefore, it is essential for authors to maintain the accessibility of their repositories.

\subsection{Future Directions}
Through analyzing the positive and negative trends present in the literature, several key knowledge gaps and promising directions that are worth further investigation in future studies are disclosed.

\subsubsection{Study the influence of encoding schemes.}
The way to encode the features could affect the internal representation of the learning models, and it has been proved in various domains that the choice of the encoding method has a critical influence on both the prediction accuracy and model training overhead~\citep{9020560, DBLP:journals/bmcbi/HeP16}. However, it can be seen in Figure~\ref{fig:encoding} that more than half of the primary studies in this SLR do not justify their choice of encoding schemes or simply use the default dataset, hindering the understanding of the effectiveness of the encoding methods. 

As such, it is considered good practice for authors in the field to provide justifications for their choice of encoding schemes in their papers, which enhance the transparency and understanding of their research. On the other hand, for future researchers, it is essential to conduct studies aiming to investigate the influence and behavior of different encoding schemes and shed light on the advantages, limitations, and optimal usage of various encoding schemes across different scenarios.

\subsubsection{Apply model-based sampling methods.} 
In Figure~\ref{fig:sampling}, it can be seen that while several sampling approaches have been employed in the studies reviewed, it is evident that a significant majority of them rely on random sampling as the primary method, which highlights a significant opportunity for future research to explore and develop more effective sampling methods. Random sampling, while straightforward, may not always yield the most informative or representative samples within the configuration space~\citep{DBLP:conf/icse/KalteneckerGSGA19, DBLP:conf/kbse/SarkarGSAC15, DBLP:conf/sigsoft/JamshidiVKS18}. By exploring alternative sampling techniques that can select samples based on specific criteria or heuristics, researchers can potentially improve the efficiency and effectiveness of their studies~\citep{DBLP:conf/wosp/PereiraA0J20}. 

Among others, addressing this gap by investigating model-based sampling methods that can provide an explicit acquisition function to reliably and accurately guide the sampling process is an important insight for future studies to pursue. For instance, BO (Bayesian Optimization) effectively explores the sample space by quantifying the uncertainty of the samples to guide the search towards potentially optimal regions~\citep{DBLP:conf/icdcs/HsuNFM18, DBLP:conf/nips/SnoekLA12, DBLP:conf/eurosys/AlabedY21}. Further, SMBO (Sequential Model-Based Optimization)~\citep{wang2021morphling} is a specific variant of BO that uses a surrogate model, such as a Gaussian process, to sequentially select new samples based on the current model's predictions and uncertainty estimates, enabling efficient exploration of the sample space.

\subsubsection{Handle sparsity and sample sparsity.}
Sparsity has been addressed by many studies as a harmful issue in the domain of data-driven performance modeling~\citep{DBLP:conf/icse/HaZ19,DBLP:conf/kbse/JamshidiSVKPA17, DBLP:conf/sigsoft/SiegmundGAK15, DBLP:conf/icse/SiegmundKKABRS12}. Yet, the results of Section~\ref{chap-review-subsec:sparsity} show that whilst several papers have dedicated efforts to address sparsity and overfitting issues, it is notable that the majority of these studies primarily focus on feature selection techniques~\citep{DBLP:conf/sc/MalikFP09, DBLP:conf/middleware/GrohmannNIKL19, DBLP:journals/corr/abs-2304-13032, DBLP:journals/pvldb/MarcusP19, DBLP:journals/tecs/TrajkovicKHZ22, DBLP:conf/splc/Acher0LBJKBP22, DBLP:conf/esem/ShuS0X20}. Indeed, reducing sparsity by eliminating insignificant parameters or deselecting non-influential features using feature selection techniques is not new in the literature. However, the consideration of sparse performance, where the distribution of samples is not continuous but grouped into clusters, has been largely ignored.

This observation highlights a promising future direction for research in the field. For example, it is promising to enhance the work of~\citet{DBLP:journals/tosem/ChengGZ23}, which is currently the state-of-the-art DNN model for performance learning, by addressing the sparse feature influence and sparse performance distribution. By considering and addressing these aspects, researchers can improve the learning capability and robustness of models, leading to enhanced accuracy and reliability in real-world applications.

\subsubsection{Employ heuristic hyperparameter tuning.}
It can be seen in Section~\ref{chap-review-subsec:hyperparameter} that while various hyperparameter tuning methods have been utilized in the studies reviewed, it is evident that a significant portion of these approaches heavily rely on human experts or manual tuning. This reliance poses several challenges, including the need for extensive domain knowledge and the associated time costs, which may limit the scalability and efficiency of the tuning process~\citep{DBLP:journals/csur/YangXLG22, DBLP:conf/sc2/FrankHLB17}. Moreover, within the existing list of hyperparameter tuning methods in the literature, most studies simply apply a brute-force approach, i.e., grid search, to tune the models, which is impractical for large models due to the combinatorial explosion of possible configurations. 

To overcome these challenges, it is promising for future research to explore and employ effective automatic and heuristic tuning methods, such as Bayesian optimization~\citep{DBLP:journals/pr/TranSWQ20, zain2022software}, which uses probabilistic models to guide the search for optimal hyperparameters, or genetic algorithms~\citep{DBLP:conf/splc/GhamiziCPT19}, which utilize principles of natural selection and evolution to select best-performing candidates. By achieving this, researchers can reduce the dependence on domain experts, save time, and enhance the overall effectiveness and efficiency of DL models, enabling researchers to focus on other critical aspects of their work.

\subsubsection{Utilize statistical and effect size test methods.}
From the analysis in Section~\ref{chap-review-subsec:statistical}, it is found that although it is worth acknowledging that a small number of studies have incorporated statistical tests and effect size measurements to assess the significance of their evaluations, it is concerning that over 93\% of the reviewed works have neglected to include these essential analyses. The omission of statistical significance testing raises concerns about the reliability and validity of the reported results. Therefore, it is vital for future studies to fill this gap. For instance, for comparisons between two paired groups of data, approaches like Wilcoxon signed-rank test~\citep{DBLP:reference/stat/ReyN11} and A$^{12}$ effect size test~\citep{vargha2000critique} can be applied, while for comparisons between more than 2 sets, Scott-Knott ESD test~\citep{DBLP:conf/sigsoft/Gong023, DBLP:conf/msr/GongC22, DBLP:journals/jss/ZhuYZZ21}, which classify data into groups according to their differences in both statistical significance and effect size.

\subsubsection{Consider dynamic environments.}
Despite the encouraging finding in Section~\ref{chap-review-subsec:environment} that 29 out of the 85 reviewed studies have considered dynamic environments, it is evident that a majority, specifically 66\%, did not address the dynamic nature of software running environments. This observation highlights the need for future research to address this gap and develop methodologies that can effectively handle performance data across multiple environments, such as multi-task learning~\citep{DBLP:conf/iclr/YangH17, DBLP:journals/corr/abs-2106-02716, DBLP:conf/eurosys/AlabedY21, DBLP:conf/icml/StandleyZCGMS20}, which learns the common representations between multiple tasks simultaneously; meta-learning~\citep{DBLP:conf/icml/YaoWHL19, DBLP:journals/air/VilaltaD02, DBLP:journals/pami/HospedalesAMS22, DBLP:conf/icse/ChaiZSG22, DBLP:conf/sigir/ZhangFW00LZ20}, 
which pretrain a set of model parameters via meta-training and can adapt quickly or unseen tasks or environments; or transfer learning~\citep{ DBLP:conf/icccnt/KumarMBCA20, DBLP:conf/noms/SanzEJ22, DBLP:conf/bigdataconf/GhoshalWPS19, icse2020-empirical, DBLP:conf/im/JohnssonMS19}, 
which seek to leverage the common information gained from related environments, enabling models to adapt and generalize to the target environments.

\subsubsection{Provide open-access repository.}
Section~\ref{chap-review-subsec:repo} reveals an absence of a public repository in 78\% of the reviewed papers, which indicates a substantial missed opportunity for paper replication, knowledge sharing, and the advancement of the research community. 

Consequently, establishing a public repository for source codes and datasets is a crucial future direction, which promotes open science principles and benefits future studies from the knowledge generated by previous research~\citep{DBLP:journals/tosem/LiuGXLGY22}. Furthermore, it is crucial for researchers to maintain the availability of their repositories and enable a sustainable research environment.

\subsection{Extracted Objectives}
\label{chap-review:objective}
In the previous sections, the good and bad trends observed from the existing literature and future directions that researchers can address have been discussed, which showcases the achievement of \textbf{objective 1}. In the subsequent phase, additional objectives will be derived from the identified knowledge gaps, and studies will be conducted to address these gaps. However, given the limitations in time, this thesis will concentrate on specific key knowledge gaps rather than attempting to cover all possible directions. Consequently, this section introduces the knowledge gaps addressed in this thesis and presents the extracted objectives and research questions to be explored.

\subsubsection{Objective 2: to study the encoding schemes.}
The best encoding method could vary with different machine/deep models and for different tasks. However, in most of the reviewed studies in the SLR, the justification of encoding scheme selection is omitted, leaving researchers with little understanding in this regard. As a result, it is crucial to understand how the encoding method performs differently for different learning models and different configurable software.

In a nutshell, to fill this knowledge gap, the objective this thesis seeks to meet is:

\begin{answerbox}
\emph{\textbf{Objective 2: }to research whether the encoding scheme is critical for performance learning and understand how to choose the most appropriate encoding scheme.}
\end{answerbox}

With this objective satisfied, the researchers would have a clear knowledge of the encoding scheme, offering researchers a faster path to select the best encoding method. Chapter~\ref{chap:encoding} will elaborate on all the details of the study.

\subsubsection{Objective 3: to solve the sparsity problem.}
In the SLR, only a number of studies have tried to solve the sparsity problem. Nevertheless, most models apply manual selection to filter the parameters and features, which is infeasible for large-scale software with thousands of configuration options. Therefore, to push the level of accuracy to the next level, this thesis seeks to develop a new method that can improve the sparse performance models from a different perspective. 

Formally, the designed objective is:

\begin{answerbox}
\emph{\textbf{Objective 3: }to design and implement a model to solve the sparsity problem in performance modeling, such that the model is more accurate than the state-of-the-art models with the same dataset.}
\end{answerbox}

By achieving the objective, the accuracy of the state of the art could be improved to a new stage. The details of how this thesis addresses this objective are presented in Chapter~\ref{chap:dal}.

\subsubsection{Objective 4: to handle dynamic environments.}
Software is often run on various hardware and software environments, yet most of the current deep learning performance models fail to adapt across environments, resulting in a waste of data from various environments. As such, it is crucial to leverage the performance data efficiently and handle the dynamic environment challenge. 

The research objective can be formally stated as follows:

\begin{answerbox}
\emph{\textbf{Objective 4:} to design and implement a multi-environment learning model that leverages performance data from other environments to improve the prediction accuracy in the target environment.}
\end{answerbox}

Through accomplishing the objective, cross-environment performance models can be enhanced, achieving higher accuracy with less data from the target environment. The detailed study is shown in Chapter~\ref{chap:meta}.

\subsection{Limitations}

The limitations of this thesis may come from three sources:

\begin{itemize}
    \item Due to the time constraints and the fast development space of the deep learning models, by the time of publication, new state-of-the-art approaches could have been invented, thus threatening the validity of the findings.
    \item The scope of this SLR is limited to six indexing services, as a great amount of time and resources are required for the data collection, data examination, and analysis processes.
    \item The quality of included studies could be one of the limitations since this study does not aim to focus on the best works, but to survey the common phenomena in the literature and, therefore, does not include any criteria to include or exclude studies based on their quality.
\end{itemize}

\subsection{Publication and Future work}
Initially, the SLR was carried out between 2015 and 2020, and a conference paper was derived from the study; however, it was not published. Subsequently, in 2024, the SLR was updated and transformed into a journal paper, which was accepted by the \textit{ACM Transactions on Software Engineering and Methodology (TOSEM)} in October 2024~\citep{DBLP:journals/corr/abs-2403-03322}.

The future work of this chapter entails expanding the time range and incorporating additional indexing services to encompass a broader selection of primary studies, particularly focusing on the most recent ones. This will involve updating the positive and negative trends derived from statistical analysis and highlighting the knowledge gaps that have been successfully bridged. Additionally, it is crucial to ensure the updated review reports any changes in trends or newly identified knowledge gaps that have emerged during the course of the study.

%% file: Chapter-survey/threats_to_validity.tex
\section{Threats to Validity}
\label{chap-reivew:threats}
When conducting a systematic literature review, it is important to consider potential threats to validity that could affect the reliability and generalizability of the findings. In the context of this survey on deep learning for performance modeling of configurable software systems, several threats to validity need to be addressed, including sampling bias, internal threats, and external threats. In this section, the threats will be discussed, and an explanation will be provided on how the survey has addressed them.

\subsection{Sampling Bias:}
Sampling bias refers to the potential threats of the results due to the selection of studies. In this survey, sampling bias could arise from the selection of primary papers from indexing services or the inclusion and exclusion of certain papers. To mitigate this threat, specific rules, procedures, and criteria have been codified in the search protocol. The development of the search protocol is based on the guidance provided by~\citet{DBLP:journals/infsof/KitchenhamBBTBL09}. Specifically, the sampling of studies is done by a comprehensive automatic search via six popular indexes in software engineering to cover a wide range of papers, then the application of domain knowledge to remove the redundant and irrelevant studies, and filtering using the carefully crafted inclusion and exclusion criteria to further remove potential bias, as summarized in Figure~\ref{fig:protocol}.

\subsection{Internal Threats:}
Internal threats to validity relate to issues within the study design or data analysis that could affect the accuracy and reliability of the results. In this survey, potential internal threats could include inconsistencies in data extraction, subjectivity in data analysis, or biased interpretation of the findings. To address these potential threats, a systematic and rigorous approach has been followed for data extraction and analysis. By clearly defining research questions and sub-questions, it is ensured that the data extraction process is focused and consistent. Additionally, multiple reviewers were involved, and two iterations of independent paper reviews were conducted among the authors, which helped to minimize biases and ensure a more comprehensive and reliable review process. Error checks and investigations were also conducted to correct any issues found during the search. Any discrepancy in the results was discussed until an agreement can be reached.

\subsection{External Threats:}
External threats to validity are usually related to the generalizability and applicability of the findings beyond the specific context of the survey. In this survey, external threats could arise from the limited time range from 2013 to 2023, or the limited scope of the papers using deep learning for performance prediction. To mitigate these potential threats, a thorough search was conducted on 948 studies from six indexing services. Clear and well-defined inclusion and exclusion criteria were established to guide the selection of papers. Based on these criteria, 85 prominent primary studies were identified and extracted for detailed analysis. By including a diverse range of studies from different indexing services, the survey can provide a broader perspective on deep learning for performance modeling. Further, the good and bad practices disclosed are justified by evidence from the literature.

%% file: Chapter-survey/conclusion.tex
\section{Chapter Summary}
\label{chap-review:conclusion}
In this chapter, to achieve objective 1 mentioned in Section~\ref{chap-intro:rq}, a systematic literature review is conducted to examine the techniques applied to each process in the pipeline of deep learning for performance modeling, covering 85 prominent work from 948 studies found on six indexing services. The results are classified and presented, and the positive and negative practices of this field are disclosed in this chapter. 

Particularly, regarding data preparation, it is observed that over half of the studies do not apply any preprocessing methods. Furthermore, the choice of encoding scheme is disregarded in a majority of the studies, and random sampling is the most commonly employed sampling technique. In the model training stage, a significant number of studies overlook addressing sparsity and overfitting issues. FNN is the most extensively studied learning algorithm, while hyperparameter tuning is predominantly conducted manually. Additionally, a notable number of studies omit to discuss the optimization method and activation function. Regarding model evaluation, the bootstrap method and MAPE (MRE) are the most widely used evaluation methods and metrics, respectively. However, more than 93\% of the studies fail to examine the statistical significance and effect size. Regarding model application, it is observed that the majority of papers solely focus on pure performance prediction tasks. Additionally, the challenge of dynamic environments is not addressed in over half of the studies, and only 22\% of the papers provide an open-access repository.



More importantly, the potential knowledge gaps in deep learning-based performance prediction are identified:

\begin{itemize}
    \item The influence of encoding schemes needs to be explored.
    \item Explore model-based sampling methods such as SMBO.
    \item Pay more attention to sample sparsity issues, for example, adaptively divide the sample space into distinct divisions and learn each one separately.
    \item Tune hyperparameters using automatic and heuristic techniques like genetic algorithms to save the tuning costs.
    \item Examine the statistical significance and effect size difference such as Wilcoxon signed-rank and A$^{12}$ effect size test to enhance the reliability of the evaluations and findings.
    \item Emphasize the challenge of dynamic environments, including workloads, software versions, and hardware configurations, to strengthen the generalizability of performance models.
    \item Provide a public repository with source codes and datasets to promote open science and facilitate replicability and reproducibility.
\end{itemize}

From these knowledge gaps, three research objectives related to encoding schemes, sparsity, and dynamic environments are extracted and addressed in this thesis. The proposed solutions to the gaps, as well as the evaluations and justifications, are given in the following Chapter~\ref{chap:encoding}, ~\ref{chap:dal} and~\ref{chap:meta}, respectively.

%% file: Chapter-encoding/chapter-encoding.tex
\chapter{On the Impact of Encoding Schemes for Performance Prediction: An Empirical Study}
\label{chap:encoding}

As shown in Chapter~\ref{chap:introduction}, learning and predicting the performance of a configurable software system using deep learning is crucial. One important engineering decision therein is how to encode the configurations into the most suitable form for the DL model. However, as disclosed in Chapter~\ref{chap:review}, despite the presence of different encoding schemes, there is still little understanding of which is better and under what circumstances, as the community often relies on some general beliefs that inform the decision in an ad-hoc manner. Thugs, the research question that still needs to be answered, as illustrated in Section~\ref{chap-intro:rq}, is:

\begin{answerbox}
\emph{\textbf{RQ2:} Which encoding scheme is better for performance learning, and under what circumstances?}
\end{answerbox}

To bridge this gap, this chapter empirically compares the widely used encoding schemes for software performance learning, namely the label, scaled label, and one-hot encoding. The study covers five systems, seven models, and three encoding schemes, leading to 105 cases of investigation. 

The key findings reveal that: (1) conducting trial-and-error to find the best encoding scheme in a case-by-case manner can be rather expensive, requiring up to 400$+$ hours on some models and systems; (2) the one-hot encoding often leads to the most accurate results while the scaled label encoding is generally weak on accuracy over different models; (3) conversely, the scaled label encoding tends to result in the fastest training time across the models/systems while the one-hot encoding is the slowest; (4) for all models studied, label and scaled label encoding often lead to relatively less biased outcomes between accuracy and training time, but the paired model varies according to the system. To promote open science, the data and code of this work can be publicly accessed at
\texttt{\textcolor{blue}{\url{https://github.com/ideas-labo/MSR2022-encoding-study}}}.

Moreover, this chapter discusses the actionable suggestions derived from the findings, hoping to provide a better understanding of this topic for the community. 

\input{Chapter-encoding/introduction}

\input{Chapter-encoding/background}

\input{Chapter-encoding/methodology}

\input{Chapter-encoding/evaluation}

\input{Chapter-encoding/suggestions.tex}

\input{Chapter-encoding/discussions}

\input{Chapter-encoding/threats_to_validity}

\input{Chapter-encoding/conclusions}

%% file: Chapter-encoding/introduction.tex
\section{Introduction}
\label{chap-encoding:introduciton}

A critical engineering decision to make in learning performance for configurable software is how to encode the configurations. In the literature, three encoding schemes are prevalent: (1) embedding the configuration options without scaling (label encoding)~\citep{DBLP:journals/corr/abs-1801-02175,DBLP:conf/sigsoft/SiegmundGAK15,DBLP:conf/sigsoft/NairMSA17}; (2) doing so with normalization (scaled label encoding)~\citep{DBLP:conf/icse/Chen19b,DBLP:journals/tse/ChenB17,DBLP:conf/icse/HaZ19} or (3) converting them into binary ones that focus on the configuration values of those options, each of which serves as a dimension (one-hot encoding)~\citep{DBLP:conf/icse/SiegmundKKABRS12,DBLP:conf/kbse/GuoCASW13,DBLP:conf/kbse/BaoLWF19}. 

\subsection{Knowledge Gap}
Existing work takes one of these three encoding schemes without systematic justification or even discussions, leaving researchers with little understanding in this regard. This is of concern, as in other domains, such as system security analysis~\citep{9020560} and medical science~\citep{DBLP:journals/bmcbi/HeP16}, it has been shown that the encoding scheme chosen can pose significant implications to the success of a machine/deep learning model. Further, choosing one in a trial-and-error manner for each case can be impractical and time-consuming, as it will be shown in Section~\ref{chap-encoding:analysis and results}. It is, therefore, crucial to understand how the encoding performs differently for learning performance of configurable software.

To provide a better understanding of \textbf{RQ2} mentioned above, this thesis conducts an empirical study that systematically compares the three encoding schemes for learning software performance and discusses the insights learned. The hope is to provide a more justified understanding towards such an engineering decision in learning software performance under different circumstances.

\subsection{Research Questions}

In order to enhance the generalizability of this study, it should be noted that machine learning models, apart from (deep) neural network, have also been considered. Consequently, this chapter covers seven widely used machine/deep learning models for learning software performance, i.e., Decision Tree (DT)~\citep{DBLP:series/smpai/RokachM14} (used by~\citep{DBLP:conf/icse/Chen19b,DBLP:journals/tse/ChenB17,DBLP:journals/corr/abs-1801-02175,DBLP:conf/kbse/GuoCASW13}), $k$-Nearest Neighbours ($k$NN)~\citep{fix1985discriminatory} (used by ~\citep{DBLP:journals/software/KalteneckerGSA20}), Kernel Ridge Regression (KRR)~\citep{vovk2013kernel} (used by~\citep{DBLP:journals/software/KalteneckerGSA20}), Linear Regression (LR)~\citep{DBLP:journals/tamm/Goldin10} (used by~\citep{DBLP:conf/icse/Chen19b,DBLP:journals/tse/ChenB17,DBLP:conf/icse/SiegmundKKABRS12}), Neural Network (NN)~\citep{wang2003artificial} (used by~\citep{DBLP:conf/icse/HaZ19, fei2016compressor}), Random Forest (RF)~\citep{DBLP:conf/icdar/Ho95} (used by~\citep{DBLP:conf/splc/ValovGC15,DBLP:conf/oopsla/QueirozBC16}), and Support Vector Regression (SVR)~\citep{cortes1995support} (used by~\citep{DBLP:conf/icse/Chen19b,DBLP:conf/splc/ValovGC15}), together with five popular real-world software systems from prior work~\citep{DBLP:journals/corr/abs-1801-02175, DBLP:journals/corr/abs-2106-02716,ChenMMO21,DBLP:journals/corr/abs-2112-07303}, covering a wide spectrum of characteristics and domains. 

To examine research question 2, this thesis proposes 4 sub-RQs to study. Naturally, the first RQ to ask is:

\begin{quotebox}
   \noindent
   \textit{\textbf{RQ2.1:} Is it practical to examine all encoding methods to find the best one under every system?}
\end{quotebox}

\textbf{RQ2.1} seeks to confirm the significance of the study: if it takes an unreasonably long time to conduct trial-and-error in a case-by-case manner, then guidelines on choosing the best encoding scheme under different circumstances become rather important. 


What to understand next is:

\begin{quotebox}
   \noindent
   \textit{\textbf{RQ2.2:} Which encoding scheme (paired with the model) helps to build a more accurate performance model?}
\end{quotebox}

Root Mean Squared Error (RMSE), which is commonly used for software performance modeling~\citep{DBLP:conf/mascots/GrohmannSELKD20,DBLP:conf/cloudcom/IorioHTA19}, is used as the metric for accuracy. In particular, to make a comparison under the best possible situation, this thesis follows the standard pipeline in software performance learning~\citep{DBLP:conf/icse/Chen19b,DBLP:journals/tse/ChenB17,DBLP:journals/corr/abs-1801-02175,DBLP:conf/sigsoft/SiegmundGAK15,DBLP:conf/sigsoft/NairMSA17} that tunes the hyperparameters of each model-encoding pair using grid search and cross-validation, which is a common way for parameter tuning~\citep{DBLP:series/lncs/Hinton12}. 

While prediction accuracy is important, the time taken for training can also become an integral factor in software performance learning. The next RQ is, therefore:

\begin{quotebox}
   \noindent
   \textit{\textbf{RQ2.3:} Which encoding scheme (paired with the model) helps train a performance model faster?}
\end{quotebox}

The training time of each model-encoding pair is examined, including all processes in the learning pipeline such as hyperparameter tuning and validation.

Since it is important to understand the relationship between accuracy and training time, in the final RQ, it is asked that:

\begin{quotebox}
   \noindent
   \textit{\textbf{RQ2.4:} What are the trade-offs between accuracy and training time when choosing the encoding and models?}
\end{quotebox}

With this, this study seeks to understand the Pareto-optimal choices that are neither the highest on accuracy nor have the fastest training time (the non-extreme points), especially those that achieve a well-balanced between accuracy and training time, i.e., the knee points.

\subsection{Contributions}

In a nutshell, this chapter shows that choosing the encoding scheme is non-trivial for learning software performance and the key findings are:

\begin{itemize}
    \item  \textbf{To RQ2.1:} Performing trial-and-error in a case-by-case manner for finding the best encoding schemes can be rather expensive under some cases, e.g., up to 400$+$ hours.
    \item  \textbf{To RQ2.2:} The one-hot and label encoding tends to be the best choice while the scaled label encoding performs generally the worst.
    \item  \textbf{To RQ2.3:} Opposed to \textbf{RQ2.2}, the scaled label encoding is generally the best choice while the one-hot encoding often exhibits the slowest training.
    \item  \textbf{To RQ2.4:} Over the models studied, the label and scaled label encoding often lead to less biased results, particularly the latter, but the paired model varies depending on the system.
\end{itemize}

Deriving from the above, actionable suggestions are provided for learning software performance under a variety of circumstances:

\begin{enumerate}
    \item When the model to be used involves RF, SVR, KRR, or NN, it is recommended to avoid trial-and-error for finding the best encoding schemes. However, this may be practical for $k$NN, DT, and LR.
    \item When the accuracy is of primary concern, 
    \begin{itemize}
        \item using deep neural network paired with one-hot encoding if all models studied are available to choose from.
        \item using one-hot encoding for deep learning (NN), lazy models ($k$NN), and kernel models (KRR and SVM).
        \item using label encoding for linear (LR) and tree models (DT and RF).
    \end{itemize}
    \item When the training time is more important,
     \begin{itemize}
        \item using linear regression paired with scaled label encoding if all models studied are available to choose from.
        \item using scaled label encoding for deep learning (NN), linear (LR), and kernel models (KRR and SVR).
        \item using label encoding for lazy ($k$NN) and tree models (DT and RF).
    \end{itemize}
    \item When a trade-off between accuracy and training time is unclear while an unbiased outcome is preferred, 
     \begin{itemize}
        \item using scaled label encoding for achieving a relatively well-balanced result if considering all models studied, but the paired model requires some effort to determine.
        \item if the model is fixed, only the kernel models (KRR and SVR) and lazy model ($k$NN) have a more balanced outcome achieved by label encoding and scaled label encoding, respectively.
    \end{itemize}
\end{enumerate}

\subsection{Chapter Outline}
The remaining of this chapter is organized as follows: Section~\ref{chap-encoding:background} introduces the background information. Section~\ref{chap-encoding:methodology} elaborates on the details of the empirical strategy. Section~\ref{chap-encoding:analysis and results} discusses the results and answers the aforementioned research questions. The insights learned and actionable suggestions are specified in Section~\ref{chap-encoding: suggestions}. Lastly, Section~\ref{chap-encoding:discussion} discusses the implications, limitations, and future work of the study, Section~\ref{chap-encoding:threats} depicts the threats to validity, and~\ref{chap-encoding:conclusions} presents the conclusion, respectively.

%% file: Chapter-encoding/background.tex
\section{Background}
\label{chap-encoding:background}


This section elaborates on the necessary background information and the motivation of the study.

\subsection{Encoding Schemes}

In machine learning, the steps involved in the automated model building form a \textbf{\textit{learning pipeline}}~\citep{mohr2021predicting}. For learning software performance, the standard learning pipeline setting consists of preprocessing, hyperparameter tuning, model training (using all configuration options), and model evaluation~\citep{DBLP:conf/icse/Chen19b,DBLP:journals/tse/ChenB17,DBLP:journals/corr/abs-1801-02175,DBLP:conf/sigsoft/SiegmundGAK15,DBLP:conf/sigsoft/NairMSA17} (see Section~\ref{chap-encoding:methodology} for details).

In all learning pipeline phases, one critical engineering decision, which this paper focuses on, is how the $\mathbf{\overline{x}}$ can be encoded. In general, existing work takes one of the following three encoding schemes:

   \textbf{Label Encoding:} This is a widely used scheme~\citep{DBLP:journals/corr/abs-1801-02175,DBLP:conf/sigsoft/SiegmundGAK15,DBLP:conf/sigsoft/NairMSA17}, where each of the configuration options occupies one dimension in $\mathbf{\overline{x}}$. Taking \textsc{MongoDB} as an example, its configuration can be represented as $\mathbf{\overline{x}}=($\texttt{cache\_size}$,$ \texttt{interval}$,$ \texttt{ssl}$,$ \texttt{data\_strategy}$)$ where \texttt{cache\_size}$=(1,10,10000)$, \texttt{interval}$=(1,2,3,4)$, \texttt{ssl}$=(0,1)$ and \texttt{data\_strategy}$=(str\_l1,$ $str\_l2,str\_l3)$. A configuration that is used as a training sample could be $(10000,2,1,1)$, where the $\texttt{data\_}$$\texttt{strategy}$ can be converted into numeric values of $(0,1,2)$.
    
     \textbf{Scaled Label Encoding:} This is a variant of the label encoding used by a state-of-the-art approach~\citep{DBLP:conf/icse/Chen19b,DBLP:journals/tse/ChenB17,DBLP:conf/icse/HaZ19}, where each configuration also takes one dimension in $\mathbf{\overline{x}}$. The only difference is that all configurations are normalized to the range between 0 and 1. The same example configuration above for label encoding would be scaled to $(1, \frac{1}{3}, 1, 0.5)$.
    
  \textbf{One-hot Encoding:} Another commonly followed scheme~\citep{DBLP:conf/icse/SiegmundKKABRS12,DBLP:conf/kbse/GuoCASW13,DBLP:conf/kbse/BaoLWF19} such that each dimension in $\mathbf{\overline{x}}$ refers to the binary form of a particular value for a configuration option. Using the above example of \textsc{MongoDB}, the representation becomes $\mathbf{\overline{x}}=($\texttt{cache\_size\_v1}$,$ \texttt{cache\_size\_v2}$,...)$. Each dimension, e.g., \texttt{cache\_size\_v1}, would have a value of 1 if it is the one that the corresponding configuration chooses, otherwise it is 0. As such, the same configuration $(10000,2,1,1)$ in the label encoding would be represented as $(0,0,1,0,1,0,0,0,1,0,1,0)$ in the one-hot encoding.


Clearly, for binary options, the three encoding methods would be identical. Hence in this chapter, the focus is on the systems that also come with complex numeric and categorical configuration options.

\subsection{Why Study Them?}

Despite the prevalence of the three encoding schemes, existing work often uses one of them without justifying their choice for learning software performance~\citep{DBLP:conf/icse/Chen19b,DBLP:journals/tse/ChenB17,DBLP:journals/corr/abs-1801-02175,DBLP:conf/sigsoft/SiegmundGAK15,DBLP:conf/sigsoft/NairMSA17,DBLP:conf/icse/SiegmundKKABRS12,DBLP:conf/kbse/GuoCASW13,DBLP:conf/kbse/BaoLWF19}, particularly relating to the accuracy and training time required for the model. Some studies have mentioned the rationals, but a common agreement on which one to use has not yet been drawn. For example, \citet{DBLP:conf/kbse/BaoLWF19} state that for categorical configuration options, e.g., \texttt{cache\_mode} with three values (\texttt{memory}, \texttt{disk}, \texttt{mixed}), the label encoding unnecessarily assume a natural ordering between the values, as they are represented as $1$, $2$, and $3$. Here, one-hot encoding should be chosen. However, \citet{DBLP:journals/jmlr/AlayaBGG19} argue that the one-hot encoding can easily suffer from the multicollinearity issue on categorical configuration options, i.e., it is difficult to handle options interaction. For numeric configuration options, label encoding may fit well, as it naturally comes with order, e.g., the \texttt{cache\_size} in \textsc{MongoDB}, which has a set of values $(1,10,10000)$. However, the values, such as the above, can be of largely different scales and thus degrade numeric stability. Indeed, using one-hot encoding could be robust to this issue, but it loses the ordinal property of the numeric configuration option~\citep{DBLP:conf/sigsoft/SiegmundGAK15}. Similarly, scaled label encoding could reduce the instability and improve the prediction performance~\citep{pan2016impact}, but it also weakens the interactions between the scaled options and the binary options (as they stay the same). Therefore, there is still no common agreement (or insights) on which encoding scheme to use under what circumstances for learning performance models.


Unlike other domains, software configuration is often highly sparse, leading to unusual data distributions. Specifically, a few configuration options could have a large influence on the software performance, while the others are trivial, which makes the decision of encoding scheme difficult. Moreover, it is often the case that the nature of every configuration option may not be fully understood, as the software may be off-the-shelf or close-sourced; hence, one may not be able to choose the right encoding based on purely theoretical understandings. As such, a high-level guideline on choosing the encoding scheme for performance modeling, which gives overall suggestions for the practitioners, is in high demand. 

The above thus motivates this empirical study, aiming to analyze the effectiveness of encoding schemes across various subject systems and machine learning models, summarizing the common behaviors of the encoding methods, and providing actionable advice based on the learning models applied as well as the requirements, e.g., accuracy and training time.


%% file: Chapter-encoding/methodology.tex
\section{Methodology}
\label{chap-encoding:methodology}

This section will discuss the methodology and experimental setup of the empirical strategy for the study.


\subsection{System and Data Selection}
The following criteria are set to select sampled data of configurable systems and their environments when comparing the three encoding schemes:

\begin{enumerate}

\item To promote the reproducibility,
the systems should be open-sourced and the data should be hosted in public repositories, e.g., GitHub.

\item The system and its environment should have been widely used and tested in existing work.

\item To ensure a case where the encoding schemes can create sufficiently different representations, the system should have at least 10\% configuration options that are not categorical/binary.

\item To promote the robustness of the experiments, the subject systems should have different proportions of configuration options that are numerical.

\item To guarantee the scale of the study, only systems with more than 5,000 configuration samples are considered.

\end{enumerate}

Systems and their data from recent studies on software configuration tuning and modeling~\citep{DBLP:journals/corr/abs-1801-02175, DBLP:journals/corr/abs-2106-02716} were shortlisted, from which five systems and their environment were identified according to the above criteria, as shown in Table~\ref{tb:subj}. The five systems contain different percentages of categorical/binary and numeric configuration options while covering five distinct domains. 

Note that since the measurement and sampling process for configurable software is usually rather expensive, e.g., \citet{DBLP:conf/icml/ZuluagaSKP13} report that the synthesis of only one software configuration design can take hours or even days, in practice it is not necessarily always possible to gather an extremely large number of data samples. Further, using the full samples for some systems with a large configuration space can easily lead to unrealistic training time for certain models, e.g., with Neural Network, it took several days to complete only one run under the learning pipeline on the full datasets of \textsc{Trimesh}. Therefore in this work, for each system, 5,000 configurations are randomly sampled from the dataset as the experiment data, which tends to be reasonable and is also a commonly used setting in previous work~\citep{DBLP:conf/kbse/DornAS20,DBLP:conf/im/JohnssonMS19, DBLP:conf/icdm/ShaoWL19, DBLP:conf/icse/Gerostathopoulos18}.

\begin{table}
  \caption{Datasets of configurable software systems used. $\lvert \mathbfcal{O} \rvert$ (C/N) denotes the number of categorical (including binary) / numerical options.}
  \footnotesize
  \setlength{\tabcolsep}{1.4mm} 
  \label{tb:subj}
  \begin{adjustbox}{width=1.1\linewidth,center}
  \begin{tabular}{ccccp{5cm}}
    \toprule
    \textbf{Dataset}&$\lvert \mathbfcal{O} \rvert$ (C/N)& \textbf{Performance} & \textbf{Description}&\textbf{Used by}\\
    \midrule
    \textsc{MongoDB} & 14/2 & runtime (ms) & NoSQL database & \citep{DBLP:journals/corr/abs-2106-02716}\\
    
    \textsc{Lrzip} & 9/3 & runtime (ms) &  compression tool & \citep{DBLP:journals/corr/abs-2106-02716}\\
    
    \textsc{Trimesh} & 9/4 & runtime (ms) & triangle meshes library& \citep{DBLP:journals/corr/abs-1801-02175}, \citep{ChenMMO21}\\

    \textsc{ExaStencils} & 4/6 & latency (ms) & stencil code generator& \citep{DBLP:journals/corr/abs-2106-02716}\\
    
    \textsc{x264} & 4/13 & energy (mW) & a video encoder & \citep{DBLP:journals/corr/abs-1801-02175}, \citep{ChenMMO21}\\
  \bottomrule
\end{tabular}
\end{adjustbox}
\end{table}

\subsection{Learning Models}
\label{chap-encoding:learning_models}


In this work, the most common learning models that are of different types are chosen as used in prior studies:

\begin{itemize}

\item\textbf{Linear Model:} This type of model builds the correlation between configuration options and performance under certain linear assumptions. 

\begin{itemize}
    \item\textbf{Linear Regression (LR):} A multi-variable linear regression model that linearly correlates the configurations and their options to make prediction. It has been used by~\citet{DBLP:conf/icse/Chen19b,DBLP:journals/tse/ChenB17,DBLP:conf/icse/SiegmundKKABRS12}. There are three hyperparameters to tune, e.g., \texttt{n\_jobs}.
\end{itemize}

\item\textbf{Deep Learning Model:} A model that is based on multiple layers of perceptrons to learn and predict the concepts. 

\begin{itemize}

\item\textbf{Neural Network (NN):} A network structure with layers of neurons and connections representing the flow of data. The weights incorporate the influences of each input unit and the interactions between them. The NN models have been shown to be successful for modeling software performance, e.g.,~\citet{DBLP:conf/icse/HaZ19, fei2016compressor}. In this work, the same network setting and hyperparameter tuning method from the work by \citet{DBLP:conf/icse/HaZ19} are utilized. 

\end{itemize}

\item\textbf{Tree Model:} This model constructs a tree-like structure with a clear decision boundary on the branches.

\begin{itemize}
    \item\textbf{Decision Tree (DT):} A regression tree model that recursively partitions the configurations space to predict the output, which is used by~\citep{DBLP:conf/icse/Chen19b,DBLP:journals/tse/ChenB17,DBLP:journals/corr/abs-1801-02175,DBLP:conf/kbse/GuoCASW13}. DT is tuned using three hyperparameters, e.g., \texttt{min\_samples\_split}.

\item\textbf{Random Forest (RF):} An ensemble of decision trees, each of which learns a subset of samples or configurations space. It is a widely used model for performance learning~\citep{DBLP:conf/splc/ValovGC15,DBLP:conf/oopsla/QueirozBC16}. There are three hyperparameters to tune, e.g., \texttt{n\_estimators}

\end{itemize}

\item\textbf{Lazy Model:} This model delays the learning until the point of prediction.

\begin{itemize}

\item\textbf{$k$-Nearest Neighbours ($k$NN):} A model that considers only already measured neighboring configurations to produce a prediction, which has been commonly used~\citep{DBLP:journals/software/KalteneckerGSA20}. It has four hyperparameters to be tuned, such as \texttt{n\_neighbors}.

\end{itemize}

\item\textbf{Kernel Model:} This model performs learning and prediction by means of a kernel function.

\begin{itemize}
\item\textbf{Kernel Ridge Regression (KRR):} A model of kernel transformation that is combined with ridge
regression, which is the $L_2$-norm regularization. It has been used by~\citet{DBLP:journals/software/KalteneckerGSA20}.  There are three hyperparameters to tune, such as \texttt{alpha}.

\item\textbf{Supportt Vector Regression (SVR):} A model that transforms the configurations space into a higher-dimensional space via the kernel function, as used by~\citet{DBLP:conf/icse/Chen19b,DBLP:conf/splc/ValovGC15}. It contains hyperparameters, e.g., \texttt{kernel\_func}.
\end{itemize}

\end{itemize}

The reasons for evaluating other machine learning models in addition to deep neural networks in this chapter are two-fold: (1) they have been widely used in previous works for configuration performance learning, therefore ensure a better generalizability of the findings about encoding schemes, and (2) they exist in standard implementation under the same and widely-used machine learning library, i.e., \texttt{Sklearn}~\citep{DBLP:journals/jmlr/PedregosaVGMTGBPWDVPCBPD11} and \texttt{Tensorflow}, which reduces the possibility of bias. Note that the aim is not to be exhaustive, but to focus on those that are the most prevalent ones such that the potential impact of this study can be maximized.

\subsection{Metrics}

Different metrics exist for measuring the accuracy of a prediction model. In this study, RMSE is used because of two reasons: (1) it is a widely used metric for performance modeling of configurable systems in prior work~\citep{DBLP:conf/mascots/GrohmannSELKD20,DBLP:conf/cloudcom/IorioHTA19}; and (2) it has been reported that RMSE can reveal the performance difference better, compared with its popular counterparts such as Mean Relative Error~\citep{chai2014root}. RMSE is calculated as:
\begin{equation}
    RMSE = \sqrt{ \frac{1}{N}\sum_{i=1}^{N} (x_{i} - \hat{x}_{i})^2}
\end{equation}
whereby $x_{i}$ and $\hat{x}_{i}$ are the actual and predicted performance value, respectively; $N$ denotes that total number of testing data samples.

As for the training time, the time taken to complete the training process is reported, including hyperparameter tuning and prepossessing as necessary.

\subsection{Learning Pipeline Setting}
\label{chap-encoding:pipeline_setting}

\begin{figure}[t!]
\centering
\includegraphics[width=0.85\columnwidth]{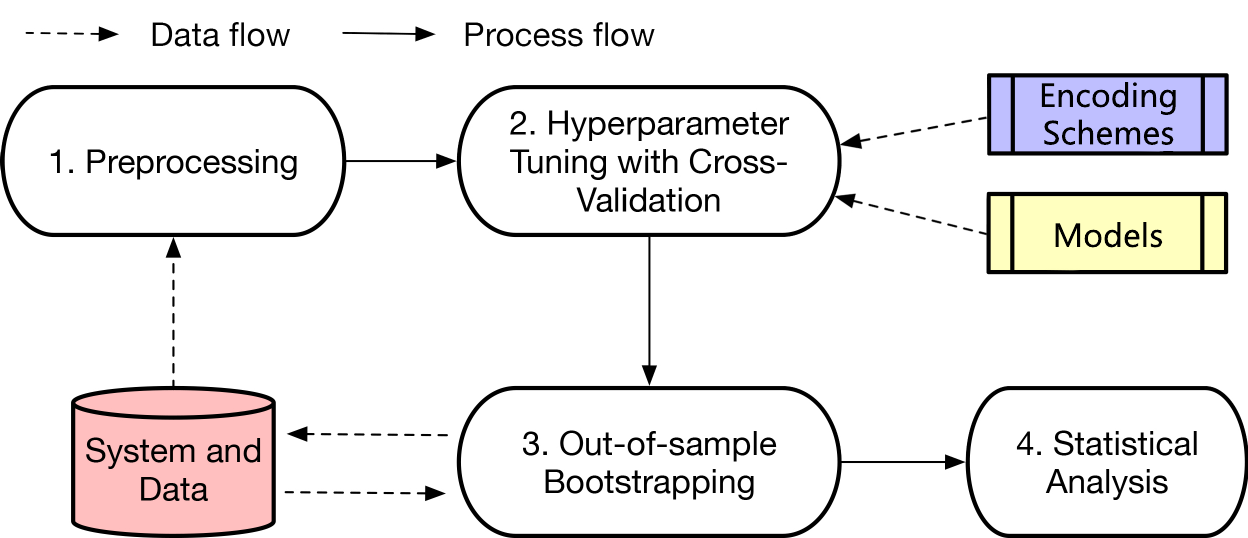}
\caption{The learning pipeline in this study.}
\label{fig:wf}
\end{figure}

As shown in Figure~\ref{fig:wf}, the standard learning pipeline setting in the empirical study has several key steps as specified below:

\begin{enumerate}
    \item\textbf{Preprocessing:} For label and one-hot encoding, the standard encoding functions from the \texttt{Sklearn} library are utilized. For the scaled label encoding, the configurations are normalized using the \textit{max-min scaling}, such that an option value $v$ is standardized as $v= {{v - v_{min}} \over {v_{max}- v_{min}}}$, where $v_{max}$ and $v_{min}$ denote the maximum and minimum bound, respectively. In this way, the values of each configuration option can be normalized within the range between $0$ and $1$. This normalization method is chosen because (1) it is computationally efficient, (2) it is easy to understand and implement, and (3) according to our systematic literature review in Section~\ref{subsec: preprocess}, max-min scaling is the most commonly-used normalization technique, used in 25 out of 32 studies. Moreover, The state-of-the-art learning pipeline is followed such that all configuration options and their values are considered in the model ~\citep{DBLP:journals/tse/ChenB17,DBLP:journals/corr/abs-1801-02175,DBLP:conf/sigsoft/SiegmundGAK15,DBLP:conf/sigsoft/NairMSA17,DBLP:conf/kbse/BaoLWF19}.

   \item\textbf{Hyperparameter Tuning:} It is not uncommon that a model comes with at least one hyperparameter~\citep{DBLP:conf/icse/LiX0WT20}. Therefore, the common practice of the pipeline for learning software performance is to tune them under all encoding schemes~\citep{DBLP:journals/tse/ChenB17,DBLP:journals/corr/abs-1801-02175,DBLP:conf/sigsoft/SiegmundGAK15,DBLP:conf/sigsoft/NairMSA17,DBLP:conf/kbse/BaoLWF19}. In this study, the \texttt{GridSearchCV} function from \texttt{Sklearn} is used, which is an exhaustive grid search that evaluates the model quality via 10-fold cross-validation on the training dataset. The one that leads to the best result is used. Note that the default values are always used as a starting point.

    \item\textbf{Bootstrapping:} To achieve a reliable conclusion, this study conducted out-of-sample bootstrap (without replacement). In particular, 90\% of the data are randomly sampled as the training dataset, those samples that were not included in the training were used as the testing samples. The process was repeated 50 times, i.e., there are 50 runs of RMSE (on the testing dataset) and training time to be reported. For each run, all encoding schemes are examined, thereby ensuring that they are evaluated under the same randomly sampled training and testing dataset.

    \item\textbf{Statistical Analysis:} To ensure statistical significance in multiple comparisons, Scott-Knott test~\citep{DBLP:journals/tse/MittasA13} is applied on all comparisons of over 50 runs and produces a score. In a nutshell, Scott-Knott sorts the list of treatments (the learning model-encoding pairs) by their median RMSE/training time. Next, it splits the list into two sub-lists with the largest expected difference~\citep{xia2018hyperparameter}. Suppose to compare \texttt{NN\_onehot}, \texttt{RF\_onehot}, and \texttt{NN\_label}, a possible split could be: $\{$\texttt{NN\_onehot}, \texttt{RF\_onehot}$\}$, $\{$\texttt{NN\_label}$\}$, with the score of 2 and 1, respectively. This means that, in the statistical sense, \texttt{NN\_onehot}~and \texttt{RF\_onehot} perform similarly, but they are significantly better than \texttt{NN\_label}. Formally, the Scott-Knott test aims to find the best split by maximizing the difference $\Delta$ in the expected mean before and after each split:
\begin{equation}
    \Delta = \frac{|l_1|}{|l|}(\overline{l_1} - \overline{l})^2 + \frac{|l_2|}{|l|}(\overline{l_2} - \overline{l})^2
\end{equation}
whereby $|l_1|$ and $|l_2|$ are the sizes of two sub-lists ($l_1$ and $l_2$) from list $l$ with a size $|l|$. $\overline{l_1}$, $\overline{l_2}$, and $\overline{l}$ denote their mean RMSE/training time values.

During the splitting, a statistical hypothesis test $H$ is applied to check if $l_1$ and $l_2$ are significantly different. This is done by using bootstrapping and $\hat{A}_{12}$~\citep{Vargha2000ACA}. If that is the case, Scott-Knott recurses on the splits. In other words, the treatments are divided into different sub-lists if both bootstrap sampling and effect size test suggest that a split is statistically significant (with a confidence level of 99\%) and not a small effect ($\hat{A}_{12} \geq 0.6$). The sub-lists are then scored based on their mean RMSE/training time. The higher the score, the better the treatment.

\end{enumerate}

Since there are five systems and environments, together with seven models and three encoding schemes, the empirical study consists of 105 cases of investigation. All the experiments were performed on a Windows 10 server with an Intel Core i5-9400 CPU 2.90GHz and 8GB RAM.

%% file: Chapter-encoding/evaluation.tex
\section{Analysis and Results}
\label{chap-encoding:analysis and results}
This section discusses the results of the empirical study with respect to the RQs.

\input{Tables/chap-encoding/rq2-plot.tex}

\begin{figure}[t]
\centering
\includegraphics[width=0.9\columnwidth]{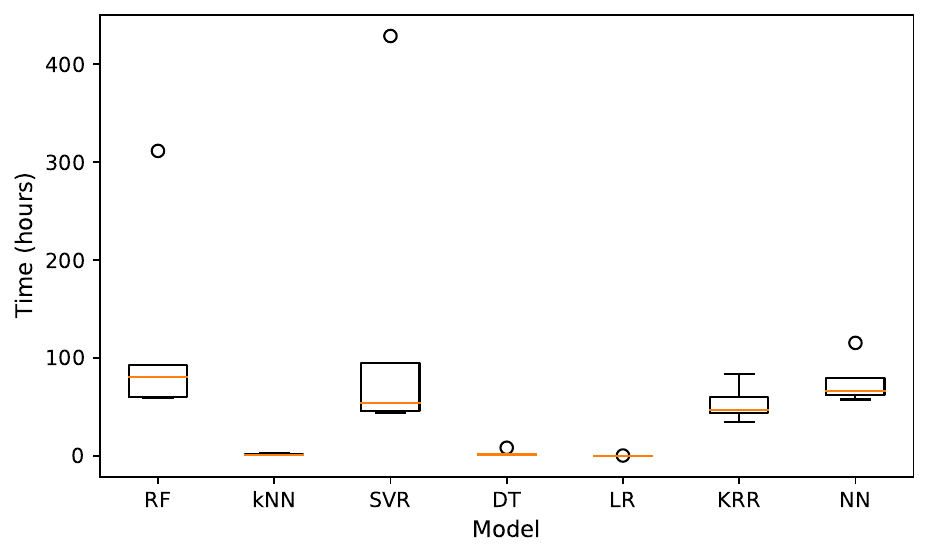}
\caption{The boxplot of the total time required for identifying the best encoding scheme (with respect to a model) over all systems studied.}
\label{fig:total-time}
\end{figure}

\subsection{RQ2.1: Cost of Trial-and-Error}
\subsubsection{Method}

To answer \textbf{RQ2.1}, for each encoding scheme, the time taken to complete all 50 runs under a model and system (including training, hyperparameter tuning, and evaluation) is recorded. To identify the best encoding scheme using trial-and-error in a case-by-case manner, the ``efforts'' required would be the total time taken for evaluating a model under all encoding schemes for a system.

\input{Tables/chap-encoding/rq3-plot.tex}

\subsubsection{Results}

Figure~\ref{fig:total-time} shows the result, from which some clear evidence can be obtained:

\begin{itemize}
    \item \textbf{Finding 1:} It can take an extremely long time to conclude which encoding scheme is better depending on the models: this is almost 100 hours (median) for RF and around 80 hours (median) for SVR in general; it can go up to 400$+$ hours on some systems. For KRR and NN, which take less time to do so, it still requires around at least two and a half days (60$+$ hours).
    \item \textbf{Finding 2:} For certain models, it may be possible to find the best encoding scheme. For example, it takes less than an hour for $k$NN, DT, and LR due to their low computational needs. Yet, whether one would be willing to spend valuable development time on this is really case-dependent.
\end{itemize}

The above confirms that finding the best encoding scheme for learning software performance can be non-trivial and the needs of the study. Therefore, for \textbf{RQ2.1}, it can be said that:

\begin{quotebox}
   \noindent
   \textit{\textbf{RQ2.1:} Depending on the model, finding the best encoding scheme using trial-and-error can be highly expensive, as it may take 60$+$ hours (median) and up to 400$+$ hours. However, for other cases, the ``effort" only needs less than one hour, which may be acceptable depending on the scenario.}
\end{quotebox}



\subsection{RQ2.2: Accuracy}
\label{sec:rq2-acc}

  

\begin{figure*}[t]
\centering

\begin{subfigure}{.33\textwidth}
  \centering
  \includegraphics[width=\linewidth]{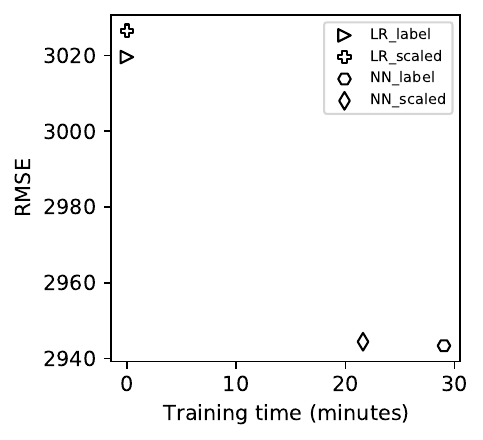}  
  \caption{\textsc{MongoDB}}
  \label{fig:tradeoff-a}
\end{subfigure}
~\hspace{-0.6cm}
\begin{subfigure}{.34\textwidth}
  \centering
  \includegraphics[width=\linewidth]{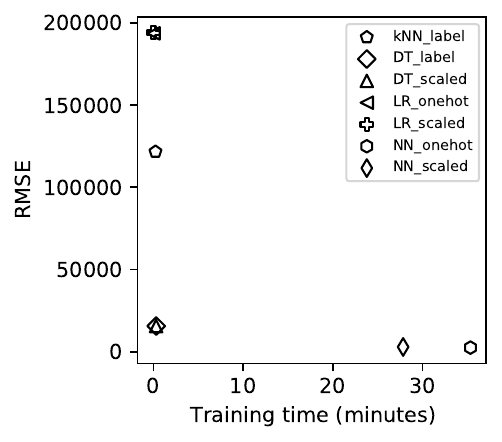} 
  \caption{\textsc{Lrzip}}
  \label{fig:tradeoff-b}
\end{subfigure}
~\hspace{-0.5cm}
\begin{subfigure}{.325\textwidth}
  \centering
  \includegraphics[width=\linewidth]{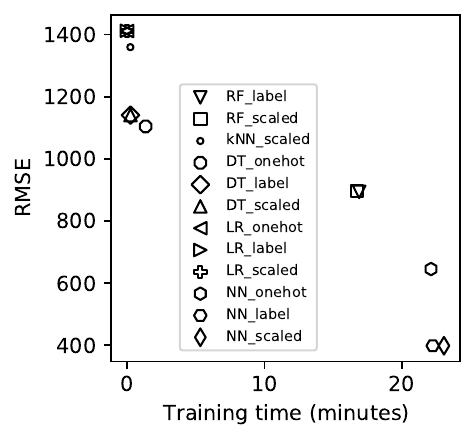} 
  \caption{\textsc{Trimesh}}
  \label{fig:tradeoff-c}
\end{subfigure}


\begin{subfigure}{.33\textwidth}
  \centering
  \includegraphics[width=\linewidth]{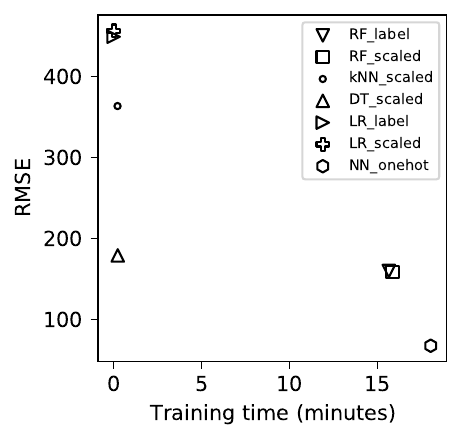}  
  \caption{\textsc{ExaStencils}}
  \label{fig:tradeoff-d}
\end{subfigure}
\begin{subfigure}{.34\textwidth}
  \centering
  \includegraphics[width=\linewidth]{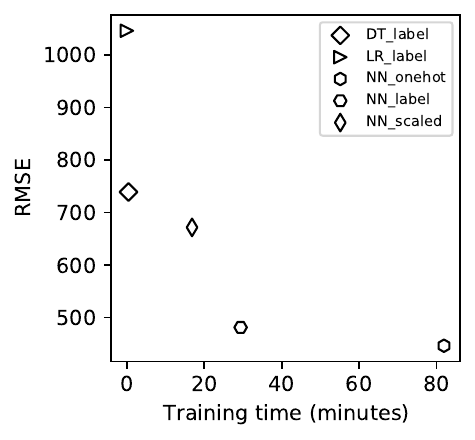} 
  \caption{\textsc{x264}}
  \label{fig:tradeoff-e}
\end{subfigure}

    \caption{The trade-off between RMSE and training time over all Pareto-optimal model-encoding pairs.}
    \label{fig:tradeoff}
\end{figure*}

\subsubsection{Method}
To study \textbf{RQ2.2}, all RMSE values for the three encoding schemes under the models and systems are compared. That is, for each subject system, there are $3 \times 7 = 21$ pairs of model-encoding (50 RMSE repeats each). To ensure statistical significance among the comparisons, Scott-Knott test is used to assign a score for each pair, hence, similar ones are clustered together (same score) while different ones can be ranked (the higher score, the better).

\subsubsection{Results}

As illustrated in Table~\ref{tb:rq2-plot}, some interesting findings are observed:

\begin{itemize}
    \item \textbf{Finding 3:} From Table~\ref{tb:rq2-plot}f, overall, label and one-hot encoding are clearly more accurate than scaled label encoding across the models, as the former two have the best total Scott-Knott scores for all models over the systems studied. Between these two, one-hot encoding tends to be slightly better across all models, as it wins on 4 models against the 3 wins by label encoding. A similar trend can be observed from Table~\ref{tb:rq2-plot}a to~\ref{tb:rq2-plot}e over the systems.

    \item \textbf{Finding 4:} For different models in Table~\ref{tb:rq2-plot}f, it is observed that one-hot encoding is the best for deep learning, lazy, and kernel models, while label encoding is preferred on linear and tree models. The same has also been registered in Table~\ref{tb:rq2-plot}a to~\ref{tb:rq2-plot}e.
    \item \textbf{Finding 5:} From Table~\ref{tb:rq2-plot}a to~\ref{tb:rq2-plot}e, NN is clearly amongst the top models on Scott-Knott score and RMSE regardless of the encoding schemes and systems. In particular, when NN is chosen, \texttt{NN\_onehot} is the best, as it has a better Scott-Knott score than the other two on 3 out of 5 systems, draw on one system and lose on the remaining one, leading to a 75\% cases of no worse outcome than \texttt{NN\_label} and \texttt{NN\_scaled}.  
\end{itemize}

To conclude, \textbf{RQ2.2} can be answered as:

\begin{quotebox}
   \noindent
   \textit{\textbf{RQ2.2:} In general, the one-hot encoding tends to have the best accuracy, and the scaled label encoding should be avoided. In particular, \texttt{NN\_onehot} is the safest option for the overall optimal accuracy among the subjects studied.}
\end{quotebox}

\subsection{RQ2.3: Training Time}
\label{sec:rq3-time}

\subsubsection{Research}
Similar to \textbf{RQ2.2}, here, the training time over 50 runs for all 21 pairs of model-encoding for each system is measured. 

\subsubsection{Results}

With Table~\ref{tb:rq3-plot}, some patterns can be observed:

\begin{itemize}
    \item \textbf{Finding 6:} Overall, from Table~\ref{tb:rq3-plot}f, label and scaled label encoding are much faster to train than their one-hot counterpart, which has never won the other two under any model across the systems. In particular, scaled label encoding appears to have the fastest training than others in general, as the former wins on 4 models, draws on one, and loses only on two. Similar results have been obtained in Table~\ref{tb:rq3-plot}a to~\ref{tb:rq3-plot}e.
    
    
    \item \textbf{Finding 7:} For different model types, in Table~\ref{tb:rq3-plot}f, the label encoding tends to be the best option in terms of training time for tree model and lazy model; the scaled label counterpart is faster on deep learning model, linear model, and kernel model. This is similar to that from Table~\ref{tb:rq3-plot}a to~\ref{tb:rq3-plot}e.
    
    \item \textbf{Finding 8:} From Table~\ref{tb:rq3-plot}a to~\ref{tb:rq3-plot}e, unexpectedly LR has the fastest training time over all systems and this model works the best with scaled label encoding since \texttt{LR\_scaled} is the fastest on 4 out of 5 systems; its difference to \texttt{LR\_label} tends to be marginal though. 
\end{itemize}

Therefore, it is said that:

\begin{quotebox}
   \noindent
   \textit{\textbf{RQ2.3:} The scaled label encoding tends to have the fastest training, while one-hot encoding takes the longest time to train. In particular, \texttt{LR\_scaled} is the best choice for the overall fastest training time over the subjects studied.}
\end{quotebox}

\subsection{RQ2.4: Trade-off Analysis}
\label{sec:rq4-tradeoff}

\subsubsection{Method}

\begin{figure}[t]
\centering

\begin{subfigure}{.31\columnwidth}
  \centering
  \includegraphics[width=\linewidth]{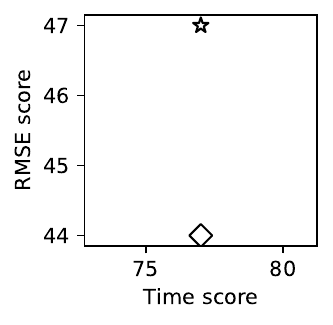}  
  \caption{\texttt{DT}}
  \label{fig:rank-a}
\end{subfigure}
\begin{subfigure}{.31\columnwidth}
  \centering
  \includegraphics[width=\linewidth]{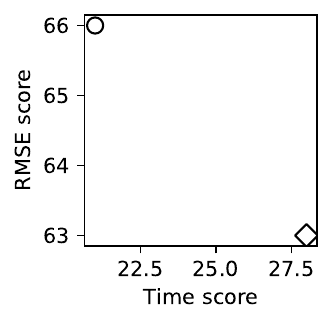} 
  \caption{\texttt{NN}}
  \label{fig:rank-b}
\end{subfigure}
\begin{subfigure}{.31\columnwidth}
  \centering
  \includegraphics[width=\linewidth]{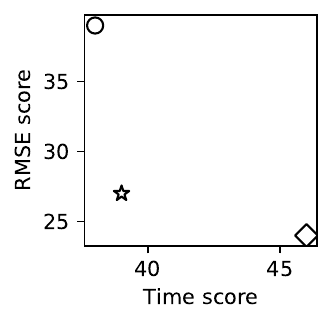} 
  \caption{\texttt{KRR}}
  \label{fig:rank-c}
\end{subfigure}


\begin{subfigure}{.30\columnwidth}
  \centering
  \includegraphics[width=\linewidth]{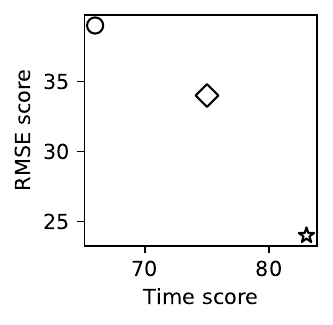}  
  \caption{\texttt{$k$NN}}
  \label{fig:rank-d}
\end{subfigure}
\begin{subfigure}{.33\columnwidth}
  \centering
  \includegraphics[width=\linewidth]{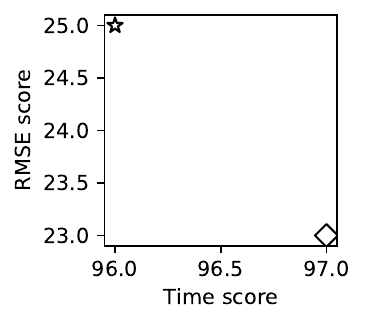} 
  \caption{\texttt{LR}}
  \label{fig:rank-e}
\end{subfigure}
\begin{subfigure}{.31\columnwidth}
  \centering
  \includegraphics[width=\linewidth]{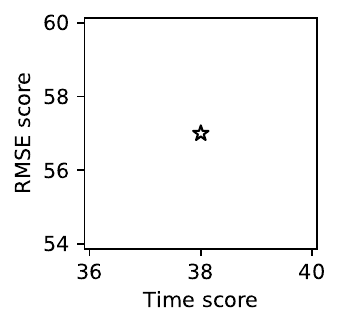}  
  \caption{\texttt{RF}}
  \label{fig:rank-f}
\end{subfigure}

\begin{subfigure}{.31\columnwidth}
  \centering
  \includegraphics[width=\linewidth]{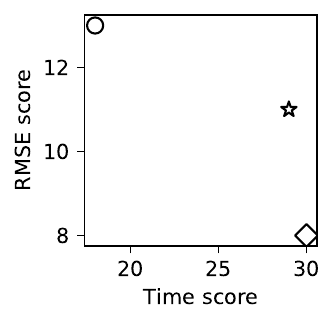} 
  \caption{\texttt{SVR}}
  \label{fig:rank-g}
\end{subfigure}
~\hspace{1cm}
\begin{subfigure}{.31\columnwidth}
  \centering
  \includegraphics[width=\linewidth]{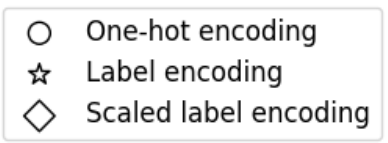} 
  ~\vspace{2cm}
  \label{fig:rank-h}
\end{subfigure}

    \caption{The total Scott-Knott scores for each model over all systems. Only the Pareto-optimal choice is presented.}
    \label{fig:rank}
\end{figure}

Understanding \textbf{RQ2.4} requires to simultaneously consider the accuracy and training time achieved by all 21 pairs of model-encoding. According to the guidance provided by ~\citet{Li2020}, for each system, the study seeks to analyze the Pareto optimal choices as those are the ones that require trade-offs. Suppose that a pair $P_x$ has $\{A_x, T_x\}$ and another $P_y$ comes with $\{A_y, T_y\}$, whereby $A_x$ and $A_y$ are their median RMSE (over 50 runs) while $T_x$ and $T_y$ are their median training time, respectively. It says $P_x$ dominates $P_y$ if $A_x \leq A_y$ and $T_x \leq T_y$ while there is either $A_x < A_y$ or $T_x < T_y$. A pair that is not dominated by any other pairs from the total set of 21 is called a Pareto-optimal pair therein. The set of all Pareto-optimal points is called the Pareto front (Figures~\ref{fig:tradeoff}). The Pareto front is also plotted with respect to the total Scott-Knott scores (over all systems) under each model (Figures~\ref{fig:rank}).

Here, a Pareto-optimal pair that has the best accuracy or the fastest training time is called a biased point (or an extreme point). Among others, this thesis is interested in the non-extreme, less biased points, especially those with a well-balanced trade-off.

\subsubsection{Results}

The results are illustrated in Figures~\ref{fig:tradeoff} and \ref{fig:rank}, from which some interesting observations can be obtained:

\begin{itemize}

    \item \textbf{Finding 9:} Over all the model-encoding pairs (Figures~\ref{fig:tradeoff}), label and scaled label encoding can more commonly lead to less biased results (non-extreme points) in the Pareto front than their one-hot counterpart. This can clearly offer more trade-off choices.

    \item \textbf{Finding 10:} In Figures~\ref{fig:tradeoff}, the scaled label encoding tends to achieve a more balanced trade-off than the others, but the paired model may vary, i.e., it can be NN, DT, or RF, across the systems. For example, it is \texttt{NN\_scaled} on \textsc{MongoDB} but becomes \texttt{DT\_scaled} on \textsc{Lrzip}.
    
    \item \textbf{Finding 11:} Kernel models like KRR and SVR have never produced Pareto-optimal outcomes over the 5 systems studied (Figures~\ref{fig:tradeoff}).
    
    \item \textbf{Finding 12:} From Figure~\ref{fig:rank}, only kernel and lazy models have relatively less biased points in their own Pareto front, which are achieved by label (Figure~\ref{fig:rank}c and~\ref{fig:rank}g) and scaled label encoding (Figure~\ref{fig:rank}d), respectively.
    
    
\end{itemize}

In summary, there are:

\begin{quotebox}
   \noindent
   \textit{\textbf{RQ2.4:} For all the model-encoding pairs, label, and scaled label encoding tend to be less biased to accuracy or training time than their one-hot counterpart. In particular, scaled label encoding can lead to relatively more balanced outcomes. However, the paired model for the above may differ depending on the system, but it would never be KRR or SVR, which produce no Pareto-optimal pairs.}
\end{quotebox}

%% file: Tables/chap-encoding/rq2-plot.tex
\begin{sidewaystable}[]
      \caption{Scott-Knott test, Med (median), and Interquartile Range (IQR) on the RMSE of all models and systems, where ``onehot'', ``label'', and ``scaled'' stand for one-hot, label and scaled label encoding, respectively. A higher score means the RMSE is lower and, therefore, better. For (a) to (e), the pairs are sorted by score, median, and then IQR. For (f), \setlength{\fboxsep}{1.5pt}\colorbox{red!10}{red} highlights the best encoding scheme for a model over all systems}
    \label{tb:rq2-plot}
  \begin{center}
    \begin{adjustbox}{max width = 0.85\textwidth}
\footnotesize
    \begin{tabular}{c@{}c@{}c}
        \begin{tabular}{lcccl}
           \cellcolor[gray]{1}\textbf{Pair} & \cellcolor[gray]{1}\textbf{Score} &     \cellcolor[gray]{1}\textbf{Med} & \cellcolor[gray]{1}\textbf{IQR} & \cellcolor[gray]{1}\\
            \hline
\texttt{NN\_label} & 10 & 2943.38 & 0.30 & \quart{6.57}{0.30}{6.73}{100.00} \\
\texttt{NN\_scaled} & 10 & 2944.45 & 0.39 & \quart{6.52}{0.39}{6.73}{100.00} \\
\texttt{NN\_onehot} & 10 & 2951.74 & 0.31 & \quart{6.64}{0.31}{6.75}{100.00} \\
\texttt{KRR\_onehot} & 9 & 3014.68 & 0.19 & \quart{6.78}{0.19}{6.89}{100.00} \\
\texttt{RF\_label} & 8 & 3019.52 & 0.20 & \quart{6.79}{0.20}{6.91}{100.00} \\
\texttt{LR\_label} & 8 & 3019.55 & 0.19 & \quart{6.80}{0.19}{6.91}{100.00} \\
\texttt{RF\_scaled} & 8 & 3019.63 & 0.20 & \quart{6.79}{0.20}{6.91}{100.00} \\
\texttt{LR\_onehot} & 8 & 3020.36 & 0.21 & \quart{6.79}{0.21}{6.91}{100.00} \\
\texttt{KRR\_label} & 8 & 3024.03 & 0.19 & \quart{6.81}{0.19}{6.92}{100.00} \\
\texttt{LR\_scaled} & 8 & 3026.49 & 0.19 & \quart{6.83}{0.19}{6.92}{100.00} \\
\texttt{KRR\_scaled} & 8 & 3028.21 & 0.19 & \quart{6.83}{0.19}{6.93}{100.00} \\
\texttt{RF\_onehot} & 7 & 3060.11 & 0.20 & \quart{6.86}{0.20}{7.00}{100.00} \\
\texttt{DT\_scaled} & 6 & 3093.12 & 0.21 & \quart{7.00}{0.21}{7.07}{100.00} \\
\texttt{DT\_onehot} & 6 & 3095.11 & 0.25 & \quart{7.00}{0.25}{7.08}{100.00} \\
\texttt{DT\_label} & 6 & 3097.61 & 0.22 & \quart{6.98}{0.22}{7.08}{100.00} \\
\texttt{$k$NN\_scaled} & 5 & 8968.09 & 1.54 & \quart{19.90}{1.54}{20.51}{100.00} \\
\texttt{$k$NN\_onehot} & 4 & 15772.52 & 0.85 & \quart{35.78}{0.85}{36.07}{100.00} \\
\texttt{$k$NN\_label} & 3 & 22114.97 & 0.99 & \quart{49.96}{0.99}{50.57}{100.00} \\
\texttt{SVR\_onehot} & 2 & 41917.30 & 5.73 & \quart{92.16}{5.73}{95.86}{100.00} \\
\texttt{SVR\_scaled} & 1 & 42228.88 & 5.85 & \quart{92.76}{5.85}{96.57}{100.00} \\
\texttt{SVR\_label} & 1 & 42229.54 & 5.86 & \quart{92.77}{5.86}{96.57}{100.00} \\
        \end{tabular} & 
        \begin{tabular}{|lcccl|}
           \cellcolor[gray]{1}\textbf{Pair} & \cellcolor[gray]{1}\textbf{Score} &     \cellcolor[gray]{1}\textbf{Med} & \cellcolor[gray]{1}\textbf{IQR} & \cellcolor[gray]{1}\\
            \hline
\texttt{NN\_onehot} & 12 & 2497.07 & 0.10 & \quart{0.61}{0.10}{0.67}{100.00} \\
\texttt{NN\_scaled} & 11 & 2887.66 & 0.20 & \quart{0.72}{0.20}{0.78}{100.00} \\
\texttt{NN\_label} & 10 & 3019.11 & 0.34 & \quart{0.71}{0.34}{0.81}{100.00} \\
\texttt{DT\_label} & 9 & 15621.69 & 5.92 & \quart{2.61}{5.92}{4.19}{100.00} \\
\texttt{DT\_scaled} & 9 & 15621.77 & 5.92 & \quart{2.61}{5.92}{4.19}{100.00} \\
\texttt{RF\_scaled} & 8 & 25417.20 & 3.37 & \quart{5.50}{3.37}{6.82}{100.00} \\
\texttt{RF\_label} & 8 & 25457.00 & 3.18 & \quart{5.70}{3.18}{6.83}{100.00} \\
\texttt{RF\_onehot} & 7 & 57173.71 & 5.31 & \quart{13.27}{5.31}{15.35}{100.00} \\
\texttt{DT\_onehot} & 7 & 59610.67 & 9.35 & \quart{10.61}{9.35}{16.00}{100.00} \\
\texttt{$k$NN\_label} & 6 & 121668.37 & 3.96 & \quart{30.25}{3.96}{32.66}{100.00} \\
\texttt{$k$NN\_scaled} & 5 & 151716.91 & 3.54 & \quart{38.37}{3.54}{40.73}{100.00} \\
\texttt{$k$NN\_onehot} & 4 & 168726.75 & 2.81 & \quart{43.77}{2.81}{45.30}{100.00} \\
\texttt{KRR\_onehot} & 3 & 192199.55 & 3.73 & \quart{49.63}{3.73}{51.60}{100.00} \\
\texttt{LR\_onehot} & 2 & 193658.25 & 3.88 & \quart{49.99}{3.88}{51.99}{100.00} \\
\texttt{LR\_scaled} & 2 & 194094.27 & 3.76 & \quart{50.36}{3.76}{52.11}{100.00} \\
\texttt{KRR\_scaled} & 2 & 194588.83 & 3.84 & \quart{50.46}{3.84}{52.24}{100.00} \\
\texttt{LR\_label} & 2 & 194648.94 & 3.79 & \quart{50.50}{3.79}{52.26}{100.00} \\
\texttt{KRR\_label} & 2 & 195128.40 & 3.88 & \quart{50.53}{3.88}{52.39}{100.00} \\
\texttt{SVR\_label} & 1 & 321782.26 & 6.84 & \quart{83.61}{6.84}{86.39}{100.00} \\
\texttt{SVR\_onehot} & 1 & 321878.92 & 6.86 & \quart{83.64}{6.86}{86.42}{100.00} \\
\texttt{SVR\_scaled} & 1 & 321923.45 & 6.86 & \quart{83.65}{6.86}{86.43}{100.00} \\
        \end{tabular} &
                \begin{tabular}{lcccl}
             \cellcolor[gray]{1}\textbf{Pair} & \cellcolor[gray]{1}\textbf{Score} &     \cellcolor[gray]{1}\textbf{Med} & \cellcolor[gray]{1}\textbf{IQR} & \cellcolor[gray]{1}
            \\
            \hline
            \texttt{NN\_scaled} & 12 & 398.39 & 8.37 & \quart{17.59}{8.37}{20.93}{100.00} \\
\texttt{NN\_label} & 12 & 398.39 & 8.37 & \quart{17.59}{8.37}{20.93}{100.00} \\
\texttt{NN\_onehot} & 11 & 645.28 & 9.91 & \quart{27.93}{9.91}{33.90}{100.00} \\
\texttt{RF\_label} & 10 & 893.39 & 9.60 & \quart{42.73}{9.60}{46.93}{100.00} \\
\texttt{RF\_scaled} & 10 & 896.18 & 9.25 & \quart{42.81}{9.25}{47.08}{100.00} \\
\texttt{RF\_onehot} & 9 & 947.02 & 11.43 & \quart{46.49}{11.43}{49.75}{100.00} \\
\texttt{DT\_onehot} & 8 & 1104.22 & 12.32 & \quart{51.41}{12.32}{58.01}{100.00} \\
\texttt{DT\_label} & 7 & 1140.59 & 17.93 & \quart{53.16}{17.93}{59.92}{100.00} \\
\texttt{DT\_scaled} & 7 & 1140.63 & 17.93 & \quart{53.16}{17.93}{59.92}{100.00} \\
\texttt{$k$NN\_onehot} & 6 & 1302.37 & 7.61 & \quart{63.67}{7.61}{68.42}{100.00} \\
\texttt{$k$NN\_scaled} & 5 & 1359.37 & 7.14 & \quart{68.06}{7.14}{71.41}{100.00} \\
\texttt{KRR\_onehot} & 4 & 1378.64 & 8.30 & \quart{68.47}{8.30}{72.42}{100.00} \\
\texttt{KRR\_scaled} & 3 & 1410.46 & 9.14 & \quart{70.32}{9.14}{74.10}{100.00} \\
\texttt{KRR\_label} & 3 & 1411.00 & 9.15 & \quart{70.34}{9.15}{74.12}{100.00} \\
\texttt{LR\_onehot} & 3 & 1411.28 & 10.07 & \quart{69.94}{10.07}{74.14}{100.00} \\
\texttt{LR\_scaled} & 3 & 1411.56 & 9.14 & \quart{70.39}{9.14}{74.15}{100.00} \\
\texttt{LR\_label} & 3 & 1412.19 & 9.08 & \quart{70.45}{9.08}{74.19}{100.00} \\
\texttt{$k$NN\_label} & 2 & 1502.74 & 10.07 & \quart{73.73}{10.07}{78.94}{100.00} \\
\texttt{SVR\_onehot} & 1 & 1521.08 & 11.72 & \quart{75.65}{11.72}{79.91}{100.00} \\
\texttt{SVR\_scaled} & 1 & 1525.33 & 11.75 & \quart{75.84}{11.75}{80.13}{100.00} \\
\texttt{SVR\_label} & 1 & 1525.57 & 11.74 & \quart{75.85}{11.74}{80.14}{100.00} \\
        \end{tabular} \\
        \\
         \textbf{\small (a). \textsc{MongoDB}} & \small \textbf{(b). \textsc{Lrzip}} & \textbf{\small (c). \textsc{Trimesh}}
        \\
        \\
        \begin{tabular}{lcccl}
            \cellcolor[gray]{1}\textbf{Pair} & \cellcolor[gray]{1}\textbf{Score} &     \cellcolor[gray]{1}\textbf{Med} & \cellcolor[gray]{1}\textbf{IQR} & \cellcolor[gray]{1}\\
            \hline
\texttt{NN\_onehot} & 14 & 67.62 & 0.88 & \quart{3.18}{0.88}{3.38}{100.00} \\
\texttt{NN\_label} & 13 & 86.18 & 0.61 & \quart{4.05}{0.61}{4.31}{100.00} \\
\texttt{NN\_scaled} & 13 & 92.37 & 1.60 & \quart{3.97}{1.60}{4.62}{100.00} \\
\texttt{RF\_scaled} & 12 & 158.61 & 4.00 & \quart{6.63}{4.00}{7.93}{100.00} \\
\texttt{RF\_label} & 12 & 159.83 & 4.00 & \quart{6.73}{4.00}{8.00}{100.00} \\
\texttt{RF\_onehot} & 12 & 164.02 & 3.99 & \quart{6.76}{3.99}{8.21}{100.00} \\
\texttt{DT\_scaled} & 11 & 179.07 & 3.46 & \quart{8.04}{3.46}{8.96}{100.00} \\
\texttt{DT\_label} & 11 & 181.96 & 3.40 & \quart{8.20}{3.40}{9.10}{100.00} \\
\texttt{DT\_onehot} & 11 & 182.42 & 3.42 & \quart{8.23}{3.42}{9.13}{100.00} \\
\texttt{$k$NN\_scaled} & 10 & 363.34 & 2.40 & \quart{17.65}{2.40}{18.18}{100.00} \\
\texttt{KRR\_onehot} & 10 & 369.58 & 0.95 & \quart{18.17}{0.95}{18.49}{100.00} \\

\texttt{$k$NN\_onehot} & 9 & 438.83 & 1.24 & \quart{21.43}{1.24}{21.95}{100.00} \\
\texttt{$k$NN\_label} & 8 & 1008.23 & 1.43 & \quart{49.62}{1.43}{50.44}{100.00} \\
\texttt{KRR\_label} & 7 & 1113.89 & 1.61 & \quart{54.82}{1.61}{55.72}{100.00} \\
\texttt{SVR\_onehot} & 6 & 1210.97 & 1.25 & \quart{59.93}{1.25}{60.58}{100.00} \\
\texttt{KRR\_scaled} & 5 & 1306.24 & 1.56 & \quart{64.59}{1.56}{65.34}{100.00} \\
\texttt{SVR\_label} & 4 & 1495.85 & 1.82 & \quart{74.05}{1.82}{74.83}{100.00} \\
\texttt{SVR\_scaled} & 3 & 1507.12 & 1.92 & \quart{74.76}{1.92}{75.39}{100.00} \\
\texttt{LR\_scaled} & 2 & 4562.24 & 1.97 & \quart{90.10}{1.97}{92.82}{100.00} \\
\texttt{LR\_label} & 2 & 4483.84 & 2.12 & \quart{90.72}{2.12}{92.45}{100.00} \\
\texttt{LR\_onehot} & 1 & 7814.81 & 99 & \quart{90.70}{9}{99.11}{100.00} \\
        \end{tabular} & 
        \begin{tabular}{|lcccl}
           \cellcolor[gray]{1}\textbf{Pair} & \cellcolor[gray]{1}\textbf{Score} &     \cellcolor[gray]{1}\textbf{Med} & \cellcolor[gray]{1}\textbf{IQR} & \cellcolor[gray]{1}\\
            \hline
\texttt{NN\_onehot} & 19 & 446.49 & 7.94 & \quart{19.60}{7.94}{22.34}{100.00} \\
\texttt{RF\_label} & 19 & 581.18 & 6.07 & \quart{27.64}{6.07}{29.07}{100.00} \\
\texttt{NN\_label} & 18 & 481.45 & 6.97 & \quart{20.77}{6.97}{24.08}{100.00} \\
\texttt{RF\_onehot} & 17 & 665.35 & 6.60 & \quart{30.65}{6.60}{33.28}{100.00} \\
\texttt{NN\_scaled} & 17 & 671.84 & 5.32 & \quart{31.75}{5.32}{33.61}{100.00} \\
\texttt{$k$NN\_onehot} & 16 & 1021.07 & 4.49 & \quart{48.97}{4.49}{51.08}{100.00} \\
\texttt{RF\_scaled} & 15 & 735.36 & 4.61 & \quart{35.10}{4.61}{36.79}{100.00} \\
\texttt{DT\_label} & 14 & 739.20 & 6.96 & \quart{34.49}{6.96}{36.98}{100.00} \\
\texttt{KRR\_onehot} & 13 & 910.47 & 4.64 & \quart{43.52}{4.64}{45.55}{100.00} \\
\texttt{DT\_onehot} & 12 & 803.96 & 8.26 & \quart{36.70}{8.26}{40.22}{100.00} \\
\texttt{DT\_scaled} & 11 & 950.32 & 6.04 & \quart{45.36}{6.04}{47.54}{100.00} \\
\texttt{LR\_label} & 10 & 1046.27 & 3.38 & \quart{50.07}{3.38}{52.34}{100.00} \\
\texttt{$k$NN\_scaled} & 9 & 939.14 & 3.79 & \quart{45.49}{3.79}{46.98}{100.00} \\
\texttt{LR\_scaled} & 8 & 1084.04 & 3.01 & \quart{53.02}{3.01}{54.23}{100.00} \\
\texttt{KRR\_label} & 7 & 1110.85 & 3.61 & \quart{53.77}{3.61}{55.57}{100.00} \\
\texttt{KRR\_scaled} & 6 & 1183.43 & 3.39 & \quart{57.17}{3.39}{59.20}{100.00} \\
\texttt{$k$NN\_label} & 5 & 1455.11 & 4.34 & \quart{70.68}{4.34}{72.79}{100.00} \\
\texttt{SVR\_label} & 4 & 1487.67 & 4.93 & \quart{72.13}{4.93}{74.42}{100.00} \\
\texttt{SVR\_onehot} & 3 & 1566.62 & 5.47 & \quart{76.28}{5.47}{78.37}{100.00} \\
\texttt{SVR\_scaled} & 2 & 1714.36 & 5.89 & \quart{83.50}{5.89}{85.76}{100.00} \\
\texttt{LR\_onehot} & 1 & 9999.99 & 17.91 & \quart{90}{17}{99}{100.00} \\
        \end{tabular}& 
        \begin{tabular}{|lcccl}
           \cellcolor[gray]{1}\textbf{Pair} & \cellcolor[gray]{1}\textbf{Total Score} &     
           & & \\
            \hline
        
\cellcolor{red!10}\texttt{NN\_onehot} & \cellcolor{red!10}66 &\cellcolor{red!10} &\cellcolor{red!10} &\cellcolor{red!10} \hspace{25 mm} \\
\texttt{NN\_label} & 63 & & & \\
\texttt{NN\_scaled} & 63 & & & \\\hline
\texttt{RF\_onehot}& 52 &&&\\
\cellcolor{red!10}\texttt{RF\_label}&\cellcolor{red!10}57 &\cellcolor{red!10}&\cellcolor{red!10}& \cellcolor{red!10}\\
\texttt{RF\_scaled}& 53 &&&\\\hline
\texttt{DT\_onehot}& 44&&&\\
\cellcolor{red!10}\texttt{DT\_label}& \cellcolor{red!10}47 &\cellcolor{red!10}&\cellcolor{red!10}&\cellcolor{red!10}\\
\texttt{DT\_scaled}& 44 &&&\\\hline
\cellcolor{red!10}\texttt{$k$NN\_onehot}& \cellcolor{red!10}39 &\cellcolor{red!10}&\cellcolor{red!10}&\cellcolor{red!10}\\
\texttt{$k$NN\_label}& 24 &&&\\
\texttt{$k$NN\_scaled}& 34 &&&\\\hline
\texttt{LR\_onehot}& 15 &&&\\
\cellcolor{red!10}\texttt{LR\_label}&\cellcolor{red!10}25 &\cellcolor{red!10}&\cellcolor{red!10}&\cellcolor{red!10}\\
\texttt{LR\_scaled}& 23 &&&\\\hline
\cellcolor{red!10}\texttt{KRR\_onehot}& \cellcolor{red!10}39 &\cellcolor{red!10}&\cellcolor{red!10}&\cellcolor{red!10}\\
\texttt{KRR\_label}& 27 &&&\\
\texttt{KRR\_scaled}& 24 &&&\\\hline
\cellcolor{red!10}\texttt{SVR\_onehot}& \cellcolor{red!10}13 &\cellcolor{red!10}&\cellcolor{red!10}&\cellcolor{red!10}\\
\texttt{SVR\_label}& 11 &&&\\
\texttt{SVR\_scaled}& 8 &&&\\
        \end{tabular}\\
        \\
      \textbf{\small (d). \textsc{ExaStencils}}  & \textbf{\small (e). \textsc{x264}}& \textbf{\small (f). Total Scott-Knott scores over all systems}\\
    \end{tabular}
  \end{adjustbox}
   \end{center}
\end{sidewaystable}

%% file: Tables/chap-encoding/rq3-plot.tex
\begin{sidewaystable}[]
  
          \caption{Scott-Knott test, Med (median), and Interquartile Range (IQR) on the training time (minutes) of all models and systems, where ``onehot'', ``label'', and ``scaled'' stand for one-hot, label and scaled label encoding, respectively. A higher score means the RMSE is lower and, therefore, better. For (a) to (e), the pairs are sorted by score, median, and then IQR. For (f), \setlength{\fboxsep}{1.5pt}\colorbox{red!10}{red} highlights the best encoding scheme for a model over all systems}
\label{tb:rq3-plot}
  \begin{center}
    \begin{adjustbox}{max width = 0.85\textwidth}
     
\footnotesize
    \begin{tabular}{c@{}c@{}c}
        \begin{tabular}{lcccl}
            \cellcolor[gray]{1}\textbf{Pair} & \cellcolor[gray]{1}\textbf{Score} &     \cellcolor[gray]{1}\textbf{Med} & \cellcolor[gray]{1}\textbf{IQR} & \cellcolor[gray]{1}\\
            \hline
\texttt{LR\_scaled}  & 19 & <0.01 & 0.00 & \quart{0.00}{0.00}{0.00}{100.00} \\
\texttt{LR\_label}  & 19 & <0.01 & 0.00 & \quart{0.00}{0.00}{0.00}{100.00} \\
\texttt{LR\_onehot}  & 18 & 0.02 & 0.01 & \quart{0.03}{0.01}{0.03}{100.00} \\
\texttt{DT\_label}  & 17 & 0.49 & 0.01 & \quart{0.75}{0.01}{0.76}{100.00} \\
\texttt{DT\_scaled}  & 16 & 0.52 & 0.01 & \quart{0.80}{0.01}{0.81}{100.00} \\
\texttt{$k$NN\_label} & 15 & 0.84 & 0.03 & \quart{1.29}{0.03}{1.31}{100.00} \\
\texttt{$k$NN\_onehot} & 14 & 0.86 & 0.02 & \quart{1.33}{0.02}{1.34}{100.00} \\
\texttt{DT\_onehot}  & 13 & 1.07 & 0.04 & \quart{1.65}{0.04}{1.67}{100.00} \\
\texttt{$k$NN\_scaled}  & 12 & 1.36 & 0.05 & \quart{2.09}{0.05}{2.12}{100.00} \\
\texttt{NN\_scaled} & 11 & 21.61 & 1.30 & \quart{33.55}{1.30}{33.74}{100.00} \\
\texttt{NN\_onehot} & 10 & 23.76 & 0.20 & \quart{36.99}{0.20}{37.09}{100.00} \\
\texttt{RF\_label} & 9 & 28.71 & 0.33 & \quart{44.70}{0.33}{44.83}{100.00} \\
\texttt{KRR\_scaled}  & 8 & 28.86 & 0.08 & \quart{45.02}{0.08}{45.06}{100.00} \\
\texttt{NN\_label} & 7 & 29.02 & 0.24 & \quart{45.23}{0.24}{45.31}{100.00} \\
\texttt{KRR\_onehot}  & 6 & 29.82 & 0.08 & \quart{46.52}{0.08}{46.57}{100.00} \\
\texttt{SVR\_label}  & 5 & 30.45 & 0.54 & \quart{47.11}{0.54}{47.55}{100.00} \\
\texttt{RF\_scaled}  & 4 & 34.20 & 0.62 & \quart{53.06}{0.62}{53.41}{100.00} \\
\texttt{SVR\_onehot} & 3 & 34.75 & 1.61 & \quart{53.53}{1.61}{54.26}{100.00} \\
\texttt{KRR\_label}  & 2 & 42.28 & 0.18 & \quart{65.94}{0.18}{66.02}{100.00} \\
\texttt{RF\_onehot} & 1 & 47.84 & 1.62 & \quart{73.81}{1.62}{74.69}{100.00} \\
\texttt{SVR\_scaled}  & 1 & 60.96 & 47.97 & \quart{48.66}{47.97}{95.18}{100.00} \\
        \end{tabular} & 
        \begin{tabular}{|lcccl|}
           \cellcolor[gray]{1}\textbf{Pair} & \cellcolor[gray]{1}\textbf{Score} &     \cellcolor[gray]{1}\textbf{Med} & \cellcolor[gray]{1}\textbf{IQR} & \cellcolor[gray]{1}\\
            \hline
\texttt{LR\_scaled}  & 20 & <0.01 & 0.00 & \quart{0.00}{0.00}{0.00}{100.00} \\
\texttt{LR\_label}  & 19 & <0.01 & 0.00 & \quart{0.00}{0.00}{0.00}{100.00} \\
\texttt{LR\_onehot}  & 18 & 0.01 & 0.01 & \quart{0.01}{0.01}{0.01}{100.00} \\
\texttt{$k$NN\_label} & 17 & 0.23 & 0.00 & \quart{0.40}{0.00}{0.40}{100.00} \\
\texttt{$k$NN\_scaled}  & 16 & 0.30 & 0.06 & \quart{0.49}{0.06}{0.52}{100.00} \\
\texttt{DT\_scaled}  & 15 & 0.30 & 0.01 & \quart{0.52}{0.01}{0.53}{100.00} \\
\texttt{DT\_label}  & 14 & 0.31 & 0.01 & \quart{0.54}{0.01}{0.55}{100.00} \\
\texttt{$k$NN\_onehot} & 13 & 0.80 & 0.03 & \quart{1.39}{0.03}{1.41}{100.00} \\
\texttt{DT\_onehot}  & 12 & 0.85 & 0.01 & \quart{1.50}{0.01}{1.51}{100.00} \\
\texttt{KRR\_label}  & 11 & 14.72 & 0.57 & \quart{25.85}{0.57}{26.03}{100.00} \\
\texttt{KRR\_scaled}  & 10 & 14.92 & 0.03 & \quart{26.36}{0.03}{26.37}{100.00} \\
\texttt{RF\_scaled}  & 9 & 17.01 & 0.48 & \quart{29.85}{0.48}{30.07}{100.00} \\
\texttt{RF\_label} & 8 & 17.15 & 0.38 & \quart{30.18}{0.38}{30.31}{100.00} \\
\texttt{SVR\_label}  & 7 & 17.73 & 0.58 & \quart{31.21}{0.58}{31.34}{100.00} \\
\texttt{SVR\_scaled}  & 6 & 18.69 & 0.35 & \quart{32.80}{0.35}{33.03}{100.00} \\
\texttt{KRR\_onehot}  & 5 & 22.22 & 1.09 & \quart{39.20}{1.09}{39.28}{100.00} \\
\texttt{SVR\_onehot} & 4 & 28.41 & 0.98 & \quart{49.43}{0.98}{50.22}{100.00} \\
\texttt{NN\_scaled} & 3 & 27.85 & 4.20 & \quart{49.03}{4.20}{49.23}{100.00} \\
\texttt{NN\_label} & 2 & 28.03 & 8.81 & \quart{47.80}{8.81}{49.54}{100.00} \\
\texttt{NN\_onehot} & 1 & 35.36 & 5.16 & \quart{60.76}{5.16}{62.51}{100.00} \\
\texttt{RF\_onehot} & 1 & 36.77 & 0.50 & \quart{64.82}{0.50}{65.00}{100.00} \\
        \end{tabular} &
                \begin{tabular}{lcccl}
             \cellcolor[gray]{1}\textbf{Pair} & \cellcolor[gray]{1}\textbf{Score} &     \cellcolor[gray]{1}\textbf{Med} & \cellcolor[gray]{1}\textbf{IQR} & \cellcolor[gray]{1}
            \\
            \hline
\texttt{LR\_scaled}  & 20 & <0.01 & 0.00 & \quart{0.00}{0.00}{0.00}{100.00} \\
\texttt{LR\_label}  & 20 & <0.01 & 0.00 & \quart{0.00}{0.00}{0.00}{100.00} \\
\texttt{LR\_onehot}  & 19 & 0.01 & 0.01 & \quart{0.01}{0.01}{0.01}{100.00} \\
\texttt{$k$NN\_label} & 18 & 0.17 & 0.00 & \quart{0.26}{0.00}{0.26}{100.00} \\
\texttt{$k$NN\_scaled}  & 17 & 0.25 & 0.02 & \quart{0.37}{0.02}{0.38}{100.00} \\
\texttt{DT\_scaled}  & 16 & 0.26 & 0.00 & \quart{0.39}{0.00}{0.39}{100.00} \\
\texttt{DT\_label}  & 15 & 0.26 & 0.00 & \quart{0.40}{0.00}{0.40}{100.00} \\
\texttt{$k$NN\_onehot} & 14 & 0.47 & 0.00 & \quart{0.72}{0.00}{0.72}{100.00} \\
\texttt{DT\_onehot}  & 13 & 1.36 & 0.03 & \quart{2.06}{0.03}{2.07}{100.00} \\
\texttt{KRR\_label}  & 12 & 13.78 & 0.03 & \quart{20.99}{0.03}{21.00}{100.00} \\
\texttt{KRR\_scaled}  & 11 & 13.75 & 0.05 & \quart{20.95}{0.05}{20.96}{100.00} \\
\texttt{KRR\_onehot}  & 10 & 14.35 & 0.50 & \quart{21.46}{0.50}{21.88}{100.00} \\
\texttt{RF\_scaled}  & 9 & 16.75 & 0.15 & \quart{25.45}{0.15}{25.53}{100.00} \\
\texttt{RF\_label} & 8 & 16.90 & 0.27 & \quart{25.63}{0.27}{25.76}{100.00} \\
\texttt{SVR\_scaled}  & 7 & 17.11 & 0.04 & \quart{26.07}{0.04}{26.09}{100.00} \\
\texttt{SVR\_label}  & 6 & 17.33 & 0.14 & \quart{26.37}{0.14}{26.41}{100.00} \\
\texttt{SVR\_onehot} & 5 & 21.01 & 0.14 & \quart{32.00}{0.14}{32.03}{100.00} \\
\texttt{NN\_label} & 4 & 22.25 & 9.76 & \quart{29.07}{9.76}{33.92}{100.00} \\
\texttt{NN\_scaled} & 3 & 23.07 & 9.06 & \quart{30.40}{9.06}{35.17}{100.00} \\
\texttt{NN\_onehot} & 2 & 22.13 & 10.26 & \quart{29.52}{10.26}{33.73}{100.00} \\
\texttt{RF\_onehot} & 1 & 63.34 & 1.15 & \quart{96.07}{1.15}{96.55}{100.00} \\
        \end{tabular} \\
        \\
         \textbf{\small (a). \textsc{MongoDB}} & \small \textbf{(b). \textsc{Lrzip}} & \textbf{\small (c). \textsc{Trimesh}}
        \\
        \\
        \begin{tabular}{lcccl}
            \cellcolor[gray]{1}\textbf{Pair} & \cellcolor[gray]{1}\textbf{Score} &     \cellcolor[gray]{1}\textbf{Med} & \cellcolor[gray]{1}\textbf{IQR} & \cellcolor[gray]{1}\\
            \hline
\texttt{LR\_scaled}  & 20 & <0.01 & 0.00 & \quart{0.00}{0.00}{0.00}{100.00} \\
\texttt{LR\_label}  & 19 & <0.01 & 0.00 & \quart{0.00}{0.00}{0.00}{100.00} \\
\texttt{LR\_onehot}  & 18 & 0.01 & 0.00 & \quart{0.01}{0.00}{0.01}{100.00} \\
\texttt{$k$NN\_label} & 17 & 0.18 & 0.00 & \quart{0.27}{0.00}{0.27}{100.00} \\
\texttt{$k$NN\_scaled}  & 16 & 0.22 & 0.02 & \quart{0.33}{0.02}{0.34}{100.00} \\
\texttt{DT\_scaled}  & 15 & 0.23 & 0.00 & \quart{0.35}{0.00}{0.35}{100.00} \\
\texttt{DT\_label}  & 14 & 0.26 & 0.01 & \quart{0.39}{0.01}{0.40}{100.00} \\
\texttt{$k$NN\_onehot} & 13 & 0.49 & 0.05 & \quart{0.71}{0.05}{0.75}{100.00} \\
\texttt{DT\_onehot}  & 12 & 0.73 & 0.01 & \quart{1.10}{0.01}{1.11}{100.00} \\
\texttt{KRR\_scaled}  & 11 & 13.96 & 0.15 & \quart{21.22}{0.15}{21.24}{100.00} \\
\texttt{KRR\_onehot}  & 10 & 15.30 & 2.39 & \quart{22.35}{2.39}{23.29}{100.00} \\
\texttt{RF\_label} & 9 & 15.64 & 0.35 & \quart{23.59}{0.35}{23.80}{100.00} \\
\texttt{RF\_scaled}  & 8 & 15.86 & 0.26 & \quart{23.93}{0.26}{24.13}{100.00} \\
\texttt{SVR\_scaled}  & 7 & 16.98 & 0.10 & \quart{25.78}{0.10}{25.84}{100.00} \\
\texttt{SVR\_label}  & 6 & 16.48 & 3.21 & \quart{25.03}{3.21}{25.07}{100.00} \\
\texttt{NN\_onehot} & 5 & 18.02 & 4.46 & \quart{25.91}{4.46}{27.42}{100.00} \\
\texttt{SVR\_onehot} & 5 & 18.50 & 0.07 & \quart{28.12}{0.07}{28.15}{100.00} \\
\texttt{KRR\_label}  & 4 & 21.01 & 21.64 & \quart{29.63}{21.64}{31.97}{100.00} \\
\texttt{NN\_scaled} & 3 & 25.70 & 19.07 & \quart{35.98}{19.07}{39.10}{100.00} \\
\texttt{NN\_label} & 2 & 29.11 & 24.35 & \quart{34.46}{24.35}{44.29}{100.00} \\
\texttt{RF\_onehot} & 1 & 32.51 & 44.27 & \quart{49.23}{44.27}{49.47}{100.00} \\
        \end{tabular} & 
        \begin{tabular}{|lcccl}
           \cellcolor[gray]{1}\textbf{Pair} & \cellcolor[gray]{1}\textbf{Score} &     \cellcolor[gray]{1}\textbf{Med} & \cellcolor[gray]{1}\textbf{IQR} & \cellcolor[gray]{1}\\
            \hline
\texttt{LR\_label}  & 19 & <0.01 & 0.00 & \quart{0.00}{0.00}{0.00}{100.00} \\
\texttt{LR\_scaled}  & 18 & <0.01 & 0.00 & \quart{0.00}{0.00}{0.00}{100.00} \\
\texttt{DT\_label}  & 17 & 0.41 & 0.00 & \quart{0.07}{0.00}{0.07}{100.00} \\
\texttt{$k$NN\_label} & 16 & 0.46 & 0.00 & \quart{0.08}{0.00}{0.08}{100.00} \\
\texttt{DT\_scaled}  & 15 & 0.48 & 0.00 & \quart{0.08}{0.00}{0.08}{100.00} \\
\texttt{$k$NN\_scaled}  & 14 & 0.48 & 0.01 & \quart{0.08}{0.01}{0.08}{100.00} \\
\texttt{LR\_onehot}  & 13 & 0.57 & 0.02 & \quart{0.09}{0.02}{0.10}{100.00} \\
\texttt{$k$NN\_onehot} & 12 & 1.98 & 0.04 & \quart{0.31}{0.04}{0.34}{100.00} \\
\texttt{DT\_onehot}  & 11 & 9.17 & 0.05 & \quart{1.53}{0.05}{1.56}{100.00} \\
\texttt{KRR\_label}  & 10 & 18.87 & 0.88 & \quart{2.42}{0.88}{3.20}{100.00} \\
\texttt{SVR\_scaled}  & 9 & 19.69 & 0.42 & \quart{2.96}{0.42}{3.34}{100.00} \\
\texttt{NN\_scaled} & 8 & 16.83 & 0.82 & \quart{2.77}{0.82}{2.86}{100.00} \\
\texttt{KRR\_onehot}  & 7 & 23.76 & 0.21 & \quart{3.98}{0.21}{4.03}{100.00} \\
\texttt{KRR\_scaled}  & 6 & 33.79 & 0.33 & \quart{5.54}{0.33}{5.73}{100.00} \\
\texttt{NN\_label} & 5 & 29.38 & 2.17 & \quart{4.00}{2.17}{4.99}{100.00} \\
\texttt{RF\_scaled}  & 5 & 30.28 & 0.07 & \quart{5.10}{0.07}{5.14}{100.00} \\
\texttt{SVR\_label}  & 5 & 35.82 & 0.15 & \quart{5.98}{0.15}{6.08}{100.00} \\
\texttt{RF\_label} & 4 & 34.27 & 3.76 & \quart{5.64}{3.76}{5.82}{100.00} \\
\texttt{NN\_onehot} & 3 & 81.93 & 4.25 & \quart{11.38}{4.25}{13.90}{100.00} \\
\texttt{RF\_onehot} & 2 & 244.98 & 10.31 & \quart{40.42}{10.31}{41.57}{100.00} \\
\texttt{SVR\_onehot} & 1 & 519.99 & 23.35 & \quart{68.66}{23.35}{88.23}{100.00} \\
        \end{tabular}& 
        \begin{tabular}{|lcccl}
           \cellcolor[gray]{1}\textbf{Pair} & \cellcolor[gray]{1}\textbf{Total Score} &     
           & & \\
            \hline
\texttt{LR\_onehot}& 86 &&&\\
\texttt{LR\_label}& 96 &&&\\
\cellcolor{red!10}\texttt{LR\_scaled}& \cellcolor{red!10}97 &\cellcolor{red!10}&\cellcolor{red!10}&\cellcolor{red!10}\\
\hline
\texttt{$k$NN\_onehot}& 66 &&&\\
\cellcolor{red!10}\texttt{$k$NN\_label}& \cellcolor{red!10}83 &\cellcolor{red!10}&\cellcolor{red!10}&\cellcolor{red!10}\\
\texttt{$k$NN\_scaled}& 75 &&&\\
\hline
\texttt{DT\_onehot}& 61 &&&\\
\cellcolor{red!10}\texttt{DT\_label}& \cellcolor{red!10}77 &\cellcolor{red!10}&\cellcolor{red!10}&\cellcolor{red!10}\\
\cellcolor{red!10}\texttt{DT\_scaled}& \cellcolor{red!10}77 &\cellcolor{red!10}&\cellcolor{red!10}&\cellcolor{red!10}\\
\hline
\texttt{KRR\_onehot}& 38 &&&\\
\texttt{KRR\_label}& 39 &&&\\
\cellcolor{red!10}\texttt{KRR\_scaled}& \cellcolor{red!10}46 &\cellcolor{red!10}&\cellcolor{red!10}&\cellcolor{red!10}\\
\hline
\texttt{SVR\_onehot}& 18 &&&\\
\texttt{SVR\_label}& 29 &&&\\
\cellcolor{red!10}\texttt{SVR\_scaled}& \cellcolor{red!10}30 &\cellcolor{red!10}&\cellcolor{red!10}&\cellcolor{red!10}\\
\hline
\texttt{RF\_onehot}& 6 &&&\\
\cellcolor{red!10}\texttt{RF\_label}& \cellcolor{red!10}38 &\cellcolor{red!10}&\cellcolor{red!10}&\cellcolor{red!10}\\
\texttt{RF\_scaled}& 35 &&&\\
\hline
\texttt{NN\_onehot} & 21 & & & \hspace{30 mm} \\
\texttt{NN\_label} & 20 & & & \\
\cellcolor{red!10}\texttt{NN\_scaled} & \cellcolor{red!10}28 & \cellcolor{red!10}&\cellcolor{red!10} &\cellcolor{red!10} \\






        \end{tabular}\\
        \\
      \textbf{\small (d). \textsc{ExaStencils}}  & \textbf{\small (e). \textsc{x264}}& \textbf{\small (f). Total Scott-Knott scores over all systems}\\
    \end{tabular}
  \end{adjustbox}
   \end{center}
\end{sidewaystable}

%% file: Chapter-encoding/suggestions.tex
\section{Answer to RQ2 and Actionable Suggestions}
\label{chap-encoding: suggestions}

By analyzing the results of the empirical study, the RQ2 proposed in Section~\ref{chap-intro:rq} can be answered by providing the following suggestions on choosing the encoding scheme for learning software performance under a variety of circumstances.

\begin{answerbox}
   
   \emph{\textbf{Suggestion 1:} When RF, SVR, KRR, or NN is to be used, the study does not recommend trial-and-error to find the best encoding scheme. However, for $k$NN, DT or LR, it may be practical to ``try them all''.}
\end{answerbox}

From \textbf{RQ2.1}, it can be rather time-consuming to compare all three encoding schemes under RF, SVR, KRR, or NN. Indeed, the ``efforts'' may be reduced if it is considered, e.g., less repeated runs or even reduced data samples. However, to provide a reliable choice, what to consider in this study is essential, and hence further reducing them may increase the instability of the result. The process can be even more expensive if different models are also to be assessed during the trial-and-error. In contrast, when $k$NN, DT, or LR is to be used, it only requires less than one hour each --- an assumption that may be more acceptable within the development lifecycle.

\begin{answerbox}

   \emph{\textbf{Suggestion 2:} 
   When the accuracy is all that matters, among all possible models studied, NN paired with one-hot encoding is recommended. When a certain model needs to be used, the study suggests avoiding scaled label encoding in general and following one-hot encoding for deep learning and kernel models; label encoding for linear and tree models.}
\end{answerbox}

Reflecting on \textbf{RQ2}, when only the accuracy is of concern, suggestions can be made for practitioners to infer the best choice of encoding schemes when experimental assessment is not possible or desirable. Among others, it is clear that NN tends to offer the best accuracy, and NN paired with one-hot encoding, i.e., \texttt{NN\_onehot}, is the most reliable choice. In contrast, scaled label encoding often performs the worst, and hence scaled label encoding can be ruled out from the suggestions.

Besides the fact that the one-hot encoding can generally lead to the best accuracy over the models, some specific patterns are observed when the model to be used is fixed: one-hot encoding for deep learning and kernel models while label encoding for linear and tree models.

\begin{answerbox}

   \emph{\textbf{Suggestion 3:} When faster training time is more preferred (e.g., the model needs to be rapidly retrained at runtime), over all models studied, it is recommended to use linear regression paired with scaled label encoding. When the model is fixed, it is suggested to adopt scaled label encoding in general (especially for deep learning, linear, and kernel models) and label encoding for tree and lazy models; one-hot encoding should be avoided.}
\end{answerbox}

Deriving from the findings for \textbf{RQ3}, if the training time is of higher importance, the suitable choice of encoding scheme in the absence of experimental evaluation can be also estimated. Over all possible models studied, linear regression is unexpectedly the fastest to train and when it is paired with scaled label encoding (\texttt{LR\_scaled}) the training is the fastest. One-hot encoding is often the slowest to train and hence can be avoided. 

Although the scaled label encoding appears to be faster to train than its label counterpart, they remain competitive. In fact, when the model to be used has been pre-defined, some common patterns are observed: the scaled label encoding is the best for deep learning, linear, and kernel models while the label encoding is preferred for tree and lazy models.

\begin{answerbox}

   \emph{\textbf{Suggestion 4:} When the preference between accuracy and training time is unclear while the unbiased outcome is preferred, over all models studied, it is recommended to use scaled label encoding, but the paired model needs some efforts to determine. it can be certainly suggested to avoid one-hot encoding and kernel model (KRR and SVR regardless of its encoding schemes). When the model is fixed to the kernel and lazy models, the label and scaled label encoding can be chosen to reduce the bias, respectively.
   }
\end{answerbox}

It is not uncommon that the preference between accuracy and training time can be unclear, and hence an unbiased outcome is important. According to the findings for \textbf{RQ4}, this needs the label and scaled label encoding. Because in this case, as it has been shown, they often lead to results that are in the middle of the Pareto front for the pairs. In particular, scaled label encoding can often lead to well-balanced results in contrast to the other, but the paired model may vary. It would also be suggested to avoid one-hot encoding and kernel model (SVR and KRR), as the former would easily bias to accuracy or training time while the latter leads to no Pareto optimal choice at all over the systems studied.

However, when the model needs to be fixed, only the kernel and lazy models can have less biased choices, which are under the label and scaled label encoding, respectively.

%% file: Chapter-encoding/discussions.tex
\section{Discussion}
\label{chap-encoding:discussion}

This section discusses a few interesting points derived from the empirical study, the limitations and publications derived from this chapter, and possible future directions.

\subsection{Practicality of Performance Models}

The performance models built can be used in different practical scenarios, under each of which the accuracy and training time can be of great importance (and thereby the choice of encoding schemes is equally crucial).

\subsubsection{Configuration debugging}

Ill-fitted Configurations can lead to bugs such that the resulting performance is dramatically worse than the expectation. Here, a performance model can help software engineers easily inspect which configuration options are likely to be the root cause of the bug and identify the potential fix~\citep{DBLP:conf/sigsoft/XuJFZPT15}. The fact that the model makes inferences without running the system can greatly improve the efficiency of the debugging process. Further, by analyzing the models, software engineers can gain a better understanding of the system's performance characteristics which helps to prevent future configuration bugs.


\subsubsection{Speed up automatic configuration tuning}

Automatic configuration tuning is necessary to optimize the performance of the software system at deployment time. However, due to the expensiveness of measuring the performance, tuning is often a slow and time-consuming process. As one resolution to that issue, the performance model can serve as the surrogate for cheap evaluation of the configuration. Indeed, there have been a few successful applications in this regard, such as those that rely on Bayesian optimization~\citep{DBLP:journals/corr/abs-1801-02175,DBLP:conf/mascots/JamshidiC16}.

\subsubsection{Runtime self-adaptation}

Self-adapting the configuration at runtime is a promising way to manage the system's performance under uncertain environments. In this context, the performance model can help to achieve the adaptation in a timely manner, as it offers a relatively cheap way to reason about the better or worse of different configurations under changing environmental conditions. From the literature on self-adaptive systems, it is not uncommon to see that the performance models are often used during the planning stage~\citep{DBLP:journals/tsc/ChenB17,DBLP:conf/wosp/0001BWY18,DBLP:journals/csur/ChenBY18,DBLP:journals/pieee/ChenBY20,ChenLiDOS22,DBLP:conf/icse/ChenB14}.


\subsection{Why Considering Different Models Beyond Deep Learning?}

Note that deep learning models, such as NN, perform overwhelmingly better than others. Yet, this study involves a diverse set of other models because, in practice, there may be other reasons that a learning model is preferred. For example, linear and tree models may be used as they are directly interpretable~\citep{DBLP:conf/icse/SiegmundKKABRS12,DBLP:conf/splc/ValovGC15}, despite that they can lead to inferior accuracy overall. Therefore, the results on the choice of encoding schemes provide evidence for a wide set of scenarios and the possibility that different models may be involved. 

The other reason for considering different models is that, although this thesis seeks to examine whether the choice of deep learning model matters when deciding what encoding schemes to use, the validity and generalizability of this empirical study could be limited due to its  scope. Hence, by considering these alternative models, the findings and recommendations offer a broader range of insights to cater to different readers and benefit the research community as a whole.

\subsection{On Interactions between Configuration Options}

The encoding schemes can serve as different ways to represent the interactions between configuration options. Since the one-hot encoding embeds the values of options as the feature dimensions and captures their interactions, it models a much more finer-grained feature space compared with that of the label and scaled-label counterparts. The results show that, indeed, such a finer-grained capture of interactions enables one-hot encoding to become the most reliable scheme across the models/software as it has the generally best accuracy. This confirms the current understanding that the interaction between configuration options is important and the way they are handled can significantly influence the accuracy~\citep{DBLP:conf/icse/SiegmundKKABRS12}. Most importantly, the findings show that it is possible to better handle the interactions at the level of encoding.

\subsection{Limitations}

There are two folds of limitations in this empirical study.
\begin{itemize}
    \item First, the number of encoding schemes examined in this study is limited. Due to time and cost budgets, this thesis has to choose three of the most common encoding methods, due to the large number of combinations with the datasets and learning models (105 cases in this study). Yet, based on the SLR in Chapter~\ref{chap:review}, the three encoding methods can cover most of the studies.
    \item Further, the range of subject systems covered in this research is restricted. For researchers on other software domains, the results could potentially be divergent. Even though, the chosen datasets have included various categories of software, and are commonly used by researchers in the field of performance learning.
\end{itemize}

\subsection{Publications and Future Work}
The empirical study on encoding methods in this chapter has been constructed into a conference paper and accepted by the \textit{IEEE/ACM Mining Software Repositories (MSR) 2022}~\citep{DBLP:conf/msr/GongC22}.

This study demonstrates considerable potential, and the possibilities for future work are vast. For instance, there is room for exploring new encoding schemes, incorporating different subject systems, and exploring novel machine and deep learning models. These endeavors would contribute to the development of more comprehensive strategies for researchers in this field when selecting encoding methods. Furthermore, given the scope for expansion, it is conceivable to extend this study into a journal-sized publication.

%% file: Chapter-encoding/threats_to_validity.tex
\section{Threats to validity}
\label{chap-encoding:threats}

Similar to many empirical studies in software engineering, the empirical study in this chapter is subject to threats to validity. Specifically, \textbf{internal threats} can be related to the configuration options used and their ranges. Indeed, a different set may lead to a different result in some cases. However, here the thesis follows what has been commonly used in state-of-the-art studies, which are representatives of the subject systems. The hyperparameter of the models to tune can also impose this threat. Ideally, widening the set of hyperparameters to tune can complement the results. Yet, considering an extensive set of hyperparameters is rather expensive, as the tuning needs to go through the full training and validation process. To mitigate such, different hyperparameters in preliminary runs have been examined to find a balance between effectiveness and overhead.

\textbf{Construct threats} to validity can be related to the metric used. While different metrics exist for measuring accuracy, here RMSE is used, which is a widely used one for learning software performance. The results are also evaluated and validated by the Scott-Knott test~\citep{DBLP:journals/tse/MittasA13}. This study also set a data samples of $5,000$, which tends to be reasonable as this is what has been commonly used in prior work~\citep{DBLP:conf/kbse/DornAS20,DBLP:conf/im/JohnssonMS19, DBLP:conf/icdm/ShaoWL19, DBLP:conf/icse/Gerostathopoulos18}. Indeed, using other metrics or different sample sizes may offer new insights, which is planned to be done in future work.

Finally, \textbf{external threats} to validity can arise from the subjects and models used. To mitigate such, five common systems that are of diverse characteristics are studied, together with seven widely-used models. This leads to a total of 105 cases of investigation. Such a setting, although not exhaustive, is not uncommon in empirical software engineering and can serve as a strong foundation to generalize the findings, especially considering that an exhaustive study of all possible models and systems is unrealistic. Yet, it is agreed that additional subjects may prove fruitful.

%% file: Chapter-encoding/conclusions.tex
\section{Conclusions}
\label{chap-encoding:conclusions}

By realizing \textbf{objective 2} in this thesis, the knowledge gap in the understating of encoding schemes for learning performance for highly configurable software is bridged. It is done by conducting a systematic empirical study, covering five systems, seven models, and three widely used encoding schemes, giving a total of 105 cases of investigation. 

First, it is proven that:

\begin{displayquote}
\textit{Choosing the encoding scheme is non-trivial for performance learning and it can be rather expensive to do it using trial-and-error in a case-by-case manner.}
\end{displayquote}

More importantly, the findings observed from the study provide actionable suggestions and a ``rule-of-thumb'' when a thorough experimental comparison is not possible or desirable. Among these, the most important ones over all models and encoding schemes are:

\begin{itemize}
    \item using neural network paired with one-hot encoding for the best accuracy.
    \item using linear regression paired with scaled label encoding for the fastest training. 
    \item using scaled label encoding for a relatively well-balanced outcome, but mind the underlying model.
\end{itemize}

Further, by answering \textbf{RQ2} in this thesis, it is hoped that this chapter can serve as a good starting point to raise awareness of the importance of choosing encoding schemes for performance learning, and the actionable suggestions are of usefulness to the practitioners in the field. Moreover, this study seeks to spark a dialog on a set of relevant future research directions in this regard. 

The limitation of this chapter is on the scale of the examined encoding methods, subject systems, and learning models. As such, the next stage on this research thread is vast, including designing specialized models paired with suitable encoding schemes or even investigating new, tailored encoding schemes derived from the findings in the paper. 

Finally, while given the actionable suggestions shown in this study, the performance prediction accuracy can be increased by choosing the right encoding scheme, yet, there are still two gaps remaining to fill, i.e., the sparsity problem and the multi-environment problem. This thesis has solved these problems by designing two state-of-the-art artifacts, which will be illustrated in Chapter~\ref{chap:dal} and~\ref{chap:meta}.

%% file: Chapter-DAL/chapter-DAL.tex
\chapter{Enhancing Performance Prediction by Handling Sparsity via Divide-and-Learn}
\label{chap:dal}

As specified in Chapter~\ref{chap:introduction}, deep learning models have been widely adopted for predicting the configuration performance of software systems. However, in Chapter~\ref{chap:review}, the systematic literature review has revealed a crucial yet unaddressed challenge, which is how to cater for the sparsity inherited from the configuration landscape: the influence of configuration options (features) and the distribution of data samples are highly sparse, which might significantly reduce the prediction accuracy of the deep performance learning models. This knowledge gap leads to the formulation of a crucial research question in this thesis:

\begin{answerbox}
\emph{\textbf{RQ3:} Can the prediction accuracy in deep configuration performance learning be improved by effectively mitigating the sparsity problem?}
\end{answerbox}

In response to this RQ, this thesis proposes a model-agnostic and sparsity-robust framework for predicting configuration performance, dubbed \Model, based on the new paradigm of dividable learning that builds a model via ``divide-and-learn''. To handle sample sparsity, the samples from the configuration landscape are divided into distant divisions, for each of which a DL local model is built, e.g., regularized Hierarchical Interaction Neural Network, to deal with the feature sparsity. A newly given configuration would then be assigned to the right model of division for the final prediction. Further, \Model~adaptively determines the optimal number of divisions required for a system and sample size without any extra training or profiling. 

Experiment results from 12 real-world systems and five sets of training data reveal that, compared with the state-of-the-art approaches, \Model~performs no worse than the best counterpart on 44 out of 60 cases (within which 31 cases are significantly better) with up to $1.61$ times improvement on accuracy; requires fewer samples to reach the same/better accuracy; and producing acceptable training overhead. In particular, the mechanism that adapted the parameter $d$ can reach the optimal value for 76.43\% of the individual runs. The result also confirms that the paradigm of dividable learning is more suitable than other similar paradigms, such as ensemble learning for predicting configuration performance. Practically, \Model~considerably improves different global models when using them as the underlying local models, which further strengthens its flexibility. To promote open science, all the data, code, and supplementary materials of this work can be accessed at the repository: \texttt{\textcolor{blue}{\url{https://github.com/ideas-labo/DaL-ext}}}.

In this chapter, the backgrounds, details, evaluations, and discussions of the \Model~framework are introduced.

\input{Chapter-DAL/introduction}

\input{Chapter-DAL/background}

\input{Chapter-DAL/framework}
\input{Chapter-DAL/setup.tex}

\input{Chapter-DAL/evaluation}
\input{Chapter-DAL/discussion}

\input{Chapter-DAL/threats_to_validity}

\input{Chapter-DAL/conclusion}

%% file: Chapter-DAL/introduction.tex
\section{Introduction}
\label{chap-dal:introduction}

Since machine learning modeling is data-driven, the characteristics and properties of the measured data for configurable software systems pose non-trivial challenges to the learning, primarily because it is known that the configuration landscapes of the systems do not follow a ``smooth'' shape~\citep{DBLP:conf/mascots/JamshidiC16}. For example, adjusting between different cache strategies can drastically influence the performance, but they are often represented as a single-digit change on the landscape~\citep{DBLP:conf/sigsoft/0001Chen21}. This leads to the notion of sparsity in two aspects:

\begin{itemize}
    \item Only a small number of configuration options can significantly influence the performance, hence there is a clear \textbf{feature sparsity} involved~\citep{DBLP:conf/nips/HuangJYCMN10, DBLP:conf/sigsoft/SiegmundGAK15, DBLP:conf/icse/HaZ19, DBLP:conf/icse/VelezJSAK21}.
    \item The samples from the configuration landscape tend to form different divisions with diverse values of performance and configuration options, especially when the training data is limited due to expensive measurement---a typical case of \textbf{sample sparsity}~\citep{DBLP:journals/corr/abs-2202-03354,DBLP:conf/icml/LiuCH20,DBLP:conf/icml/ShibagakiKHT16}. This is particularly true when not all configurations are valid~\citep{DBLP:conf/sigsoft/SiegmundGAK15}.
\end{itemize}

\subsection{Knowledge Gap}
While prior work can handle feature sparsity through tree-liked model~\citep{DBLP:conf/kbse/GuoCASW13}, feature selection~\citep{DBLP:journals/fgcs/LiLTWHQD19, DBLP:conf/mascots/GrohmannEEKKM19,DBLP:journals/tse/ChenB17}, or regularizing deep learning~\citep{DBLP:conf/icse/HaZ19, DBLP:journals/jmlr/GlorotBB11,DBLP:conf/esem/ShuS0X20,DBLP:journals/tosem/ChengGZ23}, the sample sparsity has almost been ignored, which causes a major obstacle to the effectiveness of the machine learning-based performance model. For example, it is known that sparse data samples can easily force the model to focus and memorize too much on a particular region in the landscape of configuration data, leading to a serious issue of overfitting\footnote{Overfitting means a learned model fits well with the training data but works poorly on new data.}~\citep{DBLP:journals/corr/abs-2202-03354}.

To address the above gap, this thesis proposes \Model, a framework for configuration performance learning via the concept of ``divide-and-learn''. \Model~comes under a newly proposed paradigm termed \textit{dividable learning}---the key that enables it to be naturally model-agnostic and can improve any existing models that learn configuration performance. The basic idea is that, to handle sample sparsity, the approach divides the samples (configurations and their performance) into different divisions, each of which is learned by a local model. In this way, the highly sparse samples can be split into different locally smooth regions of data samples, and hence their patterns and feature sparsity can be better captured.

\subsection{Contributions}
The contributions of this chapter are multi-fold as it introduces a novel framework that effectively tackles the longstanding challenge of sparsity in the field of deep performance learning.

First, a systematic qualitative study is conducted to comprehend the SLR in Chapter~\ref{chap:review}, which analyzes the sparsity characteristics of configuration data, including both a literature review and an empirical study, hence better motivating the needs of this research.
    
Second, by formalizing the ``divide-and-learn'' framework, several notable contributions have been achieved: 

\begin{enumerate}
    \item The formulation, on top of the regression problem of learning configuration performance, of a new classification problem without explicit labels.
    
    \item The extension of Classification and Regression Tree (\texttt{CART})~\citep{loh2011classification} as a clustering algorithm to ``divide'' the samples into different divisions with similar characteristics, for each of which a local model is built.
    
    \item Newly given configurations would be assigned into a division inferred by a Random Forest classifier~\citep{DBLP:conf/icdar/Ho95}, which is trained using the pseudo-labeled data from the \texttt{CART}. The local model of the assigned division would be used for the final prediction thereafter.

    \item Experiments show that \Model~is sensitive to its only parameter $d$, which determines the number of divisions. This chapter proposes a novel adaptive mechanism that dynamically adapts the $d$ value to an appropriate level without additional training or profiling. 

    \item To determine the optimal $d$ value in the adaptation, a new indicator $\mu$HV is proposed, extending from the standard HV that is widely used for multi-objective evaluation, which can better reflect the goodness and balance between the ability to handle sample sparsity and the amount of data for learning in the divided configuration data.
\end{enumerate}

Under 12 systems with diverse performance attributes, scales, and domains, as well as five different training sizes, this chapter evaluates \Model~against the common state-of-the-art approaches, different underlying local models, and a number of its variants. The experiment results are encouraging: compared with the state-of-the-art, this thesis demonstrates that \Model

\begin{itemize}
    \item achieves no worse accuracy on 44 out of 60 cases, with 31 of them being significantly better. The improvements can be up to $1.61$ times against the best counterpart; 
    \item uses fewer samples to reach the same/better accuracy.
    \item incurs acceptable training time considering the improvements in accuracy, while the adaptation of $d$ has negligible overhead without requiring extra training. 
\end{itemize}

Interestingly, it is also revealed that 

\begin{itemize}
    \item \Model~is model-agnostic, significantly improving the accuracy of a given local model for each division compared with using the model alone as a global model (which is used to learn the entire training dataset). However, \Model~using the hierarchical deep learning-based approach published at TOSEM'23 \citep{DBLP:journals/tosem/ChengGZ23} as the local model produces the most accurate results.
    \item Compared with ensemble learning, which is the other similar paradigm that shares information between local models, \Model, which follows the paradigm of dividable learning that completely isolates the local model, performs considerably better in dealing with the sample sparsity for configuration data with up to $28.50$ times accuracy improvement over the second-best approach depending on the local model used.
    \item The tailored \texttt{CART}, the mechanism that adapts $d$, and the proposed $\mu$HV indicators can indeed individually contribute to the effectiveness of \Model.
    \item \Model's error tends to correlate upward and quadratically with its only parameter $d$ that sets the number of divisions. Yet, the optimal $d$ value indeed varies depending on the actual systems and training/testing data. Despite such, \Model~can adapt to the optimal $d$ value in 76.43\% of the individual runs while even when it misses hit, a promising $d$ value, which leads to generally marginal accuracy degradations to that of the optimal $d$, can still be selected.
\end{itemize}

\subsection{Chapter Outline}
This chapter is organized as follows: Section~\ref{chap-dal:background} introduces the notions of sparsity in software performance learning and the motivation of the proposed framework. Section~\ref{chap-dal:framework} delineates the tailored problem formulation and the detailed designs of \Model. Section~\ref{chap-dal:setup} presents the research questions and the experiment design, followed by the analysis of results in Section~\ref{chap-dal:evaluation}. The reasons why \Model~works, its strengths, limitations, and future plans are discussed in Section~\ref{chap-dal:discussion}. Finally, Section~\ref{chap-dal:threats} and Section~\ref{chap-dal:conclusion} present the threats to validity and conclude the chapter, respectively.

%% file: Chapter-DAL/background.tex
\section{Background and Motivation}
\label{chap-dal:background}

In this section, both a literature review and an empirical study are presented to comprehensively understand the properties of sparsity in learning software performance. Besides, the formal formulation of this problem is revisited, and the key observations from the qualitative studies that motivate this work are introduced.

\subsection{Problem Formulation}
In the software engineering community, configuration performance learning has been most commonly tackled by using various machine learning models (or at least partially)~\citep{DBLP:journals/tse/YuWHH19,DBLP:conf/splc/TempleAPBJR19,DBLP:journals/tosem/ChenLBY18,DBLP:journals/corr/abs-1801-02175,DBLP:conf/esem/HanY16}. Such a data-driven process relies on observing the software’s actual behaviors and builds a statistical model to predict the configuration performance without heavy human intervention~\citep{DBLP:books/daglib/0020252}. 


Formally, modeling the performance of software with $n$ configuration options is a regression problem that builds:
\begin{equation}
    \mathcal{P} = f(\mathbfcal{S})\text{, } \mathcal{P}\in\mathbb{R}
    \label{chap-dal-eq:prob}
\end{equation}
whereby $\mathbfcal{S}$ denotes the training samples of configuration-performance pairs, such that $\mathbf{\overline{x}} \in \mathbfcal{S}$. $\mathbf{\overline{x}}$ is a configuration and $\mathbf{\overline{x}}=(x_{1},x_{2},\cdots,x_{n})$, where each configuration option $x_{i}$ is either binary or categorical/numerical. The corresponding performance is denoted as $\mathcal{P}$. 


The goal of machine learning-based modeling is to learn a regression function $f$ using all training data samples such that for newly given configurations, the predicted performance is as close to the actual performance as possible.

\subsection{Sparsity in Configuration Data: A Literature review}
\label{chap-dal-sec:review}


To confirm the known characteristics of configuration data with respect to sparsity, a literature review is first conducted for papers related to the performance of software configuration published during the last decade, i.e., from 2013 to 2023. Specifically, an automatic search is performed over six highly influential indexing services, i.e., ACM Library, IEEE Xplore, Google Scholar, ScienceDirect, SpringerLink, and Wiley Online, by using the search string below:

\begin{displayquote}
\textit{(``software" OR ``system") AND ``configuration" AND (``performance prediction" OR ``performance modeling" OR ``performance learning" OR ``configuration tuning") AND ``machine learning")} 
\end{displayquote}

The search result consists of 4,740 studies with duplication except for non-English documents. Next, the duplication is filtered out by examining the titles and eliminating clearly irrelevant documents, e.g., those about students' learning performance. This has resulted in 391 highly relevant candidate studies. Through a detailed review of the candidate studies, A set of criteria is applied to further extract a set of more representative works. In particular, a study is temporarily selected as a primary study if it meets all of the following inclusion criteria:

\begin{itemize}
\item The paper presents machine learning algorithm(s) for modeling configuration performance; This can also include search algorithm(s) for tuning the configurations to reach better performance in which some models need to be learned, e.g., different variants of Bayesian Optimization~\citep{DBLP:conf/icse/0003XC021,DBLP:journals/corr/abs-1801-02175}.
\item The paper presents a study or claims, either explicitly or implicitly, related to the sparsity of configuration data in the domain studied.
\item The goal of the paper is to predict, optimize, or analyze the performance of a software system.
\item The paper has at least one section that explicitly specifies the algorithm(s) used.
\item The paper contains quantitative experiments in the evaluation with details about how the results were obtained.

\end{itemize}

Subsequently, the following exclusion criteria are applied to the previously included studies, which would be removed if they meet any below: 
\begin{itemize}
\item The paper is not software or system engineering-related.
\item The paper is not published in a peer-reviewed public venue.
\item The paper is a survey, review, or tutorial type of work.
\item The paper is a short and work-in-progress work, i.e., shorter than 8 double-column or 15 single-column pages.
\end{itemize}

The above process was repeated for all candidate studies. Finally, 64 primary studies are identified for detailed data collection, from which the contents are carefully reviewed and statements about the sparsity nature of configuration data are extracted.

\input{Tables/chap-DAL/papers}

The selected results have been shown in Table~\ref{tb:papers}, where the key phrases are highlighted in bold\footnote{The completed list of reviewed papers can be found at: \texttt{\textcolor{blue}{\url{https://github.com/ideas-labo/DaL-ext/blob/main/supplementary_materials/SLR_full_list.xlsb}}}}. Albeit with diverse terminologies (e.g., knobs, metrics, or parameters), it can be seen that almost all papers have identified evident sparse patterns in software features related to configuration options, therefore, existing work reveals high \textbf{\textit{feature sparsity}}, which refers to the fact that only a small number of configuration options are prominent to the performance. This is inevitable as the configuration space for software systems is generally rugged with respect to the configuration options~\citep{DBLP:conf/sigsoft/0001Chen21,DBLP:conf/kbse/JamshidiSVKPA17}.

\input{Figures/chap-DAL/3d_scatter}

\subsection{Sparsity in Configuration Data: An Empirical study}
\label{chap-dal-sec:empirical}

Although the literature review has revealed evidence of sparsity in configuration data, it is difficult to ensure that all the major aspects/characteristics of sparsity in configuration data have been covered. As such, to better understand the sparse nature of configurable software systems, the data collected from the real-world systems studied in prior work is empirically analyzed. This is accomplished by reviewing the systems used in the work identified from the literature review stage and selecting the ones according to the following criteria:

\begin{itemize}
\item The system is used by more than one paper.
\item There are clear articulations on how the data is collected under the systems, e.g., how many repeated measurements.
\item The data contains neither missing measurements nor invalid configurations.
\item The data should express a full spectrum of all the valid configurations for the system.
\end{itemize}

The above criteria could still lead to many systems that are of similar categories and characteristics, but it would be unrealistic to study them all and hence further representatives need to be extracted. To that end, for each category of the software systems, (e.g., a video encoder or a compiler), the one(s) that have been overwhelmingly studied across most of the papers identified are selected.

The extraction has led to 12 real-world configurable software systems in the analysis (their details will be discussed in Section~\ref{chap-dal:setup}). Figure~\ref{fig:3d-exp} shows the projected configuration samples measured from the systems studied. Notably, it is seen that the systems all exhibit a consistent pattern---the samples tend to form different divisions with two properties:

\begin{itemize}
    \item \textbf{Property 1:} configurations in the same division share closer performance values with smoother changes but those in-between divisions exhibit drastically different performance and can change more sharply.
    \item \textbf{Property 2:} configurations in the same division can have a closer value on at least one key option than those from the different divisions.
\end{itemize}

In this regard, the values of performance and key configuration options determine the characteristics of samples. The key discovery is therefore even with the key options that are the most influential to the performance, the samples still do not exhibit a ``smooth'' distribution over the configuration landscape. Instead, they tend to be spread sparsely: those with similar characteristics can form arbitrarily different divisions, which tend to be rather distant from each other. This is a typical case of high \textbf{\textit{sample sparsity}}~\citep{DBLP:journals/corr/abs-2202-03354,DBLP:conf/icml/LiuCH20,DBLP:conf/icml/ShibagakiKHT16}. Indeed, it is difficult for precisely defining to what extent the sparsity is considered as high, but with the visualized results from the empirical study, there is strong evidence that the configuration data is certainly non-smooth and hence raising concerns of ``high sparsity smells''.

In general, such a high sample sparsity is caused by two reasons: (1) the inherited consequence of high feature sparsity and (2) the fact that not all configurations are valid because of the constraints (e.g., an option can be used only if another option has been turned on)~\citep{DBLP:conf/sigsoft/SiegmundGAK15}, thereby there are many ``empty areas'' in the configuration landscape.

\subsubsection{Discussion}

The above findings reveal the key factors to consider for the problem: when using deep learning models to learn concepts from the configuration data, a model is needed that: 

\begin{enumerate}
    \item handles the complex interactions between the configuration options with high feature sparsity while;
    \item captures the diverse characteristics of configuration samples over all divisions caused by the high sample sparsity, e.g., in Figure~\ref{fig:3d-exp}, where samples in different divisions have diverged performance ranges.
\end{enumerate}
For the former challenge, there have been some proposed approaches previously, such as \texttt{DeepPerf}~\citep{DBLP:conf/icse/HaZ19}, \texttt{Perf-AL}~\citep{DBLP:conf/esem/ShuS0X20}, and \texttt{HINNPerf}~\citep{DBLP:journals/tosem/ChengGZ23}. The latter, unfortunately, is often ignored in existing work for configuration performance learning as observed from the literature review and Table~\ref{tb:papers}, causing a major obstacle for a model to learn and generalize the data for predicting the performance of the newly given configuration. This is because those highly sparse samples increase the risk for models to overfit the training data, for instance by memorizing and biasing values in certain respective divisions~\citep{DBLP:journals/corr/abs-2202-03354}, especially considering that there are often limited samples from the configuration landscape due to the expensive measurement of configurable systems.

The above is the main motivation of this work, for which the RQ3 of this thesis is asked: how to improve the accuracy of predicting software configuration performance under such a high sample sparsity?

%% file: Tables/chap-DAL/papers.tex
\begin{footnotesize} 
~\vspace{0.2cm}
\begin{longtable}{p{0.15\textwidth}p{0.7\textwidth}p{0.15\textwidth}}

\caption{Selected papers that contain domain knowledge and explicit statements related to the characteristics of sparsity in configuration data.}.
\label{tb:papers}
\\
\toprule
\textbf{Reference} & \textbf{Description and/or definition related to sparsity} & \textbf{What is sparse?}\\
\midrule

\citep{DBLP:conf/sigmod/AkenPGZ17} & ``\textit{DBMSs can have hundreds of knobs, but \textbf{only a subset actually affect} the DBMS’s performance.}'' & Configuration knobs\\
\hline

\citep{DBLP:conf/sigsoft/SiegmundGAK15} & ``\textit{The ratio over all subject systems for OW with Plackett-Burman sampling is 0.31, indicating that \textbf{one third of the options significantly contribute} to the performance of a system.}'' & Configuration options\\
\hline

\citep{DBLP:conf/mascots/JamshidiC16} & ``\textit{More specifically, this means low-order interactions among \textbf{a few dominating factors} can explain the main changes in the response function observed in the experiments.}'' & Factors\\
\hline

\citep{DBLP:journals/corr/abs-1801-02175} & ``\textit{We know of many software options where \textbf{a small change can lead to radically different} software performance.}'' & Configuration options\\
\hline

\citep{DBLP:journals/access/ThraneZC20} & ``\textit{The satellite images offer much information, and the entirety is not necessarily relevant for radio performance prediction, especially at lower frequencies. The use of such images results ultimately in a model that is harder to train since the \textbf{latent features obtained by the CNN might be sparse}.}'' & Features\\
\hline

\citep{DBLP:journals/pvldb/MarcusP19} & ``\textit{The problem with this solution is \textbf{sparsity}:
if one has many different operator types, the vectors used to
represent them will have an increasingly \textbf{larger proportion
of zeros}.}'' & Vectors of variables\\
\hline

\citep{DBLP:conf/kbse/JamshidiSVKPA17} & ``\textit{\textbf{Only a subset of options is influential} which is largely preserved across all environment changes.}'' & Configuration options\\
\hline

\citep{DBLP:journals/tse/ChenB17} & ``\textit{Too limited inputs may not provide
enough information of relevance to the QoS (i.e., the information that drives the changes in QoS), which restricts the
model accuracy and applicability. On the other hand, too
many inputs can generate noise in the modeling, because it
introduces \textbf{irrelevant information and large redundancy} in
the inputs (i.e., the same information has been provided by
more than one selected primitives, thus it becomes noise),
this will downgrade the model accuracy and generate
unnecessary overhead.}'' & Configuration primitives and inputs\\
\hline

\citep{DBLP:conf/icse/HaZ19} & ``\textit{It has been observed that the software \textbf{performance functions are usually very sparse} (i.e. only a small number of configuration variables and their interactions have significant impact on system performance).}'' & Variables of performance function\\
\hline

\citep{DBLP:journals/pvldb/ZhouSLF20} & ``\textit{The
\textbf{performance-related features are sparsely scattered} in the
graph (e.g., for a matrix of 1000×1000, many rows only have
two 1s, meaning that the corresponding operator only has 2
directly related operators).}'' & Performance features\\
\hline

\citep{DBLP:conf/hotstorage/KanellisAV20} & ``\textit{Surprisingly, we find that with YCSB
workload-A on Cassandra, tuning \textbf{just five knobs can achieve
99\% of the performance} achieved by the best configuration
that is obtained by tuning many knobs.}'' & Configuration knobs\\
\hline
 
\citep{DBLP:conf/icsm/Ha019} & ``\textit{For software performance functions, their \textbf{Fourier coefficients are always very sparse},
i.e. most of Fourier coefficients are zeros. The reason is that only a small number of configurations have significant impact on system performance.}'' & Configuration coefficients\\
\hline

\citep{DBLP:conf/icse/0003XC021} & ``\textit{\textbf{Only a small number of optimization flags}, referred to as impactful optimizations, can have \textbf{noticeable impact} on the runtime performance of a specific program.}'' & Configuration flag\\
\hline

\citep{DBLP:conf/kdd/FekryCPRH20} & ``\textit{However, we have observed that for each workload \textbf{only a small subset of those parameters has a significant impact} on overall performance.}'' & Configuration parameters\\
\hline

\citep{DBLP:conf/middleware/GrohmannNIKL19} & ``\textit{We collect 1,040 platform metrics using the PCP monitoring tool as described in Section 3.1. 952 of these platform metrics consider the host, 88 are specific to service instances (i.e., containers) running on the host. As expected, \textbf{not all the metrics are relevant} for the machine learning model and in many cases metric preprocessing is required such that they can be useful or leveraged by the algorithm.}'' & Measured configuration metrics\\
\hline

\citep{DBLP:journals/jsa/ZhangLWWZH18} & ``\textit{\textbf{Not all the features have impact} on the performance evaluation result, we can find a way to confirm the real effective features and remove the redundancy ones.}'' & Configuration features\\
\hline

\citep{DBLP:journals/pvldb/ZhangCLWTLC22} & ``\textit{Given a limited tuning budget, tuning over the configuration space with all the knobs is inefficient. It is recommended to \textbf{preselect important knobs} to prune the configuration space.}'' & Configuration knobs\\
\hline

\citep{DBLP:conf/esem/ShuS0X20} & ``\textit{This means
that only a small number of parameters have significant impact on the model. In other words, the \textbf{parameters of the neural network could be sparse}.}'' & Configuration parameters\\
\hline

\citep{DBLP:conf/sc/XieTCCHLOVW19} & ``\textit{Our analysis shows that the write behaviors of the GPFS system are \textbf{dominated by the metadata load and load skew} within the supercomputer and the resources used in its filesystem.}'' (The metadata load and load skew here are configuration options.) & Configuration options\\
\hline

\citep{DBLP:conf/sc/MalikFP09} & ``\textit{\textbf{Not all the features actually have considerable impact} on the execution time, therefore the collected provenance data needs to be filtered and reduced.}'' & Configuration features\\
\hline

\citep{DBLP:journals/pvldb/KanellisDKMCV22} & ``\textit{However, recent studies have shown that \textbf{tuning a handful of knobs can be sufficient} to achieve near-optimal performance and significantly reducing the number of knobs can accelerate the tuning process.}'' & Configuration knobs\\
\hline

\citep{DBLP:journals/tosem/ChengGZ23} & ``\textit{One important prior knowledge is that \textbf{only a small number of configuration options and their interactions have a significant impact} on system performance, implying sparsity on the parameters of the performance model (i.e., making parameters of many insignificant options and interactions equal to zero).}'' & Configuration options\\
\hline

\citep{DBLP:conf/msr/GongC22} & ``\textit{Unlike other domains, software configuration is often highly
sparse, leading to unusual data distributions. Specifically, \textbf{a few configuration options could have large influence} on the software
performance, while the others are trivial.}'' & Configuration options\\
\hline

\citep{DBLP:journals/tecs/TrajkovicKHZ22} & ``\textit{To be able to generate a prediction model with high accuracy, we need to select features with a significant impact on the values of the targets. Including the \textbf{features that have little or no impact on the targets may result in overfitting}.}'' & Parameter features\\
\hline

\citep{DBLP:conf/splc/Acher0LBJKBP22} & ``\textit{Given these results, we can say that out of the thousands of options of Linux kernel, \textbf{only a few hundreds actually influence} its binary size.}'' & Configuration options\\


\bottomrule
\end{longtable}
\end{footnotesize}

%% file: Figures/chap-DAL/3d_scatter.tex
\begin{figure}[!t]
\centering
\footnotesize
\begin{adjustbox}{width=1.3\textwidth,center}
\begin{subfigure}{.34\columnwidth}
  \centering
  \includegraphics[width=\linewidth]{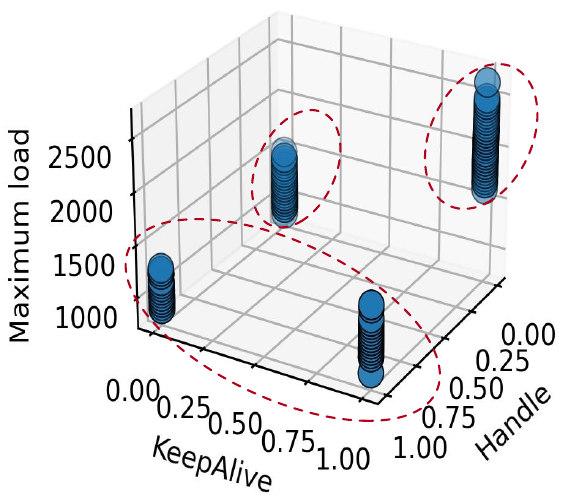} 
  \caption{\textsc{Apache}}
  \label{fig:depth-Apache}
\end{subfigure}
\hspace{-0.1cm}
\begin{subfigure}{.34\columnwidth}
  \centering
  \includegraphics[width=\linewidth]{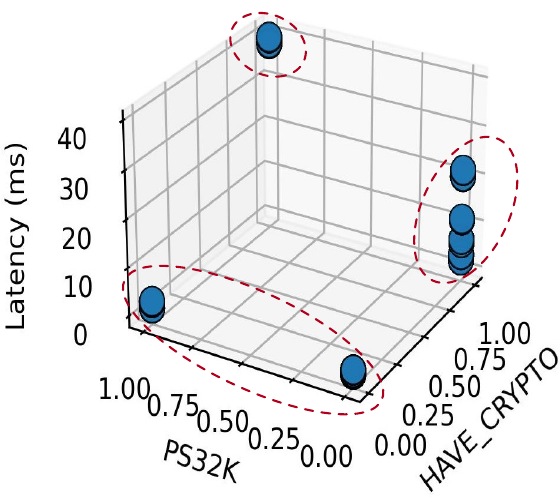} 
  \caption{\textsc{BDB-C}}
  \label{fig:depth-BDBC}
\end{subfigure}
\hspace{-0.1cm}
\begin{subfigure}{.37\columnwidth}
  \centering
  \includegraphics[width=\linewidth]{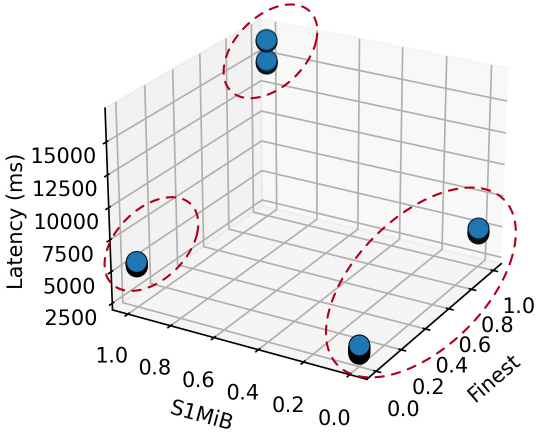} 
  \caption{\textsc{BDB-J}}
  \label{fig:depth-BDBJ}
\end{subfigure}
\hspace{-0.1cm}
\begin{subfigure}{.37\columnwidth}
  \centering
  \includegraphics[width=\linewidth]{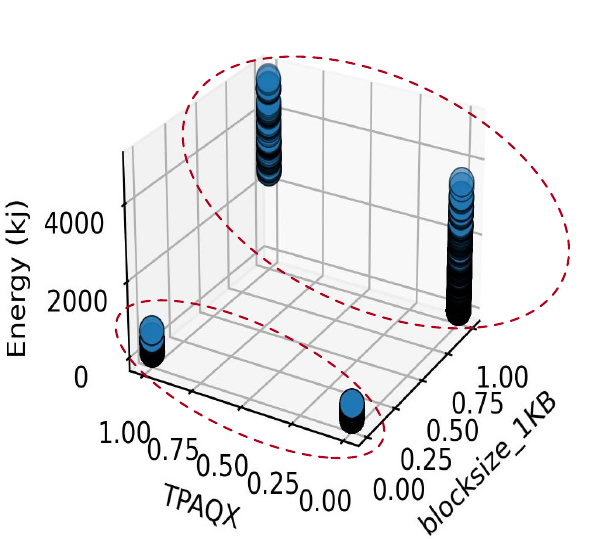}  
  \caption{\textsc{kanzi}}
  \label{fig:depth-kanzi}
\end{subfigure}
\end{adjustbox}

\begin{adjustbox}{width=1.3\textwidth,center}
\begin{subfigure}{.34\columnwidth}
  \centering
  \includegraphics[width=\linewidth]{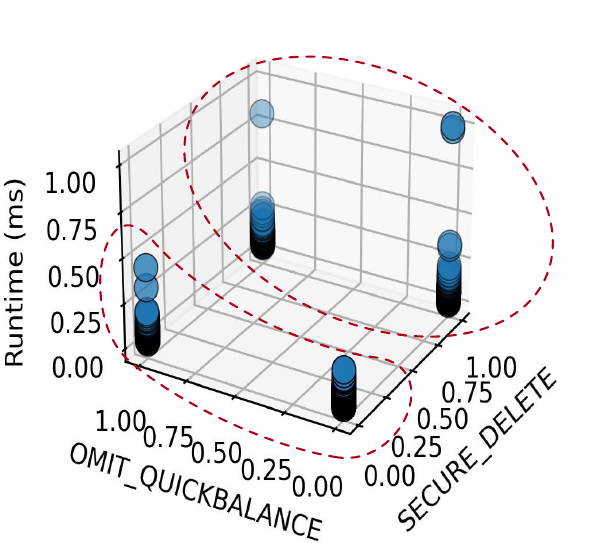} 
  \caption{\textsc{SQLite}}
  \label{fig:depth-SQLite}
\end{subfigure}
\hspace{-0.1cm}
\begin{subfigure}{.35\columnwidth}
  \centering
  \includegraphics[width=\linewidth]{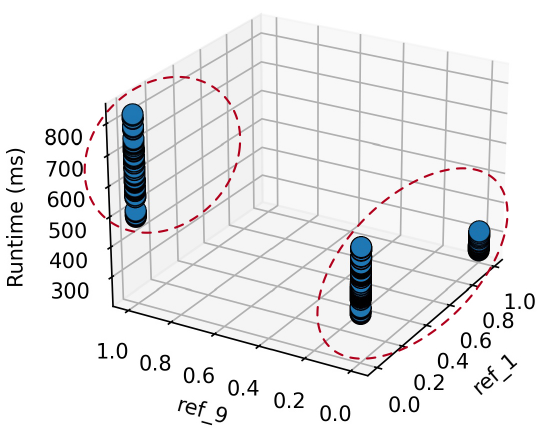} 
  \caption{\textsc{x264}}
  \label{fig:depth-x264}
\end{subfigure}
\hspace{-0.1cm}
\begin{subfigure}{.34\columnwidth}
  \centering
  \includegraphics[width=\linewidth]{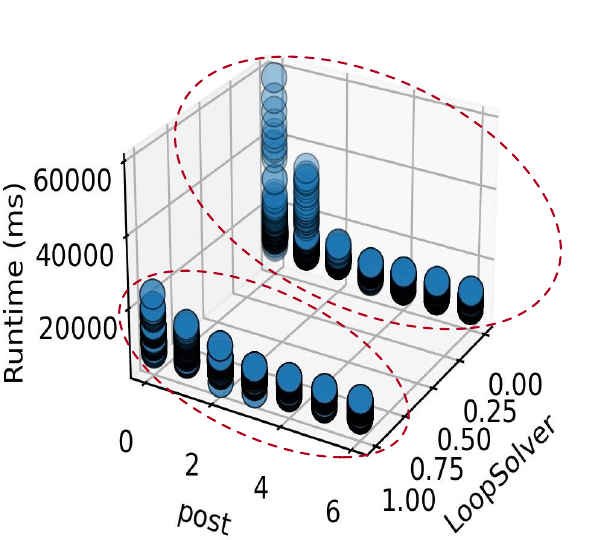}  
  \caption{\textsc{Dune MGS}}
  \label{fig:depth-Dune}
\end{subfigure}
\hspace{-0.1cm}
\begin{subfigure}{.34\columnwidth}
  \centering
  \includegraphics[width=\linewidth]{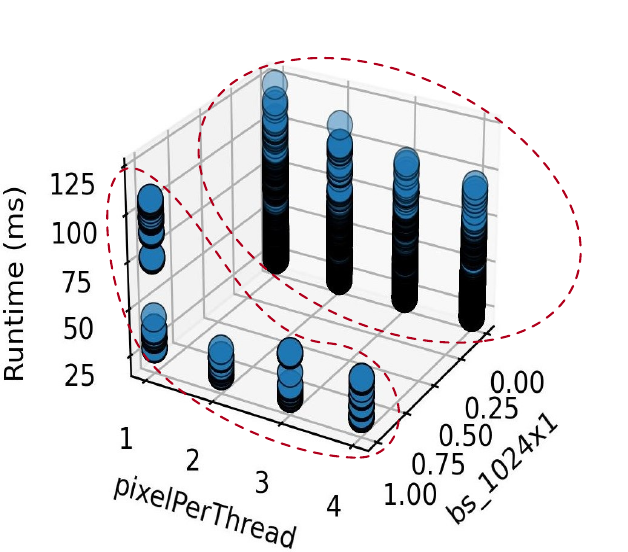} 
  \caption{\textsc{HIPA$^{cc}$}}
  \label{fig:depth-HIPAcc}
\end{subfigure}
\end{adjustbox}

\begin{adjustbox}{width=1.3\textwidth,center}
\begin{subfigure}{.37\columnwidth}
  \centering
  \includegraphics[width=\linewidth]{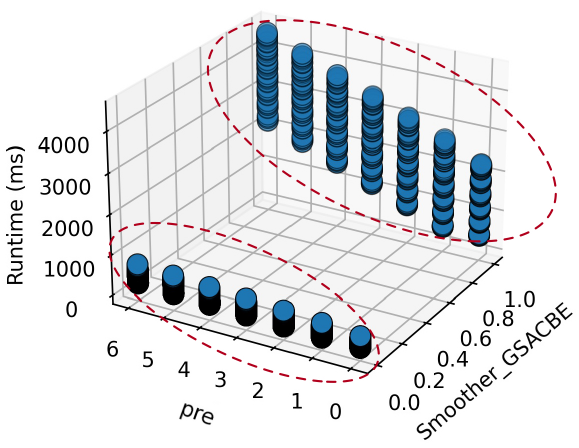} 
  \caption{\textsc{HSMGP}}
  \label{fig:depth-HSMGP}
\end{subfigure}
\hspace{-0.1cm}
\begin{subfigure}{.34\columnwidth}
  \centering
  \includegraphics[width=\linewidth]{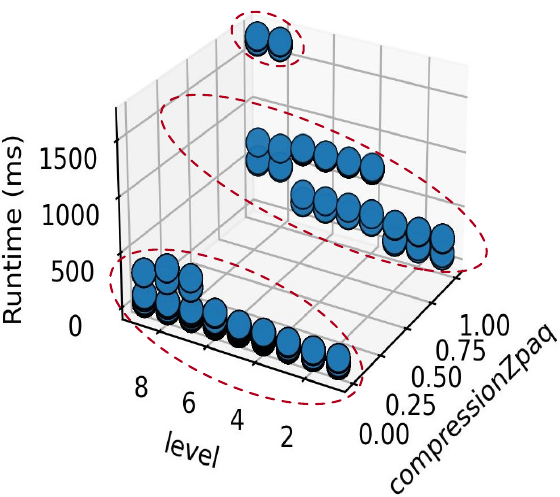} 
  \caption{\textsc{Lrzip}}
  \label{fig:depth-Lrzip}
\end{subfigure}
\hspace{-0.1cm}
\begin{subfigure}{.34\columnwidth}
  \centering
  \includegraphics[width=\linewidth]{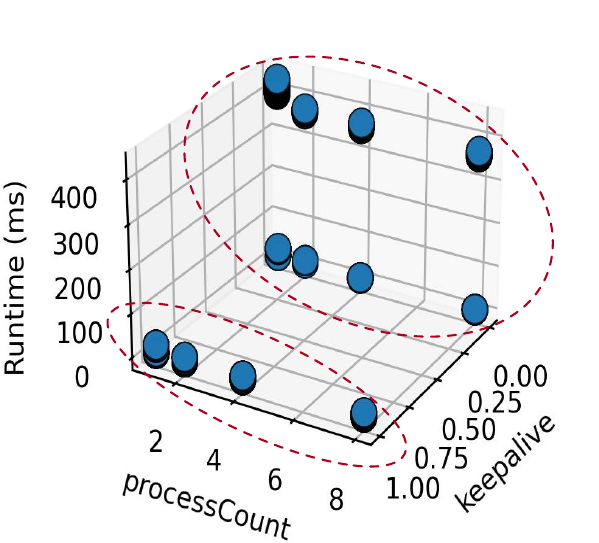} 
  \caption{\textsc{nginx}}
  \label{fig:depth-nginx}
\end{subfigure}
\hspace{-0.1cm}
\begin{subfigure}{.37\columnwidth}
  \centering
  \includegraphics[width=\linewidth]{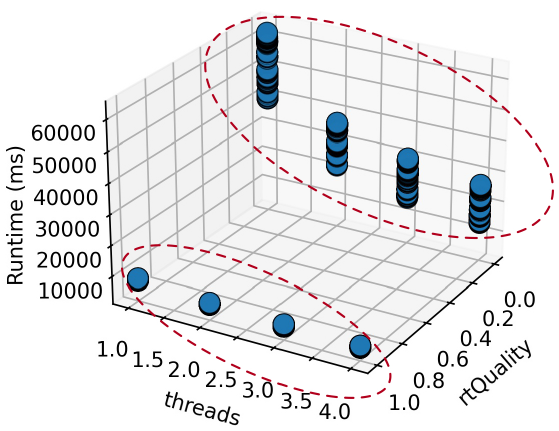}  
  \caption{\textsc{VP8}}
  \label{fig:depth-VP8}
\end{subfigure}
\end{adjustbox}


  \caption{Projection of configurations in the landscape with respect to the performance and two most important options (the divisions are circled).}
     \label{fig:3d-exp}
\end{figure}

%% file: Chapter-DAL/framework.tex
\section{Divide-and-Learn for Performance Prediction}
\label{chap-dal:framework}


Drawing on the above-mentioned observations of the configuration data, this thesis proposes \Model: an approach that enables better prediction of the software performance via ``divide-and-learn''. To mitigate the sample sparsity issue, the key idea of \Model~is that, since different divisions of configurations show drastically diverse characteristics, i.e., rather different performance values with distant values of key configuration options, the framework seeks to independently learn a local model for each of those divisions that contain \textit{locally smooth} samples, thereby the learning can be more focused on the particular characteristics exhibited from the divisions and handle the feature sparsity. Yet, this requires formulating, on top of the original regression problem of predicting the performance value, a new classification problem without explicit labels. As such, the original performance learning problem formulation (Equation~\ref{chap-dal-eq:prob}) is modified as follows: 
\begin{equation}
    \mathbfcal{D} = g(\mathbfcal{S}) 
\end{equation}
\begin{equation}
    \forall D_i \in \mathbfcal{D}\text{: } \mathcal{P} = f(D_i)\text{, } \mathcal{P}\in\mathbb{R}
\end{equation}
Overall, this framework aims to achieve three goals:

\begin{itemize}
    \item \textbf{Goal 1:} dividing the data samples into diverse yet more focused divisions $\mathbfcal{D}$ (building function $g$) and;
    \item \textbf{Goal 2:} training a dedicated local model for each division $D_i$ (building function $f$) while;
    \item \textbf{Goal 3:} assigning a newly coming configuration into the right model for prediction (using functions $g$ and $f$).
\end{itemize}

\begin{figure}[t!]
  \centering
  \includegraphics[width=0.9\columnwidth]{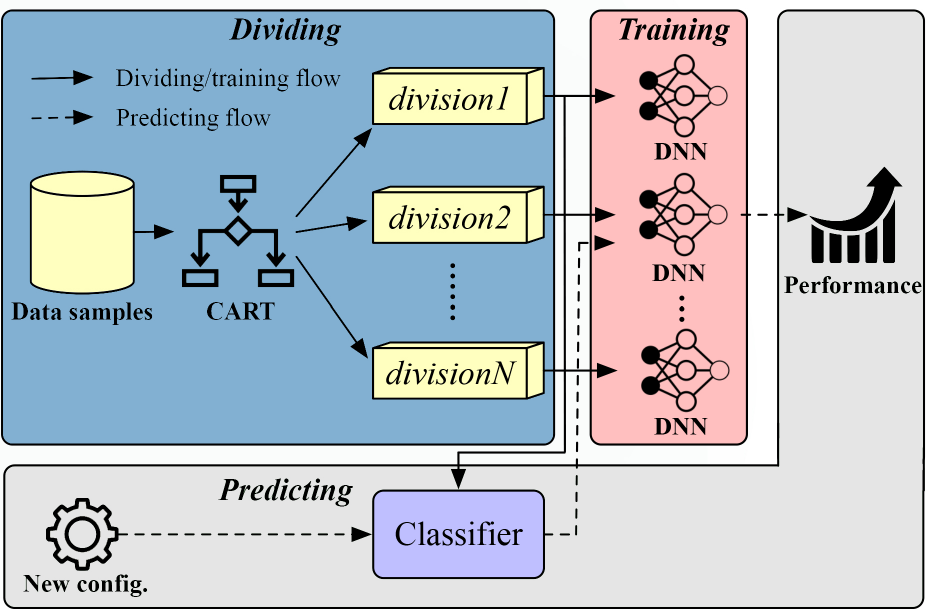}
  \caption{The architecture of \Model. Note that although the default local model in the training phase is DNN, the framework is model-agnostic and it can improve a wide range of other local machine learning models.}
  \label{fig:dal-structure}
\end{figure}

Figure~\ref{fig:dal-structure} illustrates the overall architecture of \Model, in which there are three core phases, namely \textit{Dividing}, \textit{Training}, and \textit{Predicting}. A pseudo-code can also be found in Algorithm~\ref{alg:dal-code}.

\subsection{Dividing}
\label{subsec:phase1_clustering}

The very first phase in \Model~is to appropriately divide the data into more focused divisions while doing so by considering both the configuration options and performance values. To that end, the key question to address is: how to effectively cluster the performance data with similar sample characteristics (\textbf{Goal 1})?

\subsubsection{Modifying \texttt{CART} for Generating Divisions}
\label{subsec:modeify-cart}

Indeed, for dividing the data samples, it makes sense to consider various unsupervised clustering algorithms, such as $k$-mean~\citep{macqueen1967some}, BIRCH~\citep{DBLP:conf/sigmod/ZhangRL96} or DBSCAN~\citep{DBLP:conf/kdd/EsterKSX96}. However, it is found that they are ill-suited for the problem, because:

\begin{itemize}
    \item the distance metrics are highly system-dependent. For example, depending on the number of configuration options and whether they are binary/numeric options;
    \item it is difficult to combine the configuration options and performance value with appropriate discrimination; 
    \item and clustering algorithms are often non-interpretable.
\end{itemize}

As a result, in \Model, Classification and Regression Tree (\texttt{CART}) is extended as the clustering algorithm (lines 3-11 in Algorithm~\ref{alg:dal-code}) since (1) it is simple with interpretable/analyzable structure; (2) it ranks the important options as part of training (good for dealing with the feature sparsity issue), and (3) it does not suffer the issues above~\citep{DBLP:conf/kbse/SarkarGSAC15,DBLP:journals/ese/GuoYSASVCWY18,DBLP:journals/ase/NairMSA18,DBLP:conf/icse/Chen19b,DBLP:journals/tse/ChenB17,DBLP:journals/corr/abs-1801-02175,DBLP:conf/kbse/GuoCASW13}. 

\begin{figure}[t!]
  \centering
  \includegraphics[width=0.9\columnwidth]{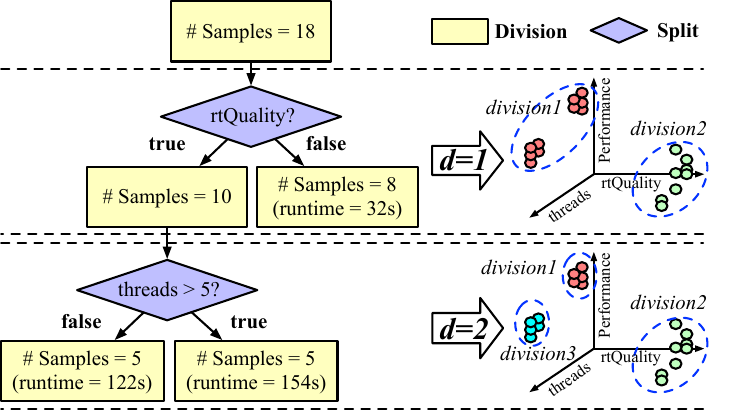}
  \caption{Projection of \texttt{CART} for \textsc{VP8} showing the possible divisions with different colors under alternative depth $d$.}
  \label{fig:DT_example}
\end{figure}

As illustrated in Figure~\ref{fig:DT_example}, \texttt{CART} is originally a supervised and binary tree-structured model, which recursively splits some, if not all, configuration options and the corresponding data samples based on tuned thresholds. A split would result in two divisions, each of which can be further split. In this work, at first, the \texttt{CART} is trained on the available samples of configurations and performance values, during which the most common mean performance of all samples is used for each division $D_i$ as the prediction~\citep{DBLP:journals/ese/GuoYSASVCWY18,DBLP:conf/kbse/GuoCASW13}:
\begin{equation}
    \overline{y}_{D_i}={{1}\over{|D_i|}} {\sum_{y_j \in D_i}y_j}
    \label{eq:seg}
\end{equation}
in which $y_j$ is a performance value. For example, Figure~\ref{fig:DT_example} shows a projected example, in which the configuration that satisfies ``\texttt{rtQuality=true}'' and ``\texttt{threads=3}'' would lead to an inferred runtime of 122 seconds, which is calculated over all the 5 samples involved using Equation~\ref{eq:seg}.

\input{Tables/chap-DAL/alg1.tex}

By choosing/ranking options that serve as the splits and tuning their thresholds, \Model~seek to minimize the following overall loss function during the \texttt{CART} training:
\begin{equation}
   \mathcal{L}= {\sum_{y_j \in D_l}{(y_j - \overline{y}_{D_l})}^2} + {\sum_{y_j \in D_r}{(y_j - \overline{y}_{D_r})}^2}
    \label{eq:loss}
\end{equation}
where $D_l$ and $D_r$ denote the left and right division from a split, respectively. This ensures that the divisions would contain data samples with similar performance values (\textbf{Property 1}) while they are formed with respect to the similar values of the key configuration options as determined by the splits/thresholds at the finest granularity (\textbf{Property 2}), i.e., the more important options would appear on the higher level of the tree with excessive splitting.

However, here \texttt{CART} is not used to generalize prediction directly on new data once it is trained, as it has been shown that the splits and simple average of performance values in the division alone can still fail to handle the complex interactions between the options, leading to insufficient accuracy~\citep{DBLP:conf/icse/HaZ19}. Further, with the loss function in Equation~\ref{eq:loss}, \texttt{CART} is prone to be overfitting\footnote{Overfitting means a learned model fits well with the training data but works poorly on new data.} especially for software quality data~\citep{DBLP:journals/ese/KhoshgoftaarA01}. This exacerbates the issue of sample sparsity~\citep{DBLP:journals/corr/abs-2202-03354} under a small amount of data samples, which is not uncommon for configurable software systems~\citep{DBLP:conf/icse/HaZ19,DBLP:conf/esem/ShuS0X20}. 

Instead, what this thesis is interested in are the (branch and/or leaf) divisions made therein (with respect to the training data), which motivate to use further dedicated and more focused local models for better generalizing to the new data (lines 6-11 in Algorithm~\ref{alg:dal-code}). As such, the final prediction is no longer a simple average, while the \texttt{CART} overfitting itself is not cared about as long as it fits the training data well. This is similar to the case of unsupervised clustering, for which the clustering is guided by implicit labels (via the loss function at Equation~\ref{eq:loss}). Specifically, \Model~extract the data samples according to the divisions made by the $d$th depth of the \texttt{CART}, including all the leaf divisions with depth smaller than $d$. An example can be seen from Figure~\ref{fig:DT_example}, where $d$ is a controllable parameter to be given. In this way, \Model~divides the data into a range of $[d+1,2^{d}]$ divisions ($d \geq 1$), each of which will be captured by a local model learned thereafter. Note that when the number of data samples in the division is less than the minimum amount required by a model, the two divisions of the same parent node are merged.

As a concrete example, from Figure~\ref{fig:DT_example}, it can be seen that there are two depths: when $d=1$ there would be two divisions (one branch and one leaf) with 10 and 8 samples, respectively; similarly, when $d=2$ there would be three leaf divisions: two of each have 5 samples, and one is the division with 8 samples from $d=1$ as it is a leaf. In this case, \texttt{CART} has detected that the \texttt{rtQuality} is a more important (binary) option to impact the performance, and hence it should be considered at a higher level in the tree. Note that for numeric options, e.g., \texttt{threads}, the threshold of splitting (\texttt{threads} $>5$) is also tuned as part of the training process of \texttt{CART}.

The above properties can be indeed confirmed in Figure~\ref{fig:3d-exp}, where the original dataset is divided into more focused divisions according to one or two most important configuration options based on the decision boundaries of \texttt{CART} and the parameter $d$. Nonetheless,  it is important to note that in some extremely sparse regions, an incorrect choice of $d$ may lead to samples with different characteristics and performance ranges being grouped together. For example, in Figure~\ref{fig:depth-nginx}, the division with \texttt{keepalive}=0 still exhibits sparse performance, indicating the need for further splitting using other decision boundaries provided by \texttt{CART}. Despite this, it is worth highlighting that even in such scenarios, the division where \texttt{keepalive}=1 displays a highly concentrated sample distribution, which can be effectively captured by the local model. As such, it is intuitive to understand that $d$ is such a critical parameter for \Model. 

Yet, manually setting the right value for it can be time-consuming and error-prone. In what follows, the role of $d$ in \Model~will be delineated, and an adaptive mechanism within \Model~that enables dynamic adaptation of the $d$ value during training will be proposed.

\subsubsection{The Role of Depth $d$ in \Model}
\label{subsubsec:play}

Since more divisions mean that the sample space is separated into more loosely related regions for dealing with the sample sparsity, one may expect that the accuracy will be improved, or at least stay similar, thereby the maximum possible $d$ from \texttt{CART} should be used in the \textit{Dividing} phase. This, however, only exists in the ``utopia case'' where there is an infinite set of configuration data samples.

In essence, with the design of \Model, the depth $d$ will manage two conflicting objectives that influence its accuracy:

\begin{enumerate}
    \item greater ability to handle sample sparsity by separating the distant samples into divisions, each of which is learned by an isolated local model;
    \item and a larger amount of data samples in each division for the local model to be able to generalize.
\end{enumerate}

Clearly, a larger $d$ may benefit the ability to handle sample sparsity, but it will inevitably reduce the data size per division for a local model to generalize since it is possible for \texttt{CART} to generate divisions with imbalanced sample sizes. From this perspective, $d$ is seen as a value that controls the trade-off between the two objectives, and neither a too small nor too large $d$ would be ideal, as the former would lose the ability to deal with sample sparsity while the latter would leave too little data for a local model to learn, hence produce negative noises that harm the overall prediction. This is the key reason that setting $d$ for a given system is crucial when using \Model. 

\begin{figure}[!t]
  \centering
   \begin{subfigure}[t]{0.4\columnwidth}
        \centering
\includegraphics[width=\columnwidth]{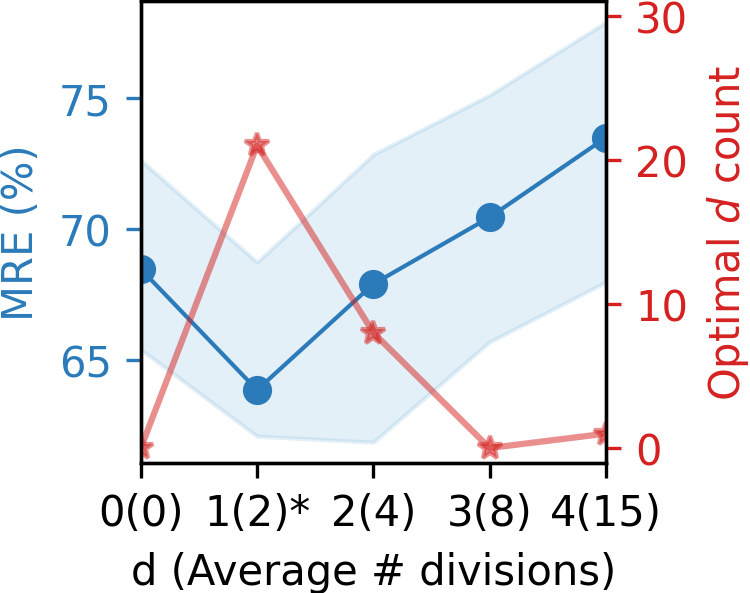}
    \subcaption{\textsc{SQLite}}
   \end{subfigure}
~\hspace{0.5cm}
      \begin{subfigure}[t]{0.39\columnwidth}
        \centering
\includegraphics[width=\columnwidth]{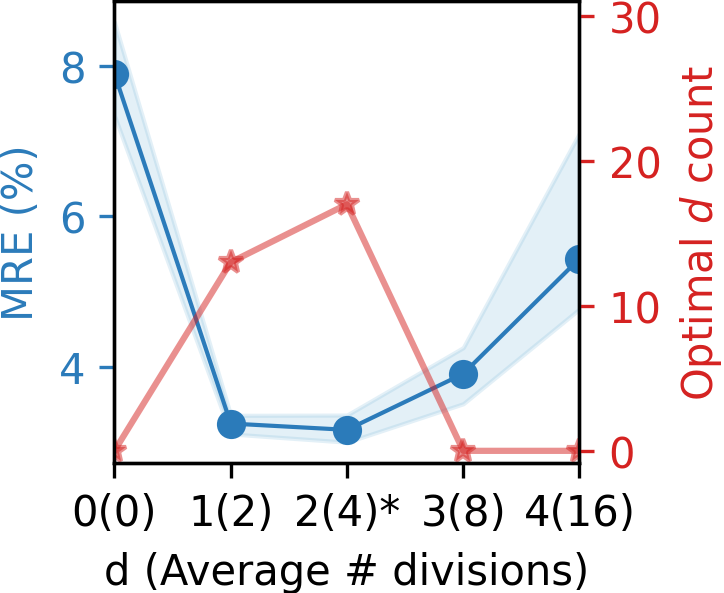}
 \subcaption{\textsc{Lrzip}}
   \end{subfigure}

    \caption{The changing optimal $d$ on \Model~depending on the software systems being modeled and the training/testing data across 30 runs.}
       \label{fig:optimal-d}
  \end{figure}

Unfortunately, finding the appropriate $d$ value is not straightforward, as it often requires repeatedly profiling different values following in a trial-and-error manner, which can be rather expensive, especially when the optimal $d$, which leads to the best mean relative error (MRE), varies depending on the system and even the training/testing data. For example, Figure~\ref{fig:optimal-d} illustrates the optimal $d$ for \Model~on two real-world software systems (the full coverage will be discussed in Section~\ref{subsec:sen}), from which it is clear that the  $d=1$ would lead to the best overall MRE for \textsc{SQLite}, but this becomes $d=2$ for \textsc{Lrzip}. If considering the individual runs that involve different training and testing configuration data, the optimal $d$ could also differ (red lines). Notably, it is seen that some $d$ values, e.g., $d=4$ for \textsc{SQLite}, can be dramatically harmful to the accuracy, leading to a result that is even worse than the case when \Model~is absent (i.e., $d=0$).

\subsubsection{Adapting the Depth $d$}
\label{subsubsec:adapting_depth}

To overcome the above, in \Model, an adaptive mechanism is designed as part of the \textit{Dividing} phase that is able to dynamically find the $d$ such that the two aforementioned objectives can be optimized and balanced (line 5 in Algorithm~\ref{alg:dal-code}).

Specifically, the following functions $h$ and $z$ are used to measure the ability to handle sample sparsity and the amount of information for a division $D_i$, respectively:
\begin{equation}
\begin{cases}
h(D_i)={1\over{|D_i|}}{\sum_{y_j \in D_i}{(y_j - \overline{y}_{D_i})}^2}\\
z(D_i)=-n_{D_i}\\
\end{cases}
\label{eq:adapt}
\end{equation}
whereby $z$ is the additive inverse of the assigned sample size, denoted as $n_{D_i}$. $h$ is basically the mean square error (or performance variance of the samples in a division) taken from the loss function (Equation~\ref{eq:loss}) that splits the divisions with respect to the important configuration options. As such, the samples in a division that are generally closer to each other in terms of the performance value, after being divided according to the importance of key configuration options, will be more likely to be beneficial for a local model to learn. Intuitively, both $h$ and $z$ need to be minimized. Given a maximum number of $d_{max}$ divisions generated by \texttt{CART} under the training data, the purpose is to find the $d$ value ($0\leq d\leq d_{max}$) that leads to the overall best and most balanced divisions (a.k.a. knee points) in the objective space of $h$ and $z$ from those generated by all possible $d$ values. In essence, from a multi-objective optimization perspective, knee points represent the solutions that, when changed, can only marginally improve one objective by largely comprising the other~\citep{DBLP:journals/tosem/ChenLBY18,DBLP:journals/tec/YuMJDLZ22}, hence achieving a well-balanced trade-off.

To this end, this thesis gains inspiration from a widely used quality indicator for evaluating multi-objective solution sets, namely Hypervolume (HV)~\citep{Zitzler1998}. In a nutshell, HV measures the volume between all nondominated points\footnote{A solution $\mathbf{\overline{a}}$ is dominated by $\mathbf{\overline{b}}$ if all objectives of $\mathbf{\overline{b}}$ are better or equivalent to those of $\mathbf{\overline{a}}$ while there is at least one objective of $\mathbf{\overline{b}}$ performs better than that of $\mathbf{\overline{a}}$. A solution is said nondominated if it cannot be dominated by any other solutions in the set.} in the objective space and a reference point (usually a nadir point); the larger the volume, the better convergence and diversity that the set achieves. The HV for a solution set $\mathbfcal{A}$ can be computed as:
\begin{equation}
	 \text{HV}(\mathbfcal{A},\mathbf{\overline{r}}) = \lambda(\bigcup_{\mathbf{\overline{a}}\in \mathbfcal{A}} \{\mathbf{\overline{x}}|\mathbf{\overline{a}} \prec \mathbf{\overline{x}} \prec \mathbf{\overline{r}}\})
	\label{eq:hv}
\end{equation}
where $\lambda$ is the Lebesgue measure~\citep{Zitzler1998}; $\mathbf{\overline{r}}$ is the reference nadir point, which is often taken as the 1.1 times of the range of the nondominated set~\citep{9252185}.~\citet{9252185} show that, with an appropriate setting of the reference point, HV can well reflect the preference for balanced/knee solutions, which fits precisely with the needs. However, the original HV cannot be directly adopted due to the fact that it can completely omit the contributions of divisions based on their relative domination relations, i.e., a division does not contribute to the HV value at all if it is dominated by the other division under a certain $d$. Indeed, this makes sense in conventional multi-objective evaluation scenarios, but in the case of dividable learning for configuration performance with \Model, the contribution of each division counts, even if it has been dominated since the local model under such a division could still impact the result when the newly given configuration falls therein. As a result, the original HV might misjudge the true effectiveness of a $d$ value.

\begin{figure*}[!t]
  \centering
  \hspace{-0.2cm}
   \begin{subfigure}[t]{0.3\textwidth}
        \centering
\includegraphics[width=\columnwidth]{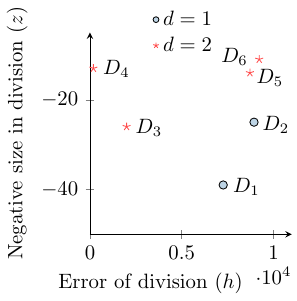}
    \subcaption{All divisions}
   \end{subfigure}
~\hfill
      \begin{subfigure}[t]{0.3\textwidth}
        \centering
\includegraphics[width=\columnwidth]{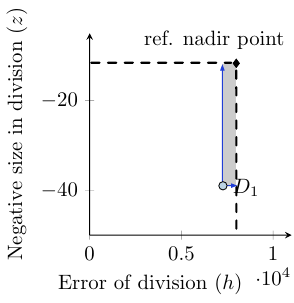}
 \subcaption{HV for $d=1$}
   \end{subfigure}
~\hfill
      \begin{subfigure}[t]{0.3\textwidth}
        \centering
\includegraphics[width=\columnwidth]{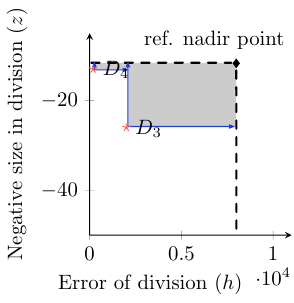}
  \subcaption{HV for $d=2$}
   \end{subfigure}
~\hfill
      \begin{subfigure}[t]{0.3\textwidth}
        \centering
\includegraphics[width=\columnwidth]{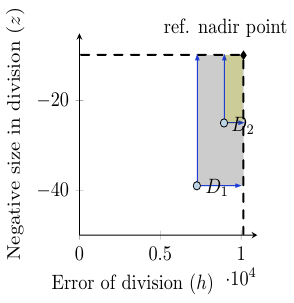}
     \subcaption{$\mu$HV for $d=1$}
   \end{subfigure}
~\hspace{0.5cm}
      \begin{subfigure}[t]{0.3\textwidth}
        \centering
\includegraphics[width=\columnwidth]{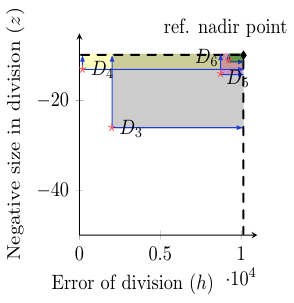}
    \subcaption{$\mu$HV for $d=2$}
   \end{subfigure}

    \caption{Comparing the differences between HV and $\mu$HV for divisions with $d=1$ and $d=2$ under system \textsc{x264}.}
       \label{fig:ahv-hv}
  \end{figure*}

Instead of leveraging the original HV, in this work, a modified HV is proposed, dubbed averaging HV ($\mu$HV), that evaluates the average quality of the HV value that takes each division into account with respect to $h$ and $z$ regardless of the domination relations---a typical specialization of a generic concept to cater for the needs. Formally, the $\mu$HV for all the divisions in a set $\mathbfcal{D}$ under a particular $d$ value is calculated as below:
\begin{equation}
\begin{aligned}
&\mu \text{HV}(\mathbfcal{D},\mathbf{\overline{r}})={1\over{|\mathbfcal{D}|}}{\sum_{D_i \in \mathbfcal{D}}{(|h_r - h(D_i)|) \times (|z_r - z(D_i)|)}} \\ 
&\text{s.t. } \mathbf{\overline{r}}=\langle h_r,z_r\rangle
\end{aligned}
\label{eq:ahv}
\end{equation}
whereby $h_r$ and $z_r$ are the objective values for mean square error and data size of the reference nadir point $\mathbf{\overline{r}}$, respectively. Since there are only two objectives to consider in the case, each individual HV value within the $\mu$HV is essentially the area between the corresponding division and the reference nadir point in the objective space. With the above, the following specializations to the original generic HV are realized:
\begin{itemize}
    \item \textbf{Specialization 1:} Through averaging the HV value for each individual division's objectives, $\mu$HV would take into account the contribution made by a division that is dominated by the other. This, conversely, is not the case in the original HV as it works only on the nondominated ones.
    \item \textbf{Specialization 2:} Unlike the original HV that uses $1.1$ times the range of the nondominated divisions as the reference nadir point, $\mu$HV takes $1.1$ times the range of each of the worst objective values\footnote{This would be $0.9$ times for the objective $z$ as it is always a negative number.}. This is necessary to account for the divisions being dominated by others.
\end{itemize}

Considering the example illustrated in Figure~\ref{fig:ahv-hv}, where (a) shows that $D_1$ is the nondominated division for $d=1$ while $D_3$ and $D_4$ are the nondominated ones for $d=2$. In (b), the original HV value equals the area between $D_1$ alone and the reference nadir point, i.e., HV$=19825.43$. In (c), the original HV value equals the non-overlapped area from $D_3$ and $D_4$ to the reference nadir point, i.e., HV$=88025.51$. In (d), $\mu$HV is the mean over the HV value of the area between $D_1$ and the reference nadir point together with that of the area between $D_2$ and the reference nadir point, i.e., $\mu$HV$=(83951.55+18216.65) /2=51084.10$. In (e), $\mu$HV is the mean over the HV value for the area of each of the divisions $D_3$, $D_4$, $D_5$, and $D_6$, i.e., $\mu$HV$=(131212.91+30873.60+5862.67+1014.70)/4=42240.97$. With the original HV, $d=2$ would be chosen while using the proposed $\mu$HV, $d=1$ would be chosen. In fact, here $d=1$ tends to be better in general, considering all the possible divisions in terms of $h$ and $z$. The actual validation also confirms that $d=1$ leads to $2.28\times$ more accurate result than that of $d=2$.  

In this example, it is possible to use $d=1$ or $d=2$ for \textsc{x264} (Figure~\ref{fig:ahv-hv}a). From Figures~\ref{fig:ahv-hv}b and~\ref{fig:ahv-hv}c, if the original HV is used directly, then there would be $\text{HV}_1=19825.43$ and $\text{HV}_2=88025.51$ for $d=1$ and $d=2$, respectively. Clearly, $d=2$ would be chosen here since $\text{HV}_1 < \text{HV}_2$. When using $\mu$HV, in Figures~\ref{fig:ahv-hv}d and~\ref{fig:ahv-hv}e, there are $\mu \text{HV}_1=51084.10$ and $\mu \text{HV}_2=42240.97$ for $d=1$ and $d=2$, respectively. As such, $d=1$ would be chosen since $\mu \text{HV}_1 > \mu \text{HV}_2$. In fact, $d=1$ is a better setting than $d=2$ as the latter contains $D_5$ and $D_6$, which are two much more inferior/less balanced divisions than those in $d=1$, leading to a higher probability of more severe negative impacts on the accuracy. Hence, $d=1$ tends to be more desired in general, considering all the possible divisions in terms of $h$ and $z$. Indeed, the experiment validation confirms that it leads to $2.28\times$ more accurate results than that of $d=2$.

The above demonstrates the limitation of the original HV for adapting $d$ in dividable learning for configuration performance: because of the way it is computed, those divisions that are dominated by the others, e.g., $D_2$ for $d=1$ and $D_5$ for $d=2$, would have no contribution to the final value. This is a serious issue for the case, given that the local model trained on those divisions is still important since \Model~naturally allows any local model to be used in isolation. For example, if upon prediction, there are newly given configurations belonging to the local models for $D_2$ or $D_5$, then these models would be used for predicting the performance independently. However, the original HV would not have been able to take that into account since the contributions of the corresponding divisions were ruled out, providing likely misleading guidance. $\mu$HV addresses such a shortcoming by averaging the individual HV value for all divisions under a $d$ value, hence taking the contribution of those being dominated by the other into account (\textbf{Specialization 1}). In this way, $\mu$HV can better assist in identifying the overall better $d$ value and its corresponding divisions with respect to $h$ and $z$. Yet, this requires changing the reference nadir point to consider not only the nondominated divisions (as in the original HV) but also anyone in the space (\textbf{Specialization 2}).

Algorithm~\ref{alg:adapt-code} illustrates the proposed adaptive mechanism with the following key steps:

\begin{enumerate}

    \item Compute the objective values of $h_i$ and $z_i$ for all division $D_i$ under the possible $d$. The set of divisions $\mathbfcal{D}_j$ and its corresponding $d_j$ are kept as a vector $\langle \mathbfcal{D}_j,d_j \rangle$.
    \item Calculate the $\mu$HV value for all the divisions under each $d$ up to the maximum $d$ possible as determined by \texttt{CART}. Here, $d=0$ is omitted since in such a case \Model~is essentially the same as the single local model.
    \item Return the $d$ with the largest $\mu$HV value.
\end{enumerate}

With such an adaptive mechanism of $d$, better and more balanced results are achieved for the aforementioned objectives while requiring neither additional training nor profiling. The computational overhead of computing $\mu$HV is also ignorable, as shown in Section~\ref{chap-dal-subsec:overhead}.

\input{Tables/chap-DAL/alg2}

\subsection{Training}

Given the divisions generated by the \textit{Dividing} phase, a local model is trained for the samples from each division identified as part of \textbf{Goal 2} (lines 13-15 in Algorithm~\ref{alg:dal-code}). 
Theoretically, they can be paired with any model given the generic concept of dividable learning. However, this chapter adopts the state-of-the-art regularized Hierarchical Interaction Neural Network~\citep{DBLP:journals/tosem/ChengGZ23} (namely \texttt{HINNPerf}) to serve as the default in this work, as Cheng \textit{et al.} provide compelling empirical evidence on its effectiveness for handling feature sparsity and the interaction therein well for configurable software, even with very small training sample size. Section~\ref{chap-dal:evaluation} reports on the use and observation of pairing \Model~with \texttt{HINNPerf}.

This chapter adopts exactly the same structure and training procedure as those used by~\citet{DBLP:journals/tosem/ChengGZ23}, hence interested readers are kindly referred to their work for the training details. Since the local models of the divisions are independent, parallel training is utilized as part of \Model. 

It is worth stressing that, as it will be shown in Section~\ref{chap-meta-subsec:local}, \Model~is a model-agnostic framework and can improve a wide range of local models compared with the case when they are used to learn the entire configuration data space.

\subsection{Predicting}

When a new configuration arrives for prediction, \Model~chooses a local model of division trained previously to infer its performance. Therefore, the question is: how to assign the new configuration to the right model (\textbf{Goal 3})? A naive solution is to directly feed the configuration into the \texttt{CART} from the \textit{Dividing} phase and check which divisions it associates with. Yet, since the performance of the new configuration is unforeseen from the \texttt{CART}'s training data, this solution requires \texttt{CART} to generalize accurately, which, as mentioned, can easily lead to poor results because \texttt{CART} is overfitting-prone when directly working on new data~\citep{DBLP:journals/ese/KhoshgoftaarA01}.

Instead, by using the divided samples from the \textit{Dividing} phase (which serves as pseudo labeled data), a Random Forest---a widely used classifier and is resilient to overfitting ~\citep{DBLP:conf/splc/ValovGC15,DBLP:conf/oopsla/QueirozBC16,DBLP:conf/icse/0003XC021}--- is trained to generalize the decision boundary and predict which division that the new configuration should be better assigned to (lines 16-22 in Algorithm~\ref{alg:dal-code}). Again, in this way, less concern is placed on the overfitting issue of \texttt{CART} as long as it matches the patterns of training data well. This has now become a typical classification problem, but there are only pseudo labels to be used in the training. Using the example from Figure~\ref{fig:DT_example} again, if $d=1$ is chosen as the best value then the configurations in the 10 sample set would have a label \textit{``division1''}; similarly, those in the 8 sample set would result in a label \textit{``division2''}.

However, one issue that was experienced is that, even with $d=1$, the sample size of the two divisions can be rather imbalanced, which severely harms the quality of the classifier trained. For example, when training \textsc{BDB-C} with 18 samples, the first split in \texttt{CART} can lead to two divisions with 14 and 4 samples, respectively.
Therefore, before training the classifier Synthetic Minority Oversampling Technique (SMOTE)~\citep{DBLP:journals/bmcbi/BlagusL13a} is used to pre-process the pseudo label data, hence the division(s) with much less data (minority) can be more repeatedly sampled.

Finally, the classifier predicts a division whose local model would infer the performance of the new configuration.

%% file: Tables/chap-DAL/alg1.tex
\begin{algorithm}[t!]
	\DontPrintSemicolon
	\footnotesize
	
	\caption{Pseudo code of \Model}
	\label{alg:dal-code}
	\KwIn{A new configuration $\mathbf{\overline{c}}$ to be predicted}
     \KwOut{The predicted performance of $\mathbf{\overline{c}}$}
	
	\If{$\mathbfcal{M} = \emptyset$}
	{
	  \tcc{\textcolor{blue}{dividing phase.}}
      $\mathbfcal{S}\leftarrow$ randomly sample a set of configurations and their performance\\
      $\mathcal{T}\leftarrow$ \textsc{trainCART($\mathbfcal{S}$)}\\
      $d=$\textsc{adaptingDepth($\mathcal{T}$)}\\
      $d'=1$\\
      \While{$d'\leq d$}{
           \If{$d'<d$}
           {
             $\mathbfcal{D}\leftarrow$ extract all the leaf divisions of samples from $\mathcal{T}$ at the $d'$th depth\\
           }
           \Else{
              $\mathbfcal{D}\leftarrow$ extract all divisions of samples from $\mathcal{T}$ at the $d'$th depth\\
           }
           $d'=d'+1$\\
      }
      
      \tcc{\textcolor{blue}{training phase.}}
      \For{$\forall D_i \in \mathbfcal{D}$}  
      {
       $\mathbfcal{M}\leftarrow$ \textsc{trainLocalModel($D_i$)}\\
    
      }
    
    }
       
    \tcc{\textcolor{blue}{predicting phase.}}
    \If{$\mathcal{F}$ has not been trained}
	{
	$\mathbfcal{U}\leftarrow$ Removing performance data and labeling the configurations based on their divisions in $\mathbfcal{D}$\\
	$\mathbfcal{U'}\leftarrow$ \textsc{SMOTE($\mathbfcal{U}$)}\\
	$\mathcal{F}\leftarrow$ \textsc{trainRandomForest($\mathbfcal{U'}$)}
	}
	
    $D_i=$  \textsc{predict($\mathcal{F}$,$\mathbf{\overline{c}}$)}\\
    $\mathcal{M}=$ get the model from $\mathbfcal{M}$ that corresponds to the predicted division $D_i$\\

    \Return \textsc{predict($\mathcal{M}$,$\mathbf{\overline{c}}$)}\\
	
\end{algorithm}

%% file: Tables/chap-DAL/alg2.tex
\begin{algorithm}[t!]
	\DontPrintSemicolon
	\footnotesize
	
	\caption{Pseudo code of \textsc{AdaptingDepth}}
	\label{alg:adapt-code}
	\KwIn{A trained CART model $\mathcal{T}$}
     \KwOut{The most balanced $d$ value}
      $d_j=1$\\
      
       \tcc{\textcolor{blue}{step 1.}}
     \While{$d_j\leq d_{max}$ from $\mathcal{T}$}{
      $d'=1$\\
      $\mathbfcal{D}_j\leftarrow \emptyset$\\
      \While{$d'\leq d_j$ from $\mathcal{T}$}{
          
           \If{$d'<d_j$}
           {
             $\mathbfcal{D}_j\leftarrow$ extract all the leaf divisions of samples from $\mathcal{T}$ at the $d'$th depth\\
           }
           \Else{
              $\mathbfcal{D}_j\leftarrow$ extract all divisions of samples from $\mathcal{T}$ at the $d'$th depth\\
           }
            $d'=d'+1$\\
      }

            \For{$\forall D_i \in \mathbfcal{D}_j$} {
             
            \If{$ \langle h_i,z_i \rangle \notin \langle \mathbf{\overline{h}},\mathbf{\overline{z}} \rangle$}{
            
            $\langle \mathbf{\overline{h}},\mathbf{\overline{z}} \rangle\leftarrow$ calculate $h_i$ and $z_i$ for $D_i$ according to Equation~\ref{eq:adapt}\\
            
            }

            }

            
                $\mathbfcal{N}\leftarrow \langle \mathbfcal{D}_j,d_j \rangle$\\
                  $d_j=d_j+1$\\
      }
      
        \tcc{\textcolor{blue}{step 2.}}
        $\mathbf{\overline{r}}=\langle h_r,z_r \rangle\leftarrow$ find the nadir reference point from $\langle \mathbf{\overline{h}},\mathbf{\overline{z}} \rangle$\\
        \For{$\forall \langle \mathbfcal{D}_j,d_j \rangle \in \mathbfcal{N}$} {
         $v_j=\mu \text{HV}(\mathbfcal{D}_j,\mathbf{\overline{r}})$ according to Equation~\ref{eq:ahv}\\
         $\mathbfcal{F}\leftarrow \langle v_j,d_j \rangle$\\
        }

      \tcc{\textcolor{blue}{step 3.}}
       \Return the $d_j$ with the largest $v_j$ such that $\langle v_j,d_j \rangle \in \mathbfcal{F}$\\

\end{algorithm}

%% file: Chapter-DAL/setup.tex
\section{Experiment Setup}
\label{chap-dal:setup}

Here, the settings of the evaluation are delineated. In this work, \Model~is implemented based on \texttt{Tensorflow} and \texttt{scikit-learn}. All experiments were carried out on a server with 64-core Intel Xeon 2.6GHz and 256G DDR RAM.

\subsection{Research Questions}

In this chapter, \Model~is assessed by answering the following research questions (RQ):

\begin{itemize}
    \item \textbf{RQ3.1:} How accurate is \Model~compared with the state-of-the-art approaches, i.e., \texttt{HINNPerf}, \texttt{DeepPerf}, \texttt{Perf-AL}, \texttt{DECART}, and \texttt{SPLConqueror}, for configuration performance learning?

    \item \textbf{RQ3.2:} To what extent \Model~can improve different generic models (i.e., \texttt{HINNPerf}, regularized Deep Neural Network, \texttt{CART}, Linear Regression, and Support Vector Regression) when they are used locally therein for predicting configuration performance?

    \item \textbf{RQ3.3:} How do \Model~perform compared with the existing ensemble approaches such as \texttt{Bagging} and \texttt{Boosting}?

    \item \textbf{RQ3.4:} What is the benefit of the components in \Model? This consists of three sub-questions:

    \begin{itemize}
        \item What is the benefit of using \texttt{CART} for dividing in \Model~over the standard clustering algorithms?
        \item What is the benefit of the mechanism that adapts $d$ compared with the variant that relies on pre-tuned fixed $d$?
        \item What is the benefit of $\mu$HV over the original HV in adapting $d$?
    \end{itemize}

    \item \textbf{RQ3.5:} What is the sensitivity of \Model~to a fixed $d$ and how well does the adaptive mechanism perform in finding the optimal $d$ for each individual run?

    \item \textbf{RQ3.6:} What is the model building time for \Model?

\end{itemize}

\textbf{RQ3.1} is asked to assess the effectiveness of \Model~under different sample sizes against the state-of-the-art approaches. Since \Model~is naturally mode-agnostic and can be paired with different local models, \textbf{RQ3.2} is studied to examine how the concept of dividable learning can benefit any given local model. Since \Model~is similar to the ensemble approaches, e.g., \texttt{Bagging}~\citep{breiman1996bagging} and \texttt{Boosting}~\citep{schapire2003boosting}, with \textbf{RQ3.3}, the target is to examine how it performs against those. \textbf{RQ3.4} is mainly an ablation analysis of \Model, consisting of confirming the necessity of using \texttt{CART} to determine the divisions over unsupervised clustering algorithms; verifying the effectiveness of adapting $d$ compared with its counterpart under a fixed best $d$ value obtained via trial-and-error; and assessing the benefit of using the newly proposed $\mu$HV in quantifying the usefulness of the divisions in a $d$ value. \textbf{RQ3.5} examines, under an extensively increased training sample size, how the depth of division ($d$) can impact the performance of \Model~and how well the mechanism that adapts $d$ can locate the optimal setting in individual runs, which is the ground truth. Finally, the overall overhead of \Model~is examined in \textbf{RQ3.6}.

\begin{table}[t!]

\centering
\footnotesize
\caption{Details of the subject systems. ($|\mathbfcal{B}|$/$|\mathbfcal{N}|$) denotes the number of binary/numerical options, and $|\mathbfcal{C}|$ denotes the number of valid configurations (full sample size).}
\begin{adjustbox}{width=\columnwidth,center}
\input{Tables/chap-DAL/subject_systems}
\end{adjustbox}
\label{tb:subject-system}
\end{table}

\subsection{Subject Systems and Sample Sizes}
\label{chap-dal-subsec:subject_system}

Leveraging the criteria and procedure mentioned in Section~\ref{chap-dal-sec:empirical}, this chapter uses the same datasets of all valid configurations from real-world systems as widely used in the literature~\citep{DBLP:conf/esem/ShuS0X20,DBLP:conf/icse/HaZ19,DBLP:conf/sigsoft/SiegmundGAK15, DBLP:journals/ese/GuoYSASVCWY18, DBLP:journals/corr/abs-2106-02716}. The 12 configurable software systems studied in this work are specified in Table~\ref{tb:subject-system}. As can be seen, these software systems come with diverse domains, scales, and performance concerns. Some of them contain only binary configuration options (e.g., \textsc{x264}) while the others involve mixed options (binary and numeric), e.g., \textsc{HSMGP}, leading to configuration data that can be more difficult to model and generalize~\citep{DBLP:conf/icse/HaZ19}.

The configuration data of all the systems are collected by prior studies using the standard benchmarks with repeated measurements~\citep{DBLP:conf/esem/ShuS0X20,DBLP:conf/icse/HaZ19,DBLP:conf/sigsoft/SiegmundGAK15, DBLP:journals/ese/GuoYSASVCWY18, DBLP:journals/corr/abs-2106-02716}. For example, the configurations of \textsc{Apache}---a popular Web server---are benchmarked using the tools \texttt{Autobench} and \texttt{Httperf}, where workloads are generated and increased until reaching the point before the server crashes, and then the maximum load is marked as the performance value~\citep{DBLP:journals/ese/GuoYSASVCWY18}. The process repeats a few times for each configuration to ensure reliability. When there exist multiple measured datasets for the same software system, the one with the largest size is used.

To ensure generalizability of the results, for each system, the protocol used by existing work~\citep{DBLP:journals/sqj/SiegmundRKKAS12,DBLP:conf/icse/HaZ19,DBLP:conf/esem/ShuS0X20} is followed to obtain five sets of training sample size in the evaluation:

\begin{itemize}
    \item \textbf{Binary systems:} $n$, $2n$, $3n$, $4n$, and $5n$ configurations and their measurements are randomly sampled, where $n$ is the number of configuration options~\citep{DBLP:conf/icse/HaZ19,DBLP:conf/esem/ShuS0X20}.
    \item \textbf{Mixed systems:} the sizes suggested by \texttt{SPLConqueror}~\citep{DBLP:journals/sqj/SiegmundRKKAS12} (a state-of-the-art approach) are leveraged depending on the amount of budget. 

\end{itemize}

The results have been illustrated in Table~\ref{tb:sizes}. All the remaining samples in the dataset are used for testing.

\input{Tables/chap-DAL/sample_sizes}

\subsection{Metric and Statistical Validation}

\subsubsection{Accuracy}

For all the experiments, mean relative error (MRE) is used as the evaluation metric for prediction accuracy, since it provides an intuitive indication of the error and has been widely used in the domain of software performance prediction~\citep{DBLP:conf/icse/HaZ19, DBLP:conf/esem/ShuS0X20, DBLP:journals/ese/GuoYSASVCWY18}. Formally, the MRE is computed as:
\begin{equation}
     MRE = {{1} \over {k}} \times {\sum^k_{t=1} {{|A_t - P_t|} \over {A_t}}} \times 100\%
\end{equation}
\noindent whereby $A_t$ and $P_t$ denote the $t$th actual and predicted performance, respectively. To mitigate bias, all experiments are repeated for 30 runs via bootstrapping without replacement.

\subsubsection{Statistical Test}

Since the evaluation commonly involves comparing more than two approaches, Scott-Knott test~\citep{DBLP:journals/tse/MittasA13} is applied to evaluate their statistical significance on the difference of MRE over 30 runs, as recommended by~\citet{DBLP:journals/tse/MittasA13}. In a nutshell, Scott-Knott sorts the list of treatments (the approaches that model the system) by their median values of the MRE. Next, it splits the list into two sub-lists with the largest expected difference~\citep{xia2018hyperparameter}. For example, when comparing $A$, $B$, and $C$, a possible split could be $\{A, B\}$, $\{C\}$, with the rank ($r$) of 1 and 2, respectively. This means that, in the statistical sense, $A$ and $B$ perform similarly, but they are significantly better than $C$. Formally, the Scott-Knott test aims to find the best split by maximizing the difference $\Delta$ in the expected mean before and after each split:
\begin{equation}
    \Delta = \frac{|l_1|}{|l|}(\overline{l_1} - \overline{l})^2 + \frac{|l_2|}{|l|}(\overline{l_2} - \overline{l})^2
\end{equation}
whereby $|l_1|$ and $|l_2|$ are the sizes of two sub-lists ($l_1$ and $l_2$) from list $l$ with a size $|l|$. $\overline{l_1}$, $\overline{l_2}$, and $\overline{l}$ denote their mean MRE.

During the splitting, a statistical hypothesis test $H$ is applied to check if $l_1$ and $l_2$ are significantly different. This is done by using bootstrapping and $\hat{A}_{12}$~\citep{Vargha2000ACA}. If that is the case, Scott-Knott recurses on the splits. In other words, the approaches are divided into different sub-lists if both bootstrap sampling and effect size test suggest that a split is statistically significant (with a confidence level of 99\%) and with a good effect $\hat{A}_{12} \geq 0.6$. The sub-lists are then ranked based on their mean MRE.

%% file: Tables/chap-DAL/subject_systems.tex
\footnotesize
\setlength{\tabcolsep}{1mm}
\begin{tabular}{cccccc}
\toprule
\textbf{System} & \textbf{$|\mathbfcal{B}|$/$|\mathbfcal{N}|$} & \textbf{Performance} & \textbf{Description} & \textbf{$|\mathbfcal{C}|$} & \textbf{Used by}\\ 
\midrule
\textsc{Apache} & 9/0  & Maximum load & Web server              & 192 & \citep{DBLP:conf/icse/HaZ19,DBLP:journals/ese/GuoYSASVCWY18,DBLP:conf/esem/ShuS0X20} \\
\textsc{BDB-C}  & 16/0 & Latency (ms) & Database (C)    & 2560 & \citep{DBLP:conf/icse/HaZ19,DBLP:journals/ese/GuoYSASVCWY18,DBLP:conf/esem/ShuS0X20} \\
\textsc{BDB-J}  & 26/0 & Latency (ms) & Database (Java) & 180 & \citep{DBLP:conf/icse/HaZ19,DBLP:journals/ese/GuoYSASVCWY18,DBLP:conf/esem/ShuS0X20} \\
\textsc{kanzi}  & 31/0 & Energy (kj) & Compression tool & 3202 & \citep{DBLP:conf/icse/WeberKSAS23, 10172849} \\
\textsc{SQLite}  & 14/0 & Runtime (ms) & Database & 1000 & \citep{DBLP:journals/tse/KrishnaNJM21, DBLP:conf/icse/HaZ19, DBLP:journals/jss/CaoBWZLZ23} \\
\textsc{x264}   & 16/0 & Runtime (ms)  & Video encoder & 1152 & \citep{DBLP:conf/icse/HaZ19,DBLP:journals/ese/GuoYSASVCWY18,DBLP:conf/esem/ShuS0X20} \\
\textsc{Dune}   & 8/3 & Runtime (ms)  & Multi-grid solver           & 2304 & \citep{DBLP:conf/icse/HaZ19,DBLP:journals/ese/GuoYSASVCWY18,DBLP:conf/esem/ShuS0X20} \\
\textsc{HIPA$^{cc}$}   & 31/2 & Runtime (ms) & Compiler                & 13485 & \citep{DBLP:conf/icse/HaZ19,DBLP:conf/sigsoft/SiegmundGAK15} \\
\textsc{HSMGP}   & 11/3 & Runtime (ms) & Compiler                & 3456 & \citep{DBLP:conf/icse/HaZ19,DBLP:conf/sigsoft/SiegmundGAK15} \\
\textsc{Lrzip}   & 9/3 & Runtime (ms) & Compression tool                & 5184 & \citep{DBLP:journals/corr/abs-2106-02716, DBLP:conf/icse/WeberKSAS23, 10172849} \\
\textsc{nginx}  & 12/2 & Runtime (ms) & Web server & 4416 & \citep{DBLP:conf/icse/WeberKSAS23, DBLP:conf/sigsoft/WangWHSSZFLSJ22} \\
\textsc{VP8}   & 9/4 & Runtime (ms) & Video encoder                & 2736 & \citep{DBLP:journals/corr/abs-2106-02716, DBLP:conf/icse/WeberKSAS23} \\
\bottomrule
\end{tabular}

%% file: Tables/chap-DAL/sample_sizes.tex
\begin{table}[t!]
\caption{The training sample sizes used. $n$ denotes the number of configuration options in a binary system.}
\centering
\begin{adjustbox}{width=0.8\columnwidth,center}
\begin{tabular}{llllll}
\toprule
\textbf{System} & \textbf{Size 1 ($S_1$)} & \textbf{Size 2 ($S_2$)} & \textbf{Size 3 ($S_3$)} & \textbf{Size 4 ($S_4$)} & \textbf{Size 5 ($S_5$)} \\ 
\midrule
\textsc{Apache} & $n$  & $2n$ & $3n$  & $4n$ & $5n$\\
\textsc{BDB-C}  & $n$  & $2n$ & $3n$  & $4n$ & $5n$\\
\textsc{BDB-J}  & $n$  & $2n$ & $3n$  & $4n$ & $5n$\\
\textsc{kanzi}  & $n$  & $2n$ & $3n$  & $4n$ & $5n$\\
\textsc{SQLite}  & $n$  & $2n$ & $3n$  & $4n$ & $5n$\\
\textsc{x264}   & $n$  & $2n$ & $3n$  & $4n$ & $5n$\\
\textsc{Dune} & 224 & 692 & 1000 & 1365 & 1612\\
\textsc{HIPA$^{cc}$}  & 261 & 528 & 736 & 1281 & 2631\\
\textsc{HSMGP} & 77 & 173 & 384 & 480 & 864\\
\textsc{Lrzip} & 127 & 295 & 386 & 485 & 907\\
\textsc{nginx}  & 228  & 468 & 814  & 1012 & 1352\\
\textsc{VP8} & 121 & 273 & 356 & 467 & 830 \\
\bottomrule
\end{tabular}
\end{adjustbox}
\label{tb:sizes}
\end{table}

%% file: Chapter-DAL/evaluation.tex
\section{Evaluation}
\label{chap-dal:evaluation}

This section presents and discusses the experimental results.

\subsection{Comparing with the State-of-the-art}
\label{subsec:RQ3.1}

\begin{table*}[t!]
\caption{The median and interquartile range of MRE, denoted as Med-IQR, for \Model~and the state-of-the-art approaches for all the subject systems and training sizes over 30 runs. For each case, \setlength{\fboxsep}{1.5pt}\colorbox{green!20}{green cells} mean \Model~has the best median MRE; or \setlength{\fboxsep}{1.5pt}\colorbox{red!20}{red cells} otherwise. The one(s) with the best rank ($r$) from the Scott-Knott test is highlighted in bold.}
\input{Tables/chap-DAL/vs_STOA_scottknott}
\label{tb:vsSOTA}
\end{table*}

To understand how \Model~performs compared with the state-of-the-art approaches, its accuracy is assessed against both the standard approaches that rely on statistical learning, i.e., \texttt{SPLConqueror}~\citep{DBLP:journals/sqj/SiegmundRKKAS12} (linear regression and sampling methods) and \texttt{DECART}~\citep{DBLP:journals/ese/GuoYSASVCWY18} (an improved \texttt{CART}), together with recent deep learning-based ones, i.e., \texttt{HINNPerf}~\citep{DBLP:journals/tosem/ChengGZ23} (a hierarchical deep neural network), \texttt{DeepPerf}~\citep{DBLP:conf/icse/HaZ19} (a single global regularized deep neural network) and \texttt{Perf-AL}~\citep{DBLP:conf/esem/ShuS0X20} (an adversarial learning method). All approaches can be used for any type of system except for \texttt{DECART}, which works on binary systems only. Following the setting used by~\citet{DBLP:journals/tosem/ChengGZ23},  \texttt{SPLConqueror}\footnote{Since \texttt{SPLConqueror} supports multiple sampling methods, the one (or combination for the mixed system) that leads to the best MRE is used.} and \texttt{DECART} use their specific sampling method while \Model, \texttt{HINNPerf}, \texttt{DeepPerf} and \texttt{Perf-AL} rely on random sampling. Since there are 12 systems and 5 sets of sample sizes each, 60 cases are obtained to compare in total. The MREs and ranks from the Scott-Knott test are reported for all cases.

To ensure consistency, the implementations published by their authors with the same parameter settings are used. The systems, training sizes, and statistical tests are explained in Section~\ref{chap-dal:setup}. All experiments are repeated for 30 runs.  

\subsubsection{Results}

The results have been illustrated in Table~\ref{tb:vsSOTA}. It can be seen that, remarkably, \Model~achieves the best median MRE on 41 out of 60 cases. In particular, \Model~considerably improves the accuracy, i.e., by up to $1.61$ times better than the second-best one on $S_4$ of \textsc{Lrzip} (5.95 vs. 15.55). The above still holds when looking into the results of the statistical test: \Model~is ranked first for 44 out of the 60 cases, in which \Model~obtain the sole best rank for 31 cases. In particular, among the 19 cases where \Model~does not achieve the best MRE, the inferiority to the best on 6 of them is actually insignificant since it is still equally ranked as the best together with the others. That is to say, \Model~is, in 44 cases, similar to (13 cases) or significantly better (31 cases) than the best state-of-the-art approaches for each specific case (which could be a different approach). Overall, \Model~obtain an average rank of 1.48---the smallest among those of the others---indicating that it is much more likely to be ranked the best in terms of MRE.

When considering different training sample sizes, it can be seen that \Model~performs generally more inferior than the others when the size is too limited, i.e., $S_1$ and $S_2$ for the binary systems. This is expected as when there are too few samples, and each local model would have a limited chance to observe the right pattern after the splitting, hence blurring its effectiveness in handling sample sparsity. However, in the other cases (especially for mixed systems that have more data even for $S_1$), \Model~needs far fewer samples to achieve the same accuracy as the best state-of-the-art. For example, on \textsc{Lrzip}, \Model~only needs 295 samples ($S_2$) to achieve an accuracy better than the accuracy of the second best model \texttt{DeepPerf} with 907 samples ($S_5$), saving 67\% effort of data measurements.

Another observation is that the improvements of \Model~is much more obvious in mixed systems than those for binary systems. This is because: (1) the binary systems have fewer training samples as they have a smaller configuration space. Therefore, the data learned by each local model is more restricted. (2) The issue of sample sparsity is more severe on mixed systems, as their configuration landscape is more complex and comes with finer granularity.

As a result, it is anticipated that the benefit of \Model~can be amplified under more complex configurable systems and/or when the size of the training data sample increases.

To summarize, \textbf{RQ3.1} can be answered as:

\begin{quotebox}
   \noindent
   \textit{\textbf{RQ3.1:} For modeling configuration performance, \Model~performs similar or significantly more accurate than the best state-of-the-art approach in 44 out of 60 cases (73\%), in which 31 cases obtain the sole best results with up to $1.61$ times improvements. It also needs fewer samples to achieve the same accuracy, and the benefits can be amplified with complex systems/more training samples.} 
\end{quotebox}

\subsection{\Model~under Different Local Models}
\label{chap-meta-subsec:local}

\begin{table*}[t!]
\centering
\caption{The median and interquartile range of MRE, denoted as Med-IQR, for \Model~under different local models against using them as the global models for all the subject systems and training sizes over 30 runs. For each case, \setlength{\fboxsep}{1.5pt}\colorbox{green!20}{green cells} mean \Model~has the best median MRE; or \setlength{\fboxsep}{1.5pt}\colorbox{red!20}{red cells} otherwise. The one(s) with the best rank ($r$) from the Scott-Knott test is highlighted in bold.}
\input{Tables/chap-DAL/compare_models}
\label{chap-dal-tb:compare_models}
\end{table*}

Since the paradigm of dividable learning is naturally agnostic to the underlying local models, the aim is to understand how well \Model~performs with different local models against their global model counterparts (i.e., using them directly to learn the entire training dataset without dividing). To that end, experiments are run on a set of global models available in \texttt{scikit-learn} that are widely used in software engineering tasks to make predictions directly~\citep{DBLP:conf/msr/GongC22}, such as regularized Deep Neural Network (\texttt{DNN}), Decision Tree (\texttt{CART}), Random Forest (\texttt{RF}), Linear Regression (\texttt{LR}), Support Vector Regression (\texttt{SVR}) and eXtreme Gradient Boosting (\texttt{XGBoost}). The same settings are used as those for \textbf{RQ3.1} and all models' hyperparameters are tuned during the training. Similarly, the ranks $r$ produced by the Scott-Knott test and the median MRE (IQR) of the evaluation results are shown.

\subsubsection{Result}

From Table~\ref{chap-dal-tb:compare_models}, it is clear that when examining each pair of the counterparts in terms of the average rank, i.e., \texttt{DaL$_X$} and \textit{X}, \Model~can indeed improve the accuracy of the local model via the concept of ``divide-and-learn''. Such an improvement is particularly obvious on some simple models, such as \texttt{LR}, which might lead to $26.08$ times better MRE on \textsc{NGINX} under $S_1$. This confirms the generality and flexibility of \Model: for example, when \texttt{LR} needs to be used as the local model for the sake of training overhead, \Model~can still significantly improve the results compared with its global counterpart. 

Interestingly, when pairing \Model~with \texttt{CART} as the local model, it remains slightly better than using \texttt{CART} as the global model alone, even though their learning procedures are similar. This is possible due to two reasons: (1) \Model~adapts different techniques to assist the learning process, i.e., min-max scalling to normalize the configuration options, and SMOTE for over-sampling, which could affect the predictions, and (2) the actual result of the computed loss can be different in \texttt{CART} when it is presented with different proportions of the data samples. Yet, the resulting MREs do not differ much, as can be seen.

It is worth noting that the default of \Model, which uses the \texttt{HINNPerf} as the local model, still performs significantly better than the others: out of the 60 cases, 25 of them come with the best MRE, and there are 29 cases with the best rank, leading to an average rank of 2.05 which is again the best of amongst the others. This aligns with the previous findings that \texttt{HINNPerf} handles the feature sparsity better~\citep{DBLP:journals/tosem/ChengGZ23}. 

Therefore, for \textbf{RQ3.2}, the answer is:

\begin{quotebox}
   \noindent
   \textit{\textbf{RQ3.2:} Thanks to the paradigm of dividable learning, \Model~is model-agnostic, being able to significantly improve a range of global models when using them as the underlying local model in configuration performance learning. It is also revealed that pairing with \Model~with \texttt{HINNPerf} can lead to considerably better accuracy in this work.}
\end{quotebox}

\subsection{Comparing with Other Ensemble Approaches}
\label{subsec:rq3.3}

Since \Model~works with multiple local models, it could naturally be compared with the ensemble learning approaches that also involve a set of local models. Here, \Model~is compared with the most common ensemble learning approaches such as \texttt{Bagging} and \texttt{Boosting}. For the simplicity of exposition, \texttt{LR} and \texttt{CART} are used as the local models for all.

Specifically, \Model~that uses \texttt{LR} (denoted as \texttt{\Model$_{LR}$}) is compared with \texttt{Bagging$_{LR}$}~\citep{breiman1996bagging}  (an aggregated ensemble of \texttt{LR} based on random data projection) and \texttt{AdaBoosting$_{LR}$}~\citep{DBLP:conf/eurocolt/FreundS95} (a sequentially learned ensemble of \texttt{LR} based on majrotiy voting). Similar settings are followed for \texttt{CART}. Yet, since \texttt{CART} is a tree-like structure, there exist more diverse variants of ensembles. Therefore, this chapter examines \Model~that uses \texttt{CART} (denoted as \texttt{\Model$_{CART}$}) against \texttt{RF} (a bagging version that trains multiple \texttt{CART} models), \texttt{XGBoost}~\citep{DBLP:conf/kdd/ChenG16} (a boosting version that builds multiple \texttt{gbtree} booster~\citep{liu2014gb}, which is a variant of \texttt{CART}), and \texttt{AdaBoosting$_{CART}$} (another boosting version of \texttt{CART} that does not rely on optimization of a loss function).

Again, the \texttt{scikit-learn} package is leveraged. The default parameter settings for all these ensemble approaches are used, which tends to be ideal for most scenarios. The other experiment settings are the same as \textbf{RQ3.1} and \textbf{RQ3.2}.

\begin{table*}[t!]
\centering
\caption{The median and interquartile range of MRE, denoted as Med-IQR, for \Model~against the other ensemble approaches for all the subject systems and training sizes over 30 runs. For each case, \setlength{\fboxsep}{1.5pt}\colorbox{green!20}{green cells} mean \Model~has the best median MRE; or \setlength{\fboxsep}{1.5pt}\colorbox{red!20}{red cells} otherwise. The one(s) with the best rank ($r$) from the Scott-Knott test is highlighted in bold.}
\input{Tables/chap-DAL/compare_ensemble}
\label{tb:ensemble}
\end{table*}

\subsubsection{Results}
The results can be seen in Table~\ref{tb:ensemble}. When using \texttt{LR} as the local model, \texttt{\Model$_{LR}$} achieves considerably better accuracy than the other ensemble learning approaches, leading to the best MRE in 55 out of 60 cases with up to $28.50$ times improvement with respect to the second best counterpart (19.17 vs. 565.62 for $S_{5}$ of \textsc{nginx}). In particular, it obtains the best Soctt-Knott ranks on 51 out of 60 cases, within which only 2 cases \texttt{\Model$_{LR}$} does not achieve the sole best rank. This has resulted in the best average rank among the counterparts, i.e., 1.2.

Since \Model~leverages \texttt{CART} to divide the data samples, one can easily expect that it would benefit less when using \texttt{CART} again as the local model. As a 
result, the relative improvement of \texttt{\Model$_{CART}$} over the counterparts becomes blurred compared with that of \texttt{\Model$_{LR}$}. However, in Table~\ref{tb:ensemble}, it is seen that \texttt{\Model$_{CART}$} still achieve highly competitive outcomes in relation to the others: amongst the 60 cases, it is ranked as the best for 28 cases, which is higher than the 24, 17, and 10 cases for \texttt{RF}, \texttt{XGBoost}, and \texttt{AdaBoosting$_{CART}$}, respectively. Notably, \texttt{\Model$_{CART}$} also more frequently achieves the best MRE. All the above has led to its best average rank of 1.85. Looking at the detailed MRE differences, compared with \texttt{RF}, the largest improvement and degradation of \texttt{\Model$_{CART}$} is $4.83\times$ (14.44 vs. 84.12 for $S_{1}$ of \textsc{nginx}) and $0.50\times$ (11.35 vs. 7.58 for $S_{3}$ of \textsc{HSMGP}), respectively; compared with \texttt{XGBoost} it is $4.20\times$ (0.94 vs. 4.89 for $S_{5}$ of \textsc{VP8}) and $1.13\times$ (14.44 vs. 6.78 for $S_{1}$ of \textsc{nginx}), respectively; in contrast to \texttt{AdaBoost$_{CART}$}, it is respectively $5.54\times$ (6.72 vs. 43.94 for $S_{3}$ of \textsc{BDB-C}) and $2.43\times$ (7.44 vs. 2.17 for $S_{2}$ of \textsc{nginx}). All the above reveals the superior benefit of \Model~against the others.

With both \texttt{LR} and \texttt{CART}, it can be seen that the benefit of \Model~is more obvious for mixed systems: this is again due to the fact that those systems often come with a more sparse and complex configuration landscape, which is precisely what \Model~can cope with.

In conclusion, the following can be stated for \textbf{RQ3.3}:

\begin{quotebox}
   \noindent
   \textit{\textbf{RQ3.3:} Compared with existing ensemble learning approaches, \Model~generally has a better ability to utilize the local models for predicting configuration performance. The benefits are particularly obvious under simple local models and complex systems.}
\end{quotebox}

\begin{table*}[t!]
\centering
\caption{The median and interquartile range of MRE, denoted as Med-IQR, for the ablation study of \Model~with respect to the effectiveness of using \texttt{CART} for dividing samples, adapting $d$, and $\mu$HV for all the subject systems and training sizes over 30 runs. For each case, \setlength{\fboxsep}{1.5pt}\colorbox{green!20}{green cells} mean \Model~has the best median MRE; or \setlength{\fboxsep}{1.5pt}\colorbox{red!20}{red cells} otherwise. The one(s) with the best rank ($r$) from the Scott-Knott test is highlighted in bold.}
\input{Tables/chap-DAL/compare_clustering}
\label{tb:clustering}
\end{table*}

\subsection{The Effectiveness of Components in \Model}
\label{subsec:component}

\subsubsection{The Benefits of \texttt{CART}}

As discussed in Section~\ref{subsec:modeify-cart}, \texttt{CART} is modified to divide the configuration data samples since its training procedure naturally fits the needs better. To confirm this point, this chapter experimentally examines its effectiveness over several alternative clustering algorithms that can serve as the replacement of \texttt{CART} in the \textit{dividing} phase of \Model, namely: \texttt{DBSCAN}~\citep{DBLP:conf/kdd/EsterKSX96} (a density-based clustering algorithm), \texttt{Agglomerative clustering}~\citep{Inchoate:Ward63} (a hierarchical clustering algorithm), and \texttt{$k$Means}~\citep{DBLP:journals/tit/Lloyd82} (a centroid-based clustering algorithm). The most common Euclidean distance is used when both the configuration options and performance values are combined as the features of clustering.

Again, the implementations from the \texttt{scikit-learn} package are directly used, and the other experiment settings are the same as the previous RQs. Among these, \texttt{DBSCAN} adaptively determines the best number of divisions, while for \texttt{Agglomerative clustering} and \texttt{$k$Means}, the division numbers are tuned for each case and use the overall best setting throughout the runs. The other parameters are set as the default values.

As can be seen from Table~\ref{tb:clustering} (left), the original design of \Model~that extends \texttt{CART} achieves significantly and overwhelmingly better results: among the 60 cases, there are 56 cases of the solely best rank and 57 cases of the best MRE with up to $5.41$ times improvement over the second-best counterpart. The main reason is that, while the clustering algorithms do divide the configuration data, they often fail to take the prediction loss into account---a key feature within the training procedure of \texttt{CART}.

\subsubsection{The Necessity of Adapting $d$}

To investigate the necessity of adapting $d$, I compare \Model~with the version from my previous FSE work~\citep{DBLP:conf/sigsoft/Gong023} where an overall best $d$ value for each system-size pair is pre-defined via profiling, denoted as \texttt{DaL}-\texttt{FSE}. Pragmatically, the procedure that one would need to perform without $d$ adaptation is as follows:

\begin{enumerate}
    \item Choose a system and a training size considered.
    \item Pick a $d=k$ where $k \in\{1,2,..,d_{max}\}$ and train \texttt{DaL}-\texttt{FSE} using the corresponding training size, where $d_{max}$ is the largest depth that the \texttt{CART} can ever reach under a training size of the system.
    \item Repeat the experiment for 30 runs under $d=k$.
    \item Calculate the MREs of \texttt{DaL}-\texttt{FSE} under $d=k$ using all the remaining data as the testing set. 
    \item Repeat from (2) if not all possible $d$ value has been examined.
    \item Calculate the Scott-Knott ranks for MRE of \texttt{DaL}-\texttt{FSE} under different $d$ values.
    \item The $d$ value with the best rank is used; if there are multiple $d$ values that lead to the best rank, the one with the best average MRE is selected therein.
    \item Repeat from (1) if not all system and size pairs have been covered.
\end{enumerate}

All other settings are the same as the previous RQs.

The comparisons between \texttt{DaL}-\texttt{FSE} and the proposed \Model~are illustrated in Table~\ref{tb:clustering} (middle). Clearly, both approaches perform very similarly across the cases. In particular, \Model~reaches identical results to that of \texttt{DaL}-\texttt{FSE}, which is pre-tuned, in 54 out of 60 cases. Interestingly, it is even possible for \Model~to achieve better accuracy, e.g., in $S_{2}$ of x264. This is because \Model~adapts $d$ for each individual run while \texttt{DaL}-\texttt{FSE} uses an overall best setting, as such \Model~will be able to control the $d$ in a finer granularity, hence further pushing the full potential of the dividable learning paradigm.

Considering the overhead required to pre-define the $d$ for \texttt{DaL}-\texttt{FSE} (more than 80 hours for selecting the best out of all possible $d$ values for each of the 60 cases) and the above competitive results of \Model, it can be said that adapting $d$ is highly necessary (and effective) to improve the efficiency of configuration performance learning, i.e., the process of $d$ adaptation only takes $\approx$5 minutes for all 30 repeated runs of the 60 cases since it does not require any extra training.

\subsubsection{The Usefulness of $\mu$HV}

Recall that Section~\ref{subsubsec:adapting_depth} has theoretically analyzed why the newly proposed indicator $\mu$HV is needed for adapting $d$. Here, to confirm its usefulness, the results of \Model~are compared with its variant where the standard HV is used instead, which is denoted as \Model$^{HV}$. To that end, the HV implementation from the Python library \texttt{pymoo} is leveraged. All other settings are the same as the previous RQs.

From Table~\ref{tb:clustering} (right), it can be seen that using $\mu$HV clearly achieves considerably better accuracy than using the standard HV: the former is overwhelmingly ranked better (47 cases) or similar (12 cases) out of the 60 cases, which is a remarkable result.

Overall, for \textbf{RQ3.4}, it is concluded that:

\begin{quotebox}
   \noindent
   \textit{\textbf{RQ3.4:}} The designed components in \Model~is effective:

   \begin{enumerate}
       \item \texttt{CART} is ranked the solely best for 93.3\% (56/60) of the cases amongst the other clustering counterparts with up to $5.41$ times MRE improvement over the second-best.
       \item Adapting $d$ achieves almost identical results to the version of \Model~that relies on a pre-tuned $d$ value while saving an extensive amount of effort.
       \item $\mu$HV is more suitable than the standard HV for determining the optimal and most balanced $d$ value on the configuration data, leading to 98\% (59/60) cases of better (47/60=78\%) or similar (12/60=20\%) accuracy.
   \end{enumerate}
   
\end{quotebox}

\subsection{Sensitivity and Adaptivity to the Depth $d$}
\label{subsec:sen}

To understand the sensitivity of \Model~to the depth, different $d$ values are examined. Since the number of divisions (and hence the possible depth) is sample size-dependent, for each system, 80\% of the full dataset is used for training and the remaining for testing. This has allowed us to achieve up to $d=4$ with 16 divisions as the maximum possible bound for all systems studied. For different $d$ values, the median MRE together over 30 runs is reported, as well as counting the number of times that a $d$ value leads to the best MRE. 

Recall that \textbf{RQ3.4} examines \Model-\texttt{FSE}, which serves as the variant that takes a generally optimal $d$ (over all repeated runs) obtained via pre-tuning. Yet, while \Model-\texttt{FSE} mimics the practical scenario, it does not cater to the optimal $d$ for each individual run in which the training/testing data differ. Therefore, here, this chapter further examines to what extent \Model~can indeed reach the optimal $d$, considering individual run under each system-size pair, which serves as the ground truth.

All other settings are the same as the previous RQs.

\begin{figure*}[!t]
    \centering
    \footnotesize
    \input{Figures/chap-DAL/Sensitivity_to_depth}
    \caption{The median MRE (\markone{0}{20}{10}{20}), its IQR (area), and the number of runs that a $d$ value leads to the best MRE (\marktwo{0}{20}{7.5}{20}) over all systems and 30 runs.}
    \label{fig:depths}
\end{figure*}

\begin{figure*}[!t]
    \centering
    \input{Figures/chap-DAL/hit-count-new}
    \caption{Hit count of reaching the optimal $d$ at each run (which leads to the best MRE) by the adaptive mechanism in \Model. The \setlength{\fboxsep}{1.5pt}\colorbox{red!20}{red bars} and \setlength{\fboxsep}{1.5pt}\colorbox{green!20}{green bars} show the miss hit counts and correct hit counts, respectively.}
    \label{fig:hit-count}
\end{figure*}
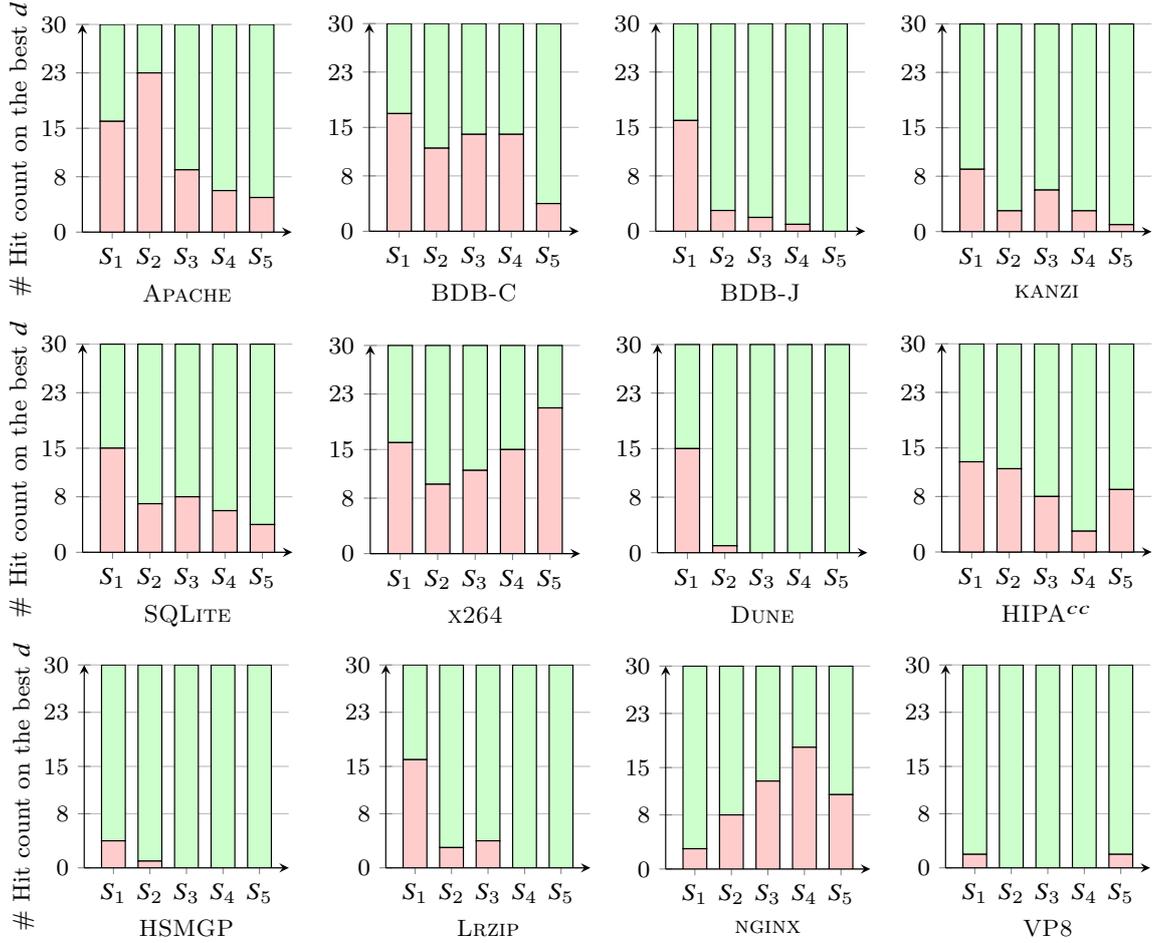

\begin{figure*}[!t]
\centering
    \input{Figures/chap-DAL/box_plot_ground_truth}
\caption{MRE difference between the $d$ adapted by \Model~and the optimal $d$ for each of the 30 runs.}
\label{fig:box-plot}
\end{figure*}
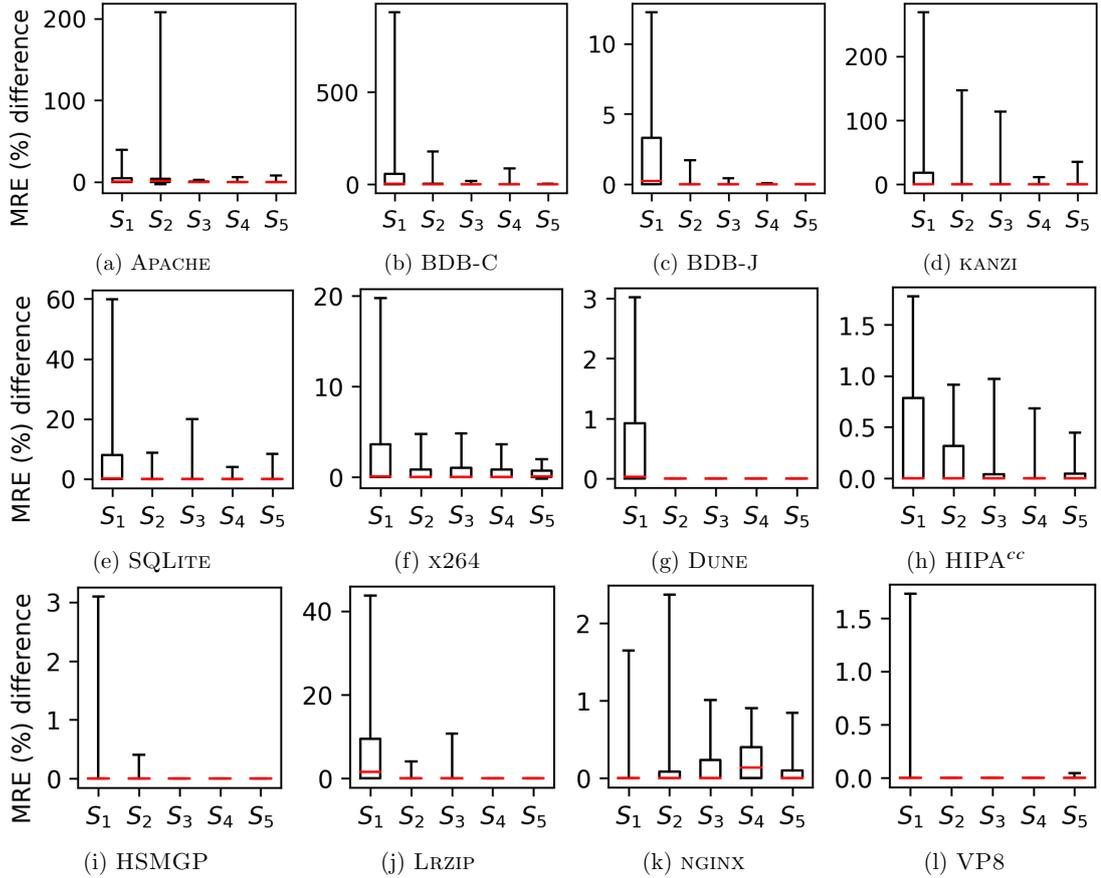

\subsubsection{Results}

In Figure~\ref{fig:depths}, the best $d$ of each subject system is marked as $*$, which has the best Scott-Knott rank, and if more than one $d$ is ranked first, the one with the smallest median MRE is marked.

As can be seen from Figure~\ref{fig:depths}, it can be seen that the correlation between the MRE (blue axis) of \Model~and $d$ value is close to quadratic: \Model~reaches its best MRE with $d=1$ (2 divisions) or $d=2$ (4 divisions). Since $d$ controls the trade-off between the ability to handle sample sparsity and ensuring sufficient data samples to train all local models, $d=1$ or $d=2$ tends to be the ``sweet points'' that reach a balance for the systems studied. After the point of $d=1$ or $d=2$, the MRE will worsen, as the local models' training size often drops dramatically. This is a clear sign that, from that point, the side-effect of having too few samples to train a local model has started to surpass the benefit that could have been brought by dealing with sample sparsity using more local models. When $d=0$, which means only one division and hence \Model~is reduced to \texttt{HINNPerf} that ignores sample sparsity, the resulted MRE is the worst on 6 out of 12 systems; a similar result can be observed for the cases when $d=4$. This suggests that neither too small $d$ (e.g., $d=0$ with only one division) nor too large $d$ (e.g., $d=4$ with up to 16 divisions, i.e., too many divisions) are ideal, which matches the theoretical analysis in Section~\ref{subsubsec:adapting_depth}.

Clearly, another important observation is that there is not a single optimal $d$ that can be applied to all systems: four of them favor $d=2$ in general while the remaining reveal that $d=1$ is the overall optimal setting. This becomes even more obvious when one takes a closer look at each individual run with different training/testing data: as from the count of optimal $d$ (red axis), for the case where $d=1$ leads to the generally best MRE, there are a considerable number of runs under which $d=2$ is the optimal, even occasionally $d=3$, e.g., for system \textsc{kanzi}. The above suggests that \Model~is highly sensitive to $d$, depending on the systems and training/testing data---the key motivation to the extension of adapting $d$ in this work.

To compare the ability of \Model~on adapting $d$ against the ground truth for each individual run, the hit/miss hit counts are plotted in Figure~\ref{fig:hit-count}. As can be seen, in the majority of the cases, \Model~can accurately adapt to the optimal $d$ of a run. There are a good amount of cases where the hit count is 100\%, and only in 5 out of 60 cases, the miss hit counts are more than 50\% of the runs. A clear pattern is also observed where the hit count often increases as the training size expands, e.g., for systems \textsc{BDB-J} and \textsc{Dune}: this is as expected as the conflicts between the ability to handle sample sparsity and the necessary amount of data to learn are relieved when the amount of data increases.

Overall, the average hit count for all the cases is 22.93 out of 30, which means in 76.43\% of the runs, \Model~can indeed adapt to the optimal $d$. For mixed systems, \Model~generally reaches the optimal $d$ more frequently, as the average hit count for all the mixed systems is 25.13 against the 20.73 for binary systems. This is because the complex configuration landscape for mixed systems can provide more information when assessing the $d$ values during its adaptation process. To understand the severity of miss hitting the optimal $d$ in a run, the MRE difference between the results of the optimal $d$ and what is achieved by \Model~under an adapted $d$ are reported. It is clear that, from Figure~\ref{fig:box-plot}, the MRE difference is marginal in most of the cases according to their distributions: only very occasionally there are one or two extreme cases of more than 50\% MRE difference, the majority of them are within 10\%. This implies that, even for the cases/runs where \Model~does miss hit, a promising $d$ value can still be chosen, leading to minimal accuracy loss.

Therefore, for \textbf{RQ3.5}, it can be concluded that:

\begin{quotebox}
   \noindent
   \textit{\textbf{RQ3.5:} The error of \Model~has a (upward) quadratic correlation to $d$, the optimal value of which varies depending on the actual systems and training/testing data. \Model~can adapt to the optimal $d$ in 76.43\% of the individual runs while even when it misses hit, a promising $d$ value that leads to generally marginal MRE degradation can still be selected.}

\end{quotebox}

\subsection{Overhead of Model Building}
\label{chap-dal-subsec:overhead}

To study \textbf{RQ3.6}, the overall time required for training and building the models is examined, including those approaches compared in \textbf{RQ3.1}.

\subsubsection{Result}

From Table~\ref{tb:max_min}, which shows the range of training time required across all 60 cases, \Model~incurs an overall overhead from 4 to 56 minutes. Yet, from the breakdown, it can be noted that the majority of the overhead comes from the \textit{training phase} that trains the local models. This is expected, as \Model~uses \texttt{HINNPerf} by default, which is a deep learning model combined with auto-encoding. In particular, the time required for adapting $d$ is rather trivial, i.e., less than a second, thanks to the fact that no extra training is needed.

Compared with other approaches, i.e., \texttt{HINNPerf} (3 to 54 minutes) and \texttt{DeepPerf} (3 to 48 minutes), \Model~have slightly higher training overhead, but the difference is merely in a matter of a few minutes, even though it trains multiple local models. This is because (1) each local model has less data to train, and (2) those local models can be trained in parallel, which speeds up the process. However, it is worth noting that, from \textbf{RQ3.1}, \Model~leads to considerably better accuracy than those two. In contrast to \texttt{Perf-AL} (a few seconds to two minutes), \Model~appears to be rather slow as the former does not use hyperparameter tuning but fixed-parameter values~\citep{DBLP:conf/esem/ShuS0X20}. Yet, as it has been shown for \textbf{RQ3.1}, \Model~achieves up to a few magnitudes of accuracy improvement. Although \texttt{SPLConqueror} and \texttt{DECART} have an overhead of less than a minute, again, their accuracy is much more inferior. Further, \texttt{SPLConqueror} requires a good selection of the sampling method(s) (which can largely incur additional overhead) while \texttt{DECART} does not work on mixed systems. 

\input{Tables/chap-DAL/min_max_time}

In summary, it can be summarized that:

\begin{quotebox}
   \noindent
   \textit{\textbf{RQ3.6:} \Model, when using \texttt{HINNPerf} as the local model, has competitive model building time with respect to \texttt{HINNPerf} and \texttt{DeepPerf}, and has higher overhead than the other state-of-the-art approaches, but this can be acceptable considering its improvement in accuracy.}
\end{quotebox}

%% file: Tables/chap-DAL/vs_STOA_scottknott.tex
\scriptsize
\begin{adjustbox}{width=1.2\textwidth,center}
\begin{tabular}{clll|ll|ll|ll|ll|ll} \toprule
\multirow{2}{*}{System} & \multicolumn{1}{c}{\multirow{2}{*}{Size}} & \multicolumn{2}{c}{\texttt{DaL}} & \multicolumn{2}{c}{\texttt{HINNPerf}} & \multicolumn{2}{c}{\texttt{DeepPerf}} & \multicolumn{2}{c}{\texttt{Perf-AL}} & \multicolumn{2}{c}{\texttt{DECART}} & \multicolumn{2}{c}{\texttt{SPLConqueror}} \\ \cline{3-14}
 & \multicolumn{1}{c}{} & $r$ & Med (IQR) & $r$ & Med (IQR) & $r$ & Med (IQR) & $r$ & Med (IQR) & $r$ & Med (IQR) & $r$ & Med (IQR) \\ \hline
\multirow{5}{*}{\textsc{Apache}} & $S_1$ & 4 & 21.86 (7.36) & 4 & 19.44 (10.53) & 3 & 20.19 (6.34) & 5 & 33.57 (11.34) & 2 & 19.44 (6.48) & \cellcolor{red!20}\textbf{1} & \cellcolor{red!20}\textbf{14.44 (0.00)} \\
 & $S_2$ & 5 & 12.89 (7.87) & \cellcolor{red!20}\textbf{1} & \cellcolor{red!20}\textbf{8.93 (3.89)} & 2 & 9.79 (4.56) & 6 & 32.97 (6.24) & 3 & 9.77 (5.36) & 4 & 13.89 (0.00) \\
 & $S_3$ & \cellcolor{green!20}\textbf{1} & \cellcolor{green!20}\textbf{6.56 (1.92)} & \textbf{1} & \textbf{7.01 (1.74)} & 2 & 7.98 (1.91) & 4 & 31.73 (4.66) & 2 & 8.07 (1.05) & 3 & 13.99 (0.00) \\
 & $S_4$ & 2 & 6.87 (1.95) & \cellcolor{red!20}\textbf{1} & \cellcolor{red!20}\textbf{6.58 (1.04)} & 2 & 6.85 (1.60) & 5 & 30.67 (6.84) & 3 & 7.47 (0.72) & 4 & 13.84 (0.00) \\
 & $S_5$ & \cellcolor{green!20}\textbf{1} & \cellcolor{green!20}\textbf{5.78 (1.08)} & \textbf{1} & \textbf{5.85 (0.99)} & 2 & 6.66 (1.61) & 5 & 30.45 (4.91) & 3 & 7.14 (0.89) & 4 & 14.07 (0.00) \\ \hline
\multirow{5}{*}{\textsc{BDB-C}} & $S_1$ & 5 & 83.60 (208.75) & 5 & 167.55 (144.47) & \cellcolor{red!20}\textbf{1} & \cellcolor{red!20}\textbf{43.66 (42.88)} & 4 & 82.38 (57.27) & 3 & 51.88 (45.91) & 2 & 86.86 (0.00) \\
 & $S_2$ & 3 & 21.92 (17.91) & 2 & 21.45 (13.61) & 2 & 18.17 (8.11) & 4 & 73.99 (22.51) & \cellcolor{red!20}\textbf{1} & \cellcolor{red!20}\textbf{13.37 (7.25)} & 5 & 162.91 (0.00) \\
 & $S_3$ & 2 & 8.31 (7.52) & \cellcolor{red!20}\textbf{1} & \cellcolor{red!20}\textbf{6.50 (9.20)} & 3 & 12.47 (4.08) & 3 & 69.74 (4.33) & \textbf{1} & \textbf{7.56 (5.07)} & 4 & 193.89 (0.00) \\
 & $S_4$ & \cellcolor{green!20}2 & \cellcolor{green!20}3.60 (4.50) & \textbf{1} & \textbf{4.86 (2.18)} & 2 & 9.06 (9.76) & 3 & 69.50 (2.49) & \textbf{1} & \textbf{5.17 (5.07)} & 4 & 231.68 (0.00) \\
 & $S_5$ & \cellcolor{green!20}\textbf{1} & \cellcolor{green!20}\textbf{1.88 (1.24)} & 3 & 4.04 (1.24) & 4 & 7.08 (8.03) & 5 & 69.29 (0.69) & 2 & 3.54 (1.92) & 6 & 296.68 (0.00) \\ \hline
\multirow{5}{*}{\textsc{BDB-J}} & $S_1$ & 3 & 5.37 (4.74) & 2 & 3.58 (1.74) & 3 & 2.98 (3.34) & 5 & 37.45 (2.45) & \cellcolor{red!20}\textbf{1} & \cellcolor{red!20}\textbf{2.36 (0.75)} & 4 & 12.29 (0.00) \\
 & $S_2$ & \cellcolor{green!20}\textbf{1} & \cellcolor{green!20}\textbf{1.83 (0.45)} & \textbf{1} & \textbf{1.96 (0.36)} & 2 & 1.91 (0.68) & 4 & 37.90 (1.67) & \textbf{1} & \textbf{1.84 (0.17)} & 3 & 24.85 (0.00) \\
 & $S_3$ & \cellcolor{green!20}\textbf{1} & \cellcolor{green!20}\textbf{1.58 (0.26)} & 2 & 1.75 (0.41) & 3 & 2.01 (1.10) & 5 & 37.19 (1.70) & \textbf{1} & \textbf{1.64 (0.24)} & 4 & 25.88 (0.00) \\
 & $S_4$ & \cellcolor{green!20}\textbf{1} & \cellcolor{green!20}\textbf{1.45 (0.22)} & 2 & 1.52 (0.24) & 3 & 1.70 (0.37) & 5 & 37.74 (2.59) & \textbf{1} & \textbf{1.47 (0.12)} & 4 & 29.01 (0.00) \\
 & $S_5$ & \textbf{1} & \textbf{1.40 (0.34)} & 2 & 1.48 (0.32) & 3 & 1.69 (0.48) & 5 & 35.76 (4.47) & \cellcolor{red!20}\textbf{1} & \cellcolor{red!20}\textbf{1.37 (0.26)} & 4 & 29.88 (0.00) \\ \hline
\multirow{5}{*}{\textsc{kanzi}} & $S_1$ & \cellcolor{green!20}\textbf{1} & \cellcolor{green!20}\textbf{196.15 (213.73)} & 2 & 262.27 (175.07) & \textbf{1} & \textbf{196.36 (122.61)} & 3 & 1434.70 (1328.18) & 2 & 276.50 (272.43) & 4 & 2528.40 (0.00) \\
 & $S_2$ & 2 & 95.31 (51.34) & \cellcolor{red!20}\textbf{1} & \cellcolor{red!20}\textbf{80.53 (51.36)} & 3 & 107.12 (43.75) & 4 & 961.11 (877.82) & 3 & 126.50 (142.62) & 4 & 1191.03 (0.00) \\
 & $S_3$ & \cellcolor{green!20}2 & \cellcolor{green!20}59.30 (38.47) & \textbf{1} & \textbf{62.20 (14.07)} & 2 & 68.62 (30.41) & 4 & 280.93 (147.45) & 3 & 74.08 (32.14) & 5 & 1059.28 (0.00) \\
 & $S_4$ & \cellcolor{green!20}\textbf{1} & \cellcolor{green!20}\textbf{42.93 (18.75)} & \textbf{1} & \textbf{46.48 (19.58)} & 2 & 62.65 (26.54) & 4 & 266.53 (179.59) & 3 & 67.83 (15.79) & 5 & 1030.25 (0.00) \\
 & $S_5$ & \textbf{1} & \textbf{35.29 (17.13)} & \cellcolor{red!20}\textbf{1} & \cellcolor{red!20}\textbf{34.27 (11.87)} & 2 & 49.42 (9.99) & 4 & 169.12 (64.60) & 3 & 67.27 (17.95) & 5 & 1007.23 (0.00) \\ \hline
\multirow{5}{*}{\textsc{SQLite}} & $S_1$ & \textbf{1} & \textbf{73.52 (20.40)} & \cellcolor{red!20}\textbf{1} & \cellcolor{red!20}\textbf{70.76 (27.86)} & \textbf{1} & \textbf{79.49 (25.36)} & 3 & 573.26 (340.38) & \textbf{1} & \textbf{74.14 (17.98)} & 2 & 157.09 (0.00) \\
 & $S_2$ & \textbf{1} & \textbf{76.41 (16.66)} & 2 & 76.69 (23.32) & 2 & 77.75 (22.47) & 3 & 386.74 (323.30) & \cellcolor{red!20}2 & \cellcolor{red!20}76.20 (22.74) & 2 & 84.41 (0.00) \\
 & $S_3$ & \textbf{1} & \textbf{71.63 (11.58)} & \cellcolor{red!20}\textbf{1} & \cellcolor{red!20}\textbf{71.02 (17.30)} & \textbf{1} & \textbf{75.01 (15.58)} & 2 & 424.32 (311.00) & \textbf{1} & \textbf{74.16 (20.12)} & \textbf{1} & \textbf{75.72 (0.00)} \\
 & $S_4$ & \cellcolor{green!20}\textbf{1} & \cellcolor{green!20}\textbf{66.94 (10.61)} & 2 & 70.84 (10.38) & 3 & 72.45 (12.11) & 4 & 363.18 (415.63) & 2 & 71.57 (14.72) & 2 & 69.61 (0.00) \\
 & $S_5$ & \cellcolor{green!20}\textbf{1} & \cellcolor{green!20}\textbf{64.20 (13.91)} & 3 & 70.15 (9.71) & 4 & 72.87 (8.86) & 5 & 212.19 (295.12) & 3 & 68.51 (8.54) & 2 & 67.78 (0.00) \\ \hline
\multirow{5}{*}{\textsc{x264}} & $S_1$ & 4 & 10.21 (4.35) & 3 & 8.44 (2.69) & 2 & 8.04 (3.06) & 5 & 37.00 (9.43) & 3 & 9.27 (2.11) & \cellcolor{red!20}\textbf{1} & \cellcolor{red!20}\textbf{7.01 (0.00)} \\
 & $S_2$ & \cellcolor{green!20}2 & \cellcolor{green!20}2.72 (1.52) & \textbf{1} & \textbf{2.77 (1.07)} & 2 & 3.17 (1.00) & 5 & 34.98 (7.81) & 3 & 6.29 (1.72) & 4 & 6.64 (0.00) \\
 & $S_3$ & 2 & 1.57 (1.52) & \cellcolor{red!20}\textbf{1} & \cellcolor{red!20}\textbf{1.37 (0.46)} & 2 & 2.23 (0.90) & 5 & 36.25 (7.73) & 3 & 4.57 (1.03) & 4 & 6.31 (0.00) \\
 & $S_4$ & 2 & 1.30 (1.17) & \cellcolor{red!20}\textbf{1} & \cellcolor{red!20}\textbf{0.81 (0.28)} & 3 & 1.74 (0.98) & 6 & 36.41 (6.47) & 4 & 3.66 (1.42) & 5 & 6.32 (0.00) \\
 & $S_5$ & 2 & 0.96 (0.79) & \cellcolor{red!20}\textbf{1} & \cellcolor{red!20}\textbf{0.58 (0.15)} & 3 & 1.44 (0.90) & 6 & 36.20 (3.75) & 4 & 2.15 (1.05) & 5 & 6.36 (0.00) \\ \hline
\multirow{5}{*}{\textsc{Dune}} & $S_1$ & 3 & 9.37 (1.15) & 2 & 7.98 (0.98) & \cellcolor{red!20}\textbf{1} & \cellcolor{red!20}\textbf{7.86 (0.93)} & 5 & 55.73 (0.12) & 6 & \texttimes & 4 & 13.05 (0.00) \\
 & $S_2$ & \cellcolor{green!20}\textbf{1} & \cellcolor{green!20}\textbf{5.95 (0.54)} & 2 & 6.15 (0.40) & \textbf{1} & \textbf{6.01 (0.21)} & 4 & 55.72 (0.53) & 5 & \texttimes & 3 & 13.09 (0.00) \\
 & $S_3$ & \cellcolor{green!20}\textbf{1} & \cellcolor{green!20}\textbf{4.81 (0.48)} & 2 & 5.48 (0.39) & 2 & 5.52 (0.34) & 4 & 55.53 (1.69) & 5 & \texttimes & 3 & 13.10 (0.00) \\
 & $S_4$ & \cellcolor{green!20}\textbf{1} & \cellcolor{green!20}\textbf{4.23 (0.35)} & 2 & 5.12 (0.38) & 3 & 5.26 (0.31) & 5 & 54.58 (1.48) & 6 & \texttimes & 4 & 13.16 (0.00) \\
 & $S_5$ & \cellcolor{green!20}\textbf{1} & \cellcolor{green!20}\textbf{3.98 (0.19)} & 2 & 4.98 (0.64) & 3 & 5.15 (0.34) & 5 & 34.53 (21.14) & 6 & \texttimes & 4 & 13.17 (0.00) \\ \hline
\multirow{5}{*}{\textsc{HIPA$^{cc}$}} & $S_1$ & \textbf{1} & \textbf{7.50 (1.42)} & \cellcolor{red!20}\textbf{1} & \cellcolor{red!20}\textbf{7.40 (1.05)} & 2 & 9.70 (1.28) & 4 & 31.99 (0.06) & 5 & \texttimes & 3 & 16.68 (0.00) \\
 & $S_2$ & \cellcolor{green!20}\textbf{1} & \cellcolor{green!20}\textbf{4.48 (0.38)} & 2 & 4.53 (0.48) & 3 & 6.89 (1.44) & 5 & 31.98 (0.06) & 6 & \texttimes & 4 & 16.96 (0.00) \\
 & $S_3$ & \cellcolor{green!20}\textbf{1} & \cellcolor{green!20}\textbf{3.43 (0.34)} & 2 & 3.67 (0.28) & 3 & 4.68 (0.90) & 5 & 31.99 (0.06) & 6 & \texttimes & 4 & 17.18 (0.00) \\
 & $S_4$ & \cellcolor{green!20}\textbf{1} & \cellcolor{green!20}\textbf{2.60 (0.15)} & 2 & 2.70 (0.18) & 3 & 3.58 (1.00) & 5 & 31.98 (0.06) & 6 & \texttimes & 4 & 17.42 (0.00) \\
 & $S_5$ & \cellcolor{green!20}\textbf{1} & \cellcolor{green!20}\textbf{2.11 (0.08)} & 2 & 2.23 (0.23) & 3 & 2.82 (0.57) & 5 & 31.97 (0.14) & 6 & \texttimes & 4 & 17.69 (0.00) \\ \hline
\multirow{5}{*}{\textsc{HSMGP}} & $S_1$ & \cellcolor{green!20}\textbf{1} & \cellcolor{green!20}\textbf{4.14 (1.44)} & 2 & 5.03 (2.56) & 4 & 7.09 (3.04) & 5 & 66.70 (1294.05) & 6 & \texttimes & 3 & 15.04 (0.00) \\
 & $S_2$ & \cellcolor{green!20}\textbf{1} & \cellcolor{green!20}\textbf{2.06 (0.19)} & 2 & 3.13 (0.49) & 3 & 3.69 (0.61) & 5 & 66.67 (0.11) & 6 & \texttimes & 4 & 22.63 (0.00) \\
 & $S_3$ & \cellcolor{green!20}\textbf{1} & \cellcolor{green!20}\textbf{1.42 (0.13)} & 2 & 2.34 (0.38) & 2 & 2.28 (0.37) & 4 & 66.63 (0.18) & 5 & \texttimes & 3 & 32.46 (0.00) \\
 & $S_4$ & \cellcolor{green!20}\textbf{1} & \cellcolor{green!20}\textbf{1.33 (0.08)} & 2 & 2.19 (0.23) & 3 & 2.23 (0.39) & 5 & 66.63 (0.14) & 6 & \texttimes & 4 & 33.74 (0.00) \\
 & $S_5$ & \cellcolor{green!20}\textbf{1} & \cellcolor{green!20}\textbf{1.19 (0.08)} & 2 & 2.09 (0.25) & 3 & 1.94 (0.63) & 5 & 66.59 (0.20) & 6 & \texttimes & 4 & 35.86 (0.00) \\ \hline
\multirow{5}{*}{\textsc{Lrzip}} & $S_1$ & \cellcolor{green!20}\textbf{1} & \cellcolor{green!20}\textbf{31.88 (10.96)} & 3 & 41.97 (12.02) & 2 & 35.40 (16.59) & 4 & 58.45 (0.12) & 5 & \texttimes & 4 & 123.70 (0.00) \\
 & $S_2$ & \cellcolor{green!20}\textbf{1} & \cellcolor{green!20}\textbf{10.51 (4.39)} & 3 & 24.10 (19.98) & 2 & 21.46 (4.53) & 4 & 58.46 (0.18) & 6 & \texttimes & 5 & 171.85 (0.00) \\
 & $S_3$ & \cellcolor{green!20}\textbf{1} & \cellcolor{green!20}\textbf{8.04 (2.37)} & 3 & 23.71 (16.49) & 2 & 18.68 (3.20) & 4 & 58.44 (0.21) & 6 & \texttimes & 5 & 177.61 (0.00) \\
 & $S_4$ & \cellcolor{green!20}\textbf{1} & \cellcolor{green!20}\textbf{5.95 (0.99)} & 3 & 19.82 (16.76) & 2 & 15.55 (1.71) & 4 & 58.48 (0.21) & 6 & \texttimes & 5 & 208.08 (0.00) \\
 & $S_5$ & \cellcolor{green!20}\textbf{1} & \cellcolor{green!20}\textbf{4.03 (0.26)} & 3 & 10.39 (2.92) & 2 & 10.17 (0.91) & 4 & 58.39 (0.21) & 6 & \texttimes & 5 & 236.48 (0.00) \\ \hline
\multirow{5}{*}{\textsc{nginx}} & $S_1$ & \cellcolor{green!20}\textbf{1} & \cellcolor{green!20}\textbf{4.11 (1.45)} & 2 & 8.08 (2.89) & 3 & 8.76 (4.10) & 5 & 1553.44 (965.63) & 6 & \texttimes & 4 & 574.23 (0.00) \\
 & $S_2$ & \cellcolor{green!20}\textbf{1} & \cellcolor{green!20}\textbf{2.60 (1.14)} & 2 & 6.95 (1.74) & 2 & 6.15 (4.17) & 4 & 1637.60 (430.94) & 5 & \texttimes & 3 & 566.42 (0.00) \\
 & $S_3$ & \cellcolor{green!20}\textbf{1} & \cellcolor{green!20}\textbf{2.14 (0.40)} & 3 & 6.46 (1.41) & 2 & 4.42 (1.44) & 5 & 1692.49 (305.78) & 6 & \texttimes & 4 & 567.41 (0.00) \\
 & $S_4$ & \cellcolor{green!20}\textbf{1} & \cellcolor{green!20}\textbf{1.99 (0.47)} & 3 & 6.01 (1.68) & 2 & 4.09 (2.27) & 5 & 1784.76 (274.62) & 6 & \texttimes & 4 & 568.88 (0.00) \\
 & $S_5$ & \cellcolor{green!20}\textbf{1} & \cellcolor{green!20}\textbf{1.98 (0.21)} & 3 & 7.27 (2.31) & 2 & 3.17 (1.72) & 5 & 1579.37 (515.54) & 6 & \texttimes & 4 & 563.91 (0.00) \\ \hline
\multirow{5}{*}{\textsc{VP8}} & $S_1$ & \cellcolor{green!20}\textbf{1} & \cellcolor{green!20}\textbf{1.56 (0.18)} & \textbf{1} & \textbf{1.72 (0.25)} & 2 & 4.68 (3.27) & 4 & 60.05 (2.03) & 5 & \texttimes & 3 & 29.39 (0.00) \\
 & $S_2$ & \cellcolor{green!20}\textbf{1} & \cellcolor{green!20}\textbf{1.15 (0.05)} & 2 & 1.27 (0.15) & 3 & 2.39 (1.21) & 5 & 60.04 (0.24) & 6 & \texttimes & 4 & 35.34 (0.00) \\
 & $S_3$ & \cellcolor{green!20}\textbf{1} & \cellcolor{green!20}\textbf{1.08 (0.04)} & 2 & 1.25 (0.14) & 3 & 1.93 (0.72) & 5 & 60.02 (0.29) & 6 & \texttimes & 4 & 36.26 (0.00) \\
 & $S_4$ & \cellcolor{green!20}\textbf{1} & \cellcolor{green!20}\textbf{0.98 (0.05)} & 2 & 1.15 (0.11) & 3 & 1.54 (0.40) & 5 & 59.99 (0.27) & 6 & \texttimes & 4 & 36.21 (0.00) \\
 & $S_5$ & \cellcolor{green!20}\textbf{1} & \cellcolor{green!20}\textbf{0.90 (0.06)} & 2 & 1.12 (0.11) & 3 & 1.45 (0.23) & 5 & 59.95 (0.68) & 6 & \texttimes & 4 & 37.14 (0.00) \\ \hline
\multicolumn{2}{c}{Average $r$} & 1.48 &  & 1.95 &  & 2.4 &  & 4.5 &  & 3.98 &  & 3.73 & 
\\
\bottomrule
\end{tabular}
\end{adjustbox}

%% file: Tables/chap-DAL/compare_models.tex
\scriptsize

\setlength{\tabcolsep}{1.2mm}
\renewcommand\arraystretch{1.1}
\begin{adjustbox}{width=1.34\textwidth,center}
\begin{tabular}{p{1cm}lp{0.2cm}p{2cm}|p{0.2cm}p{2cm}|p{0.2cm}p{2cm}|p{0.2cm}p{2cm}|p{0.2cm}p{2cm}|p{0.2cm}p{2cm}|p{0.2cm}p{2.3cm}|p{0.2cm}p{2.3cm}|p{0.2cm}p{2.1cm}|p{0.2cm}p{2.2cm}|p{0.2cm}p{2cm}|p{0.2cm}p{2cm}} \toprule
\multirow{2}{*}{System} & \multicolumn{1}{c}{\multirow{2}{*}{Size}} & \multicolumn{2}{c}{\texttt{DaL}} & \multicolumn{2}{c}{\texttt{HINNPerf}} & \multicolumn{2}{c}{\texttt{DaL$_{DNN}$}} & \multicolumn{2}{c}{\texttt{DNN}} & \multicolumn{2}{c}{\texttt{DaL$_{CART}$}} & \multicolumn{2}{c}{\texttt{CART}} & \multicolumn{2}{c}{\texttt{DaL$_{LR}$}} & \multicolumn{2}{c}{\texttt{LR}} & \multicolumn{2}{c}{\texttt{DaL$_{SVR}$}} & \multicolumn{2}{c}{\texttt{SVR}} & \multicolumn{2}{c}{\texttt{DaL$_{XGB}$}} & \multicolumn{2}{c}{\texttt{XGBoost}} \\ \cline{3-26}
 & \multicolumn{1}{c}{} & $r$ & Med (IQR) & $r$ & Med (IQR) & $r$ & Med (IQR) & $r$ & Med (IQR) & $r$ & Med (IQR) & $r$ & Med (IQR) & $r$ & Med (IQR) & $r$ & Med (IQR) & $r$ & Med (IQR) & $r$ & Med (IQR) & $r$ & Med (IQR) & $r$ & Med (IQR) \\ \hline
 
\textsc{Apache} & $S_1$ & 2 & 21.86 (7.36) & 3 & 19.44 (10.53) & \textbf{1} & \textbf{21.02 (6.64)} & 2 & 20.19 (6.34) & 2 & 21.61 (6.69) & 2 & 20.85 (12.99) & \textbf{1} & \textbf{19.71 (6.80)} & 4 & 27.72 (4.98) & 3 & 23.30 (4.08) & 3 & 22.67 (3.66) & \cellcolor{red!20}\textbf{1} & \cellcolor{red!20}\textbf{19.03 (7.50)} & \cellcolor{red!20}\textbf{1} & \cellcolor{red!20}\textbf{19.03 (7.50)} \\
 & $S_2$ & 6 & 12.89 (7.87) & \cellcolor{red!20}\textbf{1} & \cellcolor{red!20}\textbf{8.93 (3.89)} & 4 & 9.82 (6.00) & 2 & 9.79 (4.56) & 4 & 10.12 (5.05) & 3 & 10.06 (3.05) & 5 & 10.77 (7.55) & 6 & 17.61 (2.41) & 6 & 17.59 (3.76) & 6 & 20.31 (3.18) & 4 & 11.13 (4.63) & 3 & 10.70 (3.76) \\
 & $S_3$ & \cellcolor{green!20}\textbf{1} & \cellcolor{green!20}\textbf{6.56 (1.92)} & \textbf{1} & \textbf{7.01 (1.74)} & 2 & 7.17 (2.67) & 3 & 7.98 (1.91) & 4 & 9.10 (1.31) & 4 & 9.03 (1.13) & 2 & 7.42 (2.14) & 6 & 16.16 (1.41) & 5 & 15.92 (2.45) & 7 & 18.50 (1.54) & 4 & 9.33 (1.35) & 4 & 9.37 (1.36) \\
 & $S_4$ & 3 & 6.87 (1.95) & 2 & 6.58 (1.04) & 3 & 6.59 (1.77) & 3 & 6.85 (1.60) & 5 & 7.97 (1.35) & 4 & 7.81 (1.35) & \cellcolor{red!20}\textbf{1} & \cellcolor{red!20}\textbf{6.39 (1.16)} & 8 & 16.12 (0.78) & 7 & 13.89 (2.35) & 9 & 18.57 (1.58) & 5 & 8.12 (1.21) & 6 & 8.70 (1.14) \\
 & $S_5$ & 3 & 5.78 (1.08) & 3 & 5.85 (0.99) & 2 & 5.96 (2.07) & 4 & 6.66 (1.61) & 5 & 7.63 (1.00) & 5 & 7.71 (1.03) & \cellcolor{red!20}\textbf{1} & \cellcolor{red!20}\textbf{5.75 (0.55)} & 9 & 16.12 (1.37) & 8 & 12.63 (1.76) & 10 & 17.82 (1.78) & 6 & 7.89 (1.09) & 7 & 8.17 (1.69) \\ \hline
\textsc{BDB-C} & $S_1$ & 5 & 83.60 (208.75) & 5 & 167.55 (144.47) & 3 & 41.77 (32.54) & 2 & 43.66 (42.88) & \textbf{1} & \textbf{47.20 (19.75)} & \cellcolor{red!20}\textbf{1} & \cellcolor{red!20}\textbf{36.17 (9.11)} & 4 & 65.09 (37.30) & 6 & 698.12 (713.51) & 4 & 65.38 (22.26) & 5 & 154.71 (52.45) & 3 & 41.20 (57.95) & 4 & 40.05 (54.70) \\
 & $S_2$ & 4 & 21.92 (17.91) & 2 & 21.45 (13.61) & 4 & 17.22 (15.67) & 2 & 18.17 (8.11) & 3 & 16.81 (8.59) & 3 & 19.61 (7.19) & 5 & 49.04 (47.11) & 6 & 545.48 (149.70) & 5 & 70.56 (64.18) & 5 & 117.87 (49.95) & \cellcolor{red!20}2 & \cellcolor{red!20}15.83 (7.47) & \textbf{1} & \textbf{16.30 (4.99)} \\
 & $S_3$ & 5 & 8.31 (7.52) & 2 & 6.50 (9.20) & \cellcolor{red!20}3 & \cellcolor{red!20}5.74 (6.84) & 6 & 12.47 (4.08) & 5 & 6.72 (5.82) & \textbf{1} & \textbf{7.44 (3.70)} & 7 & 29.00 (78.81) & 9 & 465.51 (91.79) & 8 & 65.69 (155.89) & 7 & 97.22 (52.60) & 4 & 7.80 (5.09) & 3 & 10.23 (1.63) \\
 & $S_4$ & \cellcolor{green!20}5 & \cellcolor{green!20}3.60 (4.50) & \textbf{1} & \textbf{4.86 (2.18)} & \textbf{1} & \textbf{3.68 (2.78)} & 4 & 9.06 (9.76) & \textbf{1} & \textbf{4.15 (2.78)} & 2 & 5.74 (3.97) & 6 & 24.55 (4.18) & 8 & 441.04 (85.57) & 7 & 51.04 (29.64) & 7 & 93.39 (41.17) & 4 & 6.30 (4.63) & 3 & 8.51 (0.97) \\
 & $S_5$ & \cellcolor{green!20}\textbf{1} & \cellcolor{green!20}\textbf{1.88 (1.24)} & 3 & 4.04 (1.24) & 2 & 2.15 (1.52) & 5 & 7.08 (8.03) & 2 & 3.24 (3.09) & 2 & 3.32 (2.50) & 6 & 22.88 (3.78) & 9 & 456.02 (78.01) & 7 & 46.18 (19.08) & 8 & 105.14 (56.61) & 2 & 3.23 (2.02) & 4 & 7.50 (1.11) \\ \hline
\textsc{BDB-J} & $S_1$ & 3 & 5.37 (4.74) & 2 & 3.58 (1.74) & 2 & 3.14 (2.90) & \cellcolor{red!20}4 & \cellcolor{red!20}2.98 (3.34) & \cellcolor{red!20}\textbf{1} & \cellcolor{red!20}\textbf{2.98 (0.81)} & \textbf{1} & \textbf{3.03 (0.99)} & 5 & 5.31 (2.96) & 7 & 43.23 (6.15) & 6 & 22.77 (17.72) & 5 & 12.93 (3.37) & \textbf{1} & \textbf{3.06 (1.12)} & 4 & 7.39 (1.07) \\
 & $S_2$ & \cellcolor{green!20}\textbf{1} & \cellcolor{green!20}\textbf{1.83 (0.45)} & \textbf{1} & \textbf{1.96 (0.36)} & 4 & 1.90 (0.37) & 3 & 1.91 (0.68) & 2 & 2.05 (0.33) & \textbf{1} & \textbf{1.99 (0.29)} & 4 & 3.77 (0.49) & 8 & 37.85 (6.27) & 7 & 15.28 (12.28) & 6 & 10.54 (0.97) & 2 & 2.11 (0.17) & 5 & 6.10 (0.51) \\
 & $S_3$ & \cellcolor{green!20}\textbf{1} & \cellcolor{green!20}\textbf{1.58 (0.26)} & 2 & 1.75 (0.41) & 2 & 1.61 (0.31) & 3 & 2.01 (1.10) & 2 & 1.92 (0.27) & 2 & 1.87 (0.35) & 3 & 3.41 (0.46) & 7 & 37.61 (3.49) & 6 & 15.03 (10.36) & 5 & 11.03 (0.89) & 2 & 1.93 (0.20) & 4 & 5.71 (0.61) \\
 & $S_4$ & \cellcolor{green!20}\textbf{1} & \cellcolor{green!20}\textbf{1.45 (0.22)} & 2 & 1.52 (0.24) & 3 & 1.61 (0.29) & 5 & 1.70 (0.37) & 4 & 1.67 (0.29) & 4 & 1.69 (0.36) & 7 & 3.57 (0.66) & 10 & 37.57 (3.48) & 9 & 9.20 (5.36) & 9 & 11.25 (1.35) & 6 & 1.84 (0.23) & 8 & 5.48 (0.55) \\
 & $S_5$ & \cellcolor{green!20}\textbf{1} & \cellcolor{green!20}\textbf{1.40 (0.34)} & 2 & 1.48 (0.32) & 2 & 1.46 (0.26) & 4 & 1.69 (0.48) & 3 & 1.67 (0.43) & 3 & 1.68 (0.39) & 5 & 3.23 (0.65) & 9 & 37.48 (2.86) & 7 & 8.21 (3.87) & 8 & 11.51 (2.21) & 4 & 1.88 (0.44) & 6 & 5.34 (0.42) \\ \hline
\textsc{kanzi} & $S_1$ & 3 & 196.15 (213.73) & 5 & 262.27 (175.07) & 3 & 166.22 (122.05) & 3 & 196.36 (122.61) & 3 & 178.67 (123.16) & 3 & 163.60 (121.50) & 6 & 551.03 (1244.41) & 7 & 3106.09 (1514.60) & 4 & 253.06 (150.61) & \cellcolor{red!20}\textbf{1} & \cellcolor{red!20}\textbf{129.01 (61.27)} & 3 & 156.59 (104.87) & 2 & 190.78 (98.22) \\
 & $S_2$ & 3 & 95.31 (51.34) & \cellcolor{red!20}\textbf{1} & \cellcolor{red!20}\textbf{80.53 (51.36)} & 2 & 90.27 (63.53) & 4 & 107.12 (43.75) & \textbf{1} & \textbf{86.55 (35.63)} & \textbf{1} & \textbf{91.65 (38.71)} & 6 & 383.06 (493.33) & 7 & 1191.03 (348.39) & 5 & 190.01 (134.20) & 2 & 105.79 (34.34) & \textbf{1} & \textbf{82.53 (45.04)} & \textbf{1} & \textbf{84.19 (30.01)} \\
 & $S_3$ & \cellcolor{green!20}2 & \cellcolor{green!20}59.30 (38.47) & \textbf{1} & \textbf{62.20 (14.07)} & 3 & 64.56 (62.71) & 2 & 68.62 (30.41) & 2 & 62.50 (21.15) & 2 & 60.13 (21.42) & 6 & 177.35 (384.90) & 5 & 1058.91 (166.05) & 4 & 198.38 (146.32) & 3 & 99.24 (33.82) & 2 & 61.36 (19.58) & 2 & 65.34 (21.07) \\
 & $S_4$ & \textbf{1} & \textbf{42.93 (18.75)} & \textbf{1} & \textbf{46.48 (19.58)} & \cellcolor{red!20}\textbf{1} & \cellcolor{red!20}\textbf{41.83 (15.47)} & 3 & 62.65 (26.54) & \textbf{1} & \textbf{48.69 (10.31)} & \textbf{1} & \textbf{49.06 (14.98)} & 7 & 134.99 (522.74) & 6 & 1030.25 (180.09) & 5 & 186.29 (80.16) & 4 & 89.93 (27.29) & \textbf{1} & \textbf{48.74 (8.64)} & 2 & 54.19 (10.26) \\
 & $S_5$ & \textbf{1} & \textbf{35.29 (17.13)} & \cellcolor{red!20}\textbf{1} & \cellcolor{red!20}\textbf{34.27 (11.87)} & 4 & 38.57 (20.46) & 3 & 49.42 (9.99) & 2 & 44.15 (8.53) & 2 & 43.96 (10.22) & 7 & 605.42 (1213.79) & 7 & 1010.78 (264.34) & 6 & 152.57 (97.14) & 5 & 87.82 (16.52) & 2 & 44.66 (8.43) & 3 & 48.74 (9.81) \\ \hline
\textsc{SQLite} & $S_1$ & 4 & 73.52 (20.40) & 3 & 70.76 (27.86) & 2 & 73.62 (16.43) & 4 & 79.49 (25.36) & 5 & 82.77 (20.99) & 5 & 83.67 (21.57) & 6 & 105.75 (83.62) & 7 & 151.86 (65.00) & 3 & 79.85 (19.78) & 2 & 74.80 (19.37) & 2 & 72.10 (20.66) & \cellcolor{red!20}\textbf{1} & \cellcolor{red!20}\textbf{62.32 (11.64)} \\
 & $S_2$ & 2 & 76.41 (16.66) & 3 & 76.69 (23.32) & 4 & 77.81 (24.37) & 4 & 77.75 (22.47) & 5 & 86.76 (20.84) & 5 & 89.07 (18.89) & 6 & 161.86 (135.45) & 5 & 84.41 (23.67) & 3 & 79.52 (21.38) & 3 & 77.78 (11.49) & 5 & 86.76 (20.84) & \cellcolor{red!20}\textbf{1} & \cellcolor{red!20}\textbf{64.04 (13.07)} \\
 & $S_3$ & 2 & 71.63 (11.58) & 2 & 71.02 (17.30) & 3 & 74.86 (15.95) & 2 & 75.01 (15.58) & 4 & 87.65 (24.29) & 4 & 89.12 (23.09) & 5 & 103.75 (29.66) & 3 & 75.72 (24.83) & 3 & 80.75 (14.40) & 3 & 79.66 (10.31) & 4 & 87.65 (24.29) & \cellcolor{red!20}\textbf{1} & \cellcolor{red!20}\textbf{64.25 (11.83)} \\
 & $S_4$ & 2 & 66.94 (10.61) & 3 & 70.84 (10.38) & 3 & 70.41 (12.22) & 4 & 72.45 (12.11) & 7 & 86.47 (20.30) & 7 & 87.46 (19.46) & 7 & 82.25 (23.16) & 3 & 69.61 (13.80) & 5 & 77.23 (10.08) & 6 & 79.11 (10.34) & 7 & 86.47 (20.30) & \cellcolor{red!20}\textbf{1} & \cellcolor{red!20}\textbf{62.19 (11.78)} \\
 & $S_5$ & \textbf{1} & \textbf{64.20 (13.91)} & 2 & 70.15 (9.71) & 2 & 69.72 (12.48) & 3 & 72.87 (8.86) & 6 & 84.43 (12.41) & 6 & 84.51 (14.40) & 4 & 76.06 (16.27) & 2 & 67.78 (9.68) & 4 & 76.28 (13.41) & 5 & 79.69 (11.05) & 6 & 84.43 (12.41) & \cellcolor{red!20}\textbf{1} & \cellcolor{red!20}\textbf{63.59 (12.77)} \\ \hline
\textsc{x264} & $S_1$ & 3 & 10.21 (4.35) & 2 & 8.44 (2.69) & \cellcolor{red!20}\textbf{1} & \cellcolor{red!20}\textbf{8.04 (1.30)} & \cellcolor{red!20}\textbf{1} & \cellcolor{red!20}\textbf{8.04 (3.06)} & 3 & 10.40 (2.15) & 2 & 8.77 (2.89) & 3 & 10.43 (3.40) & 4 & 14.36 (4.34) & 4 & 14.39 (2.64) & 5 & 25.72 (2.91) & 3 & 10.59 (2.22) & 3 & 11.19 (2.06) \\
 & $S_2$ & \cellcolor{green!20}2 & \cellcolor{green!20}2.72 (1.52) & \textbf{1} & \textbf{2.77 (1.07)} & 3 & 3.34 (1.90) & 2 & 3.17 (1.00) & 5 & 6.57 (1.67) & 5 & 6.50 (1.63) & 4 & 3.51 (0.98) & 6 & 8.01 (1.23) & 8 & 9.32 (2.16) & 9 & 18.83 (5.32) & 5 & 6.66 (1.45) & 7 & 8.53 (0.86) \\
 & $S_3$ & 2 & 1.57 (1.52) & \cellcolor{red!20}\textbf{1} & \cellcolor{red!20}\textbf{1.37 (0.46)} & 2 & 1.97 (1.47) & 2 & 2.23 (0.90) & 4 & 5.01 (1.42) & 4 & 4.72 (0.73) & 3 & 2.84 (0.52) & 6 & 7.61 (0.76) & 6 & 8.25 (2.68) & 7 & 13.66 (2.32) & 4 & 4.77 (1.08) & 5 & 7.45 (0.85) \\
 & $S_4$ & 2 & 1.30 (1.17) & \cellcolor{red!20}\textbf{1} & \cellcolor{red!20}\textbf{0.81 (0.28)} & 2 & 1.39 (0.56) & 3 & 1.74 (0.98) & 5 & 3.52 (1.33) & 5 & 3.65 (1.03) & 4 & 2.60 (0.52) & 8 & 7.34 (0.39) & 6 & 7.05 (4.00) & 9 & 10.24 (1.41) & 5 & 3.52 (1.37) & 7 & 6.61 (0.73) \\
 & $S_5$ & 2 & 0.96 (0.79) & \cellcolor{red!20}\textbf{1} & \cellcolor{red!20}\textbf{0.58 (0.15)} & 2 & 1.05 (0.55) & 3 & 1.44 (0.90) & 6 & 2.54 (1.04) & 5 & 2.37 (1.11) & 4 & 2.10 (1.04) & 8 & 7.17 (0.60) & 7 & 7.72 (4.47) & 9 & 9.19 (0.76) & 5 & 2.43 (0.90) & 7 & 6.26 (0.58) \\ \hline
\textsc{Dune} & $S_1$ & 4 & 9.37 (1.15) & 2 & 7.98 (0.98) & 3 & 8.40 (1.61) & \cellcolor{red!20}\textbf{1} & \cellcolor{red!20}\textbf{7.86 (0.93)} & 4 & 9.04 (1.13) & 4 & 9.13 (0.99) & 6 & 11.21 (0.88) & 7 & 13.06 (1.90) & 8 & 14.46 (0.51) & 9 & 14.76 (0.64) & 4 & 9.16 (0.78) & 5 & 10.12 (0.71) \\
 & $S_2$ & \textbf{1} & \textbf{5.95 (0.54)} & 2 & 6.15 (0.40) & \cellcolor{red!20}\textbf{1} & \cellcolor{red!20}\textbf{5.78 (0.41)} & \textbf{1} & \textbf{6.01 (0.21)} & 3 & 6.25 (0.42) & 3 & 6.26 (0.40) & 5 & 11.13 (0.57) & 6 & 13.02 (1.06) & 6 & 13.37 (0.25) & 7 & 13.47 (0.32) & 3 & 6.33 (0.37) & 4 & 8.39 (0.26) \\
 & $S_3$ & \cellcolor{green!20}\textbf{1} & \cellcolor{green!20}\textbf{4.81 (0.48)} & 3 & 5.48 (0.39) & 2 & 5.21 (0.39) & 3 & 5.52 (0.34) & 3 & 5.48 (0.42) & 3 & 5.49 (0.37) & 5 & 11.09 (0.54) & 7 & 13.13 (0.66) & 6 & 12.95 (0.32) & 7 & 13.15 (0.41) & 3 & 5.42 (0.38) & 4 & 8.07 (0.24) \\
 & $S_4$ & \cellcolor{green!20}\textbf{1} & \cellcolor{green!20}\textbf{4.23 (0.35)} & 4 & 5.12 (0.38) & 2 & 4.79 (0.49) & 5 & 5.26 (0.31) & 3 & 5.06 (0.32) & 3 & 5.01 (0.32) & 7 & 11.04 (0.43) & 10 & 13.14 (0.55) & 8 & 12.57 (0.47) & 9 & 12.80 (0.52) & 3 & 5.02 (0.30) & 6 & 7.85 (0.26) \\
 & $S_5$ & \cellcolor{green!20}\textbf{1} & \cellcolor{green!20}\textbf{3.98 (0.19)} & 3 & 4.98 (0.64) & 2 & 4.69 (0.50) & 4 & 5.15 (0.34) & 2 & 4.69 (0.35) & 2 & 4.66 (0.36) & 6 & 10.93 (0.48) & 9 & 13.21 (0.57) & 7 & 12.30 (0.46) & 8 & 12.58 (0.69) & 2 & 4.64 (0.36) & 5 & 7.66 (0.17) \\ \hline
\textsc{HIPA$^{cc}$} & $S_1$ & \textbf{1} & \textbf{7.50 (1.42)} & \cellcolor{red!20}\textbf{1} & \cellcolor{red!20}\textbf{7.40 (1.05)} & 2 & 9.07 (1.18) & 3 & 9.70 (1.28) & 5 & 12.03 (0.93) & 5 & 12.07 (1.06) & 10 & 12.70 (1.22) & 9 & 20.49 (1.61) & 7 & 13.06 (0.51) & 8 & 14.84 (0.62) & 6 & 12.39 (1.49) & 4 & 11.19 (0.60) \\
 & $S_2$ & \cellcolor{green!20}\textbf{1} & \cellcolor{green!20}\textbf{4.48 (0.38)} & 2 & 4.53 (0.48) & 3 & 5.55 (0.60) & 4 & 6.89 (1.44) & 5 & 8.47 (1.10) & 5 & 8.20 (1.00) & 11 & 11.64 (0.35) & 10 & 19.67 (0.83) & 8 & 11.27 (0.44) & 9 & 14.52 (0.54) & 6 & 8.51 (0.97) & 7 & 9.88 (0.37) \\
 & $S_3$ & \cellcolor{green!20}\textbf{1} & \cellcolor{green!20}\textbf{3.43 (0.34)} & 2 & 3.67 (0.28) & 3 & 4.39 (0.41) & 4 & 4.68 (0.90) & 5 & 6.62 (0.80) & 5 & 6.60 (0.75) & 9 & 11.43 (0.34) & 11 & 19.35 (0.72) & 8 & 10.78 (0.28) & 10 & 14.48 (0.57) & 6 & 6.82 (0.58) & 7 & 9.41 (0.33) \\
 & $S_4$ & \cellcolor{green!20}\textbf{1} & \cellcolor{green!20}\textbf{2.60 (0.15)} & 2 & 2.70 (0.18) & 3 & 3.22 (0.35) & 4 & 3.58 (1.00) & 5 & 4.42 (0.22) & 6 & 4.31 (0.44) & 9 & 11.13 (0.25) & 11 & 18.97 (0.46) & 8 & 10.31 (0.14) & 10 & 12.06 (0.25) & 6 & 4.46 (0.29) & 7 & 8.81 (0.20) \\
 & $S_5$ & \cellcolor{green!20}\textbf{1} & \cellcolor{green!20}\textbf{2.11 (0.08)} & 2 & 2.23 (0.23) & 3 & 2.39 (0.22) & 6 & 2.82 (0.57) & 4 & 2.69 (0.15) & 4 & 2.70 (0.17) & 10 & 11.01 (0.15) & 11 & 18.81 (0.37) & 8 & 10.04 (0.09) & 9 & 10.52 (0.24) & 5 & 2.74 (0.13) & 7 & 8.30 (0.20) \\ \hline
\textsc{HSMGP} & $S_1$ & \cellcolor{green!20}\textbf{1} & \cellcolor{green!20}\textbf{4.14 (1.44)} & 2 & 5.03 (2.56) & \textbf{1} & \textbf{4.66 (1.32)} & 7 & 7.09 (3.04) & 7 & 21.65 (2.31) & 7 & 21.34 (2.41) & 3 & 7.72 (0.80) & 8 & 55.91 (19.98) & 5 & 16.00 (2.48) & 6 & 16.63 (1.38) & 7 & 20.44 (2.66) & 4 & 14.63 (1.92) \\
 & $S_2$ & \cellcolor{green!20}\textbf{1} & \cellcolor{green!20}\textbf{2.06 (0.19)} & 3 & 3.13 (0.49) & 2 & 2.66 (0.68) & 4 & 3.69 (0.61) & 8 & 15.59 (1.41) & 8 & 16.43 (1.52) & 5 & 7.35 (0.47) & 9 & 52.54 (10.70) & 7 & 15.24 (1.32) & 8 & 16.30 (2.16) & 7 & 15.21 (1.27) & 6 & 10.59 (0.88) \\
 & $S_3$ & \cellcolor{green!20}\textbf{1} & \cellcolor{green!20}\textbf{1.42 (0.13)} & 3 & 2.34 (0.38) & 2 & 1.56 (0.20) & 3 & 2.28 (0.37) & 8 & 11.35 (0.63) & 8 & 11.26 (0.98) & 4 & 7.14 (0.35) & 10 & 51.38 (9.85) & 6 & 10.05 (0.86) & 9 & 15.03 (0.66) & 7 & 11.07 (0.75) & 5 & 7.52 (0.46) \\
 & $S_4$ & \cellcolor{green!20}\textbf{1} & \cellcolor{green!20}\textbf{1.33 (0.08)} & 3 & 2.19 (0.23) & 2 & 1.49 (0.15) & 4 & 2.23 (0.39) & 9 & 10.13 (0.57) & 10 & 10.14 (0.72) & 6 & 7.16 (0.17) & 12 & 49.63 (9.43) & 7 & 8.95 (0.90) & 11 & 15.07 (0.56) & 8 & 9.94 (0.66) & 5 & 7.10 (0.38) \\
 & $S_5$ & \textbf{1} & \textbf{1.19 (0.08)} & 2 & 2.09 (0.25) & \cellcolor{red!20}\textbf{1} & \cellcolor{red!20}\textbf{1.16 (0.08)} & 3 & 1.94 (0.63) & 8 & 7.46 (0.46) & 7 & 7.43 (0.34) & 6 & 7.15 (0.20) & 10 & 49.05 (3.00) & 4 & 5.87 (0.52) & 9 & 8.60 (0.38) & 8 & 7.58 (0.44) & 5 & 6.11 (0.26) \\ \hline
\textsc{Lrzip} & $S_1$ & 5 & 31.88 (10.96) & 7 & 41.97 (12.02) & 4 & 26.40 (6.94) & 6 & 35.40 (16.59) & 2 & 17.95 (10.85) & 3 & 23.50 (10.09) & 9 & 117.37 (24.65) & 10 & 369.45 (116.78) & 8 & 59.16 (13.29) & 8 & 63.04 (23.36) & \cellcolor{red!20}\textbf{1} & \cellcolor{red!20}\textbf{16.16 (6.26)} & \textbf{1} & \textbf{16.71 (4.27)} \\
 & $S_2$ & 3 & 10.51 (4.39) & 6 & 24.10 (19.98) & 4 & 15.37 (6.04) & 5 & 21.46 (4.53) & 2 & 9.09 (2.56) & 3 & 10.47 (4.94) & 8 & 109.11 (20.26) & 9 & 315.34 (66.07) & 7 & 63.84 (11.18) & 7 & 61.82 (34.55) & \cellcolor{red!20}\textbf{1} & \cellcolor{red!20}\textbf{7.56 (2.33)} & 2 & 10.19 (1.79) \\
 & $S_3$ & 4 & 8.04 (2.37) & 7 & 23.71 (16.49) & 5 & 11.83 (4.53) & 6 & 18.68 (3.20) & 2 & 7.77 (2.09) & 4 & 8.62 (4.75) & 10 & 104.20 (17.55) & 11 & 322.55 (50.28) & 9 & 63.91 (13.58) & 8 & 66.09 (41.44) & \cellcolor{red!20}\textbf{1} & \cellcolor{red!20}\textbf{6.15 (1.03)} & 3 & 8.43 (0.69) \\
 & $S_4$ & 2 & 5.95 (0.99) & 7 & 19.82 (16.76) & 5 & 10.10 (3.23) & 6 & 15.55 (1.71) & 2 & 6.38 (2.72) & 3 & 7.09 (2.76) & 9 & 103.74 (14.29) & 10 & 323.59 (56.77) & 8 & 65.19 (12.53) & 8 & 69.07 (5.93) & \cellcolor{red!20}\textbf{1} & \cellcolor{red!20}\textbf{5.30 (0.97)} & 4 & 7.59 (0.91) \\
 & $S_5$ & 3 & 4.03 (0.26) & 8 & 10.39 (2.92) & 6 & 6.60 (1.34) & 7 & 10.17 (0.91) & 2 & 3.58 (0.97) & 4 & 4.47 (1.90) & 11 & 105.83 (7.06) & 12 & 327.20 (27.08) & 9 & 48.80 (2.86) & 10 & 51.27 (2.21) & \cellcolor{red!20}\textbf{1} & \cellcolor{red!20}\textbf{2.87 (0.27)} & 5 & 6.17 (0.43) \\ \hline
\textsc{nginx} & $S_1$ & \textbf{1} & \textbf{4.11 (1.45)} & 4 & 8.08 (2.89) & 2 & 4.95 (2.94) & 5 & 8.76 (4.10) & 6 & 14.44 (2.41) & 6 & 14.44 (2.41) & 7 & 22.02 (2.11) & 9 & 574.23 (25.93) & 8 & 456.24 (15.91) & 10 & 1603.42 (66.06) & \cellcolor{red!20}\textbf{1} & \cellcolor{red!20}\textbf{3.96 (1.59)} & 3 & 6.78 (1.92) \\
 & $S_2$ & 2 & 2.60 (1.14) & 4 & 6.95 (1.74) & 2 & 3.07 (0.94) & 4 & 6.15 (4.17) & 5 & 7.44 (3.14) & 5 & 7.44 (3.14) & 6 & 20.25 (1.60) & 8 & 568.09 (19.68) & 7 & 430.40 (22.39) & 9 & 1337.93 (148.42) & \cellcolor{red!20}\textbf{1} & \cellcolor{red!20}\textbf{1.24 (0.81)} & 3 & 4.90 (0.44) \\
 & $S_3$ & 2 & 2.14 (0.40) & 5 & 6.46 (1.41) & 3 & 2.31 (1.29) & 4 & 4.42 (1.44) & 4 & 4.81 (0.81) & 4 & 4.81 (0.81) & 6 & 19.66 (0.75) & 8 & 568.45 (16.13) & 7 & 389.40 (30.26) & 9 & 942.81 (108.67) & \cellcolor{red!20}\textbf{1} & \cellcolor{red!20}\textbf{1.03 (0.06)} & 4 & 4.56 (0.09) \\
 & $S_4$ & 2 & 1.99 (0.47) & 7 & 6.01 (1.68) & 3 & 2.16 (0.88) & 6 & 4.09 (2.27) & 4 & 4.16 (1.01) & 4 & 4.16 (1.01) & 8 & 19.74 (0.82) & 10 & 571.30 (19.18) & 9 & 364.98 (32.75) & 11 & 690.54 (133.59) & \cellcolor{red!20}\textbf{1} & \cellcolor{red!20}\textbf{1.02 (0.04)} & 5 & 4.49 (0.05) \\
 & $S_5$ & 2 & 1.98 (0.21) & 5 & 7.27 (2.31) & 2 & 1.97 (0.52) & 4 & 3.17 (1.72) & 3 & 3.33 (1.00) & 3 & 3.33 (1.00) & 6 & 19.17 (1.07) & 9 & 564.69 (23.96) & 7 & 327.92 (28.40) & 8 & 450.48 (87.27) & \cellcolor{red!20}\textbf{1} & \cellcolor{red!20}\textbf{1.00 (0.06)} & 4 & 4.44 (0.06) \\ \hline
\textsc{VP8} & $S_1$ & \cellcolor{green!20}\textbf{1} & \cellcolor{green!20}\textbf{1.56 (0.18)} & \textbf{1} & \textbf{1.72 (0.25)} & 2 & 1.59 (0.36) & 6 & 4.68 (3.27) & 3 & 2.57 (0.70) & 4 & 2.47 (0.70) & 8 & 12.72 (0.84) & 11 & 43.69 (5.81) & 9 & 13.84 (0.76) & 10 & 22.37 (3.68) & 5 & 2.70 (0.88) & 7 & 5.99 (0.31) \\
 & $S_2$ & \cellcolor{green!20}\textbf{1} & \cellcolor{green!20}\textbf{1.15 (0.05)} & 2 & 1.27 (0.15) & 4 & 1.21 (0.09) & 5 & 2.39 (1.21) & 2 & 1.24 (0.24) & 3 & 1.30 (0.22) & 8 & 12.42 (0.44) & 10 & 42.04 (3.16) & 7 & 10.30 (0.49) & 9 & 23.53 (1.46) & 3 & 1.37 (0.20) & 6 & 5.25 (0.20) \\
 & $S_3$ & \cellcolor{green!20}\textbf{1} & \cellcolor{green!20}\textbf{1.08 (0.04)} & 5 & 1.25 (0.14) & 2 & 1.12 (0.08) & 6 & 1.93 (0.72) & 3 & 1.15 (0.12) & 4 & 1.18 (0.21) & 9 & 12.41 (0.34) & 11 & 42.65 (2.99) & 8 & 10.11 (0.56) & 10 & 25.02 (1.93) & 5 & 1.24 (0.10) & 7 & 5.15 (0.15) \\
 & $S_4$ & \cellcolor{green!20}\textbf{1} & \cellcolor{green!20}\textbf{0.98 (0.05)} & 4 & 1.15 (0.11) & 3 & 1.10 (0.06) & 6 & 1.54 (0.40) & 2 & 1.07 (0.08) & 3 & 1.09 (0.08) & 9 & 12.35 (0.31) & 11 & 42.08 (2.79) & 8 & 9.81 (0.36) & 10 & 26.13 (1.14) & 5 & 1.18 (0.07) & 7 & 5.07 (0.10) \\
 & $S_5$ & \cellcolor{green!20}\textbf{1} & \cellcolor{green!20}\textbf{0.90 (0.06)} & 6 & 1.12 (0.11) & 5 & 1.07 (0.05) & 7 & 1.45 (0.23) & 2 & 0.94 (0.04) & 3 & 0.94 (0.05) & 10 & 12.41 (0.27) & 12 & 42.69 (1.94) & 9 & 9.71 (0.21) & 11 & 15.54 (0.77) & 4 & 1.05 (0.04) & 8 & 4.89 (0.09) \\ \hline
\multicolumn{2}{l}{Average $r$} & 2.05 &  & 2.9 &  & 2.65 &  & 3.9 &  & 3.77 &  & 3.85 &  & 6.1 &  & 8.12 &  & 6.52 &  & 7.33 &  & 3.63 &  & 4.28 & 
\\
\bottomrule
\end{tabular}
\end{adjustbox}


%% file: Tables/chap-DAL/compare_ensemble.tex
\scriptsize
\setlength{\tabcolsep}{1.2mm}
\begin{adjustbox}{width=1.2\textwidth,center}
\begin{tabular}{clllllll||llllllll} \toprule
\multirow{2}{*}{System} & \multicolumn{1}{c}{\multirow{2}{*}{Size}} & \multicolumn{2}{c}{\texttt{DaL$_{LR}$}} & \multicolumn{2}{c}{\texttt{Bagging$_{LR}$}} & \multicolumn{2}{c}{\texttt{AdaBoost$_{LR}$}} & \multicolumn{2}{c}{\texttt{DaL$_{CART}$}} & \multicolumn{2}{c}{\texttt{RF}} & \multicolumn{2}{c}{\texttt{XGBoost}} & \multicolumn{2}{c}{\texttt{AdaBoost$_{CART}$}} \\ \cline{3-16}
 & \multicolumn{1}{c}{} & $r$ & Med (IQR) & $r$ & Med (IQR) & $r$ & Med (IQR) & $r$ & Med (IQR) & $r$ & Med (IQR) & $r$ & Med (IQR) & $r$ & Med (IQR) \\ \hline
\multirow{5}{*}{\textsc{Apache}} & $S_1$ & \cellcolor{green!20}\textbf{1} & \cellcolor{green!20}\textbf{19.71 (6.80)} & 2 & 20.81 (4.47) & 2 & 20.44 (5.03) & 2 & 21.61 (6.69) & 3 & 21.20 (7.72) & \cellcolor{red!20}\textbf{1} & \cellcolor{red!20}\textbf{19.03 (7.50)} & 4 & 24.06 (7.47) \\
 & $S_2$ & \cellcolor{green!20}\textbf{1} & \cellcolor{green!20}\textbf{10.77 (7.55)} & 3 & 17.64 (2.12) & 2 & 16.69 (2.45) & \cellcolor{green!20}2 & \cellcolor{green!20}10.12 (5.05) & 2 & 12.02 (6.58) & \textbf{1} & \textbf{10.70 (3.76)} & 3 & 13.82 (4.42) \\
 & $S_3$ & \cellcolor{green!20}\textbf{1} & \cellcolor{green!20}\textbf{7.42 (2.14)} & 2 & 16.11 (1.34) & 2 & 16.25 (1.68) & 2 & 9.10 (1.31) & \cellcolor{red!20}\textbf{1} & \cellcolor{red!20}\textbf{7.65 (1.46)} & 2 & 9.37 (1.36) & 2 & 9.04 (1.70) \\
 & $S_4$ & \cellcolor{green!20}\textbf{1} & \cellcolor{green!20}\textbf{6.39 (1.16)} & 2 & 16.20 (0.82) & 2 & 15.96 (1.40) & 3 & 7.97 (1.35) & \cellcolor{red!20}\textbf{1} & \cellcolor{red!20}\textbf{6.63 (1.19)} & 4 & 8.70 (1.14) & 2 & 7.64 (0.99) \\
 & $S_5$ & \cellcolor{green!20}\textbf{1} & \cellcolor{green!20}\textbf{5.75 (0.55)} & 2 & 16.04 (1.37) & 2 & 16.18 (1.80) & 3 & 7.63 (1.00) & \cellcolor{red!20}\textbf{1} & \cellcolor{red!20}\textbf{6.43 (1.11)} & 4 & 8.17 (1.69) & 2 & 7.22 (0.97) \\ \hline
\multirow{5}{*}{\textsc{BDB-C}} & $S_1$ & \cellcolor{green!20}\textbf{1} & \cellcolor{green!20}\textbf{65.09 (37.30)} & 3 & 485.41 (235.62) & 2 & 453.92 (139.97) & \textbf{1} & \textbf{47.20 (19.75)} & 4 & 253.78 (93.63) & \cellcolor{red!20}2 & \cellcolor{red!20}40.05 (54.70) & 3 & 211.60 (191.00) \\
 & $S_2$ & \cellcolor{green!20}\textbf{1} & \cellcolor{green!20}\textbf{49.04 (47.11)} & 3 & 529.58 (158.69) & 2 & 565.28 (157.43) & 2 & 16.81 (8.59) & 3 & 73.41 (118.30) & \cellcolor{red!20}\textbf{1} & \cellcolor{red!20}\textbf{16.30 (4.99)} & 3 & 95.94 (19.68) \\
 & $S_3$ & \cellcolor{green!20}\textbf{1} & \cellcolor{green!20}\textbf{29.00 (78.81)} & 2 & 460.89 (90.55) & 3 & 496.74 (56.61) & \cellcolor{green!20}2 & \cellcolor{green!20}6.72 (5.82) & 3 & 18.64 (19.42) & \textbf{1} & \textbf{10.23 (1.63)} & 4 & 43.94 (7.78) \\
 & $S_4$ & \cellcolor{green!20}\textbf{1} & \cellcolor{green!20}\textbf{24.55 (4.18)} & 2 & 440.07 (77.70) & 3 & 475.09 (43.17) & \cellcolor{green!20}\textbf{1} & \cellcolor{green!20}\textbf{4.15 (2.78)} & 3 & 11.32 (6.77) & 2 & 8.51 (0.97) & 3 & 21.47 (5.45) \\
 & $S_5$ & \cellcolor{green!20}\textbf{1} & \cellcolor{green!20}\textbf{22.88 (3.78)} & 2 & 460.51 (88.56) & 3 & 473.16 (64.47) & \cellcolor{green!20}\textbf{1} & \cellcolor{green!20}\textbf{3.24 (3.09)} & 3 & 6.64 (3.35) & 2 & 7.50 (1.11) & 3 & 11.07 (2.28) \\ \hline
\multirow{5}{*}{\textsc{BDB-J}} & $S_1$ & \cellcolor{green!20}\textbf{1} & \cellcolor{green!20}\textbf{5.31 (2.96)} & 3 & 43.30 (6.44) & 2 & 40.89 (6.68) & \cellcolor{green!20}\textbf{1} & \cellcolor{green!20}\textbf{2.98 (0.81)} & 3 & 8.36 (10.48) & 2 & 7.39 (1.07) & \textbf{1} & \textbf{3.61 (2.31)} \\
 & $S_2$ & \cellcolor{green!20}\textbf{1} & \cellcolor{green!20}\textbf{3.77 (0.49)} & 2 & 37.75 (5.70) & 3 & 39.84 (6.41) & 2 & 2.05 (0.33) & \cellcolor{red!20}\textbf{1} & \cellcolor{red!20}\textbf{1.82 (0.29)} & 3 & 6.10 (0.51) & \textbf{1} & \textbf{1.85 (0.20)} \\
 & $S_3$ & \cellcolor{green!20}\textbf{1} & \cellcolor{green!20}\textbf{3.41 (0.46)} & 2 & 37.72 (3.80) & 2 & 38.85 (4.51) & 3 & 1.92 (0.27) & \cellcolor{red!20}\textbf{1} & \cellcolor{red!20}\textbf{1.57 (0.21)} & 4 & 5.71 (0.61) & 2 & 1.78 (0.25) \\
 & $S_4$ & \cellcolor{green!20}\textbf{1} & \cellcolor{green!20}\textbf{3.57 (0.66)} & 3 & 37.07 (3.03) & 2 & 36.17 (4.46) & 3 & 1.67 (0.29) & \cellcolor{red!20}\textbf{1} & \cellcolor{red!20}\textbf{1.43 (0.21)} & 4 & 5.48 (0.55) & 2 & 1.66 (0.21) \\
 & $S_5$ & \cellcolor{green!20}\textbf{1} & \cellcolor{green!20}\textbf{3.23 (0.65)} & 3 & 37.27 (3.01) & 2 & 36.52 (5.25) & 3 & 1.67 (0.43) & \cellcolor{red!20}\textbf{1} & \cellcolor{red!20}\textbf{1.40 (0.23)} & 4 & 5.34 (0.42) & 2 & 1.55 (0.29) \\ \hline
\multirow{5}{*}{\textsc{kanzi}} & $S_1$ & \cellcolor{green!20}\textbf{1} & \cellcolor{green!20}\textbf{551.03 (1244.41)} & \textbf{1} & \textbf{1158.20 (365.64)} & \textbf{1} & \textbf{1087.10 (248.90)} & \cellcolor{green!20}\textbf{1} & \cellcolor{green!20}\textbf{178.67 (123.16)} & 2 & 359.19 (147.95) & \textbf{1} & \textbf{190.78 (98.22)} & 2 & 426.88 (221.92) \\
 & $S_2$ & \cellcolor{green!20}\textbf{1} & \cellcolor{green!20}\textbf{383.06 (493.33)} & 3 & 1224.60 (377.56) & 2 & 1259.43 (331.27) & \textbf{1} & \textbf{86.55 (35.63)} & 2 & 155.50 (51.71) & \cellcolor{red!20}\textbf{1} & \cellcolor{red!20}\textbf{84.19 (30.01)} & 3 & 253.92 (74.19) \\
 & $S_3$ & \cellcolor{green!20}2 & \cellcolor{green!20}177.35 (384.90) & 3 & 1194.11 (1403.01) & \textbf{1} & \textbf{1368.20 (250.10)} & \cellcolor{green!20}\textbf{1} & \cellcolor{green!20}\textbf{62.50 (21.15)} & 2 & 87.44 (21.38) & \textbf{1} & \textbf{65.34 (21.07)} & 3 & 199.28 (45.12) \\
 & $S_4$ & \cellcolor{green!20}2 & \cellcolor{green!20}134.99 (522.74) & 3 & 1041.55 (230.02) & \textbf{1} & \textbf{1396.16 (189.09)} & \cellcolor{green!20}\textbf{1} & \cellcolor{green!20}\textbf{48.69 (10.31)} & 3 & 65.20 (18.86) & 2 & 54.19 (10.26) & 4 & 158.05 (37.81) \\
 & $S_5$ & \cellcolor{green!20}\textbf{1} & \cellcolor{green!20}\textbf{605.42 (1213.79)} & 3 & 1014.62 (255.48) & 2 & 1266.23 (184.09) & \cellcolor{green!20}\textbf{1} & \cellcolor{green!20}\textbf{44.15 (8.53)} & 3 & 52.53 (9.96) & 2 & 48.74 (9.81) & 4 & 132.03 (25.97) \\ \hline
\multirow{5}{*}{\textsc{SQLite}} & $S_1$ & 2 & 105.75 (83.62) & \cellcolor{red!20}\textbf{1} & \cellcolor{red!20}\textbf{74.67 (20.68)} & \textbf{1} & \textbf{76.22 (21.29)} & 3 & 82.77 (20.99) & \textbf{1} & \textbf{65.16 (12.06)} & \cellcolor{red!20}\textbf{1} & \cellcolor{red!20}\textbf{62.32 (11.64)} & 2 & 70.76 (20.26) \\
 & $S_2$ & 3 & 161.86 (135.45) & 2 & 95.26 (36.34) & \cellcolor{red!20}\textbf{1} & \cellcolor{red!20}\textbf{89.04 (34.14)} & 3 & 86.76 (20.84) & \textbf{1} & \textbf{69.64 (10.76)} & \cellcolor{red!20}\textbf{1} & \cellcolor{red!20}\textbf{64.04 (13.07)} & 2 & 75.59 (24.63) \\
 & $S_3$ & 3 & 103.75 (29.66) & \cellcolor{red!20}\textbf{1} & \cellcolor{red!20}\textbf{76.20 (21.41)} & 2 & 84.61 (35.20) & 3 & 87.65 (24.29) & \textbf{1} & \textbf{69.77 (13.75)} & \cellcolor{red!20}\textbf{1} & \cellcolor{red!20}\textbf{64.25 (11.83)} & 2 & 70.77 (18.66) \\
 & $S_4$ & 2 & 82.25 (23.16) & \cellcolor{red!20}\textbf{1} & \cellcolor{red!20}\textbf{68.94 (13.83)} & 2 & 82.27 (27.41) & 3 & 86.47 (20.30) & 2 & 69.29 (9.25) & \cellcolor{red!20}\textbf{1} & \cellcolor{red!20}\textbf{62.19 (11.78)} & 2 & 70.01 (16.80) \\
 & $S_5$ & 2 & 76.06 (16.27) & \cellcolor{red!20}\textbf{1} & \cellcolor{red!20}\textbf{68.03 (10.38)} & 3 & 80.61 (23.78) & 3 & 84.43 (12.41) & 2 & 70.42 (9.55) & \cellcolor{red!20}\textbf{1} & \cellcolor{red!20}\textbf{63.59 (12.77)} & 2 & 71.19 (13.68) \\ \hline
\multirow{5}{*}{\textsc{x264}} & $S_1$ & \cellcolor{green!20}\textbf{1} & \cellcolor{green!20}\textbf{10.43 (3.40)} & \textbf{1} & \textbf{10.90 (2.21)} & \textbf{1} & \textbf{11.58 (2.30)} & 2 & 10.40 (2.15) & \cellcolor{red!20}\textbf{1} & \cellcolor{red!20}\textbf{8.18 (2.78)} & 3 & 11.19 (2.06) & 4 & 12.22 (5.95) \\
 & $S_2$ & \cellcolor{green!20}\textbf{1} & \cellcolor{green!20}\textbf{3.51 (0.98)} & 2 & 8.11 (1.06) & 3 & 8.74 (1.21) & \textbf{1} & \textbf{6.57 (1.67)} & \cellcolor{red!20}\textbf{1} & \cellcolor{red!20}\textbf{6.42 (1.20)} & 3 & 8.53 (0.86) & 2 & 6.82 (1.28) \\
 & $S_3$ & \cellcolor{green!20}\textbf{1} & \cellcolor{green!20}\textbf{2.84 (0.52)} & 2 & 7.57 (0.81) & 3 & 7.91 (0.80) & \textbf{1} & \textbf{5.01 (1.42)} & \cellcolor{red!20}\textbf{1} & \cellcolor{red!20}\textbf{4.95 (0.85)} & 3 & 7.45 (0.85) & 2 & 5.59 (0.64) \\
 & $S_4$ & \cellcolor{green!20}\textbf{1} & \cellcolor{green!20}\textbf{2.60 (0.52)} & 2 & 7.27 (0.45) & 3 & 7.44 (0.37) & \cellcolor{green!20}\textbf{1} & \cellcolor{green!20}\textbf{3.52 (1.33)} & 2 & 4.04 (0.86) & 4 & 6.61 (0.73) & 3 & 4.53 (0.67) \\
 & $S_5$ & \cellcolor{green!20}\textbf{1} & \cellcolor{green!20}\textbf{2.10 (1.04)} & 2 & 7.12 (0.67) & 3 & 7.19 (0.46) & \cellcolor{green!20}\textbf{1} & \cellcolor{green!20}\textbf{2.54 (1.04)} & 2 & 2.98 (0.77) & 4 & 6.26 (0.58) & 3 & 3.55 (0.53) \\ \hline
\multirow{5}{*}{\textsc{Dune}} & $S_1$ & \cellcolor{green!20}\textbf{1} & \cellcolor{green!20}\textbf{11.21 (0.88)} & 2 & 13.01 (1.78) & 3 & 13.76 (5.73) & \cellcolor{green!20}\textbf{1} & \cellcolor{green!20}\textbf{9.04 (1.13)} & \textbf{1} & \textbf{9.27 (0.56)} & 2 & 10.12 (0.71) & \textbf{1} & \textbf{9.29 (0.65)} \\
 & $S_2$ & \cellcolor{green!20}\textbf{1} & \cellcolor{green!20}\textbf{11.13 (0.57)} & 2 & 12.94 (1.04) & 3 & 27.72 (8.04) & \cellcolor{green!20}\textbf{1} & \cellcolor{green!20}\textbf{6.25 (0.42)} & 3 & 7.27 (0.40) & 4 & 8.39 (0.26) & 2 & 6.62 (0.36) \\
 & $S_3$ & \cellcolor{green!20}\textbf{1} & \cellcolor{green!20}\textbf{11.09 (0.54)} & 2 & 13.10 (0.50) & 3 & 29.45 (5.73) & \cellcolor{green!20}\textbf{1} & \cellcolor{green!20}\textbf{5.48 (0.42)} & 3 & 6.63 (0.33) & 4 & 8.07 (0.24) & 2 & 5.79 (0.24) \\
 & $S_4$ & \cellcolor{green!20}\textbf{1} & \cellcolor{green!20}\textbf{11.04 (0.43)} & 2 & 13.15 (0.47) & 3 & 29.21 (5.15) & \cellcolor{green!20}\textbf{1} & \cellcolor{green!20}\textbf{5.06 (0.32)} & 3 & 5.97 (0.45) & 4 & 7.85 (0.26) & 2 & 5.26 (0.19) \\
 & $S_5$ & \cellcolor{green!20}\textbf{1} & \cellcolor{green!20}\textbf{10.93 (0.48)} & 2 & 13.15 (0.61) & 3 & 29.14 (3.85) & \cellcolor{green!20}\textbf{1} & \cellcolor{green!20}\textbf{4.69 (0.35)} & 3 & 5.68 (0.35) & 4 & 7.66 (0.17) & 2 & 4.90 (0.23) \\ \hline
\multirow{5}{*}{\textsc{HIPA$^{cc}$}} & $S_1$ & \cellcolor{green!20}2 & \cellcolor{green!20}12.70 (1.22) & 3 & 20.95 (699.46) & \textbf{1} & \textbf{23.50 (2.17)} & 2 & 12.03 (0.93) & \cellcolor{red!20}\textbf{1} & \cellcolor{red!20}\textbf{11.08 (0.66)} & \textbf{1} & \textbf{11.19 (0.60)} & 3 & 14.24 (0.61) \\
 & $S_2$ & \cellcolor{green!20}3 & \cellcolor{green!20}11.64 (0.35) & \textbf{1} & \textbf{19.70 (0.89)} & 2 & 21.82 (0.68) & \textbf{1} & \textbf{8.47 (1.10)} & \cellcolor{red!20}\textbf{1} & \cellcolor{red!20}\textbf{8.46 (0.70)} & 2 & 9.88 (0.37) & 3 & 10.48 (0.63) \\
 & $S_3$ & \cellcolor{green!20}\textbf{1} & \cellcolor{green!20}\textbf{11.43 (0.34)} & 2 & 19.34 (0.88) & 3 & 21.82 (0.42) & \cellcolor{green!20}\textbf{1} & \cellcolor{green!20}\textbf{6.62 (0.80)} & \textbf{1} & \textbf{6.69 (0.51)} & 3 & 9.41 (0.33) & 2 & 8.21 (0.47) \\
 & $S_4$ & \cellcolor{green!20}\textbf{1} & \cellcolor{green!20}\textbf{11.13 (0.25)} & 2 & 18.94 (0.58) & 3 & 21.75 (0.81) & \cellcolor{green!20}\textbf{1} & \cellcolor{green!20}\textbf{4.42 (0.22)} & 2 & 4.47 (0.32) & 4 & 8.81 (0.20) & 3 & 5.37 (0.24) \\
 & $S_5$ & \cellcolor{green!20}\textbf{1} & \cellcolor{green!20}\textbf{11.01 (0.15)} & 2 & 18.79 (0.35) & 3 & 21.88 (0.76) & \cellcolor{green!20}\textbf{1} & \cellcolor{green!20}\textbf{2.69 (0.15)} & \cellcolor{red!20}\textbf{1} & \cellcolor{red!20}\textbf{2.69 (0.13)} & 3 & 8.30 (0.20) & 2 & 3.10 (0.13) \\ \hline
\multirow{5}{*}{\textsc{HSMGP}} & $S_1$ & \cellcolor{green!20}\textbf{1} & \cellcolor{green!20}\textbf{7.72 (0.80)} & 2 & 55.90 (20.70) & 3 & 80.26 (38.86) & 3 & 21.65 (2.31) & 2 & 16.63 (2.10) & \cellcolor{red!20}\textbf{1} & \cellcolor{red!20}\textbf{14.63 (1.92)} & 4 & 25.42 (5.90) \\
 & $S_2$ & \cellcolor{green!20}\textbf{1} & \cellcolor{green!20}\textbf{7.35 (0.47)} & 2 & 51.49 (11.94) & 3 & 81.46 (17.14) & 3 & 15.59 (1.41) & 2 & 11.46 (1.78) & \cellcolor{red!20}\textbf{1} & \cellcolor{red!20}\textbf{10.59 (0.88)} & 4 & 18.59 (1.67) \\
 & $S_3$ & \cellcolor{green!20}\textbf{1} & \cellcolor{green!20}\textbf{7.14 (0.35)} & 2 & 51.65 (8.15) & 3 & 80.54 (19.98) & 2 & 11.35 (0.63) & \textbf{1} & \textbf{7.58 (0.67)} & \cellcolor{red!20}\textbf{1} & \cellcolor{red!20}\textbf{7.52 (0.46)} & 3 & 12.10 (0.67) \\
 & $S_4$ & \cellcolor{green!20}\textbf{1} & \cellcolor{green!20}\textbf{7.16 (0.17)} & 2 & 49.96 (10.24) & 3 & 66.32 (36.72) & 3 & 10.13 (0.57) & \cellcolor{red!20}\textbf{1} & \cellcolor{red!20}\textbf{6.88 (0.61)} & 2 & 7.10 (0.38) & 4 & 10.47 (0.65) \\
 & $S_5$ & \cellcolor{green!20}\textbf{1} & \cellcolor{green!20}\textbf{7.15 (0.20)} & 2 & 49.18 (3.13) & 3 & 85.14 (39.24) & 4 & 7.46 (0.46) & \cellcolor{red!20}\textbf{1} & \cellcolor{red!20}\textbf{5.15 (0.28)} & 2 & 6.11 (0.26) & 3 & 7.04 (0.28) \\ \hline
\multirow{5}{*}{\textsc{Lrzip}} & $S_1$ & \cellcolor{green!20}\textbf{1} & \cellcolor{green!20}\textbf{117.37 (24.65)} & 2 & 363.17 (114.48) & 3 & 475.86 (188.56) & 2 & 17.95 (10.85) & 3 & 21.71 (7.45) & \cellcolor{red!20}\textbf{1} & \cellcolor{red!20}\textbf{16.71 (4.27)} & 4 & 76.46 (23.85) \\
 & $S_2$ & \cellcolor{green!20}\textbf{1} & \cellcolor{green!20}\textbf{109.11 (20.26)} & 2 & 315.54 (57.21) & 3 & 474.88 (196.54) & \cellcolor{green!20}\textbf{1} & \cellcolor{green!20}\textbf{9.09 (2.56)} & 2 & 10.18 (2.12) & \textbf{1} & \textbf{10.19 (1.79)} & 3 & 36.47 (12.40) \\
 & $S_3$ & \cellcolor{green!20}\textbf{1} & \cellcolor{green!20}\textbf{104.20 (17.55)} & 2 & 321.82 (49.06) & 3 & 360.11 (241.44) & \cellcolor{green!20}\textbf{1} & \cellcolor{green!20}\textbf{7.77 (2.09)} & \textbf{1} & \textbf{7.90 (1.27)} & 2 & 8.43 (0.69) & 3 & 26.46 (7.48) \\
 & $S_4$ & \cellcolor{green!20}\textbf{1} & \cellcolor{green!20}\textbf{103.74 (14.29)} & 2 & 325.19 (51.13) & 3 & 346.60 (131.69) & \cellcolor{green!20}\textbf{1} & \cellcolor{green!20}\textbf{6.38 (2.72)} & \textbf{1} & \textbf{6.51 (0.69)} & 2 & 7.59 (0.91) & 3 & 19.77 (2.80) \\
 & $S_5$ & \cellcolor{green!20}\textbf{1} & \cellcolor{green!20}\textbf{105.83 (7.06)} & 2 & 324.82 (32.57) & 3 & 457.93 (299.67) & \cellcolor{green!20}\textbf{1} & \cellcolor{green!20}\textbf{3.58 (0.97)} & 2 & 3.67 (1.31) & 2 & 6.17 (0.43) & 3 & 9.37 (1.21) \\ \hline
\multirow{5}{*}{\textsc{nginx}} & $S_1$ & \cellcolor{green!20}\textbf{1} & \cellcolor{green!20}\textbf{22.02 (2.11)} & 3 & 576.17 (24.71) & 2 & 550.28 (66.10) & 3 & 14.44 (2.41) & 4 & 84.12 (150.39) & 2 & 6.78 (1.92) & \cellcolor{red!20}\textbf{1} & \cellcolor{red!20}\textbf{4.23 (1.21)} \\
 & $S_2$ & \cellcolor{green!20}\textbf{1} & \cellcolor{green!20}\textbf{20.25 (1.60)} & 2 & 568.78 (17.19) & 2 & 564.04 (37.05) & 3 & 7.44 (3.14) & 4 & 11.09 (3.33) & 2 & 4.90 (0.44) & \cellcolor{red!20}\textbf{1} & \cellcolor{red!20}\textbf{2.17 (0.48)} \\
 & $S_3$ & \cellcolor{green!20}\textbf{1} & \cellcolor{green!20}\textbf{19.66 (0.75)} & 2 & 567.70 (19.05) & 2 & 570.41 (23.98) & 2 & 4.81 (0.81) & 3 & 5.68 (1.43) & 2 & 4.56 (0.09) & \cellcolor{red!20}\textbf{1} & \cellcolor{red!20}\textbf{1.85 (0.43)} \\
 & $S_4$ & \cellcolor{green!20}\textbf{1} & \cellcolor{green!20}\textbf{19.74 (0.82)} & 2 & 566.35 (18.53) & 2 & 566.60 (22.75) & 2 & 4.16 (1.01) & 4 & 4.87 (0.63) & 3 & 4.49 (0.05) & \cellcolor{red!20}\textbf{1} & \cellcolor{red!20}\textbf{1.59 (0.28)} \\
 & $S_5$ & \cellcolor{green!20}\textbf{1} & \cellcolor{green!20}\textbf{19.17 (1.07)} & 2 & 565.62 (23.39) & 2 & 567.96 (18.51) & 2 & 3.33 (1.00) & 3 & 4.49 (0.56) & 3 & 4.44 (0.06) & \cellcolor{red!20}\textbf{1} & \cellcolor{red!20}\textbf{1.64 (0.30)} \\ \hline
\multirow{5}{*}{\textsc{VP8}} & $S_1$ & \cellcolor{green!20}\textbf{1} & \cellcolor{green!20}\textbf{12.72 (0.84)} & 3 & 45.51 (7.20) & 2 & 44.69 (5.11) & \cellcolor{green!20}\textbf{1} & \cellcolor{green!20}\textbf{2.57 (0.70)} & 3 & 3.32 (0.79) & 3 & 5.99 (0.31) & 2 & 4.42 (0.46) \\
 & $S_2$ & \cellcolor{green!20}\textbf{1} & \cellcolor{green!20}\textbf{12.42 (0.44)} & 2 & 42.04 (3.29) & 2 & 42.06 (2.51) & \cellcolor{green!20}\textbf{1} & \cellcolor{green!20}\textbf{1.24 (0.24)} & 2 & 1.45 (0.16) & 3 & 5.25 (0.20) & 2 & 1.44 (0.21) \\
 & $S_3$ & \cellcolor{green!20}\textbf{1} & \cellcolor{green!20}\textbf{12.41 (0.34)} & 2 & 42.55 (2.55) & 3 & 44.06 (2.66) & 2 & 1.15 (0.12) & 2 & 1.16 (0.16) & 3 & 5.15 (0.15) & \cellcolor{red!20}\textbf{1} & \cellcolor{red!20}\textbf{1.11 (0.09)} \\
 & $S_4$ & \cellcolor{green!20}\textbf{1} & \cellcolor{green!20}\textbf{12.35 (0.31)} & 2 & 41.90 (3.11) & 3 & 42.98 (2.47) & 3 & 1.07 (0.08) & 2 & 1.03 (0.11) & 4 & 5.07 (0.10) & \cellcolor{red!20}\textbf{1} & \cellcolor{red!20}\textbf{1.02 (0.05)} \\
 & $S_5$ & \cellcolor{green!20}\textbf{1} & \cellcolor{green!20}\textbf{12.41 (0.27)} & 2 & 42.79 (1.80) & 3 & 43.19 (2.36) & 3 & 0.94 (0.04) & \cellcolor{red!20}\textbf{1} & \cellcolor{red!20}\textbf{0.85 (0.03)} & 4 & 4.89 (0.09) & 2 & 0.88 (0.04) \\ \hline
\multicolumn{2}{c}{Average $r$} & 1.2 &  & 2.1 &  & 2.38 &  & 1.85 &  & 2 &  & 2.35 &  & 2.43 & 
\\
\bottomrule
\end{tabular}
\end{adjustbox}

%% file: Tables/chap-DAL/compare_clustering.tex
\scriptsize
\setlength{\tabcolsep}{1.2mm}
\begin{adjustbox}{width=1.3\textwidth,center}
\begin{tabular}{clllllllll||llll||llll} \toprule
\multirow{2}{*}{System} & \multicolumn{1}{c}{\multirow{2}{*}{Size}} & \multicolumn{2}{c}{\texttt{DaL}} & \multicolumn{2}{c}{\texttt{DBSCAN}} & \multicolumn{2}{c}{\texttt{Agglomerative}} & \multicolumn{2}{c}{\texttt{kMeans}} & \multicolumn{2}{c}{\texttt{DaL}} & \multicolumn{2}{c}{\texttt{DaL-FSE}} & \multicolumn{2}{c}{\texttt{DaL}} & \multicolumn{2}{c}{\texttt{DaL$^{HV}$}} \\ \cline{3-18}
 & \multicolumn{1}{c}{} & $r$ & Med (IQR) & $r$ & Med (IQR) & $r$ & Med (IQR) & $r$ & Med (IQR) & $r$ & Med (IQR) & $r$ & Med (IQR) & $r$ & Med (IQR) & $r$ & Med (IQR) \\ \hline
\multirow{5}{*}{\textsc{Apache}} & $S_1$ & \cellcolor{green!20}\textbf{1} & \cellcolor{green!20}\textbf{21.86 (7.36)} & 2 & 26.29 (15.23) & 3 & 30.99 (20.01) & 3 & 40.02 (22.13) & \cellcolor{green!20}\textbf{1} & \cellcolor{green!20}\textbf{21.86 (7.36)} & \cellcolor{red!20}\textbf{1} & \cellcolor{red!20}\textbf{21.86 (7.36)} & \cellcolor{green!20}\textbf{1} & \cellcolor{green!20}\textbf{21.86 (7.36)} & \cellcolor{red!20}\textbf{1} & \cellcolor{red!20}\textbf{21.86 (7.36)} \\
 & $S_2$ & \cellcolor{green!20}2 & \cellcolor{green!20}12.89 (7.87) & 2 & 16.24 (10.00) & \textbf{1} & \textbf{15.45 (5.76)} & \textbf{1} & \textbf{14.85 (5.21)} & \textbf{1} & \textbf{12.89 (7.87)} & \cellcolor{red!20}\textbf{1} & \cellcolor{red!20}\textbf{11.87 (9.22)} & \cellcolor{green!20}\textbf{1} & \cellcolor{green!20}\textbf{12.89 (7.87)} & \cellcolor{red!20}\textbf{1} & \cellcolor{red!20}\textbf{12.89 (7.87)} \\
 & $S_3$ & \cellcolor{green!20}\textbf{1} & \cellcolor{green!20}\textbf{6.56 (1.92)} & 3 & 13.24 (9.29) & 2 & 10.62 (3.24) & 2 & 10.60 (6.77) & \cellcolor{green!20}\textbf{1} & \cellcolor{green!20}\textbf{6.56 (1.92)} & \cellcolor{red!20}\textbf{1} & \cellcolor{red!20}\textbf{6.56 (1.92)} & \cellcolor{green!20}\textbf{1} & \cellcolor{green!20}\textbf{6.56 (1.92)} & 2 & 7.00 (2.29) \\
\bottomrule
\end{tabular}
\end{adjustbox}

%% file: Figures/chap-DAL/Sensitivity_to_depth.tex
\begin{subfigure}{.23\columnwidth}
  \centering
\includegraphics[width=\linewidth]{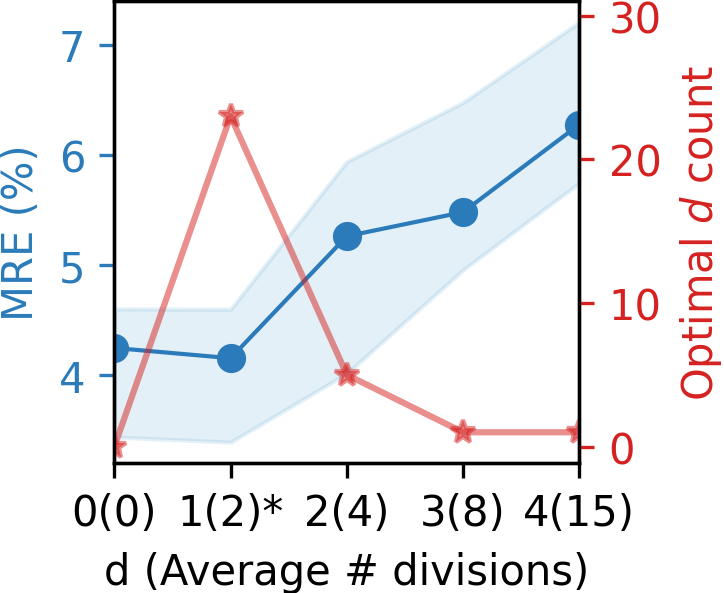}  
  \subcaption{\textsc{Apache}}
  \label{fig:sensitivity-Apache}
\end{subfigure}
~
\begin{subfigure}{.23\columnwidth}
  \centering
  \includegraphics[width=\linewidth]{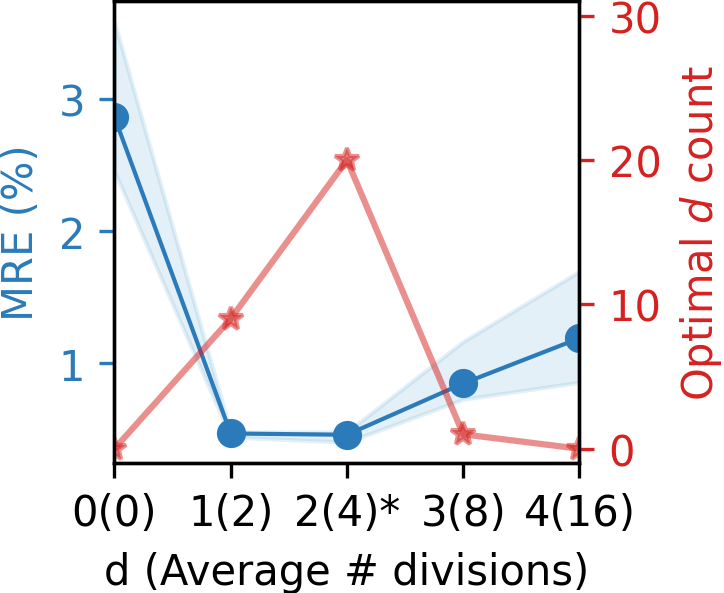} 
  \caption{\textsc{BDB-C}}
  \label{fig:sensitivity-BDBC}
\end{subfigure}
~
\begin{subfigure}{.23\columnwidth}
  \centering
  \includegraphics[width=\linewidth]{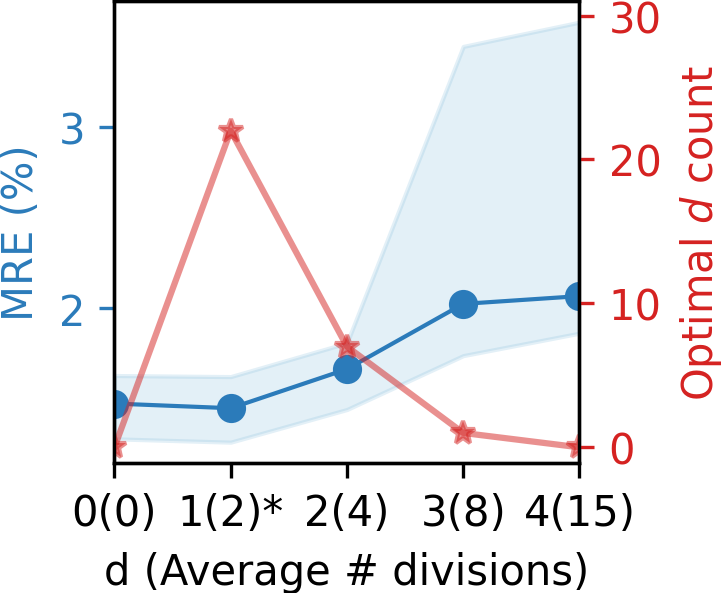} 
  \subcaption{\textsc{BDB-J}}
  \label{fig:sensitivity-BDBJ}
\end{subfigure}
~
\begin{subfigure}{.24\columnwidth}
  \centering
  \includegraphics[width=\linewidth]{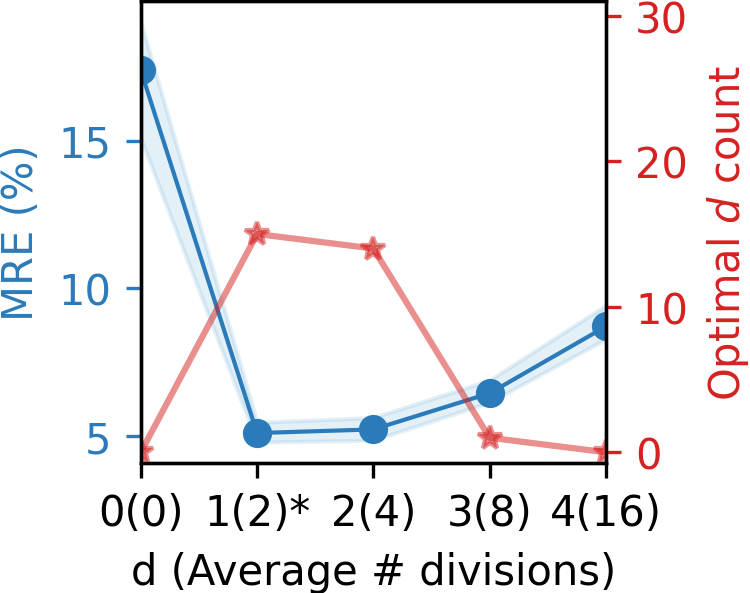}  
  \subcaption{\textsc{kanzi}}
  \label{fig:sensitivity-kanzi}
\end{subfigure}

\vspace{0.2cm}

\begin{subfigure}{.235\columnwidth}
  \centering
  \includegraphics[width=\linewidth]{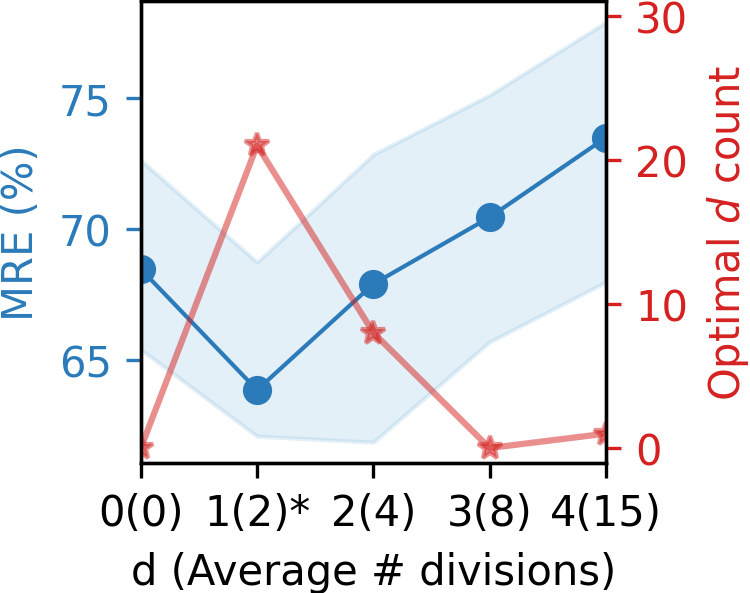} 
  \subcaption{\textsc{SQLite}}
  \label{fig:sensitivity-sqlite}
\end{subfigure}
~
\begin{subfigure}{.24\columnwidth}
  \centering
  \includegraphics[width=\linewidth]{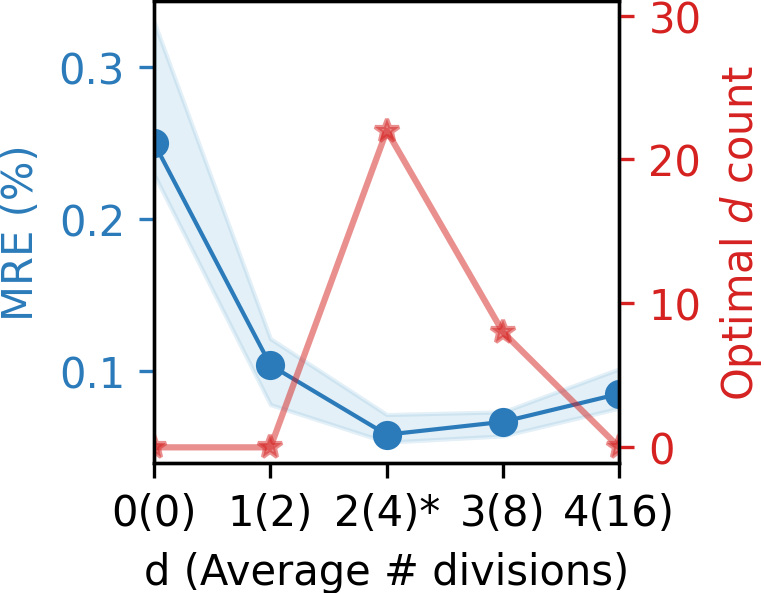}  
  \subcaption{\textsc{x264}}
  \label{fig:sensitivity-x264}
\end{subfigure}
~
\begin{subfigure}{.23\columnwidth}
  \centering
  \includegraphics[width=\linewidth]{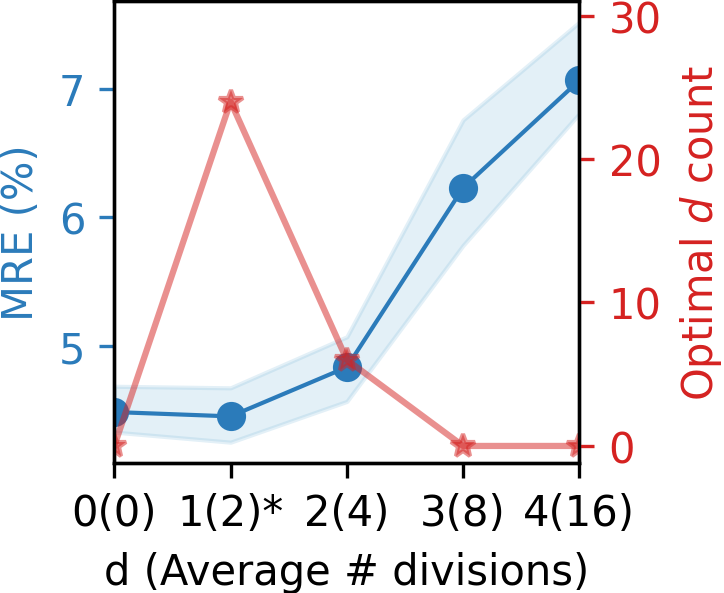} 
  \subcaption{\textsc{Dune}}
  \label{fig:sensitivity-Dune}
\end{subfigure}
~
\begin{subfigure}{.24\columnwidth}
  \centering
  \includegraphics[width=\linewidth]{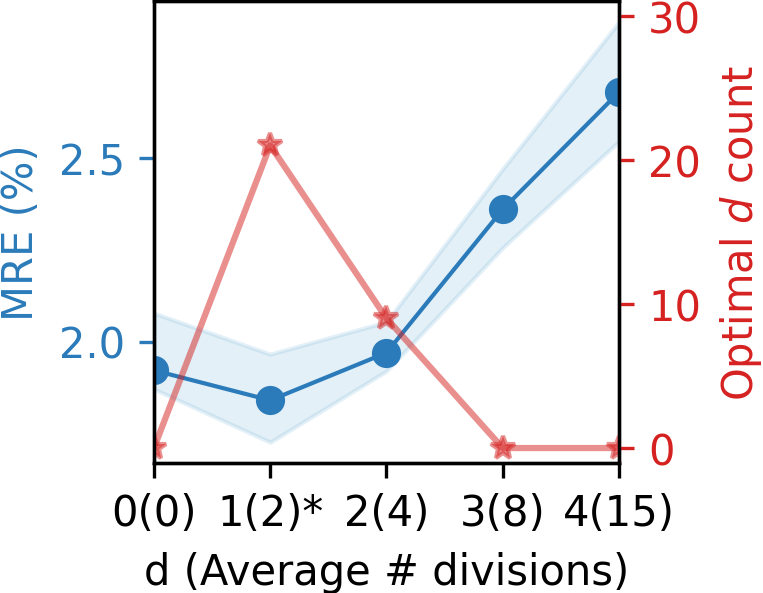} 
  \subcaption{\textsc{HIPA$^{cc}$}}
  \label{fig:sensitivity-hipacc}
\end{subfigure}

\vspace{0.2cm}

\begin{subfigure}{.24\columnwidth}
  \centering
  \includegraphics[width=\linewidth]{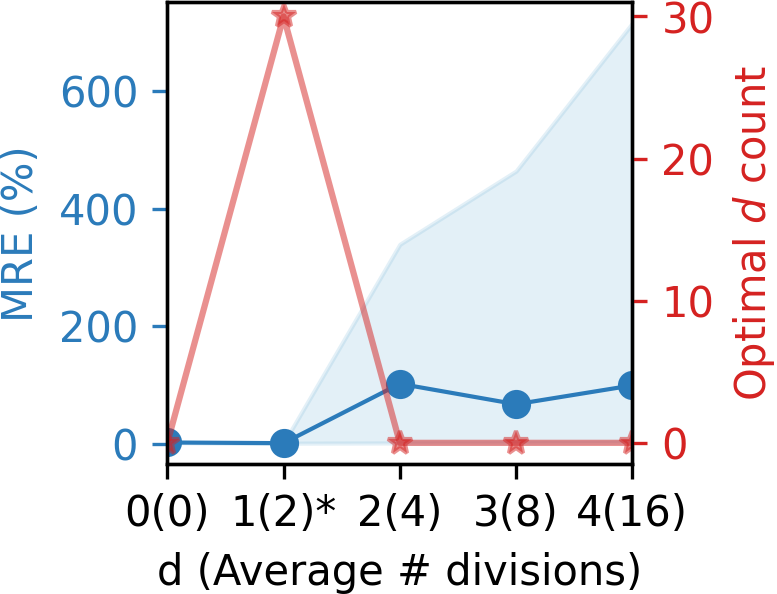} 
  \subcaption{\textsc{HSMGP}}
  \label{fig:sensitivity-Hsmgp}
\end{subfigure}
~
\begin{subfigure}{.23\columnwidth}
  \centering
  \includegraphics[width=\linewidth]{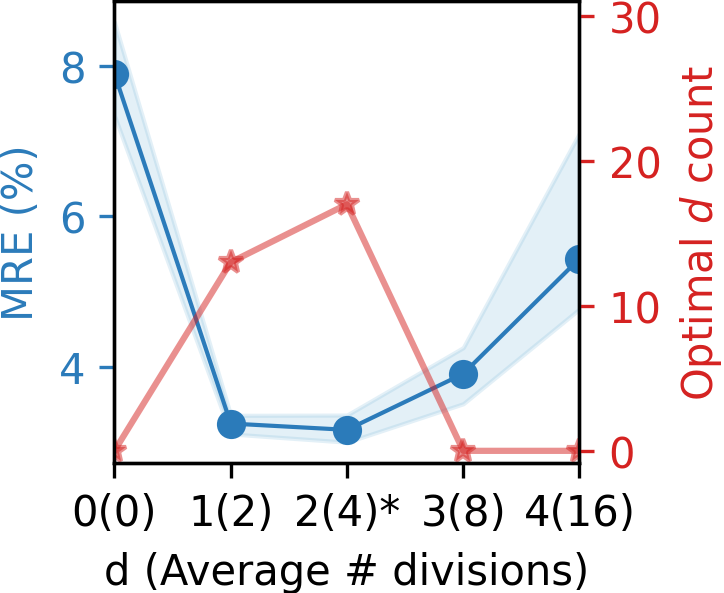} 
  \subcaption{\textsc{Lrzip}}
  \label{fig:sensitivity-Lrzip}
\end{subfigure}
~
\begin{subfigure}{.23\columnwidth}
  \centering
  \includegraphics[width=\linewidth]{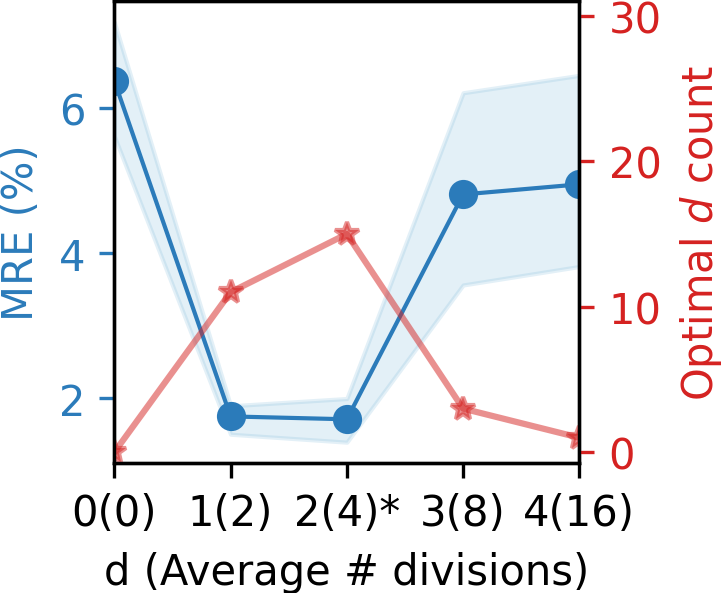} 
  \subcaption{\textsc{nginx}}
  \label{fig:sensitivity-nginx}
\end{subfigure}
~
\begin{subfigure}{.24\columnwidth}
  \centering
  \includegraphics[width=\linewidth]{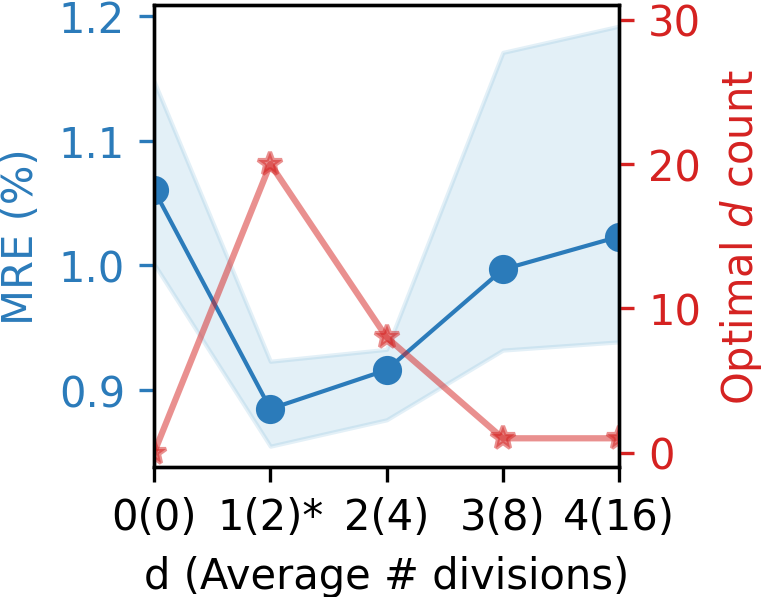} 
  \subcaption{\textsc{VP8}}
  \label{fig:sensitivity-VP8}
\end{subfigure}

%% file: Figures/chap-DAL/hit-count-new.tex
\resizebox{\textwidth}{!}{ 
\begin{subfigure}[b]{.25\textwidth}
  \centering
  \scriptsize
\begin{tikzpicture}
\begin{axis}[
    ybar stacked,
       width=4cm,
    height=4cm,
         axis x line  = bottom,
    axis y line  = left  ,
	bar width=8pt,
  enlarge x limits=0.2,
 ymin=0,
    legend style={at={(0.5,-0.20)},
      anchor=north,legend columns=-1},
    xlabel={\textsc{Apache}},
    ylabel={\# Hit count on the best $d$},
    symbolic x coords={$S_{1}$, $S_{2}$, $S_{3}$, $S_{4}$, $S_{5}$},
    xtick=data,
    ymajorgrids=true,
    ytick={0,8,15,23,30},
    y label style={at={(-0.2, 0.4)}},
    ]
\addplot+[ybar,red!20,draw=black] plot coordinates {($S_{1}$,16) ($S_{2}$,23) ($S_{3}$,9) ($S_{4}$,6) ($S_{5}$,5)};
\addplot+[ybar,green!20,draw=black] plot coordinates {($S_{1}$,14) ($S_{2}$,7) ($S_{3}$,21) ($S_{4}$,24) ($S_{5}$,25)};
\end{axis}
\end{tikzpicture}
\vspace{-0.01cm}
\end{subfigure}
~\hspace{-0.3cm}
\begin{subfigure}[b]{.15\textwidth}
  \centering
  \scriptsize
\begin{tikzpicture}
\begin{axis}[
    ybar stacked,
       width=4cm,
    height=4cm,
         axis x line  = bottom,
    axis y line  = left  ,
	bar width=8pt,
  enlarge x limits=0.2,
 ymin=0,
    legend style={at={(0.5,-0.20)},
      anchor=north,legend columns=-1},
    xlabel={\textsc{BDB-C}},
    symbolic x coords={$S_{1}$, $S_{2}$, $S_{3}$, $S_{4}$, $S_{5}$},
    xtick=data,
        ymajorgrids=true,
    ytick={0,8,15,23,30},
    y label style={at={(0.1, 0.5)}},
    ]
\addplot+[ybar,red!20,draw=black] plot coordinates {($S_{1}$,17) ($S_{2}$,12) ($S_{3}$,14) ($S_{4}$,14) ($S_{5}$,4)};
\addplot+[ybar,green!20,draw=black] plot coordinates {($S_{1}$,13) ($S_{2}$,18) ($S_{3}$,16) ($S_{4}$,16) ($S_{5}$,26)};
\end{axis}
\end{tikzpicture}
\end{subfigure}
~\hspace{0.7cm}
\begin{subfigure}[b]{.15\textwidth}
  \centering
  \scriptsize
\begin{tikzpicture}
\begin{axis}[
    ybar stacked,
       width=4cm,
    height=4cm,
         axis x line  = bottom,
    axis y line  = left  ,
	bar width=8pt,
  enlarge x limits=0.2,
 ymin=0,
    legend style={at={(0.5,-0.20)},
      anchor=north,legend columns=-1},
    xlabel={\textsc{BDB-J}},
    symbolic x coords={$S_{1}$, $S_{2}$, $S_{3}$, $S_{4}$, $S_{5}$},
    xtick=data,
    y label style={at={(0.1, 0.5)}},
        ymajorgrids=true,
    ytick={0,8,15,23,30},
    ]
\addplot+[ybar,red!20,draw=black] plot coordinates {($S_{1}$,16) ($S_{2}$,3) ($S_{3}$,2) ($S_{4}$,1) ($S_{5}$,0)};
\addplot+[ybar,green!20,draw=black] plot coordinates {($S_{1}$,14) ($S_{2}$,27) ($S_{3}$,28) ($S_{4}$,29) ($S_{5}$,30)};
\end{axis}
\end{tikzpicture}
\end{subfigure}
~\hspace{0.7cm}
\begin{subfigure}[b]{.15\textwidth}
  \centering
  \scriptsize
\begin{tikzpicture}
\begin{axis}[
    ybar stacked,
       width=4cm,
    height=4cm,
         axis x line  = bottom,
    axis y line  = left  ,
	bar width=8pt,
  enlarge x limits=0.2,
 ymin=0,
    legend style={at={(0.5,-0.20)},
      anchor=north,legend columns=-1},
    xlabel={\textsc{kanzi}},
    symbolic x coords={$S_{1}$, $S_{2}$, $S_{3}$, $S_{4}$, $S_{5}$},
    xtick=data,
    y label style={at={(0.1, 0.5)}},
        ymajorgrids=true,
    ytick={0,8,15,23,30},
    ]
\addplot+[ybar,red!20,draw=black] plot coordinates {($S_{1}$,9) ($S_{2}$,3) ($S_{3}$,6) ($S_{4}$,3) ($S_{5}$,1)};
\addplot+[ybar,green!20,draw=black] plot coordinates {($S_{1}$,21) ($S_{2}$,27) ($S_{3}$,24) ($S_{4}$,27) ($S_{5}$,29)};
\end{axis}
\end{tikzpicture}
\vspace{-0.29cm}
\end{subfigure}
}

\resizebox{\textwidth}{!}{ 
\begin{subfigure}[b]{.25\textwidth}
  \centering
  \scriptsize
\begin{tikzpicture}
\begin{axis}[
    ybar stacked,
       width=4cm,
    height=4cm,
         axis x line  = bottom,
    axis y line  = left  ,
	bar width=8pt,
  enlarge x limits=0.2,
 ymin=0,
    legend style={at={(0.5,-0.20)},
      anchor=north,legend columns=-1},
    xlabel={\textsc{SQLite}},
    ylabel={\# Hit count on the best $d$},
    symbolic x coords={$S_{1}$, $S_{2}$, $S_{3}$, $S_{4}$, $S_{5}$},
    xtick=data,
    y label style={at={(-0.2, 0.4)}},
        ymajorgrids=true,
    ytick={0,8,15,23,30},
    ]
\addplot+[ybar,red!20,draw=black] plot coordinates {($S_{1}$,15) ($S_{2}$,7) ($S_{3}$,8) ($S_{4}$,6) ($S_{5}$,4)};
\addplot+[ybar,green!20,draw=black] plot coordinates {($S_{1}$,15) ($S_{2}$,23) ($S_{3}$,22) ($S_{4}$,24) ($S_{5}$,26)};
\end{axis}
\end{tikzpicture}
\vspace{-0.05cm}
\end{subfigure}
~\hspace{-0.3cm}
\begin{subfigure}[b]{.15\textwidth}
  \centering
  \scriptsize
\begin{tikzpicture}
\begin{axis}[
    ybar stacked,
       width=4cm,
    height=4cm,
         axis x line  = bottom,
    axis y line  = left  ,
	bar width=8pt,
  enlarge x limits=0.2,
 ymin=0,
    legend style={at={(0.5,-0.20)},
      anchor=north,legend columns=-1},
    xlabel={\textsc{x264}},
    symbolic x coords={$S_{1}$, $S_{2}$, $S_{3}$, $S_{4}$, $S_{5}$},
    xtick=data,
    y label style={at={(0.1, 0.5)}},
        ymajorgrids=true,
    ytick={0,8,15,23,30},
    ]
\addplot+[ybar,red!20,draw=black] plot coordinates {($S_{1}$,16) ($S_{2}$,10) ($S_{3}$,12) ($S_{4}$,15) ($S_{5}$,21)};
\addplot+[ybar,green!20,draw=black] plot coordinates {($S_{1}$,14) ($S_{2}$,20) ($S_{3}$,18) ($S_{4}$,15) ($S_{5}$,9)};
\end{axis}
\end{tikzpicture}
\end{subfigure}
~\hspace{0.7cm}
\begin{subfigure}[b]{.1\textwidth}
  \centering
  \scriptsize
\begin{tikzpicture}
\begin{axis}[
    ybar stacked,
       width=4cm,
    height=4cm,
         axis x line  = bottom,
    axis y line  = left  ,
	bar width=8pt,
  enlarge x limits=0.2,
 ymin=0,
    legend style={at={(0.5,-0.20)},
      anchor=north,legend columns=-1},
    xlabel={\textsc{Dune}},
    symbolic x coords={$S_{1}$, $S_{2}$, $S_{3}$, $S_{4}$, $S_{5}$},
    xtick=data,
        ymajorgrids=true,
    ytick={0,8,15,23,30},
    ]
\addplot+[ybar,red!20,draw=black] plot coordinates {($S_{1}$,15) ($S_{2}$,1) ($S_{3}$,0) ($S_{4}$,0) ($S_{5}$,0)};
\addplot+[ybar,green!20,draw=black] plot coordinates {($S_{1}$,15) ($S_{2}$,29) ($S_{3}$,30) ($S_{4}$,30) ($S_{5}$,30)};
\end{axis}
\end{tikzpicture}
\end{subfigure}
~\hspace{1.4cm}
\begin{subfigure}[b]{.15\textwidth}
  \centering
  \scriptsize
\begin{tikzpicture}
\begin{axis}[
    ybar stacked,
       width=4cm,
    height=4cm,
         axis x line  = bottom,
    axis y line  = left  ,
	bar width=8pt,
  enlarge x limits=0.2,
 ymin=0,
    legend style={at={(0.5,-0.20)},
      anchor=north,legend columns=-1},
    xlabel={\textsc{HIPA$^{cc}$}},
    symbolic x coords={$S_{1}$, $S_{2}$, $S_{3}$, $S_{4}$, $S_{5}$},
    xtick=data,
    y label style={at={(0.1, 0.5)}},
        ymajorgrids=true,
    ytick={0,8,15,23,30},
    ]
\addplot+[ybar,red!20,draw=black] plot coordinates {($S_{1}$,13) ($S_{2}$,12) ($S_{3}$,8) ($S_{4}$,3) ($S_{5}$,9)};
\addplot+[ybar,green!20,draw=black] plot coordinates {($S_{1}$,17) ($S_{2}$,18) ($S_{3}$,22) ($S_{4}$,27) ($S_{5}$,21)};
\end{axis}
\end{tikzpicture}
\vspace{-0.33cm}
\end{subfigure}
}

\resizebox{\textwidth}{!}{ 
\begin{subfigure}[b]{.25\textwidth}
  \centering
  \scriptsize
\begin{tikzpicture}
\begin{axis}[
    ybar stacked,
       width=4cm,
    height=4cm,
         axis x line  = bottom,
    axis y line  = left  ,
	bar width=8pt,
  enlarge x limits=0.2,
 ymin=0,
    legend style={at={(0.5,-0.20)},
      anchor=north,legend columns=-1},
    xlabel={\textsc{HSMGP}},
    ylabel={\# Hit count on the best $d$},
    symbolic x coords={$S_{1}$, $S_{2}$, $S_{3}$, $S_{4}$, $S_{5}$},
    xtick=data,
    y label style={at={(-0.2, 0.4)}},
        ymajorgrids=true,
    ytick={0,8,15,23,30},
    ]
\addplot+[ybar,red!20,draw=black] plot coordinates {($S_{1}$,4) ($S_{2}$,1) ($S_{3}$,0) ($S_{4}$,0) ($S_{5}$,0)};
\addplot+[ybar,green!20,draw=black] plot coordinates {($S_{1}$,26) ($S_{2}$,29) ($S_{3}$,30) ($S_{4}$,30) ($S_{5}$,30)};
\end{axis}
\end{tikzpicture}
\end{subfigure}
~
\begin{subfigure}[b]{.15\textwidth}
  \centering
  \scriptsize
\begin{tikzpicture}
\begin{axis}[
    ybar stacked,
       width=4cm,
    height=4cm,
         axis x line  = bottom,
    axis y line  = left  ,
	bar width=8pt,
  enlarge x limits=0.2,
 ymin=0,
    legend style={at={(0.5,-0.20)},
      anchor=north,legend columns=-1},
    xlabel={\textsc{Lrzip}},
    symbolic x coords={$S_{1}$, $S_{2}$, $S_{3}$, $S_{4}$, $S_{5}$},
    xtick=data,
    y label style={at={(0.1, 0.5)}},
        ymajorgrids=true,
    ytick={0,8,15,23,30},
    ]
\addplot+[ybar,red!20,draw=black] plot coordinates {($S_{1}$,16) ($S_{2}$,3) ($S_{3}$,4) ($S_{4}$,0) ($S_{5}$,0)};
\addplot+[ybar,green!20,draw=black] plot coordinates {($S_{1}$,14) ($S_{2}$,27) ($S_{3}$,26) ($S_{4}$,30) ($S_{5}$,30)};
\end{axis}
\end{tikzpicture}
\end{subfigure}
~\hspace{0.7cm}
\begin{subfigure}[b]{.15\textwidth}
  \centering
  \scriptsize
\begin{tikzpicture}
\begin{axis}[
    ybar stacked,
       width=4cm,
    height=4cm,
         axis x line  = bottom,
    axis y line  = left  ,
	bar width=8pt,
  enlarge x limits=0.2,
 ymin=0,
    legend style={at={(0.5,-0.20)},
      anchor=north,legend columns=-1},
    xlabel={\textsc{nginx}},
    symbolic x coords={$S_{1}$, $S_{2}$, $S_{3}$, $S_{4}$, $S_{5}$},
    xtick=data,
    y label style={at={(0.1, 0.5)}},
        ymajorgrids=true,
    ytick={0,8,15,23,30},
    ]
\addplot+[ybar,red!20,draw=black] plot coordinates {($S_{1}$,3) ($S_{2}$,8) ($S_{3}$,13) ($S_{4}$,18) ($S_{5}$,11)};
\addplot+[ybar,green!20,draw=black] plot coordinates {($S_{1}$,27) ($S_{2}$,22) ($S_{3}$,17) ($S_{4}$,12) ($S_{5}$,19)};
\end{axis}
\end{tikzpicture}
\vspace{-0.3cm}
\end{subfigure}
~\hspace{0.7cm}
\begin{subfigure}[b]{.15\textwidth}
  \centering
  \scriptsize
\begin{tikzpicture}
\begin{axis}[
    ybar stacked,
       width=4cm,
    height=4cm,
         axis x line  = bottom,
    axis y line  = left  ,
	bar width=8pt,
  enlarge x limits=0.2,
 ymin=0,
    legend style={at={(0.5,-0.20)},
      anchor=north,legend columns=-1},
    xlabel={\textsc{VP8}},
    symbolic x coords={$S_{1}$, $S_{2}$, $S_{3}$, $S_{4}$, $S_{5}$},
    xtick=data,
    y label style={at={(0.1, 0.5)}},
        ymajorgrids=true,
    ytick={0,8,15,23,30},
    ]
\addplot+[ybar,red!20,draw=black] plot coordinates {($S_{1}$,2) ($S_{2}$,0) ($S_{3}$,0) ($S_{4}$,0) ($S_{5}$,2)};
\addplot+[ybar,green!20,draw=black] plot coordinates {($S_{1}$,28) ($S_{2}$,30) ($S_{3}$,30) ($S_{4}$,30) ($S_{5}$,28)};
\end{axis}
\end{tikzpicture}
\end{subfigure}
}

%% file: Figures/chap-DAL/box_plot_ground_truth.tex
\resizebox{\textwidth}{!}{ 
\begin{subfigure}{.355\textwidth}
  \centering
  \includegraphics[width=\linewidth]{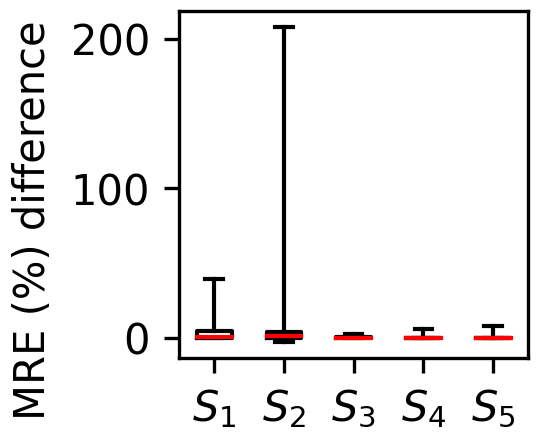} 
      \subcaption{\textsc{Apache}}
\end{subfigure}
\begin{subfigure}{.316\textwidth}
  \centering
  \includegraphics[width=\linewidth]{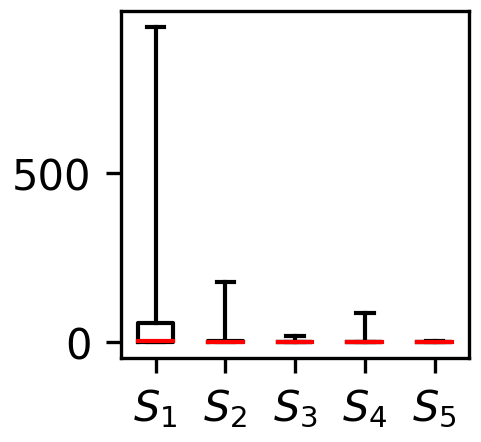} 
  \subcaption{\textsc{BDB-C}}
\end{subfigure}
\begin{subfigure}{.298\textwidth}
  \centering
  \includegraphics[width=\linewidth]{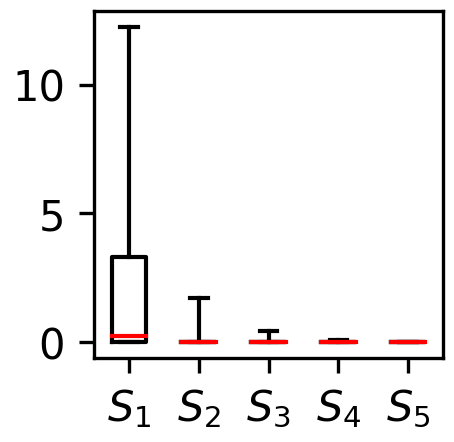} 
  \subcaption{\textsc{BDB-J}}
\end{subfigure}
\begin{subfigure}{.316\textwidth}
  \centering
  \includegraphics[width=\linewidth]{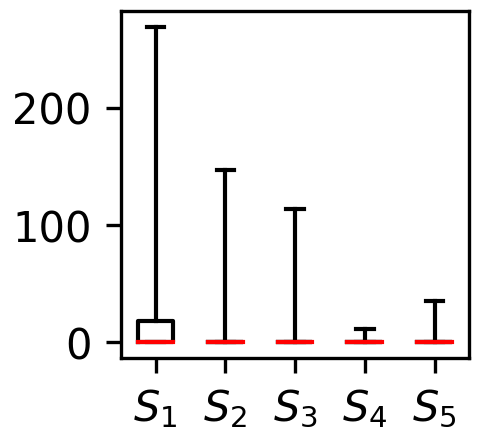} 
 \subcaption{\textsc{kanzi}}
\end{subfigure}
}

\resizebox{\textwidth}{!}{ 
\begin{subfigure}{.335\textwidth}
  \centering
  \includegraphics[width=\linewidth]{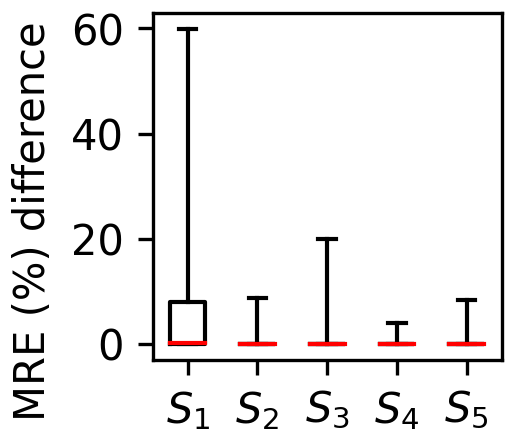} 
  \subcaption{\textsc{SQLite}}
\end{subfigure}
\begin{subfigure}{.298\textwidth}
  \centering
  \includegraphics[width=\linewidth]{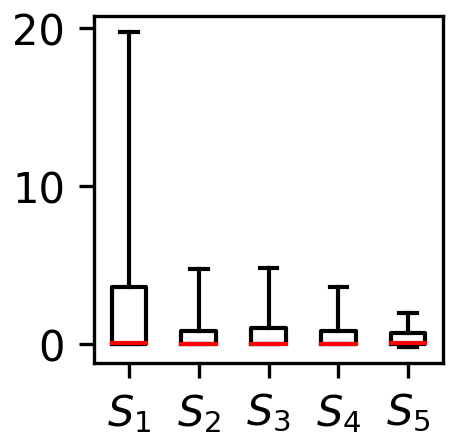} 
  \subcaption{\textsc{x264}}
\end{subfigure}
\begin{subfigure}{.282\textwidth}
  \centering
  \includegraphics[width=\linewidth]{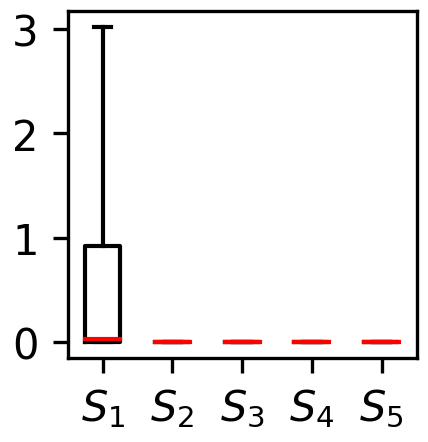} 
 \subcaption{\textsc{Dune}}
\end{subfigure}
\begin{subfigure}{.309\textwidth}
  \centering
  \includegraphics[width=\linewidth]{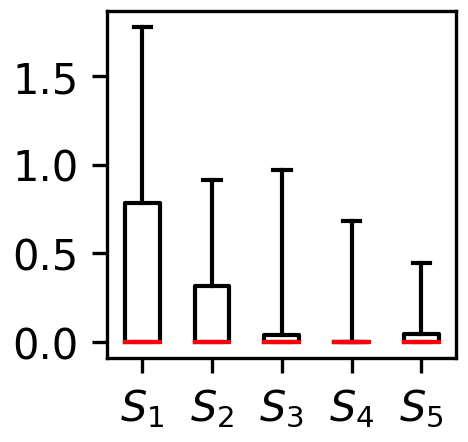} 
  \subcaption{\textsc{HIPA$^{cc}$}}
\end{subfigure}
}

\resizebox{\textwidth}{!}{ 
\begin{subfigure}{.318\textwidth}
  \centering
  \includegraphics[width=\linewidth]{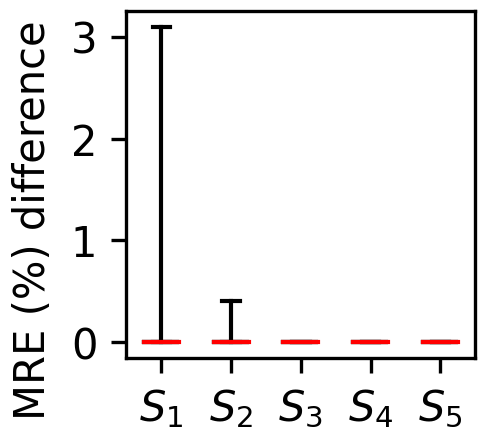} 
  \subcaption{\textsc{HSMGP}}
\end{subfigure}
\begin{subfigure}{.298\textwidth}
  \centering
  \includegraphics[width=\linewidth]{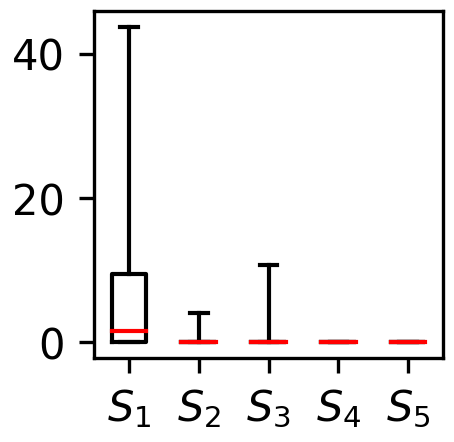} 
  \subcaption{\textsc{Lrzip}}
\end{subfigure}
\begin{subfigure}{.282\textwidth}
  \centering
  \includegraphics[width=\linewidth]{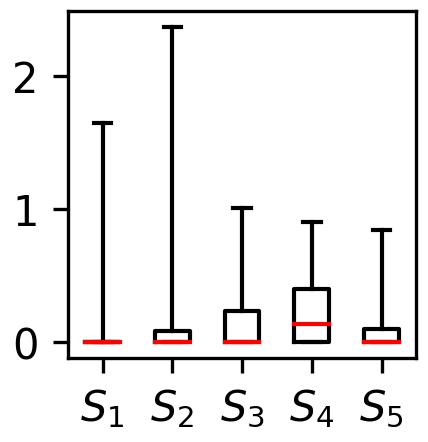} 
 \subcaption{\textsc{nginx}}
\end{subfigure}
\begin{subfigure}{.309\textwidth}
  \centering
  \includegraphics[width=\linewidth]{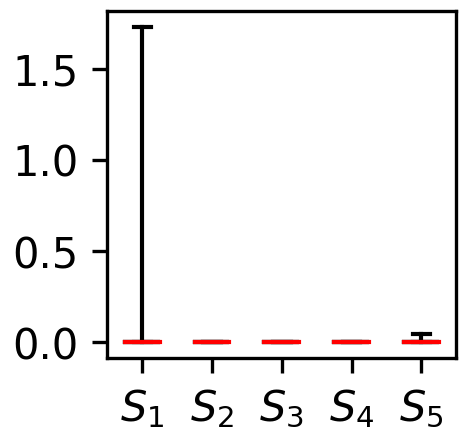} 
 \subcaption{\textsc{VP8}}
\end{subfigure}
}

%% file: Tables/chap-DAL/min_max_time.tex
\begin{table}[t!]
\caption{The overhead ranges across all systems and sizes.}
\centering
\begin{adjustbox}{width=0.85\columnwidth,center}
\begin{tabular}{lll}
\toprule
\textbf{Approach}             & \textbf{Overhead (min)}                  & \textbf{Restriction and Prerequisite}               \\
\midrule
\texttt{SPLConqueror} & 4$\times 10^{-4}$ to 5$\times 10^{-3}$ & needs to select sampling method(s) \\
\texttt{DECART} & 0.07 to 0.5 & does not work on mixed systems \\
\texttt{Perf-AL} & 0.08 to 2.4 & None \\
\texttt{DeepPerf} & 3 to 48 & None \\
\texttt{HINNPerf} & 3 to 54 & None \\
\Model & 4 to 56 & None\\
--- \Model~(\textit{adapting $d$}) & 1$\times 10^{-6}$ to 0.02 & None \\
--- \Model~(\textit{dividing}) & 9$\times 10^{-4}$ to 0.18 & None \\
--- \Model~(\textit{training}) & 3 to 53 & None \\
--- \Model~(\textit{predicting}) & 0.3 to 3 & None \\

\bottomrule
\end{tabular}
\end{adjustbox}
\label{tb:max_min}
\end{table}

%% file: Chapter-DAL/discussion.tex
\section{Discussion}
\label{chap-dal:discussion}

\begin{figure}[!t]
\centering
\footnotesize

\begin{subfigure}{.48\columnwidth}
  \centering
  \includegraphics[width=\textwidth]{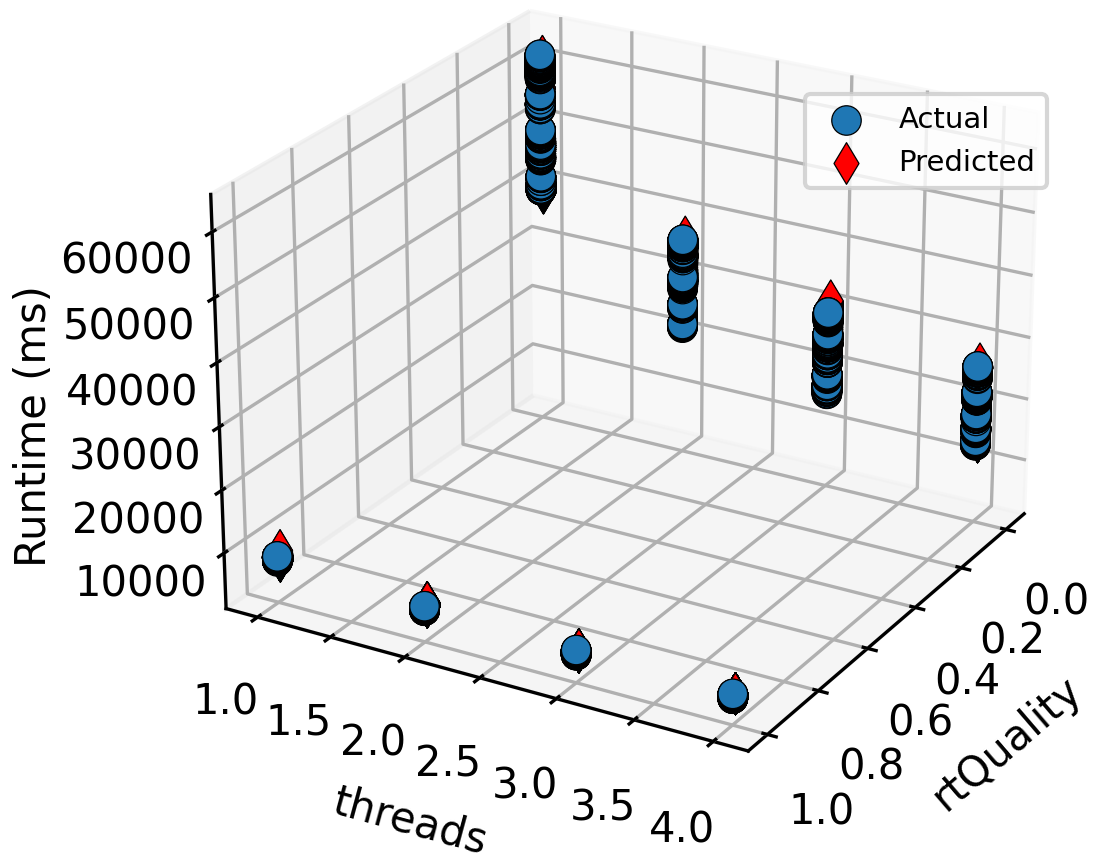} 
  \caption{\Model}
\end{subfigure}
~\hfill
\begin{subfigure}{.48\columnwidth}
  \centering
  \includegraphics[width=\textwidth]{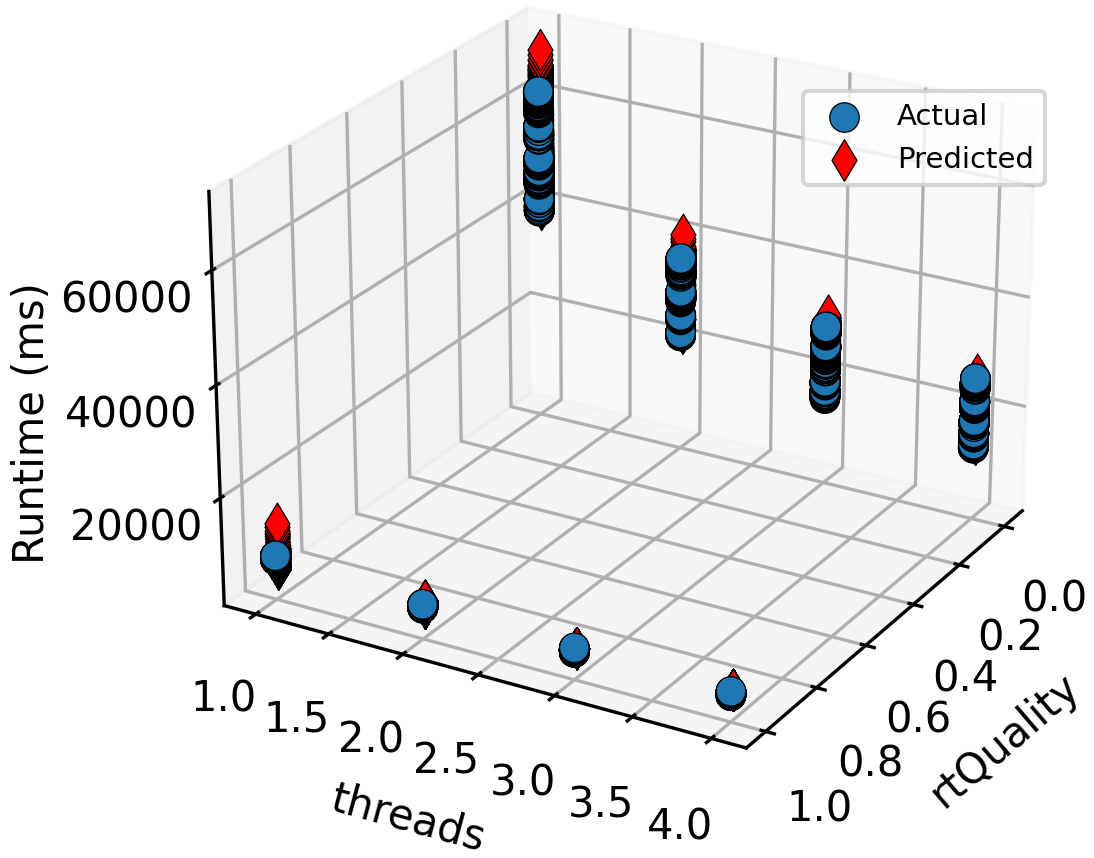} 
  \caption{\texttt{HINNPerf}}
\end{subfigure}

\begin{subfigure}{.48\columnwidth}
  \centering
  \includegraphics[width=\textwidth]{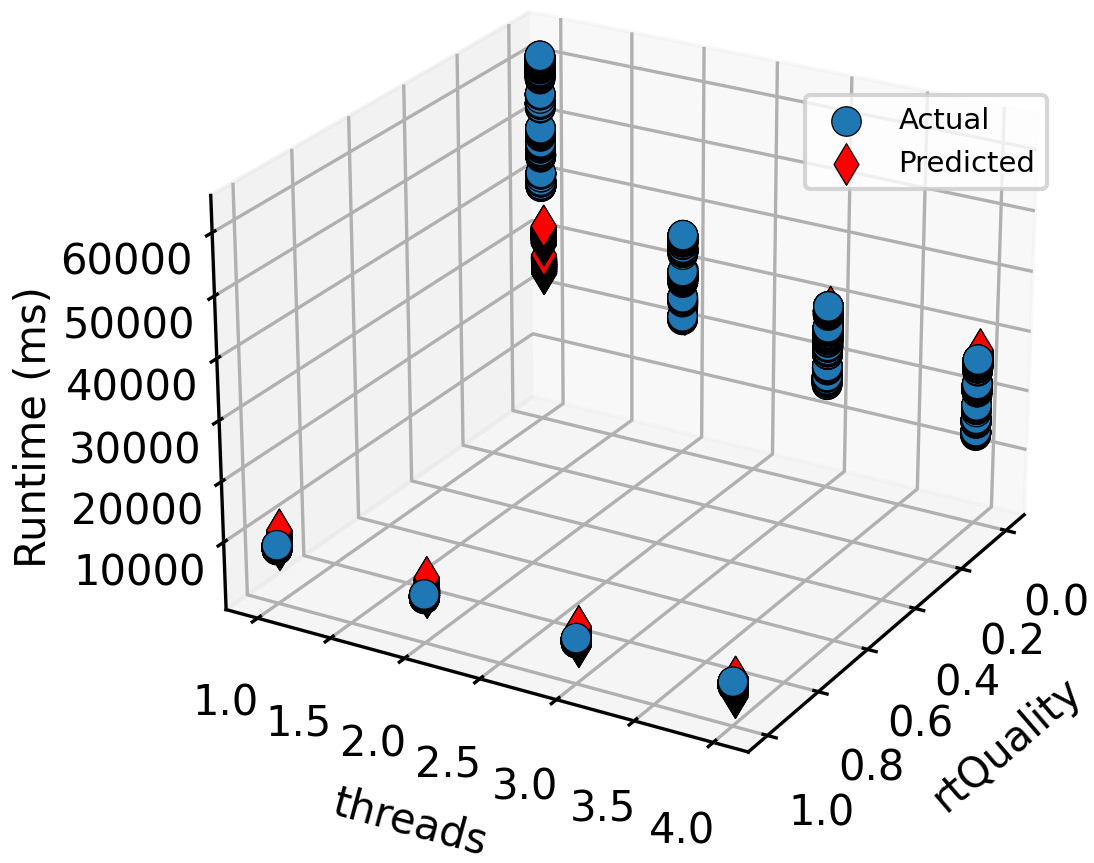} 
  \caption{\texttt{DeepPerf}}
\end{subfigure}
~\hfill
\begin{subfigure}{.48\columnwidth}
  \centering
  \includegraphics[width=\textwidth]{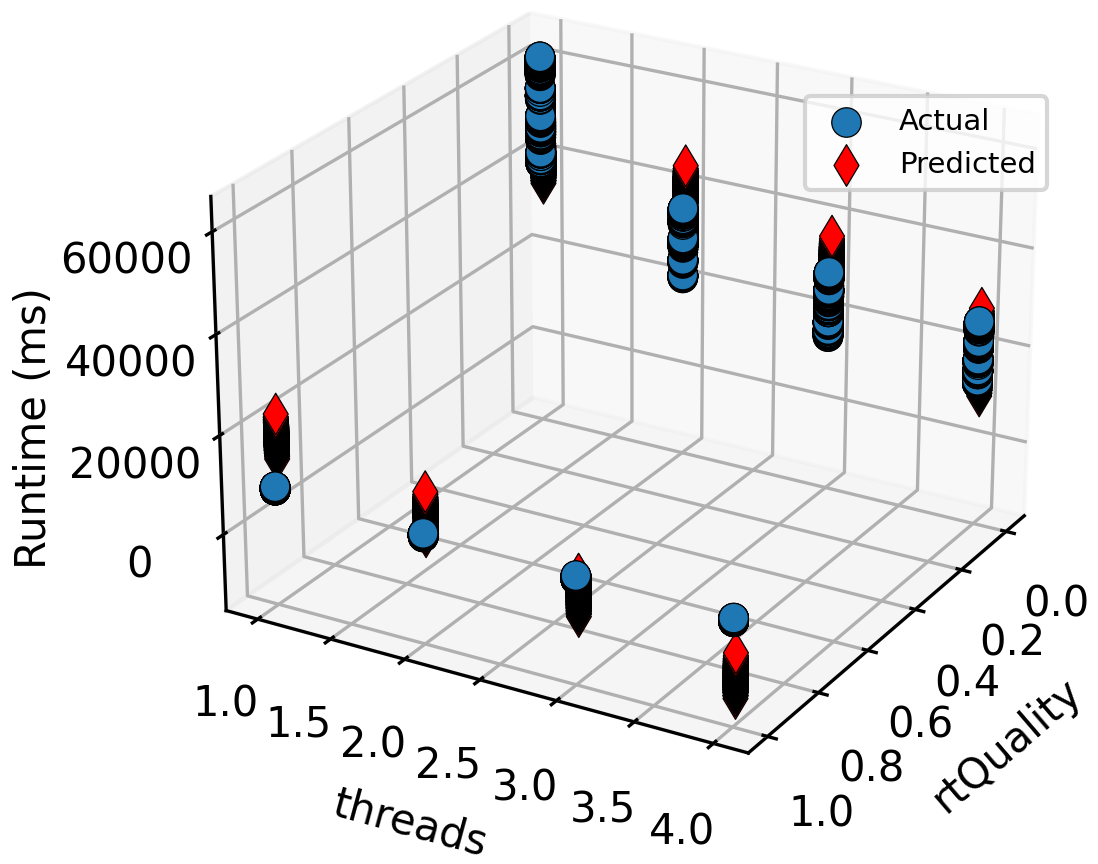} 
  \caption{\texttt{SPLConqueror}}
\end{subfigure}

\caption{Example run of the actual and predicted performance by \Model, \texttt{HINNPerf}, \texttt{DeepPerf}, and \texttt{SPLConqueror} for \textsc{VP8}.}
\label{fig:discussion_scatter}
\end{figure}

Here, a few insights and pointers observed from the experiment results are discussed, and then the answers to \textbf{RQ3} of this thesis, the limitations, the publications, and future plans are introduced.

\subsection{Why does \Model~Work?}

To provide a more detailed understanding of why \Model~performs better than the state-of-the-art approaches, Figure~\ref{fig:discussion_scatter} showcases the most common run of the predicted performance by \Model, \texttt{HINNPerf}, \texttt{DeepPerf} and \texttt{SPLConqueror} against the actual performance. Clearly, it can be noted that the sample sparsity is rather obvious where there are two distant divisions. 

Those approaches that rely on a single local model have been severely affected by such highly sparse samples: it is seen that the models try to cover points in both divisions, but fail to do so as they tend to overfit/memorize the points in one or the other. For example, on \texttt{DeepPerf} (Figure~\ref{fig:discussion_scatter}c), its predictions on some configurations, which should result in high runtime, tend to have much lower values (e.g., when \texttt{rtQuality=0} and \texttt{threads=1}) since it overfits the points with some drastically lower runtime, i.e., when \texttt{rtQuality=1}. Conversely, \texttt{HINNPerf} (Figure~\ref{fig:discussion_scatter}b) and \texttt{SPLConqueror} (Figure~\ref{fig:discussion_scatter}d) estimate high runtime on some configurations that should lead to excellent performance (e.g., \texttt{rtQuality=1} and \texttt{threads=1}), which is, again, due to the fact that it memorizes those points with much higher runtime (\texttt{rtQuality=0}). Further, it also creates additional noises that make the predictions too high against what they should be, e.g., when \texttt{rtQuality=0} and \texttt{threads=2}. \Model, in contrast, handles such a sample sparsity well as it contains different local models that particularly cater to each division identified, hence leading to high accuracy (Figure~\ref{fig:discussion_scatter}a).

\begin{figure}[!t]
  \centering
   \begin{subfigure}[t]{0.35\columnwidth}
        \centering
\includegraphics[width=\columnwidth]{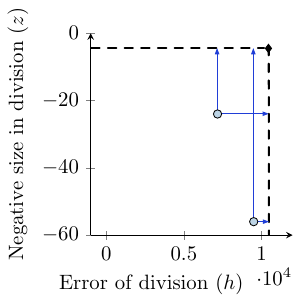}
    \subcaption{$d=1$ ($\mu$HV=$56411.36$)}
   \end{subfigure}
\hspace{0.2cm}
      \begin{subfigure}[t]{0.35\columnwidth}
        \centering
\includegraphics[width=\columnwidth]{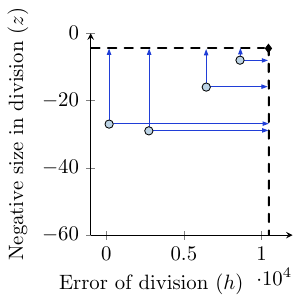}
 \subcaption{$d=2$ ($\mu$HV=$118013.45$)}
   \end{subfigure}

      \begin{subfigure}[t]{0.35\columnwidth}
        \centering
\includegraphics[width=\columnwidth]{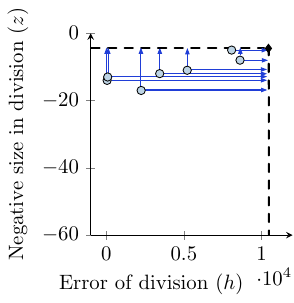}
  \subcaption{$d=3$ ($\mu$HV=$54755.69$)}
   \end{subfigure}
\hspace{0.2cm}
      \begin{subfigure}[t]{0.35\columnwidth}
        \centering
\includegraphics[width=\columnwidth]{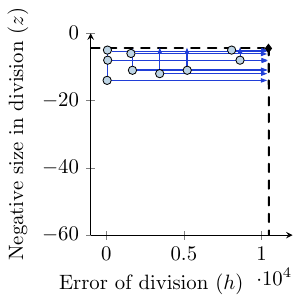}
     \subcaption{$d=4$ ($\mu$HV=$33827.60$)}
   \end{subfigure}

    \caption{The snapshot of the objective values of $h$ and $z$ on divisions when adapting $d$ for a run under system \textsc{x264}. }
       \label{fig:changing-d}
  \end{figure}

\subsection{Why Adapting $d$ Helps?}

It has been showcased that the proposed adaptive mechanism is effective in dynamically setting the suitable $d$ in \Model. In order to take a deeper look at why it helps, Figure~\ref{fig:changing-d} illustrates the process of selecting the $d$ value at a run under \textsc{x264}, where the black diamond denotes the reference point. The testing MRE for $d=1$, $d=2$, $d=3$, and $d=4$ are $0.9676$, $0.7876$, $1.7858$, and $3.1285$, respectively. Here, clear traces of how the sample size and mean square error in the divisions change along different $d$ values are observed:

\begin{itemize}
    \item The overall sample size of the divisions ($z$) decreases as the $d$ value increases since more divisions are created to divide the configuration data. This implies that there is less chance for a local model to learn and generalize.
    \item The overall mean square error of the divisions ($h$) tends to decrease with larger $d$, as the additionally created divisions contain some much-reduced errors. This means that the overall ability to handle sample sparsity is improved, which is expected because the larger $d$, the smaller the sample size. Hence, it is more likely to have more dense points in a division, leading to a lower mean square error.
\end{itemize}

The above observations match with the theoretical analysis in Section~\ref{subsubsec:adapting_depth}, and therefore adapting the $d$ value requires finding a good trade-off between $h$ and $z$. For the example in Figure~\ref{fig:changing-d}, the proposed $\mu$HV reflects the collective contributions of all possible divisions on both objectives, including those that are dominated, by calculating the average area that each $d$ value covers with respect to the reference point. In this way, it is identified that $d=2$ is the best and most balanced value, which is subsequently confirmed to be indeed the case when validating their MRE.

\subsection{Answer to RQ3}
In this chapter, a novel model named \Model~is designed and implemented to handle the sparsity problem, where a set of three key strengths can be identified through the sub-RQs.

The first strength of \Model~is that the concept of ``divide-and-learn'' under the paradigm of dividable learning, paired with an appropriately chosen local model, can handle both sample sparsity and feature sparsity well. As from Section~\ref{subsec:RQ3.1} for \textbf{RQ3.1}, this has led to better accuracy and better utilization of the sample data than the state-of-the-art approaches. 

The second strength is that while using the deep learning model \texttt{HINNPerf} has been shown to be the best option for accuracy in the experiments, one can also easily generalize it with other machine learning models, such as LR, for faster training and better interoperability, hence offering great flexibility. As such, \Model~permits software engineers to make trade-offs based on different concerns depending on the practical scenarios. In particular, as from Section~\ref{chap-meta-subsec:local} for \textbf{RQ3.2}, \Model~can additionally improve different local models compared with when they are used alone as a global model. Further, when compared with other model-agnostic frameworks from the paradigm of ensemble learning, \Model~also produces more effective and accurate predictions, as demonstrated at \textbf{RQ3.3}.

The third strength of \Model~is that it is now less parameter-dependent (apart from those of the local model) given the mechanism that adapts $d$. This, as from \textbf{RQ3.5}, is necessary, and the adaptation leads to a high hit rate with effective results. More importantly, adapting $d$ would have saved a great amount of effort to tailor and understand the approach together with its customized parameters, which is not uncommon for a model-agnostic framework~\citep{DBLP:conf/eurocolt/FreundS95}. Meanwhile, \textbf{RQ3.4} illustrates that each proposed component in \Model~can individually contribute to its superiority. 

Given these strengths of \Model, it can be concluded that \Model~satisfies \textbf{objective 3} of this thesis, which is to effectively handle the knowledge gap on sparsity in software performance learning, and thereby \textbf{RQ3} raised in Section~\ref{chap-intro:aim} can be fulfilled by: 

\begin{answerbox}
\emph{\textbf{Answer to RQ3:} the divide-and-learn framework that effectively handles both feature and sample sparsity is able to improve the performance of a number of machine learning models without any extra data, and the \Model~model (\texttt{HINNPerf} as the local learner) is able to outperform state-of-the-art performance modeling approaches.}
\end{answerbox}

\subsection{Limitations}
Three limitations can be found in the proposed artifact. Firstly, from \textbf{RQ3.6}, a limitation of \Model~is that it can incur slightly larger computational overhead: it could take a longer time to build the model than some state-of-the-art approaches. Using a typical server with a CPU of 2.6GHz and 256GB RAM, \Model~requires between 4 minutes and 56 minutes for systems with up to 33 configuration options and more than 2,000 samples. While this time effort might be relieved by implementing parallel training of the local models, the computational resources consumed are unavoidable. 

Secondly, although the only parameter $d$ of \Model~provides flexibility to adjust the number of divisions and the empirical experiments show that $d$=1 or 2 tends to be the best setting, it is not known whether there is a universal best $d$ for all the software systems. Thugs, the generalizability of \Model~might be restricted. Nonetheless, an adaptive depth algorithm is proposed in this chapter to effectively mitigate this issue, which selects the optimal $d$ based on a novel hypervolume metric $\mu$HV.

Additionally, the application scenario of this method is limited because it does not consider the impact of the software running environment. In other words, the performance model could be precise in the environment where the training data are collected, but when it comes to another hardware/network environment, the situation could vary significantly. Therefore, it lacks the ability to process dynamic-environment tasks. Hence, this thesis proposes to handle this limitation in the coming chapter (Chapter~\ref{chap:meta}).

\subsection{Publications and Future Work}
From this chapter, two papers have been derived. One paper is a conference paper that has been accepted by the \textit{ACM Joint European Software Engineering Conference and Symposium on the Foundations of Software Engineering (ESEC/FSE) 2023}~\citep{DBLP:conf/sigsoft/Gong023}, while the other is an extended journal version of the conference paper, which has been accepted by the \textit{IEEE Transactions on Software Engineering (TSE)} as a regular paper in October 2024. 

In addition, there are several directions for future studies to improve \Model. For example, in the current framework, one important assumption is that the local model for all the divisions of data is the same, while each division learns a distinct instance of the chosen deep learning model (\texttt{HINNPerf} in this chapter), as \texttt{HINNPerf} is generally the most accurate performance model in the experiments~\citep{DBLP:journals/tosem/ChengGZ23}. Indeed, this is effective in most of the cases, as it has been shown in Section~\ref{chap-dal:evaluation}. However, there are still situations where other machine learning approaches outperform \texttt{HINNPerf}. For instance, it can be seen in Table~\ref{tb:vsSOTA} that, for $S_1$ of \textsc{BDB-J} and $S_2$ of \textsc{BDB-C}, the rank and MRE of \texttt{DECART} are significantly better than the others, and correspondingly, in Table~\ref{chap-dal-tb:compare_models}, \texttt{DaL$_{CART}$}, which uses \texttt{CART} (the initial version of \texttt{DECART}) as the local model, is ranked the best. This is understandable as the sparsity and complexity of each software are different, resulting in different best local models. Therefore, there is no universal best local model for all the subject systems and all the training sample sizes, and it is crucial to combine different local models for \Model~in order to handle these situations. 

Yet, incorporating multiple local models poses a new challenge as it requires selecting the best local model, which can be resource-consuming, as the number of potential combinations is substantial due to the possible combinations of models for each division, training sample size, and software system. This challenge presents an opportunity for future studies to explore and develop efficient methods for selecting and combining different local models in \Model.

%% file: Chapter-DAL/threats_to_validity.tex
\section{Threats to Validity}
\label{chap-dal:threats}
{\textbf{Internal Threats.}} Internal threats to validity are related to the parameter values used. In this work, the same settings are used as in state-of-the-art approaches~\citep{DBLP:journals/sqj/SiegmundRKKAS12,DBLP:conf/kbse/GuoCASW13,DBLP:conf/icse/HaZ19,DBLP:conf/esem/ShuS0X20}. Automated and adaptive mechanisms are enabled for those approaches that come with parameter tuning capability, e.g., \texttt{DeepPerf} and \texttt{DBSCAN}, or the best settings are profiled in a trail-and-error manner before the experiments, e.g., for the $k$ value in $k$\texttt{Mean} clustering. If no parameter values are specified, the default settings are used. To mitigate stochastic error, the experiments are repeated for 30 runs and use the Scott-Knott test for multiple comparisons. Indeed, although this thesis has carefully ensured the fairness of comparisons, it is hard to guarantee that all compared models are in their best possible status.

{\textbf{Construct Threats.}} Threats to construct validity may lie in the metric used. In this study, MRE is chosen for two reasons: (1) it is a relative metric and hence is insensitive to the scale of the performance; (2) MRE has been recommended for performance prediction by many latest studies~\citep{DBLP:conf/icse/HaZ19, DBLP:conf/esem/ShuS0X20, DBLP:journals/ese/GuoYSASVCWY18}. To assess the cost of the modeling and training process, the time taken to build the models is measured. \Model~is also compared against state-of-the-art approaches, other ensemble learning models, and different variants thereof that are paired with different local models and come with alternative components. To confirm that adapting $d$ is indeed effective, the sensitivity of \Model~to $d$ value is verified while examining the adapted $d$ against the optimal $d$ on both the overall results and each individual run. However, programming errors or other small implementation defects are always possible.

{\textbf{External Threats.}} External validity could be raised from the subject systems and training samples used. To mitigate such, 12 commonly used subject software systems are evaluated, which are of diverse domains, scales, and performance metrics, selected from the latest studies~\citep{DBLP:journals/tse/KrishnaNJM21, DBLP:conf/icse/HaZ19,DBLP:journals/jss/CaoBWZLZ23,DBLP:conf/sigsoft/SiegmundGAK15,10172849}. Different training sample sizes have also been examined, which are determined by the number of configuration options and \texttt{SPLConqueror}~\citep{DBLP:journals/sqj/SiegmundRKKAS12}---a typical method. Nevertheless, it is agreed that using more subject systems and data sizes may be fruitful, especially for examining the adaptive mechanism for $d$.

%% file: Chapter-DAL/conclusion.tex
\section{Chapter Summary}
\label{chap-dal:conclusion}

To answer the third research question in the thesis, this chapter introduces a model-agnostic framework dubbed \Model~that effectively handles the issues of both feature and sample sparsity in configuration performance learning. The key novelty of \Model~is that it follows a new paradigm of dividable learning, in which the branches/leaves are extracted from a \texttt{CART} that divides the samples of configuration into a number of distant divisions, which is adaptively adjusted, and trains a dedicated local model for each division thereafter. Prediction of the new configuration is then made by the local model of division inferred based on a Random Forest classifier. Such a theory of ``divide-and-learn'' handles the sample sparsity while the local model used (e.g., one with regularization) deals with the feature sparsity, hence collectively addressing the overall sparsity issue in configuration data. 

By means of extensive experiments, \Model~is comprehensively evaluated on 12 real-world systems that are of diverse domains and scales, together with five sets of training data. The results show that \Model~is:

\begin{itemize}

    \item \textbf{effective} as it is competitive to the best state-of-the-art approach on 44 out of 60 cases, in which 31 of them are significantly better with up to $1.61$ times MRE improvement; the paradigm of dividable learning is also more suitable than the classic ensemble learning to handle the sample sparsity in configuration data;
    \item \textbf{efficient} since it often requires fewer samples to reach the same/better accuracy compared with the state-of-the-art approaches; the adaptation of $d$ also leads to negligible overhead as no additional training is required;
    \item \textbf{flexible} given that it considerably improves various global models when they are used as the local model therein; 
    \item \textbf{robust} because the mechanism that adapted the parameter $d$ can reach the optimal value for 76.43\% of the individual runs while a similarly promising $d$ value can be chosen even when the optimal $d$ is missed.

\end{itemize}

With the above, the chapter answers \textbf{RQ3} of this thesis by demonstrating that the proposed framework can indeed improve the performance of a flexible number of machine learning models, where in most cases, \Model~combined with \texttt{HINNPerf} outperforms the best performance prediction models.

In future work, this thesis seeks to integrate different machine learning approaches in the training phase of \Model to further strengthen its prediction accuracy. Specifically, the dedicated local model for each division can be adaptively selected based on specific characteristics of the samples in that division.

While \Model~addresses the knowledge gap regarding sparsity (discussed in Section~\ref{chap-review:discussion}), it currently operates within a fixed environment. In other words, it doesn't address the knowledge gap in predicting performance across environments. To bridge this gap, the thesis introduces another framework, presented in Chapter~\ref{chap:meta}.

%% file: Chapter-meta/chapter-meta.tex
\chapter{Improving Performance Prediction in Multiple Environments with Sequential Meta-Learning}
\label{chap:meta}

Although the \Model~framework proposed in Chapter~\ref{chap:dal} achieves solving the sparsity problem and improves the prediction accuracy, the prediction is under a fixed environment, i.e., the model can not guarantee the prediction accuracy when the environment varies, which is one of the key knowledge gaps identified in this thesis, as justified in the Section~\ref{chap-review:discussion} in Chapter~\ref{chap:review}. To address this challenge, a promising solution is to leverage knowledge from different environments. Hence, the fourth research question in this thesis is asked:

\begin{answerbox}
\emph{\textbf{RQ4:} Is it possible to improve the prediction accuracy under the target environment by utilizing data from additional environments?}
\end{answerbox}

To answer this RQ, various methods have been implemented and evaluated during my research, like transfer learning (reusing knowledge from a source environment), multi-task learning (learning and predicting multiple tasks simultaneously using a single model), and meta-learning (learning a meta-model using meta-data). Finally, a sequential meta-learning model \MetaModel is developed in the thesis, which adopts a regularized DNN as the base model, with a fast sequence selection algorithm to learn the meta tasks in an optimal order. By experiments, \MetaModel~effectively address \textbf{RQ4} by outperforming the state-of-the-art single environment performance models and other multi-environment approaches like meta-learning, transfer learning, and multi-task learning models. Note that due to the regulations on anonymous review, the code and data for this study are omitted in this thesis.

In this chapter, the backgrounds, details of the structure, evaluations, and discussions of the findings in \MetaModel~are justified.

\input{Chapter-meta/introduction}

\input{Chapter-meta/background}
\input{Chapter-meta/premises}
\input{Chapter-meta/framework}

\input{Chapter-meta/setup}
\input{Chapter-meta/evaluation}
\input{Chapter-meta/discussion}

\input{Chapter-meta/threats_to_validity}
\input{Chapter-meta/conclusion}

%% file: Chapter-meta/introduction.tex
\section{Introduction}
\label{chap-meta:introduction}

Through learning on the available data, current work has successfully leveraged deep learning to build various performance models~\citep{DBLP:journals/tosem/ChengGZ23}. Yet, those approaches mostly focus on configuration performance learning under one environment~\citep{DBLP:conf/icse/SiegmundKKABRS12,DBLP:journals/ese/GuoYSASVCWY18,DBLP:conf/icse/HaZ19,DBLP:conf/splc/ValovGC15,DBLP:conf/oopsla/QueirozBC16,DBLP:conf/icse/0003XC021,DBLP:journals/sqj/SiegmundRKKAS12}, e.g., a pre-defined workload, a fixed hardware, and a specific version.

Working on a single environment is an over-optimistic assumption, as it is not uncommon to see configurable software systems run under diverse conditions. For example, a database system may experience both read-heavy and write-heavy workload~\citep{DBLP:conf/kbse/JamshidiSVKPA17}. Similarly, the hardware between the testing and production infrastructure might be drastically different~\citep{DBLP:journals/toit/LeitnerC16,DBLP:journals/corr/BrunnertHWDHHHJ15}, especially during the modern DevOps era. The ignorance of multiple environments would inevitably harm the effectiveness of a performance model.~\citet{DBLP:conf/kbse/JamshidiSVKPA17} reveal that the accuracy of a single environment model can be severely degraded when used in a different environment. Furthermore, due to the expensive measurement of configuration, e.g., it can take hours or days to measure only a few configurations~\citep{DBLP:conf/wosp/ValovPGFC17,DBLP:conf/sigsoft/0001L21}, building a new model for every distinct environment is unrealistic. Recently, at ICSE'23,~\citet{muhlbaueranalyzing} have demonstrated that predicting under multiple environments can pose significant threats to the robustness and generalizability of performance models learned using a single environment. Through a large-scale study, they concluded that:

\begin{quote}
\textit{``Performance models based on a single workload are useless, unless the configuration options’ sensitivity to workloads is accounted for.''}
\end{quote}

However, given the importance of the multi-environment, the problem has received inadequate attention from the researchers, and it still remains one of the greatest challenges for performance learning.

\subsection{Knowledge Gap}

As demonstrated in Chapter~\ref{chap:review}, a majority of studies only assume a static environment when learning the performance model for configurable software. The failure to take multiple environments into account not only degrades accuracy but also incurs an extra overhead of model re-building, as the valuable data samples measured under different environments are wasted. This leads to a previously unaddressed problem: \textit{How to effectively leverage configurations measured in different environments for modeling configuration performance?} Yet, learning a performance model under multiple environments for configurable software is challenging due to the large variations between data measured in distinct environments. For example, several studies have revealed that varying environments can cause substantial changes in the performance distributions with non-monotonic correlations, including workloads~\citep{muhlbaueranalyzing,DBLP:conf/wcre/Chen22,DBLP:conf/wosp/PereiraA0J20}, versions~\citep{DBLP:journals/tse/MartinAPLJK22} and hardware~\citep{DBLP:conf/kbse/JamshidiSVKPA17}.

Despite being uncommon,~\citet{muhlbaueranalyzing} summarized two major categories from existing work on how multiple environments have been handled, each with its own limitations:

\begin{enumerate}
    \item \textbf{Environment as additional features~\citep{DBLP:journals/tse/ChenB17,DBLP:conf/kbse/KocMWFP21,DBLP:conf/europar/LengauerABGHKRTGKKRS14,DBLP:conf/icse/Chen19b}:} Here, the specific properties of an environment, e.g., size and job counts, are considered as model features alongside the configuration options in performance model learning. However, in this category, not only the additional measurements can be costly, but the extra dimension(s) also make the true concept more difficult to learn and generalize.

    \item \textbf{Transfer learning~\citep{ DBLP:conf/icse/JamshidiVKSK17,DBLP:journals/tse/KrishnaNJM21,DBLP:journals/tse/MartinAPLJK22,DBLP:conf/wosp/ValovPGFC17}:} Given an existing performance model trained in the source environment(s), its differences to the new environment can be learned via transfer learning models. Yet, the key shortcoming thereof is that the loss function in transfer learning is tailored to a specific target environment, hence lacking generalizability to arbitrary environments.
\end{enumerate}

The above motivates this thesis to explore the concept of meta-learning, which tries to learn the way of learning. To be specific, a meta-model is trained using several meta-tasks, which learns the best parameters to quickly fit new tasks, without any prior knowledge of the new task. 

As far as can be seen by this thesis, very few research studies have been done on meta-learning concerning performance prediction. Although it has been proved that performance tasks on different running environments are related, it is still not known whether this relationship is useful for meta-learning. Therefore, there exists a big knowledge gap that is worth examining.

\subsection{Contributions}
To address the above gap, this thesis proposes to use the concept of meta-learning---a form of machine learning that is capable of ``learning to learn'' by co-learning data across multiple environments (or tasks\footnote{In this chapter, the terms 'task' and 'environment' are used interchangeably.}), and generalize the learning to an unforeseen one. From this, the framework presented is \underline{\textbf{Se}}quential \underline{\textbf{M}}eta \underline{\textbf{P}}erformance \underline{\textbf{L}}earning (\MetaModel), which produces a general meta-model learned from data of all existing meta environments during pre-training. When needed, such a meta-model can then be quickly adapted for accurately predicting performance under a target environment via fine-tuning. What makes \MetaModel~unique is that, unlike popular general-purpose meta-learning frameworks such as \texttt{MAML}~\citep{DBLP:conf/icml/FinnAL17} and \texttt{MetaSGD}~\citep{DBLP:journals/corr/LiZCL17} which learn the data of meta environments in parallel, \MetaModel~follows sequential meta-learning that learns the datasets of meta environments one at a time in a specific order, aiming to discriminate their contributions in the meta-model built. This is a tailored design to deal with the potentially large variations between measurements in different environments from the configuration data. Moreover, it is worth noting that the default base-learner for \MetaModel~is a regularized deep neural network, yet one can easily adapt the framework to other machine learning models like \texttt{LR}.

In a nutshell, the key contributions of this chapter are:

\begin{itemize}
    \item The justification for the suitability of the concept of sequential meta-learning in \MetaModel~for learning multi-environment configuration data compared to general frameworks like \texttt{MAML} and \texttt{MetaSGD} is presented both analytically and empirically in this thesis. This assessment is based on three unique properties in \MetaModel:

    \begin{enumerate}
        \item the sequence matters;
        \item train later contributes more;
        \item and using more meta environments are beneficial.
    \end{enumerate}

    \item Drawing on those properties provided by \MetaModel, a novel method that selects the optimal sequence of learning the data of meta environments is designed, which fits the characteristics of the configuration data under multiple environments.

    \item Based on nine systems with 3-10 meta environments and five training data sizes each, \MetaModel~with DNN as the base learner is experimentally compared against 15 state-of-the-art models for single or multiple environments, taken from the software, system, and machine learning communities.
\end{itemize}

The results are encouraging: \MetaModel~significantly outperforms existing models in $89\%$ of the systems with up to $99\%$ accuracy improvement; it is also data-efficient with at most $3.86\times$ speedup.





\subsection{Chapter Outline}
This chapter is presented as the following: Section~\ref{chap-meta:background} introduces the problem of dynamic running environments in software performance learning and the formal notion of meta-learning. Section~\ref{chap-meta:premises} examines the preliminary theory of this study, and Section~\ref{chap-meta:framework} illustrates the algorithm and details of the \MetaModel~framework. Section~\ref{chap-meta:setup} presents the sub-RQs in this chapter and the evaluation design, followed by the experiment results in Section~\ref{chap-meta:evaluation}. The answer to \textbf{RQ4} of this thesis, limitations, and future plans are discussed in Section~\ref{chap-meta:discussion}. Finally, Section~\ref{chap-meta:threats} presents the threats to validity, and Section~\ref{chap-meta:conclusion} summarises the chapter.

%% file: Chapter-meta/background.tex
\section{Background}
\label{chap-meta:background}

To give a more comprehensive understanding of this chapter, this section introduces the detailed background of the single-task performance learning problem and then gives a formal definition of the most widely applied multi-environment learning approaches.

\subsection{Single Environment Configuration Performance Learning}
\label{sec:bg-single}
Learning performance for configurable software systems is commonly formulated as single-task learning that learns and generalizes data under a single environment. Formally, the aim is to build a regression model $f$ that predicts the performance $p$ of an unforeseen configuration $\mathbf{\overline{x}'}$:
\begin{equation}
\begin{aligned}
 &\text{Train: } \mathbfcal{E}_{target} \Longrightarrow f\\
 &\text{Predict: } f(\mathbf{\overline{x}'})=p_{target} \text{ | } \mathbf{\overline{x}'}\in \mathbfcal{E}_{target}
\end{aligned}
    \label{chap-meta-eq:prob}
\end{equation}
whereby $\mathbfcal{E}_{target}$ denotes the training samples of configuration-performance pairs ($\{\mathcal{C} \rightarrow \mathcal{P}\}$) measured under the target environment, such that $\mathbf{\overline{x}} \in \mathbfcal{E}_{target}=\{\mathcal{C} \rightarrow \mathcal{P}\}$. $\mathbf{\overline{x}}$ is a measured configuration and $\mathbf{\overline{x}}=(x_{1},x_{2},\cdots,x_{n})$, where there are $n$ configuration options and each option $x_{i}$ can be either binary or categorical/numerical. $f$ should be trained in such a way that its prediction ($p_{target}$) on $\mathbf{\overline{x}'}$ is as close to the actual performance as possible. Many models for learning configuration performance in a single environment have been proposed~\citep{DBLP:conf/icse/SiegmundKKABRS12,DBLP:journals/ese/GuoYSASVCWY18,DBLP:conf/icse/HaZ19,DBLP:conf/splc/ValovGC15,DBLP:conf/oopsla/QueirozBC16,DBLP:conf/icse/0003XC021,DBLP:journals/sqj/SiegmundRKKAS12}. Some state-of-the-art models, which are also compared in this chapter (Section~\ref{chap-meta:evaluation}) are explained here:

\begin{itemize}
    \item \textbf{\texttt{DeepPerf}}~\citep{DBLP:conf/icse/HaZ19}: a Deep Neural Network (DNN) model with $L_1$ regularization and fast hyperparameter tuning for sparse performance learning. 
    \item \textbf{\texttt{DECART}}~\citep{DBLP:journals/ese/GuoYSASVCWY18}: an improved regression tree model~\citep{breiman2017classification} with a specialized sampling mechanism.
    \item \textbf{\texttt{Random Forest} (\texttt{RF})}~\citep{DBLP:conf/splc/ValovGC15,DBLP:conf/oopsla/QueirozBC16,DBLP:conf/icse/0003XC021}: an ensemble of trees that tackle the feature sparsity issue.
    \item \textbf{\texttt{SPLConqueror}}~\citep{DBLP:journals/sqj/SiegmundRKKAS12}: a baseline model that relies on linear regression.
\end{itemize}

\subsection{Configuration Performance Learning with Multiple Environment Inputs}
\label{sec:bg-env}

To handle multiple environments, a natural way is to combine the environmental features with configuration options in the model. In this way, Equation (1) remains unchanged, but the configuration-performance pairs from all available environments are merged, and the environment features serve as additional inputs, denoted as $\mathcal{E}$, i.e., $\mathbfcal{E}_{target}=\{\mathcal{C} \times \mathcal{E} \rightarrow \mathcal{P}\}$. For example,~\citet{DBLP:conf/icse/Chen19b} has followed this strategy to examine various types of underlying models. The environmental features therein are the workload and request rate, \textit{etc}. More domain-specific environmental features also exist, such as those for program verification~\citep{DBLP:conf/kbse/KocMWFP21}, high-performance computing~\citep{DBLP:conf/europar/LengauerABGHKRTGKKRS14}, and cloud computing~\citep{DBLP:journals/tse/ChenB17}.

However, a major limitation thereof is that the combined input space of options and environmental features makes the training even more difficult, as not all options are environment-sensitive~\citep{LESOIL2023111671}.

\subsection{Joint Learning for Configuration Performance}
\label{sec:bg-transfer}

\subsubsection{Transfer Configuration Performance Learning}

Building performance models under multiple environments can also be handled by transfer learning, in which Equation (1) is changed to:
\begin{equation}
\begin{aligned}
    &\text{Train: } \mathbfcal{E}_{1}\cup\mathbfcal{E}_{2}...\cup\mathbfcal{E}_{m}\cup\mathbfcal{E}_{target} \Longrightarrow f\\
    &\text{Predict: } f(\mathbf{\overline{x}'})=p_{target} \text{ | } \mathbf{\overline{x}'}\in \mathbfcal{E}_{target}
\end{aligned}
    \label{eq:prob1}
\end{equation}
Here, by treating the learning of data under different environments as independent tasks, both the data from source environment(s) ($\mathbfcal{E}_{1},...,\mathbfcal{E}_{m}$) and target environment ($\mathbfcal{E}_{target}$) are jointly learned by a base learner with information sharing on, e.g., model parameters/data samples~\citep{ DBLP:conf/icse/JamshidiVKSK17,DBLP:journals/tse/KrishnaNJM21}, features~\citep{DBLP:journals/tse/MartinAPLJK22}, or prediction outputs~\citep{DBLP:conf/wosp/ValovPGFC17}. Among others, \textbf{\texttt{BEETLE}}~\citep{DBLP:journals/tse/KrishnaNJM21} is a transfer learning model for configuration performance learning. The key idea is to evaluate all the source-target pairs over the candidate environments and identify the best source as the ``bellwether'' to use. The data of the source and target environment are jointly learned by a regression tree model. \textbf{\texttt{tEAMS}}~\citep{DBLP:journals/tse/MartinAPLJK22} is a recent approach that reuses the performance model in software evolution. With feature alignment, the model trained on the data of an older version serves as the basis for learning the data under the new version.

Yet, the key limitation of transfer learning is that, regardless of how many environments serve as the source data, the loss function therein needs to be specific to the target environment, making it difficult to generalize to the other environments~\citep{torrey2010transfer}.

\subsubsection{Multi-Task Configuration Performance Learning}
To cope with the above limitation, a relevant paradigm is multi-task learning, which can also learn data from multiple environments as the sources ($\mathbfcal{E}_{1},...,\mathbfcal{E}_{m}$)~\citep{DBLP:conf/iclr/YangH17,DBLP:conf/eurosys/AlabedY21,DBLP:conf/socpros/MadhumathiS18} and has the ability to generalize the predictions for all environments learned simultaneously. Here, Equation (2) will be changed to:
\begin{equation}
\begin{aligned}
    &\text{Train: } \mathbfcal{E}_{1}\cup\mathbfcal{E}_{2}...\cup\mathbfcal{E}_{m}\cup\mathbfcal{E}_{target} \Longrightarrow f\\
    &\text{Predict: } f(\mathbf{\overline{x}'})=\{p_1,p_2,...,p_m,p_{target}\} \text{ | } \mathbf{\overline{x}'}\in \mathbfcal{E}_{target}
\end{aligned}
\label{eq:prob2}
\end{equation}
whereby $p_m$ is the predicted performance of $\mathbf{\overline{x}'}$ under the $m$th environment. There is no clear distinction between source and target environments, as the key in multi-task learning is to share information on data from whatever available environments. For example, Madhumathi and Suresh~\citep{DBLP:conf/socpros/MadhumathiS18} propose \textbf{\texttt{Multi-Output Random Forest} (\texttt{MORF})}, a multi-task learning version of Random Forest where there is one dedicated output for each environment of performance prediction.

However, the training still requires foreseeing the target environment and can suffer ``negative transfer'', as task gradients
may interfere, overcomplicating the loss function landscape~\citep{DBLP:conf/icml/StandleyZCGMS20}.

\subsection{Meta-Learning for Configuration Performance}
\label{sec:bg-meta}

Unlike transfer and multi-task learning, meta-learning seeks to learn a meta-model based on some readily available data from the known environments (a.k.a meta environments) without knowing what will be the target environment~\citep{DBLP:journals/air/VilaltaD02}. Formally, in meta-learning, the process becomes:
\begin{equation}
\begin{aligned}
    &\text{Train: } \mathbfcal{E}_{1}\cup\mathbfcal{E}_{2}...\cup\mathbfcal{E}_{m} \Longrightarrow f' \text{; } \mathbfcal{E}_{target}+f' \Longrightarrow f\\
   &\text{Predict: } f(\mathbf{\overline{x}'})=p_{target} \text{ | } \mathbf{\overline{x}'}\in \mathbfcal{E}_{target}
\end{aligned}
\label{eq:prob3}
\end{equation}
The key idea is that through learning the data of meta environments using a base learner at the meta-level, the meta-model $f'$ will obtain the ability of ``learning to learn'', enabling easier fine-tuning on the new target environment ($\mathbfcal{E}_{target}$) when it becomes available. As such, meta-learning can achieve: (1) fast adaptation; and (2) better generalization and hence a more accurate model.

Although rarely used for configuration performance learning, there exist different models for meta-learning in the machine learning community~\citep{DBLP:conf/icml/FinnAL17,DBLP:journals/corr/LiZCL17,DBLP:conf/icml/YaoWHL19,DBLP:conf/nips/ChuaLL21}, in which the meta tasks are learned simultaneously and the outcomes are combined. The most noticeable ones are \texttt{MAML}~\citep{DBLP:conf/icml/FinnAL17} and its variant \texttt{MetaSGD}~\citep{DBLP:journals/corr/LiZCL17}: \textbf{\texttt{MAML}}~\citep{DBLP:conf/icml/FinnAL17} is a state-of-the-art meta-learning framework that has been widely applied in different domains~\citep{DBLP:conf/aistats/ChenC22}, including software engineering~\citep{DBLP:conf/icse/ChaiZSG22}. \texttt{MAML}, if used for configuration performance learning, learns a set of good model parameter values on data of the meta environments in parallel, building a meta-model ($f'$), which can then be adapted to build a fine-tuned model $f$ for the target environment. In contrast, \textbf{\texttt{MetaSGD}}~\citep{DBLP:journals/corr/LiZCL17} extends the \texttt{MAML} by additionally adapting the learning rate along the meta-training, expediting the learning over \texttt{MAML}.

Nevertheless, Section~\ref{chap-meta:premises} will explain why general models like \texttt{MAML} are ill-suited to configuration performance learning and elaborate on the theory behind the proposed \Model~framework.

%% file: Chapter-meta/premises.tex
\section{The Theory behind \MetaModel}
\label{chap-meta:premises}
This section illustrates the empirical experiments and the identified premises of the proposed meta-learning model.

\begin{figure}[!t]
\centering
    \begin{subfigure}[t]{0.5\columnwidth}
    \includegraphics[width=\columnwidth]{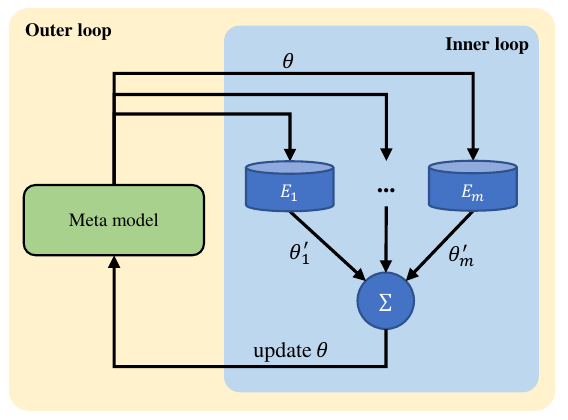}
    \subcaption{\texttt{MAML}}
    \label{subfig:maml model}
    \end{subfigure}
~\hspace{-0.4cm}
    \begin{subfigure}[t]{0.5\columnwidth}
    \includegraphics[width=\columnwidth]{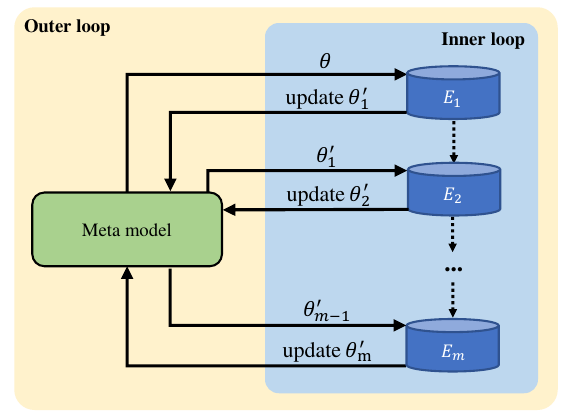}
    \subcaption{\MetaModel}
    \label{subfig:sequential model}
    \end{subfigure}
\caption{Workflow of \texttt{MAML} and the proposed \MetaModel. The meta-model can be produced by any base learner.}
\label{fig:models}
\end{figure}

In essence, the key to the success of \texttt{MAML} (and its variants) is that it can produce a set of good model parameter values for the target environment to start training, paired with any gradient descent base learner, which enables fast adaptation of the created meta-model. As in Figure~\ref{subfig:maml model}, this is achieved via iteratively updating the parameters of the meta-model ($\theta$ for the outer loop\footnote{The default iteration limit of the outer loop in \texttt{MAML} is 1.}) using those trained on the data of each meta environment individually (i.e., $\theta'_1,...,\theta'_m$ from the inner loop). In such a process, data of all meta environments are learned simultaneously, followed by a linear aggregation of their model parameter values, hence they provide equal contributions to the meta-model. This, however, is ill-suited for the performance learning of configurable software systems. Because, unlike some other domains, prior studies have revealed that the data from distinct environments exhibit drastically different correlations between performance and configuration:

\begin{itemize}

\item A study on \textsc{x264}~\citep{DBLP:conf/wosp/PereiraA0J20} shows that different video inputs can lead to varying and non-monotonic performance results.

\item \citet{DBLP:conf/wcre/Chen22} found that altered workload causes dramatic shifts in the configuration landscapes.

\item \citet{muhlbaueranalyzing} reported that varying workloads can change how configuration options affect performance, causing drastic variations among performance distributions.

\item \citet{DBLP:conf/kbse/JamshidiSVKPA17} revealed the non-linear correlation on data from different versions/hardware.
\end{itemize}

As a result, these prior observations derive the following insight: 


\begin{quotebox}
   \noindent
   \textit{\textbf{Insight of Learning Configuration Data:}} The ideal parameter values and distributions for models learned under different environments can be, unfortunately, rather different. 
\end{quotebox}

This means that treating all meta environments equally in \texttt{MAML} can force some less useful (or even misleading) meta environments to contribute, causing some model parameter values to largely deviate from the optima required for the target environment. Therefore, what is needed is a tailored model for multi-environment configuration performance learning with the following requirements:

\begin{itemize}
    \item \textbf{Requirement 1:} Configuration data of distinct meta environments should be discriminated.
    \item \textbf{Requirement 2:} Configuration data of distinct meta environments should have alterable contributions to the learning.
    \item \textbf{Requirement 3:} The valuable information of data from different meta environments should be exploited fully since measuring configurations is highly expensive~\citep{DBLP:conf/sigsoft/0001L21,DBLP:journals/tse/Nair0MSA20}.
\end{itemize}


A naive (perhaps natural) solution to the above is to consider weights between meta environments when updating the $\theta$ in \texttt{MAML}. However, this has the following shortcomings:

\begin{itemize}
\item It is difficult to precisely quantify the relative contributions between meta environments, as the model parameter values are naturally multi-dimensional. 


\item Setting/updating the weights is challenging, especially with more meta environments. 

\item It has been shown that the weights can obscure the optimization during the training~\citep{10.1145/3514233}.
\end{itemize}

Instead of weights, this thesis proposes sequential meta-learning in \MetaModel: as in Figure~\ref{subfig:sequential model}, the datasets in distinct meta environments are learned one by one in a certain order, and so does the update of meta-model's parameters $\theta$. The sequence of meta environments prioritizes their contributions to learning, thereby resolving the limitation of \texttt{MAML} for configuration performance learning without extra parameters. In what follows, the theory behind \MetaModel~and its properties are elaborated.

\subsection{The Sequence Matters}

The very first property that is intentionally designed for \MetaModel, which meets \textbf{Requirement 1}, is:

\begin{property}
\label{pro:1}
The sequence of learning the data in distinct meta environments significantly influences the meta-model built, hence leading to different adaptation effects for the target environment.
\end{property}


Given that the datasets in meta environments are learned in sequence with only one being considered each time, the model parameter values for a meta environment trained earlier would serve as the initial values of the ones that will be learned later. As such, different sequences would cause distinct orders of intermediate initialization on the model parameter values for learning some meta environments. Since the heuristic of gradient descent in training is often stochastic and the training sample size can be limited for configurable software systems, working on different sets of starting points would most likely have different results, leading to diverse meta-models in the end.

\begin{figure}[!t]
  \centering
   \begin{subfigure}[t]{\columnwidth}
        \centering
\includegraphics[width=\textwidth]{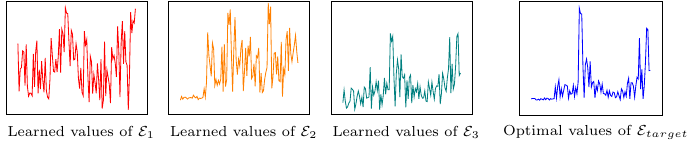}
        \subcaption{Configuration performance learning under independent single environment}
   \end{subfigure}

      \begin{subfigure}[t]{\columnwidth}
        \centering
\includegraphics[width=\textwidth]{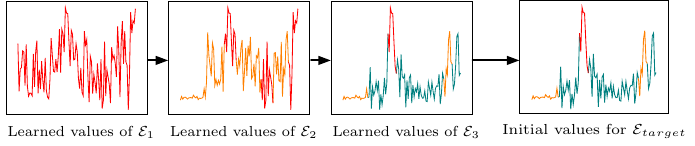}
      \subcaption{Multi-environment configuration performance learning with \MetaModel}
   \end{subfigure} 
 
    \begin{subfigure}[t]{\columnwidth}
        \centering
\includegraphics[width=\textwidth]{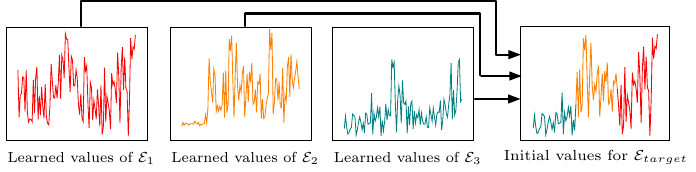}
      \subcaption{Multi-environment configuration performance learning with \texttt{MAML}}
   \end{subfigure}  
    
 \caption{Illustrating the distributions of the model parameter values in different situations under a real-world software system.}
      \label{fig:theory}
  \end{figure}

For example, Figure~\ref{fig:theory} shows an illustration of the distribution of model parameter values in different real-world software system scenarios, where the base learner is a regularized Deep Neural Network (it is best viewed in color). The x- and y-axis are model parameters and their values, respectively. Particularly, Figure~\ref{fig:theory}a illustrates the distributions of model parameter values when trained individually on data of three single environments, namely $\mathbfcal{E}_1$, $\mathbfcal{E}_2$, and $\mathbfcal{E}_3$. With \MetaModel~in this case (Figure~\ref{fig:theory}b), the model parameter values trained from $\mathbfcal{E}_1$ becomes the initial values for learning the data of $\mathbfcal{E}_2$, after which the values become the initials for learning the $\mathbfcal{E}_3$ data. Here, the resulted model parameter values when learning the data of $\mathbfcal{E}_2$, whose initialization is set by learning the data of $\mathbfcal{E}_1$, would lead to a distribution mixed from those of learning the data of each meta environment individually (Figure~\ref{fig:theory}a). As such, if $\mathbfcal{E}_1$ is swapped with $\mathbfcal{E}_3$, the resulting distribution of model parameter values would be different. \texttt{MAML}, in contrast, is insensitive to the sequence of meta environments as all of them are learned simultaneously, and the learned model parameter values are aggregated to update the meta-model, as in Figure~\ref{fig:theory}c. Figure~\ref{fig:ranking_MRE}a shows the empirical results for a randomly selected system and similar outcomes are observed for the others, where the y-axis is the testing Mean Relative Error (MRE) on $\mathbfcal{E}_{target}$; $\mathbfcal{E}_{3}$ is the most useful environments for $\mathbfcal{E}_{target}$, following by $\mathbfcal{E}_{2}$ and then $\mathbfcal{E}_{1}$. Clearly, the different sequences in the meta-learning of \MetaModel~can lead to distinct accuracy for the target environment.

\subsection{Train Later Contributes More}

To satisfy \textbf{Requirement 2}, a related property that can be derived from \textbf{Property 1} in \MetaModel~is:

\begin{property}
\label{pro:2}
If the data of meta environment $\mathbfcal{E}_{i+1}$ is learned later than that of $\mathbfcal{E}_i$, then the model parameter values of meta-model are more similar to those of learning $\mathbfcal{E}_{i+1}$ data alone than those trained under $\mathbfcal{E}_i$. Therefore, data of the more useful meta environment should be learned later.
\end{property}

With the sequential training of meta environments data in \MetaModel, it is easy to see that, if there are $n$ meta environments ($n>0$), the model parameter values learned on $\mathbfcal{E}_i$ will experience $n-i$ ($i \in [1,n]$) updates later on. Since in each update, the training tends to tune the model parameter to fit the data in the corresponding meta environment, which can be rather different to $\mathbfcal{E}_i$ due to the characteristics of configuration landscape, the distribution of model parameter values learned from $\mathbfcal{E}_i$ will be gradually overridden with more subsequent updates. As a result, the bigger the $i$, i.e., a meta environment sits at a later position of the sequence, the fewer updates to the parameters trained under $\mathbfcal{E}_i$, allowing it to preserve more model parameter values (and distribution) in the meta-model and contribute more therein than the meta environments learned earlier. Importantly, this leads to an obvious rule: the more useful meta environments---those that can provide better initial model parameter values for the target environment---should be left to later positions (the method to measure the usefulness is explained in Section~\ref{chap-meta:framework}).

\begin{figure}[!t]
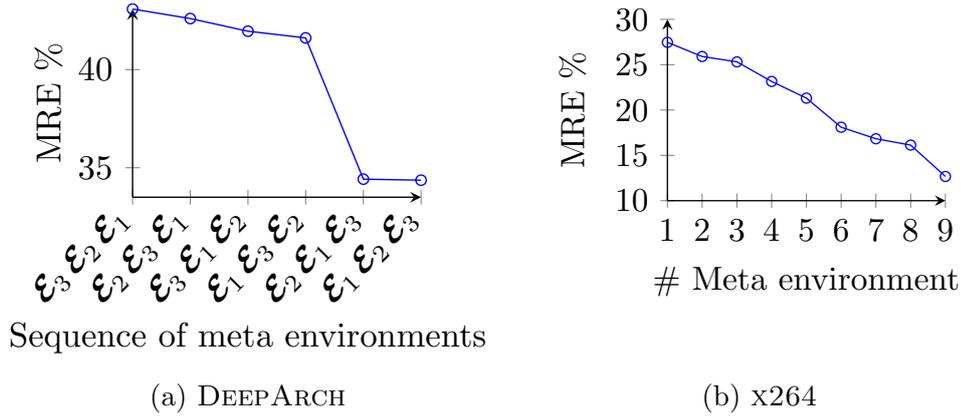

\centering
\footnotesize
 \begin{subfigure}{.45\columnwidth}
  \centering
\includestandalone[width=\columnwidth]{Figures/chap-meta/ranking_MRE} 
  \caption{\textsc{DeepArch}}
  \label{fig:sequence}
\end{subfigure}
~\hspace{0.3cm}
\begin{subfigure}{.38\columnwidth}
  \centering
\includestandalone[width=\columnwidth]{Figures/chap-meta/ranking_Scott}
~\vspace{.06\columnwidth}
\caption{\textsc{x264}}
  \label{fig:n_meta_tasks}
\end{subfigure}

\caption{Empirical results that verify the properties of real-world software systems. (a) confirms Property 1 and 2; (b) reveals Property 3.}
 \label{fig:ranking_MRE}
\end{figure}

Using the example from Figure~\ref{fig:theory}b again, clearly, the model parameter values learned on the first meta environment $\mathbfcal{E}_1$ will need to be updated via learning the $\mathbfcal{E}_2$ data and $\mathbfcal{E}_3$ data, hence it only contributes to the smallest proportion of its distribution in the meta-model. In contrast, the last meta environment $\mathbfcal{E}_3$ will contribute the most to the meta-model as most of the model parameter values learned under it will be preserved into those of the meta-model (i.e., there will be no further update). $\mathbfcal{E}_2$ contributes less than $\mathbfcal{E}_3$ but more than $\mathbfcal{E}_1$, as its resulted model parameter values are updated once. According to Figure~\ref{fig:theory}a, in this case, $\mathbfcal{E}_3$ has most parts of its parameter distribution close to that of $\mathbfcal{E}_{target}$, hence it is beneficial to leave it as the last to be learned, which preserves more of its model parameter values. Similarly, $\mathbfcal{E}_2$ tends to have more proportions of its parameter distribution closer to that of $\mathbfcal{E}_{target}$ than that of $\mathbfcal{E}_1$, hence learning its data later than that of $\mathbfcal{E}_1$ would ensure it contributes relatively more. All these can then lead to good initialization of the parameter's value in the meta-model (the rightmost in Figure~\ref{fig:theory}b) with respect to the optimal parameter distribution for $\mathbfcal{E}_{target}$ (the rightmost in Figure~\ref{fig:theory}a). Empirically, for a real-world system in Figure~\ref{fig:ranking_MRE}a, placing the more useful meta environments to the latter positions in \MetaModel~can lead to better accuracy for the target environment. Similar results have been observed in other cases.

\texttt{MAML} (Figure~\ref{fig:theory}c), in contrast, forces all meta environments to contribute equally in an aggregation to update the parameter values in the meta-model, hence for configuration data, this can easily result in an initial distribution that is highly deviated from the optimal setting of the target environment (the rightmost in Figure~\ref{fig:theory}c). This will also be experimentally justified in Section~\ref{chap-meta:evaluation}.

\subsection{More Meta Environments are Beneficial}

It is natural to ask, given \textbf{Property 1} and \textbf{Property 2}, why not use only the most useful meta environment to initialize the model parameters for the target environment? The answer is that those less useful meta environments, albeit contributing to relatively fewer proportions in the meta-model, may still provide excellent starting points for certain parts of the parameter distribution. This leads to \MetaModel's final property that fulfills \textbf{Requirement 3}:

\begin{property}
\label{pro:3}
Learning from more meta environments can help to cover the ``corner cases'' of initialing the model parameter values for the target environment. 
\end{property}

Considering the example from Figure~\ref{fig:theory}b, it is seen that, although $\mathbfcal{E}_1$ only preserves a small part of its learned parameter distribution in the meta-model, this may still be close to the optimal setting for the target environment (e.g., the peak parameter values as in Figure~\ref{fig:theory}b), hence complementary to what is missing in the contributions from $\mathbfcal{E}_2$ and $\mathbfcal{E}_3$. The same principle can be similarly applied for $\mathbfcal{E}_2$. As a random example, Figure~\ref{fig:ranking_MRE}b illustrates the sensitivity of \MetaModel's accuracy over a target environment to the number of meta environments (with the appropriate sequence). Clearly, the more meta environments, the better the accuracy---a pattern that is observed for all systems.

Noteworthily, \textbf{Property 3} may not be true for \texttt{MAML} as exploiting all meta environments equally makes it severely suffer from the side-effect of less useful meta environments, which is not uncommon with the configuration data. \MetaModel, in contrast, is able to mitigate such a side-effect by prioritizing the sequence of meta environments, thanks to \textbf{Property 1} and \textbf{Property 2}.

%% file: Figures/chap-meta/ranking_MRE.tex
\begin{tikzpicture}[]
\begin{axis}[
    ymin=33.5,
    height=3.5cm,
    width=4.5cm,
    axis x line  = bottom,
    axis y line  = left,
    xtick = {0,1,2,3,4,5},
    ytick = {30,35,40,45},
    xticklabels={ 
    $\mathbfcal{E}_3\,\mathbfcal{E}_2\,\mathbfcal{E}_1$,  $\mathbfcal{E}_2\,\mathbfcal{E}_3\,\mathbfcal{E}_1$, $\mathbfcal{E}_3\,\mathbfcal{E}_1\,\mathbfcal{E}_2$, $\mathbfcal{E}_1\,\mathbfcal{E}_3\,\mathbfcal{E}_2$, $\mathbfcal{E}_2\,\mathbfcal{E}_1\,\mathbfcal{E}_3$, $\mathbfcal{E}_1\,\mathbfcal{E}_2\,\mathbfcal{E}_3$},
    ylabel=MRE $\%$,
    xlabel=Sequence of meta environments,
    x label style={at={(0.4, -0.6)}},
    y label style={at={(-0.2, 0.5)}},
    x tick label style={rotate=45,anchor=east,font=\footnotesize},
]

\addplot+[mark=o,color=blue,mark size=1.5pt]           
coordinates                        
{                                   
  (0, 43.08265364202791)  (1, 42.59766434633329)  (2, 41.95367723155698) (3, 41.61097535228704)  (4, 34.41966822592169)  (5, 34.366265862039704)
};

\end{axis}
\end{tikzpicture}

%% file: Figures/chap-meta/ranking_Scott.tex
\begin{tikzpicture}[]
\begin{axis}[
    ymin=10,
    ymax=30,
    height=3.5cm,
    width=4.5cm,
       axis x line  = bottom,
    axis y line  = left,
    xtick = {0,1,2,3,4,5,6,7,8,9,10,11,12,13,14},
    xticklabels={1,2,3,4,5,6,7,8,9},
    ylabel=MRE $\%$,
    xlabel=\# Meta environment,
    x label style={at={(0.5,-0.30)}},
     y label style={at={(-0.25,0.5)}},
]

\addplot+[mark=o,color=blue,mark size=1.5pt]           
coordinates                         
{                                   
 (0, 27.46) (1, 25.90) (2, 25.31) (3, 23.15) (4, 21.31) (5, 18.10) (6, 16.83) (7, 16.13) (8, 12.67)};

\end{axis}
\end{tikzpicture}

%% file: Chapter-meta/framework.tex
\section{Sequential Meta-Learning for Performance Prediction}
\label{chap-meta:framework}

\begin{figure}[t!]
  \centering
  \includegraphics[width=\columnwidth]{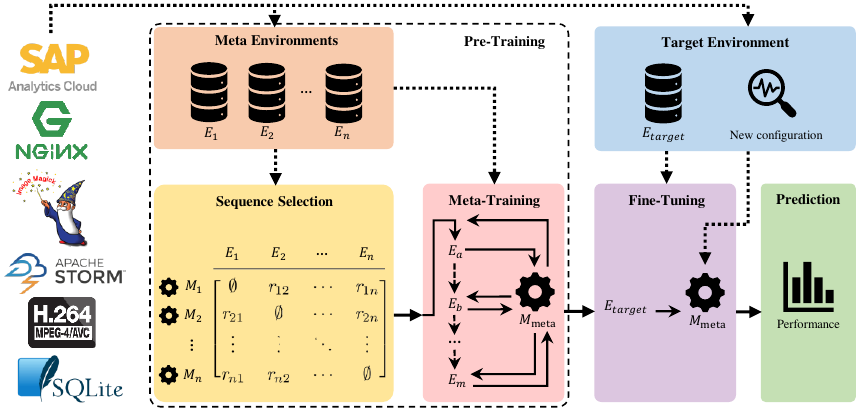}
  \caption{The architecture of \MetaModel~for learning configuration performance of a software system under multiple environments.}
  \label{chap-meta:structure}
\end{figure}


To engineer the properties from the theory behind \MetaModel~that fulfill the requirements for configuration performance learning under multiple environments, the implementation has three core components, namely, \textit{Sequence Selection}, \textit{Meta-Training}, and \textit{Fine-Tuning}. The former two resemble a pre-training process for the outer and inner loop, whereas the last is triggered when the target environment becomes available, as shown in Figure~\ref{chap-meta:structure}. These are specified below:

\begin{itemize}
    \item \textbf{Sequence Selection:} This component finds the optimal training sequence of meta environment data in \MetaModel with respect to \textbf{Property 1} and \textbf{Property 2}, considering all the available environments (\textbf{Property 3}). It deals with three low-level questions:

\begin{enumerate}
    \item How to assess the usefulness of the data of a known meta environment with respect to the unforeseen target environment?
    \item How to ensure such an assessment is reliable?
    \item How to guarantee efficiency in the assessment?
\end{enumerate}

     \item \textbf{Meta-Training:} Here, the parameters of meta-model are by sequentially learning the data of meta environments following the order provided by the \textit{Sequence Selection}.

      \item \textbf{Fine-Tuning:} Given the meta-model from the \textit{Meta-Training}, here, the aim is to update the model parameter values using measured configurations under the target environment.
    
\end{itemize}

Given the flexible nature, \MetaModel~is agnostic to the base learner that learns the meta-model, hence it can be paired with any regression learning algorithm from the literature. In the following, the above components and the pre-training process of \MetaModel~will be delineated in Algorithm~\ref{alg:pre-training}.

\subsection{Sequence Selection}

\subsubsection{How to assess the usefulness of meta environments to the target environment?}

Since the target environment is unforeseen by the time of building the meta-model, the overall Mean Relative Error (MRE) on how the single-model $\mathbfcal{M}_i$ is assessed, which is trained under an individual meta environment $\mathbfcal{E}_i$, performs when being validated across all other remaining meta environments.

Formally, MRE is a widely used scale-free metric for performance prediction~\citep{DBLP:conf/icse/HaZ19, DBLP:conf/esem/ShuS0X20, DBLP:journals/ese/GuoYSASVCWY18}:
\begin{equation}
     MRE = {{1} \over {k}} \times {\sum^k_{t=1} {{|A_t - P_t|} \over {A_t}}} \times 100\%
\end{equation}
\noindent whereby $A_t$ and $P_t$ denote the $t$th actual and predicted performance, respectively. A better overall MRE implies that the distribution of model parameter values learned from $\mathbfcal{E}_i$ 
is generally closer to all the optimal distributions required for fully learning the data of other meta environments, thereby serving as an indicator of the possible usefulness for the unforeseen target environment. As such, $\mathbfcal{E}_i$ will more likely be useful for contributing to the meta-model, enabling quicker adaptation for the unknown target environment.

\input{Tables/chap-meta/algo1}

Algorithm~\ref{alg:pre-training} (lines 2-7) demonstrates the assessment steps below:

\begin{enumerate}
    \item Train a single-model $\mathbfcal{M}_i$ using all the available data for a meta environment $\mathbfcal{E}_i$ (line 2). 
    
    \item Test single-model $\mathbfcal{M}_i$ over all data for the remaining meta environments $\mathbfcal{E}_j$. This is repeated $x$ times ($x=30$ in this chapter as it is found that using more repeats did not change the result), leading to $x$ MRE values for each tested meta environment, denoted as a vector $\mathbf{\overline{a}_{ij}}$ (line 3-7).
    \item All the $\mathbf{\overline{a}_{ij}}$ are represented in a matrix $\mathbf{A}$ below:
\input{Figures/chap-meta/matrix}

 \item Repeat from (1) till there is a single model for every meta environment.
\end{enumerate}

\subsubsection{How to ensure the reliability of assessment?}

A naive way to sort the sequence would be to directly use the overall MRE as the metric in the comparison. This, however, entails two issues:

\begin{itemize}

\item Since the overall MRE covers the accuracy tested over all the remaining meta environments, the comparisons may fail to consider statistical significance even with repeated runs.

\item Due to the residual nature of MRE, the single-model $\mathbfcal{M}_i$ has to be built by the base learner, which may not be realistic when the training is expensive (see Section~\ref{sec:efficiency}).

\end{itemize}

To address the above, in \MetaModel~Scott-Knott test~\citep{DBLP:journals/tse/MittasA13} is used to rank the MREs of all the single-models tested on a specific meta environment, then average the ranks for a single-model $\mathbfcal{M}_i$ when testing it on all the remaining meta environments. Scott-Knott sorts the list of treatments (e.g., $\mathbf{\overline{a}_{1j}},\mathbf{\overline{a}_{2j}},...,\mathbf{\overline{a}_{nj}}$ from all single-models that are tested on the $j$th meta environment) by their average MRE. Next, it splits the list into two sub-lists with the largest expected difference $\Delta$ ~\citep{xia2018hyperparameter}:
\begin{equation}
    \Delta = \frac{|l_1|}{|l|}(\overline{l_1} - \overline{l})^2 + \frac{|l_2|}{|l|}(\overline{l_2} - \overline{l})^2
\end{equation}
whereby $|l_1|$ and $|l_2|$ are the sizes of two sub-lists ($l_1$ and $l_2$) from list $l$ with a size $|l|$. $\overline{l_1}$, $\overline{l_2}$, and $\overline{l}$ denote their mean MRE. During the splitting, bootstrapping and $\hat{A}_{12}$~\citep{Vargha2000ACA} are applied to check if $l_1$ and $l_2$ are significantly different. If that is the case, Scott-Knott recurses on the splits. In other words, the models are divided into different sub-lists if both bootstrap sampling suggests that a split is significant under a confidence level of 99\% and with a good effect $\hat{A}_{12} \geq 0.6$. The sub-lists are then ranked based on their mean MRE. For example, when comparing $A$, $B$, and $C$, a possible split could be $\{A, B\}$, $\{C\}$, with the rank score of 1 and 2, respectively. Hence, statistically, it can be said that $A$ and $B$ perform similarly, but they are significantly better than $C$.

Specifically, the final rank scores are computed as in Algorithm~\ref{alg:pre-training} (lines 9-17):

\begin{enumerate}
    \item Get the repeated MREs for all single-models tested for a meta environment $\mathbfcal{E}_j$ (lines 9-10).
    \item Compute the vector of rank scores $\mathbf{\overline{r}_j}$  for all single-models when testing on $\mathbfcal{E}_j$ (line 11).
    \item Replace the tested result for $\mathbfcal{E}_j$ in matrix $\mathbf{A}$ from a vector $\mathbf{\overline{a}_{ij}}$ to a single value of rank score $r_{ij}$, forming a new matrix of ranks $\mathbf{R}$. (lines 12-13).
    \item Repeat from (1) until all tested meta environments are covered.
    \item Compute the average rank ($\mathbf{\overline{r}_i}$) over all tested meta environments for each single-model $\mathbfcal{M}_i$, and put it with the corresponding meta environment learned by $\mathbfcal{M}_i$ in $\mathbfcal{E}_{seq}$ (lines 14-16).
\end{enumerate}

The optimal sequence of training the meta environments can be found by sorting the average rank scores in $\mathbfcal{E}_{seq}$ descendingly (line 17)---the most useful meta environment is left to the last.

\subsubsection{How to guarantee efficiency?}
\label{sec:efficiency}

Ideally, the single-model $\mathbfcal{M}_i$ used when assessing the usefulness of the meta environments for a target environment should also be learned by the base learner. Yet, training some base learners, such as DNN, can be expensive, especially when training a single model for each meta environment. To ensure efficiency, in \MetaModel, linear regression is used as a surrogate of the base learner to build $\mathbfcal{M}_i$ since it has negligible overhead. 

Indeed, the MRE of linear regression may be different from that of the base learner. However, through the Scott-Knott test, the main interests are in the relative ranks between the MREs from different single models rather than their residual accuracy. As a result, simple models like linear regression can still produce similarly coarse-grained ranks to that of a complex model~\citep{DBLP:conf/sigsoft/NairMSA17}.

\subsection{Meta-Training}


The \textit{Meta-Training} in \MetaModel~learns the dataset of each meta environment sequentially according to the optimal sequence (Algorithm~\ref{alg:meta-training}). Notably, the model parameter values trained under a preceding meta-environment serve as the starting point for learning the data of the succeeding one. 

This thesis uses \texttt{DeepPerf}~\citep{DBLP:conf/icse/HaZ19}---a regularized Deep Neural Network (rDNN)---as the default base learner since it is a state-of-the-art model used for configuration performance learning. However, it is worth stressing that \MetaModel~is agnostic to the base learner, hence the choice of using \texttt{DeepPerf} is mainly due to its empirical superiority on accuracy over the others. The base learner can be replaced when other concerns, e.g., training overhead, are of higher priority.

\texttt{DeepPerf} is trained in the same process as used by~\citet{DBLP:conf/icse/HaZ19} with hyperparameter tuning for building the single-models. Interested readers are kindly referred to their work for details.

\input{Tables/chap-meta/algo2}

\subsection{Fine-Tuning}
\label{chap-meta:fine-tuning}

Since the goal is to exploit the learned model parameter values from the trained meta-model as the starting point in fine-tuning, \MetaModel~adopts the same \texttt{DeepPerf} as the base learner for the target environment. As in Algorithm~\ref{alg:fine-tuning}, the fine-tuning follows standard training of a typical machine learning model using samples for the target environment; yet, instead of using an initial model with random parameter values, the meta-model directly serves as the starting point. As a result, learning under the target environment can be greatly expedited and improved as the knowledge from meta environments should have been generalized. The training sample size from the target environment can vary, for which different sizes are examined as will be discussed in Section~\ref{chap-meta:setup}. The newly given configurations can be fed into the fine-tuned meta-model to predict their performance.

Notably, while the pre-trained meta-model demonstrates good generalizability and can be directly used for predicting configuration performance, the fine-tuning process is still necessary for the following reasons: (1) during fine-tuning, a critical step is to normalize the configuration options between 0 and 1, yet without fine-tuning, normalization would not be feasible, potentially leading to reduced prediction accuracy; (2) without fine-tuning, the meta-model can only use the default hyperparameters such as the learning rate and lambda for L1 regularization, which would also constraint its capabilities. In the subsequent experiment section, it will be further demonstrated that even with only a few fine-tuning samples, \MetaModel~is able to outperform the state-of-the-art configuration performance prediction models.

\input{Tables/chap-meta/algo3}

%% file: Tables/chap-meta/algo1.tex
\begin{algorithm}[t]

	\DontPrintSemicolon
	\footnotesize
	\caption{Sequence Selection}
	\label{alg:pre-training}
	\KwIn{Data from meta environments $\mathbfcal{E}=\{\mathbfcal{E}_1,\mathbfcal{E}_2,...,\mathbfcal{E}_n\}$}
     \KwOut{The optimal sequence $\mathbfcal{E}_{seq}$}

         \For{$\forall \mathbfcal{E}_i \in \mathbfcal{E}$} {

            $\mathcal{M}_i\leftarrow$ \textsc{train($\mathbfcal{E}_i$)}\\

               \For{$\forall \mathbfcal{E}_j \in \mathbfcal{E}$} {

                   \If{$\mathbfcal{E}_j$ is not  $\mathbfcal{E}_i$} {
                         $\mathbf{A}\leftarrow \mathbf{\overline{a}_{ij}}=$ \textsc{testMREwithRepeats($\mathcal{M}_i,\mathbfcal{E}_j$)}\\
                    } \Else{
                      $\mathbf{A}\leftarrow \emptyset$\\
                    }
               }
         }

          \For{$\forall \mathbfcal{E}_j \in \mathbfcal{E}$} {

               \For{$\forall \mathbfcal{E}_i \in \mathbfcal{E}$} {
             
                   $\mathbf{\overline{a}'_j}\leftarrow\mathbf{\overline{a}_{ij}} \in \mathbf{A}$\\
                  
               }
 $\mathbf{\overline{r}_j}=$ \textsc{scottKnottTest($\mathbf{\overline{a}'_j}$)}\\
              \For{$\forall r_{ij} \in \mathbf{\overline{r}_j}: r_{ij} \neq 0$} {
$\mathbf{R}\leftarrow r_{ij}$\\
                }        
         }

 \For{$\forall \text{row } \mathbf{\overline{r}_i} \in \mathbf{R}$} {
 
     $r_{mean}=$ averaging the rank scores $r_{ij} \in \mathbf{\overline{r}_i}$\\
               $\mathbfcal{E}_{seq}\leftarrow \{\mathbfcal{E}_i/\mathcal{M}_i,r_{mean}\}$\\
 }

    \Return $\mathbfcal{E}_{seq}\leftarrow$ \textsc{sort($\mathbfcal{E}_{seq}$)}\\   
	
\end{algorithm}

%% file: Figures/chap-meta/matrix.tex
\begin{equation}
\mathbf{A} = 
\begin{blockarray}{c cccc}
& \mathbfcal{E}_1  & \mathbfcal{E}_2 & \cdots & \mathbfcal{E}_n  \\
\cmidrule{2-5}
\begin{block}{c [cccc]}
\mathbfcal{M}_1 & \fixhd{b} \emptyset & \mathbf{\overline{a}_{12}} & \cdots & \mathbf{\overline{a}_{1n}}  \\
\mathbfcal{M}_2 & \fixhd{b} \mathbf{\overline{a}_{21}} & \emptyset & \cdots & \mathbf{\overline{a}_{2n}}  \\
\vdots & \vdots & \vdots & \ddots & \vdots  \\
\mathbfcal{M}_n & \fixhd{b} \mathbf{\overline{a}_{n1}} & \mathbf{\overline{a}_{n2}} & \cdots & \emptyset \\
\end{block}
\noalign{\vspace{-100.5ex}}
\\
\end{blockarray}
\end{equation}

%% file: Tables/chap-meta/algo2.tex
\begin{algorithm}[t]

	\DontPrintSemicolon
	\footnotesize
	
	\caption{Meta-Training}
	\label{alg:meta-training}
	\KwIn{The optimal sequence $\mathbfcal{E}_{seq}$}
     \KwOut{The meta-model $\mathbfcal{M}_{meta}$}

$\mathcal{M}_{meta}\leftarrow$ randomly initialized model \\
 \While{not done}{
         
         

          \For{$\forall \mathbfcal{E}_i \in \mathbfcal{E}_{seq}$} {
                $\mathcal{M}_{meta}\leftarrow$ \textsc{train($\mathcal{M}_{meta}, \mathbfcal{E}_i $)}\\
          }

     }

    \Return $\mathcal{M}_{meta}$\\
	
\end{algorithm}

%% file: Tables/chap-meta/algo3.tex
\begin{algorithm}[t]

	\DontPrintSemicolon
	\footnotesize
	
	\caption{Fine-Tuning}
	\label{alg:fine-tuning}
	\KwIn{The meta-model $\mathbfcal{M}_{meta}$}
     \KwOut{The fine-tuned model $\mathbfcal{M}_{tuned}$}

$\mathbfcal{E}_{target}\leftarrow$ measured data samples from the target environment \\

    \Return   $\mathcal{M}_{tuned}\leftarrow$ \textsc{train($\mathcal{M}_{meta}, \mathbfcal{E}_{target} $)}\\
	
\end{algorithm}

%% file: Chapter-meta/setup.tex
\section{Experiment Setup}
\label{chap-meta:setup}

This section delineates the settings of the experimental evaluation.

\subsection{Research Questions}

To answer the RQ4 in this thesis, a few sub-questions should be examined. First, it is natural to evaluate \MetaModel~with the single-task performance models, which are trained without extra data from other sources than the target environment. Second, it is crucial for  \MetaModel~to be compared with the multi-environment learning models to determine the effectiveness of the meta-learning framework. Besides, justifying the influence of the sequence selection algorithm for  \MetaModel~is also necessary. Since \MetaModel~defaultly uses 100\% data from the meta environments, the last sub-RQ seeks to understand the situations where fewer meta-data are available.

In summary, this chapter evaluates~\MetaModel~by answering the following sub-RQs:

\begin{itemize}
    \item \textbf{RQ4.1:} How does \MetaModel~perform compared with existing single environment models?

    \item \textbf{RQ4.2:} How does \MetaModel~perform compared with the state-of-the-art models that handle multiple environments?

    \item \textbf{RQ4.3:} How effective is the sequence selection in \MetaModel?

    \item \textbf{RQ4.4:} What is the sensitivity of \MetaModel's accuracy to the sample size used in pre-training?

    \item \textbf{RQ4.5:} What is the effectiveness of fine-tuning?
\end{itemize}

\input{Tables/chap-meta/systems.tex}

\subsection{Subject Software Systems and Environments}
\label{chap-meta-subsec:subject_system}

In the experiments, widely-used multi-environment configuration datasets collected from real-world systems\footnote{To ensure reliability, the data has been collected with repeated measurements.}~\citep{DBLP:conf/kbse/JamshidiSVKPA17, DBLP:conf/sigsoft/JamshidiVKS18, DBLP:journals/tse/KrishnaNJM21, LESOIL2023111671} are studied. These systems are selected based on the following:

\begin{itemize}
    \item To ensure diversity, systems with less than three environments are removed.
    \item In the presence of multiple datasets for the same system, the one that contains the most deviated measurements between the environments is used.
\end{itemize}

Within the identified systems, the data of environments are ruled out that (1) do not measure all valid configurations; (2) contain invalid measurements; or (3) lack detailed specifications.

As shown in Table~\ref{chap-meta-table:subject systems}, the above process leads to nine systems of diverse domains, scales, and option types, together with configurations measured in distinct valid environments of hardware, workloads, and/or versions. To further improve the external validity, the models are examined under different training sample sizes from the target environment as elaborated in Table~\ref{chap-meta-tb:sizes}. For binary systems (only binary options), five sizes as prior work~\citep{DBLP:conf/icse/HaZ19} are used: $\{x,2x,3x,4x,5x\}$ where $x$ is the number of options. For mixed systems (both binary and numeric options), five sizes as produced by the sampling method from the work of~\citet{DBLP:journals/sqj/SiegmundRKKAS12} are used. These are the common methods used in the field~\citep{DBLP:conf/icse/HaZ19,DBLP:conf/esem/ShuS0X20}. 

\input{Tables/chap-meta/sample_sizes}

\subsection{Procedure}

For each system in the experiments, the steps below are followed:

\begin{enumerate}
    \item Pick an environment as the target $\mathbfcal{E}_{target}$ and set all remaining ones as meta environments.
    \item If applicable, pre-training on all data samples of the meta environments (except for RQ4.4).
    \item Pick a training sample size $S_i$ for $\mathbfcal{E}_{target}$.
    \item Train/fine-tune a model under $S_i$ randomly selected samples and test it on all the remaining (unforeseen) samples of $\mathbfcal{E}_{target}$. 
    \item To mitigate bias, repeat (4) for 30 runs via bootstrapping without replacement\footnote{30 runs is the most common setting in the field of configuration performance learning~\citep{DBLP:conf/icse/HaZ19,DBLP:conf/esem/ShuS0X20}. This number of runs is merely a pragmatic choice given the resource constraint.}.
    \item Repeat from (3) to cover all training sample sizes of $\mathbfcal{E}_{target}$.
    \item Repeat from (1) till every environment has served as the target environment once. In this way, bias towards the prediction of a particular environment is avoided.
\end{enumerate}

As such, for each of the nine systems, there are five different training sizes and 3--10 alternative target environments, leading to 15--50 cases of comparisons. In \MetaModel, the iteration limit for the outer loop is set as 1, which is the same default for \texttt{MAML}.

\subsection{Metric}

\subsubsection{Accuracy}
As with the sequence selection, MRE is used to assess the accuracy, which is the standard practice~\citep{DBLP:conf/splc/ValovGC15,DBLP:conf/oopsla/QueirozBC16,DBLP:conf/icse/0003XC021,DBLP:conf/icse/HaZ19,DBLP:journals/ese/GuoYSASVCWY18}. Further, MRE is insensitive to the scale of performance metrics.

\subsubsection{Speedup}

As in prior work~\citep{DBLP:conf/icse/0003XC021,DBLP:conf/sigsoft/0001L21}, to assess the training speedup achieved by \MetaModel~for a system, a baseline, $b$, is taken as the smallest training size that the best
state-of-the-art counterpart consumes to achieve its best mean MRE. Then, the smallest size for \MetaModel~is found to achieve the same accuracy, denoted as $s$. The ratios, i.e., $sp={b \over s}$, are reported. Clearly, if \MetaModel~is more data-efficient, then it would be expected that $sp \geq 1 \times$. When \MetaModel~cannot achieve the same mean MRE for all sample sizes, it has $sp=N/A$.

\subsection{Statistics}

To make a fair comparison between more than two models, the Scott-Knott test~\citep{DBLP:journals/tse/MittasA13} is again used to evaluate their statistical significance on the difference of MRE over 30 runs, as recommended by~\citet{DBLP:journals/tse/MittasA13}. In a nutshell, Scott-Knott test provides the following benefits over the other statistical methods:

\begin{itemize}
    \item Unlike those parametric tests such as the t-test~\citep{student1908probable}, it has no assumption on the data distribution.
    \item It allows the assessment of more than two models as opposed to the other pair-wise nonparametric tests such as the Wilcoxon test~\citep{Wilcoxon1945IndividualCB} or Mann-Whitney U test~\citep{Mann47}.
    \item It naturally ranks the model while other multiple comparison tests, e.g., Kruskal-Wallis test~\citep{mckight2010kruskal}, only assess the differences between models without comprising the better or worse. Further, the Scott-Knott test does not need post-hoc correlation.
\end{itemize}

%% file: Tables/chap-meta/systems.tex
\begin{table}[t!]
\caption{Details of the subject systems. $|\mathbfcal{H}|$, $|\mathbfcal{W}|$, and $|\mathbfcal{V}|$ respectively denotes the number of hardware, workload, and version considered for the environments, ($|\mathbfcal{B}|$/$|\mathbfcal{N}|$) denotes the number of binary/numerical options, and $|\mathbfcal{C}|$ denotes the number of valid configurations per environment (full sample size).}
\centering
\footnotesize
\begin{adjustbox}{width=\linewidth,center}
\setlength{\tabcolsep}{1mm}
\begin{tabular}{l|lllc|ccc}
\toprule

\multirow{2}{*}{\textbf{System}} & \multirow{2}{*}{\textbf{Domain}}  & \multirow{2}{*}{\textbf{$|\mathbfcal{B}|$/$|\mathbfcal{N}|$}} & \multirow{2}{*}{\textbf{$|\mathbfcal{C}|$}}  & \multirow{2}{*}{\textbf{Used by}}    & \multicolumn{3}{c}{\textbf{Environments}}  \\

&   &  & &   & \textbf{$|\mathbfcal{H}|$} & \textbf{$|\mathbfcal{W}|$} & \textbf{$|\mathbfcal{V}|$}   \\
\midrule
\textsc{DeepArch} & DNN tool            & 12/0             & 4096  & \citep{DBLP:conf/sigsoft/JamshidiVKS18}& 3 & 1 & 1  \\
\textsc{SaC}      & Cloud tool          & 58/0             & 4999  & \citep{DBLP:journals/tse/KrishnaNJM21}       & 1 & 3 & 1    \\
\textsc{SQLite}   & DBMS                & 14/0             & 1000  & \citep{DBLP:journals/tse/KrishnaNJM21}     & 2  & 1 & 2      \\
\textsc{NGINX}      & Web server          & 16/0              & 1104    & \citep{DBLP:conf/icse/webertwins}   & 1 & 1 & 4   \\
\textsc{SPEAR}    & Audio editor        & 14/0             & 16385 & \citep{DBLP:journals/tse/KrishnaNJM21}   & 3 & 1 & 1      \\
\textsc{Storm} & Big data analyzer         & 1/11             & 2048  & \citep{DBLP:conf/sigsoft/JamshidiVKS18} & 3 & 1 & 1  \\
\textsc{ImageMagick}     & Image editor      & 0/5            & 100   & \citep{LESOIL2023111671} & 1 & 4 & 1      \\
\textsc{ExaStencils}    & Code generator      & 8/4            & 4098   & \citep{DBLP:conf/icse/webertwins} & 1  & 4 & 1     \\ 
\textsc{x264}     & Video encoder      & 11/13            & 201   & \citep{LESOIL2023111671}    & 1 & 10 & 1  \\

\bottomrule
\end{tabular}
\end{adjustbox}
\label{chap-meta-table:subject systems}
\end{table}

%% file: Tables/chap-meta/sample_sizes.tex
\begin{table}[!t]
\caption{The training sizes (for fine-tuning) when an environment is used as the target.}
\centering
\footnotesize
\begin{adjustbox}{width=0.6\linewidth,center}
\begin{tabular}{l|ccccc}
\toprule
\textbf{System} & \textbf{$S_1$} & \textbf{$S_2$} & \textbf{$S_3$} & \textbf{$S_4$} & \textbf{$S_5$} \\ 
\midrule
\textsc{DeepArch} (\textsc{DArch})  & 12  & 24 & 36 & 48 & 60\\
\textsc{SaC}  & 58 & 116 & 174 & 232 & 290\\
\textsc{SQLite}  & 14 & 28 & 42 & 56 & 70\\
\textsc{NGINX}   & 16 & 32 & 48 & 64 & 80\\
\textsc{SPEAR} & 14 & 28 & 42 & 56 & 70\\
\textsc{Storm}  & 158 & 211 & 522 & 678 & 1403\\ %
\textsc{ImageMagick} (\textsc{IM})  & 11 & 24 & 45 & 66 & 70 \\
\textsc{ExaStencils} (\textsc{ES})  & 106 & 181 & 366 & 485 & 695\\ 
\textsc{x264} & 24 & 53 & 81 & 122 & 141\\
\bottomrule
\end{tabular}
\end{adjustbox}
\label{chap-meta-tb:sizes}
\end{table}

%% file: Chapter-meta/evaluation.tex
\section{Evaluation}
\label{chap-meta:evaluation}

This section presents and discusses the experiment results following the sub-RQs.

\subsection{RQ4.1: \MetaModel~against Single Environment Models}

  \begin{figure}[!t]
  \centering
  \begin{subfigure}[t]{0.8\columnwidth}
        \centering
\includegraphics[width=\textwidth]{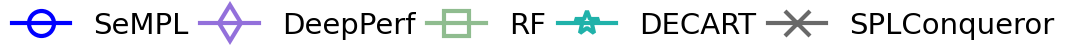}
   \end{subfigure}

   \begin{subfigure}[t]{\subfigsize\columnwidth}
        \centering
\includegraphics[width=\textwidth]{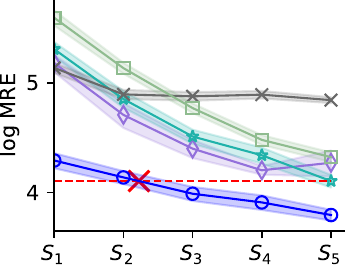}
 \subcaption*{\textsc{DeepArch}  ($sp=2.25\times$)}
   \end{subfigure}
~\hspace{-0.15cm}
\begin{subfigure}[t]{\subfigsize\columnwidth}
        \centering
\includegraphics[width=\textwidth]{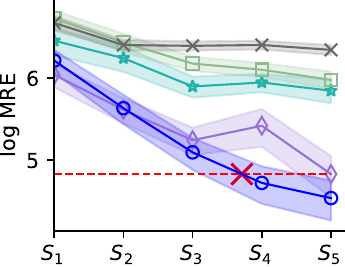}
      \subcaption*{\textsc{SaC}  ($sp=1.35\times$)}
   \end{subfigure}
~\hspace{-0.15cm}
   \begin{subfigure}[t]{\subfigsize\columnwidth}
        \centering
\includegraphics[width=\textwidth]{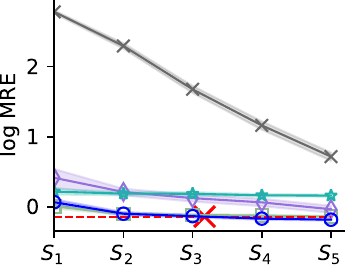}
     \subcaption*{\textsc{SQLite}  ($sp=1.58\times$)}
   \end{subfigure}

 \begin{subfigure}[t]{\subfigsize\columnwidth}
        \centering
\includegraphics[width=\textwidth]{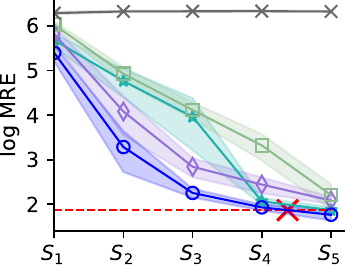}
 \subcaption*{\textsc{NGINX} ($sp=1.14\times$)}
   \end{subfigure}
~\hspace{-0.15cm}
\begin{subfigure}[t]{\subfigsize\columnwidth}
        \centering
\includegraphics[width=\textwidth]{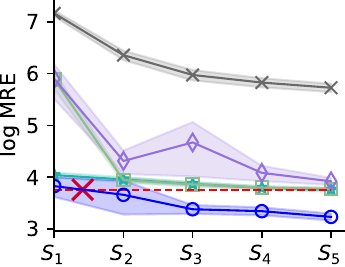}
     \subcaption*{\textsc{SPEAR} ($sp=3.54\times$)}
   \end{subfigure}
~\hspace{-0.15cm}
    \begin{subfigure}[t]{\subfigsize\columnwidth}
        \centering
\includegraphics[width=\textwidth]{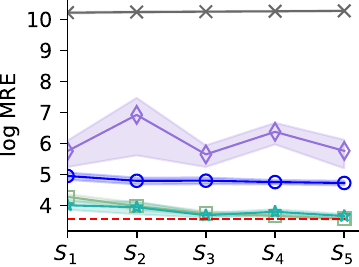}
      \subcaption*{\textsc{Storm}  ($sp=N/A$)}
   \end{subfigure}

      \begin{subfigure}[t]{\subfigsize\columnwidth}
        \centering
\includegraphics[width=\textwidth]{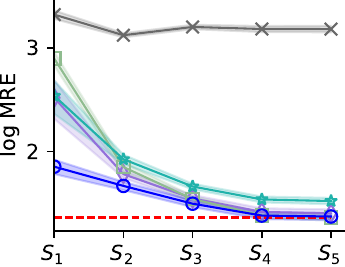}
 \subcaption*{\textsc{ImageMagick}  ($sp=N/A$)}
   \end{subfigure}
~\hspace{-0.15cm}
   \begin{subfigure}[t]{\subfigsize\columnwidth}
        \centering
\includegraphics[width=\textwidth]{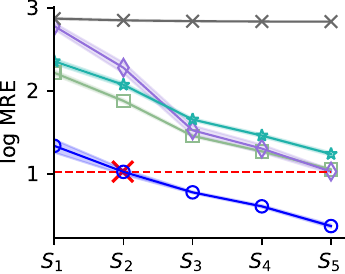}
     \subcaption*{\textsc{ExaStencils}  ($sp=3.86\times$)}
   \end{subfigure}
~\hspace{-0.15cm}
    \begin{subfigure}[t]{\subfigsize\columnwidth}
        \centering
\includegraphics[width=\textwidth]{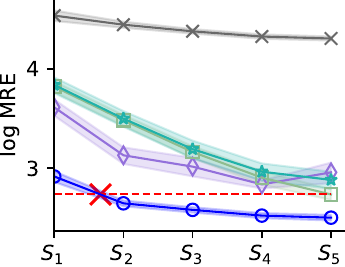}
      \subcaption*{\textsc{x264}  ($sp=3.25\times$)}
   \end{subfigure}

      \caption{\MetaModel~versus single environment models.
      For the simplicity of exposition, the log-transformed average MRE (and its standard error) of all target environments and runs is reported. For speedup ($sp={b \over s}$), \textcolor{red}{\dashed} denotes the mean MRE for $b$; \textcolor{red}{\ding{53}} indicates the point of $s$.}
      \label{fig:rq1-mean}
  \end{figure}

To answer RQ4.1, \MetaModel~is compared against four state-of-the-art single environment models, i.e., \texttt{DeepPerf}, \texttt{DECART}, \texttt{RF}, and \texttt{SPLConqueror}, as discussed in Section~\ref{sec:bg-single}. To ensure fairness, the codes published by their authors are used. All settings presented in Section~\ref{chap-meta:setup} are used.

\input{Tables/chap-meta/rq1-new}

\subsubsection{Results}

The results are visualized in Figure~\ref{fig:rq1-mean}. 
As can be seen, on nearly all systems, \MetaModel~significantly improves the MRE by up to 81\%, 88\%, 84\%, and 99\% over \texttt{DeepPerf}, \texttt{RF}, \texttt{DECART}, and \texttt{SPLConqueror}, respectively. As the default base learner for \MetaModel, \texttt{DeepPerf} might occasionally perform worse even with more data. This is because the target environments exhibit diverse data patterns, which can easily cause the hyperparameter tuning to be trapped at local optima. Such an issue can be mitigated when paired with \MetaModel~since the meta-model is fine-tuned from some good starting points of the parameters. In particular, \MetaModel~often achieves considerably better MRE in very few data samples, commonly with a significant speedup that can be up to $3.86\times$. This confirms the generalization efficiency of sequential meta-learning. Remarkably, from the average Scott-Knott ranks, \MetaModel~is the best for 8 out of 9 systems (i.e., $89\%$), confirming that it is statistically better than the single environment models for learning configuration performance, thanks to ``learning to learn".

\begin{quotebox}
   \noindent
   {\textbf{RQ4.1:} \MetaModel~performs significantly better than the state-of-the-art single environment performance models in 8 out of 9 systems with the best MRE improvement of $99\%$; it is also generally data-efficient with up to $3.86\times$ speedup.} 
\end{quotebox}

\subsection{RQ4.2: \MetaModel~against Multi-Environment Models}

      \begin{figure}[!t]
  \centering
  \begin{subfigure}[t]{0.9\columnwidth}
        \centering
\includegraphics[width=\textwidth]{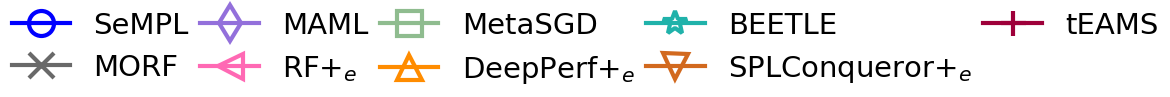}
   \end{subfigure}

   \begin{subfigure}[t]{\subfigsize\columnwidth}
        \centering
\includegraphics[width=\textwidth]{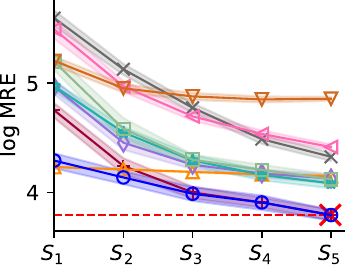}
 \subcaption*{\textsc{DArch}  ($sp=1.00\times$)}
   \end{subfigure}
~\hspace{-0.15cm}
\begin{subfigure}[t]{\subfigsize\columnwidth}
        \centering
\includegraphics[width=\textwidth]{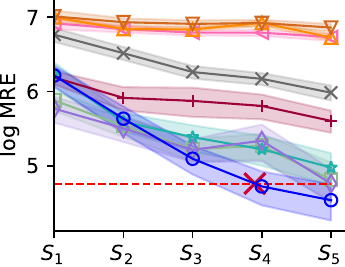}
      \subcaption*{\textsc{SaC}  ($sp=1.28\times$)}
   \end{subfigure}
~\hspace{-0.15cm}
   \begin{subfigure}[t]{\subfigsize\columnwidth}
        \centering
\includegraphics[width=\textwidth]{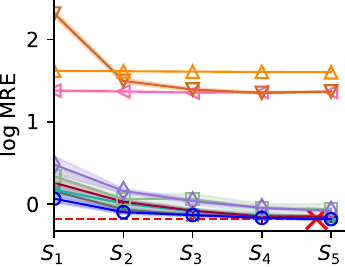}
     \subcaption*{\textsc{SQLite}  ($sp=1.04\times$)}
   \end{subfigure}

      \begin{subfigure}[t]{\subfigsize\columnwidth}
        \centering
\includegraphics[width=\textwidth]{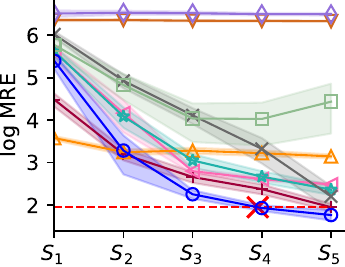}
 \subcaption*{\textsc{NGINX}  ($sp=1.27\times$)}
   \end{subfigure}
~\hspace{-0.15cm}
   \begin{subfigure}[t]{\subfigsize\columnwidth}
        \centering
\includegraphics[width=\textwidth]{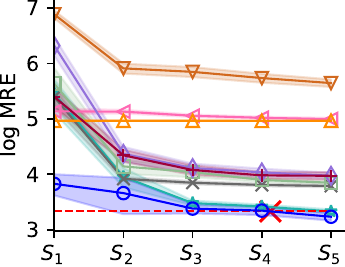}
     \subcaption*{\textsc{SPEAR}  ($sp=1.21\times$)}
   \end{subfigure}
~\hspace{-0.15cm}
    \begin{subfigure}[t]{\subfigsize\columnwidth}
        \centering
\includegraphics[width=\textwidth]{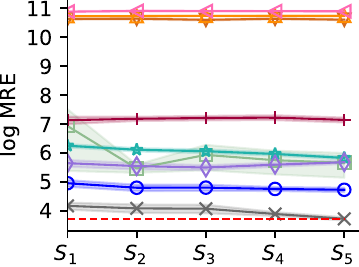}
      \subcaption*{\textsc{Storm}  ($sp=N/A$)}
   \end{subfigure}

      \begin{subfigure}[t]{\subfigsize\columnwidth}
        \centering
\includegraphics[width=\textwidth]{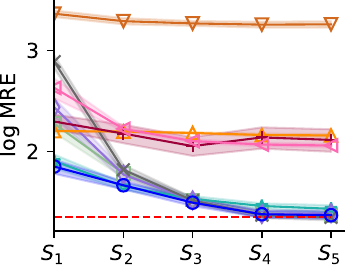}
 \subcaption*{\textsc{IM} ($sp=N/A$)}
   \end{subfigure}
~\hspace{-0.15cm}
   \begin{subfigure}[t]{\subfigsize\columnwidth}
        \centering
\includegraphics[width=\textwidth]{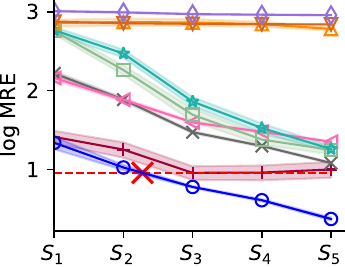}
    \subcaption*{\textsc{ES} ($sp=1.55\times$)}
   \end{subfigure}
   ~\hspace{-0.15cm}
    \begin{subfigure}[t]{\subfigsize\columnwidth}
        \centering
\includegraphics[width=\textwidth]{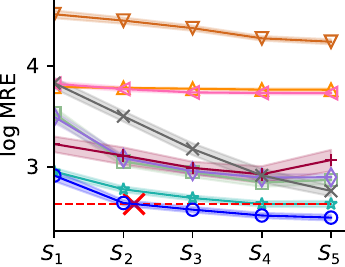}
      \subcaption*{\textsc{x264} ($sp=2.13\times$)}
   \end{subfigure}

      \caption{\MetaModel~versus multi-environment models. 
      For the simplicity of exposition, the log-transformed average MRE (and its standard error) of all target environments and runs is reported. For speedup ($sp={b \over s}$), \textcolor{red}{\dashed} denotes the mean MRE for $b$; \textcolor{red}{\ding{53}} indicates the point of $s$.}
      \label{fig:rq2-mean}
  \end{figure}

To study RQ4.2, \MetaModel~is assessed against state-of-the-art models based on transfer learning (\texttt{BEETLE} and \texttt{tEAMS}), multi-task learning (\texttt{MORF}), and meta-learning (\texttt{MAML} and \texttt{MetaSGD}), as discussed in Section~\ref{sec:bg-transfer} and~\ref{sec:bg-meta}. Again, the same source code published by their authors is used. The approaches that take environmental features as additional inputs are also examined, i.e., \texttt{DeepPerf$+_{e}$}, \texttt{RF$+_{e}$}, and \texttt{SPLConqueror$+_{e}$} (see Section~\ref{sec:bg-env}).

Since \texttt{MAML}, \texttt{MetaSGD}, \texttt{BEETLE}, and \texttt{tEAMS} are agnostic to the base learner, they are paired with \texttt{DeepPerf}, which is the same default for \MetaModel. Their pre-training data is also identical to that of \MetaModel~where applicable. All other settings are the same as \textbf{RQ4.1}.

\input{Tables/chap-meta/rq2-new}

\subsubsection{Results}

From Figure~\ref{fig:rq2-mean}, it can be seen that in general, \MetaModel~is mostly more accurate than, or at least similar to, the best state-of-the-art model that handles multiple environments. The best improvement ranges from 74\% (for \textsc{SPEAR}) to 99\% (for \textsc{NGINX}). In terms of data efficiency, \MetaModel~shows considerable speedup with up to $2.13\times$. Similar findings can be observed on the Scott-Knott test: overall, \MetaModel~is ranked the best for $89\%$ of the systems (8 out of 9 systems).

These results match with the theory: transfer learning models like \texttt{BEETLE} and \texttt{tEAMS} are restricted by their weak generalizability; while multi-task learning models like \texttt{MORF} overcomplicate the training, which makes the accuracy suffers; meta-learning models like \texttt{MAML} and \texttt{MetaSGD} fail to discriminate the contributions between different meta environments when pre-training the meta-model. All of the above shortcomings are what \MetaModel~seeks to tackle. In particular, models that take the environments as additional features generally perform badly (e.g., \texttt{DeepPerf$+_{e}$}), due to the fact that this unnecessarily increases the difficulty of training.

\begin{quotebox}
   \noindent
   {\textbf{RQ4.2:}} Compared with the state-of-the-art models that handles multiple environments, \MetaModel~is more effective on 89\% of the systems with the best MRE improvement from 74\% to 99\% while being data-efficient with mostly $sp \geq 1$ and the best speedup of $2.13\times$. In particular, this evidences that the sequential training in \MetaModel, which discriminates the contributions of each meta environment, is more beneficial for configuration performance learning than the parallel training (e.g., in \texttt{MAML}) that treats all meta environments equally.
\end{quotebox}

\begin{figure}[!t]
  \centering
\begin{subfigure}[t]{\columnwidth}
        \centering
\includegraphics[width=\textwidth]{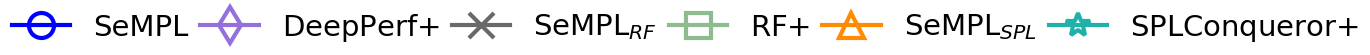}
   \end{subfigure}
  
\begin{subfigure}[t]{\subfigsize\columnwidth}
        \centering
\includegraphics[width=\textwidth]{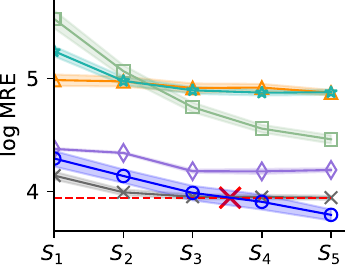}
\subcaption*{\textsc{DArch} ($sp=1.41\times$)}
   \end{subfigure}
~\hspace{-0.15cm}
    \begin{subfigure}[t]{\subfigsize\columnwidth}
        \centering
\includegraphics[width=\textwidth]{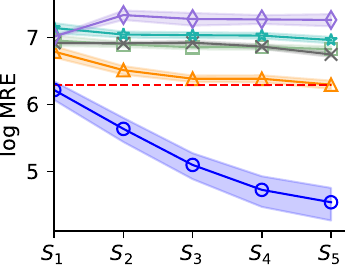}
\subcaption*{\textsc{SaC} ($sp>5.00\times$)}
   \end{subfigure}
~\hspace{-0.15cm}
   \begin{subfigure}[t]{\subfigsize\columnwidth}
        \centering
\includegraphics[width=\textwidth]{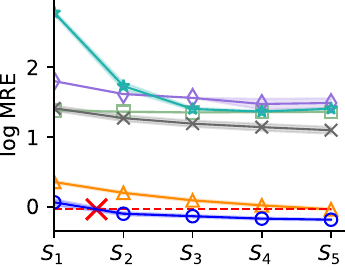}
\subcaption*{\textsc{SQLite} ($sp=3.10\times$)}
   \end{subfigure}

\begin{subfigure}[t]{\subfigsize\columnwidth}
\centering
\includegraphics[width=\textwidth]{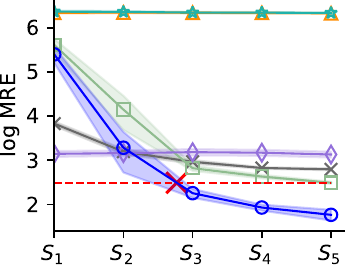}
\subcaption*{\textsc{NGINX} ($sp=1.80\times$)}
   \end{subfigure}
   ~\hspace{-0.15cm}
   \begin{subfigure}[t]{\subfigsize\columnwidth}
        \centering
\includegraphics[width=\textwidth]{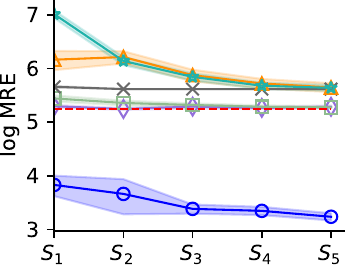}
\subcaption*{\textsc{SPEAR} ($sp>2.00\times$)}
   \end{subfigure}
~\hspace{-0.15cm}
    \begin{subfigure}[t]{\subfigsize\columnwidth}
        \centering
\includegraphics[width=\textwidth]{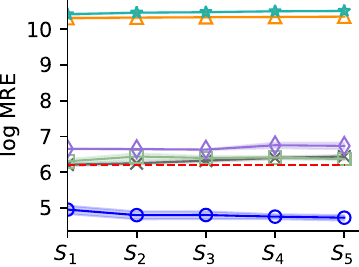}
\subcaption*{\textsc{Storm} ($sp>1.00\times$)}
   \end{subfigure}

\begin{subfigure}[t]{\subfigsize\columnwidth}
\centering
\includegraphics[width=\textwidth]{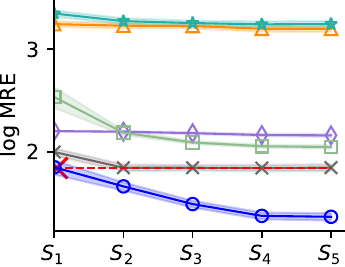}
\subcaption*{\textsc{IM} ($sp=5.72\times$)}
   \end{subfigure}
~\hspace{-0.15cm}
   \begin{subfigure}[t]{\subfigsize\columnwidth}
        \centering
\includegraphics[width=\textwidth]{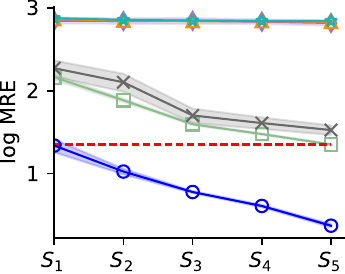}
\subcaption*{\textsc{ES} ($sp>6.56\times$)}
   \end{subfigure}
   ~\hspace{-0.15cm}
    \begin{subfigure}[t]{\subfigsize\columnwidth}
        \centering
\includegraphics[width=\textwidth]{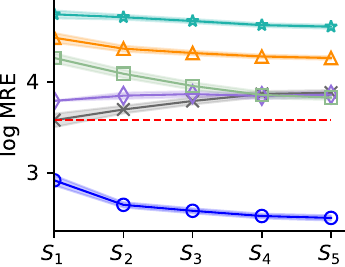}
\subcaption*{\textsc{x264} ($sp>1.00\times$)}
   \end{subfigure}
     
      \caption{Optimal sequence versus random order in \MetaModel. 
      For the simplicity of exposition, the log-transformed average MRE (and its standard error) of all target environments and runs is reported. For speedup ($sp={b \over s}$), \textcolor{red}{\dashed} denotes the mean MRE for $b$; \textcolor{red}{\ding{53}} indicates the point of $s$.}
      \label{fig:rq3-mean}
  \end{figure}

\subsection{RQ4.3: Effectiveness of Sequence Selection}

To confirm the necessity of sequence selection in \MetaModel, for RQ4.3, the other models from RQ4.1 are equipped with the sequential meta-learning process using all data of the meta environments in random orders, denoted as \texttt{DeepPerf}$+$, \texttt{RF}$+$, and \texttt{SPLConqueror}$+$. Thus, they will have access to exactly the same amount of data as \MetaModel~and the meta-model will also offer learned parameter values for fine-tuning. Note that \texttt{DECART} is not considered because its sampling mechanism requires initializing the model from scratch, which is incompatible with the concept of fine-tuning.

To eliminate the noise caused by the default base learner, this chapter also additionally pairs \MetaModel~with two alternative base learners, i.e., \texttt{RF} and \texttt{SPLConqueror}, denoted as \MetaModel$_{RF}$ and \MetaModel$_{SPL}$, respectively. All other settings are the same as the previous RQs.

\input{Tables/chap-meta/rq3-new}

\subsubsection{Result}

From Figure~\ref{fig:rq3-mean}, it is seen that \MetaModel, \MetaModel$_{RF}$ and \MetaModel$_{SPL}$ generally perform better than their counterparts on 9 (100\%), 5 (56\%), and 9 (100\%) out of 9 systems, respectively. From this, the importance of obtaining the optimal sequence for the sequential learning in \MetaModel can be confirmed, as a random order would likely amplify the side-effect of some badly performing meta environments. Clearly, using \texttt{DeepPerf} as the base learner in \MetaModel~dramatically boosts the accuracy across nearly all training sizes on 8 out of 9 systems with the best speedup of more than $6.56\times$.

Interestingly, \MetaModel$_{RF}$ might be exacerbated with more training data, e.g., for \textsc{x264}. This is because the base learner, \texttt{RF}, suffers from the issue of overfitting, which, when updating the meta-models over multiple meta environments, can be harmful due to the cumulatively overfitted model parameters. It is also the reason that \MetaModel$_{RF}$ does not perform as well as \texttt{RF$+$} on some systems.

\begin{quotebox}
   \noindent
   \textbf{RQ4.3:} The sequence selection helps to improve the results on between 56\% and 100\% of the systems based on the base learner. Notably, pairing \MetaModel~with \texttt{DeepPerf} achieves the best results overall and leads to considerable speedup.
\end{quotebox}

\subsection{RQ4.4: Sensitivity to the Pre-Training Size}

Previous sub-RQs use full data samples in pre-training, for RQ4.4, the MRE at $0\%, 25\%,...,100\%$ of the full samples for meta environments is reported, where $0\%$ means no meta-learning.

\subsubsection{Results}

\begin{figure}[!t]
\centering
    \begin{subfigure}[t]{0.2124\columnwidth}
    \includegraphics[width=\columnwidth]{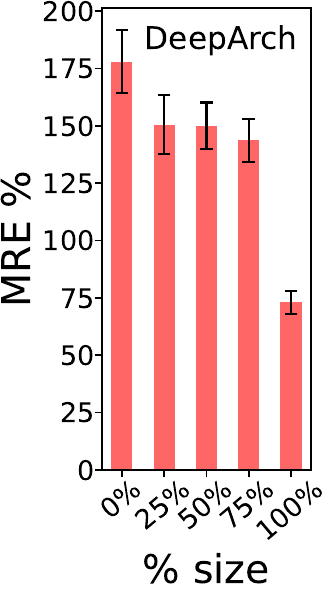}
    \end{subfigure}
 ~\hspace{-0.3cm}
    \begin{subfigure}[t]{0.1854\columnwidth}
    \includegraphics[width=\columnwidth]{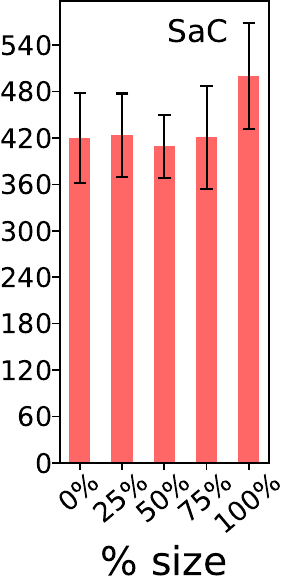}
    \end{subfigure}
 ~\hspace{-0.3cm}
    \begin{subfigure}[t]{0.1908\columnwidth}
    \includegraphics[width=\columnwidth]{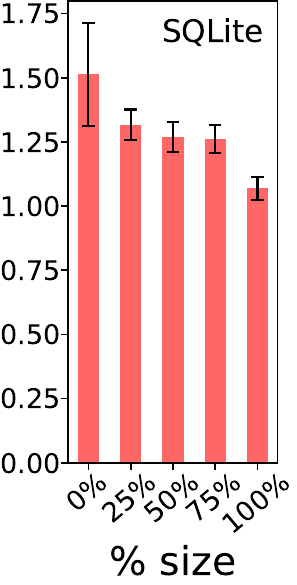}
    \end{subfigure}
 ~\hspace{-0.3cm}
     \begin{subfigure}[t]{0.1854\columnwidth}
    \includegraphics[width=\columnwidth]{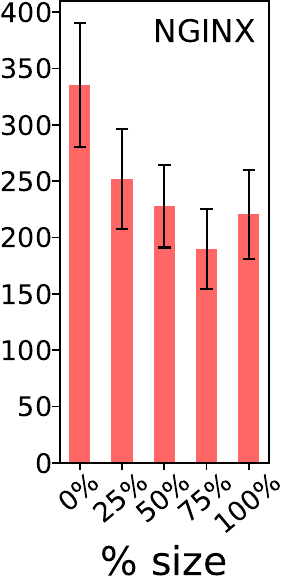}
    \end{subfigure}
 ~\hspace{-0.3cm}
     \begin{subfigure}[t]{0.1854\columnwidth}
    \includegraphics[width=\columnwidth]{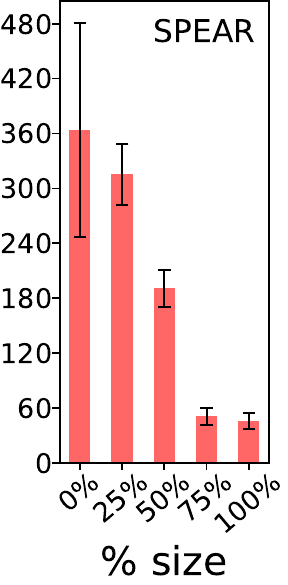}
    \end{subfigure}

     \begin{subfigure}[t]{0.18486\columnwidth}
    \includegraphics[width=\columnwidth]{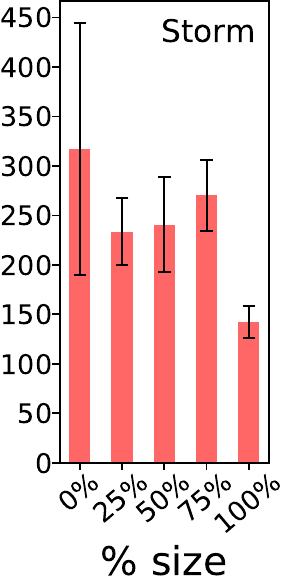}
    \end{subfigure}
 ~\hspace{-0.3cm}
     \begin{subfigure}[t]{0.1746\columnwidth}
    \includegraphics[width=\columnwidth]{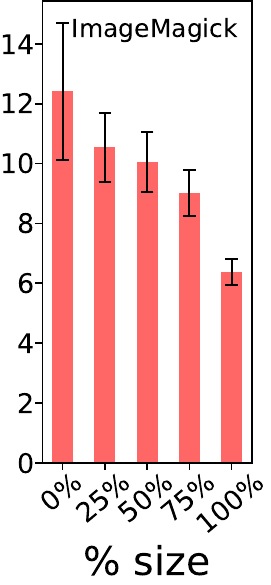}
    \end{subfigure}
 ~\hspace{-0.3cm}
     \begin{subfigure}[t]{0.1746\columnwidth}
    \includegraphics[width=\columnwidth]{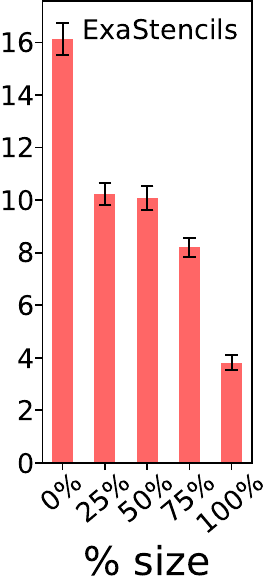}
    \end{subfigure}
 ~\hspace{-0.3cm}
     \begin{subfigure}[t]{0.1746\columnwidth}
    \includegraphics[width=\columnwidth]{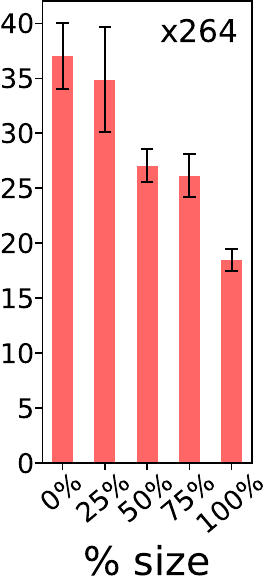}
    \end{subfigure}
    
\caption{Mean MRE and standard error under different percentages of pre-training sizes over 30 runs.}
\label{fig:meta sizes}
\end{figure}

In Figure~\ref{fig:meta sizes}, it can be noted that there is generally a monotonic correlation between the sample size of meta environments and the accuracy: the MRE decreases with more data samples used in the pre-training of \MetaModel. In particular, the improvement occurs from $0\%$ data onward in 8 out of 9 systems, suggesting the importance of meta-learning. The only exception is \textsc{SaC}, which makes the model suffer the ``curse of dimensionality'' since it has $58$ configuration options.

\begin{quotebox}
   \noindent
    {\textbf{RQ4.4:}} While more data samples in the pre-training can be beneficial to \MetaModel, the biggest improvement often occurs as soon as the meta-learning is used (from $0\%$). 
\end{quotebox}

%% file: Tables/chap-meta/rq1-new.tex
\begin{table}[!t]
\caption{Average Scott-Knott ranks for \MetaModel~versus single environment models over all cases/runs. \setlength{\fboxsep}{1.5pt}\colorbox{steel!20}{blue} is the best. Detailed data in Appendix~\ref{appendix-chap-meta-tb:rank_rq1}.}
\centering
\footnotesize
\begin{adjustbox}{width=\linewidth,center}
\begin{tabular}{l|ccccccccc}
\toprule
\textbf{Model} & \rotatebox[origin=c]{0}{\textsc{DArch}} & \rotatebox[origin=c]{0}{\textsc{SaC}} & \rotatebox[origin=c]{0}{\textsc{SQLite}} & \rotatebox[origin=c]{0}{\textsc{NGINX}} & \rotatebox[origin=c]{0}{\textsc{SPEAR}} & \rotatebox[origin=c]{0}{\textsc{Storm}} & \rotatebox[origin=c]{0}{\textsc{IM}} & \rotatebox[origin=c]{0}{\textsc{ES}} & \rotatebox[origin=c]{0}{\textsc{x264}}\\ 
\midrule
\MetaModel  & \cellcolor{steel!20}\textbf{1.0}  & \cellcolor{steel!20}\textbf{1.53}  & \cellcolor{steel!20}\textbf{1.15}  & \cellcolor{steel!20}\textbf{1.2}  & \cellcolor{steel!20}\textbf{1.07}  & 3.2  & \cellcolor{steel!20}\textbf{1.3}  & \cellcolor{steel!20}\textbf{1.0}  & \cellcolor{steel!20}\textbf{1.02}  \\ 
\texttt{DeepPerf}  & 2.33  & 2.27  & 2.8  & 2.25  & 2.87  & 3.53  & 1.9  & 3.05  & 2.2  \\ 
\texttt{RF}  & 3.8  & 2.47  & 1.4  & 3.1  & 2.0  & \cellcolor{steel!20}\textbf{1.13}  & 2.1  & 2.1  & 2.54  \\ 
\texttt{DECART}  & 2.73  & 1.67  & 3.15  & 2.15  & 1.87  & 1.73  & 3.1  & 3.35  & 2.9  \\ 
\texttt{SPLConqueror}  & 3.73  & 3.6  & 4.4  & 4.2  & 3.87  & 4.87  & 4.3  & 4.75  & 4.04  \\ 

\bottomrule
\end{tabular}
\end{adjustbox}
\label{chap-meta-tb:rank_rq1}
\end{table}

%% file: Tables/chap-meta/rq2-new.tex
\begin{table}[!t]
\caption{Average Scott-Knott ranks for \MetaModel~versus multi-environment models over all cases/runs. \setlength{\fboxsep}{1.5pt}\colorbox{steel!20}{blue} is the best. Detailed data in Appendix~\ref{appendix-chap-meta-tb:rank_rq2}.}
\centering
\footnotesize
\begin{adjustbox}{width=\linewidth,center}
\begin{tabular}{l|ccccccccc}
\toprule
\textbf{Model} & \rotatebox[origin=c]{0}{\textsc{DArch}} & \rotatebox[origin=c]{0}{\textsc{SaC}} & \rotatebox[origin=c]{0}{\textsc{SQLite}} & \rotatebox[origin=c]{0}{\textsc{NGINX}} & \rotatebox[origin=c]{0}{\textsc{SPEAR}} & \rotatebox[origin=c]{0}{\textsc{Storm}} & \rotatebox[origin=c]{0}{\textsc{IM}} & \rotatebox[origin=c]{0}{\textsc{ES}} & \rotatebox[origin=c]{0}{\textsc{x264}}\\ 
\midrule
\MetaModel  & \cellcolor{steel!20}\textbf{1.67}  & \cellcolor{steel!20}\textbf{1.2}  & \cellcolor{steel!20}\textbf{1.2}  & \cellcolor{steel!20}\textbf{1.55}  & \cellcolor{steel!20}\textbf{1.6}  & 2.73  & \cellcolor{steel!20}\textbf{1.9}  & \cellcolor{steel!20}\textbf{1.15}  & \cellcolor{steel!20}\textbf{1.48}  \\ 
\texttt{MAML}  & 3.53  & 2.67  & 3.95  & 7.0  & 4.6  & 3.73  & 2.8  & 7.65  & 4.16  \\ 
\texttt{MetaSGD}  & 4.2  & 1.93  & 3.4  & 4.65  & 3.87  & 3.4  & 2.5  & 4.0  & 3.9  \\ 
\texttt{BEETLE}  & 3.67  & 2.4  & 2.25  & 3.1  & 2.07  & 3.4  & 2.5  & 4.75  & 2.04  \\ 
\texttt{tEAMS}  & 1.8  & 3.47  & 2.55  & 1.85  & 4.67  & 4.87  & 4.4  & 1.85  & 4.2  \\ 
\texttt{MORF}  & 5.73  & 2.4  & 1.85  & 4.25  & 3.07  & \cellcolor{steel!20}\textbf{1.0}  & 3.25  & 2.95  & 4.78  \\ 
\texttt{RF$+_{e}$}  & 5.93  & 5.07  & 5.15  & 2.8  & 4.47  & 7.73  & 4.7  & 3.65  & 4.18  \\ 
\texttt{DeepPerf$+_{e}$}  & 2.8  & 5.8  & 6.5  & 3.0  & 4.6  & 7.07  & 3.7  & 6.3  & 4.06  \\ 
\texttt{SPLConqueror$+_{e}$}  & 6.13  & 6.4  & 6.0  & 6.0  & 7.13  & 6.4  & 6.7  & 6.4  & 7.2  \\

\bottomrule
\end{tabular}
\end{adjustbox}
\label{chap-meta-tb:rank_rq2}
\end{table}

%% file: Tables/chap-meta/rq3-new.tex
\begin{table}[!t]
\caption{Average Scott-Knott ranks for optimal sequence versus random order in \MetaModel~over all cases/runs. \setlength{\fboxsep}{1.5pt}\colorbox{steel!20}{blue} is the best. Detailed data in Appendix~\ref{appendix-chap-meta-tb:rank_rq3}.}
\centering
\footnotesize
\begin{adjustbox}{width=\linewidth,center}
\begin{tabular}{l|ccccccccc}
\toprule
\textbf{Model} & \rotatebox[origin=c]{0}{\textsc{DArch}} & \rotatebox[origin=c]{0}{\textsc{SaC}} & \rotatebox[origin=c]{0}{\textsc{SQLite}} & \rotatebox[origin=c]{0}{\textsc{NGINX}} & \rotatebox[origin=c]{0}{\textsc{SPEAR}} & \rotatebox[origin=c]{0}{\textsc{Storm}} & \rotatebox[origin=c]{0}{\textsc{IM}} & \rotatebox[origin=c]{0}{\textsc{ES}} & \rotatebox[origin=c]{0}{\textsc{x264}}\\ 
\midrule
\MetaModel  & 1.53  & \cellcolor{steel!20}\textbf{1.0}  & \cellcolor{steel!20}\textbf{1.0}  & \cellcolor{steel!20}\textbf{1.6}  & \cellcolor{steel!20}\textbf{1.73}  & \cellcolor{steel!20}\textbf{1.33}  & \cellcolor{steel!20}\textbf{1.3}  & \cellcolor{steel!20}\textbf{1.0}  & \cellcolor{steel!20}\textbf{1.18}  \\ 
\texttt{DeepPerf$+$}  & 2.4  & 5.27  & 4.65  & 2.75  & 2.2  & 3.4  & 2.55  & 3.95  & 2.76  \\ 
\texttt{\MetaModel$_{RF}$}  & \cellcolor{steel!20}\textbf{1.47}  & 3.53  & 3.85  & 2.4  & 2.87  & 2.2  & 2.7  & 2.45  & 2.62  \\ 
\texttt{RF$+$}  & 4.4  & 3.07  & 3.45  & 2.65  & 2.47  & 2.67  & 3.2  & 2.15  & 3.6  \\ 
\texttt{\MetaModel$_{SPL}$}  & 4.33  & 2.2  & 1.95  & 4.9  & 4.33  & 5.07  & 5.05  & 4.15  & 4.62  \\ 
\texttt{SPLConqueror$+$}  & 4.8  & 4.73  & 4.8  & 5.4  & 4.73  & 5.4  & 5.25  & 4.35  & 5.32  \\

\bottomrule
\end{tabular}
\end{adjustbox}
 \label{chap-meta-tb:rank_rq3}
\end{table}

%% file: Chapter-meta/discussion.tex
\section{Discussion}
\label{chap-meta:discussion}

This section discusses the specialty, applications, and overhead of \MetaModel, and then illustrates the answer to \textbf{RQ4}, the limitations, publications, and future work of this chapter.

\subsection{What Make \MetaModel~Special for Configuration Performance Learning?}

The proposed sequential pre-training is a specialization of meta-learning for the unique characteristics of configuration data between different environments (as discussed in Section~\ref{chap-meta:premises}). As such, \MetaModel~is specifically designed based on my understanding of the software engineering task at hand.

Indeed, it is impractical to claim that \MetaModel~can completely eliminate the potentially misleading meta environments, nor is it feasible to do so since this may incur significant overhead. However, with the sequential pre-training, \MetaModel~is able to at least mitigate the effect of those misleading meta environments by placing them earlier in the sequence. This, as it has been shown, can obtain significant improvement in accuracy over classic frameworks like \texttt{MAML}, which treats all the meta environments equally. As discussed in Section~\ref{chap-meta:premises}, even though certain meta environments may be misleading overall, some of their parameter distribution could still fit well with those needed for the target environment. Hence, mitigating their effect could be more suitable than their complete omission.

\subsection{Practical Application and Overhead of \MetaModel}

\MetaModel~was designed under the assumption that there are available data measured under different environments of a running configurable software system, which is not uncommon. Therefore, the applications of \MetaModel~in real-world cases are straightforward. Thanks to the sequential pre-training, the contributions of meta environments can be discriminated with no manual selection required. Hence, the more useful ones will contribute more to the learning. Of course, like any learning models, software engineers would need to decide on the training sample size to be measured for the target environment, together with whether the default base learner in \MetaModel, i.e., \texttt{DeepPerf} that favors accuracy, is suitable for the specific scenario as the training of the base learner would constitute the majority of the \MetaModel's overhead during pre-training.

On a machine with CPU 2GHz and 16GB RAM, \MetaModel~only needs 2 and 60 seconds to find the optimal sequence of learning meta environments on a system with 3 and 10 meta environments, respectively, thanks to the linear regression. Of course, the overhead can increase with more meta environments, but the magnitude is still acceptable since the pre-training is an offline process.

\subsection{Answer to RQ4}

Drawing from the answers to the four sub-RQs and the discussions above, it can be concluded that \textbf{objective 4} in this thesis is accomplished, where \MetaModel~can leverage prior knowledge studied from the meta-environments via sequential meta-learning, and with rDNN as the base learner, its accuracy is better than the state-of-the-art models in most the experiments. Thereafter, the final RQ in this thesis can be addressed as:

\begin{answerbox}
\emph{\textbf{Answer to RQ4:} This thesis successfully improves the prediction accuracy under multiple environments by \MetaModel, a sequential meta-learning model with efficient sequence selection proposed in this chapter.}
\end{answerbox}

\subsection{Limitations}

The potential limitations in this study could be in three folds. 
\begin{itemize}
    \item During the course of \MetaModel, the most recent deep learning models, such as \Model~and \texttt{HINNPerf}, were not available in published form. Consequently, these models were not included in the examination conducted in this thesis, resulting in limited validation of this chapter.
    \item The optimal sequence for meta-learning is environment-specific, which means it needs to be rerun for any new target environment. However, the sequence selection process is very fast, with 2-60 seconds for 2-9 meta-environments (thanks to \texttt{LR}).
    \item The prediction accuracy of \MetaModel~is generally not as good as the results of tree-based models like \texttt{RF} for \textsc{Storm}. This is due to a combined effect of the highly sparse samples (because of the system’s nature) and more diverse environments (only hardware changes).
    \item When meta-data from the meta-environments is not enough, the performance of \MetaModel~might be degraded. This is a natural problem for all the multi-environment learning approaches, yet, as it has been demonstrated in RQ4.4, even just using 25\% of meta-training size, \MetaModel~can improve the accuracy in 8 out of 9 systems compared to the cases without meta-learning.
\end{itemize}

Hence, although \MetaModel~is constrained in a few ways, it is still evidently effective, and there could be future studies based on \MetaModel.

\subsection{Publication and Future Work}
\label{chap-meta:future_work}
A conference paper has been extracted from this chapter and has been accepted by the \textit{The ACM International Conference on the Foundations of Software Engineering (FSE) 2024}, which is a top-tier conference in the field of Software Engineering.

The first future work to do for this study is to explore new base learning models, such as the recently published paper in TOSEM'23~\citep{DBLP:journals/tosem/ChengGZ23}, since it achieves better results than the DNN used in this chapter. A possible challenge is which model parameters should be updated in the meta-learning process, and the way of pre-training these parameters using meta-data, as the model structure of this paper is more complicated than traditional MLP.

In addition, in order to mitigate the sparsity of performance functions for some software systems like \textsc{Storm}, it is worth trying to combine \MetaModel~with sparsity handling techniques, for example, simply applying the ``divide-and-learn'' framework from Chapter~\ref{chap:dal}.

Further, it is also worth exploring incremental learning or lifelong learning approaches to strengthen the adaptability and generalizability of SeMPL. For example,~\citet{ChenLiDOS22} proposes \texttt{LiDOS}, which employs the idea of lifelong learning to continuously adapt to the changing environment for configuration tuning of self-adaptive systems. Moreover,~\citet{DBLP:conf/aspdac/WangHFHSYZL22} leverage incremental learning to solve the problem of catastrophic forgetting, where the model weights that are related to the previous samples are frozen/preserved and the rest are updated incrementally. With similar techniques as the above, it is promising to keep the most important information from the previous environments while learning new knowledge from upcoming environments, thereby improving the effectiveness of \MetaModel.

%% file: Chapter-meta/threats_to_validity.tex
\section{Threats to Validity}
\label{chap-meta:threats}

{\textbf{Internal threats}} is concerned with the parameter settings, e.g., the training size. This is mitigated by following the same settings used in the state-of-the-art studies~\citep{DBLP:journals/sqj/SiegmundRKKAS12, DBLP:conf/kbse/GuoCASW13, DBLP:conf/icse/HaZ19, DBLP:conf/esem/ShuS0X20}. This chapter also examined the sensitivity of \MetaModel~to the pre-training sample size in RQ4.4. \MetaModel~is tested with \texttt{DeepPerf} as the base learner, which can be flexibly replaced. Of course, as in RQ4.3, replacing \texttt{DeepPerf} could lead to different results but this is up to the software engineers to decide if other factors, such as training efficiency, are more important. The use of linear regression as a surrogate of the base learner might also raise threats to internal validity since it might not be perfectly accurate. However, the results show this still leads to significant improvement compared with the state-of-the-art.

{\textbf{Construct threats}} are mainly related to the evaluation metric. In this chapter, MRE is used as the main accuracy metric as it is less sensitive to the scales in different environments and systems while being the most common metric from the literature of configuration performance learning~\citep{DBLP:conf/icse/HaZ19, DBLP:conf/esem/ShuS0X20, DBLP:journals/ese/GuoYSASVCWY18}. To measure the speedup, the same protocol is followed as used in prior work~\citep{DBLP:conf/icse/0003XC021,DBLP:conf/sigsoft/0001L21} in the field.

{\textbf{External threats}} could lie in the systems studied including the environments. To mitigate such, this thesis selects the datasets from a diverse set of systems that are collected by the leading researchers~\citep{DBLP:journals/tse/KrishnaNJM21, DBLP:conf/sigsoft/JamshidiVKS18, LESOIL2023111671}. For each of these, all the eligible environments for evaluation are selected. The experiments are also repeated 30 times, which tends to be sufficient to reduce randomness. The Scott-Knott test is adopted to ensure the statistical significance of the results. However, it is worth acknowledging that using more systems and environments might be beneficial.

%% file: Chapter-meta/conclusion.tex
\section{Chapter Summary}
\label{chap-meta:conclusion}
To deal with multiple environments when predicting performance for configurable software systems (the last research objective in the thesis), this chapter proposes a new category of framework that leverages meta-learning and deep learning, dubbed \MetaModel. What makes \MetaModel~unique is that it learns the meta environments, one at a time, in a specific order according to the likely usefulness of the meta environment to the unforeseen target environment. Such sequential learning, unlike the existing parallel learning, naturally allows the pre-training to discriminate the contributions between meta environments, thereby handling the largely deviated samples from different environments---a key characteristic of the configuration data. Further, although the base learner for \MetaModel~is a regularized deep neural network, it can be applied with different types of models.


Extensive evaluations of \MetaModel~on 9 real-world multi-environment datasets of diverse domains and scales and the comparisons with the SOTA performance learning models demonstrate that:

\begin{itemize}
    \item In 8 out of 9 systems, \MetaModel~demonstrates significantly superior performance compared to state-of-the-art single environment performance models, achieving a remarkable improvement of $99\%$ in terms of MRE. Additionally, \MetaModel~exhibits notable data efficiency, delivering a speedup of up to $3.86$ times.

    \item When compared to state-of-the-art models designed to handle multiple environments, \MetaModel~ranks first in 89\% of the cases, achieving the best improvement in MRE ranging from 74\% to 99\%. Moreover, \MetaModel~demonstrates data efficiency in most systems with a speedup greater than or equal to 1, and achieves the highest $sp$ of $2.13$ times. 

    \item By comparing with random sequences in the meta-training phase, it is proven that the sequence selection algorithm in \MetaModel~is able to improve the performance of a large number of systems, ranging from 56\% to 100\% based on the base learning model.

    \item Although \MetaModel~put some requirements on pre-training data, it is shown that starting from 25\% data from the meta-environments, \MetaModel~improves the accuracy in 8 out of 9 subject systems.
\end{itemize}

As such, \textbf{RQ4} of this thesis can be answered that: by leveraging knowledge from related environments in an optimized sequence, the proposed \MetaModel~model can outperform both the SOTA single-environment and multi-environment performance models in the majority of cases.

In future research, it is promising to explore the use of the latest deep learning models as the base learner to further enhance performance. Additionally, investigating techniques to handle sparse systems and accommodate diverse environments with lifelong learning and incremental learning could offer opportunities for further improvements in the proposed approach.

%% file: Chapter-discussion/chapter-discussion.tex
\chapter{Discussion}
\label{chap:discussion}
The primary aim of this thesis is to push the boundaries of deep learning performance models by enhancing their accuracy. In pursuit of this aim, four specific goals were set out, as outlined in Section~\ref{chap-intro:objectives}. While these chapters have effectively addressed the research objectives, the extent to which the aim of this thesis has been accomplished has not been evaluated. Therefore, this chapter serves as a venue for comprehensive discussion and analysis of the outcomes of this thesis.

By pursuing the research aim and objectives, this thesis has resulted in five technical articles submitted or published in top Software Engineering proceedings, i.e., a systematic literature review (Chapter~\ref{chap:review}), an empirical study (Chapter~\ref{chap:encoding}), and three papers about deep learning models for performance prediction (two from Chapter~\ref{chap:dal} and one from Chapter~\ref{chap:meta}). While the contributions of each study have been summarized in corresponding chapters, they have not been synthesized and discussed as a whole. Therefore, in this chapter, the contributions to the existing literature will be introduced.

Moreover, this thesis not only contributes theoretical knowledge but also provides practical insights for researchers and practitioners. The implications and significance of the studies will be thoroughly justified in this chapter.

Additionally, although the previous chapters have highlighted the limitations specific to each study, the overall constraints and weaknesses in the design and implementation of this thesis have not been acknowledged. This chapter will provide a comprehensive overview of these limitations.

In summary, this chapter aims to discuss the accomplishments of the research objectives, outline the contributions to the field, justify the implications and significance of the studies, and address the limitations and weaknesses of this thesis as a whole.

\section{Has the Research Aim Been Accomplished?}

Recall from Section~\ref{chap-intro:aim}, the \textbf{aim} of this thesis is: 

\begin{displayquote}
\textit{To identify and overcome current limitations of performance modeling using deep learning, enable more accurate, robust, and reliable performance predictions for configurable software systems, thereby pushing the boundaries of what can be achieved in this domain.}
\end{displayquote}

To achieve this aim, four key objectives were established, each of which plays a crucial role in fulfilling the aim. Specifically, \textbf{objective 1}, aiming at the systematic review and categorization of the current literature, directly supports the aim by identifying the existing limitations and knowledge gaps in performance modeling using deep learning, which is essential for overcoming the identified limitations and advancing the field. \textbf{Objective 2}, focusing on investigating the impact of encoding schemes, contributes to the aim by fulfilling one of the knowledge gaps in understanding the behavior of encoding schemes under different conditions, which enables researchers to make informed decisions to optimize model accuracy. Moreover, \textbf{objecttive 3} seeks to design a deep learning-based model that effectively tackles sparsity---the limitation identified from the SLR, thereby improving the reliability and accuracy of predictions that coincide with the aim. Lastly, in \textbf{objective 4}, the target of enabling more accurate performance predictions in dynamic environments is realized by designing a DL-based model that can adapt and generalize across multiple environments, which is a critical constraint for configurable software systems.


In the process of fulfilling the aim of this thesis, the satisfaction of the four key objectives is evident in the preceding chapters. In particular, Chapter~\ref{chap:review} presents a systematic literature review that surveys 85 latest deep learning studies in software performance modeling, providing taxonomies, positive trends, and negative trends of techniques used in the deep learning pipeline, and directions that need to be addressed by future studies, thereby, \textbf{objective 1} is addressed. 

Then, Chapter~\ref{chap:encoding} achieves \textbf{objective 2} by performing a comprehensive empirical study on the influence of encoding schemes. Through this investigation, the chapter presents valuable insights into the behavior of different encoding approaches on performance modeling under different situations, demonstrates the necessity and expensiveness of searching for the best encoding scheme, and gives actionable suggestions on choosing the most suitable encoding method with different requirements, which enables the readers to optimize their encoding strategies and create more explainable and reliable performance models. 

Additionally, the invention of \Model, which effectively tackles the sparsity problem by a 'divide-and-learn' Framework combined with Hierarchical Interaction Neural Network, is discussed in Chapter~\ref{chap:dal}. Experiments on 12 software systems in various domains and scales with 5 training sizes have demonstrated the remarkable accuracy improvements of \Model, thereby pushing the boundaries of performance modeling using deep learning in static-environment scenarios and fulfilling \textbf{objective 3}.

Finally, Chapter~\ref{chap:meta} introduces \MetaModel---a novel sequential meta-learning framework combined with DNN to tackle the multi-environment problem in performance modeling, which demonstrates the ability to leverage the hidden information in different meta-environments in a sequential training process, with an efficient sequence selection algorithm. In evaluating 9 subject systems with 5 training sizes, \MetaModel~is better than the existing models in $89\%$ of the systems with up to $99\%$ accuracy improvement; it is also data-efficient with at most $3.86$ times speedup. Therefore, \textbf{objective 4} is reached. 

Collectively, these chapters provide compelling evidence of the successful fulfillment of the four key objectives, and the achievement of the research aim. Formally, it can be concluded that:
\begin{answerbox}
    \textit{The current knowledge gaps in the literature have been identified in the SLR, and by investigating the encoding schemes, resolving the sparsity problem, and addressing the multi-environment performance learning problem, the prediction accuracy of deep learning performance models for highly configurable software has been pushed to the next level.}
\end{answerbox}

\section{What Does This Thesis Add to the Literature?}

Throughout the progression of this thesis, a series of significant achievements have been accomplished, as outlined in the previous chapters. Nonetheless, there is a lack of alignment between these achievements and the existing literature on performance prediction, including beyond the domain of deep learning. In this section, a comprehensive discussion on the introduced contributions of this thesis to the existing literature is presented.


\subsection{A Survey on Deep Learning for Performance Modeling}
Although a number of surveys have been conducted in the literature on performance modeling, this thesis stands out as the very initial one to undertake a systematic literature review of deep learning techniques employed in performance prediction of highly configurable software.

In particular, there are a few surveys related to the topic of performance modeling. Among others,~\citet{DBLP:journals/pe/BalsamoPI03} conduct a review of performance prediction using Queuing Network, and~\citet{DBLP:journals/tse/BalsamoMIS04} review how performance model can be used to help software development.~\citet{DBLP:journals/sigmetrics/NambiarKBSD16} emphasize the significance of performance modeling in SE tasks and calls for further research to enhance the predictive capabilities of performance models, and~\citet{DBLP:conf/wosp/PereiraA0J20} seek to understand to what extent are sampling strategies sensitive to performance prediction of configurable systems. A recent work~\citep{DBLP:journals/spe/HanYY23} classifies software performance learning studies into 6 categories, provides a mapping of them within these categories, and highlights potential weaknesses of the literature. Further, surveys on performance modeling for specific categories of configurable software systems exist, for example, there are reviews on the performance models for distributed systems~\citep{DBLP:conf/cisis/PllanaBB07,DBLP:journals/tpds/Lopez-NovoaMM15} and~\citet{DBLP:journals/tjs/Flores-Contreras21} survey the performance prediction methods for parallel applications published between 2005 and 2020. Similarly,~\citet{DBLP:conf/sc2/FrankHLB17} review 34 studies on performance modeling for multi-core systems, and performance learning and tuning techniques for mobile application systems are reviewed by~\citet{DBLP:journals/tse/HortKSH22}. However, the above work does not focus on how deep learning can be used in the performance model building for configurable software in general. 

On the other hand, several surveys have been conducted on the application of deep learning in software engineering. For example,~\citet{DBLP:journals/csur/YangXLG22} investigates the deep learning-related techniques for software engineering tasks, including the DL models, data preprocessing methods, SE tasks, and model optimization methods, while an SLR on deep learning is also conducted by~\citet{DBLP:journals/tosem/WatsonCNMP22}, which provides a research roadmap and guidelines for future exploration in the context of software engineering.~\citet{DBLP:journals/tosem/LiuGXLGY22} investigates the reproducibility and applicability of DL studies in SE, and observes that a significant number of them have overlooked this challenge. A novel work~\citep{DBLP:journals/tse/WangHGGZFSLZN23} explores the use of ML/DL techniques in various SE tasks and the challenges and differences between ML and DL. Yet, these studies do not focus on a specific SE task like performance learning, but rather provide a general review of software engineering. 

Moreover, an SLR by~\citet{DBLP:journals/jss/PereiraAMJBV21} explores the application objectives, sampling, learning and measuring methods applied, and evaluation methods related to machine learning models for performance modeling, which is the most similar to our work. However, the focus of this study is on deep learning models, which is a specific type of machine learning, and cover a wider range of techniques such as preprocessing, hyperparameter tuning, and sparsity handling methods. Besides, they did not discuss and justify the good and bad practices in the field.

In summary, the survey in this thesis contributes to the existing literature on the following points:

\begin{itemize}
    \item This SLR is the very early one that focuses explicitly on deep learning-based performance modeling for configurable software, which has been an emergent and powerful approach over the past decades.
    \item The review covers aspects that have not been summarized before, such as the processes of data preprocessing, encoding, sparsity handling, hyper-parameter tuning, statistical test and effect size test, dynamic environments, and public repository.
    \item The study discloses and justifies the positive and negative practices, which have not been revealed previously, with a summary of knowledge gaps to shed light on future works in the field.
\end{itemize}


\subsection{An Empirical Study on Encoding Schemes for Performance Modeling}
While there are only a few studies that justify the rationale for choosing the encoding scheme in the field of performance learning, the importance of choosing the encoding schemes for building machine learning models has been discussed frequently in other domains. For example, \citet{9020560} compares the most common encoding schemes for predicting security events using logs. The result shows that it is considerably harmful to encode the representation without systematic justification. Similarly, \citet{DBLP:journals/bmcbi/HeP16} study the effect of two encoding schemes for genetic trait prediction, and a thorough analysis of the encoding schemes has been provided, together with which could be better under what cases. 

However, those findings cannot be directly applied in the context of software performance learning, due to two of its properties:

\begin{itemize}
\item Sampling from the configurable systems is rather expensive~\citep{DBLP:conf/icml/ZuluagaSKP13,DBLP:journals/corr/abs-1801-02175,DBLP:conf/mascots/JamshidiC16, DBLP:conf/icse/HaZ19}, thus the sample size is often relatively smaller.
\item Software configuration is often sparse, as it has been shown in Chapter~\ref{chap:review} and~\ref{chap:dal}. This is because options like \texttt{cache}, when enabled, can create significant implications to the performance, but such a change is merely represented as a one-bit difference in the model. Therefore, the distribution of the data samples can be intrinsically different from the other domains.
\end{itemize}

Therefore, this thesis is one of the earliest ones in the literature to conduct an empirical study on the influence of encoding schemes for machine and deep performance models. Most importantly, this work demonstrates the importance and expansiveness of choosing the encoding scheme, and provides an in-depth analysis of the influence of encoding schemes, together with insights and suggestions under different circumstances.

\subsection{A State-of-the-art Performance Model for Addressing Sparsity in Performance Learning}
In the existing literature on performance prediction, a variety of approaches have been utilized, including analytical models, traditional data-driven models, ensemble learning frameworks, and emergent deep learning models. Yet, a novel model \Model~that leverages the concept of ``divide-and-learn'' and deep learning is proposed in this thesis, which remarkably addresses the sparsity issues and achieves state-of-the-art accuracy as shown in Chapter~\ref{chap:dal}. However, there is a lack of comprehensive analysis and discussion over the current literature.

To bridge this gap, this section introduces the existing performance models, and discusses the advantages of \Model~over them.

\subsubsection{Analytical models} 

Predicting software performance can be done by analyzing the code structure and architecture of the systems~\citep{di2004compositional,DBLP:conf/icse/VelezJSAK21}. For example,~\citet{di2004compositional} apply queuing network to model the latency of requests processed by the software.~\citet{DBLP:conf/icse/VelezJSAK21} use local measurements and
dynamic taint analysis to build a model that can predict performance for part of the configuration code. 

However, analytical models require full understanding and access to the software's internal states, which may not always be possible/feasible.  Further, due to the theoretical assumption or limitations of the measurement tool, analytical models are often restricted to certain performance metrics, most commonly time-related ones such as runtime and latency. \Model~is not restricted to those scenarios as it is a data-driven approach while still being designed to cater to the unique sparse nature of configuration data.

\subsubsection{Statistical learning-based models} 

Data-driven learning has relied on various statistical models, such as linear regressions~\citep{DBLP:journals/sqj/SiegmundRKKAS12,DBLP:conf/sigsoft/SiegmundGAK15,DBLP:journals/tc/SunSZZC20,DBLP:journals/is/KangKSGL20}, tree-liked model~\citep{DBLP:conf/icdcs/HsuNFM18,DBLP:conf/kbse/SarkarGSAC15,DBLP:journals/corr/abs-1801-02175}, fourier-learning models~\citep{zhang2015performance,DBLP:conf/icsm/Ha019}, and even transfer learning~\citep{DBLP:conf/sigsoft/JamshidiVKS18,DBLP:journals/tse/KrishnaNJM21,DBLP:conf/kbse/JamshidiSVKPA17}, \textit{etc}. Among others, \texttt{SPLConqueror}~\citep{DBLP:journals/sqj/SiegmundRKKAS12} utilizes linear regression
combined with different sampling methods and a step-wise feature selection to capture the interactions between configuration options. \texttt{DECART}~\citep{DBLP:journals/ese/GuoYSASVCWY18} is an improved CART with an efficient sampling method~\citep{zhang2015performance}.~\citet{DBLP:conf/mascots/JamshidiC16} use Gaussian Process (GP) to model configuration performance, which is also gradually updated via Bayesian Optimization. However, recent work reveals that those approaches do not work well with small datasets~\citep{DBLP:conf/icse/HaZ19}, which is rather common for configurable software systems due to their expensive measurements. This is a consequence of not fully handling the sparsity in configuration data. Further, they come with various restrictions and prerequisites, e.g., \texttt{DECART} does not work on mixed systems while \texttt{SPLConqueror} needs an extensive selection of the right sampling method(s); GP has also been proved to struggle in modeling the sparse configuration data~\citep{DBLP:journals/corr/abs-1801-02175}.

In contrast, it is shown that \Model~produces significantly more accurate results while not limited to those restrictions. In addition, the paradigm of dividable learning underpins \Model~prevent it from suffering the particular pitfalls of a single machine learning algorithm, as it can be paired with different local models.

\subsubsection{Deep learning-based models}

A variety of studies apply deep neural networks with multiple layers to predict configuration performance~\citep{DBLP:conf/icse/HaZ19, DBLP:conf/esem/ShuS0X20,DBLP:conf/sbac-pad/NemirovskyAMNUC17,app11083706,DBLP:journals/concurrency/FalchE17,DBLP:conf/iccad/KimMMSR17,DBLP:conf/sc/MaratheAJBTKYRG17,DBLP:conf/im/JohnssonMS19,DBLP:journals/jsa/ZhangLWWZH18}. Among others, \texttt{HINNPerf} uses embedding to encodes the configuration into a latent space after which a deep neural network, combined with the hierarchical regularization, is applied to handle feature sparsity. \texttt{DeepPerf}~\citep{DBLP:conf/icse/HaZ19} is a state-of-the-art DNN model with $L_{1}$ regularization to mitigate feature sparsity for any configurable systems, and it can be more accurate than many other existing approaches. The most recently proposed \texttt{Perf-AL}~\citep{DBLP:conf/esem/ShuS0X20} relied on adversarial learning that consists of a generative network to predict the performance and a discriminator network to distinguish the predictions and the actual labels. Nevertheless, existing deep learning approaches capture only the feature sparsity while ignoring the sample sparsity, causing severe risks of overfitting even with regularization in place. 

Compared with those, it has been demonstrated that, by additionally capturing sample sparsity, \Model~is able to improve the accuracy considerably with better efficiency and acceptable overhead.

\subsubsection{Ensemble models}

Similar to the paradigm of dividable learning, ensemble learning also leads to different frameworks that can be paired with different local models, some of which have already been adopted for configuration performance learning. For example, Chen and Bahsoon~\citep{DBLP:journals/tse/ChenB17} propose an ensemble approach, paired with feature selection for mitigating feature sparsity, to model software performance. Other ensemble learning such as Bagging~\citep{breiman1996bagging} and Boosting~\citep{schapire2003boosting} can also be similarly applied. However, those ensemble learning approaches allow different local models to share information at one or more phases of the learning pipeline:

\begin{itemize}
    \item Local models can be trained on the same data~\citep{DBLP:journals/tse/ChenB17}; or data that has been processed sequentially~\citep{schapire2003boosting}.
    \item Local models can make predictions collectively as opposed to individually~\citep{breiman1996bagging}.
\end{itemize}

While information sharing may be useful when different models learn data with high similarity; they might not be effective in coping with highly sparse data. \Model, in contrast, produces isolated local models without sharing knowledge at all phases, including training and prediction. This has been shown to be more suitable for dealing with the sample sparsity exhibited in configuration data, preventing overfitting and memorizing largely spread data samples.

\subsubsection{Hybrid models}

The analytical models can be combined with data-driven ones to form a hybrid model~\citep{DBLP:conf/wosp/HanYP21,didona2015enhancing,DBLP:conf/icse/WeberAS21}. Among others,~\citet{didona2015enhancing} use linear regression and $k$NN to learn certain components of a queuing network. Conversely,~\citet{DBLP:conf/icse/WeberAS21} propose to learn the performance of systems based on the parsed source codes from the system to the function level. 

\Model~is seen as being complementary to those hybrid models due to its flexibility in selecting the local model: when needed, the local models can be replaced with hybrid ones, making itself a hybrid variant. In case the internal structure of the system is unknown, \Model~can also work in its default as a purely data-driven approach.


\subsection{A State-of-the-art Framework for Multi-environment Performance Learning}
Although various approaches have been applied to deal with performance data from multiple environments, they all suffer from certain limitations. In this thesis, a sequential meta-learning and deep learning-based framework---\MetaModel~is newly proposed to address this issue. 

While Chapter~\ref{chap:meta} provides an overview of the multi-environment learning models in general domains, it currently lacks a direct comparative analysis of \MetaModel~comparing to the other joint learning models for performance modeling. To bridge this gap, this section further discusses the existing studies in the literature, clarifies their strengths and shortcomings, and highlights the advantages of \MetaModel~over them.

\subsubsection{Single environment performance models}
As explained in the systematic literature review in this thesis (Chapter~\ref{chap:review}), the performance learning problem is often formed as a single environment problem~\citep{DBLP:journals/ese/GuoYSASVCWY18,DBLP:conf/icse/HaZ19,DBLP:conf/splc/ValovGC15,DBLP:conf/oopsla/QueirozBC16,DBLP:conf/icse/0003XC021,DBLP:journals/tosem/ChengGZ23}, where the influence of environment factors is not considered, such as the inputs/workloads of the software, leading to a series accuracy degradation~\citep{LESOIL2023111671}. 

Although it is possible to consider the environment as an extra feature, as used by~\citet{DBLP:conf/icse/Chen19b}, this is not a universal solution, as the relationship between the environment feature and the original configurations is often complicated, thugs making the modeling of performance even more difficult. Moreover, it has been proved in the evaluations in Section~\ref{chap-meta:evaluation} that simply combining the state-of-the-art models with environment features tends to lead to a worse outcome. Therefore, even though \Model~has outstanding accuracy in single-task performance learning, it can not handle the environmental issues. 

On the other hand, \MetaModel~is able to leverage the concept of ``learning to learn'', where a meta-model is pre-trained with performance data from other environments via sequential meta-training, leading to more accurate predictions.

\subsubsection{Transfer learning for multi-environment performance prediction}
Transfer learning, which seeks to transfer knowledge from one task (source task) to the other (target task) to improve the target task accuracy~\citep{DBLP:journals/tse/KrishnaNJM21}, has been a widely used approach in performance modeling, e.g., \textbf{\texttt{BEETLE}}~\citep{DBLP:journals/tse/KrishnaNJM21}, which elaborates all the available environments to search for the best one to learn jointly using regression trees, and \textbf{\texttt{tEAMS}}~\citep{DBLP:journals/tse/MartinAPLJK22}, which utilize predictions in the source environments as enhanced features and train Gradient Boosting Trees for prediction. However, these models might not generalize well due to their specific assumption on the software system~\citep{torrey2010transfer}. 

Meanwhile, meta-learning models, which aim to pre-train a meta-model from the data in the existing environments, do not require any foreknowledge of the specific target environment, thereby often showcasing better generalization abilities. Specifically, supported by experiments in Section~\ref{chap-meta:discussion}, \MetaModel~can maintain good learning outcomes across a range of software domains, and may require less reliance on domain-specific settings.

Moreover, in contrast to conventional meta-learning models, \MetaModel~adopts a sequential approach to learning meta-data, which allows it to discern and prioritize the learning of ``good'' environments, resulting in state-of-the-art prediction accuracy. This is also proven to be useful in the evaluations involving meta-learning models from other domains, such as \texttt{MAML}~\citep{DBLP:conf/icml/FinnAL17}, as can be found in Section~\ref{chap-meta:discussion}.

\section{What Can Be Taken Away from This Thesis?}
With the above contributions to the current literature of performance prediction using deep learning models, this thesis offers a list of recommendations and suggestions for both researchers and practitioners, which will be delineated in this section.

\subsection{For Researchers}
For researchers in the domain of performance prediction, especially those subject to the method of deep learning, the following takeaways are provided:

\subsubsection{A list of positive and negative trends}
In the SLR presented in Chapter~\ref{chap:review}, the taxonomy, positive and negative practices in each process of the deep learning pipeline are summarized, which could be a good reference to take away when conducting studies in this domain. 

For instance, the SLR highlights a positive trend where various pre-processing techniques are applied to enhance the quality and suitability of data for deep learning models, which should be retained and further explored in future research endeavors. On the other hand, a concerning pattern is revealed, where a majority of studies neglect the promotion of open science by failing to provide source codes and datasets, which should be addressed and improved upon.

By leveraging the information provided in this SLR, researchers can develop a comprehensive understanding of the current techniques and practices in performance prediction using deep learning models, and can make more informed decisions when designing and conducting their own research.

\subsubsection{A collection of knowledge gaps and future directions}
 As specified in Section~\ref{chap-review:discussion}, an examination of knowledge gaps in the current literature is presented to researchers, such as the absence of model-based sampling methods to enhance the quality of samples, and the demand for efficient hyperparameter tuning algorithms to handle the large search space for deep learning models. 
 
 Moreover, although this thesis has addressed a few of the knowledge gaps, there are still future directions in the addressed problems that are promising to make contributions to the current literature for researchers to pursue, e.g., combining the sequential meta-learning framework with the latest performance learners, such as \texttt{HINNPerf}.

\subsubsection{Actionable suggestions on choosing the encoding scheme}
To bridge the knowledge gap pertaining to encoding schemes, this thesis performs an empirical study to examine the impacts of three widely used encoding schemes for software performance learning, namely label encoding, scaled label encoding, and one-hot encoding, which showcases the significance of carefully selecting an appropriate encoding scheme for deep performance learning.

Most importantly, a list of actionable suggestions for choosing the most suitable encoding scheme under various scenarios is provided in Section~\ref{chap-encoding: suggestions}, enabling a deeper understanding of encoding schemes and facilitating improved decision-making processes for researchers. For example, it is recommended to apply one-hot encoding to get more accurate but slower predictions. 

\subsection{For Practitioners}
For practitioners who seek to apply performance models in various software engineering tasks, e.g., configuration design, performance optimization, and performance testing, this thesis also offers an abundance of insights and recommendations.

\subsubsection{A framework to handle sparse software}
Firstly, by undertaking a systematic study that synthesizes the observations of sparsity in the primary studies, this thesis formalizes the problems of feature sparsity and sample sparsity, and demonstrates the importance of handling this problem.

Then, this thesis offers an innovative framework \Model, whose key idea is to learn the performance data with similar characteristics using a local deep learner, thereby addressing the sparse sample distribution. Combined with a state-of-the-art HINN model with regularization techniques to mitigate the sparse feature influence, \Model~performs significantly better than the best existing performance model. Moreover, Section~\ref{chap-meta-subsec:local} has demonstrated the ability of \Model~to enhance different models when used as the local model. 

Therefore, for practitioners, \Model~will serve as a valuable tool when dealing with complex and sparse software systems, and its robustness and flexibility will make it compatible with any other performance models when employed as the local model.

\subsubsection{A framework to handle dynamic environments}
In real-world scenarios, the environmental conditions of performance modeling tasks are often dynamic, making it difficult to capture the performance behavior. For an example of the video encoder \textsc{x264}, even with the same configuration, the performance could be completely different under two distinct input videos (i.e., workloads), which is hard for single-task performance models like \Model~to capture. 

To tackle this issue, practitioners can employ the sequential meta-learning framework, named \MetaModel, as proposed in this thesis. The framework capitalizes on meta-data from known environments in a sequential manner, facilitating adept adaptation of a deep neural network to the target environment. Taking the example of \textsc{x264}, \MetaModel~employs a process wherein it ranks all existing data under diverse workloads based on their similarity to the target workload. Subsequently, a DNN meta-model is pre-trained in an order that the most similar workloads are trained at the final steps, which enables rapid adaptation to the target input and enhances the accuracy of performance predictions.

With the above, the readers of this thesis can benefit from a collection of useful knowledge and artifacts. However, the limitations of these studies are worth noting.




\section{What Are the Limitations of This Thesis?}
The limitations and threats to the validity of each study in this thesis have been thoroughly addressed and justified in their respective chapters. However, it is essential to recognize and discuss the limitations of the thesis as a whole. This section aims to justify and explore the potential limitations and threats to validity that may impact the outcomes of the research.

\subsection{Time Constraints}
The initiation of my doctoral study occurred in January 2020, concluding in December 2023, thereby spanning an approximate duration of four years. Although this time range is not short, it is important to recognize that the research conducted for this thesis still had certain limitations. 

Specifically, as a beginner researcher, a substantial portion of this period was dedicated to the learning and practicing of research skills, such as reading, data analysis, and writing. Therein, the initial year mainly focused on conducting a comprehensive systematic literature review, while the following year was dedicated to exploring various methods to address identified knowledge gaps. Unfortunately, none of these endeavors produced substantial results until March 2022, when my study on encoding schemes was published at MSR'22. Subsequently, in February 2023, I received acceptance for my work on handling sparsity using \Model~at ESEC/FSE'23. In the final year, three additional pieces of work were developed, including a journal version of my FSE'23 publication, a conference paper on the meta-learning framework of \MetaModel, and an updated journal version of the SLR, all of which have been submitted as outlined in Chapter~\ref{chap:introduction}.

Nonetheless, this period is not enough to address all knowledge gaps identified in the SLR, such as the exploration of sampling methods, and the future directions of the current studies. Given more time, the outcomes of this thesis could have been further enhanced.

\subsection{Scope Limitations}
The scope of this thesis primarily focuses on exploring the deep learning techniques for performance modeling. Although DL models have been demonstrated to be widely used and accurate within the past decade, they have inherent limitations, particularly in terms of their training overhead and explainability. Consequently, the resulting performance models often exhibit high training costs and low explainability, which can hinder their practicability, reliability, and trustworthiness.

Moreover, it is important to acknowledge that there exist numerous state-of-the-art performance models that are beyond the scope of deep learning methodologies, for instance, \texttt{SPLConqueror}, \texttt{DECART} and \texttt{RF}, which have been introduced and evaluated in multiple chapters of this thesis. Yet, due to time and resource constraints, these models lie beyond the scope of the SLR in this thesis. Therefore, expanding the scope to enhance and improve these techniques presents a valuable opportunity for future research endeavors.

\subsection{Methodological Limitations}
As shown in Chapter~\ref{chap:methodology}, the conduct of this thesis follows the Design Science Research Methodology, which aims at designing, developing, and validating artifacts to address specific problems and solve existing practices. Through the conduct of this thesis, a collection of artifacts and publications has been procured; however, it is important to acknowledge certain limitations within the methodology.

One key limitation is the iterative nature of the DSRM process, which, although beneficial for refining the artifact, can also be time-consuming. This iterative aspect may exacerbate the time constraints inherent in a thesis, posing challenges in achieving comprehensive results within the available timeframe.

Moreover, with the rapid pace of development in deep learning models, there exists a risk that the designed artifact may become outdated by the time of publication. For instance, during the study of \Model~in this thesis, which employs deep learning models as the local model, the state-of-the-art deep learning model transitioned from \texttt{DeepPerf} to \texttt{HINNPerf}. Consequently, I had to restart the evaluation process to accommodate the new techniques, which resulted in additional time requirements.

These limitations highlight the need for careful consideration and adaptation within the DSRM framework, particularly when addressing dynamic fields such as deep learning.


\subsection{Data Limitations}
Another possible constraint in this thesis is that the datasets utilized are sourced solely from existing literature, which introduces a lack of transparency in the data collection process. This absence of transparency can lead to potential bias and noise in the data, thereby hindering the reliability and validity of the evaluations and findings. While these datasets have been employed in state-of-the-art studies within the field, it is important to acknowledge their limitations. To address this concern, one feasible solution is to collect and build original datasets by benchmarking software systems and strive to eliminate any bias and randomness in this data collection process.

Additionally, to align with existing studies, the training sample sizes in this thesis are set to be the same as those utilized in state-of-the-art studies like~\citep{DBLP:conf/icse/HaZ19}. For instance, in the case of binary configuration software, the sample sizes range from n to 5n, where n denotes the number of features. Similarly, for categorical, numeric, or mixed software, the sample sizes are determined by the hybrid sampling method employed by~\citet{DBLP:conf/sigsoft/SiegmundGAK15}. However, it is necessary to acknowledge that this choice imposes a potential limitation on this thesis. In future studies, it is recommended to incorporate a wider range of experimental setups in order to offer more comprehensive and reliable evaluations. 




%% file: Chapter-conclusion/chapter-conclusion.tex
\chapter{Conclusion}
\label{chap:conclusion}
Performance stands as a crucial non-functional aspect of software, exerting a direct influence on the user experience. Therefore, performance modeling becomes a crucial task in software engineering, as inappropriate configurations often lead to software bugs and issues. With the increasing complexity of modern software systems, traditional analytical models prove inadequate in capturing the intricate correlations involved. Consequently, this thesis concerns the utilization of deep learning models in performance prediction tasks, which offers the capability to handle the complexities and challenges of performance modeling, providing more accurate and reliable predictions.

In this chapter, the research objectives of this thesis are revisited, the contributions resulting from the objectives are summarized, and the limitations and future work of the studies are discussed. 

\section{Research Storyline}
First of all, following the Design Science Research Methodology, the first activity of this thesis is to identify the aims and objectives. Through primary reading and studies, it is found that few literature reviews exist exclusively on deep learning approaches for performance modeling. Therefore, the \emph{\textbf{first research objective}} seeks to disclose the behaviors of the current state-of-the-art research works, and identify the knowledge gaps for future studies. By doing a \emph{\textbf{systematic literature review}} on the recent data-driven performance models, the first objective is fulfilled, and three additional objectives and research questions are raised, which are about the encoding schemes, the sparsity problem, the dynamic-environment problem, respectively.

Subsequently, the \emph{\textbf{second research objective}} is to assess the importance of choosing the encoding scheme for performance models, and understand which encoding scheme is better and under what conditions. To that end, this thesis conducts an \emph{\textbf{empirical study}}, covering 3 widely used encoding methods for performance learning, 5 real-world subject systems, and 7 prominent machine/deep learning models (leading to 105 cases of investigation). The results demonstrate that choosing the encoding scheme is indeed crucial for performance prediction, and 4 actionable suggestions for researchers to find the most appropriate encoding method are provided.

Thereafter, the \emph{\textbf{third objective}} is to mitigate the problem of feature and sample sparsity in performance data, which may severely decrease the accuracy of deep learning models. To bridge this gap, the thesis develops a \emph{\textbf{`divide-and-learn' framework}} that has three stages. First of all, the training samples are clustered into subsets based on a CART regressor, which can capture the characteristics of each divide of samples well. Then, a deep HINN model is built to learn each subset of samples. Finally, each testing sample is classified into a local HINN using a random forest classifier, and the performance is predicted using the assigned local model. The divide-and-learn framework with HINN as a learning model is named \Model. In the evaluation experiments of 12 software systems, the model has outperformed the state-of-the-art performance prediction models in 44 out of 60 cases, with up to 161\% improvements in accuracy. Hence, the objective 3 is accomplished. 

While \Model~can accurately predict the software performance under a given environment, it can not work in a dynamic environment, where the workload, hardware, and software settings might change. Therefore, the \emph{\textbf{last research objective}} seeks to improve the prediction accuracy by leveraging data from different environments. In this thesis, a \emph{\textbf{sequential meta-learning model}} dubbed \MetaModel~is designed and implemented with a DNN model to resolve this issue. The key observation in this thesis is that by training the meta-environments in a proper sequence, the positive impacts can be accumulated while the environment-specific parameters are tuned in each meta-training epoch. Then, a fast sequence selection algorithm is proposed by this study, which evaluates the single-environment meta-models using an SGD-based linear model, and then decides the final sequence based on the rankings of the single-environment models. In this way, the factorial search space is reduced to quadratic, and the evaluation overhead is largely saved. In the evaluating experiments, \MetaModel~outperforms the state-of-the-art single-task and multi-task learning models in 89\% of the cases with up to $99\%$ accuracy improvement. Therefore, the last research objective is achieved. 

\section{RQs, Contributions and Limitations Revisited}

To make a straightforward revisit of this thesis, Table~\ref{chap-conc-tb:summary} provides a summary of the research questions, the contributions made, and their respective limitations.

\begin{longtable}{|p{0.91\columnwidth}|p{0.09\columnwidth}|}
\caption{Summaries of the RQs, answers, contributions, limitations, and the related chapters in this thesis.}
\label{chap-conc-tb:summary} \\
\hline
\rowcolor[HTML]{C0C0C0} RQ, answer, contribution and limitation                        & Chapter \\ \hline
\textbf{RQ1:} What are the potential gaps to bridge for more accurate performance modeling using deep learning?
\begin{itemize}
\item \textbf{Answer:} Details of the good and bad trends have been summarised in Section~\ref{chap-review:discussion}. Most importantly, the potential knowledge gaps in data-driven performance prediction found in the thesis are: (1) only a few studies justify the selection of encoding scheme, which could be dangerous for the community; (2) sparsity is still a critical limitation in the development of performance models; and (3) most studies only consider the accuracy of modeling under a single environment, ignoring the dynamism of the running environment.

\item \textbf{Contribution:} A systematic literature review that investigates each process of the deep learning pipeline for performance prediction, covering six indexing services, 52 venues and 85 primary studies. Through this rigorous examination, the study provides a taxonomy of the deep learning pipeline, identifies both positive and negative practices, and offers insights into future directions. A journal paper on this study has been accepted by one of the top SE journals: 

    \emph{\underline{Jingzhi Gong}, Tao Chen, Deep Configuration Performance Learning: A Systematic Survey and Taxonomy, \textit{ACM Transactions on Software Engineering and Methodology (TOSEM)}.}

\item \textbf{Limitations:} (1) The SLR could be out-of-date by the publication due to the high development speed of deep learning; (2) It is possible that not all relevant papers were included in the search process; (3) the quality of the selected papers are limited since this study aims to provide an overview of the field, rather than finding the best methods.
\end{itemize}

&
  Chapter \ref{chap:review} \\ \hline
\textbf{RQ2:} Which encoding scheme is better for performance learning, and under what circumstances?
\begin{itemize}
\item \textbf{Answer:} Choosing the encoding scheme is non-trivial for performance learning and it can be rather expensive to do it using trial-and-error in a case-by-case manner. Details of the suggestions on choosing the encoding scheme can be found in Section~\ref{chap-encoding: suggestions}.

\item \textbf{Contribution:} An empirical study that systematically compared the widely used encoding schemes for software performance learning, namely the label, scaled label, and one-hot encoding. The study covers five systems, seven models, and three encoding schemes, leading to 105 cases of investigation, and 4 actionable suggestions for researchers in the field. A paper of this chapter has been published:

\emph{\underline{Jingzhi Gong} and Tao Chen. 2022. Does Configuration Encoding Matter in Learning Software Performance? An Empirical Study on Encoding Schemes. \textit{In 19th IEEE/ACM International Conference on Mining Software Repositories (MSR'22)},
Pittsburgh, PA, USA, May 23-24, 2022. ACM, 482–494.}

\item \textbf{Limitations:} The number of encoding schemes and subject systems examined in this study is limited. Due to time and cost budgets, this thesis has to choose three of the most common encoding methods and five widely utilized software systems in the related studies.
\end{itemize}

&
  Chapter \ref{chap:encoding} \\ \hline
\textbf{RQ3:} Is it possible to improve the prediction accuracy by mitigating the sparsity problem specialises deep learning?
\begin{itemize}
\item \textbf{Answer:} Yes. The thesis proposes a `divide-and-learn' framework that divides the training dataset into subsets and learns each set separately using a deep learning model, in which way, the feature and sample sparsity in each subset are both reduced, and the overall performance of the model is improved.

\item \textbf{Contribution:} A novel model named \Model, that effectively handles the issues of both feature and sample sparsity in software performance prediction, which is able to outperform state-of-the-art performance modeling approaches in 44 out of 60 experiments. A conference paper of \Model~has been published, and the extended journal paper has been accepted by one of the top-tier journals, as shown below:

\emph{\underline{Jingzhi Gong}, Tao Chen and Rami Bahsoon, Dividable Configuration Performance Learning, \textit{IEEE Transactions on Software Engineering (TSE)}. }

\emph{\underline{Jingzhi Gong} and Tao Chen. 2023. Predicting Software Performance with Divide-and-Learn. \textit{In Proceedings of the 31st ACM Joint European Software Engineering Conference and Symposium on the Foundations of Software Engineering (ESEC/FSE’23)}, December 3–9, 2023, San Francisco, CA, USA., 13 pages. }

\item \textbf{Limitations:} (1) Higher training overhead compared to traditional machine learning models such as \texttt{RF} and \texttt{DT}, but similar to other deep learning models; (2) Limited generalizability due to the parameter $d$, which controls the number of divisions, and needs to be carefully selected; (3) Lack of consideration for environmental factors like software inputs, which may impact the accuracy when environment changes.
\end{itemize}

&
  Chapter \ref{chap:dal} \\ \hline
  
\textbf{RQ4:} Is it possible to improve prediction accuracy by tackling the dynamic environment problem with deep learning techniques?
\begin{itemize}
\item \textbf{Answer:} Yes. This thesis has successfully addressed the performance modeling problem under multiple environments by proposing a sequential meta-learning model with a fast sequence selection method and a regularized deep neural network learner.

\item \textbf{Contribution:} A meta-learning approach dubbed \MetaModel~that learns the meta tasks sequentially, such that the common feature representations are incrementally perfected, while the environment-specific representations are fitted quickly during the fine-tuning process, which effectively addresses the problem of multi-environment performance prediction. The proposed model is better than the state-of-the-art single-task and multi-task models in 89\% of the cases, and the conference paper for this study has been published:

\emph{\underline{Jingzhi Gong} and Tao Chen, Predicting Configuration Performance in Multiple Environments with Sequential Meta-Learning, \textit{The ACM Joint European Software Engineering Conference and Symposium on the Foundations of Software Engineering (FSE'24)}, July 15-19, 2024, Porto de Galinhas, Brazil, 24 pages.}

\item \textbf{Limitations:} (1) the latest single-environment performance models like \texttt{HINNPerf} is not included in this study due to time constraints; (2) The optimal sequence for meta-learning is specific to each environment, which implies that it needs to be re-selected whenever a new target environment is encountered; (3) The performance of \MetaModel~on certain systems like \textsc{Storm} is not state-of-the-art; (4) The accuracy of the predictions may be hindered when there is an insufficient amount of meta-data available.
\end{itemize}

&
  Chapter \ref{chap:meta} \\ \hline

\end{longtable}

\section{Future Work}
Through the conduction of this thesis, several promising directions have been identified that are worth exploring in future studies. In a nutshell, they can be summarized as follows.

\begin{itemize}
\item For the research outcomes of this thesis, the future work can include:
    \begin{itemize}
        \item Expanding the time range and incorporating additional indexing services to encompass a broader selection of primary studies for the SLR.
        \item Investigating more encoding schemes, different subject systems, and new deep learning models in the empirical study on encoding schemes, and preparing a journal paper.
        \item Enhancing \Model~by exploring the selection and utilization of mixed local learners, instead of employing the same DNN for all divisions.
        \item Exploring the combination of \MetaModel~with sparsity handling, lifelong learning, or incremental learning techniques to further enhance the prediction accuracy for highly sparse systems, such as \textsc{Storm}.
    \end{itemize}
\item Regarding the knowledge gaps that have not been addressed in this thesis, there are several future practices to consider:
    \begin{itemize}
    \item Explore the use of model-based sampling methods to improve the quality of training samples, thereby enhancing the effectiveness of deep learning models.
    \item Investigate automatic and efficient hyperparameter tuning algorithms to reduce the training overhead associated with deep learning models.
    \item Examine the statistical significance and effect size differences of the experiment results to enhance the validity of the research findings.
    \item Promote open science within the research community by providing and maintaining source codes and datasets, facilitating further advancements and cooperation in the field.
    \end{itemize}

\end{itemize}

\section{Summary}

In this thesis, the primary aim is to push the research community to a higher level in terms of prediction accuracy in performance modeling using deep learning. To achieve this aim, four research objectives and corresponding research questions are established based on the primary study and systematic literature review. Subsequently, comprehensive studies are conducted following the design science research methodology, which encompasses six sequential activities for the design and development of data-driven artifacts.

In the end, this thesis successfully accomplishes the research aim and all of the defined objectives, resulting in two technical papers published in the top software engineering conferences and three innovative works under review. Particularly, the contributions made in this thesis include the identification and addressing of current knowledge gaps in the literature, the investigation of encoding schemes, the resolution of the sparsity issues, and the handling of the multi-environment performance learning problem. Furthermore, this thesis offers a collection of recommendations that can be taken away by readers, including a list of positive and negative trends and future directions in deep learning for performance learning, a set of actionable suggestions on choosing the encoding scheme, and two practical frameworks that can be readily used with deep learning approaches to solve real-world problems. However, limitations exist in this thesis, including the constraints of time, data, and scope, which suggest a vast amount of future work that can be done. 

Through this thesis, the key knowledge gaps identified in the community are fulfilled, and the prediction accuracy of deep learning performance models for highly configurable software has been significantly improved, thereby pushing the boundaries of deep performance learning to the next level.

%% file: Appendix/chap-meta-rq1-full.tex
\chapter{Detailed Data for Table~\ref{chap-meta-tb:rank_rq1}}
\label{appendix-chap-meta-tb:rank_rq1}

\textbf{This appendix presents the detailed data of all subject systems for Table~\ref{chap-meta-tb:rank_rq1}, where \setlength{\fboxsep}{1.5pt}\colorbox{steel!20}{blue} cells stand for the best rank.}

\begin{table}[h]
\caption{The Scott-Knott rank ($r$), mean MRE, and standard error of the mean (SEM) of all target environments and training sizes for \textsc{DeepArch}.}
\centering
\footnotesize
\begin{adjustbox}{width=\linewidth,center}


\end{adjustbox}
\end{table}

\begin{table}[h]
\caption{The Scott-Knott rank ($r$), mean MRE, and standard error of the mean (SEM) of all target environments and training sizes for \textsc{SaC}.}
\centering
\footnotesize
\begin{adjustbox}{width=\linewidth,center}

\end{adjustbox}
\end{table}

\begin{table}[h]
\caption{The Scott-Knott rank ($r$), mean MRE, and standard error of the mean (SEM) of all target environments and training sizes for \textsc{SQlite}.}
\centering
\footnotesize
\begin{adjustbox}{width=\linewidth,center}

\end{adjustbox}
\end{table}

\begin{table}[h]
\caption{The Scott-Knott rank ($r$), mean MRE, and standard error of the mean (SEM) of all target environments and training sizes for \textsc{NGINX}.}
\centering
\footnotesize
\begin{adjustbox}{width=\linewidth,center}

\end{adjustbox}
\end{table}

\begin{table}[h]
\caption{The Scott-Knott rank ($r$), mean MRE, and standard error of the mean (SEM) of all target environments and training sizes for \textsc{SPEAR}.}
\centering
\footnotesize
\begin{adjustbox}{width=\linewidth,center}

\end{adjustbox}
\end{table}

\begin{table}[h]
\caption{The Scott-Knott rank ($r$), mean MRE, and standard error of the mean (SEM) of all target environments and training sizes for \textsc{Storm}.}
\centering
\footnotesize
\begin{adjustbox}{width=\linewidth,center}

\end{adjustbox}
\end{table}

\begin{table}[h]
\caption{The Scott-Knott rank ($r$), mean MRE, and standard error of the mean (SEM) of all target environments and training sizes for \textsc{ImageMagick}.}
\centering
\footnotesize
\begin{adjustbox}{width=\linewidth,center}

\end{adjustbox}
\end{table}

\begin{table}[h]
\caption{The Scott-Knott rank ($r$), mean MRE, and standard error of the mean (SEM) of all target environments and training sizes for \textsc{ExaStencils}.}
\centering
\footnotesize
\begin{adjustbox}{width=\linewidth,center}

\end{adjustbox}
\end{table}

\begin{table}[h]
\caption{The Scott-Knott rank ($r$), mean MRE, and standard error of the mean (SEM) of all target environments and training sizes for \textsc{x264}.}
\centering
\footnotesize
\begin{adjustbox}{width=\linewidth,center}

\end{adjustbox}
\end{table}

%% file: Appendix/chap-meta-rq2-full.tex
\chapter{Detailed Data for Table~\ref{chap-meta-tb:rank_rq2}}
\label{appendix-chap-meta-tb:rank_rq2}

\textbf{This appendix presents the detailed data of all subject systems for Table~\ref{chap-meta-tb:rank_rq2}, where \setlength{\fboxsep}{1.5pt}\colorbox{steel!20}{blue} cells stand for the best rank.}

\begin{table}[h]
\caption{The Scott-Knott rank ($r$), mean MRE, and standard error of the mean (SEM) of all target environments and training sizes for \textsc{DeepArch}.}
\centering
\footnotesize
\begin{adjustbox}{width=\linewidth,center}
\begin{tabular}{p{0.15cm}cp{0.05cm}p{1cm}p{0.05cm}p{1.3cm}p{0.05cm}p{1.3cm}p{0.05cm}p{1cm}p{0.05cm}p{1cm}p{0.05cm}p{1.3cm}p{0.05cm}p{1.3cm}p{0.05cm}p{1cm}p{0.05cm}p{1cm}}
\toprule
\multirow{2}{*}{Size} & \multirow{2}{*}{$\mathbfcal{T}_{target}$} & \multicolumn{2}{c}{\Model} & \multicolumn{2}{c}{\texttt{MAML}} & \multicolumn{2}{c}{\texttt{MetaSGD}} & \multicolumn{2}{c}{\texttt{Beetle}} & \multicolumn{2}{c}{\texttt{tEAMS}} & \multicolumn{2}{c}{\texttt{MORF}} & \multicolumn{2}{c}{\texttt{RF$+_{e}$}} & \multicolumn{2}{c}{\texttt{DeepPerf$+_{e}$}} & \multicolumn{2}{c}{\texttt{SPLConqueor$+_{e}$}} \\ \cline{3-20} 
 &  & $r$ & MRE (SEM) & $r$ & MRE (SEM) & $r$ & MRE (SEM) & $r$ & MRE (SEM) & $r$ & MRE (SEM) & $r$ & MRE (SEM) & $r$ & MRE (SEM) & $r$ & MRE (SEM) & $r$ & MRE (SEM) \\

\hline \multirow{3}{*}{$S_{1}$} & 1 & \cellcolor{steel!20}\textbf{1} & \cellcolor{steel!20}\textbf{82.5 (5.4)} & 3 & 139.9 (8.8) & 4 & 163.3 (10.1) & 2 & 120.2 (5.1) & 2 & 120.3 (6.9) & 6 & 226.5 (14.7) & 7 & 267.9 (22.0) & \cellcolor{steel!20}\textbf{1} & \cellcolor{steel!20}\textbf{83.8 (0.1)} & 5 & 206.2 (8.9) \\
 & 2 & 2 & 76.6 (5.8) & 5 & 172.1 (11.1) & 7 & 217.3 (14.1) & 4 & 152.3 (6.5) & 3 & 139.5 (9.5) & 8 & 342.1 (25.5) & 6 & 189.4 (14.3) & \cellcolor{steel!20}\textbf{1} & \cellcolor{steel!20}\textbf{56.6 (0.0)} & 3 & 142.8 (5.1) \\
 & 3 & \cellcolor{steel!20}\textbf{1} & \cellcolor{steel!20}\textbf{60.4 (3.8)} & 4 & 118.9 (8.5) & 5 & 161.3 (15.7) & 5 & 161.2 (7.8) & 3 & 87.8 (7.0) & 7 & 239.5 (17.4) & 8 & 271.7 (23.6) & 2 & 66.4 (0.1) & 6 & 197.5 (8.2) \\
\hline \multirow{3}{*}{$S_{2}$} & 1 & \cellcolor{steel!20}\textbf{1} & \cellcolor{steel!20}\textbf{62.7 (3.6)} & 4 & 86.8 (5.8) & 4 & 88.7 (6.3) & 3 & 84.4 (4.2) & 2 & 77.3 (4.9) & 5 & 144.6 (8.2) & 6 & 156.3 (8.7) & 3 & 83.1 (0.1) & 6 & 159.6 (6.3) \\
\bottomrule
\end{tabular}
\end{adjustbox}
\end{table}

\begin{table}[h]
\caption{The Scott-Knott rank ($r$), mean MRE, and standard error of the mean (SEM) of all target environments and training sizes for \textsc{SaC}.}
\centering
\footnotesize
\begin{adjustbox}{width=\linewidth,center}
\begin{tabular}{p{0.15cm}cp{0.05cm}lp{0.05cm}lp{0.05cm}lp{0.05cm}lp{0.05cm}lp{0.05cm}lp{0.05cm}lp{0.05cm}lp{0.05cm}l}
\toprule
\multirow{2}{*}{Size} & \multirow{2}{*}{$\mathbfcal{T}_{target}$} & \multicolumn{2}{c}{\Model} & \multicolumn{2}{c}{\texttt{MAML}} & \multicolumn{2}{c}{\texttt{MetaSGD}} & \multicolumn{2}{c}{\texttt{Beetle}} & \multicolumn{2}{c}{\texttt{tEAMS}} & \multicolumn{2}{c}{\texttt{MORF}} & \multicolumn{2}{c}{\texttt{RF$+_{e}$}} & \multicolumn{2}{c}{\texttt{DeepPerf$+_{e}$}} & \multicolumn{2}{c}{\texttt{SPLConqueor$+_{e}$}} \\ \cline{3-20} 
 &  & $r$ & MRE (SEM) & $r$ & MRE (SEM) & $r$ & MRE (SEM) & $r$ & MRE (SEM) & $r$ & MRE (SEM) & $r$ & MRE (SEM) & $r$ & MRE (SEM) & $r$ & MRE (SEM) & $r$ & MRE (SEM) \\

\hline \multirow{3}{*}{$S_{1}$} & 1 & 2 & 1353.0 (191.7) & \cellcolor{steel!20}\textbf{1} & \cellcolor{steel!20}\textbf{775.0 (148.5)} & \cellcolor{steel!20}\textbf{1} & \cellcolor{steel!20}\textbf{909.4 (124.4)} & 2 & 1358.2 (237.9) & 2 & 1317.1 (98.8) & 3 & 2437.0 (231.2) & 3 & 2611.2 (218.0) & 4 & 2811.5 (50.3) & 3 & 2568.8 (245.6) \\
 & 2 & 2 & 105.4 (8.4) & 2 & 105.1 (8.1) & \cellcolor{steel!20}\textbf{1} & \cellcolor{steel!20}\textbf{91.7 (9.1)} & \cellcolor{steel!20}\textbf{1} & \cellcolor{steel!20}\textbf{88.1 (7.1)} & \cellcolor{steel!20}\textbf{1} & \cellcolor{steel!20}\textbf{80.9 (16.3)} & 2 & 113.9 (8.5) & 3 & 191.0 (5.1) & 4 & 207.1 (1.0) & 5 & 346.5 (12.3) \\
 & 3 & \cellcolor{steel!20}\textbf{1} & \cellcolor{steel!20}\textbf{41.9 (6.7)} & 3 & 81.4 (8.9) & 3 & 73.9 (9.6) & 3 & 80.5 (13.3) & 2 & 62.7 (12.8) & \cellcolor{steel!20}\textbf{1} & \cellcolor{steel!20}\textbf{47.4 (3.1)} & 4 & 224.5 (5.2) & 5 & 265.0 (0.6) & 6 & 405.1 (18.0) \\
\bottomrule
\end{tabular}
\end{adjustbox}
\end{table}

\begin{table}[h]
\caption{The Scott-Knott rank ($r$), mean MRE, and standard error of the mean (SEM) of all target environments and training sizes for \textsc{SQlite}.}
\centering
\footnotesize
\begin{adjustbox}{width=\linewidth,center}
\begin{tabular}{p{0.15cm}cp{0.05cm}lp{0.05cm}lp{0.05cm}lp{0.05cm}lp{0.05cm}lp{0.05cm}lp{0.05cm}lp{0.05cm}lp{0.05cm}l}
\toprule
\multirow{2}{*}{Size} & \multirow{2}{*}{$\mathbfcal{T}_{target}$} & \multicolumn{2}{c}{\Model} & \multicolumn{2}{c}{\texttt{MAML}} & \multicolumn{2}{c}{\texttt{MetaSGD}} & \multicolumn{2}{c}{\texttt{Beetle}} & \multicolumn{2}{c}{\texttt{tEAMS}} & \multicolumn{2}{c}{\texttt{MORF}} & \multicolumn{2}{c}{\texttt{RF$+_{e}$}} & \multicolumn{2}{c}{\texttt{DeepPerf$+_{e}$}} & \multicolumn{2}{c}{\texttt{SPLConqueor$+_{e}$}} \\ \cline{3-20} 
 &  & $r$ & MRE (SEM) & $r$ & MRE (SEM) & $r$ & MRE (SEM) & $r$ & MRE (SEM) & $r$ & MRE (SEM) & $r$ & MRE (SEM) & $r$ & MRE (SEM) & $r$ & MRE (SEM) & $r$ & MRE (SEM) \\

 \hline \multirow{4}{*}{$S_{1}$} & 1 & \cellcolor{steel!20}\textbf{1} & \cellcolor{steel!20}\textbf{1.0 (0.0)} & 5 & 1.5 (0.1) & 3 & 1.2 (0.1) & 4 & 1.2 (0.1) & 4 & 1.2 (0.0) & 2 & 1.1 (0.0) & 6 & 3.7 (0.0) & 7 & 4.5 (0.0) & 8 & 9.6 (0.4) \\
 & 2 & \cellcolor{steel!20}\textbf{1} & \cellcolor{steel!20}\textbf{1.0 (0.0)} & 4 & 1.3 (0.0) & 2 & 1.0 (0.1) & 3 & 1.1 (0.1) & 4 & 1.3 (0.2) & 4 & 1.2 (0.0) & 5 & 7.8 (0.1) & 6 & 10.7 (0.0) & 7 & 12.2 (0.4) \\
 & 3 & \cellcolor{steel!20}\textbf{1} & \cellcolor{steel!20}\textbf{1.1 (0.1)} & 5 & 2.0 (0.5) & 4 & 1.6 (0.3) & 2 & 1.1 (0.1) & 3 & 1.3 (0.1) & 2 & 1.2 (0.0) & 5 & 2.3 (0.0) & 6 & 2.6 (0.0) & 7 & 9.4 (0.4) \\
 & 4 & \cellcolor{steel!20}\textbf{1} & \cellcolor{steel!20}\textbf{1.2 (0.1)} & 4 & 1.7 (0.1) & 5 & 1.8 (0.3) & 2 & 1.3 (0.1) & 3 & 1.4 (0.0) & \cellcolor{steel!20}\textbf{1} & \cellcolor{steel!20}\textbf{1.2 (0.0)} & 6 & 2.2 (0.0) & 7 & 2.5 (0.0) & 8 & 9.4 (0.4) \\
\hline \multirow{4}{*}{$S_{2}$} & 1 & 2 & 0.9 (0.0) & 5 & 1.2 (0.0) & 4 & 1.1 (0.1) & 3 & 1.0 (0.0) & 3 & 1.0 (0.0) & \cellcolor{steel!20}\textbf{1} & \cellcolor{steel!20}\textbf{0.9 (0.0)} & 6 & 3.6 (0.0) & 8 & 4.5 (0.0) & 7 & 4.0 (0.1) \\
\bottomrule
\end{tabular}
\end{adjustbox}
\end{table}

\begin{table}[h]
\caption{The Scott-Knott rank ($r$), mean MRE, and standard error of the mean (SEM) of all target environments and training sizes for \textsc{NGINX}.}
\centering
\footnotesize
\begin{adjustbox}{width=\linewidth,center}
\begin{tabular}{p{0.15cm}cp{0.05cm}lp{0.05cm}lp{0.05cm}lp{0.05cm}lp{0.05cm}lp{0.05cm}lp{0.05cm}lp{0.05cm}lp{0.05cm}l}
\toprule
\multirow{2}{*}{Size} & \multirow{2}{*}{$\mathbfcal{T}_{target}$} & \multicolumn{2}{c}{\Model} & \multicolumn{2}{c}{\texttt{MAML}} & \multicolumn{2}{c}{\texttt{MetaSGD}} & \multicolumn{2}{c}{\texttt{Beetle}} & \multicolumn{2}{c}{\texttt{tEAMS}} & \multicolumn{2}{c}{\texttt{MORF}} & \multicolumn{2}{c}{\texttt{RF$+_{e}$}} & \multicolumn{2}{c}{\texttt{DeepPerf$+_{e}$}} & \multicolumn{2}{c}{\texttt{SPLConqueor$+_{e}$}} \\ \cline{3-20} 
 &  & $r$ & MRE (SEM) & $r$ & MRE (SEM) & $r$ & MRE (SEM) & $r$ & MRE (SEM) & $r$ & MRE (SEM) & $r$ & MRE (SEM) & $r$ & MRE (SEM) & $r$ & MRE (SEM) & $r$ & MRE (SEM) \\

 \hline \multirow{4}{*}{$S_{1}$} & 1 & 3 & 130.2 (23.3) & 8 & 578.2 (43.5) & 5 & 209.2 (21.4) & 5 & 207.2 (27.4) & 2 & 101.7 (15.8) & 6 & 298.1 (22.4) & 4 & 165.9 (22.6) & \cellcolor{steel!20}\textbf{1} & \cellcolor{steel!20}\textbf{50.5 (1.0)} & 7 & 422.7 (16.5) \\
 & 2 & 3 & 186.4 (34.2) & 7 & 619.7 (46.2) & 4 & 264.5 (44.9) & 4 & 256.2 (35.1) & 2 & 105.4 (17.7) & 5 & 357.2 (25.7) & 4 & 229.1 (33.7) & \cellcolor{steel!20}\textbf{1} & \cellcolor{steel!20}\textbf{26.9 (1.9)} & 6 & 520.7 (15.0) \\
 & 3 & 3 & 313.4 (48.7) & 6 & 745.1 (55.3) & 3 & 341.8 (52.3) & 3 & 321.0 (36.1) & 2 & 82.4 (12.4) & 4 & 496.2 (35.3) & 3 & 346.9 (51.1) & \cellcolor{steel!20}\textbf{1} & \cellcolor{steel!20}\textbf{31.6 (1.7)} & 5 & 680.3 (15.0) \\
 & 4 & 3 & 252.7 (51.3) & 7 & 743.6 (55.2) & 5 & 468.7 (78.7) & 4 & 319.6 (38.6) & 2 & 65.2 (9.0) & 5 & 496.4 (35.4) & 4 & 345.7 (51.3) & \cellcolor{steel!20}\textbf{1} & \cellcolor{steel!20}\textbf{34.3 (2.3)} & 6 & 679.6 (15.0) \\
\bottomrule
\end{tabular}
\end{adjustbox}
\end{table}

\begin{table}[h]
\caption{The Scott-Knott rank ($r$), mean MRE, and standard error of the mean (SEM) of all target environments and training sizes for \textsc{SPEAR}.}
\centering
\footnotesize
\begin{adjustbox}{width=\linewidth,center}
\begin{tabular}{p{0.15cm}cp{0.05cm}lp{0.05cm}lp{0.05cm}lp{0.05cm}lp{0.05cm}lp{0.05cm}lp{0.05cm}lp{0.05cm}lp{0.05cm}l}
\toprule
\multirow{2}{*}{Size} & \multirow{2}{*}{$\mathbfcal{T}_{target}$} & \multicolumn{2}{c}{\Model} & \multicolumn{2}{c}{\texttt{MAML}} & \multicolumn{2}{c}{\texttt{MetaSGD}} & \multicolumn{2}{c}{\texttt{Beetle}} & \multicolumn{2}{c}{\texttt{tEAMS}} & \multicolumn{2}{c}{\texttt{MORF}} & \multicolumn{2}{c}{\texttt{RF$+_{e}$}} & \multicolumn{2}{c}{\texttt{DeepPerf$+_{e}$}} & \multicolumn{2}{c}{\texttt{SPLConqueor$+_{e}$}} \\ \cline{3-20} 
 &  & $r$ & MRE (SEM) & $r$ & MRE (SEM) & $r$ & MRE (SEM) & $r$ & MRE (SEM) & $r$ & MRE (SEM) & $r$ & MRE (SEM) & $r$ & MRE (SEM) & $r$ & MRE (SEM) & $r$ & MRE (SEM) \\

 \hline \multirow{3}{*}{$S_{1}$} & 4 & 2 & 44.5 (8.8) & 6 & 116.3 (13.6) & 4 & 56.7 (5.5) & 3 & 50.1 (3.0) & 6 & 105.0 (10.7) & 5 & 74.6 (8.0) & \cellcolor{steel!20}\textbf{1} & \cellcolor{steel!20}\textbf{34.3 (0.8)} & 2 & 42.9 (0.0) & 7 & 191.1 (14.1) \\
 & 5 & \cellcolor{steel!20}\textbf{1} & \cellcolor{steel!20}\textbf{59.3 (15.3)} & 5 & 1149.8 (202.1) & 4 & 589.6 (207.0) & 4 & 595.6 (194.2) & 3 & 378.0 (72.2) & 3 & 405.4 (75.3) & 3 & 357.5 (18.6) & 2 & 304.5 (0.5) & 6 & 1955.3 (133.9) \\
 & 6 & \cellcolor{steel!20}\textbf{1} & \cellcolor{steel!20}\textbf{34.1 (2.1)} & 6 & 401.3 (63.5) & 5 & 192.6 (60.6) & 2 & 57.0 (6.4) & 5 & 181.3 (31.6) & 5 & 179.8 (30.3) & 4 & 117.3 (7.2) & 3 & 82.6 (0.1) & 7 & 784.7 (55.0) \\
\hline \multirow{3}{*}{$S_{2}$} & 4 & 2 & 34.2 (1.3) & 5 & 50.3 (3.3) & 4 & 45.1 (2.4) & 4 & 45.6 (2.5) & 6 & 57.7 (4.3) & 3 & 42.3 (2.1) & \cellcolor{steel!20}\textbf{1} & \cellcolor{steel!20}\textbf{32.7 (0.6)} & 3 & 42.9 (0.0) & 7 & 71.5 (6.9) \\
 & 5 & \cellcolor{steel!20}\textbf{1} & \cellcolor{steel!20}\textbf{30.1 (2.6)} & 4 & 124.1 (21.5) & 3 & 83.6 (14.0) & 3 & 76.8 (21.6) & 4 & 110.1 (15.6) & 2 & 66.2 (1.8) & 6 & 357.0 (11.5) & 5 & 304.5 (0.4) & 7 & 746.9 (64.6) \\
\bottomrule
\end{tabular}
\end{adjustbox}
\end{table}

\begin{table}[h]
\caption{The Scott-Knott rank ($r$), mean MRE, and standard error of the mean (SEM) of all target environments and training sizes for \textsc{Storm}.}
\centering
\footnotesize
\begin{adjustbox}{width=\linewidth,center}
\begin{tabular}{p{0.15cm}cp{0.05cm}lp{0.05cm}lp{0.05cm}lp{0.05cm}lp{0.05cm}lp{0.05cm}lp{0.05cm}lp{0.05cm}lp{0.05cm}l}
\toprule
\multirow{2}{*}{Size} & \multirow{2}{*}{$\mathbfcal{T}_{target}$} & \multicolumn{2}{c}{\Model} & \multicolumn{2}{c}{\texttt{MAML}} & \multicolumn{2}{c}{\texttt{MetaSGD}} & \multicolumn{2}{c}{\texttt{Beetle}} & \multicolumn{2}{c}{\texttt{tEAMS}} & \multicolumn{2}{c}{\texttt{MORF}} & \multicolumn{2}{c}{\texttt{RF$+_{e}$}} & \multicolumn{2}{c}{\texttt{DeepPerf$+_{e}$}} & \multicolumn{2}{c}{\texttt{SPLConqueor$+_{e}$}} \\ \cline{3-20} 
 &  & $r$ & MRE (SEM) & $r$ & MRE (SEM) & $r$ & MRE (SEM) & $r$ & MRE (SEM) & $r$ & MRE (SEM) & $r$ & MRE (SEM) & $r$ & MRE (SEM) & $r$ & MRE (SEM) & $r$ & MRE (SEM) \\

\hline \multirow{3}{*}{$S_{1}$} & 1 & 2 & 31.2 (1.7) & 3 & 281.7 (41.9) & 5 & 2369.9 (2105.5) & 4 & 340.1 (25.9) & 5 & 2369.2 (190.0) & \cellcolor{steel!20}\textbf{1} & \cellcolor{steel!20}\textbf{21.8 (2.0)} & 8 & 36239.4 (740.3) & 6 & 9416.9 (53.5) & 7 & 28147.9 (679.9) \\
 & 2 & 4 & 144.5 (29.1) & 5 & 208.0 (31.8) & 3 & 100.3 (26.5) & 2 & 21.6 (1.4) & 6 & 681.1 (138.0) & \cellcolor{steel!20}\textbf{1} & \cellcolor{steel!20}\textbf{14.3 (1.3)} & 9 & 71222.3 (1237.7) & 8 & 68082.6 (261.6) & 7 & 56364.5 (1363.3) \\
 & 3 & 2 & 251.3 (18.3) & 3 & 363.2 (24.9) & 4 & 608.4 (294.7) & 5 & 1197.5 (61.3) & 4 & 715.8 (149.0) & \cellcolor{steel!20}\textbf{1} & \cellcolor{steel!20}\textbf{159.0 (20.2)} & 7 & 52452.0 (1447.2) & 8 & 59449.7 (248.0) & 6 & 39741.6 (1437.2) \\
\hline \multirow{3}{*}{$S_{2}$} & 1 & 2 & 31.4 (2.6) & 3 & 303.0 (37.3) & 3 & 339.8 (53.9) & 3 & 297.0 (12.7) & 4 & 2790.9 (159.2) & \cellcolor{steel!20}\textbf{1} & \cellcolor{steel!20}\textbf{14.2 (0.3)} & 7 & 36884.2 (477.3) & 5 & 9879.8 (138.0) & 6 & 28602.0 (652.3) \\
\bottomrule
\end{tabular}
\end{adjustbox}
\end{table}

\begin{table}[h]
\caption{The Scott-Knott rank ($r$), mean MRE, and standard error of the mean (SEM) of all target environments and training sizes for \textsc{ImageMagick}.}
\centering
\footnotesize
\begin{adjustbox}{width=\linewidth,center}
\begin{tabular}{p{0.15cm}cp{0.05cm}lp{0.05cm}lp{0.05cm}lp{0.05cm}lp{0.05cm}lp{0.05cm}lp{0.05cm}lp{0.05cm}lp{0.05cm}l}
\toprule
\multirow{2}{*}{Size} & \multirow{2}{*}{$\mathbfcal{T}_{target}$} & \multicolumn{2}{c}{\Model} & \multicolumn{2}{c}{\texttt{MAML}} & \multicolumn{2}{c}{\texttt{MetaSGD}} & \multicolumn{2}{c}{\texttt{Beetle}} & \multicolumn{2}{c}{\texttt{tEAMS}} & \multicolumn{2}{c}{\texttt{MORF}} & \multicolumn{2}{c}{\texttt{RF$+_{e}$}} & \multicolumn{2}{c}{\texttt{DeepPerf$+_{e}$}} & \multicolumn{2}{c}{\texttt{SPLConqueor$+_{e}$}} \\ \cline{3-20} 
 &  & $r$ & MRE (SEM) & $r$ & MRE (SEM) & $r$ & MRE (SEM) & $r$ & MRE (SEM) & $r$ & MRE (SEM) & $r$ & MRE (SEM) & $r$ & MRE (SEM) & $r$ & MRE (SEM) & $r$ & MRE (SEM) \\

\hline \multirow{4}{*}{$S_{1}$} & 1 & \cellcolor{steel!20}\textbf{1} & \cellcolor{steel!20}\textbf{5.3 (0.3)} & 4 & 11.2 (1.1) & 3 & 9.6 (0.7) & 2 & 6.8 (0.5) & 3 & 9.3 (0.6) & 6 & 17.5 (1.1) & 6 & 16.7 (1.4) & 5 & 14.0 (0.0) & 7 & 32.6 (0.9) \\
 & 2 & 2 & 6.0 (0.2) & 5 & 11.2 (1.1) & 4 & 9.6 (0.9) & 3 & 7.6 (0.6) & 5 & 10.5 (0.8) & 6 & 19.9 (1.3) & 5 & 11.3 (1.3) & \cellcolor{steel!20}\textbf{1} & \cellcolor{steel!20}\textbf{4.7 (0.0)} & 7 & 28.0 (0.9) \\
 & 3 & 2 & 6.9 (0.7) & 5 & 11.9 (1.1) & 4 & 10.2 (0.8) & \cellcolor{steel!20}\textbf{1} & \cellcolor{steel!20}\textbf{5.8 (0.3)} & 3 & 9.2 (0.5) & 8 & 17.9 (1.1) & 7 & 16.3 (1.4) & 6 & 13.8 (0.0) & 9 & 27.4 (1.0) \\
 & 4 & 3 & 7.3 (0.6) & 5 & 11.5 (1.0) & 4 & 9.5 (0.9) & 2 & 6.6 (0.4) & 5 & 10.7 (1.0) & 6 & 16.8 (1.1) & 5 & 11.3 (1.4) & \cellcolor{steel!20}\textbf{1} & \cellcolor{steel!20}\textbf{3.8 (0.0)} & 7 & 27.8 (0.9) \\
\hline \multirow{4}{*}{$S_{2}$} & 1 & \cellcolor{steel!20}\textbf{1} & \cellcolor{steel!20}\textbf{4.5 (0.1)} & 3 & 5.6 (0.3) & 4 & 6.0 (0.2) & 2 & 4.8 (0.1) & 5 & 8.7 (0.8) & 4 & 6.0 (0.3) & 6 & 12.2 (0.2) & 7 & 14.0 (0.1) & 8 & 30.6 (0.9) \\
\bottomrule
\end{tabular}
\end{adjustbox}
\end{table}

\begin{table}[h]
\caption{The Scott-Knott rank ($r$), mean MRE, and standard error of the mean (SEM) of all target environments and training sizes for \textsc{ExaStencils}.}
\centering
\footnotesize
\begin{adjustbox}{width=\linewidth,center}
\begin{tabular}{p{0.15cm}cp{0.05cm}lp{0.05cm}lp{0.05cm}lp{0.05cm}lp{0.05cm}lp{0.05cm}lp{0.05cm}lp{0.05cm}lp{0.05cm}l}
\toprule
\multirow{2}{*}{Size} & \multirow{2}{*}{$\mathbfcal{T}_{target}$} & \multicolumn{2}{c}{\Model} & \multicolumn{2}{c}{\texttt{MAML}} & \multicolumn{2}{c}{\texttt{MetaSGD}} & \multicolumn{2}{c}{\texttt{Beetle}} & \multicolumn{2}{c}{\texttt{tEAMS}} & \multicolumn{2}{c}{\texttt{MORF}} & \multicolumn{2}{c}{\texttt{RF$+_{e}$}} & \multicolumn{2}{c}{\texttt{DeepPerf$+_{e}$}} & \multicolumn{2}{c}{\texttt{SPLConqueor$+_{e}$}} \\ \cline{3-20} 
 &  & $r$ & MRE (SEM) & $r$ & MRE (SEM) & $r$ & MRE (SEM) & $r$ & MRE (SEM) & $r$ & MRE (SEM) & $r$ & MRE (SEM) & $r$ & MRE (SEM) & $r$ & MRE (SEM) & $r$ & MRE (SEM) \\

\hline \multirow{4}{*}{$S_{1}$} & 1 & \cellcolor{steel!20}\textbf{1} & \cellcolor{steel!20}\textbf{3.1 (0.0)} & 8 & 18.9 (0.1) & 6 & 14.8 (0.7) & 5 & 13.4 (0.6) & 2 & 6.0 (0.2) & 4 & 8.4 (0.2) & 3 & 8.1 (0.3) & 6 & 15.4 (0.6) & 7 & 17.0 (0.2) \\
 & 2 & 2 & 4.3 (0.5) & 8 & 19.8 (0.2) & 5 & 14.6 (0.5) & 7 & 17.3 (0.7) & \cellcolor{steel!20}\textbf{1} & \cellcolor{steel!20}\textbf{3.5 (0.4)} & 4 & 9.1 (0.2) & 3 & 8.8 (0.3) & 6 & 15.8 (0.6) & 7 & 17.3 (0.2) \\
 & 3 & 2 & 4.4 (0.4) & 8 & 20.7 (0.2) & 6 & 16.8 (0.8) & 5 & 15.6 (0.6) & \cellcolor{steel!20}\textbf{1} & \cellcolor{steel!20}\textbf{3.6 (0.4)} & 4 & 9.4 (0.3) & 3 & 8.9 (0.3) & 7 & 18.9 (0.9) & 7 & 18.3 (0.2) \\
 & 4 & \cellcolor{steel!20}\textbf{1} & \cellcolor{steel!20}\textbf{3.4 (0.3)} & 7 & 21.7 (0.2) & 4 & 16.7 (0.7) & 4 & 17.4 (0.7) & \cellcolor{steel!20}\textbf{1} & \cellcolor{steel!20}\textbf{3.4 (0.4)} & 3 & 9.8 (0.3) & 2 & 9.0 (0.3) & 6 & 19.5 (1.0) & 5 & 18.2 (0.2) \\
\hline \multirow{4}{*}{$S_{2}$} & 1 & \cellcolor{steel!20}\textbf{1} & \cellcolor{steel!20}\textbf{2.7 (0.0)} & 8 & 18.7 (0.2) & 4 & 8.9 (0.6) & 5 & 9.7 (0.5) & 2 & 5.4 (0.3) & 3 & 6.1 (0.1) & 3 & 6.2 (0.1) & 6 & 15.7 (0.7) & 7 & 16.7 (0.1) \\
\bottomrule
\end{tabular}
\end{adjustbox}
\end{table}

\begin{table}[h]
\caption{The Scott-Knott rank ($r$), mean MRE, and standard error of the mean (SEM) of all target environments and training sizes for \textsc{x264}.}
\centering
\footnotesize
\begin{adjustbox}{width=\linewidth,center}
\begin{tabular}{p{0.15cm}cp{0.05cm}lp{0.05cm}lp{0.05cm}lp{0.05cm}lp{0.05cm}lp{0.05cm}lp{0.05cm}lp{0.05cm}lp{0.05cm}l}
\toprule
\multirow{2}{*}{Size} & \multirow{2}{*}{$\mathbfcal{T}_{target}$} & \multicolumn{2}{c}{\Model} & \multicolumn{2}{c}{\texttt{MAML}} & \multicolumn{2}{c}{\texttt{MetaSGD}} & \multicolumn{2}{c}{\texttt{Beetle}} & \multicolumn{2}{c}{\texttt{tEAMS}} & \multicolumn{2}{c}{\texttt{MORF}} & \multicolumn{2}{c}{\texttt{RF$+_{e}$}} & \multicolumn{2}{c}{\texttt{DeepPerf$+_{e}$}} & \multicolumn{2}{c}{\texttt{SPLConqueor$+_{e}$}} \\ \cline{3-20} 
 &  & $r$ & MRE (SEM) & $r$ & MRE (SEM) & $r$ & MRE (SEM) & $r$ & MRE (SEM) & $r$ & MRE (SEM) & $r$ & MRE (SEM) & $r$ & MRE (SEM) & $r$ & MRE (SEM) & $r$ & MRE (SEM) \\

\hline \multirow{10}{*}{$S_{1}$} & 1 & 2 & 26.5 (0.8) & 4 & 42.3 (2.5) & 4 & 42.8 (2.6) & \cellcolor{steel!20}\textbf{1} & \cellcolor{steel!20}\textbf{25.7 (0.6)} & 3 & 37.1 (2.3) & 5 & 60.6 (3.3) & 6 & 82.0 (2.4) & 7 & 85.5 (0.5) & 8 & 133.4 (4.9) \\
 & 2 & 2 & 16.3 (0.8) & 6 & 37.4 (2.3) & 7 & 40.4 (2.6) & 4 & 19.6 (0.6) & 5 & 24.9 (1.6) & 8 & 53.6 (2.8) & 3 & 18.4 (0.6) & \cellcolor{steel!20}\textbf{1} & \cellcolor{steel!20}\textbf{10.7 (0.1)} & 9 & 64.7 (2.9) \\
 & 3 & 2 & 19.3 (0.9) & 6 & 37.6 (2.6) & 6 & 39.3 (2.7) & 3 & 20.9 (0.6) & 5 & 31.0 (3.0) & 7 & 54.4 (2.8) & 4 & 24.1 (0.9) & \cellcolor{steel!20}\textbf{1} & \cellcolor{steel!20}\textbf{15.6 (0.1)} & 8 & 77.7 (3.3) \\
 & 4 & 2 & 25.1 (1.4) & 4 & 38.4 (2.5) & 5 & 42.2 (3.0) & 2 & 25.3 (1.0) & 3 & 26.0 (2.6) & 6 & 53.7 (3.1) & \cellcolor{steel!20}\textbf{1} & \cellcolor{steel!20}\textbf{24.4 (0.4)} & 3 & 27.2 (0.0) & 6 & 53.2 (1.8) \\
 & 5 & 3 & 17.6 (0.9) & 5 & 31.5 (1.6) & 5 & 33.8 (2.5) & 2 & 16.8 (0.5) & 4 & 22.9 (1.5) & 6 & 42.1 (2.2) & 2 & 16.4 (0.6) & \cellcolor{steel!20}\textbf{1} & \cellcolor{steel!20}\textbf{11.6 (0.0)} & 7 & 56.9 (2.3) \\
\bottomrule
\end{tabular}
\end{adjustbox}
\end{table}

%% file: Appendix/chap-meta-rq3-full.tex
\chapter{Detailed Data for Table~\ref{chap-meta-tb:rank_rq3}}
\label{appendix-chap-meta-tb:rank_rq3}

\textbf{This appendix presents the detailed data of all subject systems for Table~\ref{chap-meta-tb:rank_rq3}, where \setlength{\fboxsep}{1.5pt}\colorbox{steel!20}{blue} cells stand for the best rank.}

\begin{table}[h]
\caption{The Scott-Knott rank ($r$), mean MRE, and standard error of the mean (SEM) of all target environments and training sizes for \textsc{DeepArch}.}
\centering
\footnotesize
\begin{adjustbox}{width=\linewidth,center}

\begin{tabular}{ccllllllllllll}
\toprule

\multirow{2}{*}{Size} & \multirow{2}{*}{$\mathbfcal{T}_{target}$} & \multicolumn{2}{c}{\Model} & \multicolumn{2}{c}{\texttt{DeepPerf$+$}} & \multicolumn{2}{c}{\Model$_{RF}$} & \multicolumn{2}{c}{\texttt{RF$+$}} & \multicolumn{2}{c}{\Model$_{SPL}$} & \multicolumn{2}{c}{\texttt{SPLConqueror$+$}} \\  \cline{3-14} 
 &  & $r$ & MRE (SEM) & $r$ & MRE (SEM) & $r$ & MRE (SEM) & $r$ & MRE (SEM) & $r$ & MRE (SEM) & $r$ & MRE (SEM) \\

\hline \multirow{3}{*}{$S_{1}$} & 1 & 2 & 82.5 (5.4) & 3 & 95.3 (1.6) & \cellcolor{steel!20}\textbf{1} & \cellcolor{steel!20}\textbf{68.2 (1.2)} & 6 & 273.4 (21.1) & 4 & 117.4 (4.5) & 5 & 215.7 (10.3) \\
 & 2 & \cellcolor{steel!20}\textbf{1} & \cellcolor{steel!20}\textbf{76.6 (5.8)} & \cellcolor{steel!20}\textbf{1} & \cellcolor{steel!20}\textbf{72.6 (0.0)} & \cellcolor{steel!20}\textbf{1} & \cellcolor{steel!20}\textbf{72.0 (3.2)} & 3 & 196.2 (15.4) & 3 & 189.3 (10.7) & 2 & 147.9 (5.9) \\
 & 3 & 2 & 60.4 (3.8) & 3 & 71.1 (0.4) & \cellcolor{steel!20}\textbf{1} & \cellcolor{steel!20}\textbf{48.5 (2.1)} & 6 & 283.7 (22.6) & 4 & 132.5 (6.9) & 5 & 202.9 (8.6) \\
\hline \multirow{3}{*}{$S_{2}$} & 1 & \cellcolor{steel!20}\textbf{1} & \cellcolor{steel!20}\textbf{62.7 (3.6)} & 2 & 95.7 (1.5) & \cellcolor{steel!20}\textbf{1} & \cellcolor{steel!20}\textbf{60.7 (0.8)} & 4 & 171.1 (9.7) & 3 & 104.2 (3.6) & 4 & 164.7 (6.3) \\
 & 2 & 2 & 74.5 (4.9) & 2 & 71.6 (0.0) & \cellcolor{steel!20}\textbf{1} & \cellcolor{steel!20}\textbf{62.0 (1.9)} & 4 & 129.0 (6.0) & 5 & 198.1 (9.3) & 3 & 117.7 (3.8) \\
 & 3 & 2 & 50.8 (2.7) & 3 & 63.3 (0.5) & \cellcolor{steel!20}\textbf{1} & \cellcolor{steel!20}\textbf{40.1 (1.4)} & 6 & 175.0 (9.4) & 4 & 131.5 (6.1) & 5 & 153.5 (5.3) \\
\bottomrule
\end{tabular}
\end{adjustbox}
\end{table}

\begin{table}[h]
\caption{The Scott-Knott rank ($r$), mean MRE, and standard error of the mean (SEM) of all target environments and training sizes for \textsc{SaC}.}
\centering
\footnotesize
\begin{adjustbox}{width=\linewidth,center}
\begin{tabular}{ccllllllllllll}
\toprule
\multirow{2}{*}{Size} & \multirow{2}{*}{$\mathbfcal{T}_{target}$} & \multicolumn{2}{c}{\Model} & \multicolumn{2}{c}{\texttt{DeepPerf$+$}} & \multicolumn{2}{c}{\Model$_{RF}$} & \multicolumn{2}{c}{\texttt{RF$+$}} & \multicolumn{2}{c}{\Model$_{SPL}$} & \multicolumn{2}{c}{\texttt{SPLConqueror$+$}} \\  \cline{3-14} 
 &  & $r$ & MRE (SEM) & $r$ & MRE (SEM) & $r$ & MRE (SEM) & $r$ & MRE (SEM) & $r$ & MRE (SEM) & $r$ & MRE (SEM) \\

\hline \multirow{3}{*}{$S_{1}$} & 1 & \cellcolor{steel!20}\textbf{1} & \cellcolor{steel!20}\textbf{1353.0 (191.7)} & 4 & 2918.7 (47.2) & 3 & 2615.0 (43.8) & 3 & 2720.2 (234.1) & 2 & 2174.6 (200.8) & 4 & 3079.0 (260.1) \\
 & 2 & \cellcolor{steel!20}\textbf{1} & \cellcolor{steel!20}\textbf{105.4 (8.4)} & 2 & 245.6 (1.4) & 3 & 253.5 (1.7) & 2 & 241.5 (6.5) & 4 & 367.6 (17.5) & 4 & 359.4 (13.5) \\
 & 3 & \cellcolor{steel!20}\textbf{1} & \cellcolor{steel!20}\textbf{41.9 (6.7)} & 4 & 177.9 (0.3) & 3 & 166.1 (0.6) & 5 & 185.4 (5.4) & 2 & 122.1 (4.6) & 6 & 352.0 (18.7) \\
\hline \multirow{3}{*}{$S_{2}$} & 1 & \cellcolor{steel!20}\textbf{1} & \cellcolor{steel!20}\textbf{754.1 (143.1)} & 5 & 3874.7 (264.3) & 3 & 2598.9 (60.4) & 3 & 2579.9 (111.2) & 2 & 1646.5 (112.4) & 4 & 2840.7 (128.4) \\
 & 2 & \cellcolor{steel!20}\textbf{1} & \cellcolor{steel!20}\textbf{59.1 (4.0)} & 5 & 375.8 (23.4) & 3 & 251.6 (1.9) & 2 & 223.1 (2.8) & 3 & 257.0 (10.4) & 4 & 320.5 (12.1) \\
 & 3 & \cellcolor{steel!20}\textbf{1} & \cellcolor{steel!20}\textbf{26.9 (2.4)} & 6 & 350.2 (37.7) & 4 & 165.3 (0.6) & 3 & 163.2 (2.6) & 2 & 115.8 (5.5) & 5 & 277.9 (12.4) \\
\bottomrule
\end{tabular}
\end{adjustbox}
\end{table}

\begin{table}[h]
\caption{The Scott-Knott rank ($r$), mean MRE, and standard error of the mean (SEM) of all target environments and training sizes for \textsc{SQlite}.}
\centering
\footnotesize
\begin{adjustbox}{width=\linewidth,center}
\begin{tabular}{ccllllllllllll}
\toprule
\multirow{2}{*}{Size} & \multirow{2}{*}{$\mathbfcal{T}_{target}$} & \multicolumn{2}{c}{\Model} & \multicolumn{2}{c}{\texttt{DeepPerf$+$}} & \multicolumn{2}{c}{\Model$_{RF}$} & \multicolumn{2}{c}{\texttt{RF$+$}} & \multicolumn{2}{c}{\Model$_{SPL}$} & \multicolumn{2}{c}{\texttt{SPLConqueror$+$}} \\  \cline{3-14} 
 &  & $r$ & MRE (SEM) & $r$ & MRE (SEM) & $r$ & MRE (SEM) & $r$ & MRE (SEM) & $r$ & MRE (SEM) & $r$ & MRE (SEM) \\

\hline \multirow{4}{*}{$S_{1}$} & 1 & \cellcolor{steel!20}\textbf{1} & \cellcolor{steel!20}\textbf{1.0 (0.0)} & 5 & 5.5 (0.0) & 4 & 4.5 (0.1) & 3 & 3.7 (0.0) & 2 & 1.6 (0.0) & 6 & 15.5 (0.5) \\
 & 2 & \cellcolor{steel!20}\textbf{1} & \cellcolor{steel!20}\textbf{1.0 (0.0)} & 5 & 11.7 (0.0) & 3 & 4.8 (0.2) & 4 & 7.9 (0.0) & 2 & 1.7 (0.0) & 6 & 17.8 (0.6) \\
 & 3 & \cellcolor{steel!20}\textbf{1} & \cellcolor{steel!20}\textbf{1.1 (0.1)} & 5 & 3.6 (0.0) & 4 & 3.1 (0.1) & 3 & 2.2 (0.0) & 2 & 1.2 (0.0) & 6 & 15.6 (0.5) \\
 & 4 & \cellcolor{steel!20}\textbf{1} & \cellcolor{steel!20}\textbf{1.2 (0.1)} & 3 & 3.5 (0.0) & 4 & 4.0 (0.2) & 2 & 2.1 (0.0) & \cellcolor{steel!20}\textbf{1} & \cellcolor{steel!20}\textbf{1.2 (0.0)} & 5 & 15.6 (0.5) \\
\hline \multirow{4}{*}{$S_{2}$} & 1 & \cellcolor{steel!20}\textbf{1} & \cellcolor{steel!20}\textbf{0.9 (0.0)} & 5 & 4.5 (0.0) & 4 & 4.1 (0.1) & 3 & 3.6 (0.0) & 2 & 1.4 (0.0) & 6 & 5.2 (0.3) \\
 & 2 & \cellcolor{steel!20}\textbf{1} & \cellcolor{steel!20}\textbf{0.8 (0.0)} & 6 & 10.6 (0.0) & 3 & 4.2 (0.1) & 4 & 7.9 (0.0) & 2 & 1.3 (0.0) & 5 & 8.4 (0.2) \\
\bottomrule
\end{tabular}
\end{adjustbox}
\end{table}

\begin{table}[h]
\caption{The Scott-Knott rank ($r$), mean MRE, and standard error of the mean (SEM) of all target environments and training sizes for \textsc{NGINX}.}
\centering
\footnotesize
\begin{adjustbox}{width=\linewidth,center}
\begin{tabular}{ccllllllllllll}
\toprule
\multirow{2}{*}{Size} & \multirow{2}{*}{$\mathbfcal{T}_{target}$} & \multicolumn{2}{c}{\Model} & \multicolumn{2}{c}{\texttt{DeepPerf$+$}} & \multicolumn{2}{c}{\Model$_{RF}$} & \multicolumn{2}{c}{\texttt{RF$+$}} & \multicolumn{2}{c}{\Model$_{SPL}$} & \multicolumn{2}{c}{\texttt{SPLConqueror$+$}} \\  \cline{3-14} 
 &  & $r$ & MRE (SEM) & $r$ & MRE (SEM) & $r$ & MRE (SEM) & $r$ & MRE (SEM) & $r$ & MRE (SEM) & $r$ & MRE (SEM) \\

\hline \multirow{4}{*}{$S_{1}$} & 1 & 3 & 130.2 (23.3) & \cellcolor{steel!20}\textbf{1} & \cellcolor{steel!20}\textbf{24.2 (1.7)} & 2 & 61.7 (2.8) & 4 & 347.5 (51.2) & 5 & 649.2 (9.0) & 6 & 682.8 (15.3) \\
 & 2 & 3 & 186.4 (34.2) & \cellcolor{steel!20}\textbf{1} & \cellcolor{steel!20}\textbf{25.2 (1.5)} & 2 & 62.0 (2.7) & 4 & 347.9 (51.1) & 5 & 649.8 (9.1) & 6 & 683.6 (15.4) \\
 & 3 & 4 & 313.4 (48.7) & \cellcolor{steel!20}\textbf{1} & \cellcolor{steel!20}\textbf{18.8 (2.0)} & 2 & 36.0 (2.0) & 3 & 229.7 (33.6) & 5 & 502.7 (9.6) & 6 & 525.3 (15.1) \\
 & 4 & 3 & 252.7 (51.3) & \cellcolor{steel!20}\textbf{1} & \cellcolor{steel!20}\textbf{24.8 (1.0)} & \cellcolor{steel!20}\textbf{1} & \cellcolor{steel!20}\textbf{24.5 (1.5)} & 2 & 167.5 (22.6) & 5 & 456.8 (13.2) & 4 & 427.4 (16.2) \\
\hline \multirow{4}{*}{$S_{2}$} & 1 & 2 & 26.4 (3.2) & \cellcolor{steel!20}\textbf{1} & \cellcolor{steel!20}\textbf{23.5 (1.8)} & 3 & 33.3 (1.9) & 4 & 79.9 (30.0) & 5 & 645.7 (7.9) & 6 & 668.7 (9.4) \\
 & 2 & \cellcolor{steel!20}\textbf{1} & \cellcolor{steel!20}\textbf{11.7 (1.2)} & 2 & 27.8 (1.7) & 3 & 33.2 (1.9) & 4 & 80.1 (30.0) & 5 & 646.8 (8.0) & 6 & 669.9 (9.5) \\
\bottomrule
\end{tabular}
\end{adjustbox}
\end{table}

\begin{table}[h]
\caption{The Scott-Knott rank ($r$), mean MRE, and standard error of the mean (SEM) of all target environments and training sizes for \textsc{SPEAR}.}
\centering
\footnotesize
\begin{adjustbox}{width=\linewidth,center}
\begin{tabular}{ccllllllllllll}
\toprule
\multirow{2}{*}{Size} & \multirow{2}{*}{$\mathbfcal{T}_{target}$} & \multicolumn{2}{c}{\Model} & \multicolumn{2}{c}{\texttt{DeepPerf$+$}} & \multicolumn{2}{c}{\Model$_{RF}$} & \multicolumn{2}{c}{\texttt{RF$+$}} & \multicolumn{2}{c}{\Model$_{SPL}$} & \multicolumn{2}{c}{\texttt{SPLConqueror$+$}} \\  \cline{3-14} 
 &  & $r$ & MRE (SEM) & $r$ & MRE (SEM) & $r$ & MRE (SEM) & $r$ & MRE (SEM) & $r$ & MRE (SEM) & $r$ & MRE (SEM) \\

\hline \multirow{3}{*}{$S_{1}$} & 4 & 3 & 44.5 (8.8) & 2 & 34.4 (1.6) & \cellcolor{steel!20}\textbf{1} & \cellcolor{steel!20}\textbf{15.2 (0.6)} & 2 & 34.4 (1.6) & 3 & 51.9 (5.1) & 4 & 208.1 (14.6) \\
 & 5 & \cellcolor{steel!20}\textbf{1} & \cellcolor{steel!20}\textbf{59.3 (15.3)} & 2 & 396.1 (1.8) & 4 & 617.8 (10.5) & 3 & 486.1 (27.5) & 5 & 1065.9 (186.3) & 6 & 2224.9 (158.8) \\
 & 6 & \cellcolor{steel!20}\textbf{1} & \cellcolor{steel!20}\textbf{34.1 (2.1)} & 2 & 167.7 (11.8) & 3 & 226.4 (2.9) & 2 & 171.5 (10.8) & 4 & 306.3 (60.8) & 5 & 893.4 (63.6) \\
\hline \multirow{3}{*}{$S_{2}$} & 4 & 3 & 34.2 (1.3) & 2 & 28.3 (0.5) & \cellcolor{steel!20}\textbf{1} & \cellcolor{steel!20}\textbf{13.8 (0.1)} & 3 & 33.0 (1.3) & 4 & 54.1 (4.7) & 5 & 87.5 (7.2) \\
 & 5 & \cellcolor{steel!20}\textbf{1} & \cellcolor{steel!20}\textbf{30.1 (2.6)} & 2 & 398.6 (6.0) & 4 & 591.3 (0.2) & 3 & 446.4 (19.3) & 6 & 1110.9 (116.7) & 5 & 938.8 (69.8) \\
 & 6 & \cellcolor{steel!20}\textbf{1} & \cellcolor{steel!20}\textbf{52.4 (33.0)} & 2 & 141.2 (2.6) & 4 & 218.1 (0.1) & 3 & 156.3 (7.5) & 5 & 332.2 (49.3) & 5 & 364.4 (28.6) \\
\bottomrule
\end{tabular}
\end{adjustbox}
\end{table}

\begin{table}[h]
\caption{The Scott-Knott rank ($r$), mean MRE, and standard error of the mean (SEM) of all target environments and training sizes for \textsc{Storm}.}
\centering
\footnotesize
\begin{adjustbox}{width=\linewidth,center}
\begin{tabular}{ccllllllllllll}
\toprule
\multirow{2}{*}{Size} & \multirow{2}{*}{$\mathbfcal{T}_{target}$} & \multicolumn{2}{c}{\Model} & \multicolumn{2}{c}{\texttt{DeepPerf$+$}} & \multicolumn{2}{c}{\Model$_{RF}$} & \multicolumn{2}{c}{\texttt{RF$+$}} & \multicolumn{2}{c}{\Model$_{SPL}$} & \multicolumn{2}{c}{\texttt{SPLConqueror$+$}} \\  \cline{3-14} 
 &  & $r$ & MRE (SEM) & $r$ & MRE (SEM) & $r$ & MRE (SEM) & $r$ & MRE (SEM) & $r$ & MRE (SEM) & $r$ & MRE (SEM) \\

\hline \multirow{3}{*}{$S_{1}$} & 1 & \cellcolor{steel!20}\textbf{1} & \cellcolor{steel!20}\textbf{31.2 (1.7)} & 3 & 225.7 (5.5) & 4 & 934.4 (21.1) & 2 & 157.4 (39.3) & 6 & 62365.6 (1192.9) & 5 & 21531.3 (558.9) \\
 & 2 & 2 & 144.5 (29.1) & 4 & 529.8 (8.1) & \cellcolor{steel!20}\textbf{1} & \cellcolor{steel!20}\textbf{63.9 (0.6)} & 3 & 373.6 (54.5) & 5 & 21690.5 (474.8) & 6 & 43178.7 (853.9) \\
 & 3 & \cellcolor{steel!20}\textbf{1} & \cellcolor{steel!20}\textbf{251.3 (18.3)} & 4 & 1578.8 (10.4) & 2 & 500.8 (34.4) & 3 & 1113.8 (16.8) & 5 & 6480.9 (356.7) & 6 & 36052.8 (728.3) \\
\hline \multirow{3}{*}{$S_{2}$} & 1 & \cellcolor{steel!20}\textbf{1} & \cellcolor{steel!20}\textbf{31.4 (2.6)} & 2 & 228.2 (6.7) & 3 & 963.0 (23.5) & 2 & 255.1 (62.3) & 5 & 63627.1 (992.0) & 4 & 23140.7 (483.9) \\
 & 2 & 2 & 119.0 (21.9) & 3 & 530.1 (8.4) & \cellcolor{steel!20}\textbf{1} & \cellcolor{steel!20}\textbf{64.0 (0.8)} & 3 & 506.6 (90.3) & 4 & 21993.3 (374.0) & 5 & 44703.3 (631.2) \\
 & 3 & \cellcolor{steel!20}\textbf{1} & \cellcolor{steel!20}\textbf{214.9 (15.2)} & 4 & 1560.1 (11.6) & 2 & 532.1 (32.8) & 3 & 1121.7 (23.6) & 5 & 6101.7 (264.1) & 6 & 37365.1 (484.6) \\
\bottomrule
\end{tabular}
\end{adjustbox}
\end{table}

\begin{table}[h]
\caption{The Scott-Knott rank ($r$), mean MRE, and standard error of the mean (SEM) of all target environments and training sizes for \textsc{ImageMagick}.}
\centering
\footnotesize
\begin{adjustbox}{width=\linewidth,center}
\begin{tabular}{ccllllllllllll}
\toprule
\multirow{2}{*}{Size} & \multirow{2}{*}{$\mathbfcal{T}_{target}$} & \multicolumn{2}{c}{\Model} & \multicolumn{2}{c}{\texttt{DeepPerf$+$}} & \multicolumn{2}{c}{\Model$_{RF}$} & \multicolumn{2}{c}{\texttt{RF$+$}} & \multicolumn{2}{c}{\Model$_{SPL}$} & \multicolumn{2}{c}{\texttt{SPLConqueror$+$}} \\  \cline{3-14} 
 &  & $r$ & MRE (SEM) & $r$ & MRE (SEM) & $r$ & MRE (SEM) & $r$ & MRE (SEM) & $r$ & MRE (SEM) & $r$ & MRE (SEM) \\

\hline \multirow{4}{*}{$S_{1}$} & 1 & 2 & 5.3 (0.3) & \cellcolor{steel!20}\textbf{1} & \cellcolor{steel!20}\textbf{3.7 (0.0)} & 3 & 6.0 (0.2) & 4 & 9.9 (1.3) & 5 & 25.7 (0.7) & 6 & 27.3 (0.8) \\
 & 2 & \cellcolor{steel!20}\textbf{1} & \cellcolor{steel!20}\textbf{6.0 (0.2)} & 3 & 13.8 (0.0) & 2 & 8.3 (0.2) & 4 & 15.5 (1.2) & 5 & 26.1 (0.7) & 5 & 27.0 (0.9) \\
 & 3 & 2 & 6.9 (0.7) & \cellcolor{steel!20}\textbf{1} & \cellcolor{steel!20}\textbf{4.7 (0.0)} & 2 & 6.8 (0.3) & 3 & 10.1 (1.2) & 4 & 25.6 (0.7) & 5 & 27.6 (0.8) \\
 & 4 & \cellcolor{steel!20}\textbf{1} & \cellcolor{steel!20}\textbf{7.3 (0.6)} & 3 & 14.0 (0.0) & 2 & 8.5 (0.4) & 4 & 15.1 (1.3) & 5 & 25.3 (0.6) & 6 & 32.0 (0.8) \\
\hline \multirow{4}{*}{$S_{2}$} & 1 & 2 & 4.5 (0.1) & \cellcolor{steel!20}\textbf{1} & \cellcolor{steel!20}\textbf{3.7 (0.0)} & 3 & 5.0 (0.2) & 4 & 5.6 (0.2) & 5 & 25.1 (0.6) & 5 & 25.1 (0.7) \\
 & 2 & \cellcolor{steel!20}\textbf{1} & \cellcolor{steel!20}\textbf{5.6 (0.2)} & 4 & 13.7 (0.1) & 2 & 7.1 (0.2) & 3 & 11.9 (0.3) & 6 & 26.1 (0.7) & 5 & 24.8 (0.8) \\
\bottomrule
\end{tabular}
\end{adjustbox}
\end{table}

\begin{table}[h]
\caption{The Scott-Knott rank ($r$), mean MRE, and standard error of the mean (SEM) of all target environments and training sizes for \textsc{ExaStencils}.}
\centering
\footnotesize
\begin{adjustbox}{width=\linewidth,center}
\begin{tabular}{ccllllllllllll}
\toprule
\multirow{2}{*}{Size} & \multirow{2}{*}{$\mathbfcal{T}_{target}$} & \multicolumn{2}{c}{\Model} & \multicolumn{2}{c}{\texttt{DeepPerf$+$}} & \multicolumn{2}{c}{\Model$_{RF}$} & \multicolumn{2}{c}{\texttt{RF$+$}} & \multicolumn{2}{c}{\Model$_{SPL}$} & \multicolumn{2}{c}{\texttt{SPLConqueror$+$}} \\  \cline{3-14} 
 &  & $r$ & MRE (SEM) & $r$ & MRE (SEM) & $r$ & MRE (SEM) & $r$ & MRE (SEM) & $r$ & MRE (SEM) & $r$ & MRE (SEM) \\

\hline \multirow{4}{*}{$S_{1}$} & 1 & \cellcolor{steel!20}\textbf{1} & \cellcolor{steel!20}\textbf{3.1 (0.0)} & 3 & 16.8 (0.3) & 2 & 8.7 (0.9) & 2 & 9.0 (0.3) & 5 & 18.9 (0.2) & 4 & 18.2 (0.2) \\
 & 2 & \cellcolor{steel!20}\textbf{1} & \cellcolor{steel!20}\textbf{4.3 (0.5)} & 3 & 18.7 (0.7) & 2 & 9.4 (1.1) & 2 & 8.9 (0.3) & 3 & 18.3 (0.2) & 3 & 18.3 (0.2) \\
 & 3 & \cellcolor{steel!20}\textbf{1} & \cellcolor{steel!20}\textbf{4.4 (0.4)} & 3 & 16.5 (0.6) & 2 & 9.5 (1.0) & 2 & 8.8 (0.3) & 3 & 16.6 (0.2) & 4 & 17.4 (0.2) \\
 & 4 & \cellcolor{steel!20}\textbf{1} & \cellcolor{steel!20}\textbf{3.4 (0.3)} & 5 & 16.7 (0.6) & 3 & 11.2 (0.8) & 2 & 8.1 (0.3) & 4 & 16.2 (0.2) & 5 & 17.1 (0.2) \\
\hline \multirow{4}{*}{$S_{2}$} & 1 & \cellcolor{steel!20}\textbf{1} & \cellcolor{steel!20}\textbf{2.7 (0.0)} & 4 & 17.5 (0.5) & 3 & 7.6 (0.9) & 2 & 6.8 (0.2) & 5 & 18.7 (0.1) & 4 & 17.9 (0.1) \\
 & 2 & \cellcolor{steel!20}\textbf{1} & \cellcolor{steel!20}\textbf{2.9 (0.2)} & 4 & 18.4 (0.6) & 3 & 7.5 (1.0) & 2 & 6.7 (0.2) & 4 & 18.1 (0.1) & 4 & 18.0 (0.1) \\
\bottomrule
\end{tabular}
\end{adjustbox}
\end{table}

\begin{table}[h]
\caption{The Scott-Knott rank ($r$), mean MRE, and standard error of the mean (SEM) of all target environments and training sizes for \textsc{x264}.}
\centering
\footnotesize
\begin{adjustbox}{width=\linewidth,center}
\begin{tabular}{ccllllllllllll}
\toprule
\multirow{2}{*}{Size} & \multirow{2}{*}{$\mathbfcal{T}_{target}$} & \multicolumn{2}{c}{\Model} & \multicolumn{2}{c}{\texttt{DeepPerf$+$}} & \multicolumn{2}{c}{\Model$_{RF}$} & \multicolumn{2}{c}{\texttt{RF$+$}} & \multicolumn{2}{c}{\Model$_{SPL}$} & \multicolumn{2}{c}{\texttt{SPLConqueror$+$}} \\  \cline{3-14} 
 &  & $r$ & MRE (SEM) & $r$ & MRE (SEM) & $r$ & MRE (SEM) & $r$ & MRE (SEM) & $r$ & MRE (SEM) & $r$ & MRE (SEM) \\

\hline \multirow{10}{*}{$S_{1}$} & 1 & \cellcolor{steel!20}\textbf{1} & \cellcolor{steel!20}\textbf{26.5 (0.8)} & 4 & 200.5 (0.4) & 2 & 95.6 (6.8) & 5 & 254.9 (9.0) & 3 & 122.6 (7.9) & 6 & 317.7 (12.8) \\
 & 2 & \cellcolor{steel!20}\textbf{1} & \cellcolor{steel!20}\textbf{16.3 (0.8)} & 2 & 24.2 (0.1) & 3 & 31.4 (1.9) & 4 & 36.8 (2.0) & 6 & 108.8 (5.7) & 5 & 74.0 (3.1) \\
 & 3 & \cellcolor{steel!20}\textbf{1} & \cellcolor{steel!20}\textbf{19.3 (0.9)} & 2 & 24.3 (0.1) & 2 & 23.9 (2.2) & 3 & 37.7 (2.1) & 5 & 107.4 (5.4) & 4 & 74.6 (3.1) \\
 & 4 & 2 & 25.1 (1.4) & \cellcolor{steel!20}\textbf{1} & \cellcolor{steel!20}\textbf{12.8 (0.1)} & 3 & 35.3 (1.9) & 4 & 38.4 (2.7) & 6 & 112.6 (6.5) & 5 & 81.2 (3.8) \\
 & 5 & \cellcolor{steel!20}\textbf{1} & \cellcolor{steel!20}\textbf{17.6 (0.9)} & 3 & 31.3 (0.1) & 2 & 28.6 (2.6) & 4 & 67.4 (3.9) & 5 & 80.5 (4.6) & 6 & 115.4 (5.1) \\
 & 6 & 2 & 15.3 (1.2) & \cellcolor{steel!20}\textbf{1} & \cellcolor{steel!20}\textbf{11.6 (0.0)} & 3 & 18.7 (2.5) & 4 & 37.9 (2.6) & 5 & 75.9 (3.2) & 5 & 77.7 (3.4) \\
\bottomrule
\end{tabular}
\end{adjustbox}
\end{table}